\documentclass[10pt]{extbook}
\usepackage{etex}
\usepackage{index}
\usepackage[french,english]{babel} 
\usepackage[utf8]{inputenc}    
\usepackage{amssymb,amstext,amscd, bbding}
\usepackage{latexsym} 
\usepackage{amsmath}
\numberwithin{equation}{section}
\usepackage{graphicx}
\usepackage{pstricks}
\usepackage{verbatim}
\usepackage{moreverb}
\usepackage{fancyhdr}
\usepackage{tabularx}
\usepackage{rotating}
\usepackage{url}
\usepackage{colortbl}
\usepackage{epsfig}
\usepackage{psfrag}
\usepackage{pstricks}
\usepackage{listings}
\usepackage{caption}
\usepackage{hhline}
\usepackage{booktabs}
\usepackage{multirow}
\usepackage{tocloft}
\usepackage[rflt]{floatflt}
\usepackage{helvet}
\usepackage{wrapfig}
\usepackage[final]{pdfpages}
\usepackage{geometry}
\usepackage{bbold}
\usepackage{bm}
\usepackage{bbm}
\usepackage{lscape}
\usepackage[toc,page]{appendix}
\usepackage[nottoc,notlot,notlof]{tocbibind}
\usepackage{chngcntr}

\usepackage{textcomp}
\usepackage{gensymb}
\usepackage{lscape} % for landscape orientation
\usepackage{subfig}
\usepackage{palatino}

\usepackage{enumitem}
\usepackage{wasysym}
\usepackage{mathtools}	

\DeclareUnicodeCharacter{00A0}{~}

\usepackage{algorithm}
\usepackage{algorithmic}
\usepackage{multirow}
\usepackage{multicol}
\usepackage{listings}
\usepackage[pdfpagelabels]{hyperref}
\hypersetup{colorlinks=true}
\usepackage{minitoc}

\usepackage{pifont}

\usepackage{array}
\definecolor{gray}{rgb}{0.4,0.4,0.4}
\definecolor{darkblue}{rgb}{0.0,0.0,0.6}
\definecolor{cyan}{rgb}{0.0,0.6,0.6}

\lstdefinelanguage{XML}
{
  morestring=[b]",
  morestring=[s]{>}{<},
  morecomment=[s]{<?}{?>},
  stringstyle=\color{black},
  identifierstyle=\color{darkblue},
  keywordstyle=\color{cyan},
  morekeywords={xmlns,version,type}% list your attributes here
}

\newcommand{\be}{\begin{equation}}
\newcommand{\ee}{\end{equation}}
\newcommand{\beq}{\begin{equation}}
\newcommand{\eeq}{\end{equation}}
\newcommand{\bea}{\begin{eqnarray}}
\newcommand{\eea}{\end{eqnarray}}
\newcommand{\non}{\nonumber}
\newcommand{\rr}{\mathrm}
\newcommand{\dd}{\mathrm{d}}
\newcommand{\om}{\omega}
\newcommand{\df}{\partial}
\newcommand{\bPPN}{\beta_\rr{PPN}}
\newcommand{\gPPN}{\gamma_\rr{PPN}}

\newcommand{\mA}{m_{\rr A}}
\newcommand{\RA}{R_{\rr A}}
\newcommand{\RL}{L}
\newcommand{\WT}{R_{\rr w}}
\newcommand{\rhoW}{\rho_{\rr w}}
\newcommand{\rhoV}{\rho_{\rr v}}
\newcommand{\rhoATM}{\rho_{\rr {atm}}}
\newcommand{\rhoA}{\rho_{\rr {A}}}

\newcommand{\Veff}{V_{\rr {eff}}}
\newcommand{\phibg}{\phi_{\rr {bg}}}
\newcommand{\mbg}{m_{\rr {bg}}}
\newcommand{\GeV}{\ \rr {GeV}}
\newcommand{\mpl}{m_\rr{pl}}
\newcommand{\GN}{G_\rr{N}}
\newcommand{\phib}{\phi_\rr{b}}
\newcommand{\rs}{r_\rr{s}}
\newcommand{\Mp}{M_\rr{pl}}

\newcommand{\bg}{\bar\gamma}   

\newcommand{\vv}{\ding{51}}
\newcommand{\xxx}{\ding{55}} 	%\XSolidBrush

\title{Hunting modifications of gravity: from the lab to inflation via compact objects\\~scrazenfreuzenfeu\\}
\author{Sandrine Schl\"ogel\\
University of Namur \\
Université Catholique de Louvain}
\renewcommand{\chaptermark}[1]{\markboth{\small\textsc{#1}}{}}

\counterwithin{figure}{chapter}
\counterwithin{table}{chapter}

\begin{document}
\sloppy	  

%\mathtoolsset{showonlyrefs}
    
\frontmatter
%\maketitle
\pagenumbering{Alph}
%\setcounter{page}{}
%*********************** TITLEPAGE ***************************
\thispagestyle{empty}

\begin{minipage}[t]{0.3\textwidth}
  \raisebox{\dimexpr 0.6\baselineskip-\height}{\includegraphics*[width=2cm]{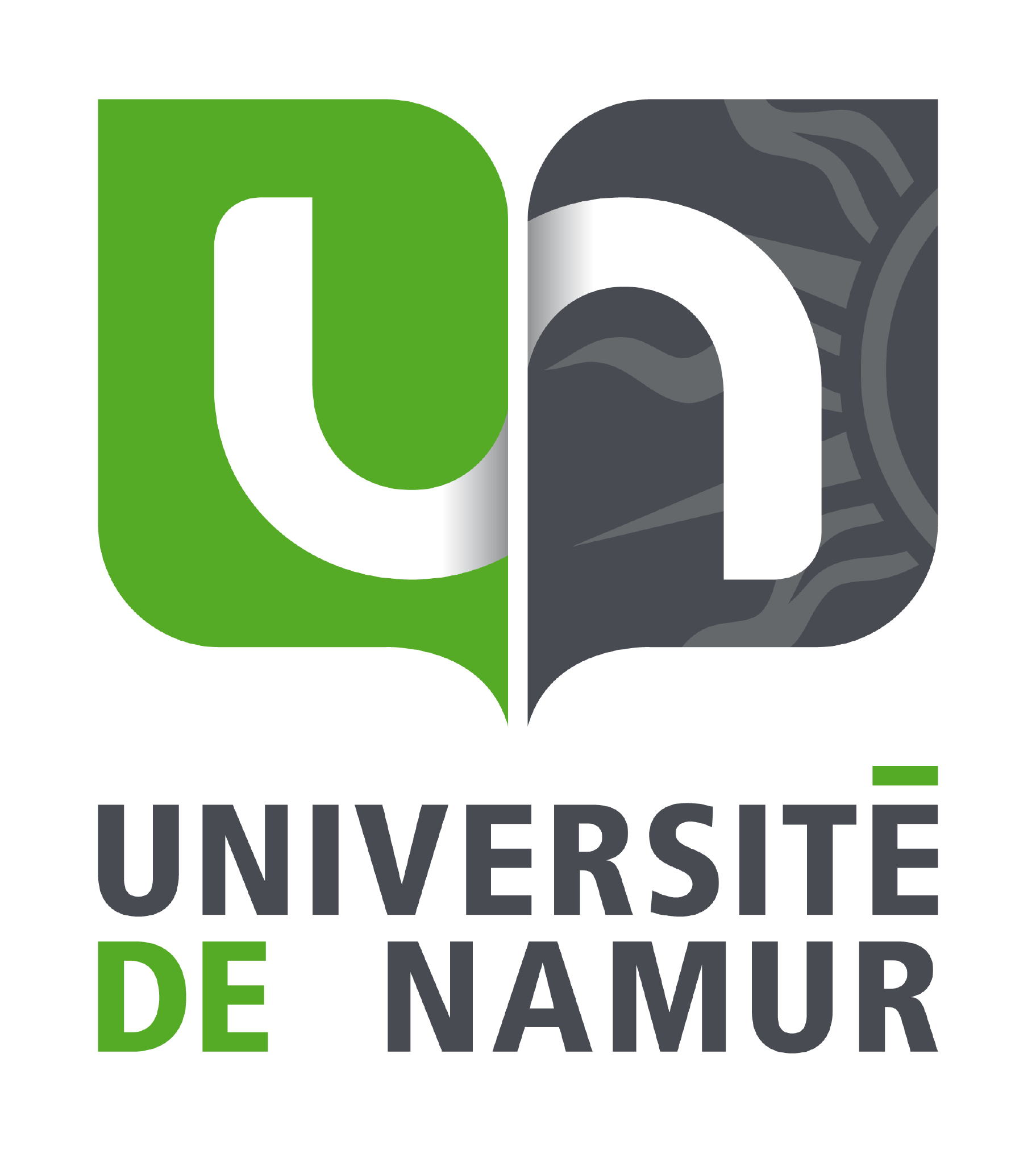}}
\end{minipage}
\hfill\begin{minipage}[t]{0.65\textwidth}
  \flushright \textbf{UNIVERSIT\'E DE NAMUR} \\
  \ \hrulefill \\
  \vspace{0.1cm}
  FACULT\'E DES SCIENCES\\
  \vspace{0.1cm}
  D\'EPARTEMENT DE MATH\'EMATIQUE
\end{minipage}
% 
% \vspace{0.6cm}
% 
% % \begin{minipage}[t]{0.65\textwidth}
% %   \vspace{0.05cm}
% %   \flushleft \textbf{UNIVERSIT\'E CATHOLIQUE DE LOUVAIN}\\
% %   \ \hrulefill \\
% %   \vspace{0.1cm}
% %   SECTEUR DES SCIENCES ET TECHNOLOGIES\\
% %   \vspace{0.1cm}
% %   INSTITUT DE RECHERCHE EN MATH\'EMATIQUE ET PHYSIQUE
% % \end{minipage}
% % \hfill
% % \begin{minipage}[t]{0.3\textwidth}
% %   \flushright
% %   \raisebox{\dimexpr 0.6\baselineskip-\height}{\includegraphics*[width=1.7cm]{UCL.jpg}}
% % \end{minipage}
% 

\vspace{3cm}
\begin{center}  
\Huge{\bf{Hunting modifications of gravity:\\~\\from the lab to cosmology \\ via compact objects}}
\end{center}
%\vspace{0.8cm}
\vspace{2cm}
\begin{flushright}
Th\`ese pr\'esent\'ee par\\
{\bf{Sandrine Schlögel}}\\
 pour l'obtention du grade\\
 de Docteur en Sciences
\end{flushright}
%\vspace{0.8cm}
\vspace{1cm}

Composition du Jury: \\

\noindent 
Ruth {\sc Durrer}\\
André {\sc Füzfa} (Promoteur)\\
Anne {\sc Lemaître} (Pr\'esident du Jury)\\
David {\sc F. Mota}\\
Christophe {\sc Ringeval} (Promoteur)\\
\vspace{3cm}

\begin{minipage}[t]{0.25\textwidth}
  \flushleft
 \raisebox{\dimexpr 0.6\baselineskip-\height}{\includegraphics*[width=2cm]{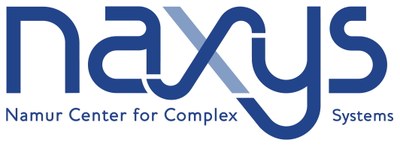}}
\end{minipage}
\begin{minipage}[t]{0.4\textwidth}
 \begin{center}
    Octobre 2016
  \end{center}
\end{minipage}\hfill
\begin{minipage}[t]{0.25\textwidth}
  \vspace{-0.7cm}
  \flushright
 \raisebox{\dimexpr 0.6\baselineskip-\height}{\includegraphics*[width=2cm]{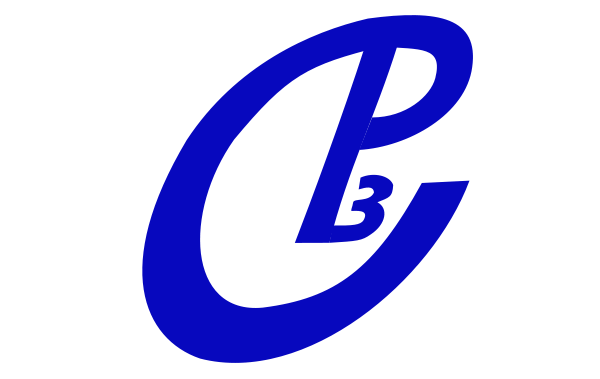}}
\end{minipage}

%%%%%%%%%%%%%%%%%%%%%%%%%%%%%%%%%%%%%%%%%%%%%%%%%%%%%%%%%%%%%%%%%%%%%%%
\newpage
%\vspace{2cm}
\thispagestyle{empty}
$\\$
\vspace{15cm}
\begin{center}
  \copyright Presses universitaires de Namur \& Sandrine Schlögel\\
  Rempart de la Vierge, 13\\
  B-5000 Namur (Belgique)\\
\end{center}

\begin{center}
  Toute reproduction d'un extrait quelconque de ce livre,\\
  hors des limites restrictives pr\'evues par la loi,\\
  par quelque proc\'ed\'e que ce soit, et notamment par photocopie ou
  scanner,\\
  est strictement interdite pour tous pays.\\
\end{center}

\begin{center}
  Imprim\'e en Belgique\\
\end{center}

\begin{center}
  %ISBN :  978-2-87037-842-7\\
  ISBN :  978-2-87037-939-4\\
  %D\'ep\^ot l\'egal: D / 2014 / 1881/ 32\\
  D\'ep\^ot l\'egal: D / 2016/ 1881/27\
\end{center}

\begin{center}
  Universit\'e de Namur\\
  Facult\'e des Sciences\\
  rue de Bruxelles, 61, B-5000 Namur (Belgique)
\end{center}
\newpage
%\

\thispagestyle{empty}
%\vspace*{0.2cm}
%\setborder[\hsize,0.2cm,\smallskipamount]\border{
%\begin{center}
%        Facult\'{e}s Universitaires Notre-Dame de la Paix\\
%        Facult\'{e} des Sciences\\
%        Rue de Bruxelles, 61, B-5000 Namur, Belgium \\
%      \end{center}

\begin{center}
  {\bf{Hunting modifications of gravity: \\from the lab to cosmology via compact objects}}\\
  by Sandrine Schlögel
\end{center}
{\bf{Abstract:}} 
Modifications of gravity have been considered to model the primordial inflation and the late-time cosmic acceleration. Provided that modified gravity models do not suffer from theoretical instabilities, they must be confronted with observations, not only at the cosmological scales, but also with the local tests of gravity, in the lab and in the Solar System, as well as at the astrophysical scales. Considering in particular sub-classes of the Horndeski gravity, we study their observational predictions at different scales. 
In order to pass the local tests of gravity while allowing for long-range interactions in cosmology, Horndeski gravity exhibits screening mechanisms, among them the chameleon. The chameleon screening mechanism has been tested recently using atom interferometry in a vacuum chamber. Numerical simulations are provided in this thesis in order to refine the analytical predictions. 
At the astrophysical scale, Horndeski gravity predicts a variation of the gravitational coupling inside compact stars. Focusing on Higgs inflation, we discuss to what extent the Higgs vacuum expectation value varies inside stars and conclude whether the effect is detectable in gravitational and nuclear physics.
Finally, the covariant Galileon model exhibits non-linearities in the scalar field kinetic term such that it might pass the local tests of gravity thanks to the Vainshtein screening mechanism. We discuss if a sub-class of the covariant Galileon theory dubbed the Fab Four model leads to a viable inflationary phase and provide combined analysis with neutron stars and Solar System observables.

\vspace{0.5cm}
\raggedbottom

\begin{center}
  {\bf{\`{A} la poursuite de modifications de la gravitation:} \\
  \bf{du laboratoire à la cosmologie en passant par les objets compacts}}\\
  par Sandrine Schlögel
\end{center}
{\bf{R\'esum\'e~:}} 
Dans le cadre de la cosmologie moderne, des modifications de la théorie d'Einstein ont été étudiées pour expliquer l'inflation primordiale et l'accélération actuelle de l'expansion de l'Univers. Pourvu que ces modèles de gravitation modifiée soient bien posés théoriquement, ils sont confrontés aux observations, non seulement en cosmologie, mais aussi en laboratoire et dans le Système Solaire ainsi qu'en astrophysique. Nous étudions dans cette thèse les prédictions de quelques sous-classes de la théorie d'Horndeski à différentes échelles. 
Ces théories font en général appel à des mécanismes d'écrantage afin de modéliser des interactions à longue portée tout en étant conformes aux contraintes observationnelles aux échelles locales. Parmi ces mécanismes d'écrantage, le caméléon a été récemment testé en laboratoire grâce à l'interférométrie atomique. Des simulations numériques de cette expérience sont développées dans cette thèse afin de raffiner les prédictions dérivées analytiquement. 
Aux échelles astrophysiques, la théorie d'Horndeski prévoit une variation du couplage gravitationnel à l'intérieur des objets compacts. Nous nous attardons dans cette thèse sur la 'Higgs inflation' en particulier. Nous discutons les variations de la valeur moyenne dans le vide du champ de Higgs prédites par ce modèle afin d'établir si celles-ci sont détectables dans des objets astrophysiques. 
Finalement, nous étudions une sous-classe du modèle du 'Galileon covariant', le modèle des 'Fab Four', qui fait appel au mécanisme d'écrantage de Vainshtein pour passer les contraintes locales. Nous analysons si le modèle des 'Fab Four' peut donner lieu à une phase d'inflation conforme aux observations. Nous étudions également les prédictions de ce dernier modèle aux échelles astrophysiques, en particulier dans les étoiles à neutron et le Système Solaire, et mettont en évidence les déviations du couplage gravitationnel prédites par les 'Fab Four'. 
%\vspace{0.5cm}
%\raggedbottom
~\\~\\~\\
\noindent Ph.D. thesis in Physics\\
Date: October 7th, 2016\\
Department of Mathematics \\
Advisors: André {\sc{Füzfa}} (UNamur), Christophe {\sc{Ringeval}} (UCLouvain) 
%}

\thispagestyle{empty}
\ \\

\newpage
%\strut %pour mettre une page blanche
%\newpage

\pagenumbering{roman}
\setcounter{page}{1}
\setcounter{minitocdepth}{2}

\thispagestyle{empty}%Pas de numerotation
%Remerciements

\chapter*{Thank you...} % Main chapter title
\markboth{\textsc{Thank you...}}{\textsc{Thank you...}}

\lhead{Chapter 1. \emph{Awesome chapter 1}} % This is for the header on each page - perhaps a shortened title

%----------------------------------------------------------------------------------------
\begin{flushright}
 \textit{"Des chercheurs qui cherchent,\\
 On en trouve.\\
 Des chercheurs qui trouvent,\\
 On en cherche.\\  
 Inventrices et vous inventeurs, \\
 De tous les pays unissez-vous.\\
 L'homme n'a pas été au bout\\
 De tous ses possibles."}\\
 \vspace{0.5cm}
 Julos Beaucarne\\
 \vspace{1.2cm}
 \textit{"Chercher n'est pas une chose\\
 et trouver une autre,\\
 mais le gain de la recherche,\\
 c'est la recherche elle-même."}\\
 \vspace{0.5cm}
 St-Grégoire de Nysse
 \\ \vspace{0.5cm}
\end{flushright}

\noindent
I like to say that my PhD thesis was most of the time a solitary work, but that I did it not alone.
I met a lot of people during those four years. Some of them taught me the job of researcher, others allowed me to take distance from my research topics (breaks are important too) and some of them have finally become friends. 

I would like to thank first Ruth Durrer and David F. Mota for having accepted to be the rapporteurs of this thesis. Their reflections, suggestions and the ensuing discussion allowed me to bring a fresh perspective to my thesis. Merci aussi à Ruth de m'avoir accueillie dans le groupe de recherche à Genève pendant quelques mois. Même si notre projet de recherche n'aura donné lieu à aucun résultat publiable, certains fruits de ce travail se trouvent entre les lignes de cette thèse. 

Merci à Anne Lemaître d'avoir présidé ce jury... et d'avoir repéré des fautes d'orthographe et d'anglais qui m'avaient échappées ! 

Cette thèse n'aurait pas pu voir le jour sans mes deux directeurs de thèse. Merci à André Füzfa de m'avoir guidée dans la recherche en relativité générale et en cosmologie. Je le remercie pour la confiance qu'il m'a accordée tout au long de ces quatres années, non seulement dans mon travail de recherche, mais aussi lors d'expériences de vulgarisation scientifique. Enfin, je lui suis reconnaissante de m'avoir permis de m'essayer à l'astronomie. Je pense que nous nous souviendrons pendant longtemps de ces nuits passées en compagine de Jean-Pol, Eve, Chapi Chapo et bien d'autres, les pieds glacés guettant l'amas de la ruche (M44 pour les intimes) entre deux nuages en partageant un \textit{Pick Up!}. Merci à Christophe Ringeval pour la relecture de ce manuscrit alors que ce dernier était encore à l'état d'ébauche. Ses remarques m'auront permis d'acquérir une plus grande rigueur et d'approfondir mes idées.

Je tiens aussi à remercier l'ensemble des chercheurs avec qui j'ai eu la chance de collaborer, de près ou de loin, durant ma thèse. Merci à Max Rinaldi de m'avoir accompagnée dans l'apprentissage du métier de chercheur. Merci aussi de m'avoir incitée à découvrir d'autres univers de recherche, d'abord en me poussant à aller à Genève, ensuite en m'invitant à Trento.

Merci à Sébastien Clesse d'avoir donné une nouvelle impulsion au "groupe cosmo" de Namur. Je me souviendrai non seulement de l'élevage de caméléons lancé ensemble, mais aussi des différentes activités du Printemps des Sciences, comme les contes d'Alice et d'Umitou, ou encore le spectacle du Pr Big et du Dr Bang. 

Thanks to Karel Van Acoleyen for the guidance of my first steps in modified gravity when I was still a master student. The tools I developed then have been proven very useful all along my thesis. Thanks also for having accepted to be part of my "comité d'accompagnement" during the last four years.

Merci à Jean-Philippe Bruneton pour nos discussions scientifiques annuelles autour d'une bonne bière, à Paris ou à Namur... c'est également ainsi qu'avance la science. Merci encore à Aurélien Hees et Olivier Minazzoli pour cette dernière année de collaboration. Si le Higgs ne nous a pas encore révélé toutes ses subtilités, c'est sans doute que nous n'en avons pas encore assez discuté autour d'un verre. Thanks to Holger Müller to have contacted me during the Marcel Grossmann meeting. It was really great to have the oppotunity to discuss with experimentalists about chameleons.

Nevertheless, my PhD thesis also strongly benefits from what I did with the university in addition to scientific research. I have already mentionned scientific outreach but that's not all! 

Le voyage en Inde organisé par l'ONG de l'Université de Namur m'aura permis de me donner un second souffle au milieu de ces quatre années, de voir "autre chose" (par exemple, découvrir le jack fruit). Merci à tout le groupe Dinde ainsi qu'à ceux qui nous ont aidés à préparer ce voyage. In particular, I thank John for the corrections of some selected parts of this thesis. 

Thanks to all the researchers I met during my research stays at the University of Geneva. Thanks for having adopted me as a member of the research group for a few months. I not only appreciated the discussions about science, but also about our respective cultures, for instance smashing the traditional Genevan marmite together or celebrating the German (and Belgian!) St. Nikolaus. 

Thanks to all my office mates for one week, one month, one year (Aurélie, Marco and Giovanni, Alexis, Ioannis, Marie-Hélène and Adrien). I strongly enjoyed to share time with you and I took advantage of the opportunity to learn some words in Italian (I would not mention which ones here)! 

As I said, at the end of those four years, some of my colleagues have become friends. Merci à Evallou et Marou pour les Megagigantorigolades et autres Poussinmontagnes\footnote{Pour des informations complémentaires, voir par exemple \cite{ponti}.}. Merci aussi pour ce merveilleux pélerinage cosmologique à Rome à l'occasion du Marcel Gross'blague. Merci à Alexis pour le voyage sur la Lune. Nos débats passionnés sur la politique et la philosophie des sciences qui nous ont occupés de longues heures, me manqueront très certainement ! Thanks also to Yérali, Delphinou, Mara, Vivi,... for the great breaks spent in their company.

Merci à mes très-proches, mes amis de longue date et ma famille pour leur soutien ; pour leur compréhension, car ils se sont toujours intéressés à ce que je faisais ; et pour leur incompréhension, car le monde de la recherche est un monde à part qui a ses propres règles, parfois incongrues. Merci à Manu, pour sa présence de tous les jours. Cette thèse lui doit beaucoup, à commencer par la qualité des figures et de la mise en page de ce manuscrit que j'aurais eu bien du mal à faire toute seule.

Thanks to you who reads this thesis...

\cleardoublepage	

%\dominitoc
% \renewcommand\contentsname{\fontfamily{palatino}\selectfont
% {Contents}
% }
%\renewcommand{}{\markboth{\small\textsc{Contents}}}
%\renewcommand{\contentsname}{Contents}
\thispagestyle{empty}

\renewcommand{\cftmarktoc}{\markboth{\small\textsc{Contents}}{\small\textsc{Contents}}}

\tableofcontents

\newpage
\mainmatter

\renewcommand{\chaptermark}[1]{\markboth{\small\textsc{Chapter \thechapter.\ #1}}{}}
\cleardoublepage

\chapter*{Introduction} % Main chapter title
\markboth{\textsc{Introduction}}{\textsc{Introduction}}

\label{Intro} % For referencing the chapter elsewhere, use \ref{Chapter1} 

\lhead{Chapter 1. \emph{Awesome chapter 1}} % This is for the header on each page - perhaps a shortened title

%----------------------------------------------------------------------------------------
Over the last century, general relativity has emerged as the theory of gravitation. It not only allowed the explanation of phenomena that are not in the Newton’s theory of gravitation, but also has opened up new perspectives in predictions, especially in astrophysics and cosmology. Indeed, when the gravitational field is very strong, i.e. in compact objects like black holes and neutron stars, relativistic effects are expected. The existence of compact objects is favored by the observations today, using either electromagnetic or gravitational wave signals.

During the last decades, the observations of the sky at large scales have made the study of precision cosmology possible. According to our current understanding, the Universe has a history and is expanding over time. Only $5\%$ of its matter-energy content is described by the standard model of particle physics. Around $68\%$ of the Universe is responsible for the late-time cosmic acceleration which can be modelled thanks to the cosmological constant $\Lambda$. The remaining $27\%$ of the matter-energy content of the Universe is composed of cold (i.e. non-relativistic) dark matter (CDM) whose nature is still unknown. The $\Lambda-$CDM concordance picture reproduces the current observations.

Despite of the successes of general relativity and the standard model of particle physics, there are at least two reasons to look for theories beyond this paradigm:
\begin{enumerate}
 \item At high energy and small distances, quantum effects are expected in general relativity, like other fundamental interactions. Since general relativity is not renormalizable, its quantization is therefore problematic, it should be considered rather as an effective classical theory of gravity emerging from its quantum counterpart. A viable quantum theory of gravitation should first be well-defined and second, conform to current observations. Compact objects and cosmology provide the most promising observational tests of quantum gravity. 
 \item In cosmology the nature of dark matter and the late-time cosmic acceleration is still debated. Today research for dark matter has shifted to particle physics. The detection of a new particle could reveal an extension of the standard model of particle physics although the distribution of dark matter in the sky should also be explained. The cosmological constant raises some theoretical issues, in particular why is its value fine-tuned to such a tiny value while still non-zero? Finally, within the $\Lambda-$CDM paradigm, the initial conditions in the early Universe appear to be fine-tuned. 
 An exponentially accelerated phase of expansion in the early Universe referred to as primordial inflation, around $10^{-43}~\rr{s}$ after the Big Bang, is able to give an explanation of the current observations. The late-time cosmic acceleration and inflation can be modelled either by a) invoking new particles and fields beyond the standard model of particle physics, the scalar fields being privileged for the sake of simplicity, or by b) calling on modifications of gravity. 
\end{enumerate}

Since the conception of general relativity, modifications and extensions of the Einstein’s theory have been extensively studied, devoted to quantum gravity, the explanation of observations in cosmology or the unification of the fundamental interactions. However, modified gravity is challenging from the theoretical point of view since general relativity has a privileged status as highlighted in Chap.~\ref{chap:math_fundations}: this is the only well-defined theory (for instance it exhibits second-order equations of motion) which guarantees that the physical laws are valid in all coordinate systems in a four dimensional spacetime. In particular, this result is based on the fact that the metric is the only field describing gravity.

Alternative theories must thus first be well-defined and then compared to present observations in order to establish whether they are viable. In Chap.~\ref{chap:test_GR}, the current observational tests of gravitation are briefly reviewed. We propose to classify them depending on the tested regime given by the strength of the gravitational field and the presence of energy-momentum sources. Today, general relativity has been tested with very good accuracy in the vacuum, either in the weak field regime, i.e. in the Solar System and in the lab, or in the strong field regime, thanks to indirect and direct detection of gravitational waves coming from binary systems.

The presence of sources renders the tests of general relativity trickier since there are uncertainties about the modeling of the sources. This is the case for neutron stars and in cosmology where the cosmological fluid can be considered as a source, possibly of unknown nature like dark energy, dark matter or scalar fields responsible for primordial inflation. Meanwhile the current cosmic expansion and primordial inflation could also reveal modifications of gravity at large scales. In Chap.~\ref{chap:MG}, alternative models to general relativity are introduced and classified depending on which underlying assumptions of general relativity they violate. In particular, we focus on alternatives to general relativity invoking a scalar field counterpart to the metric in order to describe gravity, i.e. scalar-tensor theories. Initially, those theories were devoted to  testing the constancy of gravitational coupling in spacetime. Today, scalar-tensor theories have been found to give rise to models for inflation and late-time cosmic expansion in agreement with cosmological observations. Meanwhile, they must be confronted with other observations, like experiments on the Earth as well as observations in the Solar System and around compact objects, in order to conclude if they are viable alternatives to general relativity.

%In 1961, Brans and Dicke proposed to explain the smallness of the gravitational constant by promoting it as a scalar field which evolves in spacetime. The Brans-Dicke theory boosted the tests of general relativity, notably in the Solar System. Since then, scalar-tensor theories have been found to give rise to models for inflation and current cosmic expansion. However such theories must be confronted to Solar System observations as well as laboratory experiments in order to conclude if they are viable.

The first scalar-tensor theory we focus on, is the chameleon model. This model was first proposed by Khoury and Weltman in 2004 \cite{KhouryWeltmanPRL, KhouryWeltmanPRD} in order to reproduce the current cosmic acceleration and to pass the weak-field constraints. This model is based on a screening mechanism: the chameleon effective mass is small in low density environment, the scalar field mediating long range interaction like the late-time cosmic acceleration, while the chameleon becomes massive in high density environment, the effects are so short ranged that they are difficult to measure.

In Chap.~\ref{chap:chameleon}, we briefly comment the current status of the chameleon model. It has been tested in cosmology, in the Solar System, and recently in astrophysics; a part of the parameter space of the chameleon model remains unconstrained. The most promising test today comes from lab experiments using atom interferometry, as first proposed by Burrage, Copeland and Hinds in 2014 \cite{Burrage}. Indeed, if the atom interferometer is placed inside a vacuum chamber in the presence of a test mass, the acceleration induced by the chameleon field can be determined with very good accuracy by measuring the interference fringes. A first experiment has been developed in Berkeley. It shows that almost all the chameleon parameter space should be reachable with this experimental set-up in the near future. However, the forecasts for the acceleration due to the chameleon were computed only analytically, neglecting the effects of the vacuum chamber geometry. In Chap.~\ref{chap:chameleon}, we provide numerical simulations for the on-going experiment in Berkeley in order to refine the analytical constraints and to take into account the effects of the chamber geometry.

In general, scalars mediating gravity are not assumed to be detected in nature as yet and are only considered as hypothetical degrees of freedom. In 2012, the first elementary scalar field, the Higgs field, was detected in the Large Hadron Collider \cite{Aad:2012tfa, Chatrchyan:2012xdj}. The question then arises, could the Higgs field be a partner of the Einstein metric for describing gravity? In 2008, Bezrukov and Shaposhnikov propose a model where the Higgs field is the inflaton, provided it is nonminimally coupled to gravity, within the framework of modified gravity \cite{Bezrukov:2008ej}. This model reviewed in Chap.~\ref{chap:Higgs} is favored by the latest cosmological observations.

Because of the nonminimal coupling of the Higgs field to gravity, the distribution of the Higgs field around compact objects in the presence of matter is expected to be non trivial, contrary to general relativity (i.e. with a minimally coupled Higgs field) where the Higgs field has settled to its vacuum expectation value (vev) everywhere. If the distribution is not trivial, it implies that the vev varies in spacetime and the Higgs field is not necessarily settled to it, for instance inside compact objects. Because of these variations, the question arises if this model is able to pass current constraints in the Solar System and if the nuclear physics inside neutron stars is affected. These questions are discussed in Chap.~\ref{chap:Higgs}.

In 1974, Horndeski derived the most general Lagrangian for the extension of general relativity invoking an additional scalar field \cite{horndeski1974}. Scalar-tensor theory is one particular case in the general class highlighted by Horndeski. In Chap.~\ref{chap:FabFour}, we focus on another theory dubbed the Fab Four model in reference to the four general Lagrangian terms arising in Horndeski gravity, which were rediscovered in 2012 by Copeland, Padilla and Saffin \cite{Copeland:2012qf}. This model appears to be well posed from the theoretical point of view. In Chap.~\ref{chap:FabFour}, the predictions of two of the Fab Four are discussed for primordial inflation as well as in the Solar System and around compact objects.

Finally, we draw some conclusions and perspectives in Part~\ref{CCL}.

%We will explore if it is possible to propose well-defined modifications of gravity which are able to model either the current cosmic acceleration or primordial inflation unless additional matter-energy sources. In particular those theories must pass the Solar System tests. 

\phantomsection
\addcontentsline{toc}{chapter}{Introduction}

\renewcommand{\chaptermark}[1]{\markboth{\small\textsc{Chapter \thechapter.\ #1}}{}}
\cleardoublepage

\chapter*{Conventions} % Main chapter title
\markboth{\textsc{Conventions}}{\textsc{Conventions}}

\label{Convention} % For referencing the chapter elsewhere, use \ref{Chapter1} 

\lhead{Conventions. \emph{Conventions}} % This is for the header on each page - perhaps a shortened title

%----------------------------------------------------------------------------------------
% The units system by default is the natural one, i.e. $\hbar=c=1$ and so, all physical density can be expressed in
% GeV with the right power according what they represent. In this unit system, the Newton's gravitational constant is defined by
% \bea
% G_\rr{N}\equiv\frac{1}{m_\rr{p}^2},
% \eea
% while the factor $\kappa$ which enters in the action for GR is
% \bea
% \kappa=\frac{8\pi}{m_\rr{p}^2}=\frac{1}{M_\rr{p}^2}.
% \eea
\textbf{\textit{Unit system,}} We use natural units in which $c=\hbar=1$ such that  all quantities are expressed in powers of GeV.
The Newton's constant is given by $\GN=1/\mpl^2$, $\mpl$ being the Planck mass. 

\textbf{\textit{Notations,}} The most used notations are reported in Tab.~\ref{tab:notations}. Furthermore, the Greek indices refer to spacetime coordinates, $x^{\mu}$ with $\mu=0,1,2,3$, while the Latin ones to space coordinates, $x^\rr{i}$ with $i=1,2,3$. The covariant and partial derivatives for any function or tensor $\mathbf{T}$ are respectively denoted by,
\bea
  \nabla_\mu \mathbf{T}=\mathbf{T}_{;\mu},\\
  \df_\mu \mathbf{T}=\mathbf{T}_{,\mu}.
\eea
% We remind that the covariant derivative of a tensor $\mathbf{T}$ is given by, 
% \be
% \nabla_\lambda T^{\nu_\rr{1}...\nu_\rr{q}}_{\mu_\rr{1}...\mu_\rr{p}}=
% \partial_\lambda T^{\nu_\rr{1}...\nu_\rr{q}}_{\mu_\rr{1}...\mu_\rr{p}}
% + \Gamma^{\nu_\rr{1}}_{\lambda\rho} T^{\rho...\nu_\rr{q}}_{\mu_\rr{1}...\mu_\rr{p}}...
% +\Gamma^{\nu_\rr{q}}_{\lambda\rho} T^{\nu_\rr{1}...\rho}_{\mu_\rr{1}...\mu_\rr{p}}
% -\Gamma^{\rho}_{\lambda\mu_\rr{1}} T^{\nu_\rr{1}...\nu_\rr{q}}_{\rho...\mu_\rr{p}}
% -\Gamma^{\rho}_{\lambda\mu_\rr{p}} T^{\nu_\rr{1}...\nu_\rr{q}}_{\mu_\rr{1}...\rho}.
%\ee
Time and radial  derivatives are respectively denoted by the dot $\dd f/\dd t\equiv \dot{f}$ and the prime $\dd f/\dd r\equiv f'$. 
Vector fields are in bold characters.

\textbf{\textit{Differential geometry,}} We follow the convention of Misner \textit{et al.}'s reference book \cite{Misner:1974qy}. In particular, we adopt the Einstein's implicit summation for repeated indices and the mostly plus signature for the metric $(-,~+,~+,~+)$. Following this convention, the covariant derivative of a tensor $\mathbf{T}$ is given by, 
\be
\nabla_\lambda T^{\nu_\rr{1}...\nu_\rr{q}}_{\mu_\rr{1}...\mu_\rr{p}}=
\partial_\lambda T^{\nu_\rr{1}...\nu_\rr{q}}_{\mu_\rr{1}...\mu_\rr{p}}
+ \Gamma^{\nu_\rr{1}}_{\lambda\rho} T^{\rho...\nu_\rr{q}}_{\mu_\rr{1}...\mu_\rr{p}}...
+\Gamma^{\nu_\rr{q}}_{\lambda\rho} T^{\nu_\rr{1}...\rho}_{\mu_\rr{1}...\mu_\rr{p}}
-\Gamma^{\rho}_{\lambda\mu_\rr{1}} T^{\nu_\rr{1}...\nu_\rr{q}}_{\rho...\mu_\rr{p}}
-\Gamma^{\rho}_{\lambda\mu_\rr{p}} T^{\nu_\rr{1}...\nu_\rr{q}}_{\mu_\rr{1}...\rho}.
\ee
The components of the Levi-Civita connection are,
\bea \label{eq:levi-civita_compo}
%\Gamma^\rho_{\mu\nu}
\begin{Bmatrix}
 \rho
 \\ \mu\nu
\end{Bmatrix}=\frac{1}{2}g^{\lambda\rho}\left(\partial_\mu g_{\lambda\nu}+\partial_\nu g_{\lambda\mu}
-\partial_\lambda g_{\mu\nu}\right),
\eea
the Riemann tensor,
\bea
R^\alpha_{\mu\beta\nu}=\left(\df_\beta\Gamma^\alpha_{\mu\nu}
+\Gamma^\alpha_{\sigma\beta}\Gamma^\sigma_{\mu\nu}\right)-(\beta\leftrightarrow\nu),
\eea
the Ricci tensor,
\bea
R_{\mu\nu}=R^\alpha_{\mu\alpha\nu},
\eea
and the Ricci scalar $R=R^\mu_\mu$. The Einstein tensor is $G_{\mu\nu}=R_{\mu\nu}-\frac{1}{2} g_{\mu\nu} R$ and the Einstein equations read,
\be
  G_{\mu\nu}=8\pi\GN T_{\mu\nu}.
\ee

\textbf{\textit{Symbols and acronyms,}} The list of symbols and acronyms are reported in Tabs.~\ref{tab:notations} and~\ref{tab:acronyms}. 

\begin{table*} 
  \begin{center}
    \begin{tabular}{ |l|l| } 
  \hline
  %\multicolumn{2}{|c|}{Team sheet} \\
  %\hline
  Symbols & Signification \\
  \hline
  $a(t)$ & scale factor \\
  $\beta_\rr{PPN}$, $\gamma_\rr{PPN}$ & post-Newtonian parameters $\beta$, $\gamma$  \\
  $c$, $G_\rr{N}$ & speed of light, Newton's constant \\
  %$\gamma_\rr{PPN}$ & $\gamma$ post-newtonian parameter \\
  %$e_\mu$ & basis vectors \\
  $E_\rr{b}$ & binding energy \\
  $\epsilon, \eta$ & slow-roll parameters \\
  $\eta_{\mu\nu}$ & Minkowski metric  \\
  $g_{\mu\nu}$, $g$ & metric tensor, determinant of $g_{\mu\nu}$ \\
  $h_{\mu\nu}$ & perturbations of the metric \\
  %$\eta_\rr{UFF}$ & experimental parameter for the Universality of Free Fall \\
  $\Gamma^\rho_{\mu\nu}$ & connection coefficients \\
  $G_{\mu\nu}$ & Einstein tensor \\
  $G$ & bare gravitational constant \\
  $\mathcal{G}$ & Gauss-Bonnet combination \\
  $H$, $H_0$ & Hubble parameter, Hubble constant \\
  %$H_0$ & Hubble constant \\
  $\kappa\equiv{8\pi}/{\mpl^2}$ & parameter (action) \\
  $\Lambda$ & cosmological constant \\
  $L$, $\mathcal{L}$ & Lagrangian, Lagrangian density \\
  $\lambda_\rr{sm}$, $v$ & SM Higgs self-interaction coupling and vev \\
  $\nu(r)$, $\lambda(r)$ & metric fields (Schwarzschild gauge) \\
  %$\mathcal{L}$ & Lagrangian density \\
  %$\mathcal{L}_\xi$ & Lie derivative \\
  $\mathcal{M}$ & manifold \\
  %$\nu(r)$ & metric field (Schwarzschild gauge) \\
  $\mpl$, $M_\rr{pl}=\mpl/\sqrt{8\pi}$ & Planck mass, reduced Planck mass \\
  %$M_\rr{pl}=\mpl/\sqrt{8\pi}$ & normalized Planck mass \\
  $m_\rr{i}$, $m_\rr{g}$ & inertial mass, gravitational mass \\
  $M_\odot$, $R_\odot$ & Solar mass, Solar radius \\
  %$\nu$ & frequency \\
  $N(t)$ & number of e-folds \\
  $n_\rr{s}, r$ & spectral index, tensor-to-scalar ratio (inflation) \\
  $\Omega(x)$ & conformal factor \\
  $\dd\Omega^2\equiv \dd\theta^2+\sin^2\theta \dd\varphi^2 $ & infinitesimal solid angle \\ %(spherical coordinates) \\
  $\Phi$, $U\equiv-\Phi/\GN$ & Newtonian gravitational potential \\
  $\phi, \pi$ & scalar fields \\
  $\psi_\rr{M}$ & matter fields \\
  $q (t)$,~$\Omega (t)$ & deceleration parameter, density parameters \\ %(cosmology) 
  $\rho$, $p$ & energy density, pressure \\
  %$q$ & deceleration parameter (cosmology) \\
  %$\rho$ & energy density \\
  $\mathcal{R}$, $r_\rr{s}$ & radius of compact objects, Schwarzschild radius \\
  %$R_\odot$ & Solar radius \\
  %$r_\rr{s}$ & Schwarzschild radius \\
  %$R$ & Ricci scalar \\
  $R^\mu_{\nu\lambda\kappa}$, $R_{\mu\nu}$, $R$ & Riemann tensor, Ricci tensor, Ricci scalar \\
  %$R_{\mu\nu\lambda\kappa}$ & Riemann tensor \\
  $s$ & compactness \\
  $t$, $0$ subscript & cosmic time, (cosmology) for today \\
  $T_{\mu\nu}$ & stress-energy-momentum tensor \\
  $\mathbf{u}$ & four-velocity \\
  %$v$ & velocity \\
  $V(\phi)$ & potential \\
  $w$ & equation of state \\
  $\xi$ & nonminimal coupling constant in Higgs inflation \\
  %$\mathbf{\xi}$ & Killing vector \\
  $z$ & redshift \\
  %$\square\equiv\nabla_\mu\nabla^\mu$ & d'Alembertian \\
  %$\equiv$ & definition \\
  %$x^{\mu}$ & spacetime coordinates ($\mu=1,~2,~3,~4$) \\
  %$x^\rr{i}$ & spatial coordinates ($i=1,~2,~3$) \\
  %greek indices & spacetime indices \\
  %latin indices & space indices \\
  %$\nabla$ & covariant derivative \\
  %$\df$ & partial derivative \\
  %$0$ subscript & (cosmology) for today \\
  \hline
\end{tabular}
\caption{List of symbols appearing in this thesis and their signification.}
\label{tab:notations}
\end{center}
\end{table*}

\begin{table*}
  \begin{center}
    \begin{tabular}{ |l|l| } 
  \hline
  %\multicolumn{2}{|c|}{Team sheet} \\
  %\hline
  Acronyms & Signification \\
  \hline
  BAO & Baryon Acoustic Oscillations \\
  BBN & Big Bang Nucleosynthesis \\
  BH & Black Hole \\
  BVP & Boundary Value Problem \\
  CDM & Cold Dark Matter \\
  C.L. & Confidence Level \\
  CMB & Cosmic Microwave Background \\
  DE & Dark Energy \\
  DM & Dark Matter \\
  EF & Einstein frame \\
  EH & Einstein-Hilbert \\
  EoS & Equation of State \\
  FLRW & Friedmann-Lemaître-Robertson-Walker \\
  GR & General Relativity \\
  GW & Gravitational Wave \\
  %HI & Higgs inflation \\
  %ISW & Integrated Sachs-Wolfe \\
  IVP & Initial Value Problem \\
  JF & Jordan frame \\
  LHC & Large Hadron Collider \\
  LLI & Local Lorentz Invariance \\
  LPI & Local Position Invariance \\
  LSS & Large-Scale Structure \\
  MOND & MOdified Newton Dynamics \\
  NS & Neutron Star \\
  PN & Post-Newtonian \\
  PPN & Parametrized Post-Newtonian \\
  QCD & Quantum Chromodynamics \\
  %QED & Quantum Electrodynamics \\
  SgrA* & Sagittarius A* \\
  SEP & Strong Equivalence Principle \\
  SM & Standard Model of particle physics \\
  SN & Supernovae \\
  SR & Special Relativity \\
  STT & Scalar-Tensor theories \\
  TOV & Tolmann-Oppenheimer-Volkoff \\
  UFF & Universality of Free Fall \\
  vev & Vacuum Expectation Value \\
  WEP & Weak Equivalence Principle \\
  \hline
\end{tabular}
\caption{List of acronyms appearing in this thesis and their signification.}
\label{tab:acronyms}
\end{center}
\end{table*}

\phantomsection
\addcontentsline{toc}{chapter}{Conventions}

\part{General context}

\renewcommand{\chaptermark}[1]{\markboth{\small\textsc{Chapter \thechapter.\ #1}}{}}
\cleardoublepage

\chapter{The Theory of General Relativity} % Main chapter title

\label{chap:math_fundations} % For referencing the chapter elsewhere, use \ref{Chapter1} 

\lhead{Chapter 1. \emph{Awesome chapter 1}} % This is for the header on each page - perhaps a shortened title

%----------------------------------------------------------------------------------------

This first chapter is devoted to General Relativity (GR) and its foundations, the action formalism being privileged. The theoretical pillars of GR, among them Lorentz invariance and causality, locality, general covariance as well as second-order equations of motion are introduced. Then the Lovelock theorem is formulated in order to highlight the particular status of GR in 4-dimensional spacetime. Finally, the equivalence principles are discussed in the light of the previous analysis of GR.  

\section{The Mach principle} \label{sec:mach}
The concepts of space and time dramatically change from the Newton's theory to the Einstein's one, revealing two philosophical stances. In the XVIIIth century (see also \cite{rovelli2004quantum}), Clarke (and Newton after him) claimed that \textit{"space exists as a substance or as an absolute being and absolute motion is present"}  \cite{Clarcke_Leibniz} while Leibniz (and Descartes before him) maintained that \textit{"the space is constituted only in relation to co-existent things allowing for relativism in motions only"}  \cite{Clarcke_Leibniz}. During the foundations of GR, Einstein was inspired by the Mach's principle\footnote{We would like to point out here that many different formulations of the Mach principle have been proposed in the literature (see \cite{rovelli2004quantum} for a summary), some of them being fulfilled by GR. In this manuscript, we will follow Brans and Dicke's formulation \cite{Brans:1961sx}.} \cite{Mach}, which is descended from the Leibniz's point of view, stating that \cite{Brans:1961sx},
\begin{quote}
 MACH PRINCIPLE - \textit{"The geometrical and the inertial properties of space are meaningless for an empty space, [...] the physical properties of space have their origin in the matter contained therein and [...] the only meaningful notion of a particle is motion relative to other matter in the Universe."}
\end{quote}
In GR, matter affects the gravitational field according to the Mach principle \cite{rovelli2004quantum}. Also position and motion are fully relational in the sense that they are not determined with respect to a fixed non-dynamical background like in Newton's theory \cite{Gaul:1999ys}. The local inertial frame is even fully determined by the dynamical fields \cite{rovelli2004quantum}.  However, it is clear that GR does not implement the Mach's principle entirely since it admits many vacuum solutions, like the Schwarzschild, Kerr and de Sitter ones. Those aspects will be further discussed in the rest of this chapter. The Mach's point of view on gravity has not only inspired GR but also ways to test GR as well as modified gravity, as we will see in Chap.~\ref{chap:test_GR} and \ref{chap:MG}.

\section{General picture of General Relativity} \label{sec:GR}
Even if GR and the Newton's theory of gravity do not follow from the same philosophical stances, GR must reduce to the Newton's laws in the non-relativistic limit where the gravitational field is weak and velocities are small $v\ll c$, $c$ being the speed of light, according to the \textbf{correspondence principle}. In this section, the mathematical aspects of GR are briefly reviewed as well as their weak-field Newtonian counterparts.

\subsection{Field equations}
GR is a classical field theory which can be formulated with an action principle. The most general action is divided into a geometrical part known as the Einstein-Hilbert (EH) action $S_\rr{EH}$ and in the matter action $S_\rr{M}$,
\bea
S&=&S_\rr{EH}+S_\rr{M}
\\&=&\frac{1}{2\kappa}\int \dd^4 x \sqrt{-g}\left(R-2\Lambda\right) + S_\rr{M}\left[\psi_\rr{M};g_{\mu\nu}\right],
\eea
with $\kappa\equiv8\pi/\mpl^2$, $\mpl$ being the Planck mass; $g_{\mu\nu}$ the metric and $g$ its determinant; $R$ the Ricci scalar; $\Lambda$ the cosmological constant; and $\psi_\rr{M}$ the matter fields. The equations of motion or Einstein equations are then derived from the action variation with respect to $g_{\mu\nu}$ or equivalently by the \textbf{Euler-Lagrange equations}\footnote{This Ricci scalar is function of second-order derivatives of the metric (see also Secs.~\ref{sec:Ostro} and~\ref{sec:ccl_lovelock} for a discussion), leading to the definition of the Euler-Lagrange equations given by Eq.~\eqref{eq:Euler_Lagrange_2nd_gmunu}.},
\be \label{eq:Euler_Lagrange_2nd_gmunu}
  \frac{\df}{\df x^\rho}\left[\frac{\partial L}{\partial g_{\mu\nu,\,\rho}}-\frac{\df}{\df x^\lambda}\left(\frac{\partial L}{\partial g_{\mu\nu,\,\rho\lambda}}\right)\right]-\frac{\partial L}{\partial g_{\mu\nu}}=0.
\ee
Reminding the following relations,
\bea
\delta g^{\mu\nu}&=&-g^{\alpha\mu}g^{\beta\nu}\delta g_{\alpha\beta},\label{eq:variation_metric}\\
\frac{1}{\sqrt{-g}}\frac{\delta \sqrt{-g}}{\delta g^{\mu\nu}}&=&-\frac{1}{2} 
g_{\mu\nu}, \label{eq:variation_det-metric}\\
{\frac{\delta R}{\delta g^{\mu\nu}}}&=& R_{\mu\nu} + g_{\mu\nu}\square 
-\nabla_{\mu}\nabla_{\nu}, \label{eq:variation_scalR}
\eea
the Einstein equations read,
\bea
G_{\mu\nu}+\Lambda g_{\mu\nu}=\kappa T^{\left(\rr{M}\right)}_{\mu\nu},
\label{eq:Einstein}
\eea
with the stress-energy-momentum tensor $T^{\left(\rr{M}\right)}_{\mu\nu}$\footnote{Following standard practice, we will abbreviate "stress-energy-momentum tensor" as "stress-energy tensor".},
\bea \label{eq:def_T}
T^{\left(\rr{M}\right)}_{\mu\nu}\equiv-\frac{2}{\sqrt{-g}}\frac{\delta S_\rr{M}}{\delta 
g^{\mu\nu}}.
\eea
The Einstein equations are of second-order in the metric and their solutions determine the metric field $g_{\mu\nu}$ up to a diffeomorphism\footnote{Notice that the metric is not determined univocally because of the diffeomorphism-invariance of GR. This is because metric fields are potentials rather than observables (see also Sec.~\ref{sec:gen_cova} for a discussion). Einstein was troubled with this characteristic of GR during its conception (see e.g. \cite{rovelli2004quantum, Norton1993}).}.
The Einstein equations describe the dynamics of the spacetime predicted by GR, that is how the spacetime is curved (left-hand side of Eq.~\eqref{eq:Einstein}) depending on the matter-energy composition of the spacetime (right-hand side of Eq.~\eqref{eq:Einstein}). According to the correspondence principle, GR reduces to the Newton's theory in the non-relativistic limit if,
\be
  g_{\mu\nu}=\eta_{\mu\nu}+h_{\mu\nu},
\ee
with $\eta_{\mu\nu}$ the Minkowski metric and $h_{\mu\nu}$ the perturbation ($h\ll1$). So, the Einstein equations generalize the Poisson equation\footnote{If $\Lambda\neq0$, it is rather a Helmholtz equation.} of the classical mechanics,
\be \label{eq:poisson}
  \nabla^2 \Phi=4\pi G_\rr{N} \rho,
\ee
with $\Phi=-2 h_{00}$, the Newtonian gravitational potential. Indeed, the left-hand side of Eq.~\eqref{eq:poisson} is related to the second-order derivative of $\Phi$ (more precisely $R_{00}=\nabla^2 \Phi$ \cite{Misner:1974qy}) while the right-hand side is related to the energy distribution in space. 

The conservation of the stress-energy tensor\footnote{The conservation of $T_{\mu\nu}$ can be also derived from the Noether theorem, as shown in App.~\ref{app:gen_cova_math}},
%\textbf{les equations de conservation peuvent etre obtenues independamment grace au théoreme de noether et à l’invariance par transformations de Poincaré locales},
\bea
\nabla^\mu T^{\left(\rr{M}\right)}_{\mu\nu}=0,
\eea
generalizes the Eulerian equations of hydrodynamics in the case of a perfect fluid (see Sec.~\ref{sec:PNformalism}) to curved spacetime.
On the other hand, the Einstein tensor is automatically conserved $\nabla^\mu G_{\mu\nu}=0$ according to the second Bianchi identity,
\bea
\left(\nabla_\kappa R\right)^\rho_{\lambda\mu\nu}+\left(\nabla_\nu R\right)^\rho_{\lambda\kappa\mu}+
\left(\nabla_\mu R\right)^\rho_{\lambda\nu\kappa}=0,
\label{eq:Bianchi}
\eea
by contracting twice Eq.~\eqref{eq:Bianchi} provided that the affine connection $\Gamma$ defined as the covariant derivative of basis vectors $e_\mu$,
\be \label{eq:affine_connection}
  \nabla_\mu e_\nu\equiv\Gamma^{\lambda}_{\mu\nu} e_\lambda. 
\ee
is the so-called Levi-Civita one, i.e. is only determined by the metric.

\subsection{The Levi-Civita connection}
\label{sec:add_math}
%In order to recover Special Relativity (SR) in flat spacetime, geometrical structure in GR is pseudo-Riemannian or Lorentzian. Indeed at every point of spacetime, a locally inertial frame where the metric is diagonal exists because of general covariance and the laws of SR thus apply. Moreover such a pseudo-Riemannian spacetime ensures the light-cone causal structure. 
The metric has non-trivial properties in GR: it is symmetric $g_{\mu\nu}=g_{\nu\mu}$ such that it has 10 degrees of freedom (provided that the spacetime is four-dimensional), and its covariant derivative vanishes $\nabla^\mu g_{\mu\nu}=0$, the parallel transport preserving thus distances and angles, so that the second Bianchi identity \eqref{eq:Bianchi} holds.

In general, the affine connection \eqref{eq:affine_connection} has $4^3=64$ degrees of freedom in a four-dimensional spacetime, raising the question where these degrees of freedom come from. In GR, it reduces to the \textbf{Levi-Civita connection} which is fully determined by the metric and has no independent degrees of freedom. In particular, the path along which the particles are freely falling i.e. the geodesics (see Sec.~\ref{sec:geodesics}), is also fully determined by the metric. 

In the case of the general affine connection, the \textbf{non-metricity} tensor $Q_{\lambda\mu\nu}$,
\be
\nabla_\lambda g_{\mu\nu}=Q_{\lambda\mu\nu},
\ee
is responsible for $4\times10=40$ degrees of freedom, assuming $g_{\mu\nu}=g_{\nu\mu}$. The remaining 24 degrees of freedom are related to the antisymmetric part of the connection ($4\times6=24$ degrees of freedom) given by the \textbf{torsion tensor}, %$T^\lambda_{\mu\nu}$, 
\be \label{eq:torsion}
T^\lambda_{\mu\nu}\equiv\Gamma^{\lambda}_{\mu\nu}-\Gamma^{\lambda}_{\nu\mu}.
\ee

We can derive the components of the general affine connection $\Gamma^\lambda_{\mu\nu}$ by computing the combination,
\bea
  \nabla_\mu g_{\alpha\beta}+\nabla_\alpha g_{\beta\mu}-\nabla_\beta g_{\mu\alpha}&=&
  Q_{\mu\alpha\beta}+Q_{\alpha\beta\mu}-Q_{\beta\mu\alpha}\\
  &=&\df_\mu g_{\alpha\beta}+\df_\alpha g_{\beta\mu}-\df_\beta g_{\mu\alpha}\non\\
  &&+\left(\Gamma^\lambda_{\beta\alpha}-\Gamma^\lambda_{\alpha\beta}\right) g_{\lambda\mu}
  +\left(\Gamma^\lambda_{\beta\mu}-\Gamma^\lambda_{\mu\beta}\right) g_{\lambda\alpha} \non\\
  &&-\left(\Gamma^\lambda_{\mu\alpha}+\Gamma^\lambda_{\alpha\mu}\right) g_{\lambda\beta}.
\eea
Using the definitions of the Levi-Civita connection \eqref{eq:levi-civita_compo} as well as the definition of the torsion \eqref{eq:torsion}, we obtain,
\bea
  Q_{\mu\alpha\beta}+Q_{\alpha\beta\mu}-Q_{\beta\mu\alpha}
  &=& 2 \begin{Bmatrix}
 \	\beta\,\mu\,\alpha
  \end{Bmatrix}
  +T^\lambda_{\beta\alpha}g_{\lambda\mu}+T^\lambda_{\beta\mu}g_{\lambda\alpha}-T^\lambda_{\alpha\mu}g_{\lambda\beta}\non\\
  &&-2 \Gamma_{\beta\mu\alpha},
\eea
such that the components of the general affine connection read,
\bea \label{eq:gen_connection}
\Gamma^\lambda_{\mu\nu}=
\begin{Bmatrix}
 \lambda
 \\ \mu\nu
\end{Bmatrix}
+ K^\lambda_{\mu\nu}+\frac{1}{2} \left(Q^\lambda_{\,\mu\nu}-Q_{\mu\nu}^{\,\,\,\lambda}-Q_{\nu\,\,\,\mu}^{\,\lambda}\right),
\eea
where $K^\lambda_{\mu\nu}$ is the contortion tensor,
\be
K^\lambda_{\mu\nu}=\frac{1}{2}\left(T^\lambda_{\,\mu\nu}+T_{\mu\,\,\,\nu}^{\,\lambda}+T_{\nu\,\,\,\mu}^{\,\lambda}\right).
\ee
% 
% The affine connection can be generally treated as an independent field. Applying the variational principle with respect to $\Gamma$ to $S_\rr{EH}$ gives rise to the so-called Palatini equation \cite{Palatini},
% \be
%   \nabla_\alpha \left(\sqrt{-g} g^{\mu\alpha}\right) \delta^\nu_\lambda -
%   \nabla_\lambda \left(\sqrt{-g} g^{\mu\nu}\right) +\sqrt{-g} T^{\mu\nu}_\lambda=0,
% \ee
% provided that $S_\rr{M}$ is independent on $\Gamma$.
% The Palatini formalism  highlights that assuming $T^{\lambda}_{\mu\nu}=0$ and $Q_{\lambda\mu\nu}=0$, the connection is the Levi-Civita one.

\subsection{Geodesics equations} \label{sec:geodesics}
The Einstein equations only determine the spacetime dynamics. The motion of a body in spacetime derives from the so-called geodesics equations which generalize the Newton's second law for the gravitational force ($\mathbf{F_\rr{g}}=-m\
\mathbf{\nabla}\Phi$, $m$ being the test mass), 
\be \label{eq:Newton2}
  \mathbf{\nabla}\Phi= -\frac{\dd \mathbf{p}}{\dd t},
\ee
with the momentum $\mathbf{p}$.
%, the second equality being only valid in the limit of a constant mass and assuming the Galileo's equivalence principle. 
%In the case of a gravitational force (in the limit of a constant mass and assuming the Galileo's equivalence principle), Eq.~\eqref{eq:Newton2} becomes,
% \be
%   \frac{\dd^2 \mathbf{x}}{\dd t^2}=-\mathbf{\nabla}\Phi.
% \ee

In GR, because of the spacetime curvature, the notions of straight line and parallelism are adapted by introducing the parallel transport along a curve. A freely falling body takes the shortest path, and moves along the so-called \textbf{geodesics}. The general definition of geodesics states that they are the curves whose tangent vector $\mathbf{V}$ is parallel propagated along itself \cite{wald}, that is satisfying,
\be
  \nabla_\mathbf{V} \mathbf{V}=0.
\ee 
%the covariant derivative requiring the connection $\Gamma$,
In terms of spacetime components, the geodesics equations read,
\bea
  \frac{\dd^2 x^\mu}{\dd \lambda^2}+\Gamma^{\mu}_{\nu\rho}\frac{\dd x^{\nu}}{\dd \lambda}\frac{\dd x^{\rho}}{\dd \lambda}=0,
\eea
with $\lambda$ the affine parameter and  $\Gamma^{\mu}_{\nu\rho}$ either the affine connection or the Levi-Civita one\footnote{The torsion does not affect the geodesics because of the symmetry over the indices $\nu$ and $\rho$, such that only the non-metricity makes the affine connection different than the Levi-Civita one in this case \cite{Misner:1974qy}.}.
The geodesics equations can also be derived from the action variation by extremizing the infinitesimal path length $\dd s$,
\be \label{eq:action_geod}
S_\rr{geo}=\int \dd s = \int \sqrt{-g_{\mu\nu} \dd x^\mu \dd x^\nu}
=\int \sqrt{-g_{\mu\nu} \frac{\dd x^\mu}{\dd\lambda} \frac{\dd x^\nu}{\dd\lambda}} \dd\lambda,
\ee
where $\lambda$ is the affine parameter.  

Both the geodesics and the Einstein equations implement the GR theory: on the one hand, spacetime curvature depends on the presence of matter-energy and on the other hand, body motion depends on spacetime curvature. 
If the motion is considered for test particles with negligible mass, one can solve first Einstein equations to determine spacetime shape and then solve geodesics (which correspond to the conservation of $T_{\mu\nu}$ in this case) to characterize the geometry. However, if one considers a large number of test particles whose contribution to curvature cannot be neglected, the problem is hard to handle since no clear starting point exists. This is the reason why a mean field approach is used in general, for instance by assuming a fluid approximation (see Chap.~\ref{chap:test_GR}).

\section{The underlying assumptions of GR}
After reviewing the usual point of view on GR, we will question in this section what its underlying assumptions are in order to understand why the EH action is so special at some point\footnote{The requirement of a Lagrangian formulation is also not trivial \cite{Durrer:2007re, Uzan:2010pm} since the evolution of every functions appearing in the Lagrangian must be determined self-consistently via the equations of motion.}. Those assumptions are not trivial and have inspired theories of gravity beyond GR.
%The pillars of GR, which are Lorentz invariance and causality, locality, general covariance as well as second-order equations of motion, are commonly accepted today.

\subsection{Lorentz invariance and causality} \label{sec:causality}
%\subsubsection{Causality}
In classical mechanics, equations of motion are of second order (see Sec.~\ref{sec:Ostro} for a discussion). Initial conditions for the position and the velocity then univocally determine how the system evolves - at least locally - in both direction in time according to the Cauchy theorem. 

In Special Relativity (SR), Lorentz invariance implies that the maximal speed of the information propagation, corresponds to the speed of light in the vacuum\footnote{Note that Lorentz-invariance does not always ensures the absence of superluminal motion (see \cite{Durrer:2007re, Bruneton:2006gf}).} $c=1$ and that there exists no closed timelike curve in spacetime. An event is said to be causally connected to another one if and only if points of the spacetime can be joined by non-spacelike curves \cite{hawking1973large}. 
%Moreover \tcb{Lorentz-invariance also implies that the only absolute values are true constants} (see also Sec.~\ref{sec:varying_cst}) \cite{Durrer:2007re}. 
Because of the Lorentzian signature of the metric, the time coordinate has a privileged status \cite{Bruneton:2006gf}. Hence, considering a four dimensional spacetime with one time dimension, the equations of motion are hyperbolic. The Cauchy problem is not necessarily well-defined globally in GR due to the fact that Einstein equations are hyperbolic and non-linear. Whether it can be done or not must be decided on a case by case basis \cite{Friedrich:2000qv}. In GR, even if the spacetime signature is Lorentzian such that SR can be recovered locally (see also Sec.~\ref{sec:EP}), there exist solutions of the Einstein equations with closed timelike curves, for instance the Gödel spacetime \cite{RevModPhys.21.447}, the question of causality being thus not trivial.

Three requirements allow to define \textbf{causality} in GR \cite{Bruneton:2006gf},
\begin{enumerate}
 \item \textit{Global chronology}, No global chronology exists in relativity since any field defines its own chronology locally on the spacetime manifold. However, in order to impose a global chronology, the spacetime is (in general) required to be \textbf{globally hyperbolic} \cite{Bruneton:2006gf}, that is it can be decomposed into the three space components and one time component, a procedure called 3+1 decomposition. In order to define this notion in more detail, let us introduce the concept of Cauchy surface \cite{gourgoulhon}, 
  \begin{quote}
  CAUCHY SURFACE - \textit{"a Cauchy surface is a spacelike hypersurface $\Sigma$ in [the manifold] $\mathcal{M}$ such that each causal (i.e. timelike or null) curve without end point intersects $\Sigma$ once and only once." }
  \end{quote}
  A spacetime equipped of a metric $(\mathcal{M},~g)$  that admits a Cauchy surface is by definition globally hyperbolic and satisfies the 3+1 decomposition. Most of the relevant spacetimes for cosmology and astrophysics admits this property, the G\"odel spacetime being an exception (see e.g. \cite{hawking1973large}). The topology of such a spacetime $\mathcal{M}$ is necessarily $\Sigma\times\mathbb{R}$. 
 \item \textit{No superluminal motion}, In GR, no signal propagates faster than the graviton speed (which corresponds to $c$) since superluminality would break the equivalence of all inertial frames  \cite{Durrer:2007re} (see also Sec.~\ref{sec:gen_cova}). In some cases, for instance when the dominant energy condition reading,
 \be
  T_{\mu\nu} t^\mu t^\nu \geq 0 \hspace{1cm} \text{and} \hspace{1cm} T_{\mu\nu} T^{\nu}_\lambda t^\mu t^\lambda \leq 0,
 \ee
 with $t^\mu$ any timelike vector\footnote{This condition is equivalent to $\rho>|p|$ for a perfect fluid given by Eq.~\eqref{eq:perfect_fluid}.}, is not satisfied, GR admits solutions with superluminal motion and closed loops, like the wormholes. 
 \item \textit{Cauchy problem}, The Cauchy problem is well-posed for globally hyperbolic spacetime in the absence of superluminality (see \cite{hawking1973large, wald} for a rigorous treatment of this question), provided that the gauge freedom  due to general covariance  (see Sec.~\ref{sec:gen_cova} for the definition) is fixed.
\end{enumerate}

\subsection{Locality} \label{sec:locality}
In classical field theory, \textbf{locality} refers to the fact that interactions at one point of the spacetime depend only on the infinitesimal vicinity of this point. In non-local theories, the dynamics of a field at the spacetime point $x$ is not only determined by its neighborhood $x+\delta x$, but also by the values of the field in a region of spacetime possibly infinite. As an example, in case of time non-localities, fields can exhibit memory effects \cite{Mitsou:2015yfa}.
Mathematically, it means that the field dynamics is given by integro-differential equations rather than differential ones.

Locality is preserved in GR since equations of motion are differential equations which derive from a Lagrangian formulation and are identified to the Euler-Lagrange equations. Indeed, the Lagrangian formalism is unable to account directly for either non-conservative interaction or causal history-dependence processes, that is for non-local interactions, since they are time-symmetric and necessarily energy conserving provided that $\df \mathcal{L}/\df t=0$ \cite{Galley:2012hx, 2014arXiv1412.3082G}. However, in some cases, it is possible to formulate the Lagrangian in such a way that non-conservative forces are included, their equations of motion being given by the Euler-Lagrange equations \cite{jose1998classical}. 

While causality and locality are sometimes confused, it is possible to build non-local theories which preserve causality (see e.g. \cite{Tsamis:2014hra}). In this case, the equations of motion cannot derive from an action and an arrow of time exists. 
%Microcausality rather than causality is usually assumed. It means that the (anti-)commutation relations of observables which are spacelike do not affect each other. 

\subsection{General covariance}\label{sec:gen_cova}
%\footnote{This section is inspired from \cite{rovelli2004quantum, Bertschinger}}.
%\subsubsection{Definitions and generalities}
Einstein defines general covariance by the following statement \cite{Einstein1916}:
\begin{quote}
  GENERAL COVARIANCE - \textit{"All physical laws have to be expressed by equations that are valid in all coordinate systems, i.~e., which are covariant under arbitrary substitutions  (or generally covariant)".
  %$x^{\mu}\longrightarrow y^\mu=y^\mu \left(x^\alpha\right)$
  }
\end{quote}
Contrary to the Newton's second law where fictious forces have to be invoked in non-inertial frame, Einstein equations are now valid in all coordinate systems, gravity acting like a fictious force in GR. Indeed, at each point of the spacetime, there exists a frame where gravity is vanishing and the laws of SR thus apply. 
%\textbf{Indeed since GR is a metric theory which can be diagonalized locally, the existence of such inertial frame is guaranteed (for instance this is no more the case in a bimetric theory where there is an ambiguity in the locally inertial frame). In a word GR has to be diffeomorphism-invariant.}
%in order to guarantee that such a transformation to locally flat spacetime does exist. 

As any dynamical field theory based on tensorial quantities, GR can be formulated in such a way that it is invariant under coordinate transformation $x^\mu\longrightarrow y^\mu=y^\mu\left(x^\alpha\right)$, that is under \textbf{passive diffeomorphism}, the diffeomorphism being defined as \cite{Gaul:1999ys},
\begin{quote}
  DIFFEOMORPHISM - \textit{"An infinitely differentiable $(C^\infty)$ map between manifolds that is one-to-one, onto and has a $C^\infty$ inverse."}
\end{quote}
In GR, the spacetime is a differential manifold and passive diffeomorphism corresponds to a mapping between two differential charts on the manifold. In particular, the EH action in the presence of the cosmological constant is invariant under coordinate transformation since it involves only scalars: the Ricci scalar, the cosmological constant and the volume element $\sqrt{-g}\, \dd^4 x$.  Indeed, the volume element transforms under a change of coordinates ${x^\mu\longrightarrow y^\mu=y^\mu\left(x^\alpha\right)}$ as,
\bea
  \dd^4 y&=&\dd^4 x \left(\det\frac{\df y}{\df x}\right),\\
  \sqrt{-g(y)}\equiv\sqrt{-\tilde{g}}&=&\sqrt{-\det \tilde{g}_{\mu\nu}}, \\ 
  \det \tilde{g}_{\mu\nu}&=&\det\left(\frac{\df x^\alpha}{\df y^\mu}\frac{\df x^\beta}{\df y^\nu} {g}_{\alpha\beta}\right),\\
  &=& \det\left({g}_{\alpha\beta}\right) \det\left(\frac{\df x^\alpha}{\df y^\mu}\frac{\df x^\beta}{\df y^\nu}\right),\\
  \Rightarrow \hspace{1cm} \dd^4 y \sqrt{-\tilde{g}}&=& \dd^4 x \left(\det\frac{\df y}{\df x}\right) 
  \sqrt{-g \left(\det\frac{\df x}{\df y}\right)^2}\\
  &=&\dd^4 x \sqrt{-g }
\eea
where $\left({\df y}/{\df x}\right)$ is the Jacobian matrix of the transformation and the property of determinant $\det(AB)=\det (A) \det (B)$, $A$ and $B$ being two matrices, is used.
It results that the EH action is invariant under change of coordinates,
\be \label{eq:diffeo_EH}
  S\propto \int \dd^4 x ~\sqrt{-g}~ \left[R(x)-2 \Lambda\right] = \int \dd^4 y~ \sqrt{-\tilde{g}(y)}~ \left[\tilde{R}(y)-2\tilde{\Lambda}\right] \propto \tilde{S},
\ee
where the tilde denotes quantities expressed in the coordinate system $y^\mu$.

However, since position and motion are fully relational in GR according to the Mach principle, GR is also invariant under \textbf{active diffeomorphism}: if all physical dynamical objects are shifted at once on the spacetime manifold $\mathcal{M}$ (without change of coordinate system), nothing is generated but an equivalent mathematical description \cite{Gaul:1999ys}. 
From the mathematical point of view, active diffeomorphism is a smooth displacement of any dynamical fields along an integral curve of vector field $\mathbf{\xi}\in T_P(\mathcal{M})$, $T_P(\mathcal{M})$ being the tangent space to the spacetime manifold at point $P$. Such a transformation is generated by the \textbf{pushforward}  $\left(\phi^* \mathbf{\xi}\right):T_P(\mathcal{M})\longrightarrow T_{\phi(P)}(\mathcal{M})$ which carries the tangent vectors $\mathbf{\xi}$ along the $C^\infty$ map $\phi:\mathcal{M} \longrightarrow \mathcal{M}$ between two tangent spaces of the spacetime manifold,  
\be
    \left(\phi^* \mathbf{\xi}\right)(f)=\mathbf{\xi}\left(f \circ\phi\right),
\ee
the smooth function $f:\mathcal{M} \longrightarrow \mathbb{R}$ being "pushed forward" to  $f\circ\phi:\mathcal{M} \longrightarrow \mathbb{R}$ by composing $f$ with $\phi$. 

In GR, general covariance does not only refer to passive diffeomorphism invariance but also on active diffeomorphism invariance which is made possible by the \textbf{lack of prior geometry}. By prior geometry, one means  \cite{Misner:1974qy},
\begin{quote}
 PRIOR GEOMETRY - \textit{"Any aspect of the spacetime geometry that is fixed immutably, i.e. that cannot be changed by changing the distribution of the gravitational sources."}  
\end{quote}
The lack of prior geometry implies that the spacetime geometry is purely dynamical and that gravity is entirely described in terms of the geometry, the active diffeomorphism invariance being rendered equivalent to the passive one, revealing an additional symmetry \cite{Bertschinger}. As an example, the Nordstr\o{}m theory where $g_{\mu\nu}=\varphi^2 \eta_{\mu\nu}$ with $\varphi$ a scalar field, admits a prior geometry since the Minkowski spacetime is fixed a priori.
%For example, quantum electrodynamics (QED) is not generally covariant since dynamical fields propagate on Minkowski spacetime which constitues a prior geometry, breaking the active diffeomorphism invariance???.  In GR, no aspect of the geometry of spacetime is fixed immutably \cite{Misner:1974qy}.
Contrary to the passive diffeomorphism invariance, the active one is far from obvious considering the EH action since the Lagrangian does not only depend on coordinates, but also on the metric field which is affected by the pushforward.

In App.~\ref{app:gen_cova_math}, we show that the invariance of tensor fields under active diffeomorphism in GR implies that the second Bianchi identity holds, as long as the connection is Levi-Civita. As a result, a shift of the metric field does not affect the Einstein equations which only determine the spacetime geometry \cite{Misner:1974qy}.

Finally, general covariance implies that rods and clocks measurements depend on the reference frame where the observer is located because of the gauge freedom of the metric. The interpretation of the measurements is thus much more tricky in GR than in classical mechanics where the spacetime is in addition euclidean. It is thus crucial to work with \textbf{gauge-invariant} quantities, that is quantities which do not depend on the coordinate system. 
%In analogy to the gauge invariance of classical electromagnetism, metric is defined up to a gauge transformation due to the diffeomorphism invariance. 
Over the ten degrees of freedom of the metric, four are gauge degrees of freedom and must be fixed by the four Bianchi identities, the six remaining ones being dynamical. It results that, in the Hamiltonian formulation of GR, Einstein equations reduce to four elliptic constraint equations and six hyperbolic Hamilton equations.
%As an example, in a static and spherically symmetric spacetime, several coordinates systems exist like the Schwarzschild, isotropic or Finkelstein ones. Once the coordinate system is defined, the gauge is fixed.  
This is the reason why metric fields are not observables, since they are defined up to the gauge transformation. %\textbf{COORDINATE SYSTEM FIX THE GAUGE???}
However, some quantities are gauge-invariant like the proper time $\dd\tau$. They constitute the useful quantities to be computed in order to confront the theory with the observations. 
%Four degrees of freedom have to be fixed thanks to the four Bianchi identities since the metric is a rank-2 symmetric tensor with ten independent components and six of the ten degrees of freedom are determined by the Einstein equations. 

\subsection{Second order equations of motion} \label{sec:Ostro}
Since the Newton's theory, equations of motion are of second order, two initial conditions determining the solution univocally (see also Sec.~\ref{sec:causality}). Actually, the laws of physics must involve no more than second-order time derivatives of the fundamental dynamical variables or generalized coordinates $q^i$  in order to preserve the stability of the solution \cite{Woodard:2015zca} as stated by the \textbf{Ostrogradsky theorem} \cite{Ostro}. 
%The only assumption is the nondegeneracy of the system. 
%If it is not so, a linear instability arises in the Hamiltonian in such a way that it cannot be eliminated by partial integration.

For the sake of simplicity\footnote{This result is so general that applies to all classical field theory.}, let us introduce the Ostrogradsky's result in classical mechanics for a point particle in one dimension. In general, the Lagrangian $L=T-V$ with $T$ and $V$ the kinetic and potential energies, depends upon the point particle position and its derivative, $L(q,~\dd{q}/\dd t\equiv\dot{q})$ with a quadratic dependence on $\dot{q}$ coming from the kinetic term. In this case, the equation of motion is derived from the usual Euler-Lagrange equation and is of second order, $\ddot{q}=\mathcal{F}(q,~\dot{q})$, provided that the system is \textbf{nondegenerate}, i.~e. ${\partial L}/{\partial{q^{(n)}}}$ depends on up to ${q^{(n)}}$ with $n$ the order of time derivative. The evolution of $q(t)$ is then univocally determined by two initial conditions, $q_0$ and $\dot{q}_0$. In this case, the phase space transformation $(q,~\dot{q})\longleftrightarrow(Q,~P)$ with $Q$ and $P$ the two canonical coordinates, is invertible,
\be
  Q\equiv q \hspace{1cm} \text{and} \hspace{1cm} P\equiv\frac{\partial L}{\partial \dot{q}},
\ee
%since two initial conditions must be fixed. 
since $P$ can be solved for determining $\dot{q}=v(Q,~P)$. 
% \be
%   \left.\df L/\df \dot{q}\right|_{\begin{cases}
%                                    q&=&Q\\
%                                    \dot{q}&=&v(Q,~P)
%                                   \end{cases}}=P
% \ee

The canonical Hamiltonian is given by the Legendre transformation of the Lagrangian,
\bea
  H&=& \dot{q}\frac{\partial L}{\partial \dot{q}} - L(q,~\dot{q}),\\
   &=& P v(Q,~P)-L\left[Q,~v(Q,~P)\right],
\eea
the time evolution of the canonical coordinates being given by the Hamilton's equations.

Let us now assume a Lagrangian with second order derivative $L(q,\dot{q},\ddot{q})$.
Since the Euler-Lagrange reads now,
\be \label{eq:Euler_Lagrange_2nd}
\frac{\partial L}{\partial q}-\frac{\dd}{\dd t}\frac{\partial L}{\partial \dot{q}}+\frac{\dd^2}{\dd t^2}\frac{\partial L}{\partial \ddot{q}}=0,
\ee
and because of the assumption of nondegeneracy i.~e. ${\partial L}/{\partial \ddot{q}}$ depends upon $\ddot{q}$, the equation of motion is of fourth order $q^{\rr{(4)}}=\mathcal{F}\left(q,~\dot{q},~\ddot{q},~q^{\rr{(3)}}\right)$. It means that four initial conditions must be fixed in order to get the solution $\left(q_0,~\dot{q}_0,~,\ddot{q}_0,~,q^\rr{(3)}_0\right)$ and four canonical variables have to be defined, for instance,
\bea
Q_{1}\equiv q, &\hspace{2cm}& P_1\equiv \frac{\partial L}{\partial \dot{q}}-\frac{\dd}{\dd t}\frac{\partial L}{\partial \ddot{q}}, \\
Q_2\equiv\dot{q}, &\hspace{2cm}& P_2\equiv \frac{\partial L}{\partial \ddot{q}}.
\eea
The assumption of nondegeneracy guarantees that the system is invertible, so $P_2$ can be inverted in order to determine $\ddot{q}=a(Q_1, Q_2, P_2)$. Only three of the four canonical coordinates are needed since $L(q,\dot{q},\ddot{q})$ only depends on three phase space coordinates.
The Hamiltonian is then derived by the usual Legendre transformation,
\bea
H(Q_1,Q_2,P_1,P_2)&\equiv&\sum_{i=1}^2 q^{\rr{(i)}} P_i -L(q,\dot{q},\ddot{q})\\
&=& P_1 Q_2 + P_2\, a(Q_1,Q_2,P_2)\non\\ \label{eq:hamil_4order}
&&\hspace{1.5cm}-L[Q_1,Q_2,a(Q_1,Q_2,P_2)].
\eea
As for the previous case, the time evolution is given by the Hamilton's equations.

However, the Hamiltonian \eqref{eq:hamil_4order} is ill-defined because of the linear term in $P_1$. Indeed, whereas $P_2$ is constrained by $(q,~\dot{q},~\ddot{q})$, there is no constraint among the element of $P_1$ due to the fact that there are three phase space coordinates for four canonical variables. It results that $P_1$ can take any values and that the Hamiltonian can take arbitrary positive or negative values, leading to the so-called \textbf{Ostrogradsky instability} \cite{Motohashi:2014opa}. If the system is free, it is not pathological. However, as soon as it is interacting with a 'normal' system with positive energy, the total system will lower its energy \cite{Durrer:2007re} and will quickly develop into excitation of positive and negative degrees of freedom \cite{Motohashi:2014opa}
even if the energy is conserved (the Hamiltonian being constant provided that the system is autonomous, i.~e. $\df L/\df t=0$).
%, dynamical variables which experience the instability will carry both positive and negative creation and annihilation operators and the system can decay in a collection of positive and negative energy excitations (see \cite{Woodard:2015zca} for details). Moreover, as soon as the system is interacting, 

In general, the Ostrogradsky's result constitutes a no-go theorem in the sense that equations of motion up to more than the second order leads to an instability in the theory assuming the nondegeneracy of the system. In the case of a degenerate system, then $\ddot{q}$ can be integrated out and the Ostrogradsky's instability is evaded (see Chap.~\ref{chap:MG}). 

In GR, the Lagrangian density of the EH action is function of up to second order derivatives of $g_{\mu\nu}$. It results that the equations of motion could be of fourth order in $g_{\mu\nu}$. Necessary and sufficient conditions for these Euler-Lagrange equations to be of second order are given by the Lovelock theorem.

\section{The Lovelock theorem} \label{sec:ccl_lovelock}
The underlying assumptions of GR developed in the last sections, i.e. the general covariance and the second-order equations of motion, are summarized in the \textbf{Lovelock theorem} \cite{Lovelock1969, Berti:2015itd},
\begin{quote}
 LOVELOCK THEOREM - \textit{"In four spacetime dimensions the only divergence free symmetric rank-2 tensor} (general covariance) \textit{constructed solely from the metric $g_{\mu\nu}$} (lack of prior geometry and the Levi-Civita connection) \textit{and its derivative up to second differential order} (second-order equations of motion)\textit{, is the Einstein tensor plus a cosmological term"}.
\end{quote}
%Causality and locality are not explicitly mentioned even if they are of course primeval GR ingredients.

Mathematically the Lovelock theorem implies that if the action is assumed to depend only on $g_{\mu\nu}$ up to second order derivative,
%\footnote{Remind that the Ricci scalar is built on second order partial derivative of $g_{\mu\nu}$.},
\be \label{eq:action_lovelock}
S=\int \dd^4 x L\left(g_{\mu\nu};~g_{\mu\nu,\rho};~g_{\mu\nu,\rho\sigma}\right),
\ee
the equations of motion $E^{\mu\nu}$ reading (see Eq.~\eqref{eq:Euler_Lagrange_2nd}),
\be
E^{\mu\nu}\left[L\right]=\frac{\df}{\df x^\rho}\left[\frac{\partial L}{\partial g_{\mu\nu,\,\rho}}-\frac{\df}{\df x^\lambda}\left(\frac{\partial L}{\partial g_{\mu\nu,\,\rho\lambda}}\right)\right]-\frac{\partial L}{\partial g_{\mu\nu}},
\ee
then the only second order equations of motion in $D=4$ (assuming the Levi-Civita connection) correspond to the Einstein equations with the cosmological constant \cite{Lovelock1969},
\be
\frac{E^{\mu\nu}}{\sqrt{-g}}=\alpha \mpl^2 \left(R^{\mu\nu}-\frac{1}{2}\,g^{\mu\nu}\,R\right)+\Lambda\,g_{\mu\nu},
\ee
with $\alpha$ and $\Lambda$ two arbitrary constants. 

This result is false for $D>4$ since in that case, the Lagrangian density with the Gauss-Bonnet term $\mathcal{G}$ defined as,
\be
  \mathcal{L}=\sqrt{-g}~\mathcal{G}
  =\sqrt{-g}\left(  R_{\mu\nu\alpha\beta}R^{\mu\nu\alpha\beta}-4 R_{\mu\nu} R^{\mu\nu}+R^2\right),
\ee
gives also rise to second-order equations of motion \cite{Lovelock1969}. Indeed, for $D=4$, the Gauss-Bonnet term is related to the Euler characteristic $\chi$ which is a topological invariant of the spacetime manifold $\mathcal{M}$,
\be
  \int \dd^4 x \sqrt{-g}~ \mathcal{G} =\frac{1}{8\pi^2} \chi(\mathcal{M}), 
\ee
such that the Gauss-Bonnet term is not dynamical.
%In higher dimensions, the Gauss-Bonnet term is no longer a topological invariant and becomes dynamical.

% 
% As a result the Lovelock theorem \cite{Lovelock1971} states that in four dimensions ($D=4$) for a vanishing torsion and a metric connection, only three geometrical terms are well-defined at the action level: the Ricci scalar $R$, the cosmological constant $\Lambda$ and the Gauss-Bonnet combination $\mathcal{G}$, 
% Since the Gauss-Bonnet term is a topological invariant in $D=4$, it can be thus integrated out. In $D=2$, the topology of manifolds $\mathcal{M}$ are classified thanks to the \textbf{Euler characteristic} $\chi_\rr{E} (\mathcal{M})$ which is intuitively related to the number of holes in the space considered. As an example the Euler characteristic for a polyhedron $P$ is defined by
% \be
% \chi_\rr{E}(P)=V-E+F,
% \ee
% where $V$ is the number of vertices, $E$ of edges and $F$ of faces. Any polyhedrons deformable into a sphere verify $\chi_\rr{E}=2$ while into a torus, $\chi_\rr{E}=0$. More generally, we can show that, for a $g-$holes torus $\mathcal{M}_\rr{g}$, 
% \be
%   \chi_\rr{E}(\mathcal{M}_\rr{g})=2-2 g.
% \ee
% The Gauss-Bonnet theorem implies that ($D=2$),
% \be
%   \chi_\rr{E} (\mathcal{M})=\frac{1}{4\pi}\int R,
% \ee
% with $R$ the curvature. This results was generalized by Chern for $D=2n$, 
% \be
%   \chi_\rr{E} (\mathcal{M})=\frac{1}{32\pi^2}\int \mathcal{G} \hspace{1.5cm} (D=4).
% \ee
% The Gauss-Bonnet term is thus a topological invariant, that does not depend upon the metric structure and only contributes to boundary terms. 

\section{The equivalence principles} \label{sec:EP}
From a phenomenological point of view, equivalence principles are fundamental in GR. Thanks to our current understanding, gravitation seems to be different than the three others fundamental interactions since it couples to test-particles and fields universally at all scales.
%, from the photons to the black holes. 
%Moreover, every particles and fields are identically charged under the gravitational interaction. 
%The definition of the so-called equivalence principles is one of the fundamental ideas questioned in this thesis.
The classical mechanics had already called on the Galileo's equivalence principle, usually referred to as the \textbf{universality of free fall} (UFF). Indeed two concepts of mass, apparently not related to each other, are invoked in classical mechanics, the inertial $m_\rr{i}$ and the gravitational $m_\rr{g}$ ones. In particular the acceleration inside a gravitational field,
\be
  \mathbf{a}=-\frac{m_\rr{g}}{m_\rr{i}}\mathbf{\nabla}\Phi,
\ee
is independent of the composition and the amplitude of the involved test mass provided that $m_\rr{i}=m_\rr{g}$. 

In GR the UFF derives from a novel formulation of the equivalence principle dubbed the \textbf{weak equivalence principle} (WEP)\footnote{The definitions of the different equivalence principles in GR differ from an author to another one. Some authors distinguish the weak and Einstein equivalence principles (see e.g. \cite{Carroll:2004st} where the WEP guarantees the UFF while the Einstein equivalence principle is related to the existence of local inertial frame). Throughout this thesis, we will consider that they overlap.} \cite{Carroll:2004st},
\begin{quote}
  WEAK EQUIVALENCE PRINCIPLE - \textit{"It is impossible to detect the existence of a gravitational field by means of nongravitational experiments, at least locally in small enough regions of spacetime, where the gravitational field is homogeneous and there is no tidal effect. In the presence of an arbitrary gravitational field, it is possible to find out a local inertial frame where the physical laws are those of SR."}
\end{quote}
Thus gravitation does not universally couple to the rest mass only, but also to the energy and momentum, photons being also affected by the gravitational field in GR. This first version of the equivalence principle is related to pseudo-Riemannian nature of spacetime. Because of the existence of a metric on the spacetime manifold, there exists a set of differential charts which are compatible with each other such that diffeomorphism-invariance is guaranteed. 
%Indeed if a metric is defined on a spacetime manifold then the diffeomorphism invariance   since this invariance guarantees that an inertial frame where the laws of SR are valid, exists. \textbf{en fait, cela vient du caractère riemanienn (existence d’une métrique). Si tu as une métrique c’est que tu es dans une variété différentiable et que tu as l’invariance par difféom. (changement de cartes)}
The WEP is explicitly assumed in the Lovelock theorem since $g_{\mu\nu}$ is the only tensor appearing in the Lagrangian \eqref{eq:action_lovelock}. The UFF is thus guaranteed whatever objects, gravitationally bounded or not.

In addition to the WEP, GR implements an even stronger version of the equivalence principle as first noticed by Dicke \cite{Brans:2008zz} when he looked after the possibility of testing GR. The \textbf{strong equivalence principle} (SEP) states that \cite{Carroll:2004st},
\begin{quote}
 STRONG EQUIVALENCE PRINCIPLE - \textit{It is impossible to detect the existence of a gravitational field by means of local experiments, either gravitational or nongravitational. }
\end{quote}
This means that the gravitational binding energy contributes equally to gravitational and inertial mass. The effect of a violation of the SEP would be especially sensible in compact objects (see Sec.~\ref{sec:schwa_in}) where the gravitational self-binding energy is non-negligible. Indeed, GR predicts that compact objects fall in the same way as light particles like photons. Moreover, the fluid approach for describing the matter fields is also questionable since the bunch of test particles composing the fluid self-gravitates and backreacts on spacetime curvature, contrary to point particles.
In summary, the SEP implies that the only effect of gravity is the gravity acceleration which is universal. Mathematically,
the metric only mediates gravity and the affine connection is the Levi-Civita one, the second Bianchi identity being guaranteed.

% Three distinct tests are generally distinguished \cite{Will1993} in order to test the equivalence principle: the universality of free fall (UFF), the local position invariance (LPI) and the local Lorentz invariance (LLI). The first one is related to the WEP while the two others are to the SEP. 
The mathematical formulation of the WEP and the SEP can be identified at the level of the action. Matter fields $\psi_\rr{M}$ are universally (and only) coupled to the metric $g_{\mu\nu}$. Hence, gravitation acts universally on all matter contained inside the Universe, so the WEP is guaranteed. Since the gravitational coupling is given by the Newton's constant (or the Planck mass) which does not vary in spacetime, therefore guaranteeing the constancy of the gravitational binding-energy, according to the SEP. 
%Second the EH action is built on scalar quantities which guarantees that it is Lorentz invariant. It results that there is no preferred frame, the causal structure is well-posed and the speed of light is constant.
%\textbf{objection: on peut avoir des théories sans le SEP mais avec invariance de Lorentz… comme les STT ;)}

All these assumptions are questionable from the theoretical point of view since
the equivalence principles could be broken at some point, leading to variations of the gravitational coupling into spacetime or depending on the relative velocity between observers, or even to a breaking of the universality of the gravitational coupling to all the matter-energy components. In addition, GR is a classical theory of gravity which is not compatible with quantum mechanics. Indeed, gravity in GR is only defined locally according to the differential geometry formulation while quantum mechanics is non-local in the sense that it calls on wave function. In that sense, the equivalence principles should be violated at some point.
However, as we will see in Chap.~\ref{chap:test_GR}, the equivalence principles are tested with very good accuracy today, GR being seemingly well-formulated.

%\textbf{j’ajouterais ici que la physique classique est locale alors que la mécanique quantique est non-locale, mais pas au sens où les equations du champ seraient integro-differentielles, mais bien au sens ou l’objet principal est un champ de probabilité. La non localité existe donc dans la nature et on sait donc que le principe d’équivalence (ou plutôt celui de localité) ne peut pas être rigoureusement correct.}

%This will be further discussed in Secs.~\ref{sec:tests_EP} and \ref{sec:EP_revisited}.

\section{Conclusion} \label{sec:chap1ccl}
In this chapter, we proposed an interpretative framework of GR, highlighting the particular status of the Einstein's theory of gravitation. Provided that the connection is the Levi-Civita one and the spacetime is four-dimensional, the most general second-order equations of motion are the Einstein equations in the presence of the cosmological constant. In addition, GR preserves locality and causality (up to some point, for instance assuming the dominant energy condition) and is generally covariant. 

At the end of this first chapter, we formulate the following conjecture: in four dimensions, the SEP is valid only if gravitation is mediated by a metric (without prior geometry) and only one metric, with the Levi-Civita connection, equations of motion being of second order (such that the Cauchy problem may be well-posed).

Provided this framework, we will first question how to test GR observationally and experimentally, either the dynamics predicted by the field equations and the geodesics or the equivalence principles. Testing GR is not trivial since the derivation of the observables is tricky because of general covariance.

Moreover, from the theoretical point of view, the underlying assumptions of GR are also questionable since GR has limitations: it is a classical theory of gravity which is expected to break down at high energy scale (usually the cut-off scale of GR is assumed to be the Planck scale). However, some issues appearing in cosmology today could also come from modifications of GR at the low energy scale, in particular due to the difficulty to give a physical interpretation to the cosmological constant. 

In Chap.~\ref{chap:MG}, the interpretative framework presented in this chapter will allow one to classify theories of gravity beyond GR in order to establish which assumptions of GR they violate.

\renewcommand{\chaptermark}[1]{\markboth{\small\textsc{Chapter \thechapter.\ #1}}{}}
\cleardoublepage

\chapter{General Relativity under scrutiny} % Main chapter title

\label{chap:test_GR} % For referencing the chapter elsewhere, use \ref{Chapter1} 

\lhead{Chapter 2. \emph{Awesome chapter 2}} % This is for the header on each page - perhaps a shortened title

%----------------------------------------------------------------------------------------
% \begin{quote}
% \textit{"Des chercheurs qui cherchent, on en trouve,\\
% des chercheurs qui trouvent, on en cherche."}
% \end{quote}

% After reviewing why GR is particular from the theoretical point of view, the current and future tests of gravitation either experimental or observational ones, are briefly reviewed.
% We start from the tests of the equivalence principles. Then we introduce the weak-field tests of gravity realized in laboratory and in the Solar System, which are now of amazing precision. However those tests only probe the regime where the relativistic effects appear to be small. Astrophysical tests enable strong field tests through the study of the dynamics of compact binaries and the gravitational waves (GWs) detection. Finally predictions of $\Lambda-$Cold Dark Matter ($\Lambda-$CDM) concordance picture in cosmology are reviewed.  Some questions raise by cosmological observations like the nature of Dark Matter (DM) and of the current cosmic acceleration as well as the fine-tuning of initial conditions are then discussed.

In this chapter, we question how far GR is tested by the current experiments and observations. Indeed, GR has been tested in the regime where the gravitational field is weak like in the lab, in the solar system and in cosmology, either where it is strong, for instance neutron stars (NSs) and black holes (BHs). The recent direct detection of gravitational waves (GWs) enables one to test also the radiative regime. 
Moreover, testing GR in the vacuum enables one to test directly the spacetime dynamics whereas in the presence of sources, the modeling of the stress-energy tensor is also tested. In particular, NSs and cosmology belong to this regime, the sources being relativistic or not. At the end of this chapter, we propose a classification of the GR tests depending on the presence and the properties of the sources. Even if no strong deviation from GR has been highlighted up to now it enables one to highlight which regimes have been tested.

%\textbf{TO BE READ \cite{PhysRevLett.116.221101}}

%\textbf{note perso: a ma connaissance, la plupart des tests de la RG ont été faits avec des énergies de liaison et des systèmes composites, ainsi que des situations où c’est principalement la rest mass qui prédomine. Une exception serait les tests de la RG avec les étoiles à neutrons dans lesquelles la matière est relativiste et bien sûr la cosmo où visiblement on n’a pas de la dust p=0… (DE)}

\section{Tests of the equivalence principles} \label{sec:tests_EP}
Following \cite{Will1993}, the equivalence principles are tested at three different levels: the UFF, the local position invariance (LPI) and the local Lorentz invariance (LLI). These three statements are tested at different scales, from the lab to cosmology. In this section current constraints are briefly reviewed.
\subsection{The test of the Universality of Free Fall}
GR predicts the UFF for any composition, mass and gravitational binding energy of the test body (see Chap.~\ref{chap:math_fundations}). Experimentally, any deviations from the UFF for two bodies inside the same gravitational field are parametrized by the Eötvös parameter, 
%Torsion balance experiment with 2 test masses gives best constraints parametrized by $\eta$, 
\be
  \eta_\rr{UFF}\equiv2\frac{\left| a_1-a_2 \right|}{\left| a_1 + a_2 \right|},
\ee
with $a_1$ and $a_2$, the acceleration of the first and the second body respectively. If $\eta_\rr{UFF}\neq 0$, then at least one kind of energy contributes differently for the inertial and the gravitational mass \cite{Will1993}. 

The best constraint on the UFF at the Solar System scale is due to the Lunar Laser Ranging experiment \cite{Williams:2004qba}, 
\be
  \boxed{\eta_\rr{UFF}=(-1.0\pm1.4)\times10^{-13} \hspace{2cm} \text{(gravitationally bounded objects)}}
\ee
It reveals that the Moon and the Earth, differing in composition and gravitational binding energy, fall in the gravitational field of the Sun by the same way to very good accuracy. The distance of the Moon from the Earth has been measured thanks to a laser beam for several decades, a reflector being planted on the Moon during Apollo space missions, and is currently known at the cm level. Notice that the SEP is tested rather than the WEP in this case because the Moon and the Earth are gravitationally bounded objects. %\textbf{These are also passive tests: you probe the free fall in an exterior field and not the gravitational interactions between both masses}.

The UFF has been tested in labs by many sophisticated Eötvös-type experiments, using torsion balance for instance (see \cite{Will} for more details). The principle of modern torsion balance experiments is the following: two bodies of different compositions are connected by a rod and suspended by a wire. If the gravitational acceleration of the two bodies differ and if this difference has a component perpendicular to the suspension wire, then a torque is induced on the wire. Current best constraints have been obtained by \cite{Schlamminger:2007ht} with, 
\be
  \boxed{\eta_\rr{UFF}=(0.3\pm 1.8)\times 10^{-13} \hspace{1.7cm} \text{(non-gravitationally bounded objects)}}
\ee
Eötvös experiments test the WEP since the test mass is not gravitationally bounded, the masses being only bounded by other interactions.

The UFF has been also tested at the atomic level where quantum mechanics comes into play. In 1975, Colella, Overhauser and Werner proposed to test the UFF with neutron interferometry \cite{PhysRevLett.34.1472}, the interferometer being tilted with respect to the Earth gravitational acceleration such that the neutrons are in free fall in the external gravitational field of the Earth. Other experiments are based on atom interferometry: atoms are cooled thanks to laser beams and are then trapped in a precise location. Placing atoms in different atomic levels, differences in acceleration due to the Earth are measured with very good accuracy (see e.g. \cite{RIS_0}). The same principle can be used for testing the difference in acceleration for different isotopic species \cite{Fray:2004}, for different spins states \cite{Tarallo:2014oaa}, etc. For instance, the best bounds on the Eötvös parameter for $^{87}$Rb and $^{39}$K atoms read \cite{PhysRevLett.112.203002},
\be
  \boxed{\eta_\rr{UFF}=(0.3\pm 5.4)\times 10^{-7} \hspace{2cm} \text{(quantum objects)}}
\ee
%\textbf{you should mention the COW (Colella-Overhauser-Werner) experiment. In the 1970s they have measured the effect of the earth gravitational field on neutrons in neutron interferometry. J’ai un bouquin de Werner la dessus dans mon bureau.}

A similar experiment is described in much more detail in Chap.~\ref{chap:chameleon} where numerical simulations are provided for an atom interferometry experiment testing the acceleration due to a scalar field possibly responsible for the current cosmic acceleration. However, the presence of a fifth force is tested in this case (see also Sec.~\ref{sec:fifth_force}) rather than the UFF.

\subsection{The Local Lorentz Invariance}\label{sec:LLI}
LLI is one of the cornerstone in SR, and thus in GR and standard model of particle physics (SM). In GR, the WEP guarantees that it is always possible to find a frame at each point of the spacetime where the laws of SR are valid. Active Lorentz invariance\footnote{The passive one is always fulfilled provided that the equations of motion are tensorial (a local Lorentz transformation is a subgroup of the general coordinates transformations) \cite{Mattingly:2005re} (see also Sec.~\ref{sec:gen_cova}).} is maybe not an exact symmetry at all energies; either the Lorentz symmetry is broken since there exists a preferred frame determined by other field(s) than the metric, for instance a vector field \cite{Jacobson:2000xp}; or it is deformed such as the Lorentz transformation from one frame to another is modified.

Observing a Lorentz violation implies that observables differ depending on the velocity of the frame. Two laws are thus tested: the constancy of the speed of light $c$, the best bound being \cite{PhysRevLett.110.200401},
\be
  \boxed{\frac{\delta c}{c}\lesssim 10^{-14}}
\ee 
as well as the vacuum dispersion relation of SR, $E^2=mc^2+p^2c^4$ which can have higher order terms (see e.g. \cite{Mattingly:2005re}). Several formalisms and parametrizations have been proposed, at the classical level like the $c^2$ formalism or at the quantum one like the Standard Model Extension (see e.g. \cite{Mattingly:2005re, Will}). 
%\textbf{peut-être faut-il développer ici les formulations mathématiques de l’invariance de Lorentz brièvement. Côté passif il y a le caractère tensoriel. il y a aussi la construction du lagrangien, l’invariance de la théorie par diffeomorphisme qui garantit donc la construction possible d’un fibré des repères. Bref, expliquer les différentes façons de mathématiser ce principe.}

Many experiments have been performed at different scales, none of them highlighting a violation of the LLI (see \cite{Mattingly:2005re, Kostelecky:2008ts}). Most of them are realized in conditions where the gravitational effects can be neglected. However, gravitational tests exist too, using the Cosmic Microwave Background (CMB) power spectrum and the polarization of  GWs for instance (see also Secs.~\ref{sec:inflation} and \ref{sec:GW} respectively). The direct detection of GWs from coalescing BHs (see Sec.~\ref{sec:GW}) will allow one to improve the constraints on the LLI in the gravity sector \cite{Kostelecky:2016kfm}.

\subsection{The Local Position Invariance I: gravitational redshift experiments}
%\textbf{là c’est plus chaud mathématiquement car selon moi ça touche au caractère fonctionnel de la physique classique (les mesures en un point ont un sens). Cela s’oppose à la vision de la physique quantique où seul l’intégrale a un sens et donc une formulation à l’aide des distributions.}

LPI states that the measure of observables does not depend on the position (in space and in time) where it takes place. The \textbf{gravitational redshift} experiment,  
%has two rather different parts: the position in space given by the gravitational redshift and the position in time, that is the question of the variation of fundamental constants.
which consists of measuring the relative difference in frequency $\Delta\nu/\nu$ of two identical frequency clocks placed at two different positions in a static gravitational field, directly tests the LPI. GR predicts that, up to first order,
\be
\frac{\Delta \nu}{\nu}=\left(1+\alpha\right)\frac{\Delta U}{c^2},
\ee
where $\Delta U$ is the difference between the gravitational fields at the two different places and $\alpha=0$ according to GR. 

This experiment was first realized by Pound, Rebka and Snider in 1959-1960 at Harvard University \cite{PhysRevLett.4.337, PhysRevLett.13.539}. They studied the redshift of a photon emitted upwards from the basements of a tower at Harvard to a receiving atom at the top of the building. Since the receiving atom is moving downward, this experiment combines the gravitational redshift and the Doppler redshift predicted by SR. Because of the Mössbauer effect, i.e. the resonant absorption of a photon by an atomic nuclei bounded in a solid, the receiving atom absorbs the photon only if the energy of the photon exactly corresponds to the transition between two atomic energy levels. Using this method, they determined $\alpha$ at the 1\% level \cite{PhysRevLett.13.539}.
The current best constraint was obtained in 1976 by GRAVITY PROBE-A \cite{PhysRevLett.45.2081},
\be
  \boxed{|\alpha|<2\times 10^{-4} \hspace{2cm} \text{(gravitational redshift)}}
\ee
They measured the difference in frequency of two hydrogen maser clocks, one being aloft in a spacecraft 10,000 km away from the Earth and the second one staying on the ground. 

LPI has been also tested by \textbf{null-redshift experiment} where the difference in frequency between two different clocks at different places in a gravitational field is measured. In this case, the effect of the clocks structure is also tested implying a potential violation of the UFF. Best constraints have been obtained at the Solar System scale by comparing the frequency of atomic clocks in the time-varying gravitational potential of the Earth due to its orbital motion around the Sun \cite{PhysRevD.65.081101}.
The measure of $\alpha$ should be improved by the future space missions like the Galileo 5 and 6 satellites ($|\alpha|<4\times 10^{-5}$) \cite{Delva:2015kta} as well as the ACES space mission \cite{ACES}. 
In addition, the test of the gravitational redshift would also be explored in the strong field regime, for example by looking at the bodies orbiting around the central BH of the Milky Way, Sagittarius A* (SgrA*) \cite{Meyer2012}. 

Constraining the variation of the fundamental constants is a second test of the LPI since it would imply a violation of the LPI: depending on where and when the experiment is performed, the measure of the observables would differ. We discuss this second test in Sec.~\ref{sec:varying_cst}. In the following, we will refer to the LPI either for gravitational redshift experiments or the constancy of fundamental constants.

\section{The weak field regime}\label{sec:PPN}
The weak field regime describes gravitational systems where the linearization of GR is valid since the gravitational field is weak. The (quasi-)stationarity or slow motion regime is also implicitly assumed, in the sense that the massive bodies motion is slow compared to $c$. 
We have already explored the test of the equivalence principles in this regime (see Sec.~\ref{sec:tests_EP}). We will now turn to the test of the dynamics predicted by GR either the Einstein or the geodesics equations.

The dynamics predicted by GR has been already tested a lot, for instance with the perihelion precession of Mercury, the deflection of light by the Sun or the Lense-Thirring effect for spinning objects in orbit, for citing only some of them. In order to take into account all the possible deviations from GR, the parametrization of the deviations from the predictions is required. This is the reason why post-Newtonian (PN) formalism \cite{Eddington} has been developed and extensively used, especially in the Solar system and for binary pulsar.
%However, even in the weak field regime, we will see in the following that models beyond GR which can not easily be parametrized by the approaches presented in this chapter exist.
\subsection{The Post-Newtonian formalism} \label{sec:PNformalism}
The PN formalism is devoted to testing gravity in the weak field regime and, in particular, in the Solar System where the modeling of the Sun and its flock of planets would require many-body system formulated in GR.
%while not impossible (for instance thanks to the metric for many-body system developed by Einstein, Infeld and Hoffmann \cite{EinsteinInfeldHoffmann}) \textbf{attention il y a des techniques pour le gravitational N-Body problem, par exemple les équations de Einstein Infeld Hoffman en champ faible}
Assuming the spherical symmetry with the Sun at the center, the expansion of the Einstein equations order by order makes the study of the Solar System possible. 

This phenomenological approach consists of expanding all the possible terms of $g_{\mu\nu}$ and $T_{\mu\nu}$ in the weak field and slow motion regime. The PN formalism is perfectly suitable for the Solar System since, 
\begin{enumerate}
 \item \textit{The gravitational field is weak}: $|\Phi|/c^2\lesssim 10^{-5}$ in the Solar System. The upper limit is at the center of the Sun where $\Phi_{\odot}/c^2\lesssim 10^{-5}$ while for the Earth $\Phi_{\oplus}/c^2\lesssim 10^{-10}$ \cite{Will1993}. The strength of a gravitational field is further defined by the \textbf{compactness} $s$ (in spherical symmetry),
 \be \label{eq:compactness}
  s=\frac{2 \GN M}{\mathcal{R}}=2|\Phi|\equiv\frac{r_\rr{s}}{\mathcal{R}},
 \ee
 with $M$ the mass of the object, $\mathcal{R}$ either its radius or its characteristic scale (see Secs.~\ref{sec:strong_field} and \ref{sec:cosmo}) and $r_\rr{s}=2 \GN M$ its \textbf{Schwarzschild radius}, 
 \item \textit{The matter generating the Solar System gravity is in slow motion} compared to the Solar System center of mass, $v/c \ll 1$ (more precisely $v^2 / c^2 \lesssim 10^{-7}$),
 \item \textit{The energy density is much larger than the pressure} $\rho c^2\gg p$: For the sake of simplicity the stress-energy tensor is usually assumed to be a perfect and non viscous fluid,
 \be \label{eq:perfect_fluid}
  T^{\mu\nu}=\left(\epsilon + p\right)u^\mu u^\nu + p~g^{\mu\nu}
 \ee
 with $u^\mu$, the 4-velocity of the perfect fluid, $\epsilon=\rho c^2\left(1+\Pi\right)$, $\epsilon$, $\rho$, $\Pi$ and $p$ being the energy density, the rest-mass energy density, the specific energy density which takes into account other forms of energy than the rest mass one, and the pressure respectively.
\end{enumerate}
These assumptions are no longer valid in the strong field regime which requires other parametrizations of the metric (see Sec.~\ref{sec:strong_field}). Moreover, even for the Solar System, the PN formalism is valid in the limit where the rest of the Universe does not affect it \cite{Misner:1974qy}.

As stated before, the fields are expanded on the Minkowski background which constitutes the asymptotic solution at spatial infinity. The metric expansion is given by,
\be
g_{\mu\nu} (x)=\eta_{\mu\nu}+h_{\mu\nu}(x),
\ee
where $h_{\mu\nu}\ll 1$ is developed order by order and can be considered as a field propagating on the Minkowski background where $\eta_{\mu\nu}$ allows for raising/lowering the indices. The \textbf{Newtonian limit} is obtained assuming the metric expansion (see also the correspondence principle in Sec.~\ref{sec:GR}),
\be
g_{00}=-\left(1+2\frac{\Phi}{c^2}\right),\hspace{1.5cm} g_{i0}=0,\hspace{1.5cm} g_{ij}=\delta_{ij}.
\ee
This limit is referred to as the \textbf{first order} approximation. The stress-energy tensor must be expanded in the same way, for instance assuming a perfect fluid with $T^{00}=\rho$, $T^{0i}=\rho v^i$ and $T^{ij}=\rho v^i v^j +p \delta^{ij}$. Up to the first order the conservation of $T^{\mu\nu}$ leads to the Eulerian equations of hydrodynamics,
\bea
  \frac{\df \rho}{\df t}+\mathbf{\nabla}\cdot \left(\rho \mathbf{v}\right)=0,
  \label{eq:continuity}
  \\
  \rho \frac{\dd \mathbf{v}}{\dd t}=\rho \mathbf{\nabla} U- \mathbf{\nabla} p,
  \label{eq:fluid_mecha}
\eea
with $\dd /\dd t=\df/\df t+\mathbf{v}\cdot \mathbf{\nabla} $. Indeed up to first order,
\be \label{eq:conservation_Newton}
  \nabla_\mu T^{\mu\nu}=0 \hspace{1cm} \Rightarrow \hspace{1cm} \df_\mu T^{\mu\nu} +\Gamma^\nu_{00} T^{00} \simeq0,
\ee
with $\Gamma^{k}_{00}=a^k=-\nabla U$ by the Newtonian limit of the geodesics equations, $a^k=\dd^2 x^k/\dd t^2$ being the acceleration and $U\equiv-\Phi/\GN$ the gravitational potential\footnote{In the PN literature, the gravitational potential is referred to as $U\equiv-\Phi/\GN$ rather than to $\Phi$ \cite{Misner:1974qy}}. The component $\nu=0$ and $\nu=i$ of Eq.~\eqref{eq:conservation_Newton} reduced to Eqs.~\eqref{eq:continuity} and \eqref{eq:fluid_mecha} respectively using the definitions of the total derivative, $a^k$ as well as the relation $\df_i(\rho v^i v^j)=\mathbf{\nabla}\cdot(\rho \mathbf{v}) {v}^j+\mathbf{v}\cdot \mathbf{\nabla} v^j$.

The study of GR requires the \textbf{second order} approximation or the 1PN order. The order of smallness of physical quantities $\epsilon_\rr{PN}$ intervening in the equations of motion are evaluated with respect to the gravitational potential $U$. Keeping in mind that quantities appearing in the Newton's theory like %$\mathbf{\nabla}$, 
$v$, $\rho$ and $\eta_{\mu\nu}$ are of 0PN order or $\epsilon_\rr{PN}$, terms arising in the perturbation theory like $h_{\mu\nu}$ and $\df/\df t$ are of the order 1PN. Comparing terms appearing in the Eulerian equations of hydrodynamics as well as the definition of the components of $T_{\mu\nu}$ for a perfect fluid (see \cite{Will1993} for details), the bookkeeping of the order of smallness reads\footnote{The cosmological constant does not intervene in the calculations because of its far too low value compared to other terms of the Einstein equations.},
\be \label{eq:bookkeping}
  U\sim v^2 \sim \frac{p}{\rho}\sim \epsilon^2_\rr{PN},
\ee
and the time derivative which is vanishing in the Newtonian limit, is now of order $\epsilon_\rr{PN}$. The expansion of the metric to the 2PN order implies that the expansion of each component is determined up to,
\bea
g_{00}&=&\eta_{00}+h_{00}^{(2)}+h_{00}^{(4)}+...\simeq-1+2 U +h_{00}^{(4)},\\
g_{i0}&=&\eta_{i0}+h_{i0}^{(3)}+h_{i0}^{(5)}+...\simeq h_{i0}^{(3)},\\
g_{ij}&=&\eta_{ij}+h_{ij}^{(2)}+h_{ij}^{(4)}+...\simeq \delta_{ij} + h_{ij}^{(2)}.
\eea
The relevant even/odd terms in the expansion depend on their change of sign under time reversal: terms whose total $v$'s and $\partial/\partial t$'s are odd like $g_{i0}$ change sign under time reversal contrary to $g_{00}$ and $g_{ij}$\footnote{This is true up to $\epsilon^{5}_\rr{PN}$ where other effects like radiation damping come into play.} \cite{Misner:1974qy}. 

The stress-energy tensor has to be expanded by the same way. Its expansion requires the definition of some potentials (see also App.~\ref{sec:PPN_BD} for an example), among them the Newtonian potential \cite{Will1993},
\be \label{eq:U_PPN}
U \left(\mathbf{x},\,t\right)\equiv -\frac{\Phi  \left(\mathbf{x},\,t\right)}{G_\rr{N}} \equiv \int \dd^3 x'\frac{\rho\left(\mathbf{x},\,t\right)}{\left|\mathbf{x'}-\mathbf{x}\right|}.
\ee
Einstein equations are then solved order by order.

\subsection{The Parametrized Post-Newtonian formalism}\label{sec:PPNformalism}
The PN expansion is dubbed the \textbf{parametrized Post-Newtonian} (PPN) formalism when the metric is parametrized in the most general way, including ten parameters depending on ten potentials defined similarly to the Newtonian potential \eqref{eq:U_PPN} (see \cite{Nutku1969, Will1993, Will} for the whole expansion and technical details). The PPN parameters can be directly constrained from the observations in a model-independent way. We will consider only two PPN parameters in this thesis, namely $\gamma_\rr{PPN}$ and $\beta_\rr{PPN}$. In the Standard PN gauge they read \cite{Will1993},
\be \label{eq:metric_PPN}
  g_{00}=-1+2\,\bar G U-2\,\beta_\rr{PPN}\,\bar G^2 U^2, \hspace{2cm} g_{ij}=\delta_{ij}\left(1+2\,\gamma_\rr{PPN}\,\bar G U\right),
\ee
with $\gamma_\rr{PPN}=\beta_\rr{PPN}=1$ according to GR ($c=1$), $\bar G$ being the measured gravitational constant. Best constraints today,
\bea \label{eq:Cassini}
  \boxed{|\gamma_\rr{PPN}-1|<2.3\times 10^{-5}, \hspace{2cm}
  |\beta_\rr{PPN}-1|<7\times 10^{-5}}
\eea
were obtained by the Cassini spacecraft thanks to Shapiro effect \cite{Cassini} and by studying orbital effects in planetary ephemerids \cite{Fienga:2011qh} respectively. For the additional PPN parameters, the reader is reported to \cite{Will}.
The constraints on PPN parameters should be improved in the near future thanks to space missions like GAIA (see for example \cite{IAU:6911396} and \cite{Hees:2015ixa}) and BepiColombo \cite{PhysRevD.66.082001} which should enable to constrain them up to $\gamma_\rr{PPN}-1\sim 10^{-6}$ and $\beta_\rr{PPN}-1\sim 10^{-6}$.

The PPN formalism applies to all theories which possess one metric describing spacetime provided this metric satisfies the WEP (see e.g. \cite{Misner:1974qy}). In this case the asymptotic behavior of metric fields is polynomial. As an example, if the metric has a  massive scalar field counterpart, the gravitational Newtonian potential is of Yukawa type (see also Sec.~\ref{sec:fifth_force}). The asymptotic behavior of the metric field is then exponentially decreasing and the PPN formalism is thus not valid. However, if its mass is sufficiently small, we can consider that it contributes to higher order terms and neglect it in the PPN expansion as for the cosmological constant in GR. Other metric parametrizations have been proposed in order to generalize the PPN parametrization for extended theories of gravity, like the Parametrized Post-Einsteinian formalism \cite{Jaekel:2005qe} which takes into account more generalizations of GR.  In addition, simulations of observables for general modifications of gravity have been developed \cite{Hees:2012nb}.

%(see \cite{Capozziello:2004sm} for a discussion})

%\textbf{The PPN formalism has been extensively used in the literature, in particular in order to constraint extended models of gravity. However this parametrization is limited to metric theories, that is where the spacetime possesses a metric which satisfies the weak equivalence principle ???. Alternative theories of gravity which does not satisfy those statements require other metric parametrization. }

%\textbf{you should mention some limitations of the PPN. According to me, it is convenient for theories with a WEP but not a SEP, and for theories which predicts gamma and beta constant in space and time, as well as non Yukawa interactions}

\section[The strong field regime]{The strong field regime}
\label{sec:strong_field}
In the limit where $U\sim v^2 \sim p/\rho \sim \epsilon^2_\rr{PN}\ll 1$ is no more valid everywhere in the gravitational system, strong field regime effects come into play and the linearization of GR, i.e. the 1PN approximation, is no longer appropriate \cite{Will}.
This is the case for compact objects ($s\sim 1$), the most compact objects predicted by GR being BHs with $s\sim1$ (see the definition of the compactness \eqref{eq:compactness}). The strong field regime also applies for less compact objects like NSs, $s=0.2-0.4$, and white dwarf, $s=10^{-2}$. In comparison, $s=10^{-6}$ for the Sun and $s=10^{-10}$ for the Earth. 

In the case where orbital velocity in binary systems is very large ($v\sim c$), relativistic effects have to be taken into account. Binary pulsar does not belong to this regime so that a kind of PN approximation might work \cite{Will}. This is no longer true for binary systems of BHs which require other tools like numerical relativity simulations.

Compact objects also enable one to test the SEP because of their non-negligible binding energy.
In this section, the GR solution for spacetime around a compact object is briefly reviewed and the current and future tests of GR in the strong field regime are briefly discussed, focusing on BHs, GWs and NSs.

%\textbf{actually it is s>1 if you assume R to be the radius of the body. s=1 for the event horizon which is the physical yet non material limit of the hole QUESTION: cela a du sens de définir la compacité à l'intérieur de l'horizon?}

\subsection{The Schwarzschild solution}\label{sec:GR_compact}
Studying spacetime inside and around compact objects, the most simple spacetime symmetry is the spherical one.
%even if it appears that most of the astrophysical objects rotate (usually slowly enough for assuming the quasi-stationarity). 
%In this last case, the spacetime is rather axisymmetric. 
In this case, the spacetime geometry  surrounded compact objects, i.e. in the vacuum, is the Schwarzschild one whether the star is static, vibrating or collapsing, according to the \textbf{Birkhoff theorem} \cite{birkhoff23}\footnote{This theorem was actually discovered and published two years earlier by Jebsen \cite{Jebsen1921, VojeJohansen:2005nd}},
%\textbf{J’ai en tête une formulation plus forte du théorème de birkhoff : en GR, la seule solution du VIDE en symétrie sphérique est celle de Schwarzschild. Mais celle-là est compatible également avec l’interdiction d’existence d’ondes grav en symétrie sphérique},
\begin{quote}
  BIRKHOFF THEOREM - \textit{"All spherically symmetric solutions of Einstein equations in the vacuum must be static and asymptotically flat (in the absence of a cosmological constant), that is a piece of the Schwarzschild geometry." } \cite{Clifton:2011jh, Misner:1974qy} 
\end{quote}
In particular, this theorem implies that far from the compact objects, their gravitational influence is negligible so that the spacetime is asymptotically flat at spatial infinity (neglecting the cosmological constant). The most general metric for a static and spherically symmetric spacetime given here in the \textbf{Schwarzschild coordinates}\footnote{Note that Schwarzschild coordinates apply for any spherical system and do not imply the Schwarzschild solution which is only valid in the vacuum.}, is,
\be \label{eq:metric_schwa}
  \dd s^2=-\rr{e}^{2\nu(r)}\,\dd t^2+\rr{e}^{2\lambda(r)}\,\dd r^2+r^2\,\dd\Omega^2,
\ee
where $\nu$ and $\lambda$ are both metric fields which have to be determined by solving the Einstein equations, and $\dd\Omega^2\equiv\dd\theta^2+\sin^2\theta~ \dd \varphi^2$ is the infinitesimal solid angle. The influence of asymptotically expanding spacetime, taken into account in the Schwarzschild-de Sitter spacetime solution, is neglected in Eq.~\eqref{eq:metric_schwa}. 
%\textbf{on peut en effet avoir une solution matchant une expansion cosmo comme dans Schwarzschild-de Sitter avec un terme en $H^2r^2$ (solution de Gautreau Mc-Vittie)}

The solution of the Einstein equations in the vacuum for the metric \eqref{eq:metric_schwa} is the Schwarzschild solution \cite{Schwarzschild:1916uq},
\be \label{eq:Schwa_metric}
\dd s^2=-\left(1-\frac{r_\rr{s}}{r}\right)\dd t^2+\left(1-\frac{r_\rr{s}}{r}\right)^{-1}\dd r^2+r^2\,\dd\Omega^2,
\ee
with $r_\rr{s}$ the Schwarzschild radius. 
Two singularities appear in Eq.~\eqref{eq:Schwa_metric}. The first one at $r=r_\rr{s}$ is only a  coordinate system singularity\footnote{Change of coordinates to more involved coordinate systems like Eddington-Finkelstein or Kruskal-Szekeres ones, are non-singular in $r=r_\rr{s}$.} while the singularity at $r=0$ is a true one in the sense that the spacetime curvature becomes infinite. In order to check if the singularities are physical, gauge-invariant observable for the spacetime curvature must be derived, that is
%Because of gauge invariance (see Sec.~\ref{sec:gen_cova}), 
the Kretschmann invariant $\Xi$, 
\be \label{eq:Kretschmann}
  \Xi=R_{\mu\nu\rho\sigma}R^{\mu\nu\rho\sigma}=\frac{48 M^2}{r^6},
\ee
confirming that the only physical singularity is at $r=0$. 

BH solution implies the existence of an \textbf{event horizon} in $r=\rs$ within which nothing can escape and such as no event inside the horizon affects the dynamics outside. Singularities could be created during gravitational collapse without the formation of a horizon \cite{JacobsonLectures}, such that the singularity in $r=0$ would be "naked". Penrose conjectured that appearance of such a "naked" singularity is forbidden because it would be causally disconnected from the exterior of the event horizon \cite{PhysRevLett.14.57}. This conjecture is referred to as the \textbf{cosmic censorship}. 

The Schwarzschild solution is sufficient for describing spacetime in the vacuum, i.e. in the absence of matter. In particular, it enables one to study static BHs where only gravity comes into play. Note that most of the BHs rotate (usually slowly enough for assuming the quasi-stationarity), the spacetime being then axisymmetric rather than spherically symmetric \cite{GourgoulhonLect}. A metric is stationary if all its components are time-independent or equivalently if it possesses a timelike Killing vector. If the spacetime is static, then there exists also a time reflection symmetry \cite{Teukolsky:2014vca}. Hence, rotating BHs are stationary and modeled by the Kerr metric while non rotating ones are static and modeled by the Schwarzschild metric. The asymptotic flatness has to be imposed for recovering the Minkowski solution at spatial infinity according to the Birkhoff theorem.

\subsection{The uniqueness theorems for black hole solution}
Chandrasekhar wrote \cite{chandrasekhar1983mathematical},
\begin{quote}
 BLACK HOLE - \textit{"The black holes of nature are the most perfect macroscopic objects there are in the Universe: the only elements in their construction are our concepts of space and time. And since the general theory of relativity provides only a single unique family solutions for their descriptions, they are the simplest objects as well."}
\end{quote}
From the classical point of view, BHs observations only probe spacetime curvature effects, without any prior knowledge on the matter coupling to gravity. As we will see in the following, GR in the vacuum is directly tested by the recent direct detection of GWs by LIGO \cite{PhysRevLett.116.061102}.

In 1967, Israel proved that the only static asymptotically flat solution of the Einstein equations with a regular horizon is the Schwarzschild one in the absence of BH electric charge. This is the beginning of a serie of \textbf{uniqueness theorems} (see e.g. the Robinson's contribution of \cite{Uniqueness}) guaranteeing that there is a very limited family of stationary, asymptotically flat BH solutions in Einstein-Maxwell's theory: the unique spacetime solutions are the Kerr \cite{Kerr} and Schwarzschild metrics for stationary and static spacetimes respectively (or the Kerr-Newman \cite{Kerr_Newman} and the Reissner-Nordstr\"om ones \cite{Reissner, Nordstrom} in the presence of an electric field). The mass, the angular momentum and the electric charge are the only three parameters for describing all BHs in nature, all the other properties in the previous stages of the life of the star being not relevant anymore \cite{thorne1994black}. In particular, all stellar properties like deviation from spherical symmetry and magnetic field are not relevant anymore during the collapse of a star since the gravitational field decouples from its matter source in the late stages of collapse \cite{Teukolsky:2014vca}.

However, the uniqueness theorems assume that there is no additional scalar, vector or spinor field degrees of freedom and that no naked singularity exists\footnote{This last assumption which should be unnecessary \cite{Teukolsky:2014vca} and is still a limitation of the uniqueness theorem.}.  \textbf{The no-hair conjecture} states that BHs are completely specified by giving their mass, angular momentum as well as electric and magnetic charges, the "hair" being fields associated with stationary BHs apart from the gravitational and the electromagnetic ones. It has been proven for particular cases only (see e.g. \cite{PhysRevD.5.1239, PhysRevD.51.R6608}). Basically no-hair theorem guarantees that the scalar field is constant outside the horizon, i.e. $|\nabla\phi|=0\,\forall r>r_\rr{H}$, $r_\rr{H}$ being the horizon radius, such that the scalar field is settled to its asymptotic value outside the horizon \cite{Weinberg:2001gc}. The proofs of no-hair theorems for more sophisticated models are still under investigation (see e.g.~\cite{Berti:2015itd}).
A lot of models beyond GR violate this theorem \cite{Chrusciel:2012jk} \footnote{For instance hairy BHs are predicted in the presence of non-Abelian gauge fields, like in the Einstein-Yangs-Mills theory where the solution is static and has vanishing Yang-Mills charges whereas it is not characterized by its total mass. However, physical observables remain identical (see e.g. \cite{Volkov:1998cc}).}.

%The gravitational collapse of a star to a BH raises the question of the fate of  Some physicists, Novikov among them, showed that  According to the Price's expression \cite{thorne1994black} \textit{"Whatever can be radiated} [by GWs]\textit{, is radiated"}. This is a famous \textbf{no-hair theorem}, which proves that the mass, the angular momentum and the electric charge are the only three parameters for describing all BHs in nature, all the other properties in the previous stages of the life of the star being not relevant anymore. 

\subsection{Tests in the vacuum}
In this section, the current and future observations in the strong field regime appearing in the vacuum are briefly discussed. Those observations directly probe the spacetime properties since they are performed in the absence of matter.

\subsubsection{Isolated black holes}
Because of the uniqueness theorems, any deviation from the Kerr-Newman family of solutions would invalidate GR. In order to test the dynamics predicted by GR, parametrization of generic spacetimes would be a very useful tool, similarly to PPN expansion in the weak-field regime. However, no unique reference metric exists in the strong-field regime like the Minkowski spacetime in the weak-field one. Some attempts have been developed \cite{PhysRevD.83.124015} (see also \cite{Berti:2015itd} for a summary).  

Experimental tests have been proposed, notably by measuring the dynamics of orbiting objects like pulsars around BHs \cite{Sadeghian:2011ub}, for instance around SgrA* thanks to the telescope General Relativity Analysis via Vlt InTerferometrY (GRAVITY) \cite{2011Msngr.143...16E}. Any deviation from the timelike geodesics of the Kerr spacetime would be an evidence for physics beyond GR.
In the forthcoming decades, the radio telescopes Five hundred meter Aperture Spherical Telescope (FAST) and Square Kilometer Array (SKA) will discover most of the active pulsars beamed toward us in the Milky Way \cite{Kramer:2004hd}, so that some binary systems of pulsar-BH or pulsar orbiting around SgrA* should be detected.

Some further tests of GR exist too: are all compact objects with a mass $m\gtrsim3 M_\odot$ BHs? Do all BHs have a horizon?
Observations in the electromagnetic spectrum are tricky but GWs modes detection should give rise to precision test.

\subsubsection{Binary pulsar}
\textbf{Pulsars} are rotating NSs which emit radio waves due to their intense magnetic field, around $10^8\,$T. Even if they are compact objects, \textbf{binary pulsar systems} only probe GR in the vacuum provided that both compact objects can be considered as "point" masses without complicated tidal effects. Indeed, if the equivalence principles are fully satisfied, the only way to detect gravitational effects is via tides which generate GWs. In order to be detectable, GWs must be generated either by the coalescence of two compact objects either by isolated not perfectly symmetric NSs. Because of the emission of GWs, the orbital period of the binary system decreases over time. In the case of binary pulsar, the orbital velocity is relatively small $v/c\sim 10^{-3}\ll 1$ such that the PN formalism at leading order is still valid and the orbital period changes at an effectively constant rate \cite{PhysRevLett.116.221101}. 

In 1974, Hulse and Taylor discovered the binary pulsar PSR B1913+16 which enables to test the strong field as well as the radiative regime of GR for the first time.
Computation of 3 over the 5 post-keplerian parameters leads to a self-consistent estimation of the 2 remaining parameters, i.e. the mass of the pulsar and of its companion \cite{Will}. In addition, the variations of the orbital period in time $\dot{P}_\rr{b}$ due to the emission of GWs were measured during 30 years (the coalescence process lasts around $10^7$ years in the case of binary pulsar systems \cite{PhysRevLett.116.221101}), yielding   \cite{Damour, Weisberg:2010zz},
\be
  \frac{\dot{P}_\rr{b}^{\rr{obs}}}{\dot{P}_\rr{b}^\rr{GR}}=1.0013\pm 0.0021,
\ee
GR predictions being tested at $10^{-3}$ level. Other binary systems have been studied since then, giving rise to even better tests of gravity (see e.g. \cite{Kramer:2006nb, Antoniadis:2013pzd}).

Binary pulsar systems are today able to test GR, either by measuring the emission rate of GWs over decades or by studying the nonradiative strong-gravity effects \cite{Damour} by testing the SEP (see Sec.~\ref{sec:spont_scala}). In the future, tests should be improved by detecting a lot of binary systems.
%Pulsar timing: REF REVIEW \cite{Stairs:2003eg}

\subsubsection{Direct detection of gravitational waves}\label{sec:GW}
The recent direct detection by LIGO of GWs coming from the coalescence of two binary BHs systems \cite{PhysRevLett.116.061102, PhysRevLett.116.241103} enables one to probe the large velocity and highly non-linear regime of GR \cite{PhysRevLett.116.221101}. This regime not only requires the PN formalism but also numerical relativity simulations \cite{Gair:2012nm} in order to take into account the full non-linearities of GR.

The coalescence process of two BHs is divided into three parts: the inspiral phase during which BHs spiral together on nearly circular orbit; the merger phase where the relative velocity is close to the speed of light, $v/c\sim1/3$ and the oscillation frequencies of the emitted GWs are very specific \cite{Berti:2015itd}; and the ring-downs where any remaining deformity of the final single BH is dissipated in GWs.  At the end of the coalescence process, the final BH remnant must settle down to a stable stage, satisfying the Kerr solution according to the uniqueness theorems.

The coalescence process as a whole is found to be in agreement with predictions of Einstein equations in the vacuum. According to the observations during the inspiral phase, i.e.  using the low-frequency of the signal, the estimated masses of the primary and secondary BHs are given by $m_1=39^{+6}_{-4} m_\odot$ and $m_2=32^{+4}_{-5} m_\odot$ for the first event \cite{PhysRevLett.116.061102} and $m_1=14.2^{+8.3}_{-3.7} m_\odot$ and $m_2=7.5^{+2.3}_{-2.3} m_\odot$ for the second one \cite{PhysRevLett.116.241103}. Moreover, those results are consistent with the estimated mass and the dimensionless spin of the final BH as predicted from the inspiral phase and inferred from the merger and ringdowns phases. 
%The frequencies of quasi-normal modes GWs emitted during the ringdowns phase enable one to test the no-hair theorem provided that more than one mode is detected, which has not been the case yet.

LIGO has given the best upper bound on the graviton mass in the dynamical regime, $m_\rr{g}<1.2\times10^{-22}$~eV at $90\%$ Confidence Level (C.L.) \cite{PhysRevLett.116.221101}, a result which provides constraints on modifications of gravity predicting a massive graviton. Indeed, such theories predict that the massive graviton propagates at a frequency dependent speed \cite{PhysRevLett.116.061102}.

Up to now, LIGO has not provided constraints on the polarization states of GWs. Once the other ground based detectors like Advanced Virgo, Kagra and LIGO-India, will be operational, it would be possible to measure the polarization, such that models beyond GR which predict other polarization modes that the quadrupole one (the only mode predicted by GR) could be ruled out. 

%Prediction of GR in terms of GW phasing of compact binaries as they inspiral as well as the oscillation frequencies they produce as they merge are very specific \cite{Berti:2015itd}, so that if deviations are observed, it could perhaps be a smoking-gun test of GR. More generally, the type of gravitational radiation which would be detected, could be a very stringent test of GR.
The advantages of GW astronomy with respect to optical astronomy as well as astroparticle physics, are multiple: their signal is very clean since they are not affected by the presence of matter or electromagnetic fields when they are emitted and as a result, they do not suffer from the uncertainty on the astrophysical matter like NSs (see Sec.~\ref{sec:NS}). Moreover, GWs enable one to probe some astrophysical phenomenons in the absence of any other signal, for instance  BH binary system.
However, the inspiral and merger processes are intrinsically transient. 
%Hence the evolution over time which is really powerful for binary pulsar, is not for BHs coalescence. 
The intrinsic feebleness of the signal-to-noise ratio of GWs detection is a second drawback since GWs detection requires complex data analysis for extracting the signal. 

\subsubsection{GWs detection in the future}
Experiments like LIGO are dedicated to the coalescence of NSs and stellar BHs at late time, for a redshift $z\sim 1$, since they are able to detect signals from deca- to hecto-Hz \cite{Yunes:2013dva} (the frequency of the first signal detected by LIGO is $\sim35-150$ Hz). The space mission eLISA (see \cite{AmaroSeoane:2012km} for the scientific review of the mission) should be launched in the horizon 2034. It would be rather dedicated to the detection of GWs coming from supermassive BH binary system, up to a redshift of $z\sim 10$ \cite{Gair:2012nm}. Because of the very large size of its arms, around $10^6$ km, such an experiment can detect GWs from $10^{-5}$ to 1 Hz \cite{Yunes:2013dva}. The recent results from LISA-Pathfinder experiment confirmed that the sensitivity of eLISA is reachable \cite{PhysRevLett.116.231101}.

%Careful analysis of GW signals should provide new tests of GR, hopefully in a near future, since predictions of alternatives to GR could differ from GR ones \cite{Clifton:2011jh}, notably in terms of GW polarity and speed. The presence of additional fields in addition to the metric for describing gravity could also be probed (see \cite{Gair:2012nm} and \cite{Yunes:2013dva} for reviews focusing on space and ground-based missions respectively). 

\subsection{Schwarzschild interior solution} \label{sec:schwa_in}
Computing the GR solution for a compact star interior
%that is the solution of the Einstein equations inside the compact stars, 
involves the knowledge of the fluid composing the compact stars. In this case, not only the curvature effects are probed but also the composition of the matter sources as well as its coupling to the curvature. As we will see, GR is not tested directly in this case. As for the vacuum solution, we restrict the discussion to the static and spherically symmetric spacetime with the metric ansatz \eqref{eq:metric_schwa}.

The most general stress-energy tensor associated to a spherical distribution of matter bounded by gravitation is locally anisotropic \cite{Lemaitre19333} such that the radial $p_\rr{r}$ and tangential $p_\rr{t}$ pressures are independent. In the standard perfect fluid limit\footnote{We emphasize here that perfect fluid is for sure a strong assumption which is only justified by the sake of simplicity. Realistic description of the fluid composing compact objects requires more involved equations of state (see Fig.~\ref{fig:mass_radius}).}, i.e. $p_\rr{r}=p_\rr{t}=p$ \eqref{eq:perfect_fluid},
%One more equation is needed describing the stellar hydrodynamics.
% \be
% T^{\mu\nu}=\left(\rho+p\right) u^\mu u^\nu + p g^{\mu\nu},
% \ee
the so-called \textbf{Tolman-Oppenheimer-Volkoff (TOV) equation} generalizes the Euler equation of fluid dynamics and is derived from the conservation of the stress-energy tensor, $\nabla_{\alpha}T^{\alpha\beta}=0$ for $\beta=r$ ($\Gamma^0_{0r}=\nu'$), 
\bea 
  \nabla_{\alpha}T^{\alpha}_r&=&\frac{\dd p}{\dd r} + \Gamma^\alpha_{r \alpha} p-\Gamma^\alpha_{r \beta} T^\beta_\alpha=0, \\
  \frac{\dd p}{\dd r}&=&-\nu'\left(p+\rho\right),\label{eq:TOV}\\
  &=&-(p+\rho)\frac{2 m(r)+\kappa r^3 p}{2r\left[r-2 m(r)\right]},
\eea
where the second equality derives from the Einstein equations (see e.g. \cite{wald} for the detailed calculations) and $m(r)$ is the mass function of the compact object,
\bea \label{eq:mass_fct}
  m(r)&=&4\pi \int_0^r \dd r'\, T^0_0\, r'^2,\\
  &=&4\pi \int_0^r \dd r'\, \rho(r')\, r'^2. 
\eea
Then the solution of the TOV equation requires an equation of state (EoS) for the star interior $p=p(\rho)$. 

%\textbf{remarque importante: 2.3.5 n’est pas la forme la plus générale de TOV en symétrie sphérique. En effet, la symétrie sphérique permet à $T_{rr}$ (pression radiale) d’etre diffférente de la pression tangentielle $T_{\theta\theta}=T_{\phi\phi}$. C’est le cas par exemple avec un champ scalaire ou plein d’autres champs. Lemaître donne d’ailleurs en 1931 une solution d’équilibre soutenue purement tangentiellement dans laquelle R peut descendre jusque rS}

%The \textbf{total mass} of a compact object is thus given by $m(\mathcal{R})$ \textbf{je ne suis pas d’accord: tu dois intégrer jusqu’à l’infini. La masse est une quantité globale, topologique, reliée à la décroissance asymptotique du champ de gravité. Elle contient notamment la backreaction provenant de l’énergie du champ gravitationnel généré et éventuellement les contributions d’autres champs (scalaires ou autres)}.

Starting from the definition of the mass function \eqref{eq:mass_fct} and assuming a static and spherically symmetric spacetime, the total mass of the matter distribution is referred to as the \textbf{Arnowitt Deser Misner (ADM) mass},
\be \label{eq:ADMmass}
  m_\rr{ADM}=4\pi \int_0^\infty \dd r\, \rho(r)\, r^2. 
\ee
However, the common sense of the mass that is the density inside a proper volume element $\sqrt{-^{(3)}g}\dd^3 x={e}^\lambda \, r^2 \dd r \dd \theta \dd\varphi$, is rather referred to as the \textbf{proper mass} in GR \cite{wald},
\be
M_\rr{pr}=\int \dd^3 x  \sqrt{-^{(3)}g} \rho =4\pi \int_0^\infty \dd r \, \rho(r) \, \rr{e}^\lambda \, r^2,
\ee
the difference between the proper and the total mass being interpreted as the \textbf{gravitational binding energy},
\be
E_\rr{b}=M_\rr{pr}-m \hspace{0.3cm} > \hspace{0.3cm} 0.
\ee

Assuming \textbf{top-hat density profile} inside the star,
\be
\rho(r) = \left\{\begin{array}{ll}
 \rho_0 & \mbox{ if $r\leq \mathcal{R}$},\\
 0 & \mbox{ otherwise},
 \end{array}\right.
\ee
the TOV equation admits an analytical solution, imposing $p(r=\mathcal{R})=0$ \cite{Schwarzschild1916b, Tolman1939},
%\textbf{es-tu sûre de ta ref? Je crois que seule la solution de Schwarzschild extérieure est dans le papier de 1916. En effet, dans les années 20, Einstein, Lemaître et d’autre ont travaillé sur ce qu’on a appelé la « catastrophe d’Hadamard » du nom du matheux français qui a embêté Einstein avec la singularité de Schwarzschild. La solution de Schwarzschild intérieure vient peu après (par Eddington je crois) et plus tard on aura le Buchdal theorem qui dit que $R>=9/8 r_S$ avec des pressions isotropes},
\be \label{eq:pressureGR}
  p(r)=\rho_0 \frac{\sqrt{1-s}-\sqrt{1-\frac{s^3 r^2}{\mathcal{R}^2}}}{\sqrt{1-\frac{s^3 r^2}{\mathcal{R}^2}}-3\sqrt{1-s}}.
\ee
The central pressure $p_\rr{c}=p(r=0)$ predicted by GR,
\be \label{eq:lim_pressureGR}
  p_\rr{c}=\rho_0 \frac{\sqrt{1-s}-1} {1-3\sqrt{1-s}}.
\ee
becomes infinite for,
\be
  1-3\sqrt{1-s}=0 \hspace{0.5cm} \Leftrightarrow \hspace{0.5cm}
  1-s=\frac{1}{9} \hspace{0.5cm} \Leftrightarrow \hspace{0.5cm}
  s=\frac{8}{9},
\ee
or equivalently for the critical mass $M_\rr{cr}=(4/9)\mpl^2 \mathcal{R}$ (see Eq.~\eqref{eq:compactness}) assuming uniform density stars whatever EoS \cite{PhysRev.116.1027}\footnote{The last two assumptions are not restrictive actually \cite{wald}.}. Hence, observations of stars with $s>8/9$ might reveal either the existence of anisotropic stars with $p_\rr{t}\gtrsim p_\rr{r}$ \cite{Fuzfa:2001jj} or a deviation from GR.
%\textbf{c’est faux: il te suffit de diminuer la pression radiale et d’avoir une pression tangentielle plus forte… c’est toujours en RG ça s’appelle des anisotropic stars, c’était le sujet de mon premier papier…}
% 
% In the case of rotating objects, the spacetime is axisymmetric and the solution of the Einstein equations  in the vacuum is the \textbf{Kerr metric}. This metric is not only parametrized by the mass of the compact object, but also by its angular momentum. However, since compact stars appear to rotate slowly the slowly rotating approximation where spin corrections are assumed to be small perturbations on the static and spherically symmetric background, is commonly used. In the most general case where an electric field surrounds the compact object, the \textbf{Kerr-Newman metric} is solution and depends on the mass, the angular momentum and the electric charge of the object while, if the object is not rotating, the \textbf{Reissner-Nordstr\"om} metric is solution. However, this solution is in general not useful for astrophysics since the electric field tends to decrease fast if only it exists. 

\subsection{Tests in the presence of relativistic matter: neutron stars}\label{sec:NS}
%In the perfect fluid approximation $p_\rr{r}=p_\rr{t}=p$, if the mass of the star is upper than the critical mass $M_\rr{cr}$, the gravitational collapse should be completed and the star should form a Schwarzschild BH. This is the reason why the mass-radius diagram and the related EoS are so important for studying compact objects like NSs (see Sec.~\ref{sec:NS}).

\begin{figure} 
  \begin{center}
    \includegraphics[width=0.9\textwidth]{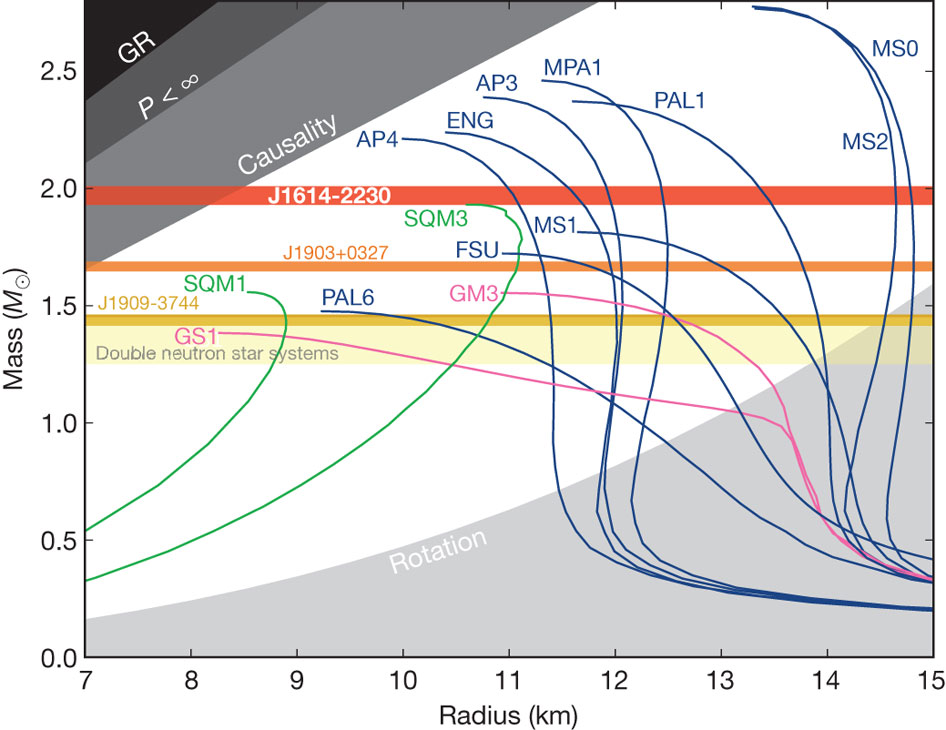}
    \caption{NS mass-radius diagram for various EoS (see \cite{Demorest:2010bx} for details). The horizontal bands in red, orange and dark yellow show the current observational constraints from several pulsars while in light yellow for double NS binaries. Any EoS line which does not intersect the red band ($m_\rr{NS}=1.97\pm 0.04 M_\odot$) is ruled out. Reprinted from ~\protect\cite{Demorest:2010bx} .}
    \label{fig:mass_radius}
    \centering
  \end{center}
\end{figure}

When the mass of a star is sufficiently large, that is when it reaches the \textbf{Chandrasekhar limit} of $m\gtrsim1.44\, M_\rr{\odot}$, then electron degeneracy pressure due to the Pauli principle is not sufficient anymore for counterbalancing the gravitational collapse and the star ends up exploding in a supernova (SN) Ib/c or II. The SN remnant can be either a NS or a stellar BH, depending on the mass and the metallicity, i.e. the presence of atomic elements other than hydrogen. The maximal NS mass which has been detected to date, is $2.01\pm0.04\, M_\rr{\odot}$ \cite{Demorest:2010bx} (see Fig.~\ref{fig:mass_radius}). 

In order to be able to test GR with NSs, the EoS should be determined. However, the EoS inside the core of NSs where the density becomes supranuclear around $10^{15}\rr{g/cm^3}$, is still largely unknown. The NSs crust is mainly composed of neutrons with electrons and protons while the density in the inner core is so high that it is constituted by a quark-gluon plasma which requires lattice Quantum Chromodynamics (QCD) computations in order to be simulated (see e.g. \cite{Chamel:2013efa}). Depending on the EoS, the maximum mass and compactness predicted by GR are different \cite{Berti:2015itd}. The mass-radius diagram represented in Fig.~\ref{fig:mass_radius} enables one to represent the different predictions of GR depending on the EoS. If the EoS predicts a maximal mass smaller than the maximal mass detected, i.e. $2.01\pm0.04\, M_\rr{\odot}$, it is rejected (see e.g. \cite{Wex:2014nva}). 

Hence, the difficulty of testing GR thanks to isolated NSs arises from the fact that the EoS of high-density matter is degenerated with strong-gravity effects \cite{Berti:2015itd}. In order to tackle this problem, some strategies have been invoked using almost EoS-independent relations between macroscopic observable properties of NSs \cite{ChamelDelsate} (see e.g. \cite{Yagi:2013bca}) like the "I-Love-Q" relation, which is a universal relation between the moment of inertia of the NS $I$, the Love numbers which measure the tidal deformability and the quadrupole moment of the NS $Q$ (see \cite{Berti:2015itd} and references therein for details). More recently a universal relation between $I$ and the compactness has also been highlighted \cite{Breu:2016ufb}. With such a universal relation, gravity can be tested in the strong-field regime without any prior knowledge of the EoS and astrophysical observations enable one to constrain nuclear physics up to a very large density.

Another axis of research consists of determining the EoS of NSs thanks to \textbf{astereoseismology} by observing characteristic NSs oscillation frequency or quasi-normal modes (see \cite{Kokkotas:1999bd} for a review). Those oscillations are responsible for the emission of GWs. As an example, the measure of the frequency and the damping time due to GW emission of one particular mode (called the $f-$mode) would give rise to both constraints on mass and radius of the NS up to at least $10\%$ accuracy and could be detected by LIGO up to $20~$ Mpc\footnote{1 pc is the distance at which 1 AU (the averaged distance from the Earth to the Sun) subtends an angle of one arcsecond i.~e. $1\,\text{Mpc}\sim 3\times 10^{22} \text{m}$.} for supermassive NSs according to \cite{Surace:2015ppq}. 

Observations of glitches, i.~e. sudden changes in the pulsar rotation rate, thanks to X-ray astronomy \cite{1969Natur.222..228R, 2011MNRAS.414.1679E} also shed light on the pulsars EoS. GWs should be also emitted by those instabilities. Glitches of a few minutes have been observed in some pulsars while the relaxation to the initial rotation period appears to take a few years. This means that the core of NSs has to be made of a non viscous liquid like helium-superfluid \cite{citeulike:12284426}. Two X-rays telescopes, CHANDRA and XMM-Newton, observed the evolution of the temperature in time at the surface of quite young NSs in SN remnants. Their observations have confirmed the presence of \textbf{superfluid} in NSs (see e.g. \cite{2011MNRAS.412L.108S}).

\section{Cosmology}\label{sec:cosmo}
In this section dedicated to cosmology, the solution of the Einstein equations assuming the cosmological principle is briefly reviewed. Similarly to NSs, cosmology not only probes the spacetime curvature because of the presence of a source, that is the cosmological fluid. Depending on the era in the Universe history, the source is either relativistic or not. According to the $\Lambda-$CDM (Cold Dark Matter) concordance model, the Universe is composed of matter, radiation, Dark Matter (DM) and Dark Energy (DE). Because of the presence of those sources, GR is not tested directly in cosmology.

\subsection{The cosmological principle}
According to the current observations, the Universe is \textbf{isotropic} at cosmological scales, i.e. above 100 Mpc. By extending the Copernican principle stating that \textit{"we have no privileged place in the world"} to all observers at cosmological scales, it results that the Universe is supposed to be \textbf{homogeneous} at a given cosmic time.
%Moreover, isotropy is supposed to apply to all observers at rest with respect to the cosmological fluid, so the Universe is assumed to be \textbf{homogeneous} at a given cosmic time. 

Such a maximally symmetric assumption allows one to predict background cosmology. However, whereas the isotropy is in agreement with the observations, for instance with CMB and galaxy surveys, homogeneity is difficult to test since it requires observations on spatial hypersurfaces \cite{Maartens:2011yx}. 

In order to explain the growth of cosmic structure appearing at smaller scales, the cosmological principle is relaxed to \textbf{statistically isotropy and homogeneity}. Several observations today enable one to test GR at the perturbative level and a possible departure from statistical isotropy is still under investigation (see e.g. \cite{Schwarz:2015cma}). Most of the cosmological observations today rely on statistical analysis.

\subsection{The {$\Lambda-$}CDM concordance picture}
%% Version electronique
%\subsection{The \texorpdfstring{$\Lambda-$}{TEXT}CDM concordance picture}

The metric for a \textbf{maximally symmetric spacetime}, i.e. satisfying the cosmological principle, or Friedmann-Lemaître-Robertson-Walker (FLRW) spacetime reads,
%\footnote{We use here the synchronous time coordinate $t$ and the metric is written in the synchronous gauge. We further assume that the spacetime is spherically symmetric.},
\be \label{eq:metric_FLRW}
 \dd s^2=-\dd t^2+a^2(t)\left(\frac{\dd r^2}{1-kr^2}+r^2 \dd\Omega^2\right),
\ee
where $t$ is the cosmic time that is the proper time of a comoving observer, $r$ is the radial coordinate and $k=[-1~;0~;+1]$ the curvature parameter depending on the spacetime geometry (hyperbolic, flat or spherical respectively). The \textbf{scale factor} $a(t)$ is the only metric field to be determined by solving the Friedmann-Lemaître equations,
\bea
  H^2&=&\frac{\kappa}{3} \sum_i \rho_i + \frac{\Lambda}{3}-\frac{k}{a^2} \label{eq:FR1},\\
  \frac{\ddot{a}}{a}&=&-\frac{\kappa}{6} \sum_i \rho_i \left(1+3 w_i\right) +\frac{\Lambda}{3}\label{eq:FRacc},
\eea
with $H=\dot{a}/a$, the \textbf{Hubble parameter}\footnote{In general, the dimensionless Hubble parameter $h$ yielding $H_0=h\times 100~ \text{km~Mpc}^{-1}~\text{s}^{-1}$ where the subscript $0$ refers to parameters evaluated today, is compared to the observations.}. The stress-energy tensor is assumed to be a perfect fluid\footnote{As before, this assumption is only justified by the sake of simplicity and is not valid during all the Universe history.} whose species (DM, dust and radiation) is labeled by $i$. In cosmology, the EoS $w$ is assumed to be time-independent and barotropic in the most simple case\footnote{In general the EoS is time-dependent. Different parametrizations of the EoS exist like the polytropic one $p=K\rho^{(n+1)/n}$ with $K$ a constant and $n$ the polytropic index, and the generalized Chaplygin gas one $p=-A\rho^\alpha$ with $A$ a positive constant and $0<\alpha\lesssim1$ \cite{Bento:2002ps}.},
\be
  w=\frac{p}{\rho},
\ee
with $w=-1,~1/3,~0$ for a Universe dominated by $\Lambda$, radiation and matter respectively. Assuming a barotropic EoS, the conservation of $T_{\mu\nu}$ gives the evolution of the density during the Universe expansion,
\be \label{eq:conservation_energy_cosmo}
  \rho(a)\propto a^{-3\left(1+w\right)}.
\ee

Current observations give rise to stringent constraints on background cosmology. In order to confront the theory with them, Friedmann-Lemaître Eqs.~\eqref{eq:FR1} and \eqref{eq:FRacc} have to be written in terms of dimensionless quantities,
\bea
  1&=&\sum_i \Omega_i+\Omega_\Lambda+\Omega_k,\\
  q&=&\frac{1}{2}\sum_i\Omega_i\left(1+3\,w_i\right)-\Omega_\Lambda.
\eea
where $\Omega_i(t)=\left[2\kappa/3 H^2 (t)\right]\rho_i(t)\equiv\rho_i(t)/\rho_\rr{c}(t)$ with $\rho_i$ the energy density of the component $i$ and $\rho_\rr{c}$ the critical density, the label $i$ denoting baryonic matter, DM and radiation (photons and neutrinos); $\Omega_\Lambda\equiv\Lambda/(3\,H^2)$ is the density parameter corresponding to the cosmological constant term is the density parameters;  $\Omega_k\equiv-k/(a\,H)^2$ is the density parameter corresponding to the curvature term; and  $q=-\ddot{a}a/\dot{a}^2$ is the \textbf{deceleration parameter},

Best bounds on density parameters have been obtained by the Planck satellite probing the CMB \cite{Ade:2015xua} by analyzing the (relative) amplitudes and the positions of the acoustic peaks in the power spectrum of CMB temperature anisotropies. The best-fit of cosmological parameters yields
\be
\boxed{\Omega_{\rr{k},0}<0.005, \hspace{0.7cm} \Omega_{\Lambda,0}=0.68, \hspace{0.7cm} \Omega_{\rr{m},0}=0.05 \hspace{0.7cm} \text{and} \hspace{0.7cm} \Omega_{\rr{DM},0}=0.27,}
\ee
the radiation being negligible today. The \textbf{Hubble constant} is given by $H_0=67.80\pm0.77~\rr{km~Mpc^{-1}~s^{-1}}$ \cite{Ade:2015xua}. This is the so-called \textbf{$\Lambda-$CDM concordance picture} which is in agreement with all the current observations today (see e.g. \cite{Kowalski:2008ez, Ade:2015rim}). 

Other observations than the CMB ones enable one to probe the cosmological parameters, among them large galaxy surveys like SDSS-II, BOSS and 6dFGS, which probe the formation of large scale structure (LSS), i.e. groups or filaments of galaxy clusters, during the matter era, by measuring the matter power spectrum. Baryons leave imprints on the matter power spectrum because of their interactions to photons in the early Universe. 
Before recombination, photons and baryons formed a single fluid because of the Thomson and Coulomb interactions. At the decoupling, photons freely move across spacetime while baryons remained at rest and were attracted by DM gravitational potential wells (see e.g. \cite{Hu} for details). The distance traveled by the baryons-photons sound-waves from Big Bang to the last scattering surface, of about 150~Mpc, is still observable in the matter power spectrum, through the so-called \textbf{baryon acoustic oscillations} (BAO) \cite{Eisenstein:2005su}. BAO is thus a standard ruler enabling to infer cosmological parameters at different redshifts thanks to galaxy surveys, independently of other observations \cite{Anderson:2013zyy}.

\begin{figure}[!tbp]
  \centering
  \subfloat[The present DE fraction $\Omega_\rr{\Lambda}$ vs. matter fraction $\Omega_\rr{m}$ assuming the $\Lambda-$CDM model. Predictions for a flat Universe is denoted by the diagonal dashed line and is strongly favored by CMB+BAO data.]{\includegraphics[width=0.37\textwidth]{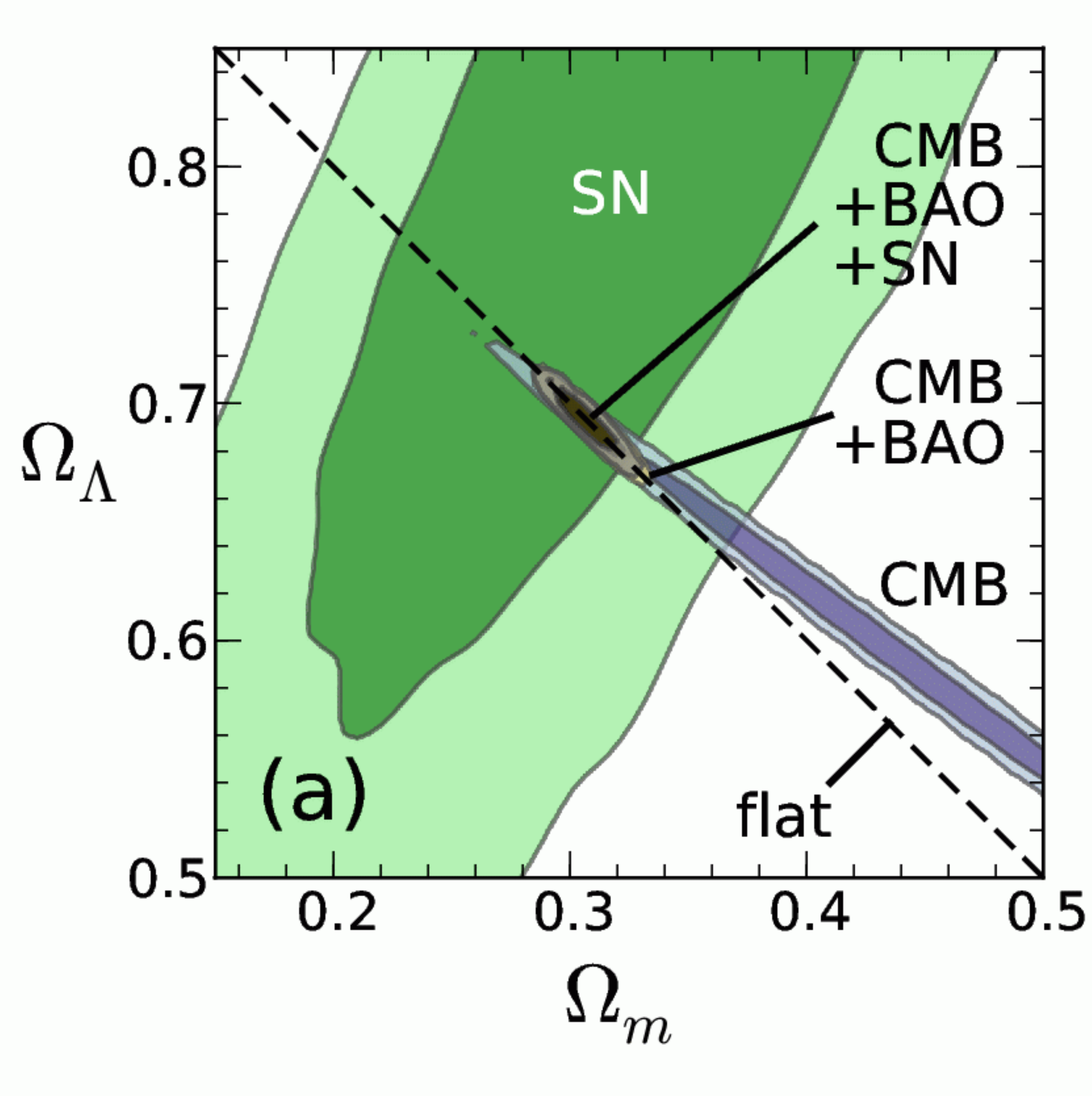}\label{fig:OmL_vs_OmM}}
  \hfill
  \subfloat[The DE EoS $w$ vs. $\Omega_\rr{m}$, assuming a constant value for $w$. The dashed contours show the $1~\sigma$ and $2~\sigma$ C.L. regions for the combination of WMAP and BAO data.]{\includegraphics[width=0.37\textwidth]{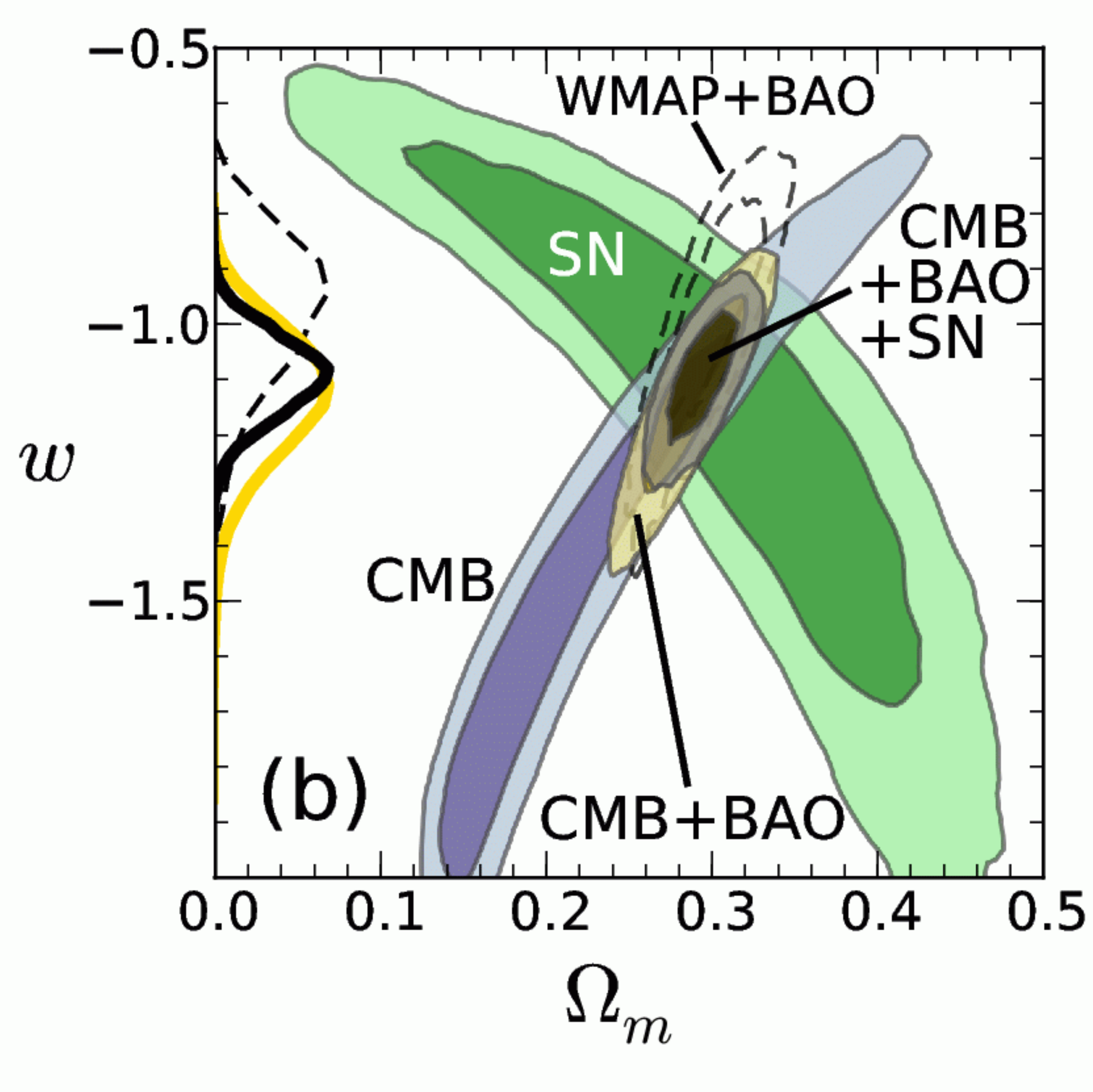}\label{fig:w_vs_OmM}}
  \hfill
  \subfloat[Constraint on the two parameters of the DE model with a time-dependent EoS, for $w(z=0.5)$.]{\includegraphics[width=0.37\textwidth]{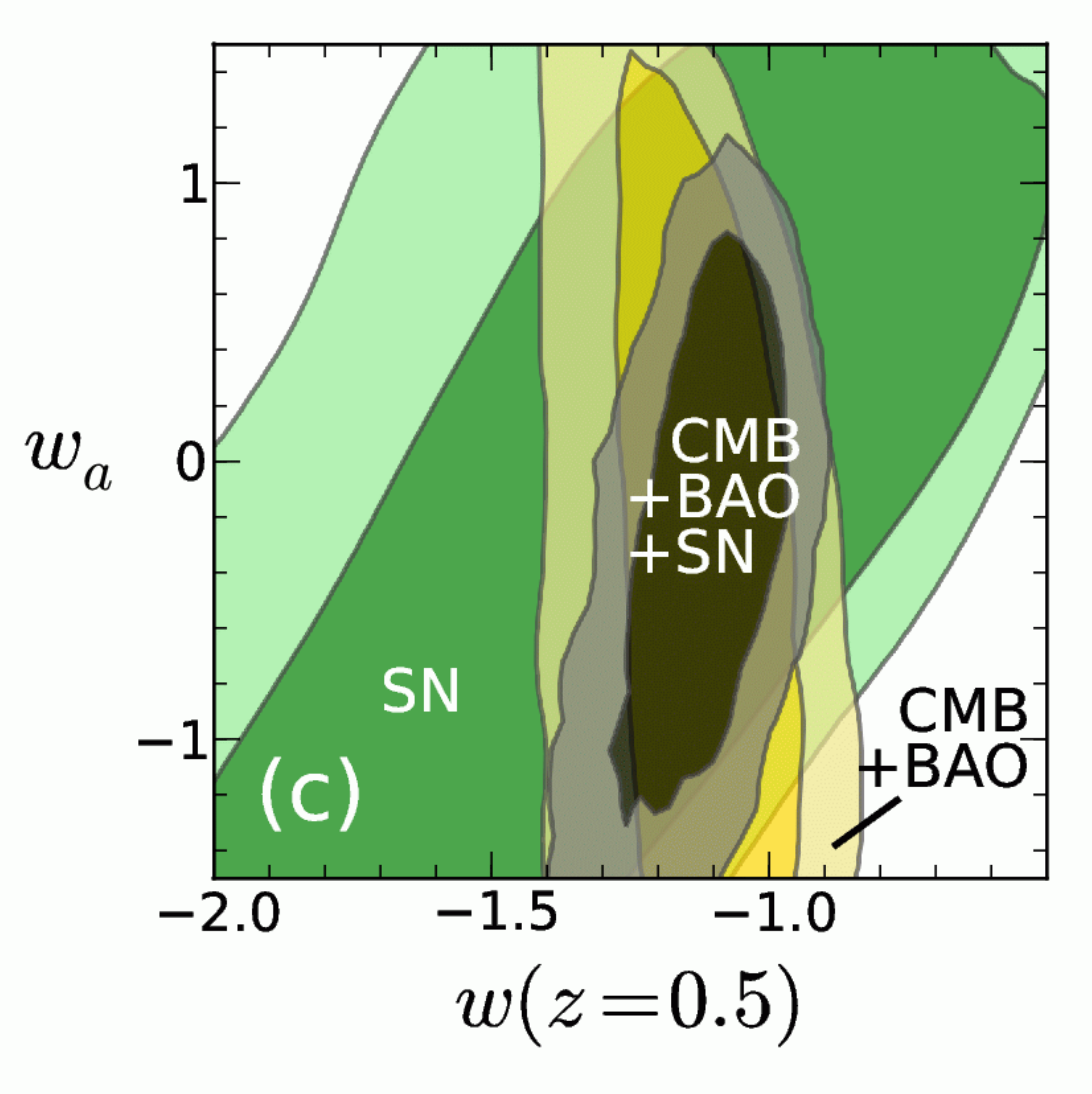}\label{fig:wa_vs_w}}
  \caption{\protect Constraints on the present matter fraction $\Omega_\rr{m}$ (DM+ baryonic matter) and the DE parameters ($\Omega_\rr{\Lambda}$, $w$ and $w_{a}= -\dd w/\dd a$). Dark and light shaded regions indicate $1~\sigma$ and $2~\sigma$ C.L. respectively. "CMB" is Planck+WMAP Polarization, "BAO" is the combination of SDSS-II, BOSS and 6dFGS, and "SN" is Union 2. Reprinted from \cite{Mortonson:2013zfa}.}
\end{figure}

In Fig.~\ref{fig:OmL_vs_OmM} the combined constraint on the cosmological parameters $\Omega_\rr{m,~0}$ and $\Omega_\rr{\Lambda,~0}$ obtained by the Wilkinson Microwave Anisotropy Probe (WMAP) and Planck satellites as well as galaxy surveys is represented, assuming the $\Lambda-$CDM model. The prediction of a flat Universe is strongly favored. The observations of SN Ia (see Sec.~\ref{sec:DE}) are also represented.  

In summary, according to our current understanding, the nature of around $95\%$ of the matter-energy content of the Universe is still an open debate: first the cold DM, i.e. non-relativistic matter ($w=0$) which does not interact with electromagnetic field, and second the DE which is responsible for the current cosmic acceleration. Today DE is compatible with the cosmological constant $\Lambda$ ($w=-1$).

% \begin{figure}
% \begin{center}
%   \includegraphics[scale=0.5]{./chapters/chapter2/}
%   \caption{\protect Reprinted from \cite{Mortonson:2013zfa}.}
%   \centering
% \label{fig:combined_cosmo}
% \end{center}
% \end{figure}

\subsection{The nature of Dark Matter}
In 1933, Zwicky suggested the existence of DM. He measured the velocity of galaxies inside the Coma cluster \cite{1933AcHPh...6..110Z} and compared the underlying total mass of the Coma cluster to the visible matter, concluding that there is a missing mass dubbed DM. Further strong evidences have been obtained in the 1970s by looking at the \textbf{galaxy rotation curves} \cite{1970ApJ...159..379R}. Typical velocity distribution does not correspond to visible disk velocity. The presence of a DM halo at the galactic scale must be invoked in order to fulfill the observations. 

The evidences for DM are now numerous, at the galactic and cosmological scales. According to our current understanding, temperature fluctuations in the CMB correspond to DM over- and under-densities, which acts as seeds for the formation of LSS hosting galaxies, from the redshift $z\approx20-30$\footnote{Rigorously, galaxies are biased tracer of DM.}.  Without DM, the formation of LSS would be not efficient since primordial fluctuations would vanish because of the diffusion damping. At the scale of clusters of galaxy, gravitational lensing reveals that DM has to be invoked for reconstructing the gravitational potential well \cite{Markevitch:2003at, Clowe:2006eq, Massey:2007wb}. 

%\tcb{ As for the CMB observations, the statistical analysis of the matter correlation function and the corresponding power spectrum are a powerful tool today for probing late-time Universe. Linear perturbations, considering in first approximation only DM, is a predictive theory up to small scales, where non-linearities become important. For instance, $\Omega_\Lambda$ can be traced thanks to galaxy surveys since it is responsible for a shift of the overall power spectrum: because of the accelerating expansion of the background Universe, the accretion of matter tends to slow down.}

The $\Lambda-$CDM model does not tell anything about the nature of the DM, the observations only revealing that DM has to be (predominantly) cold or non-relativistic. Strictly speaking, we observe evidences for an extra hidden mass, which is not observed at any wavelengths of electromagnetic signals. Although the evidences of DM come from astronomy and cosmology, most of experimental effort has now been shifted to particle and astroparticle physics. The DM could be a new particle beyond the SM. Several candidates exist, some of them related to issues appearing in the SM: WIMPs and gravitinos (related to supersymmetry), axions (also devoted to the strong CP problem), sterile neutrinos (possibly ruled out \cite{Schneider:2016uqi}), primordial BHs \cite{Carr01081974, 1975ApJ...201....1C}, their predicted mass range being in agreement with BHs detected by LIGO \cite{Bird:2016dcv, Clesse:2016vqa}\footnote{Primordial BHs in this mass range are possibly ruled out according to \cite{Ricotti:2007au}, but this result is still controversial.}, etc. No evidence for new particles has been obtained so far, either by direct or indirect detection. In the presence of a DM signal, one should still justify the observed abundance of DM on cosmological scales.

Alternative models have been proposed, some of them in the framework of modified gravity. For instance, the MOdified Newton Dynamics (MOND) theory \cite{1983ApJ...270..365M, Famaey:2011kh} or its covariant version TeVeS \cite{Bekenstein:2004ne}, are able to reproduce the galaxy curves whereas they do not fit the CMB observations nor the gravitational lensing ones by invoking baryonic matter only \cite{Angus:2006ev, Clowe:2006eq}. Others like the Chaplygin gas \cite{Bento:2002ps} where DM and DE are both described by a single fluid, are able to reproduce the growth of large-scale structure. 
%The major challenge for modified gravity models is to reproduce all the observations at once. 

\subsection{The current accelerated expansion} \label{sec:DE}
The first evidence for the acceleration of the spacetime expansion came from the observation of distant galaxies hosting SN Ia by Riess et al. \cite{Riess} and Perlmutter et al. \cite{Perlmutter:1998np}. They measured the luminosity distance $D_\rr{L}$ and the redshift $z$ defined as,
\be
  1+z\equiv \frac{\lambda_\rr{obs}}{\lambda_\rr{e}}=\frac{a_0}{a(t)},
\ee
with $\lambda_\rr{e}$, $\lambda_\rr{obs}$, the emitted and the observed wavelengths respectively, the last equality being valid for recessional redshift only, neglecting the local Doppler effects. The relation $D_\rr{L}$-$z$ enables one to probe the expansion up to $z\sim2$, highlighting the current expansion. 

% \textbf{attention: à la base le premier papier de l’équipe de Perlmutter (1998) concluait à un modèle sans Lambda mais ouvert ; les premiers à avoir déduit que $\Lambda\ne 0$ sont l’équipe de Riess, peu de temps après l’équipe de Perlmutter a corrigé ses data et conclut aussi à $\Lambda\ne 0$. C’est manifeste dans leurs papiers.}
%\textbf{independent observations, among them the CMB and the BAO, are all in favor of $\Lambda$  (see \cite{lrr-2001-1} for a review). }

However, the nature of the acceleration is still an open debate. The cosmological constant predicts that the EoS is constant $w=-1$, while more sophisticated models invoking a scalar field for instance, may predict variable EoS, generally parametrized up to first order by,
\be
  w(z)=w_{0}+w_a(1-a),
\ee
with $w_{0}=w(z=0)$ and $w_{a}=-\dd w/\dd a$. Current bounds on the plane $w_0-w_a$ obtained by combining CMB+BAO+SN data are reported in Figs.~\ref{fig:w_vs_OmM} and \ref{fig:wa_vs_w}. There are compatible with $w_0=-1$ and $w_a=0$.

Even if $\Lambda$ is able to reproduce current observations, it leads to a theoretical issue named the \textbf{cosmological constant problem} \cite{RevModPhys.61.1}. 
The cosmological constant has the same properties as the vacuum energy of quantum mechanics. Indeed, in order to preserve Lorentz invariance, $T^\rr{(vac)}_{\mu\nu}\propto g_{\mu\nu}$  (see Eq.~\eqref{eq:perfect_fluid}) \cite{weinberg2008cosmology} such that there exists no preferred direction. It results that $p^\rr{(vac)}=-\rho^\rr{(vac)}$ yielding,
\be
  T^\rr{(vac)}_{\mu\nu}= - \rho^\rr{(vac)} g_{\mu\nu},
\ee
where $\rho^\rr{(vac)}$ is constant (see Eq.~\eqref{eq:conservation_energy_cosmo}).
A cosmological constant enters in the Einstein equations \eqref{eq:Einstein} exactly in the same way as the contribution of the vacuum energy \cite{Durrer:2007re}, leading to,
\be
  \rho^\rr{(vac)}=-\frac{\Lambda}{\kappa}.
\ee

%In order to be compatible with the SM, the vacuum energy must be at least of the order of the electroweak scale \cite{peter2009primordial}, 
In general, the cut-off scale for GR is given by the Planck scale $E=\mpl$ such that \cite{Durrer:2007re},
\be
  \rho^\rr{(vac)}\sim\mpl^4\sim10^{76} \rr{GeV}^4,
\ee
while the observed value of the cosmological constant yields,
\be
  \rho_\Lambda\sim\left(10^{-12}\right)^4\sim10^{-48} \rr{GeV}^4.
\ee
It results that a cancellation of the vacuum energy of around 120 orders of magnitude is required for explaining the cosmological observations. This is the so-called \textbf{fine-tuning problem}. 

In addition the \textbf{Weinberg no-go theorem} \cite{RevModPhys.61.1} states on very general grounds like the Poincaré invariance at the level of spacetime curvature and fields, that no dynamical adjustment mechanisms could be used to solve the fine-tuning problem \cite{Padilla:2010tj}. 

The second issue arises when looking at the Universe history: the acceleration of the expansion started around $z\sim 0.6$ ($\rho_{\Lambda,~0}\simeq\rho_\rr{DM,~0}$), which is referred as \textbf{the coincidence problem}. In order to justify $\Lambda$, either anthropic or multiverse arguments have been invoked (see e.g. \cite{carr2007universe}). 

In order to face the coincidence issue, it has been proposed that the cosmic acceleration is due to a dynamical scalar field, possibly massive, rather than the cosmological constant. Since this scalar field has an exotic EoS, it has been dubbed Dark Energy. 
We focus here on one of the simplest models of DE, quintessence \cite{Amendola}. It invokes a scalar field $\phi$ minimally coupled to the metric, i.e. there is no modifications of the Einstein's theory, such that the isotropy is not broken,  
\be \label{eq:S_SF}
    \mathcal{L}=\frac{R}{2\kappa}-\frac{\mpl^2}{2}\left(\partial\varphi\right)^2+V(\varphi)+\mathcal{L}_\rr{M},
\ee
with $\phi=\mpl\,\varphi$ such that $\varphi$ is dimensionless, and $V(\varphi)$ the potential of runaway type. 
%the tracker potential, ensuring that the scalar field density follows that of the dominant matter component.
Assuming that the Universe is flat and dominated by DE $T_{\mu\nu}^{\rr{(M)}}\simeq0$, the Friedmann-Lemaître equations read,
\bea \label{eq:FR1_SF}
  H^2&=&\frac{\kappa}{3}\rho_\phi,\\ \label{eq:FR2_SF}
  \frac{\ddot{a}}{a}&=&-\frac{\kappa}{6}\rho_\phi(1+3 w_\phi),
\eea
where $w_\phi$ is the EoS of the scalar field,
\bea \label{eq:EoS_SF}
  w_\phi=\frac{\rho_\phi}{p_\phi} \hspace{1cm} \text{with} \hspace{1cm} \left\{
    \begin{array}{ll}
        \rho_\phi=\frac{1}{2}\dot{\phi}^2+V(\phi), \label{eq:rho_phi}\\
        p_\phi=\frac{1}{2}\dot{\phi}^2-V(\phi).
    \end{array}
\right.
\eea
while the Klein-Gordon equation,
\be \label{eq:KG_SF}
  \ddot{\phi}+3H\dot{\phi}+\frac{\dd V}{\dd \phi}=0.
\ee
The condition for the acceleration of the expansion is thus given by Eq.~\eqref{eq:FR2_SF}\footnote{We assume that $\rho_\phi>0$, so there is no violation of the weak energy condition $T_{\mu\nu} t^\mu t^\nu \geq 0$, $t^\mu$ being any timelike vector.},
\be \label{eq:condition_acc}
  \ddot{a}>0\hspace{0.5cm}\longleftrightarrow\hspace{0.5cm}w_\phi<-\frac{1}{3}.
\ee
In the case of the cosmological constant ($V(\phi)\propto \Lambda$ and no kinetic term), $w_\phi=-1$ and the solution of Eq.~\eqref{eq:FR1_SF} is de Sitter,
\be
  a(t)\propto \exp{(\sqrt{\Lambda} t)}.
\ee
Quintessence is able to solve the fine-tuning problem of the initial conditions, for instance if an attractor solution exists. Indeed, for any initial conditions of the scalar field in the early Universe, the scalar field converges to the path given by its attractor by rolling down its potential \cite{PhysRevLett.82.896}.
%provided that its effective mass $m_\phi^2\equiv V_{,\phi\phi}$ is sufficiently small, i.e. $|m_\phi|\lesssim 10^{-33}$~eV 
%since it is (almost) insensitive to the initial conditions. 
In addition, quintessence may not suffer from the coincidence problem since it calls on a dynamical mechanism provided that the energy scale of the scalar field today $m_\phi^2\equiv V_{,\phi\phi}$ is sufficiently small, i.e. $|m_\phi|\lesssim 10^{-33}$~eV \cite{Amendola}.
However, quintessence models rely on the existence of a potential whose parameters must be fixed, introducing additional parameters in the theory. 
%require a strong fine-tuning of the potential parameters in order to predict that the acceleration of the expansion started around $z\sim0.6$, the coincidence problem being thus not evaded. 

Scalar fields with a non-standard kinetic term like phantom ($w<-1$) and k-essence, are also able to reproduce the late-time acceleration, but phantom usually suffers from instability because it is a ghost (it has negative kinetic energy density such that its energy density grows with expansion, its quantum vacuum being possibly unstable, see e.g. \cite{Amendola}) while k-essence violates causality \cite{Durrer:2007re, Bonvin:2006vc}.

Infrared modifications of GR (see also Chap.~\ref{chap:MG}) could also be responsible for the acceleration appearing at large scales, for instance by introducing a scalar field nonminimally coupled to the metric (see Sec.~\ref{sec:STT} and \cite{Copeland:2006wr, Amendola} for reviews). 
%and the equations of motion becomes more complicated. 
The challenge in this latter case, is that those models must pass the local tests of gravity like the PPN parameters in the Solar system (see Sec.~\ref{sec:PPN}). In order to fulfill this requirement, modified gravity models invoke screening mechanisms (see Sec.~\ref{sec:screening}), like the chameleon model which nevertheless appears to be fine-tuned (see Chap.~\ref{chap:chameleon}). %Moreover modified gravity models must not suffer from instabilities (see Sec.~\ref{sec:MG_issues}).

DE and modified gravity models both predict dynamical EoS ($w_a\neq0$), a prediction which is constrained today using the combined CMB+BAO+SN data sets (see Fig.~\ref{fig:wa_vs_w}). Those bounds should be improved by future observations, in particular large galaxy surveys, for instance thanks to the Euclid satellite \cite{Amendola:2016saw}.

Note that, within the general relativistic framework, the current acceleration could be due to the backreaction, i.e. the effect of deviations from exact homogeneity and isotropy coming from the nonlinear growth of matter density perturbations, on the average expansion  (see \cite{Buchert:2011sx} for a review). Indeed, the timescales at which the cosmic acceleration and the structure formation started, are similar (around $10^{10}$ years) \cite{Buchert:2011sx}. If the non-perturbative effect of the backreaction is so large that it can explain the cosmic acceleration, it could solve the fine-tuning and coincidence problems at once. However, the effect has not been quantified yet in a fully realistic way.

\subsection{Fine-tuning of the initial conditions}
\label{sec:inflation}
Within the $\Lambda-$CDM concordance model, initial conditions in the early Universe appear to be fine-tuned given the current observations, raising the question of an underlying mechanism.
%Let's first introduce the \textbf{Hubble horizon}.
\begin{enumerate}
 \item \textit{\textbf{The horizon problem}}: Since temperature anisotropies of CMB are so tiny, it suggests that the different patches in the sky were in causal contact or inside the so-called Hubble radius $H^{-1}$ before the recombination, so that the thermalisation of the Universe is effective. However, it should not be the case if we assume that the $\Lambda-$CDM model is valid up to the Planck scale. Because of the finite speed of photons, the distance that the photons travel from the early Universe until recombination, corresponds to only 1 deg angular separation in the sky today. How can we explain the temperature isotropy in the CMB for regions much more separated in the sky today? 
 \item \textit{\textbf{The flatness problem}}: Why does the Universe appear to be so flat today ($\Omega_{k,\,0}<0.005$) although to do so, it has to be even flatter in the past? Initial conditions would be incredibly  fine-tuned to $\Omega_k=0$, for instance $\Omega_k<10^{-10}$ at the Big Bang nucleosynthesis (BBN)\footnote{The BBN is the stage in  the early Universe when temperature and density conditions were such that, during a brief epoch, nuclear reactions were effective in building complex light nuclei, D, $^3$He, $^4$He and $^7$Li (see e.g. \cite{Cyburt:2015mya}). By measuring the relic abundances of these nuclei, the physical conditions at BBN are predicted by GR and SM.}, which is in addition an unstable point between an open and a closed Universe.
 \item \textit{\textbf{Adiabatic (or curvature) primeval fluctuations}}: By definition, adiabatic density fluctuations are identical for each species (photons, baryons and DM) since there is no contribution of entropy ($S=0$). By opposition, the entropy (or isocurvature) ones are generated in case of entropy inhomogeneities, assuming vanishing spatial curvature (see e.g.~\cite{peter2009primordial}). Only adiabatic fluctuations have been detected so far \cite{Ade:2015lrj}, isocurvature ones being thus negligible in the early Universe. The mechanism generating the initial fluctuations, must thus (mostly) generate adiabatic perturbations.
 \item \textit{\textbf{The topological defects problem}}: The breaking of the gauge group of a Grand Unified Theory to the gauge group of the SM in the early Universe results in a serie of phase transitions induced by spontaneous symmetry breaking (see e.g. \cite{PhysRevLett.43.1365, PhysRevD.21.3295}). Such a process implies the formation of topological defects like monopoles. The annihilation rate of these topological defects is found to be very slow and since they are not observed, it means that a mechanism may be responsible for their disappearance \cite{1978PhLB...79..239Z}.
\end{enumerate}

Up to now, the most powerful mechanism solving all these problems at once, is the \textbf{primordial inflation}. This is an almost de Sitter phase in the early Universe, differing in DE since the predicted acceleration is much more larger and inflation requires a \textbf{graceful exit}, i.e. inflation must end.
The huge acceleration of the expansion in the early Universe explains at once why it appears locally flat today even if it was not the case before inflation and why non causally connected patches in the sky exhibit the same physical properties since they have been in causal contact before inflation. Moreover, the topological defects are diluted due to the huge expansion. The magnitude of the expansion is given by the \textbf{number of e-folds} $N$,
\be \label{eq:e_fold}
  N(t)=\ln\frac{a(t)}{a_\rr{i}},
\ee
with $a_\rr{i}$ the scale factor at the onset of inflation. $N(t_\rr{end})\gtrsim60-70$ solves at once the horizon and the flatness problems.

\begin{figure}
\begin{center}
  \includegraphics[scale=0.5]{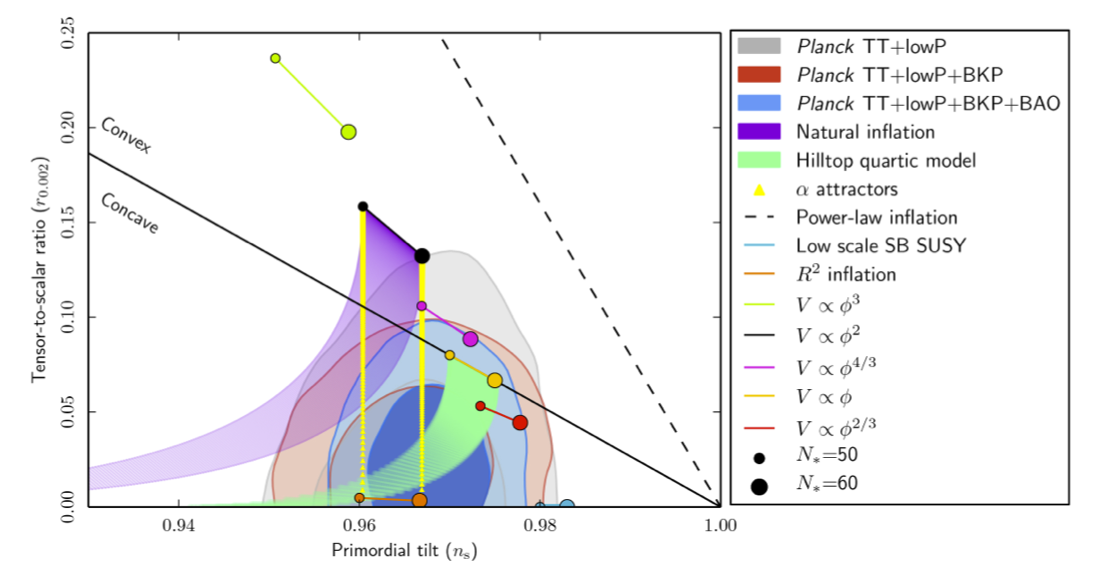}
  \caption{\protect Predictions of some inflationary models compared to the Planck observations in terms of $n_\rr{s}$ and $r$. Dark and light shaded regions indicate $1~\sigma$ and $2~\sigma$ C.L. respectively. Power-law inflation is strongly disfavored while Starobinsky inflation is favored, in the absence of tensor modes detection \cite{Ade:2015lrj}. Moreover, in the absence of non-gaussianities, one field inflation is most likely than multifields ones.}
  \centering
\label{fig:planck_inflation}
\end{center}
\end{figure}

The most simple inflationary models, first built by
%Starobinsky \cite{1980PhLB...91...99S}, 
Guth \cite{1981PhRvD..23..347G} and Linde \cite{1982PhLB..108..389L}\footnote{There are other precursory papers, among them \cite{BROUT197878, Sato01071981, 1980PhLB...91...99S}. In particular, Starobinsky proposed a model relying on a modifications of the EH action giving rise to an inflationary model which is still viable today (see also Fig.~\ref{fig:planck_inflation} and discussion in Chap.~\ref{chap:Higgs}).}, 
call on one scalar field, similarly to DE (see Eqs.\eqref{eq:S_SF} to \eqref{eq:KG_SF}). Assuming the first \textbf{slow-roll condition} (see Eqs.~\eqref{eq:condition_acc} and \eqref{eq:EoS_SF}), 
\be \label{eq:slow_roll1}
  \dot{\phi}^2\ll V(\phi),
\ee
it results that the scalar field starts to roll slowly down its potential and that the inflation naturally ends when it oscillates around its minimum. This latter phase is called the \textbf{reheating}. The inflationary phase also has to be sufficiently long, providing a second slow-roll condition,
\be \label{eq:slow_roll2}
  \left|\ddot{\phi}\right| \ll \left|\frac{\dd V}{\dd \phi}\right|.
\ee
Both slow-roll conditions \eqref{eq:slow_roll1} and \eqref{eq:slow_roll2} are usually quantified by the dimensionless \textbf{slow-roll parameters} using Eqs.~\eqref{eq:FR1_SF}, \eqref{eq:FR2_SF},~\eqref{eq:EoS_SF}, 
%depending on the potential, 
\bea
  \epsilon_\rr{V}\equiv\frac{\Mp^2}{2} \left(\frac{\partial V / \partial \varphi}{V}\right)^2 \ll1,\hspace{1.2cm}
  \left|\eta_\rr{V}\right|=\Mp^2\left|\frac{\partial^2 V / \partial \varphi^2}{V}\right| \ll 1,
\eea
with $\Mp=1/\sqrt{\kappa}$ the reduced Planck mass.
% or, equivalently, on the Hubble parameter,
% \bea
%   \epsilon_\rr{H}\equiv
%   \left|\eta_\rr{H}\right|\equiv
% \eea
Slow-roll parameters usually constrain background inflation as well as the CMB power spectrum, even if they appear to be restrictive in some cases \cite{Clesse:2011jt}.

Primordial scalar and metric fluctuations are generated in the early Universe, according to quantum mechanics (see e.g. \cite{durrer2008cosmic, mukhanov2005physical, peter2009primordial} for technical details). In particular, the scalar field fluctuates around its averaged value at very small scales. However, because of the huge expansion, microscopic regions are stretched so fast that they became larger than the size of the Hubble radius and are then frozen. As a consequence, there are scalar field fluctuations on super-horizon scales at the end of inflation. The scalar field decayed then into particles during reheating. Assuming one-field inflation, the adiabatic initial fluctuations are thus explained since all species fluctuations derive identically from the scalar field ones (provided constant branching ratios), 
\be
  \frac{\delta \rho_\phi}{\rho_\phi}\propto\frac{\delta \rho_\rr{f}}{\rho_\rr{f}},
\ee
with $\rho_\phi$, the density of the scalar field and $\rho_\rr{f}$, the density of each species labeled by $f$. On the contrary, multifield inflation could generate entropy modes.
% REMARK: either entropy+adiabatic modes, either SVT decomposition

Because of the quantum nature of the primordial fluctuations, statistical properties only can be derived from the distribution of the temperature fluctuations in the CMB and of matter fluctuations at smaller redshift. In the most simple case, where only one scalar field in slow-roll is assumed, the distribution of the two-point correlation function of $\delta\rho/\rho$ is Gaussian and the two point-correlation function describes all the statistical properties.

However, if \textbf{non-gaussianities} are detected, the three- and four-point correlation functions contain additional information. The corresponding parameters are the bispectrum $f^\rr{loc}_\rr{NL}$ and the trispectrum amplitude $g^\rr{loc}_\rr{NL}$, the current bounds reading \cite{Ade:2015ava},
\be
 f^\rr{loc}_\rr{NL}=0.8\pm 5.0 \hspace{1.5cm}  g^\rr{loc}_\rr{NL}=-9.0\pm 7.0\times 10^{-4}.
\ee
Non-gaussianities are thus negligible even if they could be produced not only during inflation but also during (p)reheating, by cosmic strings or astrophysical processes. It results that single-field inflationary models are favored.  

Statistical properties are usually derived in the Fourier space by computing the perturbations of the metric decomposed in scalar, vector and tensor perturbations.
The predicted curvature power spectrum ${P_\zeta}(k)$\footnote{The curvature $\zeta$ is used here for describing the scalar perturbations because it is gauge-invariant.} reads,
\be
  k^3{P_\zeta}(k)\equiv{A_\rr{s}}\left(\frac{k}{k_*}\right)^{n_s-1},
\ee
with $A_\rr{s}$ the amplitude of the scalar power spectrum measured at the pivot scale $k_*=0.05 \, \rr{Mpc}^{-1}$ and $n_s$ the \textbf{scalar spectral index} measuring the departure from scale-invariance. Planck $1~\sigma$ constraints today yield \cite{Ade:2015lrj},
\be
 \boxed{\ln\left(10^{10}~A_\rr{s}\right)=3.089 \pm0.036, \hspace{2cm} n_\rr{s}=0.9655\pm 0.0062.}
\ee
The fact that the $\mathcal{P_\zeta}(k)\equiv k^3 P_\zeta(k)/({2\pi^2})$ is \textbf{almost scale invariant} ($n_\rr{s}\simeq1$), has been probed by the large angular scales of the CMB temperature fluctuations, that is the \textbf{Sachs-Wolfe plateau} since, on super-Hubble scales, perturbations are almost constant (up to the Integrated Sachs-Wolfe effect, i.e. how the presence of evolving gravitational potential wells affect the temperature of the CMB photons along their line-of-sight). Indeed, the fluctuations which were super-Hubble at the last scattering surface, were frozen and thus initial conditions are directly probed today by measuring the tilt of $\mathcal{P_\zeta}(k)$.

Perturbations of the metric are not only scalar. Vector perturbations are negligible since they decrease during the expansion, except if they are sourced (locally), for instance by a magnetic field. Tensor modes are related to the generation of primordial GWs during inflation. The \textbf{tensor-to-scalar ratio} $r$ \cite{Ade:2015lrj},
\be
r\equiv\frac{\mathcal{P}_{t}}{\mathcal{P}_\zeta},
\ee
with $\mathcal{P}_\rr{t}$ the power spectrum of tensor perturbations in the metric, has been constrained recently by the observation of the polarization modes of the CMB: only tensor perturbations generate B-modes while E-modes are generated by both tensor and scalar perturbations. The measure of $r$ enables to fix the energy scale of inflation. Up to now, the combined constraints coming from Planck satellite (operating in the range 30-353~GHz) as well as the ground-based telescopes Background Imaging of Cosmic Extragalactic Polarization 2 (BICEP2) and the Keck Array (operating at 150~GHz) give,
\be
 \boxed{r<0.10 \hspace{3cm} (95\% \text{C.L.}).}
\ee

Assuming the slow-roll conditions \eqref{eq:slow_roll1} and \eqref{eq:slow_roll2}, the analytical expression for the power spectrum at the pivot scale $k=k_*$ reads (see e.g. \cite{peter2009primordial}),
\bea 
  \mathcal{P_\zeta}(k_*)\simeq\frac{1}{\pi}\frac{H^2_*}{\Mp^2\epsilon_\rr{V,\,*}}.
  %\frac{1}{\pi}\frac{H^2}{\Mp^2\epsilon_\rr{V}}
  %\Bigg[1-2\left(2C+1\right)\epsilon_\rr{V}+2C\left(\epsilon_\rr{V}-\eta_\rr{V}\r%ight)
  %\Bigg. \non\\ &&\qquad\qquad\qquad\Bigg.
  %+2\left(\eta_\rr{V}-3\epsilon_\rr{V}\right)\ln\left(\frac{k}{aH}\right)\Bigg],\\
  %&\simeq& \frac{1}{\pi}\frac{H^2}{\Mp^2\epsilon_\rr{V}},
  \label{eq:power_spectrum_slow_roll}
\eea
%with $C=\gamma_\rr{E}+\ln 2-2$, $\gamma_\rr{E}\sim0.577$ being the Euler constant.
In the slow-roll conditions the parameters $n_\rr{s}$ and $r$ can be determined by expanding the primordial power spectra $\mathcal{P_\zeta}$ and $\mathcal{P_\rr{t}}$. To first order they read \cite{Liddle:1994dx},
\be \label{eq:inflation_param}
  n_\rr{s}=1-6\epsilon_\rr{V,*}+2\eta_\rr{V,*}, \hspace{1.7cm}
  r=16\epsilon_\rr{V,*},
\ee
where the asterisk denotes the pivot scale at which $n_\rr{s}$ and $r$ are evaluated.

Numerous inflationary models exist and constraints coming from Planck + BICEP2 + Keck Array \cite{Ade:2015tva} allow to rule out some of them. Their results are reported in Fig.~\ref{fig:planck_inflation}. Bayesian inference analysis has been provided using one of the most favored models as prior, the Higgs inflation \cite{Martin:2013tda, Martin:2013nzq} which is equivalent to the Starobinsky model (see also Chap.~\ref{chap:Higgs}). Future space and ground based missions like BICEP3, LiteBIRD \cite{Matsumura:2013aja}, COrE+ \cite{core} and PRISM \cite{Andre:2013afa} should enable one to constrain $r$ up to $10^{-3}$.

Viable alternatives to inflation exist too, among them ekyroptic scenarios \cite{Steinhardt:2001st}, string gas cosmology and matter bounces \cite{Brandenberger:2009jq, Brandenberger:2016vhg}. None of them solves all the problems exposed above at once, so inflation is generally considered as the best explanation today even if it remains an effective model valid up to high enough energy scale $E$, usually $E\simeq \mpl$ being assumed.

\section{Conclusion} \label{sec:CCL_chap2}

\begin{figure} 
  \begin{center}
    \includegraphics[width=0.7\textwidth]{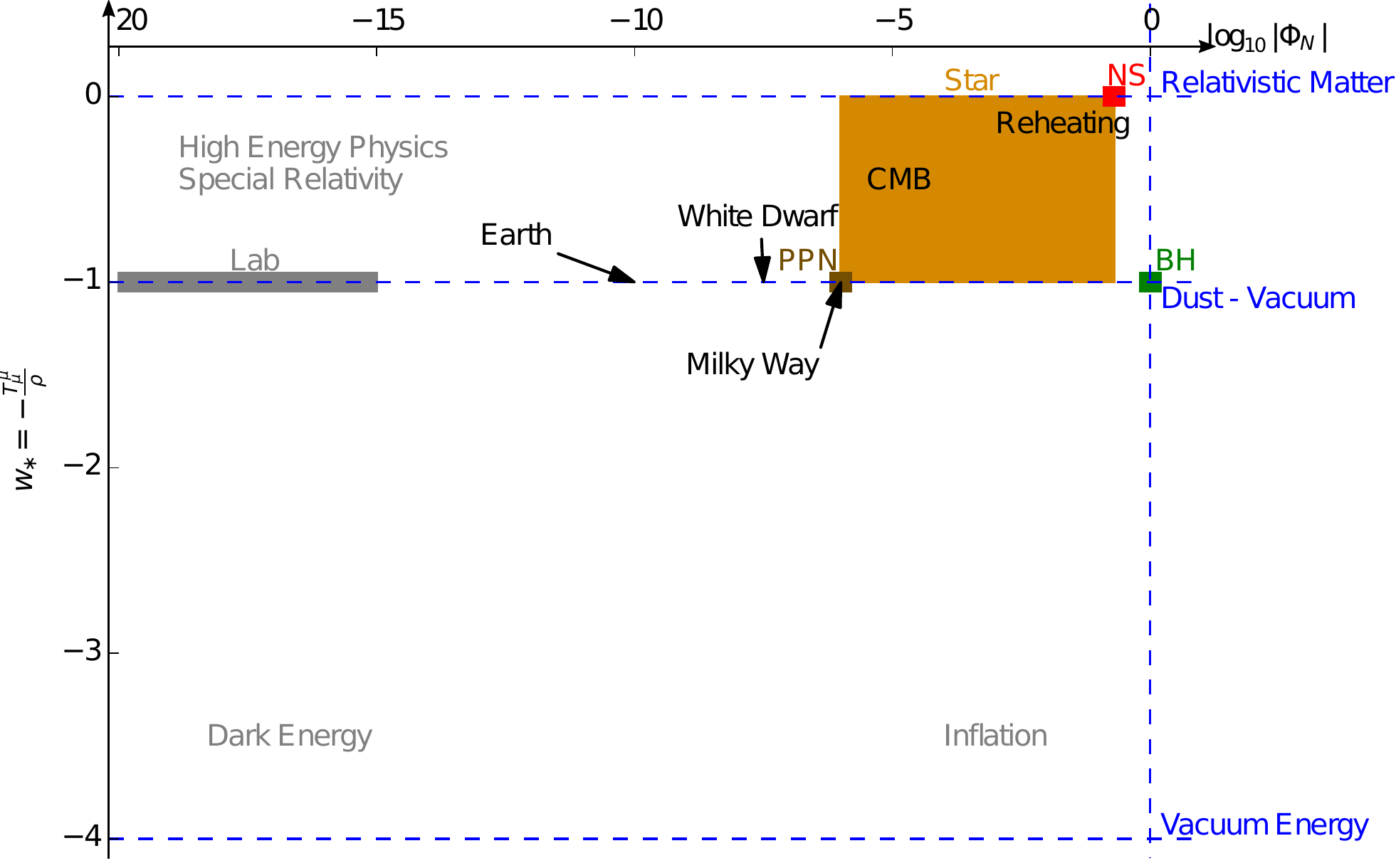}
    \caption{Classification of the tests of GR depending on the strength of the Newtonian potential as well as the nature of the sources given by $w_*=-{T^\mu_\mu}/{\rho}$. The best bounds on GR have been obtained in the vacuum, i.e. when the gravitational source has negligible pressure, while the modeling of the sources, for instance NSs and DE, is still under progress, the bounds being therefore less stringent.}
    \label{fig:chap2_recap}
    \centering
  \end{center}
\end{figure}

In this chapter, we discussed some tests of GR. In Fig.~\ref{fig:chap2_recap}, we propose to classify them depending on the strength of the gravitational field given by the Newtonian potential $|\Phi|$, as well as the nature of the gravitational source.

In spherically symmetric spacetime, the compactness \eqref{eq:compactness} provides a natural scale for the strength of the gravitational field. In cosmology, such a parameter cannot be defined. Nevertheless, Baker et al. \cite{Baker:2014zba} proposed a definition of the Newtonian potential $|\Phi|$ for cosmology. For that reason, the tests of GR in Fig.~\ref{fig:chap2_recap} are classified as a function of this latter parameter.

In order to define the nature of the source, we define the parameter $w_*$,
\be
  w_*=-\frac{T^\mu_\mu}{\rho}=3w-1.
\ee
assuming a perfect fluid \eqref{eq:perfect_fluid}. The parameter $w_*$ is equal to $0$ for relativistic sources like NSs, $-1$ in the absence of sources (or sources with negligible pressure), for instance BHs and the Sun, and $-4$ for vacuum energy sources like DE. Only the diagonal terms of $T_{\mu\nu}$, i.e. the mass-energy density $\rho$ and the isotropic pressure $p$, are taken into account in the definition of $w_*$. Relaxing the assumption of perfect fluid, there exist also phenomena invoking the off-diagonal terms, i.e. the momentum transfer and the shear stress. In particular, the classification of the tests proposed in Fig.~\ref{fig:chap2_recap} does not take into account gravitomagnetic effects, that is the contribution of moving and rotating material sources to the gravitational field, which are predicted in GR (see e.g.\cite{Will}). As an example, according to GR, the rotation of a massive body is dragging the local inertial frames of reference around it such that the orbits of moving objects around it are affected. This is the Lense-Thirring effect.

According to the classification of tests represented in Fig.~\ref{fig:chap2_recap}, best bounds on GR have been obtained in the vacuum. In the presence of sources, GR cannot be directly tested. In the case of NSs, there exists still an uncertainty about the EoS in the core of the stars while the nature of DE and the inflaton is still debated. 
%It results that modifications of gravity should not appear in the vacuum where GR seems to be tested with very good accuracy, but rather in the presence of sources.

In order to complete this chapter, we point out that some issues arising in the SM could also be related to cosmology. Why are the neutrinos massive? What is the origin of matter-anti matter asymmetry? and are the Higgs sector and the cosmological evolution connected? For instance, the existence of mass varying neutrinos could explain the late-time cosmic acceleration \cite{Fardon:2003eh}. In Chap.~\ref{chap:Higgs}, the possible relation between the Higgs field and gravitation will be further developed.

In Chap.~\ref{chap:MG}, the possibility to modify GR is further explored from the theoretical and phenomenological point of views. In the rest of this thesis, we will focus on the phenomenological predictions of some modifications of gravity, keeping in mind the classification of GR tests represented in Fig.~\ref{fig:chap2_recap}.

%\textbf{très bonne idée ça… les problèmes peuvent être reliés… je pense que tu dois étoffer ce paragraphe en ce sens (par exemple le problème de la masse des neutrinos a donné plusieurs modeles de DE dont les mass varying neutrinos)…}

%Since this thesis consists of apply combined constraints from different gravitational regime to modified gravity, Fig.~\ref{fig:chap2_recap} can be used as a guide for comparing gravitational regimes. Some of them appear to be close to each other while it does not seem so a priori.

%While we highlighted why GR is so special from a theoretical point of view in Chap.~\ref{chap:math_fundations}, this second chapter, we review the wide kinds of gravitational systems where GR predictions are in agreement with the current observations. 
%we showed that observations today are such that modifications of GR have to be tiny.

%\textbf{réflexion perso: en fait, si les problèmes de la cosmo sont imputables à la RG (modification de la RG) alors cela veut dire que les modifs sont dues au couplage à la matière (on n’est pas dans le vide) et éventuellement aussi à faible densité (sinon on aurait peut-être déjà vu qqch avec les tests étoiles à neutrons ). Dans le « vide », ça a l’air rapé (système solaire, etc.)}

\renewcommand{\chaptermark}[1]{\markboth{\small\textsc{Chapter \thechapter.\ #1}}{}}
\cleardoublepage

\chapter{Looking beyond General Relativity: Modified Gravity} % Main chapter title

\label{chap:MG} % For referencing the chapter elsewhere, use \ref{Chapter1} 

\lhead{Chapter 3. \emph{Awesome chapter 3}} % This is for the header on each page - perhaps a shortened title

%----------------------------------------------------------------------------------------

In Chap.~\ref{chap:math_fundations}, we introduce the underlying assumptions of GR which seem to indicate that GR has a privileged status. However, there are at least two motivations for studying theories of gravitation beyond GR. 
First,  GR is a classical theory which does not include quantum effects since it is not renormalizable. Second, in cosmology, the current cosmic acceleration is of unknown nature and could be due to modifications of gravity (see Sec.~\ref{sec:DE} for a discussion). In addition, the initial conditions of the Universe appear to be fine-tuned in the $\Lambda$-CDM concordance model. Primordial inflation is able to solve this problem and inflationary models may rely on modifications of gravity. 

\section{Beyond the Lovelock Theorem:\\ Modified Gravity} \label{sec:beyond_lovelock}
\subsection{Classification of Modified Gravity models}
The Lovelock theorem (see Sec.~\ref{sec:ccl_lovelock}) restricts rather drastically, the possibilities of building theories of gravity beyond GR. At least one of the assumptions of the theorem has to be broken:
\begin{itemize}
 \item \textit{\textbf{The number of dimensions (higher than 4)}}: This idea has been widely explored since the Kaluza-Klein theory,
 % developed in the 1920s
 giving rise to string and braneworld theories, as well as the Dvali-Gabadadze-Porrati model (DGP). Higher than four dimensions theories are devoted to the unification of the fundamental interactions (for instance, Kaluza-Klein and string theory), quantum gravity (string theory and braneworld) as well as phenomenological considerations like the late-time cosmic acceleration (DGP). When compactified, such theories generally exhibit additional degrees of freedom. As an example, the Kaluza-Klein theory is an attempt to unify gravitation and electromagnetism by generalizing GR to 5 dimensions \cite{Kaluza1921, 1926ZPhy...37..895K, 1926Natur.118..516K}. The metric in 5 dimensions $g^{(5)}_{AB}$ (with 15 independent components since it is symmetric) is decomposed in the 4-dimensional metric field $g_{\mu\nu}$, a vector field $A_\mu$ and a scalar $\varphi$ (see e.g. \cite{peter2009primordial}),
 \bea
  g^{(5)}_{AB}=\begin{pmatrix}
  g_{\mu\nu}+\rr{e}^{2\varphi} A_{\mu} A_{\nu} & \rr{e}^{2\varphi} A_{\mu} \\
  \rr{e}^{2\varphi} A_{\nu} & \rr{e}^{2\varphi}
 \end{pmatrix}.
 \eea
 When the fifth dimension is compactified, that is the cylinder condition $\df_y=0$ applies such that the fifth dimension is ignored, the corresponding action for the equations of motion in four dimensions, reads (see e.g. \cite{peter2009primordial} for the detailed calculations),
 \be
  S=\frac{1}{2\kappa}\int \dd^4 x \sqrt{-g} ~\rr{e}^{2\varphi} \left(R-\frac{\rr{e}^{2\varphi}}{4} F_{\mu\nu} F^{\mu\nu}\right),
 \ee
 with $F_{\mu\nu}=\df_\mu A_{\nu}-\df_\nu A_{\mu}$ the Faraday tensor. As a result, when the Kaluza-Klein theory is compactified, it reduces to a theory of gravitation in four dimensions where the metric has a scalar and a vector counterparts.
 Another example is Lovelock gravity \cite{Lovelock1971} which is an extension of the Lovelock theorem introduced in Sec.~\ref{sec:ccl_lovelock} to higher dimensions where the Gauss-Bonnet term is not trivial anymore.
 \item \textit{\textbf{Additional degree(s) of freedom (not only the spin-2 metric), whether it is scalar, vector or tensor field(s), dynamical or not}}: Many theories have been proposed with additional degrees of freedom, either by adding scalar, vector or tensor components, or by making the connexion dynamical, i.e. the assumption on the Levi-Civita connection is relaxed. Those additional degrees of freedom are justified by the compactification of the higher-dimensional theories of gravity like Kaluza-Klein and superstring theory \cite{Lidsey:1999mc}. Additional degrees of freedom also enable one to test phenomenological predictions, for instance, do GWs propagate at the speed of light?, and is the gravitational coupling $G$ constant in spacetime?   
 %Some of them suffers from instabilities like ghosts (massive gravity, DGP,...) (see also Sec.~\ref{sec:MG_issues})  
 Depending on the way they are formulated, theories with additional degrees of freedom imply violations of the WEP and/or the SEP. 
 Additional scalar fields (Horndeski gravity) lead to the LPI breaking (see Sec.~\ref{sec:varying_cst}), and possibly the UFF if the theory is formulated in such a way that the WEP is violated (see Sec.~\ref{sec:EP_revisited} for a discussion). Additional vector and tensor field(s) (TeVeS, Einstein-\AE{}ther, massive gravity,...) (usually) imply Lorentz-violation, breaking thus the LLI, in addition to the LPI. In some cases, general covariance may also be violated (see below). 
 \item \textit{\textbf{Equations of motion of higher than second order}}: We have already mentioned in Sec.~\ref{sec:Ostro} that equations of motion of higher than second order lead to the Ostrogradsky instability, excepted if the system is degenerate. Some modified gravity theories avoid this instability, for instance $f(R)$ theories \cite{DeFelice:2010aj, Sotiriou:2008rp},
 \be \label{eq:fR}
  S_{f(R)}=\int \dd^4 x \sqrt{-g} f(R) + S_\rr{M}\left[\psi_\rr{M};~g_{\mu\nu}\right],
 \ee
 which are found to be equivalent to scalar-tensor theory (STT) (see Sec.~\ref{sec:equiv_HI_andStaro} for an example).
 %\tcb{at least in Jordan frame. Through conformal transformation, you retrieve dynamical terms at the level of the action.}
 \item \textit{\textbf{Give up general covariance}}: As stated in Sec.~\ref{sec:gen_cova}, general covariance covers the diffeomorphism-invariance and the lack of prior geometry. Some modified gravity theories violate the diffeomorphism-invariance like massive gravity \cite{lrr-2014-7}, others require prior geometry, among them the Nordstr\o{}m and Rosen's bimetric theories. 
 %\item Give up the WEP
 \item \textit{\textbf{Give up Lorentz invariance and/or causality}}: Superluminal motion may be allowed either in Lorentz-violating theories at the action level, or by modifying the term responsible for the propagation velocity in the dispersion relation \cite{Bruneton:2006gf}. In some cases, the causality is violated. However, such models may rely on higher than second-order equations of motion and therefore may suffer from instabilities. Hence, the Cauchy equation could be not well-posed.
 In this case, there is no guarantee that either equations of motion admit a solution or this solution is unique. 
 \item \textit{\textbf{Give up locality}}: Some modified gravity models introducing terms like $f\left(R/\square\right)$ \cite{Woodard:2014iga} or $R (1/\square^2) R$ at the action level \cite{Maggiore:2014sia}, are non-local. Indeed, the operator $1/\square$ is the inversed d'Alembertian operator which is computed by the retarded Green's function (advanced Green's function are avoided in order to preserve causality). It results that non-local effects arise (see also Sec.~\ref{sec:locality}). Non-local models could explain the late-time cosmic acceleration \cite{Woodard:2014iga} and have been studied as effective theories for quantum gravity \cite{Hamber:2005dw}.
\end{itemize}
%Given the assumption(s) which are violated, the SEP and/or the WEP are violated.
According to the conjecture that we formulated at the end of Chap.~\ref{chap:math_fundations}, the SEP is violated in most of the cases introduced above. Higher dimension theories usually give rise to additional degrees of freedom when they are compactified, such that they violate the SEP. Non-local theories of gravity should avoid our conjecture. Depending on the way the modified gravity theories are formulated, violations of the WEP may arise (see Sec.~\ref{sec:EP_revisited} for the case of STT).

\begin{figure} 
  \begin{center}
    \includegraphics[width=0.95\textwidth]{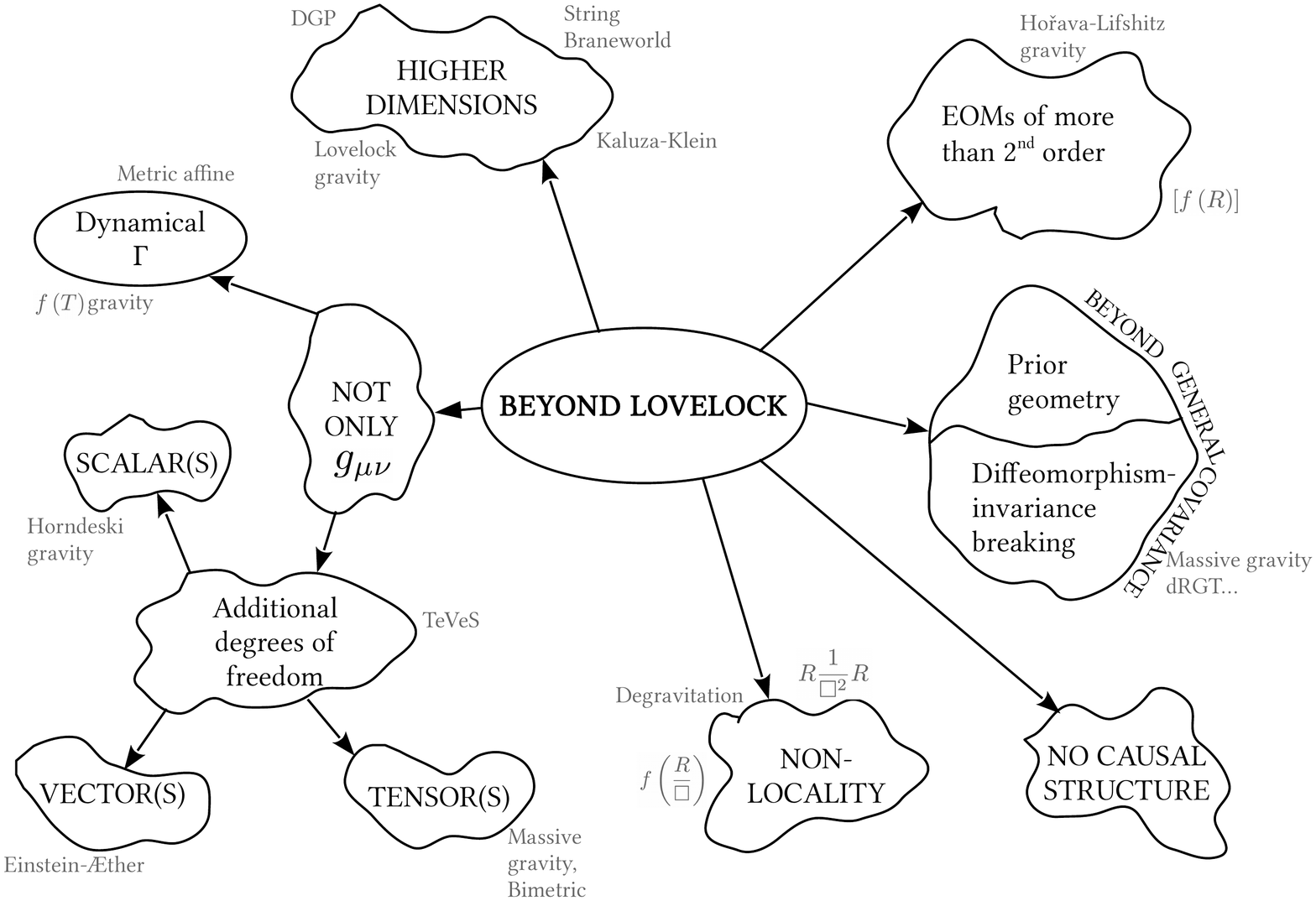}
    \caption{Axis of research of modified gravity, depending on the assumptions of the Lovelock theorem which are relaxed (inspired from T.~Baker 2013, E.~Berti \textit{et al.} \cite{Berti:2015itd} and the gravity apple tree proposed by M.~E.~Aldama \cite{Aldama:2015bxa}).}
    \label{fig:beyond_lovelock}
    \centering
  \end{center}
\end{figure}

In Fig.~\ref{fig:beyond_lovelock}, the directions for going beyond the Lovelock theorem are summarized. Some examples of the models corresponding to the violated assumptions of the Lovelock theorem are indicated.
A non-exhaustive list of modified gravity models and their characteristics is reported on Tab.~\ref{tab:MG_models}\footnote{This classification is inspired from T.~Baker \cite{bakerMG}, E. Berti \textit{et. al} \cite{Berti:2015itd} and the gravity apple tree proposed by M.~E.~Aldama \cite{Aldama:2015bxa}. 
}. Although all those models excepted Lovelock gravity predict a violation of the SEP, the WEP can be also violated depending the way the theory is formulated. 

The motivations for all these modified gravity models are different. The nature of late-time cosmic acceleration can be explained by DGP, Horndeski gravity and non-local theories for instance. MOND and TeVeS are phenomenological models explaining the matter galactic curves. Inflationary models are often built thanks to scalar field(s) with a large variety of potentials, using Horndeski models for example. We already point out here the large phenomenology of Horndeski gravity which encompasses the generalized STT, $f(R)$, the covariant Galileons, the Fab Four, K-mouflage, in order to cite only some of them. Finally, higher dimension models like string theory, braneworld scenarios, Lovelock gravity, etc. are attempts to unify the four interactions and to renormalize gravity. Quantum gravity has also been investigated in order to solve the problem of BHs and the Big Bang singularities (see Ho\v{r}ava-Lifshitz for instance).

%\textcolor{red}{Weinberg's theorem: the unique low energy field theory of spin-2 particles respecting Lorentz invariance is GR \cite{WeinbergT}.???}

%!!!!!!!!!!!Since the LPI is valid for all metric theory of gravity, a detection of $\alpha\neq 0$ would be in favor of non-metric theory \cite{Will1993} (see Chap.~\ref{chap:MG}). 

Before confronting the models with the observations, they have to be viable from the theoretical point of view, i.e. they must not suffer from instabilities like ghosts, as further developed in the next section. 

\newcolumntype{C}{>{\centering\arraybackslash\let\newline}m{4em}}
%\newcolumntype{Q}{>{\centering\arraybackslash\let\newline}m{1.5em}}
\newcolumntype{Q}{>{\centering\arraybackslash\let\newline}p{2.5em}}
\newcolumntype{M}{>{\centering\arraybackslash\let\newline}m{2.8em}}

\begin{landscape}
\begin{table}[]
\centering
%\caption{My caption}
%\label{my-label}
\begin{tabular}{|C|Q|Q|Q|M|Q|Q|C|C|Q|C|}
\hline
\multicolumn{1}{|c|}{MG Theories}                    & \multicolumn{3}{c|}{Equivalence principles}                                     & Add. \newline dofs  & \multicolumn{2}{c|}{General covariance}                                   & \multicolumn{1}{c|}{Causality} & \multicolumn{1}{c|}{Locality} & \multicolumn{1}{c|}{Higher  dim.} & 2nd order \newline EOMs \\ \hline
\multicolumn{1}{|l|}{}                               & \multicolumn{1}{Q|}{UFF} & \multicolumn{1}{Q|}{LLI} & \multicolumn{1}{Q|}{LPI}  & \multicolumn{1}{l|}{}          & \multicolumn{1}{C|}{No prior geom.} & \multicolumn{1}{C|}{Diffeo.-invar.} & \multicolumn{1}{l|}{}          & \multicolumn{1}{l|}{}         & \multicolumn{1}{l|}{}            & \multicolumn{1}{l|}{}               \\ \hline
\multicolumn{1}{|l|}{GR}                             & \multicolumn{1}{c|}{\vv} & \multicolumn{1}{c|}{\vv} & \multicolumn{1}{c|}{\vv}  & \multicolumn{1}{c|}{\xxx}      & \multicolumn{1}{c|}{\vv}            & \multicolumn{1}{c|}{\vv}            & \multicolumn{1}{c|}{\vv}       & \multicolumn{1}{c|}{\vv}      & \multicolumn{1}{c|}{\xxx}        & \multicolumn{1}{c|}{\vv}            \\ \hline
\multicolumn{1}{|l|}{Horndeski gravity}              & \multicolumn{1}{l|}{}    & \multicolumn{1}{l|}{}    & \multicolumn{1}{l|}{}     & \multicolumn{1}{l|}{}          & \multicolumn{1}{l|}{}               & \multicolumn{1}{l|}{}               & \multicolumn{1}{l|}{}          & \multicolumn{1}{l|}{}         & \multicolumn{1}{l|}{}            & \multicolumn{1}{l|}{}               \\ 
\multicolumn{1}{|l|}{\hspace{0.2 cm}Brans-Dicke}   & \multicolumn{1}{c|}{\vv} & \multicolumn{1}{c|}{\vv} & \multicolumn{1}{c|}{\xxx} & \multicolumn{1}{c|}{S}         & \multicolumn{1}{c|}{\vv}            & \multicolumn{1}{c|}{\vv}            & \multicolumn{1}{c|}{\vv}       & \multicolumn{1}{c|}{\vv}      & \multicolumn{1}{c|}{\xxx}        & \multicolumn{1}{c|}{\vv}            \\ 
\multicolumn{1}{|l|}{\hspace{0.2 cm}Scalar tensor} & \multicolumn{1}{c|}{\vv/\xxx} & \multicolumn{1}{c|}{\vv} & \multicolumn{1}{c|}{\xxx} & \multicolumn{1}{c|}{S}         & \multicolumn{1}{c|}{\vv}            & \multicolumn{1}{c|}{\vv}            & \multicolumn{1}{c|}{\vv}       & \multicolumn{1}{c|}{\vv}      & \multicolumn{1}{c|}{\xxx}        & \multicolumn{1}{c|}{\vv}            \\ 
\multicolumn{1}{|l|}{\hspace{0.2 cm}Fab Four}      & \multicolumn{1}{c|}{\vv} & \multicolumn{1}{c|}{\vv} & \multicolumn{1}{c|}{\xxx} & \multicolumn{1}{c|}{S}         & \multicolumn{1}{c|}{\vv}            & \multicolumn{1}{c|}{\vv}            & \multicolumn{1}{c|}{\vv}       & \multicolumn{1}{c|}{\vv}      & \multicolumn{1}{c|}{\xxx}        & \multicolumn{1}{c|}{\vv}            \\ 
\multicolumn{1}{|l|}{\hspace{0.2 cm}$f(R)$}        & \multicolumn{1}{c|}{\vv} & \multicolumn{1}{c|}{\vv} & \multicolumn{1}{c|}{\xxx} & \multicolumn{1}{c|}{S}         & \multicolumn{1}{c|}{\vv}            & \multicolumn{1}{c|}{\vv}            & \multicolumn{1}{c|}{\vv}       & \multicolumn{1}{c|}{\vv}      & \multicolumn{1}{c|}{\xxx}        & \multicolumn{1}{c|}{\vv}            \\ \hline
\multicolumn{1}{|l|}{Einstein-\AE{}ther}           & \multicolumn{1}{c|}{\vv}    & \multicolumn{1}{c|}{\xxx} & \multicolumn{1}{c|}{\xxx} & \multicolumn{1}{c|}{S/V}   & \multicolumn{1}{c|}{\vv}            & \multicolumn{1}{c|}{}               & \multicolumn{1}{c|}{\xxx ?} & \multicolumn{1}{c|}{\vv}  & \multicolumn{1}{c|}{\xxx}  & \multicolumn{1}{c|}{\xxx} \\ 
\multicolumn{1}{|l|}{Ho\v{r}ava-Lifshitz}          & \multicolumn{1}{c|}{\xxx ?} & \multicolumn{1}{c|}{\xxx} & \multicolumn{1}{c|}{\xxx} & \multicolumn{1}{c|}{V}     & \multicolumn{1}{c|}{\vv}            & \multicolumn{1}{c|}{\xxx}           & \multicolumn{1}{c|}{\xxx}   & \multicolumn{1}{c|}{\vv}  & \multicolumn{1}{c|}{\vv} & \multicolumn{1}{c|}{\vv}  \\ \hline
\multicolumn{1}{|l|}{Massive gravity}                & \multicolumn{1}{c|}{\vv}    & \multicolumn{1}{c|}{\xxx} & \multicolumn{1}{c|}{\xxx} & \multicolumn{1}{c|}{S/V/T} & \multicolumn{1}{c|}{\vv}            & \multicolumn{1}{c|}{\xxx ?}         & \multicolumn{1}{c|}{}       & \multicolumn{1}{c|}{}     & \multicolumn{1}{c|}{}     & \multicolumn{1}{c|}{}     \\ 
\multicolumn{1}{|l|}{Bigravity}                      & \multicolumn{1}{c|}{}       & \multicolumn{1}{c|}{}     & \multicolumn{1}{c|}{}     & \multicolumn{1}{c|}{T}     & \multicolumn{1}{c|}{\xxx ?}         & \multicolumn{1}{c|}{\vv}            & \multicolumn{1}{c|}{}       & \multicolumn{1}{c|}{\vv}  & \multicolumn{1}{c|}{\xxx}  & \multicolumn{1}{c|}{}     \\ \hline
\multicolumn{1}{|l|}{TeVeS}                          & \multicolumn{1}{c|}{\vv}    & \multicolumn{1}{c|}{\xxx} & \multicolumn{1}{c|}{\xxx} & \multicolumn{1}{c|}{S/V/T} & \multicolumn{1}{c|}{}               & \multicolumn{1}{c|}{}               & \multicolumn{1}{c|}{}       & \multicolumn{1}{c|}{\vv}  & \multicolumn{1}{c|}{}     & \multicolumn{1}{c|}{}     \\ \hline
\multicolumn{1}{|l|}{$f(R/\square)$}                 & \multicolumn{1}{c|}{}       & \multicolumn{1}{c|}{}     & \multicolumn{1}{c|}{}     & \multicolumn{1}{c|}{}      & \multicolumn{1}{c|}{}               & \multicolumn{1}{c|}{}               & \multicolumn{1}{c|}{\vv}    & \multicolumn{1}{c|}{\xxx} & \multicolumn{1}{c|}{}     & \multicolumn{1}{c|}{}     \\ \hline
%\multicolumn{1}{|l|}{$R (1/\square) R$}              & \multicolumn{1}{c|}{}       & \multicolumn{1}{c|}{}     & \multicolumn{1}{c|}{}     & \multicolumn{1}{c|}{}      & \multicolumn{1}{c|}{}               & \multicolumn{1}{c|}{}               & \multicolumn{1}{c|}{\vv}    & \multicolumn{1}{c|}{\xxx} & \multicolumn{1}{c|}{}     & \multicolumn{1}{c|}{}     \\ \hline
\multicolumn{1}{|l|}{Kaluza-Klein}                            & \multicolumn{1}{c|}{}       & \multicolumn{1}{c|}{}     & \multicolumn{1}{c|}{}     & \multicolumn{1}{c|}{}      & \multicolumn{1}{c|}{}               & \multicolumn{1}{c|}{}               & \multicolumn{1}{c|}{\vv}    & \multicolumn{1}{c|}{\vv}  & \multicolumn{1}{c|}{\vv} & \multicolumn{1}{c|}{\vv}  \\ 
\multicolumn{1}{|l|}{DGP}                            & \multicolumn{1}{c|}{}       & \multicolumn{1}{c|}{}     & \multicolumn{1}{c|}{}     & \multicolumn{1}{c|}{}      & \multicolumn{1}{c|}{}               & \multicolumn{1}{c|}{}               & \multicolumn{1}{c|}{\vv}    & \multicolumn{1}{c|}{\vv}  & \multicolumn{1}{c|}{\vv} & \multicolumn{1}{c|}{\vv}  \\ 
\multicolumn{1}{|l|}{Lovelock gravity}               & \multicolumn{1}{c|}{\vv}    & \multicolumn{1}{c|}{\vv}  & \multicolumn{1}{c|}{\vv}  & \multicolumn{1}{c|}{\xxx}  & \multicolumn{1}{c|}{\vv}            & \multicolumn{1}{c|}{\vv}            & \multicolumn{1}{c|}{\vv}    & \multicolumn{1}{c|}{\vv}  & \multicolumn{1}{c|}{\vv} & \multicolumn{1}{c|}{\vv}  \\ \hline
\multicolumn{1}{|l|}{$f(T)$}                         & \multicolumn{1}{c|}{\vv}    & \multicolumn{1}{c|}{\xxx} & \multicolumn{1}{c|}{\xxx} & \multicolumn{1}{c|}{}      & \multicolumn{1}{c|}{}               & \multicolumn{1}{c|}{}               & \multicolumn{1}{c|}{}       & \multicolumn{1}{c|}{}     & \multicolumn{1}{c|}{}     & \multicolumn{1}{c|}{}     \\ 
\multicolumn{1}{|l|}{Metric affine}                                        & \multicolumn{1}{c|}{?}                           & \multicolumn{1}{c|}{?}                         & \multicolumn{1}{c|}{?}                         & \multicolumn{1}{c|}{$\Gamma$}                   & \multicolumn{1}{c|}{\vv}                                 & \multicolumn{1}{c|}{\vv}                                 & \multicolumn{1}{c|}{\vv}                         & \multicolumn{1}{c|}{\vv}                       & \multicolumn{1}{c|}{\xxx}                       & \multicolumn{1}{c|}{\xxx}                      \\ \hline
\end{tabular}
\caption{Comparison of a non-exhaustive list of modified gravity (MG) models depending on which assumption of the Lovelock theorem is violated as well as on the consequences in terms of equivalence principles (see also Chap.~\ref{chap:math_fundations}). Abbreviations: dof=degree of freedom, geom.=geometry, Diffeo. invar.= Diffeomorphism-invariance, S=scalar, V=vector, T=tensor, PS=pseudo-scalar.}
\label{tab:MG_models}
\end{table}
\end{landscape}

\subsection{Some issues and challenges of Modified Gravity models} \label{sec:MG_issues}
While building modified gravity models, the first question raised is whether the theory is well-posed or whether it suffers from some instabilities, like the Ostrogradsky one. As an example, let us consider an alternative to GR with an additional scalar field, $\mathcal{L}=\mathcal{L}(\phi, \partial_\mu\phi)$. The stability of the solution $\phi(x,~t)$ is established by computing perturbations $\delta\phi(x,~t)$ around the background solution $\phib (t)$, 
\be
  \delta\phi(x,~t)=\phi(x,~t)-\phib (t),
\ee
up to second order. If the perturbation modes are decaying, then the theory is stable. 
%The perturbations typically give,
% \be
%   \mathcal{L}^{(2)}\subset Z(\phib)\delta\dot{\phi}^2-V(\phib)\left|\nabla\delta\phi\right|^2-M^2(\phib)\delta\phi.
% \ee

A common instability appearing in modified gravity is the \textbf{ghost}, especially in theories which attempt to reproduce the late-time cosmic acceleration. In this case, the kinetic term has the "wrong" sign, that is the opposite of the canonical one. From the classical point of view, it means that its kinetic energy is increasing (instead of decreasing) when it climbs up its potential. It is generally accepted that one cannot make sense of such a theory, at least at the quantum level \cite{Durrer:2007re}\footnote{If gravitation is considered as a low energy effective theory, then it is not necessary to care about its quantization. However, in quantum gravity, the quantization of the theory must be well-defined.}. From the quantum field theory point of view, ghosts can carry out negative energy eigenvalues (if unitarity is imposed \cite{Clifton:2011jh}). If the ghost is coupled to conventional matter field, it generates instabilities because of the possible creation of ghost-non-ghost pairs  in the vacuum (see also Sec.~\ref{sec:Ostro}). Several ways to "exorcise" the ghost have been explored (see e.g. \cite{Clifton:2011jh}). Other instabilities exist like tachyons, where the perturbations of the degrees of freedom have negative effective mass $m^2(\phi)<0$, and the \textbf{Laplacian instability} where the perturbations propagate with a negative squared speed \cite{jose}. In all cases, the quantization of the theory is not well-defined and the energy functional is not bounded from below \cite{Durrer:2007re}.
%\tcb{voir ma remarque ci-dessous. Si la gravité n’est pas une interaction fondamentale mais une force effective, plus besoin de la quantifier. Tu dois toujours la cohabiter avec la mécanique quantique mais on s’en tape de sa quantification. La question de l’extension de la RG est alors reliée aux anomalies (DE DM inflation) uniquement. C’est un peu comme avec la constante cosmologique: faut-il résoudre à la fois le problème du fine-tuning et de la coincidence? }

Another pathology of some modified gravity models is the so-called  \textbf{strong coupling}.
When the coupling constants of gravitational fields to matter ones are too strong, there exists a scale $\Lambda$ where the perturbative quantum field theory approach breaks down for the Minkowski background. It results that the theory can be non renormalizable and, if $\Lambda$ is too low, the theory is ill-defined at large scale. For instance, this is problematic for DGP where $\Lambda\sim 1000~$km \cite{Clifton:2011jh}.
% It is crucial to combine constraints on modified gravity models at different scales, in order to see where they are fruitful and where they fail. Indeed, alternatives are often motivated by lack of explanation power of GR for a particular phenonmenon (could the problem of singularity vanish by writing a model which is regular at the origin? could be the flatness of the Universe explained by invoking a mechanism predicting that the Universe is (almost) flat?) without considering additional constraints coming from other gravitational processes.

%\tcb{Today there is no modified gravity model able to pass all the tests of gravity introduced in Chap.~\ref{chap:test_GR}, contrary to GR. The PPN parameters provide stringent constraints in the weak-field limit and in the vacuum. Future observations are also promising in order to discriminate between viable models and the others, and require the developpement of new tools, either to perform model-dependent calculations or to develop parametrizations like the post-Einsteinian parametrization for cosmology. The recent detection of astrophysical GWs should shed light on gravitation since GR and its alternatives do not predict the same polarisation modes which should be measured in the future by the network of ground based detectors of GWs (see Sec.~\ref{sec:GW}). Nevertheless, modified gravity enables to extend the phenomenology as well as the theoretical framework developed in GR. It results that GR success can be better understood.  }

\subsection{Screening mechanisms}\label{sec:screening}
In the case where modified gravity is devoted to the explanation of the observations in cosmology like the late-time cosmic acceleration (see Sec.~\ref{sec:DE}), they have to pass the stringent constraints of the weak-field regime, that is the PPN parameters in the Solar system (see Sec.~\ref{sec:PPN}) and the tests of the equivalence principles (see Sec.~\ref{sec:tests_EP}), in order to be viable.

Therefore a common feature of viable modified gravity models is the so-called screening mechanism, that is a mechanism suppressing the modified gravity effects in local environments. Let us consider the sketchy general action for modified gravity models with an extra scalar field\footnote{This action is written in the Einstein frame where there is no explicit coupling to the Ricci scalar (see Sec.~\ref{app:EF} for a formal definition).}, 
\be
\mathcal{L}\supset p(\phi, ~X)-\frac{m^2~(\phi)}{2}\phi^2-g W(\phi) T,
\ee
where $X=(1/2)\left(\df \phi\right)^2$, $p(\phi, ~X)$ is a nonminimal coupling function of the kinetic term, $m(\phi)$, the effective mass of the scalar field, $g$ the coupling between the scalar field and matter, $W(\phi)$ a free function of $\phi$, and $T$, the trace of $T_{\mu\nu}$. Three screening mechanisms exist, coming from different terms at the action level:
\begin{itemize}
 \item \textit{Non-linearities in the kinetic term $p(\phi,~X)$}: First invoked by Vainshtein in the framework of massive gravity \cite{Vainshtein:1972sx}, the so-called \textbf{Vainshtein mechanism} arises from the non-linearities of the extra degrees of freedom in the kinetic term (coming from higher derivative for instance). Because of the strong kinetic self-coupling, extra degrees of freedom may be hidden and may almost not propagate (see \cite{Babichev:2013usa} and references therein). Inside the so-called Vainshtein radius, GR is restored because of the non-linearities, while the linear solution is recovered at large scales, reproducing DE phenomenology. Several modified gravity models exhibit such mechanism, either in the framework of massive gravity (DGP, bigravity,~etc) or of STT (Galileons and k-essence), the latter case being referred to as the \textbf{k-mouflage} \cite{Babichev:2009ee}.
 %\tcb{n’y aurait-il pas un parallèle intéressant entre le mécanisme de Vainshtein et celui  de la liberté asymptotique des quarks?}
 \item \textit{Large effective mass $m(\phi)$}: In the case of the \textbf{chameleon} model \cite{KhouryWeltmanPRL,KhouryWeltmanPRD}, the effective potential $V_\rr{eff}$ is defined by,
 \be
  \square\phi\equiv\frac{\dd V_\rr{eff}}{\dd \phi},
 \ee
 and the corresponding effective mass $m(\phi)=\dd^2 V_\rr{eff}/\dd\phi^2$, depends on the density of the environment. In sparse environment, the scalar field has a low mass and is thus able to mediate long-range force, while it acquires a mass in dense environment. We will detail this mechanism in Chap.~\ref{chap:chameleon}. 
 \item \textit{Small coupling $g$}: In the symmetron model \cite{Hinterbichler, Hinterbichler2}, 
 %the strength of the matter coupling $g$ which depends on the effective potential minimum, vary with the environment. 
 the screening mechanism relies on the symmetry breaking of the effective potential. The coupling $g$ is related to the vacuum expectation value (vev) of the scalar field which varies with the environment. In sparse environment, the vev is vanishing and the scalar field acts as a cosmological constant, while the symmetry is restored when the environment is dense, resulting in a non-negligible coupling between the scalar field and matter. 
\end{itemize}
In general, screening mechanisms are studied in the (quasi-)static limit in a spherically symmetric spacetime with a massive object, the Sun for instance, at the center, assuming a Minkowski background.
% This is the reason why some of the modified gravity models exhibit a \textbf{screening mechanism}, like the chameleon (see Chap.~\ref{chap:chameleon}), the Vainshtein or the K-mouflage. In the case of the chameleon potential, the effective mass of the scalar field adapts with the environment: at large density, it is massive and at low ones, it is not. However, more involved screening mechanisms exist

\section{Scalar-tensor theories} \label{sec:STT}
In the rest of this thesis we will focus on theories where an additional scalar field is the counterpart of the Einstein metric for describing gravity. Among them, STT were first proposed by Jordan in 1955 \cite{Jordan1955}, and rediscovered independently by Brans and Dicke in 1961 \cite{Brans:1961sx}. Since then, they have been studied extensively. On the one hand, STT are one of the most simple extensions of GR in the sense that it invokes only one additional degree of freedom, possibly justified by theories of gravity in higher-dimensions (see the discussion of Sec.~\ref{sec:beyond_lovelock}). On the other hand, they have a rich phenomenology, in particular around compact objects and in cosmology. In the next section, the mathematical formulations of STT, in the so-called Jordan and Einstein frames, are introduced.

\subsection{The Jordan frame}\label{sec:eom_JF}
% In this section, the equations of motion for a general STT written in the Jordan frame are derived, assuming a standard kinetic term (see App.~\ref{sec:eom_fabfour} for an example of a non standard one), first in the absence of $S_M\left[\psi_\rr{M};g_{\mu\nu}\right]$. 
In the so-called Jordan frame, the nonminimal coupling between the scalar field and the Ricci scalar is explicit. In the absence of matter, the action reads,
\bea \label{eq:actionSTTwithoutMAT}
S_\rr{JF}=\int \dd^4 x \, \sqrt{-g} \, \left[ \frac{F(\phi)}{2\kappa}R - Z(\phi) 
\left(\partial\phi\right)^2  - V\left(\phi\right) \right], 
%+ S_\rr{M}\left[\psi_\rr{M};g_{\mu\nu}\right],
\eea
where $F\left(\phi\right)$ and $Z(\phi)$ are the nonminimal coupling functions\footnote{\label{footnote1}The function $Z(\phi)$ can be reabsorbed in a scalar field redefinition such that $Z(\phi)$=1. However, we keep 3 independent functions $F(\phi)$, $Z(\phi)$ and $V(\phi)$ here in order to reuse the definition of the equations of motion in the rest of the thesis.}, $R$ the scalar 
curvature, $V\left(\phi\right)$, a generic potential. 
The modified Einstein equations are then given by the variation of the action with respect to $g_{\mu\nu}$ (see Eqs.~\eqref{eq:variation_det-metric} and \eqref{eq:variation_scalR}), assuming vanishing boundary terms\footnote{In the presence of nontrivial topology in space, such contributions may be physically relevant for some symmetries, e.g. the supersymmetry, leading to 
quantization rules on some parameters at the quantum level \cite{Govaerts2008}.},
\bea \label{eq:BD_Einstein_withoutMAT}
\left(G_{\mu\nu}+g_{\mu\nu}\square-\nabla_{\mu}\nabla_\nu\right)F\left(\phi\right)=\non
\qquad\qquad\qquad\qquad\qquad\qquad\\ 
\kappa \left\lbrace %T_{\mu\nu}^{(\rr{M})} 
Z(\phi)\left[2~\partial_{\mu}\phi \, \partial_{\nu}\phi - g_{\mu\nu}\left(\partial\phi\right)^2 \right] 
-  g_{\mu\nu}  V\left(\phi\right)
 \right\rbrace,
\eea
while the Klein-Gordon equation derives from the variation of the scalar field,
\bea \label{eq:KGtensor}
2~Z(\phi)\square\phi+{\frac{\partial F}{\partial \phi}} \frac{R}{2\kappa}=
-\frac{\df Z}{\df \phi}\left(\partial\phi\right)^2+{\frac{\partial V}{\partial \phi}}.
\eea

Including the contribution of matter $S_M\left[\psi_\rr{M};g_{\mu\nu}\right]$,
\bea \label{eq:actionSTT}
S_\rr{JF}=\int \dd^4 x \, \sqrt{-g} \, \left[ \frac{F(\phi)}{2\kappa}R - Z(\phi) 
\left(\partial\phi\right)^2  - V\left(\phi\right) \right]
+ S_\rr{M}\left[\psi_\rr{M};g_{\mu\nu}\right],
\eea
the modified Einstein equations read,
\bea \label{eq:BD_Einstein}
\left(G_{\mu\nu}+g_{\mu\nu}\square-\nabla_{\mu}\nabla_\nu\right)F\left(\phi\right)=\non
\qquad\qquad\qquad\qquad\qquad\qquad\\ 
\kappa \left\lbrace T_{\mu\nu}^{(\rr{M})} 
+Z(\phi)\left[2~\partial_{\mu}\phi \, \partial_{\nu}\phi - g_{\mu\nu}\left(\partial\phi\right)^2 \right] 
-  g_{\mu\nu}  V\left(\phi\right)
 \right\rbrace,
\eea
with the stress-energy tensor defined by,
\bea
T^{\left(\rr{M}\right)}_{\mu\nu}=-\frac{2}{\sqrt{-g}}{\frac{\delta S_\rr{M}}{\delta 
g_{\mu\nu}}}.
\eea
The Klein-Gordon equation remains unchanged.

In the Jordan frame, the stress-energy tensor is conserved. Indeed, by computing $\nabla^{\mu}$[Eq.~\eqref{eq:BD_Einstein}] where $Z(\phi)=1$\textsuperscript{\ref{footnote1}} and using the second Bianchi identity, the left-hand side reads,
\bea
  G_{\mu\nu}\frac{\dd F}{\dd \phi}\nabla^{\mu}\phi+\nabla_{\nu}\square F -\square\nabla_{\nu} F
  &=&G_{\mu\nu}\frac{\dd F}{\dd \phi}\nabla^{\mu}\phi+\left[\nabla_\nu,~\nabla_\alpha\right]\nabla^\alpha F, \qquad\\
  &=&\left(R_{\mu\nu}-\frac{1}{2}R g_{\mu\nu}\right)\frac{\dd F}{\dd \phi}\nabla^{\mu}\phi \non\\
  &&\qquad\qquad-R_{\alpha\nu}\nabla^\alpha \phi \frac{\dd F}{\dd \phi},\\
  &=&-\frac{R}{2}\frac{\dd F}{\dd \phi} \nabla_{\nu}\phi.
  \label{eq:left}
\eea
Between the first and the second equality, we used the relation for the commutator $[\cdot,~\cdot]$ of two covariant derivatives (see e.g.\cite{Carroll:2004st}),
\be
  [\nabla_\mu,~\nabla_\nu]V^\rho=R^\rho_{\sigma\mu\nu}V^\sigma,
\ee
with $V^\rho$ a vector field ($[\nabla_\mu,~\nabla_\nu]\phi=0$), assuming a vanishing torsion. The right-hand side reads,
\bea
  &&\kappa\left[\nabla^{\mu} T^{\left(\rr{M}\right)}_{\mu\nu}+2\square\phi\nabla_{\nu}\phi+2\nabla^{\mu}\phi \nabla_{\mu}\nabla_{\nu}\phi - 2\nabla^{\alpha}\phi\nabla_{\nu}\nabla_\alpha\phi-
  \frac{\dd V}{\dd \phi}\nabla_{\nu}\phi\right]\non\\
  &&\hspace{3.5cm}=\kappa\left[\nabla^{\mu} T^{\left(\rr{M}\right)}_{\mu\nu}+\left(2\square\phi-\frac{\dd V}{\dd \phi}\right)\nabla_\nu\phi\right],\\
  &&\hspace{3.5cm}=\kappa\left[\nabla^{\mu} T^{\left(\rr{M}\right)}_{\mu\nu}-\frac{\dd F}{\dd \phi}\frac{R}{2\kappa}\nabla_\nu\phi\right], \qquad\label{eq:right}
\eea
using the Klein-Gordon equation \eqref{eq:KGtensor}. Comparing Eqs.~\eqref{eq:left} and \eqref{eq:right} we conclude,
\be
  \nabla^{\mu} T^{\left(\rr{M}\right)}_{\mu\nu}=0.
\ee
%The SEP is violated since the gravitational constant is varying (see also Sec.~\ref{sec:varying_cst}).

\subsection{The Einstein frame}\label{app:EF}
Starting from the action in the Jordan frame \eqref{eq:actionSTT} where the function $Z(\phi)$ has been absorbed into the kinetic term,
\bea \label{eq:actionSTT_EF}
S=\frac{1}{2\kappa}\int \dd^4 x \, \sqrt{-g} \, \left[ F(\phi)R - g^{\mu\nu}\partial_\mu\phi \partial_\nu\phi  - V\left(\phi\right) \right]
+ S_\rr{M}\left[\psi_\rr{M};g_{\mu\nu}\right],
\eea
it is possible to rewrite it in such a way that it looks like GR
by performing a conformal transformation,
\bea \label{eq:conformal_metric_g}
  g_{\mu\nu}\longrightarrow\tilde{g}_{\mu\nu}=\Omega^2(\phi)\,g_{\mu\nu}
  \hspace{1cm}\Rightarrow \hspace{1cm} \left\{
    \begin{array}{ll}
        \sqrt{-\tilde{g}}=\Omega^4\sqrt{-g} \\
        \tilde{g}^{\mu\nu}=\Omega^{-2}{g}^{\mu\nu},
    \end{array}
\right.
\label{eq:conformal_det_g}
\eea
with the conformal factor,
\be \label{eq:Om_F}
  \Omega^2(\phi)=F(\phi).
\ee
The conformal transformation of the Ricci scalar reads (see e.g. \cite{Carroll:2004st}),
\bea \label{eq:Ric_conformal_transfo}
  R=\Omega^2 \tilde{R}+ 6\, \tilde{g}^{\alpha\beta} \Omega \left(\tilde{\nabla}_{\alpha}\tilde{\nabla}_{\beta}\Omega\right)
  -12\, \tilde{g}^{\alpha\beta} \left(\tilde{\nabla}_{\alpha}\Omega\right) \left(\tilde{\nabla}_{\beta}\Omega\right).
\eea
The action \eqref{eq:actionSTT_EF} transforms then as,
\bea \label{eq:intermediaire}
  S_\rr{EF}=\frac{1}{2\kappa}\int \dd^4 \tilde{x} \, \sqrt{-\tilde{g}} \, 
  \left[\tilde{R}+6F^{-1/2} \tilde{g}^{\alpha\beta}\left(\tilde{\nabla}_\alpha\tilde{\nabla}_\beta F^{1/2}\right)\right. \qquad\qquad\non\\\left.
  -12 F^{-1} \tilde{g}^{\alpha\beta}\tilde{\nabla}_\alpha F^{1/2} \tilde{\nabla}_\beta F^{1/2}
  -F^{-1}\tilde{g}^{\alpha\beta} \df_\alpha \phi \df_\beta \phi - U 
  \right] \non\\
  + S_\rr{M}\left[\psi_\rr{M};{g}_{\mu\nu}=A^2\tilde{g}_{\mu\nu}\right],
\eea
with
\be \label{eq:tututile}
  U\equiv F^{-2} V(\phi) \hspace{1cm} \text{and} \hspace{1cm} A\equiv\Omega^{-1}\left(\phi\right).
\ee
By computing,
\bea
  \tilde{\nabla}_\alpha F^{1/2}&=&\frac{1}{2 F^{1/2}}\frac{\dd F}{\dd \phi} \tilde{\nabla}_\alpha \phi, \\
  \tilde{\nabla}_\alpha\tilde{\nabla}_\beta F^{1/2}&=&\frac{1}{2 F^{1/2}}\left[
  \tilde{\nabla}_\alpha\tilde{\nabla}_\beta \phi \frac{\dd F}{\dd \phi} 
  + \frac{\dd^2 F}{\dd \phi^2} \df_\alpha \phi \df_\beta \phi
  - \frac{1}{2 F} \left(\frac{\dd F}{\dd \phi}\right)^2 \df_\alpha \phi \df_\beta \phi
  \right], \non\\
\eea
and by integrating by parts,
\be
  \frac{3}{F}\frac{\dd F}{\dd \phi} \tilde{\square}\phi
  =\frac{3}{F}\left[\frac{1}{F} \left(\frac{\dd F}{\dd \phi}\right)^2-\frac{\dd^2 F}{\dd \phi^2}\right] 
  \left(\tilde{\df} \phi\right)^2,
\ee
the action \eqref{eq:intermediaire} is finally formulated in the Einstein frame,
\bea \label{eq:action_EF_gen}
S_\rr{EF}&=&\frac{1}{{2\kappa}}\int \dd^4 \tilde{x} \, \sqrt{-\tilde{g}} \, \left[\tilde{R} - 2\left(\tilde{\partial}\sigma\right)^2  - U\left(\sigma\right) \right]
\non\\&&\qquad\qquad\qquad\qquad\qquad
+ S_\rr{M}\left[\psi_\rr{M};{g}_{\mu\nu}=A^2(\sigma)\,\tilde{g}_{\mu\nu}\right],
\eea
where the tildes denote Einstein frame quantities and $\sigma$ is defined by,
\bea
  \left(\frac{\tilde{\df}\sigma}{\tilde{\df}\phi}\right)^2=
  \frac{3}{4}\left(\frac{\df \ln F}{\df \phi}\right)^2+\frac{1}{2 F}.
\eea

In the Einstein frame, the action \eqref{eq:action_EF_gen} looks like the EH one with a minimally coupled scalar field $\sigma$, the coupling of the scalar field to matter $A(\sigma)$ appearing in $S_\rr{M}$. The modified Einstein equations then read (see Eq.~\eqref{eq:BD_Einstein_withoutMAT}),
\bea \label{eq:einstein_EF_gen}
\tilde{G}_{\mu\nu}=
\kappa \tilde{T}_{\mu\nu}^{(\rr{M})}+ 2~\tilde{\partial}_{\mu}\sigma \, \tilde{\partial}_{\nu}\sigma -  \tilde{g}_{\mu\nu} 
\left[\left(\tilde{\partial}\sigma\right)^2 + \frac{U\left(\sigma\right)}{2}\right],
\eea
the stress-energy tensor in the Einstein frame being defined as,
\be
  \tilde{T}^{\left(\rr{M}\right)}_{\mu\nu}=-\frac{2}{\sqrt{-\tilde{g}}}{\frac{\delta S_\rr{M}}{\delta 
  \tilde{g}_{\mu\nu}}}.
\ee
The Klein-Gordon equation yields,
\bea \label{eq:KG_EF_gen}
\tilde{\square}\sigma=\frac{1}{4}\frac{\df U}{\df \sigma}-\frac{\kappa}{2}\alpha(\sigma)\,\tilde{T}^{\left(\rr{M}\right)},
\eea
with $\alpha$, the nonminimal coupling function,
\be
  \alpha(\sigma)=\frac{\dd \ln A}{\dd \sigma},
\ee
and $\tilde{T}^{\left(\rr{M}\right)}$, the trace of $\tilde{T}^{\left(\rr{M}\right)}_{\mu\nu}$. 

Because of the nonminimal coupling in $S_\rr{M}$, the stress-energy tensor in the Einstein frame $\tilde{T}^\rr{(M)}_{\mu\nu}\equiv A^4 {T}^\rr{(M)}_{\mu\nu}$ is not conserved.  Indeed, we can show it explicitly by applying the Bianchi identity to Eq.~\eqref{eq:einstein_EF_gen},
\bea
  \tilde{\nabla}^\mu \tilde{T}_{\mu\nu}^{(\rr{M})}
  &=&\frac{1}{\kappa}\Bigg[-2\left(\tilde{\square}\sigma \tilde{\df}_\nu \sigma + \tilde{\df}_\beta\sigma \tilde{\nabla}^\beta\tilde{\nabla}_\nu \sigma\right)
  \Bigg. \non\\&& \qquad\qquad \Bigg.
  +g_{\mu\nu}\left(2\tilde{\df}_\beta\sigma \tilde{\nabla}^{\mu}\tilde{\nabla}^{\beta}\sigma+\frac{1}{2}\frac{\df U}{\df \sigma}\tilde{\df}^\mu \sigma\right)\Bigg],\\
  &=&\frac{1}{\kappa}\left\{-2\left[\tilde{\df}_\nu\sigma \left(\frac{1}{4}\frac{\df U}{\df \sigma}-\frac{\kappa}{2}\alpha(\sigma)\,\tilde{T}^{(\rr{M})}\right)+ \tilde{\df}_\beta\sigma \tilde{\nabla}^\beta\tilde{\nabla}_\nu \sigma\right]\right. \non\\
  &&\qquad\qquad\left.
  +2\tilde{\df}_\beta\sigma \tilde{\nabla}_{\nu}\tilde{\nabla}^{\beta}\sigma+\frac{1}{2}\frac{\df U}{\df \sigma}\tilde{\df}_\nu \sigma\right\},\\
  &=&\alpha \tilde{T}^{(\rr{M})} \tilde{\df}_\nu\sigma,
\eea
using the Klein-Gordon equation \eqref{eq:KG_EF_gen} for the second equality and $[\nabla_\mu,~\nabla_\nu]\sigma=0$ for the third one.

\subsection{Discussion about the frames} \label{sec:frame_discussion}
In this section, we briefly discuss the equivalence between the two formulations of STT. As we will see, the phenomenology predicted by STT does not depend on the frame. This is a mere change of variables \cite{esposito}. 

In the Jordan frame, the effective gravitational coupling defined as,
\be \label{eq:def_G_EF}
  G_\rr{eff}\left(x^\mu\right)=\frac{G}{F\left(\phi\right)},
\ee
with $G=\mpl^{-2}$ the bare gravitational coupling that is the parameter appearing in the action, is varying in spacetime. On the other hand, matter is minimally coupled to gravity such that the definition of the lengths and times measured by rods and clocks, which are made of matter, does not depend on the scalar field \cite{esposito}. 
%This is only true for laboratory-sized objects with negligible binding energy. Indeed, in that case, the variation of $G_\rr{eff}$ affect the measures.
%In this sense, the interpretation of observable is intuitive in the Jordan frame.

In the Einstein frame, the kinetic terms for the metric and the scalar fields are separated whereas the matter is directly coupled to the scalar field via the coupling function $A(\phi)$. As a result, the units system has to be re-calibrated since the rods and clocks are made of matter \cite{larena}. On the other hand the effective coupling is not varying $\tilde{G}_\rr{eff}=G$.

In physics, observables are defined by dimensionless quantities. Indeed, such quantities do not depend on the spacetime coordinates $\{x^\mu\}$ nor on a unit system. Observables  are frame-invariant since diffeomorphism-invariance is preserved in STT. As an example, let us consider the inertial mass $m_\rr{i}$ and define the corresponding observable,
\be \label{eq:obs_mass}
  \frac{m_\rr{i}}{\mpl}=m_\rr{i}\sqrt{G_\rr{eff}}.
\ee
In order to compute $m_\rr{i}$, we consider the action for a point-particle starting with the Jordan frame,
\bea
  S_\rr{pp}&=&-\int m_\rr{i} \dd s, \\
  &=&-\int m_\rr{i} \sqrt{-g_{\mu\nu}\dd x^\mu \dd x^\nu},\\
  &=&-\int m_\rr{i} A(\phi) \sqrt{-\tilde{g}_{\mu\nu}\dd x^\mu \dd x^\nu},\\
  &=&-\int \underbrace{m_\rr{i} A(\phi)}_{\equiv \tilde{m}_\rr{i}(\phi)} \dd \tilde{s} \label{eq:def_mi_EF}
\eea
where we used Eq.~\eqref{eq:conformal_metric_g} with $A(\phi)=\Omega^{-1} (\phi)$, the tilde denoting quantities expressed in the Einstein frame.
As a result, the inertial mass measured in the Einstein frame $\tilde{m}_\rr{i}(\phi)$ is found to vary in spacetime, even for laboratory-size, non self-gravitating objects \cite{esposito}.
Therefore, the ratio of inertial mass in the Jordan and the Einstein frames is varying,
\be
  \frac{{m}_\rr{i}(\phi)}{\tilde{m}_\rr{i}(\phi)}=A^{-1}(\phi),
\ee
whereas the ratio of two inertial masses labeled by the subscript $1$ and $2$ does not depend on the frame,
\be
  \frac{\tilde{m}_\rr{i,1}(\phi)}{\tilde{m}_\rr{i,2}(\phi)}=\frac{A(\phi){m}_\rr{i,1}}{A(\phi){m}_\rr{i,2}}
  =\frac{{m}_\rr{i,1}}{{m}_\rr{i,2}},
\ee
provided that the matter fields are universally coupled to the scalar field, i.e. $A(\phi)$ does not depend on the matter species. This question is further discussed in Sec.~\ref{sec:EP_revisited}. 

Moreover, using Eqs.~\eqref{eq:def_G_EF} and \eqref{eq:def_mi_EF}, the measure of the observable $m_\rr{i} \sqrt{G_\rr{eff}}$ \eqref{eq:obs_mass} does not depend on the frame,
\be
  \frac{m_\rr{i} \sqrt{G_\rr{eff}}}{ \tilde{m}_\rr{i} \sqrt{\tilde{G}_\rr{eff}}}
  =\frac{m_\rr{i} \sqrt{\frac{G}{F(\phi)}}}{ {m}_\rr{i} A(\phi) \sqrt{{G}}}
  =\frac{m_\rr{i} A(\phi) \sqrt{{G}}}{ {m}_\rr{i} A(\phi) \sqrt{{G}}}
  =1,
\ee
since $F(\phi)=\Omega^2(\phi)=A^{-2}(\phi)$ (see Eqs.~\eqref{eq:Om_F} and \eqref{eq:tututile}). In conclusion, there is an equivalence between the variation of the inertial mass in the Einstein frame and the variation of $G_\rr{eff}$ in the Jordan frame.

The calculations of other observables have been widely discussed in the literature (see e.g. \cite{Flanagan:2004bz} and references therein), for instance in cosmology \cite{Hees, Chiba:2013mha}. It shows that observables, i.e. dimensionless quantities, like the redshift, are frame-invariant while dimensional ones like the Hubble parameter are not \cite{Chiba:2013mha}. 

In conclusion, the computation of observables gives the same result in both frames. In the Jordan frame, they are obtained as in GR but solving the equations of motion is trickier since the limit to GR is not obvious (in fact it is even singular, see Sec.~\ref{sec:modelSTT} for the example of the Brans-Dicke theory). In the Einstein frame, computation of observables requires to take into account the rescaling of the metric and of the units system, because of the direct coupling of the scalar field to matter $A(\phi)$. However, in the Einstein frame, the equations of motion are written as in GR.

\subsection{The equivalence principles}
In the STT model formulated by Brans and Dicke in 1961, the scalar field involved in addition to the metric, is related to the Newton's constant $\GN$. Brans and Dicke were motivated by the Mach's principle as stated in Sec.~\ref{sec:mach}\footnote{Brans and Dicke were also inspired from the Dirac's Large Number Hypothesis, $1/\GN\propto M/R$ with $M$ the mass of the Universe and $R$ the Hubble radius. If $M/R$ varies with time then $\GN$ does \cite{Brans:2008zz}.}. Indeed, according to the Mach principle, the inertial mass of an object is related to its acceleration with respect to the local distribution of matter in the Universe. The dimensionless mass ratio $m_\rr{i}\sqrt{G_\rr{eff}}$ (see Sec.~\ref{sec:frame_discussion}) should then depend upon the matter distribution in the Universe, considering a variation in spacetime of the inertial mass $m_\rr{i}$ or of the effective gravitational coupling $G_\rr{eff}$. This is the reason why Brans and Dicke questioned the constancy of the gravitational "constant" (or equivalently of the inertial masses) and assumed that it could be a function of the matter distribution of the Universe. In order to formulate this statement mathematically, the gravitational constant has to be promoted as a scalar field, such that the SEP is violated. In more sophisticated STT, the WEP can be also violated.

\subsubsection{The Brans-Dicke theory} \label{sec:modelSTT}
Brans and Dicke initially considered the action \cite{Brans:1961sx}, 
\bea \label{BD_action}
  S=\frac{1}{2\kappa}\int \dd^4 x \, \sqrt{-g} \, \left[ \Phi R - \frac{\omega}{\Phi} 
  \left(\partial\Phi\right)^2 \right] 
  + S_\rr{M}\left[\Psi_\rr{M};g_{\mu\nu}\right],
\eea
where $\Phi$ is the dimensionless scalar field and $\omega$, a constant parameter. 
%\tcb{si je me rappelle bien omega n’est pas constant au début dans le papier original, ils le posent très vite à une constante avec l’argument qui est en gros: lorsque c’est constant les effets non locaux sont aussi importants que les effets locaux… }

The equations of motion for the Brans-Dicke theory, that is the modified Einstein equations, the Klein-Gordon equation and the conservation of $T^{(\rr{M})}_{\mu\nu}$ are given by (see Sec.~\ref{sec:eom_JF} with $F(\Phi)=\Phi$, $Z(\Phi)=2\kappa\omega/\Phi$ and $V(\Phi)=0$ for the calculations),
\bea
  \Phi G_{\mu\nu}&=&{\kappa}T^{(\rr{M})}_{\mu\nu}+\nabla_{\mu}\nabla_{\nu}\Phi-g_{\mu\nu}\square\Phi \non\\
   &&\hspace{1cm}+\frac{\omega}{\Phi}\left[\nabla_{\mu}\Phi\nabla_{\nu}\Phi-\frac{1}{2}g_{\mu\nu}(\partial\Phi)^{2}\right], \\
   \square\Phi&=&\frac{\kappa}{(2\omega+3)} T^{(\rr{M})}, \label{eq:BD_KG} \\
  \nabla^\mu T^{(\rr{M})}_{\mu\nu}&=&0.
\eea
%In this so-called \textbf{Jordan frame}, the nonminimal coupling between the scalar field and the metric is explicit (\tcb{bof: ce couplage n’apparait pas dans le terme de matiere. Il est explicite au niveau du scalaire de courbure ok mais pas de la metrique…}) while $T_{\mu\nu}$ is conserved. In particular, 
GR is recovered by imposing $\omega\longrightarrow\infty$, which means that STT is indistinguishable from GR if $\omega$ becomes unnaturally large. Indeed, in the limit where $\omega\longrightarrow\infty$, the Klein-Gordon equation reads $\square \Phi=0$, i.e. the scalar field is no more coupled to matter and GR is recovered (see also Eq.~\eqref{geff} for a second proof that GR is recovered when $\omega\longrightarrow\infty$).
%\tcb{(attention que cette limite est subtile et ne saute pas aux yeux dans les EOM. Tu devrais étayer un peu ici.)}

\subsubsection{The Local Position Invariance II: Varying fundamental constants} \label{sec:varying_cst}
The LPI ensures that the measure of the observables does not depend on the spacetime position in GR. However, if the fundamental constants, either in GR or in SM, vary in spacetime, then measure of the observables would also vary in space and in time.

The fundamental constants of a physical theory are defined as any parameter that cannot be explained by this theory \cite{Uzan:2010pm}. On the contrary, the other constants can be expressed in terms of the fundamental ones. In Sec.~\ref{sec:LLI}, we have already mentioned that the constancy of the speed of light has been questioned within the framework of Lorentz violation. 
Actually the constancy of the fundamental constants is tightly linked to the violation of the equivalence principles: in the particular case of $\GN$ only the SEP is violated since it is related to the gravitational interaction only while for other constants like $c$, both SEP and WEP are in general violated \cite{Uzan:2010pm}. It has been widely explored along the last decades (see \cite{Uzan:2010pm} for a review) for several fundamental constants, in particular $\GN$, $c$ and the fine-structure constant for electromagnetism $\alpha_\rr{EM}$. 
% It is interesting to notice that the definition of the constant may depend on the theory: for instance, the mass of the electron is a constant in the framework of the classical mechanics while it is no more the case in the SM where the mass of the electron is given by the Higgs vev and the Yukawa coupling between the Higgs field and the electron. 

A fundamental constant has by definition no dynamical equation which predicts its evolution. In order to implement the variation of fundamental "constants" mathematically in a consistent way, fundamental "constants" have to be promoted as dynamical scalar fields at the level of the action such that equations of motion for the scalar field, can be derived from a variational principle. At the end of the day, some constants will depend on the value of the scalar fields and a set of new fundamental constants must be defined. 

In the case of Brans-Dicke theory \eqref{BD_action}, the effective gravitational "constant" $G_\rr{eff}$\footnote{In Chaps.~\ref{chap:chameleon}, \ref{chap:Higgs} and \ref{chap:FabFour}, we will use $\GN$ rather than $G$ for the bare gravitational constant. However, we must keep in mind that it is a misuse of language.} is,
\be
  G_\rr{eff}\left(x^\mu\right)=\frac{G}{\Phi\left(x^\mu\right)},
\ee
and its distribution in spacetime is determined by the Klein-Gordon equation \eqref{eq:BD_KG}. 
Both time and spatial variations of $G_\rr{eff}$ are then predicted, depending on the spacetime symmetry (for instance only variations in space are predicted in a static spacetime). 
%As an example, the spatial fluctuations of the scalar field could leave imprints on the CMB power spectrum. 
$G_\rr{eff}$ has been measured at different redshifts, constraining its variation in time and, for some cases like for the CMB last scattering surface, also in space. Among the phenomenons probing the variation of the gravitational effective coupling $G_\rr{eff}$, we point out the Solar system tests, the stellar physics and cosmology, particularly at the BBN and the CMB last scattering surface (see \cite{Uzan:2010pm} and references therein for experimental bounds).

However, we have to keep in mind that $G_\rr{eff}$ does not correspond to the gravitational constant as measured by torsion balance or Cavendish experiments which requires further computations (see e.g. \cite{Uzan:2010pm}). Indeed, the effective Newton's constant is computed by expanding the conformal factor $A(\phi)=F^{-1/2}(\phi)$ in the weak-field approximation. In App.~\ref{sec:PPN_BD}, the Newton's constant as measured by Cavendish experiment $G_\rr{Cav}$ for the Brans-Dicke theory is obtained \eqref{geff} and reads,
\bea
  G_\rr{Cav}= {2 G\over \Phi_{0}}{\om +2\over 2\om +3},
\eea
with $\Phi_0$ the asymptotic value of $\Phi$, far away from the gravitational source.

\subsubsection{The equivalence principles revisited}\label{sec:EP_revisited}
In the 1960s, Dicke noticed that only the WEP had been tested through experiments of the UFF involving weakly self-gravitating bodies. The case of strongly gravitating bodies like stars where the gravitational binding energy contributes largely to the total mass, had never been studied. The Brans-Dicke theory \eqref{BD_action} predicts violation of the SEP while the WEP is not violated.

In the Brans-Dicke theory, the scalar field $\Phi$ is nonminimally coupled to gravity in such a way that it introduces a variation of the effective gravitational "constant" only. Indeed, matter fields are universally coupled to the metric $g_{\mu\nu}$. It results that there exists a reference frame locally where the gravitational effects are vanishing for all matter species such that the SR laws applies. 
However, in the case of strongly gravitating objects, the gravitational binding energy which is affected by the variation of the gravitational strength, contributes largely to the inertial mass, the SEP being violated.

This variation of the gravitational mass $m_\rr{g}$ due to the scalar field was further formalized by Nordtvedt \cite{Will},
\be
 m_\rr{g}= m_\rr{i} +\eta \frac{E_\rr{b}}{c^2},
\ee
with $E_\rr{b}$ the Newtonian gravitational binding energy \cite{PhysRev.169.1014, PhysRev.169.1017, PhysRev.170.1186}. The best constraint has been obtained by the Lunar Laser Ranging experiment, $|\eta|<(4.4\pm4.5)\times 10^{-4}$ \cite{Williams:2005rv}.
The \textbf{Nordtvedt effect} affects also the weakly gravitating bodies, but it cannot be detected experimentally, the sensitivity of the experimental set-up being far too low.   

Considering now general STT formulated in the Einstein frame as introduced in Sec.~\ref{app:EF}, violations of the WEP and the SEP may arise depending on the way matter fields $\psi_\rr{M}$ are coupled to the gravitational fields in $\mathcal{L}_\rr{M}\left[\psi_\rr{M},~g^{\mu\nu},~\phi\right]$. In the case of a conformal coupling to matter $A(\phi)$ which is universal for all matter fields (see Eq.~\eqref{eq:action_EF_gen}), only the SEP is violated since the effect of the gravitational field is vanishing locally (see the discussion above about the Brans-Dicke theory). However, if the metric does not couple universally to matter fields, i.e. $g_{\mu\nu}=A^{(i)}(\phi)\tilde{g}_{\mu\nu}$, the superscript $(i)$ denoting different matter species, for instance photons and electrons, then the WEP is also violated. Indeed, it is not possible to find out a reference frame locally where the gravitational effects are vanishing for all the matter species at once.

One example of such STT is the Abnormally Weighting Hypothesis (AWE) \cite{Alimi:2008ee} where three different geometries are assumed, describing gravitation ($g_{\mu\nu}$), the matter of the SM ($A^2(\varphi)_\rr{M}g_{\mu\nu}$) and the matter-energy content in the dark sector ($A^2(\varphi)_\rr{AWE}~g_{\mu\nu}$),
\bea
  S=\frac{1}{2\kappa}\int \dd^4 x \sqrt{-g} \left(R-2~g^{\mu\nu}\df_\mu\varphi\df_\nu\varphi\right)
   \non\qquad\qquad\qquad\qquad\qquad\qquad\\
  +S_\rr{M}\left[\psi_\rr{M},~A_\rr{M}^2(\varphi)~g_{\mu\nu}\right]+S_\rr{AWE}\left[\psi_\rr{AWE},~A_\rr{AWE}^2(\varphi)~g_{\mu\nu}\right].
\eea

In general, the SEP is violated provided that the modified gravity model exhibits two different geometries in a single model, one of them describing gravitation and the other, the geometry in which matter plays out its dynamics \cite{Bekenstein:1992pj}. In order to predict a violation of the WEP, different geometries describe the matter dynamics like for AWE. The conformal transformation between two geometries is the easiest. However, more sophisticated transformations exist like the disformal one \cite{Bekenstein:1992pj},
\be
  g_{\mu\nu}=A(\phi,~X)\tilde{g}_{\mu\nu}+B(\phi,~X)\df_\mu\phi\,\df_\nu\phi,
\ee
where the disformal functions $A$ and $B$ depend on the scalar field $\phi$ and the metric $\tilde{g}^{\mu\nu}$ via the kinetic term $X=\tilde{g}^{\mu\nu} \df_{\mu} \phi \df_{\nu} \phi$. The disformal transformation between two geometries may lead to violations of the SEP or the WEP depending on the functions $A$ and $B$.

%\tcb{attention, c’est vrai tant que tu as un couplage conforme du champ scalaire à la métrique. Si ce n’est plus le cas, tu violes aussi le WEP… Un autre cas est la théorie AWE qui est une théorie bi-métrique toutes deux conformément reliées. Kaluza-Klein est un autre exemple…}
%We could think that STT violate the general covariance by invoking a prior geometry. Actually, it is not the case since the field equations for the metric involves the scalar field, and vice versa \cite{Will}.

In summary, GR satisfies both the WEP and the SEP, and thus predicts no violation of the UFF, the LPI nor the LLI. In STT, the SEP is violated whereas the WEP can be violated, depending on the way matter fields are coupled to the scalar field. In this last case, the UFF is violated too. The LLI is usually satisfied in STT since, even if the asymptotic value of the scalar field is varying depending of the location of the frame (the LPI is violated), the metric and the scalar fields are Lorentz invariant asymptotically\footnote{The LLI can be violated in STT in a particular case. If the asymptotic value of the scalar field is varying on cosmological time, for instance because of a de Sitter background, then variations of the rate of the scalar field (and thus of the effective gravitational coupling) are generated. As a result, the local physics may depend on the velocity of the reference frame \cite{Will}.}
For more involved modified gravity models invoking vector, tensor or prior geometry (see Tab.~\ref{tab:MG_models}), both LPI and LLI are generally not satisfied.

\subsection{Current status of scalar-tensor theories from the observations}
In this section we report the current observational constraints on general STT. Indeed, the Brans-Dicke theory passes the PPN constraints only for very large values of the $\omega$ parameter (see Sec.~\ref{sec:PPN_STT}), such that it is now disfavored. In the last decades, the phenomenology predicted by more general STT, in particular in the presence of a potential, has been extensively explored. Today, the STT are also devoted to provide alternative to BHs or to model accelerated expansion of the Universe (either in the early or in the late-time Universe). In order to be viable, they must pass Solar System constraints.

\subsubsection{The PPN parameters} \label{sec:PPN_STT}
The PPN parameters  (see Sec.~\ref{sec:PPN} for a general introduction) enable one to constrain the Brans-Dicke theory in the Solar System. The calculations of the parameters are reported in App.~\ref{sec:PPN_BD} for the Brans-Dicke theory \eqref{BD_action} with $\omega=\omega(\Phi)$. They read,
\be
  \boxed{|\gPPN-1|=\frac{1}{\om+2}, \hspace{2cm} \bPPN-1={1\over (2\om+3)^{2}(2\om+4)}\frac{\dd\om}{\dd\Phi},}
\ee
the lower bound on the Brans-Dicke parameter $\omega$ yielding $\omega\gtrsim4.3\times10^4$ according to Eqs.~\eqref{eq:Cassini}. It results that the Brans-Dicke is indistinguishable from GR in the Solar system. 

In principle, the PPN parameters do not allow one to constrain general STT in the presence of a potential. However, as we will see in the rest of this thesis, PPN parameters are used provided that the potential can be considered as higher order terms in the Einstein equations, as the cosmological constant in GR. In order to test STT in the presence of a potential, the fifth force formalism has been developed.

\subsubsection{Fifth force formalism}\label{sec:fifth_force}
\begin{figure} 
  \begin{center}
    \includegraphics[scale=0.5]{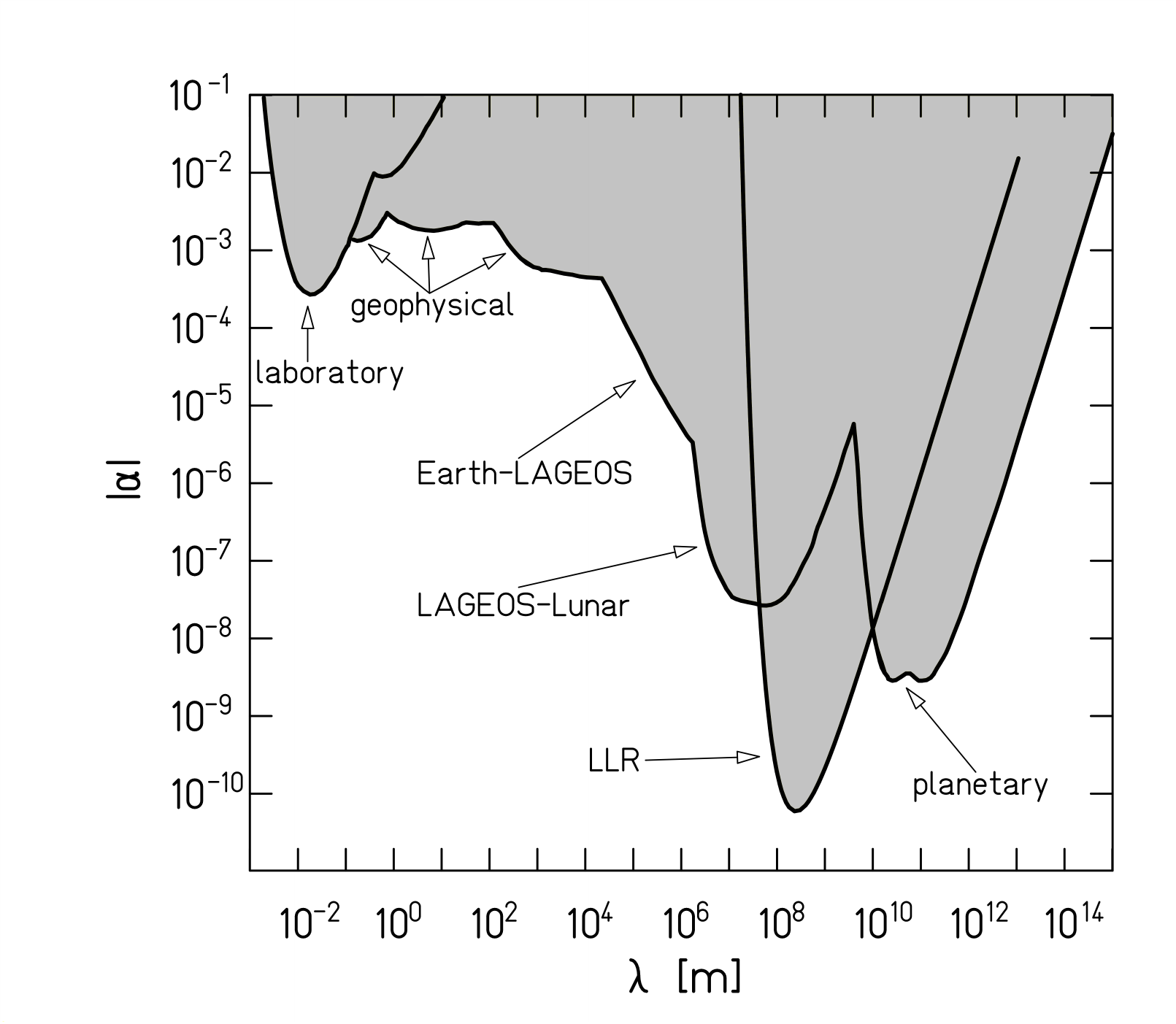}
    \caption{Current bounds on fifth force depending on the length of interaction $\lambda$ (for $\lambda>1$~cm) and its strength $\alpha$. Regions excluded at $95\%$ C.L. are shaded. Reprinted from \protect{\cite{Adelberger:2003zx}}.}
    \label{fig:fifth_force}
    \centering
  \end{center}
\end{figure}

By analogy with particle physics, low mass scalar field coupled to gravity could mediate a fifth force of Yukawa-type. Such an interaction implies a deviation from the Newtonian gravitational potential $V(r)$ in $1/r$ in the weak-field regime (see \cite{Fischbach} for a review).  
The fifth force formalism enables to parametrize the deviation of Yukawa type,
\be \label{eq:fifthforce}
  V(r)=-G_\rr{N}\frac{m_1 m_2}{r}\left(1+\alpha \rr{e}^{-r/\lambda}\right),
\ee
with $\alpha$, the dimensionless strength of interaction parameter and $\lambda$, the length scale \cite{Adelberger:2003zx}. The wide variety of scales where gravity can be tested is characterized as a function of $\lambda$ and $\alpha$. The remaining viable space parameters is constrained by various experiments depending on the probed scale of interaction \cite{Adelberger, Upadhye, Kapner}, as reported on Fig.~\ref{fig:fifth_force}. 
Actually, the detection of a fifth force would correspond to a violation of the UFF in the particular case where a low mass particle would be responsible for the deviation. 
%has to be constrained at all different scales of interaction, i.e. all ranges . It means that the fifth force tests involve a lot of different phenomenons, appearing at different scales, from LLR and planetary orbits to Casimir effect. For the very large ranges and the very short ones, the constraints are much more involved.  
 
% In Chap.~\ref{chap:chameleon}, we will study a modified gravity model where a fifth-force is mediated by an additional scalar field, the chameleon, which is a counterpart to the metric for describing gravity. The experiment of interest is similar to those used for studying the presence of the fifth force. In the following, we will review experimental tests of chameleon for short range interaction performed in labs.

\subsubsection{Strong field regime: spontaneous scalarization and particlelike solutions}
\label{sec:spont_scala}
If STT beyond the Brans-Dicke theory pass Solar System constraints, then the question arises if they are able to pass strong field regime bounds. Following \cite{PhysRevLett.70.2220} it is convenient to consider the Einstein frame (see  \cite{Salgado:1998sg} for an analysis in the Jordan frame) with the coupling function $A(\varphi)$ (see Eq.~\eqref{eq:action_EF_gen}) parametrized as,
\bea
  \ln A(\varphi)\simeq \ln A(\varphi_0)+\alpha_0(\varphi-\varphi_0)+\frac{1}{2}\beta_0(\varphi-\varphi_0)^2+...,
\eea
with $\varphi_0$ the background value of the scalar field imposed by the cosmological evolution which is usually assumed to be zero.  $\alpha_0$ and $\beta_0$ are related to the PPN parameters (see e.g. \cite{PhysRevLett.70.2220}). In particular, GR predicts $\alpha_0=\beta_0=0$ while the Brans-Dicke theory $\alpha_0^2=(2\omega_\rr{BD}+3)^{-1}$ and $\beta_0=0$. Considering the PPN parameters today gives $\alpha_0^2<10^{-5}$ \cite{Freire:2012mg} while $\beta_0$ is poorly constrained. It results that the scalar field must be close to the minimum of the coupling function $\dd A/\dd \varphi\simeq0$, i.e. the GR attractor. 

Considering now strong field experiments, NSs appear to be the most promising candidates for confronting STT with observations, in particular compact binaries.
Indeed, most of STT admit the same BH solutions as in GR, the no-hair theorem applying to a large class of modified gravity theories \cite{Sotiriou:2011dz}, whereas the dynamics of BHs can differ in some particular cases (see e.g. \cite{Berti:2015itd, Silva:2014fca} and references therein). 

In the 1990s, Damour and Esposito-Farèse \cite{PhysRevLett.70.2220, Damour:1996ke} found a non-perturbative effect (see Sec.~\ref{sec:NS}), named \textbf{spontaneous scalarization} such that the deviation from GR can be of the order one for NSs for parameters passing the Solar System tests. 
For stars with a baryonic mass above the critical value $m_\rr{bar}^{cr}$, the GR solution is less favored energetically than the solution with $\varphi\neq0$ since the ADM mass \eqref{eq:ADMmass} (assuming the baryonic matter and the scalar field contributions to $T_{\mu\nu}$) is smaller for the solution with $\varphi\neq0$. Spontaneous scalarization is usually parametrized by the \textbf{scalar charge} $\alpha_\rr{s}$ given by the asymptotic behavior of the scalar field,
\be \label{eq:scalar_charge}
  \varphi=\varphi_0+\frac{G\alpha_\rr{s}}{r},
\ee
with $G$ the bare gravitational constant.
% \be
%   m_\rr{ADM}=\frac{1}{2}\lim_{r\longrightarrow \infty} r \left(\rr{e}^{2\lambda}-1\right),
% \ee
% which is a "measure of the departure of the metric (at large distances) from the Minkowski metric" \cite{Salgado:1998sg},

Spontaneous scalarization has been dubbed by analogy with the spontaneous magnetization in ferromagnets below the Curie temperature \cite{Damour:1996ke}. The theory predicts that spontaneous scalarization cannot occur for $\beta_0 \gtrsim -4.35$ assuming spherically symmetric NSs \cite{PhysRevLett.70.2220}. The best bounds on $\beta_0$ have been obtained using compact binary observations (see Sec.~\ref{sec:NS}) and show that STT are forbidden if $\beta_0 < -5$  whatever $\alpha_0$ \cite{Freire:2012mg, Wex:2014nva}, such that $\beta_0$ is tightly constrained by compact binary observations. However, the effect of spontaneous scalarization in anisotropic NSs is found to increase when the tangential pressure is larger than the radial one \cite{Silva:2014fca}. Finally, the mass-radius diagram of stars (see Sec.~\ref{sec:NS}) is significantly affected by spontaneous scalarization \cite{Berti:2015itd}. 
%Finally the stability of the solution has been studied by \cite{Harada:1997mr} (see also \cite{Berti:2015itd}).

\paragraph*{}
\textbf{Particlelike solutions} are another example of non-perturbative effect arising in the strong field regime. Those solutions are globally regular (i.e. they exhibit no singularity), asymptotically flat (Minkowski spacetime is recovered at spatial infinity) and are of finite energy. In the best case scenario, they are also stable under perturbations. Particlelike solutions allow to regularize singular solutions appearing in spherical symmetry, for instance BH.

As an example, boson stars are particlelike solutions appearing in the framework of GR. They are compact objects made of a self-gravitating massive complex scalar field \cite{PhysRev.187.1767, PhysRevLett.57.2485}. Assuming static and spherically symmetric spacetime with the complex scalar field minimally coupled to gravity, there exists a continuous family of particlelike solutions for the scalar field depending on their potential parameters. Boson stars are BHs mimickers whereas the prediction in terms of accretion disk or emission of GWs are different (see \cite{Schunck:2003kk, Liebling:2012fv} for reviews). 

A second particlelike solution was discovered by Bartnik and Mc Kinnon when the $\rr{SU}(2)$ Yang-Mills theory is coupled to gravity. Whereas there exists no particlelike solution for gravity nor for the Yangs-Mills theory in spherical symmetry, the gauge field solutions in spherically symmetric spacetime are particlelike \cite{Bartnik}. However, those solutions were found to be unstable with respect to small spherically symmetric perturbations (see \cite{Volkov:1998cc} and references therein).

\subsubsection{Cosmology: inflation and late-time cosmic acceleration}
In order to build viable models for inflation and current cosmic acceleration, a potential term modeling the self-interaction of the scalar field is introduced in STT,
\bea \label{eq:STTwithpot}
  S=\int \dd^4 x \, \sqrt{-g} \, \left[ F(\phi)R - Z(\phi) 
  \left(\partial\phi\right)^2  - V\left(\phi\right) \right] 
  + S_\rr{M}\left[g_{\mu\nu},\, \Psi_\rr{M}\right].
\eea
If the STT is justified by the explanation of the late-time cosmic acceleration, it generally calls on a screening mechanism (see Sec.~\ref{sec:screening}) in order to pass the Solar System constraints while mediating long range effect in cosmology. STT also allow to build viable inflationary models, among them the Starobinsky model (see Fig.~\ref{fig:planck_inflation}). This is a particular example of $f(R)$ models \eqref{eq:fR},
\be
  S=\int \dd^4 x \sqrt{-g} \left(R+\frac{R^2}{6 M^2}\right),
\ee
which is equivalent to a STT in the presence of a potential (see Sec.~\ref{sec:equiv_HI_andStaro}).

\section[Beyond scalar-tensor theories: Horndeski gravity]{Beyond scalar-tensor theories: \\Horndeski gravity}
\label{sec:horndeski}
In 1974, Horndeski generalized the Lovelock theorem to models where the metric $g_{\mu\nu}$ has a scalar counterpart for describing gravity, imposing that the equations of motion are up to second order \cite{horndeski1974}.
On the other hand, the Galileon theory was proposed in 2008 \cite{Nicolis:2008in} (while it was actually discovered in 1992 by \cite{Fairlie:1992he, Fairlie:2011md}),
\be
S_\pi=S_\rr{EH} + \int \dd^4x \sqrt{-g} \sum_{i=1}^5 c_i \mathcal{L}^{(i)}_\pi +S_\rr{M}\left[\psi_\rr{M};~g_{\mu\nu}\right],
\ee
with,
\bea
  \mathcal{L}_1&=&\pi, \\
  \mathcal{L}_2&=&\left(\nabla\pi\right)^2, \\
  \mathcal{L}_3&=&\square\pi\left(\nabla\pi\right)^2, \\
  \mathcal{L}_4&=& \left(\square \pi\right)^2 \left(\pi_{;\mu}\,\pi^{;\mu}\right)
- 2 \left(\square \pi\right)\left(\pi_{;\mu}\,\pi^{;\mu\nu}\,\pi_{;\nu}\right)
\nonumber \\
&& - \left(\pi_{;\mu\nu}\,\pi^{;\mu\nu}\right) \left(\pi_{;\rho}\,\pi^{;\rho}\right)
+2 \left(\pi_{;\mu}\pi^{;\mu\nu}\,\pi_{;\nu\rho}\,\pi^{;\rho}\right), \\
 \mathcal{L}_5&=& \left(\square \pi\right)^3 \left(\pi_{;\mu}\,\pi^{;\mu}\right)
- 3 \left(\square \pi\right)^2\left(\pi_{;\mu}\,\pi^{;\mu\nu}\,\pi_{;\nu}\right)
- 3 \left(\square \pi\right) \left(\pi_{;\mu\nu}\,\pi^{;\mu\nu}\right) \left(\pi_{;\rho}\,\pi^{;\rho}\right)
\nonumber \\
&& +6 \left(\square \pi\right)\left(\pi_{;\mu}\pi^{;\mu\nu}\,\pi_{;\nu\rho}\,\pi^{;\rho}\right)
+2 \left(\pi_{;\mu}^{\hphantom{;\mu}\nu}\,\pi_{;\nu}^{\hphantom{;\nu}\rho}\,\pi_{;\rho}^{\hphantom{;\rho}\mu}\right) \left(\pi_{;\lambda}\,\pi^{;\lambda}\right) \nonumber \\
&& +3 \left(\pi_{;\mu\nu}\,\pi^{;\mu\nu}\right)
\left(\pi_{;\rho}\,\pi^{;\rho\lambda}\,\pi_{;\lambda}\right)
-6 \left(\pi_{;\mu}\,\pi^{;\mu\nu}\,\pi_{;\nu\rho}\,\pi^{;\rho\lambda}\,\pi_{;\lambda}\right),
\eea
the semi-colon denoting a covariant derivative. This model is an effective theory in a four-dimensional Minkowski background with second-order equations of motion and thus provides the well-defined modifications of gravity in the low energy limit.  Various modified gravity models are particular cases of the Galileon effective theory, among them the DGP model \cite{Dvali:2000hr} and the Lovelock gravity \cite{VanAcoleyen:2011mj} (see also \cite{deRham:2012az}).
The resulting theory is dubbed Galileon because of the Galilean shift symmetry,
\be
  \df_\mu\pi\longrightarrow \df_\mu\pi+b_\mu \hspace{1cm}\text{and}\hspace{1cm}
  \pi\longrightarrow\pi+c+b_\mu x^\mu,
\ee
with $c$ and $b_\mu$ arbitrary constant and vector field respectively, in flat spacetime.

In order to study the predictions of the Galileon theory in a curved spacetime, the covariant version of the Galileon theory, i.e. the most general theory where a scalar field is coupled to gravity with at most second order equations of motion, was formulated \cite{Deffayet:2009mn, Deffayet:2009wt, Kobayashi:2011nu},
\be
  \mathcal{L}=\sum_{i=2}^5 \mathcal{L}_i,
\ee
with,
\begin{eqnarray} \label{eq:horn1}
{\cal L}_2&=&K(\phi, X),
\\
{\cal L}_3&=&-G_3(\phi, X)\square\phi,
\end{eqnarray}
where $K$ and $G_3$ are generic functions of $\phi$ and $X\equiv-\partial_\mu\phi \partial^\mu\phi/2$.
Higher-order Galileons read,
\begin{eqnarray}
{\cal L}_4&=&G_{4}(\phi, X)R+G_{4X}\left[
\left(\square\phi\right)^2-\left(\nabla_\mu\nabla_\nu\phi\right)^2
\right],
\\
{\cal L}_5&=&G_5(\phi, X) G_{\mu\nu}\nabla^\mu\nabla^\nu\phi
\nonumber\\&&
-\frac{G_{5X}}{6}\Bigl[
\left(\square\phi\right)^3
-3\left(\square\phi\right)\left(\nabla_\mu\nabla_\nu\phi\right)^2
+2\left(\nabla_\mu\nabla_\nu\phi\right)^3
\Bigr],\,\,\,\,\,\, \label{eq:horn_end}
\end{eqnarray}
where $(\nabla_\mu\nabla_\nu\phi)^2=\nabla_\mu\nabla_\nu\phi\nabla^\mu\nabla^\nu\phi$,
$(\nabla_\mu\nabla_\nu\phi)^3=\nabla_\mu\nabla_\nu\phi\nabla^\nu\nabla^\lambda\phi\nabla_\lambda\nabla^\mu\phi$,
and $G_{iX}=\partial G_i/\partial X$.
It appears that the covariant Galileon model is equivalent to the Horndeski model  \cite{Kobayashi:2011nu}.

Besides the usual STT and the $f(R)$ models, Horndeski gravity also includes all the nonminimal derivative couplings of the scalar field to gravity like k-mouflage (see Sec.~\ref{sec:screening}) \cite{Deffayet:2009wt, Deffayet:2011gz, Deffayet:2009mn}. The EH action is also included by construction in the covariant version of the Galileon model. 
If equations of motion of more than second-order are allowed provided that they avoid the Ostrogradsky instability (see Sec.~\ref{sec:Ostro}), then the class of well-posed models is extended and is referred to as the beyond Horndeski theory \cite{Zumalacarregui:2013pma, Gleyzes:2014dya}.

Since the additional degree of freedom is scalar, it generally preserves the LLI (see the discussion in Sec.~\ref{sec:EP_revisited}) as well as the diffeomorphism-invariance. In some sense, Horndeski gravity is thus the minimal extension of GR since only the LPI is violated (see also Tab.~\ref{tab:MG_models}). %\tcb{what about WEP and SEP?}

Because of the non-linearities appearing in the scalar field kinetic term, the Horndeski models may exhibit the Vainshtein screening mechanism (see Sec.~\ref{sec:screening}). The Vainshtein mechanism makes possible to build viable cosmological models, for the late-time acceleration and inflation, with sufficiently small effects at local scales to evade Solar System constraints. 
In addition, inflationary phase can be generated by the non-linearities appearing in the kinetic term of the scalar field without the introduction of a potential term for the scalar field.

As a result, the Galileon model (or equivalently Horndeski gravity) has opened the way to new models for cosmology \cite{Chow:2009fm, Charmousis:2011bf, Kobayashi:2011nu, Tsujikawa:2012mk} (see also Chap.~\ref{chap:FabFour}).

\section{Summary of the thesis}

In the rest of this thesis, some modified gravity models are studied from the phenomenological point of view (assuming that there are well-posed) at different scales: in the lab, in the Solar System, around compact objects and at cosmological scales.

In Sec.~\ref{sec:screening} screening mechanisms have been introduced, among them the chameleon model. This is an example of STT, usually written in the Einstein frame \eqref{eq:action_EF_gen}. 
%The chameleon model is able to reproduce the current cosmic acceleration in the presence of a potential whereas it is able to pass Solar System constraint because of its screening mechanism for some potential and nonminimal coupling functions and parameters.
Due to the explicit coupling of the scalar field to matter $A(\phi)$, the chameleon acquires an effective mass (see Sec.~\ref{sec:screening}) varying as a function of the density. In relatively high density environment like in the Solar System or inside stars, the chameleon has a large effective mass such that it mediates a short-ranged fifth force (see the length of interaction introduced in Sec.~\ref{sec:fifth_force}) while in sparse environment, its effective mass is small, so that the chameleon is able to mediate long-ranged fifth force, that is the current cosmic acceleration.

Since its formulation by Khoury and Weltman \cite{KhouryWeltmanPRD, KhouryWeltmanPRL}, the chameleon model has been widely studied from the phenomenological point of view. Depending on the parameters of the potential and the nonminimal coupling function, the chameleon model can reproduce the current cosmic acceleration while it passes the current constraints in the Solar System. However, a part of this parameter space remains unconstrained. In Chap.~\ref{chap:chameleon} we focus on a lab experiment which appears to be the most promising probe of the chameleon model today. In 2012, Burrage, Copeland and Hinds proposed an atom interferometry experiment where the atom interferometer is placed inside a vacuum chamber in the presence of a test mass \cite{Burrage}. While the test mass is screened, the atoms are not, due to their small size and mass. They are thus sensitive to the chameleon field and the measure of interference fringes enable one to measure the additional acceleration due to the chameleon field.

The experiment has been performed at Berkeley in 2015 \cite{khoury}. Analytical forecasts have been provided \cite{khoury, Burrage}, relying on restrictive assumptions like negligible effects of the vacuum chamber wall. In this thesis we provide the full numerical solutions of the Klein-Gordon equation for a spherical vacuum chamber. This numerical method allows one to refine the analytical constraints and to analyze the effects of the chamber geometry.

In Chap.~\ref{chap:Higgs}, we study a second STT where the scalar field is identified to the \textbf{Brout-Englert-Higgs field}\footnote{In the following of this thesis, we will rather refer to the Higgs field.}.
Since its discovery in 2012 \cite{Aad:2012tfa, Chatrchyan:2012xdj}, the Higgs field is the first elementary scalar particle ever detected such that the existence of elementary scalar fields is not hypothetical anymore. The Higgs field has a crucial role in the SM because it is responsible to the mass generation of elementary particles relying on the spontaneous symmetry breaking of the SM gauge symmetry. Whereas SM is a quantum theory, the spontaneous symmetry breaking is a classical mechanism and could possibly be related to cosmology. 

The question thus arises if the Higgs field could play a role in cosmology, for instance during inflation. In 2008, Bezrukov and Shaposhnikov highlighted that the Higgs field could be the inflaton, provided that it is nonminimally coupled to gravity \cite{Bezrukov:2008ej}. This model is still favored by Planck+Keck+BICEP2 data (see the Starobinsky model\footnote{The Higgs inflation is equivalent to the Starobinsky model as highlighted in Sec.~\ref{sec:equiv_HI_andStaro}.} in Fig.~\ref{fig:planck_inflation}). 

In this thesis, we focus on the predictions of Higgs inflation around compact objects. Because of the nonminimal coupling, the distribution of the Higgs field in spherically symmetric spacetime is expected to be non-trivial, possibly leading to deviations from GR predictions. Moreover, variations of the Higgs vev could induce modifications in the nuclear processes inside neutron stars. Those questions are discussed in Chap.~\ref{chap:Higgs} and highlights the existence of a novel particlelike solution (see Sec.~\ref{sec:spont_scala}) for STT like the Higgs inflation.

In Chap.~\ref{chap:FabFour}, we focus on a more sophisticated model, the Fab Four dubbed in reference to the four general Lagrangians appearing in Horndeski gravity (see Sec.~\ref{sec:horndeski}) which may escape the Weinberg no-go theorem in order to solve the cosmological constant problem (see Sec.~\ref{sec:DE}). We study the phenomenology predicted by two of the four Lagrangians for inflation, in the Solar System and around compact objects. The Fab Two are found to be able to reproduce the inflationary phase without any potential, because of the nonminimal derivative coupling between the scalar field counterpart to the metric and the Einstein tensor, depending on the nonminimal coupling parameter. The Fab Two model predicts that compact objects are spontaneously scalarized (see Sec.~\ref{sec:spont_scala}). Eventually some observables in the Solar System are computed numerically since the PPN parameters do not allow one to derive any constraint on the Fab Two model due to the presence of the nonminimal derivative coupling.

%\tcb{retravailler 3.3 à la lumière de tout ce que tu as écris dans les trois premiers chapitres. Où nous conduis-tu? Tu ne parles pas de tous tes résultats, ni de l’ordre dans lesquels tu vas les montrer. Tu as bien aborder les éléments théoriques à la base des modèles que tu as regardés mais ils sont mis plic ploc. Tu dois terminer par une lecture globale de tes résultats (qui sont tres différents) à la lumière de ta belle présentation de la gravitation modifiée. 3.3 doit être un paragraphe que le lecteur peut relire pour avoir un vrai résumé de ta contribution scientifique à la lueur de la littérature abondante sur le sujet.}

% We restrict the discussion here to classical theory of gravity even if there exist several reasons why quantum gravity would be interesting, like the problem of singularities and of the renormalizability of GR.
% 
% 
% because of the spontaneous symmetry breaking, the Higgs vev obviously change from $<H>=0$ to $<H>=v$ with $v=246$ GeV during the history of the universe, so that the mass of the elementary particles also change. A mechanism - even questionable - is invoked here which explains the variation of the Higgs vev and, by the way, the fact that elementary particles acquires a mass at the electroweak energy scale.  
% 
% In this thesis, we will constrain two of them: generalized scalar-tensor theories in the presence of a potential and the Fab Four model. In some point, this is the easiest way to extend GR in $D=4$.

\part{Combined constraints on modified gravity}

\renewcommand{\chaptermark}[1]{\markboth{\small\textsc{Chapter \thechapter.\ #1}}{}}
\cleardoublepage

\chapter{Probing the chameleon model with atom-interferometry} % Main chapter title

\label{chap:chameleon} % For referencing the chapter elsewhere, use \ref{Chapter1} 

\lhead{Chapter 4. \emph{Awesome chapter 4}} % This is for the header on each page - perhaps a shortened title

%----------------------------------------------------------------------------------------
%% Version electronique
\begin{center}
\textit{based on}\\
\end{center}
\begin{center}
S.~Schlögel, S.~Clesse, A.~Füzfa, 
\\\textit{Probing Modified Gravity with Atom Interferometry: \\
a Numerical Approach},\\
\href{http://journals.aps.org/prd/abstract/10.1103/PhysRevD.93.104036}{Phys. Rev. D 93, 104036 (2016)}, \href{http://arxiv.org/abs/1507.03081}{\texttt{arXiv:1507.03081}}\\
\end{center}

\vspace{1cm}

\noindent
In this chapter we focus on the chameleon model which exhibits the eponymous screening mechanism introduced in Sec.~\ref{sec:screening}. 
%Initially motivated by the fact that Brans-Dicke theory cannot pass the Solar System whereas it can explain the current cosmic acceleration the chameleon fulfilled both conditions thanks to its screening mechanism. 
In sparse environment the chameleon behaves as a free field, allowing for the cosmic acceleration whereas in dense environment, it becomes massive, therefore possibly passing the Solar System constraints. After a brief introduction of  the chameleon models, we review the current bounds on its parameter space, the constraints coming from the cosmological and astrophysical observations as well as from experiments. Then we focus on an atom-interferometry experiment recently proposed by \cite{Burrage}. We refine the constraints derived in \cite{Burrage, khoury} providing numerical profiles of the chameleon field and of the induced acceleration on atoms. We establish that the near future atom-interferometry experiments could be able to rule out the chameleon parameter space up to the Planck scale. 

\section{The chameleon models}\label{sec2}
Chameleon models have been first proposed by Khoury and Weltman \cite{KhouryWeltmanPRL, KhouryWeltmanPRD}. They are generally formulated in the Einstein frame (see Sec.~\ref{app:EF}), 
\bea \label{eq:action_chamel}
S=\int \dd^4x \sqrt{-g} \left[\frac{R}{2\kappa}-\frac{1}{2} \left(\partial\phi\right)^2-V(\phi)\right]
+ S_\rr{M}\left[\psi_\rr{M};\tilde{g}_{\mu\nu}=A^{2}\left(\phi\right)g_{\mu\nu}\right],
\eea
%with $V(\phi)$ and $A(\phi)$ the general potential and coupling functions, while $\tilde{g}_{\mu\nu}$ the Jordan frame metric\footnote{In this chapter 
the tilde denoting Jordan frame quantities in this chapter.
Chameleon models were initially justified by the fact that quintessence is able to model cosmic acceleration (see Sec.~\ref{sec:DE}) provided that the coupling of the scalar field to matter is extremely small in order to pass local tests of gravity. Similarly to tracking quintessence the typical chameleon potential is of the runaway type, that is a monotically decreasing function satisfying the tracker condition defined by,
\bea
  \frac{V_{,\phi\phi} V}{V^2_{,\phi}}>1,
\eea
and diverging at some finite value of $\phi=\phi_*$ (in the following $\phi_*=0$). Other potential functions have been proposed (see e.g. \cite{Gubser, Mota:2010uy}). 
Contrary to quintessence models, the chameleon field exhibits a coupling to matter possibly strong,
\bea
  A(\phi)=\exp{\frac{\beta\phi}{\Mp}}\simeq\left(1+\frac{\beta\phi}{\Mp}\right),
\eea
where $\beta$ can be of order unity (or even larger in the strongly coupled case \cite{Mota, MotaShaw}), the chameleon field being then allowed to pass local tests of gravity.
% $\beta=\Mp/M$

The effective dynamics is driven by the effective potential $V_\rr{eff}$ defined by, 
\bea \label{eq:KG_chamel}
\square\phi\equiv\frac{\dd V_\rr{eff}}{\dd \phi}, \hspace{1.7cm}\frac{\dd V_\rr{eff}}{\dd \phi}=\frac{{\dd V}}{{\dd \phi}}-T\frac{{\dd \ln A}}{{\dd \phi}},
\eea
where $T$ is the trace of the stress-energy tensor in the Einstein frame $T_{\mu\nu}$, which is related to its Jordan frame counterpart $\tilde{T}$ by,
\be
  T=A^4(\phi) \tilde{T}.
\ee
Assuming a perfect fluid 
%We introduce $\tilde{T}_{\mu \nu}$ the stress-energy tensor for a perfect fluid in the Jordan frame in order to consider conserved quantities, i.e. $\tilde{\nabla}_\mu \tilde{T}^{\mu\nu}=0$, 
the energy density $\rho$ and the pressure $p$ in the Einstein frame read \cite{Damour:1993id},
\bea
\rho&=&A^4\left(\phi\right)\tilde{\rho},
\label{rho}
\\
p&=&A^4\left(\phi\right)\tilde{p}.
\eea
In the rest of this chapter we will consider the Jordan frame energy tensor only since $\tilde{\nabla}_\mu \tilde{T}^{\mu\nu}=0$ and the tilde are dropped.

\begin{figure}[!htb]
    \centering
    \includegraphics[width=0.6\textwidth]{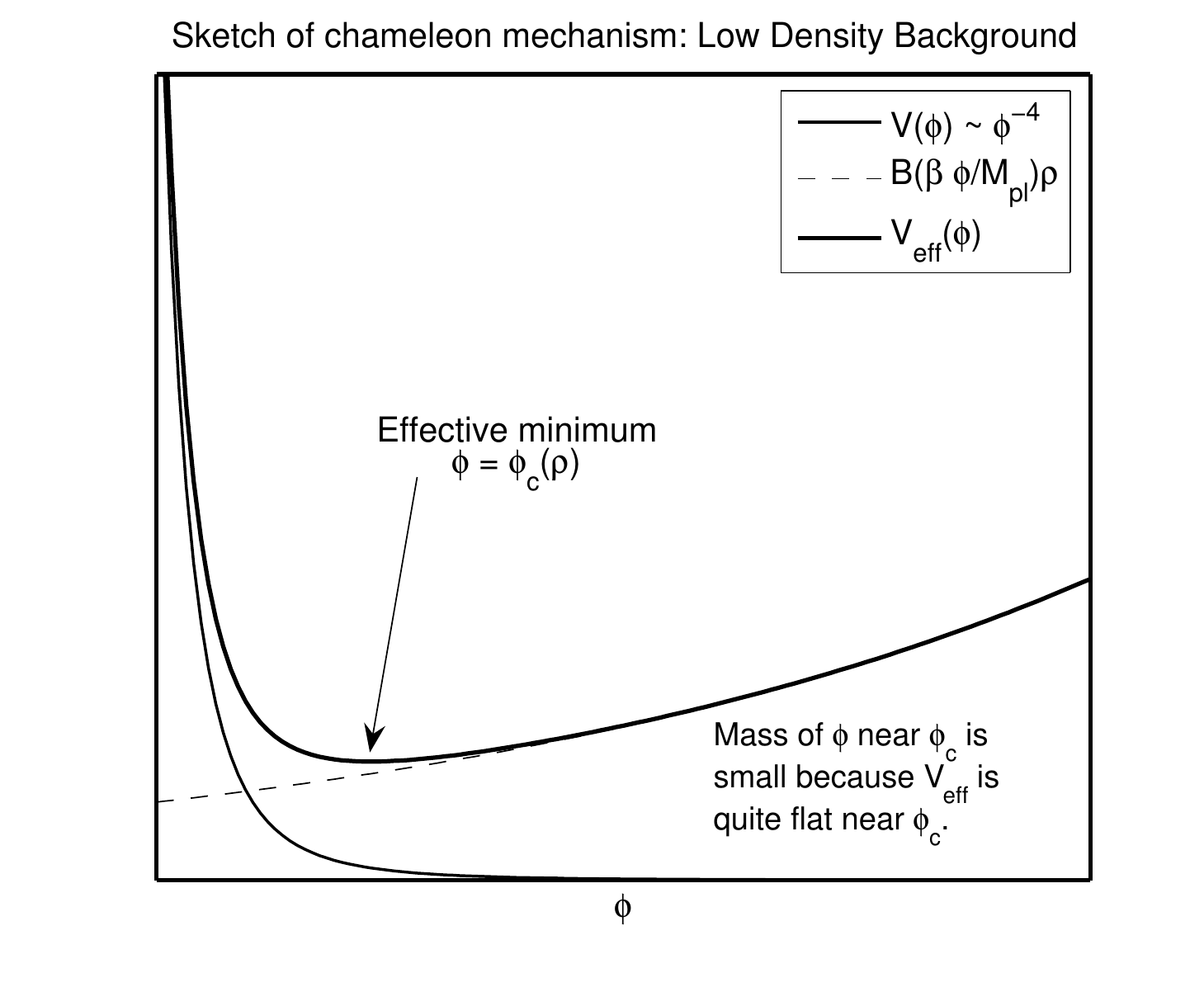}
    \caption{Effective potential, given by the runaway potential and the coupling to matter according to Eq.~\eqref{eq:KG_chamel} for the chameleon field in sparse environment. The effective chameleon mass is small since $V_\rr{eff}$ is shallow around its minimum, allowing the chameleon to drive the current cosmic acceleration. Reprinted from \cite{MotaShaw}.}
    \label{fig:sparse_chamel}
    \includegraphics[width=0.6\textwidth]{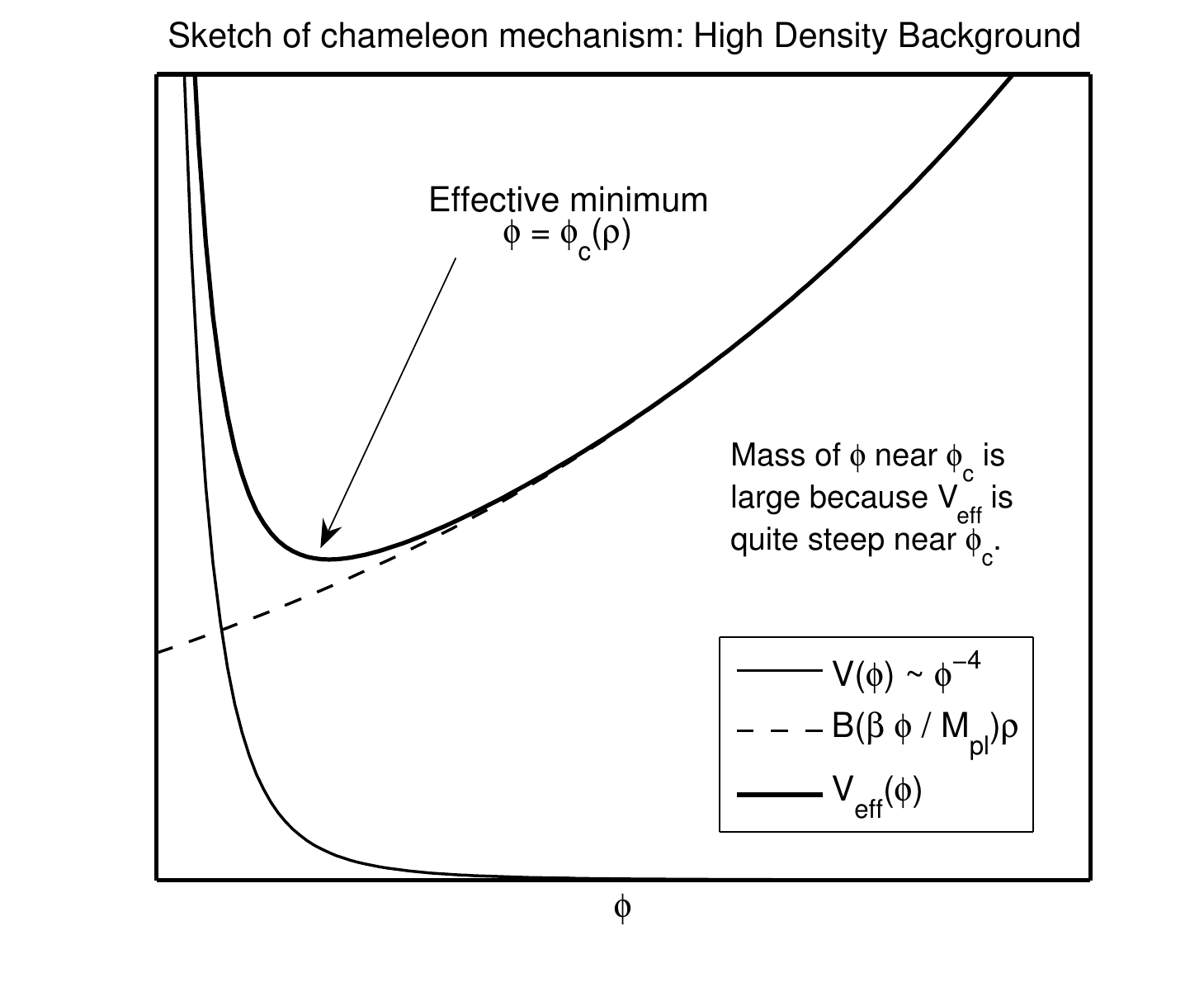}
    \caption{Effective potential, given by the runaway potential and the coupling to matter according to Eq.~\eqref{eq:KG_chamel} for the chameleon field in dense environment. The effective chameleon mass is large since $V_\rr{eff}$ is steep around its minimum, allowing the chameleon to pass local tests of gravity, for instance in the Solar system. Reprinted from \cite{MotaShaw}.} 
    \label{fig:dense_chamel}
\end{figure}

Provided $\beta>0$, the effective potential has a minimum $\phi_\rr{min}$ and an effective mass $m_{\rr{min}}^2\equiv\left.\dd^2 V_\rr{eff}/\dd\phi^2\right|_{\phi=\phi_\rr{min}}$ (or equivalently the Compton wavelength $\lambda_\rr{C}=m_\rr{min}^{-1}$)  which depend on $\rho$ (see Figs.~\ref{fig:sparse_chamel} and \ref{fig:dense_chamel}): in dense (sparse) environment, $\phi_\rr{min}$ (denoted $\phi_\rr{c}$ in the figures) is small (large) while $m_\rr{min}$ is large (small). It results that in dense environment, the chameleon is decaying rapidly since its Compton wavelength is small while it mediates long range force in sparse environment. This is the reason why this screening mechanism has been named the chameleon (see Sec.~\ref{sec:screening}).

\subsection{The original chameleon model}
Originally Khoury and Weltman considered the model for the Ratra-Peebles potential and the exponential coupling function,
\be
  V(\phi)=\frac{\Lambda^{4+\alpha}}{\phi^\alpha}, \hspace{1.5cm} A(\phi)={\rm{e}}^{\frac{\phi}{M}},
\ee
with
\be \label{eq:betaM}
  M=\frac{\Mp}{\beta}.
\ee
Assuming $A(\phi)\simeq1$ and $\alpha>0$, the minimum of the effective potential as well as the effective mass are respectively given by,
\bea
  &&\hspace*{-0.5cm}\left.\frac{\dd V_\rr{eff}}{\dd \phi}\right|_{\phi=\phi_\rr{min}}=0=-\frac{\alpha \Lambda^{\alpha+4}}{\phi_\rr{min}^{\alpha+4}}+\frac{\rho}{M}\hspace{0.2cm} \Rightarrow \hspace{0.2cm}
  \phi_\rr{min}=\left(\frac{\alpha \Lambda^{\alpha+4}M}{{\rho}}\right)^{\frac{1}{\alpha+1}}, \hspace{1.5cm}
  \label{eq:phimin}
  \\
  &&\hspace*{-0.5cm} m_{\rr{min}}^2=\left.\frac{\dd^2 V_\rr{eff}}{\dd \phi^2}\right|_{\phi=\phi_\rr{min}} 
  =\alpha  (1+ \alpha) \frac{\Lambda^{\alpha+4}}{\phi_\rr{min}^{\alpha+2}}
  \non\\
  &&   \qquad\qquad \Rightarrow \hspace{0.2cm}
  m_{\rr{min}}^2=\alpha (1+ \alpha) \Lambda^{4+\alpha}  \left( \frac{ \rho}{\alpha M \Lambda^{4+\alpha}} \right)^{\frac{2+\alpha}{1+\alpha}}.
  \label{eq:m_eff_chamel}
\eea

We first derive the order of magnitude required for the parameters in order to explain the current cosmic acceleration. Following \cite{Zhang:2016njn}, we only consider the condition on the cosmological parameters today,
\be \label{eq:cosmo_model1}
  \left.\frac{\rho_{\Lambda,0}}{\rho_\rr{m,0}}\right|_\rr{obs}=2.15.
\ee
Assuming that the chameleon field is at the minimum of its effective potential today\footnote{In their original paper \cite{Brax:2004qh} Brax et al. showed that the chameleon exhibits an attractor behavior such as it remains very close to the minimum of its effective potential all along the cosmic history, the model being mainly insensitive to the scalar field initial conditions. Background cosmological constraints are then fulfilled provided that the chameleon has already settled to its minimum by the onset of BBN, $\phi_\rr{BBN}\lesssim0.1 \beta^{-1}\Mp$, ensuring small mass variations. This results have been extended for large couplings by \cite{MotaShaw}. Numerical computations allowed to establish that the chameleon model exhibits an attractor mechanism. While it is very efficient in the early Universe, where the effective potential is narrow since the density is high, it can be not strong enough at late time, especially for small $\beta$ values. The attractor can be reached for a large span of initial conditions.}, thus running a cosmological constant, yields,
\be
  \rho_{\Lambda,0}=V\left[\phi_\rr{min}\left(\rho_\infty\right)\right]=\frac{\Lambda^{\alpha+4}}{\phi_\rr{min}^\alpha}
  =\frac{\rho_\infty \phi_\infty}{\alpha M},
\ee
where the subscript $\infty$ refers to background values with $\phi_\infty=\phi_\rr{min}(\rho_\infty)$ and $\rho_\infty$ being either the cosmological matter density identified to $\rho_\rr{m,0}\sim 10^{-47}~\rr{GeV^4}$ either the galactic background density $\rho_\rr{gal}\sim 10^5 \times \rho_\rr{m,0}$. Assuming $\rho_\infty=\rho_\rr{m,0}$ we obtain,
\be \label{eq:chamel1_cosmo}
  \frac{\rho_{\Lambda,0}}{\rho_\rr{m,0}}=\frac{\phi_\infty}{\alpha M}
  =\frac{1}{\alpha M}\left(\frac{\alpha M \Lambda^{\alpha+4}}{\rho_\rr{m,0}}\right)
  ^{\frac{1}{\alpha+1}}.
  %\lesssim 2.4\times 10^{-11},
\ee
It is possible to rewrite this equation in order to find out an explicit relation between $\alpha$ and $\Lambda$ using Eq.~\eqref{eq:betaM},
\bea
  \hspace*{-0.5cm}\Lambda^{\frac{\alpha+4}{\alpha+1}}&=&\rho_\rr{\Lambda,0}\left(\rho_\rr{m,0}\right)^{-\frac{\alpha}{\alpha+1}} \left(\alpha M\right)^{\frac{\alpha}{\alpha+1}}, \\
  \hspace*{-0.5cm}\log \Lambda&=&\frac{1}{\alpha+4}\left[\left(\alpha+1\right)\log\rho_\rr{\Lambda,0}-\alpha \log \rho_\rr{m,0}+\alpha\log\left(\frac{\alpha \mpl}{\sqrt{8\pi}\beta}\right)\right], \\
  &=&\frac{1}{\alpha+4}\left[\log\rho_\rr{\Lambda,0}+\alpha \log \mpl +\alpha\log\left(\frac{\rho_{\Lambda,0}}{\rho_\rr{m,0}}\frac{\alpha}{\sqrt{8\pi}\beta}\right)\right],
\eea
the last term of the last equality being negligible for $\beta\sim1$ and $\alpha\sim 1$. In this case, the equation reduces to,
\be \label{eq:quintcosmo}
\log \frac{\Lambda}{1~\rr{GeV}}\approx\frac{19\alpha-47}{4+\alpha},
\ee
corresponding to the relation for quintessence as found by \cite{Schimd}. In summary the original chameleon model is able to reproduce the cosmic acceleration provided that Eq.~\eqref{eq:chamel1_cosmo} is fulfilled or equivalently the potential parameters obey to Eq.~\eqref{eq:quintcosmo}.
Notice that Hees and Füzfa analyzed the likelihood of SN Ia data and obtained the same results \cite{Hees}: the relation $\Lambda-\alpha$ \eqref{eq:quintcosmo} is not weakly affected by the nonminimal coupling even if the latter contributes non-negligibly to the luminosity distance $D_\rr{L}$ measurement where $D_\rr{L}\simeq z /H (v/c\ll 1)$ with $z$ the redshift and $H$ the Hubble parameter. 

However, the original chameleon model is not able to pass the local tests of gravity for the corresponding parameters \cite{Hees, Zhang:2016njn}, as revealed by the computations of the PPN parameters. The Brans-Dicke formulas are not useful here since the potential cannot be neglected in the Klein-Gordon equation (even if the potential is negligible in the Einstein equations assuming that it contributes to higher order terms as for the cosmological constant in GR). As for the Brans-Dicke theory (see App.~\ref{sec:PPN_BD}), the Klein-Gordon equation must first be solved. Following \cite{Hees}, the numerical simulations are in good agreement with the analytical solutions for $\phi$ outside the Sun whereas deviations arise inside the Sun.
Since the PPN parameters require the solution for the scalar field outside the Sun only, the analytical treatment of the solution is valid (see also App.~\ref{sec5} for the analytical calculations). Then, it is possible to derive the PN expansion for the metric in the Einstein frame as for the Brans-Dicke theory (see App.~\ref{sec:PPN_BD}). In order to obtain the PPN parameters, the metric must be transformed in the Jordan frame by a conformal rescaling (see Sec.~\ref{app:EF}). We report the reader to \cite{Zhang:2016njn} for the detailed calculations.

Eventually, the PPN parameters for the original chameleon read \cite{Zhang:2016njn},
\bea
  \gamma_\rr{PPN}-1&=&-\frac{2\phi_\infty}{M\Phi}, \\
  \beta_\rr{PPN}-1&=&-\frac{3}{4(\alpha+1) \Phi}\left(\frac{\phi_\infty}{\Mp}\right)^2,
\eea
where the gravitational potential $\Phi\simeq 2.12\times 10^{-6}$ for the Sun. The constraints on $\gamma_\rr{PPN}$ are much more powerful than on $\beta_\rr{PPN}$ since,
\be
  \left|\beta_\rr{PPN}-1\right|=\frac{3\Phi}{16 \left(\alpha+1\right)}\left(\frac{M}{\Mp}\right)^2 \left(\gamma_\rr{PPN}-1\right)^2
  \ll \left|\gamma_\rr{PPN}-1\right|.
\ee
Indeed for $M\lesssim \Mp$ and $\alpha\sim \mathcal{O}(1)$, $\left|\beta_\rr{PPN}-1\right|\sim 10^{-16}$, thus far below the current constraints on $\beta_\rr{PPN}$ (see Sec.~\ref{sec:PPNformalism}). 
Thus we focus on $\gamma_\rr{PPN}$ constraints from Cassini probe given by Eq.~\eqref{eq:Cassini} yielding,
\be \label{eq:PPN_gen_chamel}
  \frac{\phi_\infty}{M}
  =\left(\frac{\alpha \Lambda^{\alpha+4}}{\rho_\infty M^\alpha}\right)^{\frac{1}{\alpha+1}}
  \lesssim 2.4 \times 10^{-11}.
\ee

Combining the condition for the cosmic acceleration \eqref{eq:chamel1_cosmo} and the $\gamma_\rr{PPN}$ \eqref{eq:PPN_gen_chamel}, we obtain for $\alpha\sim1$,
\be
  \frac{\rho_{\Lambda,0}}{\rho_\rr{m,0}}\lesssim 2.4\times 10^{-11},
\ee
which is incompatible with the cosmological observations, the original chameleon model being thus ruled out.

%This lower bound is valid for both chameleon models since only the derivative of the potential appears in the Klein-Gordon equation and the potential is neglected in the modified Einstein equations. However this is not the same in cosmology where the potential term cannot be neglected anymore.

\subsection{The exponential chameleon}
In order to pass the local tests of gravity in the Solar System while explaining the current cosmic acceleration, we must rather assume the exponential potential\footnote{Following Brax et al. \cite{Brax:2004qh}, the choice of this potential function is justified by two sufficient conditions: the potential is of runaway form and it diverges at some finite value, for instance $\phi=0$. In addition the potential is flat and of order unity for the current value of the scalar field, ensuring the late-time cosmic acceleration.},
\be
  V(\phi)=\Lambda^4\exp\left(\frac{\Lambda^\alpha}{\phi^\alpha}\right)\simeq\Lambda^4\left(1+\frac{\Lambda^{\alpha}}{\phi^\alpha}\right).
\ee
The Klein-Gordon equation is exactly the same, the previous PPN analysis being thus still valid, while the cosmological constraint on the exponential chameleon differs from the original one due to the additional constant in the potential. Indeed, the minimum of the potential now reads,
\be
  \rho_{\Lambda,0}=V\left[\phi_\rr{min}\left(\rho_\infty\right)\right]=\Lambda^4+\frac{\rho_\infty \phi_\infty}{\alpha M},
\ee
where the second term in the right-hand side is negligible since,
\be
  \frac{\rho_{\Lambda,0}}{\rho_\rr{m,0}}=2.15
  =\frac{\Lambda^4}{\rho_\rr{m,0}}+\underbrace{\frac{\phi_\infty(\rho_\rr{m,0})}{\alpha M}}_{\lesssim 2.4\times 10^{-11}},
\ee
by using \eqref{eq:cosmo_model1} and \eqref{eq:PPN_gen_chamel}, for $\alpha\sim1$ and $\rho_\infty=\rho_\rr{m,0}$. It results that, for the exponential chameleon,
\be
  \Lambda=\rho_{\Lambda,0}^{1/4}\simeq 2~\rr{meV},
\ee
in order to explain the current cosmic acceleration.
%, the effective mass of the chameleon being $m_\rr{min}(\phi_0)\sim H_0$ today. 
The PPN parameters \eqref{eq:PPN_gen_chamel} allow one to further constrain the parameter space $\alpha-\rr{M}$,
\bea
  \log\left(\frac{\alpha \Lambda^{\alpha+4}}{M^\alpha \rho_\infty}\right)&\lesssim& -10.6 ~(\alpha+1),\\
  \alpha\left(\log M-\log \Lambda -10.6\right)-\log \alpha& \gtrsim & 10.6 - \log \frac{\rho_\infty}{\rho_{\Lambda,0}}.
\eea
The viable parameter space $\alpha-M$ is plotted in Fig.~\ref{fig:beta_n} for $\Lambda=2.4~$meV and $\rho_\infty=\rho_\rr{m,0}$ (the choice $\rho_\infty=\rho_\rr{gal}$ is more conservative). The combination of the constraints on PPN parameters and background cosmology enables one to rule out the exponential chameleon model for small $\alpha$ values, the constraint being more stringent for large $\beta$ values. As we will see in the following additional tests on the exponential chameleon model are much more stringent than those coming from the Solar System observations today.

Further observables signatures for the chameleon model have been investigated (see \cite{Joyce:2014kja} for a review of modified gravity models, among them the chameleon). Since the chameleon couples to the trace of $T_{\mu\nu}$ no significant effect is expected during the radiation era, provided that the chameleon is not coupled to photons\footnote{ Further generalizations of the chameleon model were proposed, notably introducing a coupling function $A_\gamma^2(\phi)$ of the chameleon to photons,
\be
  S=\int \dd^4x \sqrt{-g} \left[\frac{R}{2\kappa}-\frac{1}{2} \left(\partial\phi\right)^2-V(\phi)-\frac{1}{4} A_\gamma^2(\phi) F^{\mu\nu} F_{\mu\nu}\right]
  + S_\rr{M}\left[\psi_\rr{M};\tilde{g}_{\mu\nu}=A^{2}\left(\phi\right)g_{\mu\nu}\right],
  %S\supset-\frac{1}{4}\int \dd^4x \sqrt{-g} A^2(\phi) F^{\mu\nu} F_{\mu\nu},
\ee
with $F_{\mu\nu}$ the Faraday tensor, leading to a variation of the fine-structure constant \cite{Brax:2007ak}.

Assuming a coupling of the chameleon field to photons ($\beta_\gamma$) enables one to test the chameleon models with other experimental setup. Several experiments have put constraints on $\beta_\gamma$ for a given $\beta$ (for different $\alpha$). Among them, CHameleon Afterglow SEarch-GammeV \cite{Steffen:2010ep} and Axion DM eXperiment \cite{ADMX} where the chameleon is tested using a laser beam in a vacuum chamber and in a microwave cavity respectively thanks to an intense magnetic field in both cases,  and, more recently by CERN Axion Telescope, a telescope which detects soft X-ray coming from the Sun and possibly produced by the chameleon field \cite{CAST}. Those constraints read $\beta_\gamma\lesssim10^{11}$ for the range $1<\beta\lesssim10^6$ ($\Lambda$ being of the order of the cosmological constant), a result being mostly independent of $\alpha$ \cite{CAST}. 
}.
Moreover, the range of its interaction is always much smaller than the horizon scale so no super-horizon effect is expected \cite{Brax:2005ew}. 

Chameleon field should leave imprints during the structure formation in the matter era, especially when the coupling to matter $\beta$ is large. The growth of matter fluctuations has been studied in a serie of papers, in the linear regime \cite{Brax:2005ew, Gannouji:2010fc, Mota:2010uy, Hojjati:2015ojt} and in the non-linear one (see \cite{Brax:2013mua} and references therein). It was found that first halos to form in chameleon cosmology are significantly more concentrated than according to the $\Lambda-$CDM concordance picture and matter collapses earlier to form structure \cite{Brax:2005ew}, the linear approximation thus fails at larger scale than for the $\Lambda-$CDM model. The main effects appear in the non-linear scales where the density contrast of matter is found to increase anomalously \cite{Brax:2013mua}, the matter power spectrum being altered. The deviations are $\lesssim~10\%$ in the non-linear part of the power spectrum, which is hard to detect today. 
The observations of LSS, for instance by the Euclid satellite, should enable one to improve the current constraints on the chameleon model in the near future \cite{Amendola:2016saw}. 
%The variation of the effective EoS along the cosmic history $w_\rr{eff}$ was found to be potentially measurable by Euclid \cite{Amendola:2012ys}. 

\section{Current constraints on chameleon}
\label{sec:chamel_constraints}
\begin{figure}[!htb]
    \centering
    \includegraphics[width=0.6\textwidth]{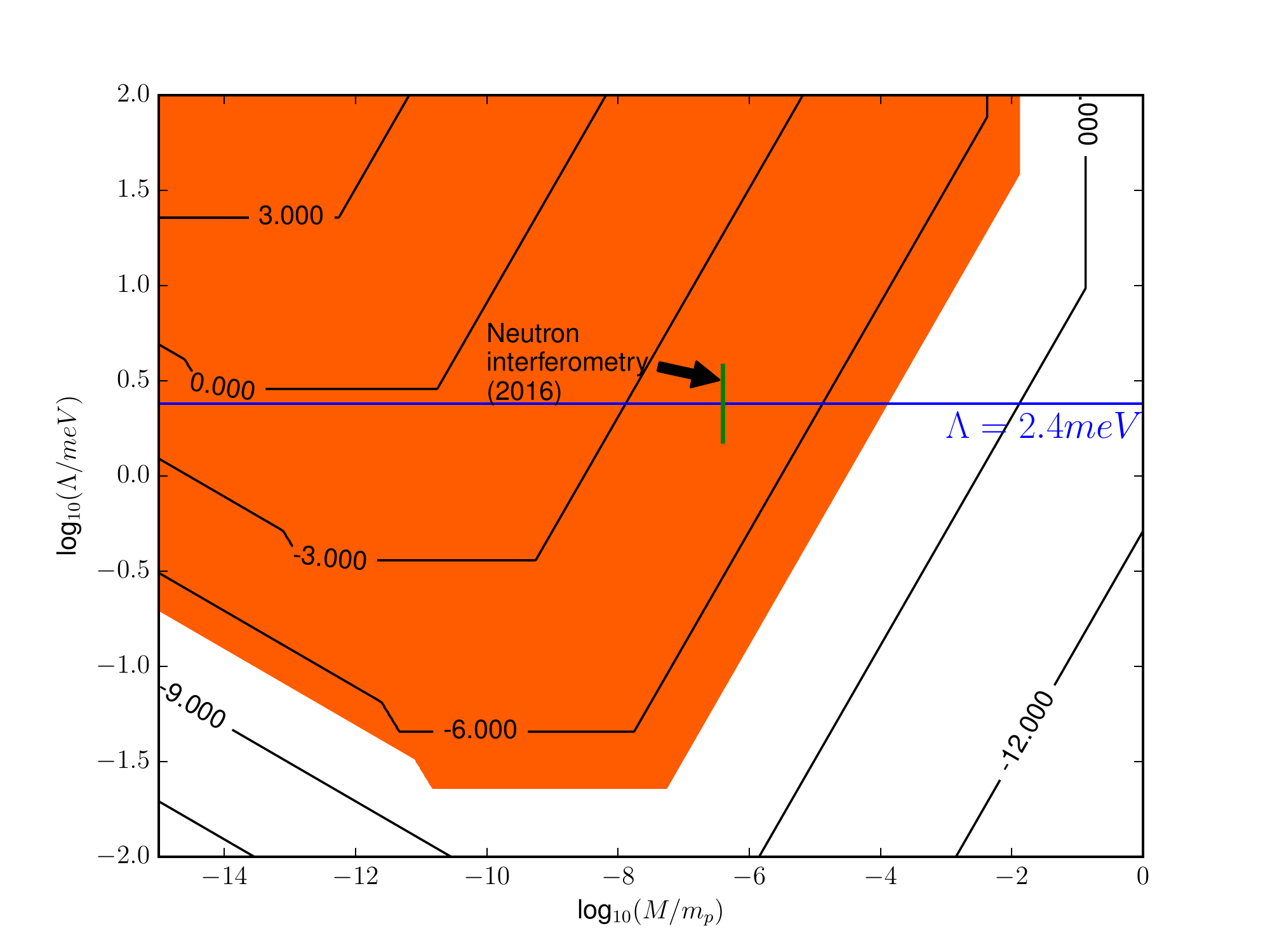}
    \caption{Exclusion contours for the chameleon parameters $M$ and $\Lambda$ ($\alpha=1$ being fixed) for atom interferometry (in orange) and neutron interferometry (on the left of the green line), the blue line referring to the cosmological constant $\Lambda=2.4~$meV. The contour lines refer to the logarithm of the normalized chameleon acceleration $a_\phi/g$, the current constraint being given by Eq.~\eqref{eq:acc_exp} with $a_\phi/g\sim 10^{-7}$.}
    \label{fig:exclusion}
    \includegraphics[width=0.6\textwidth]{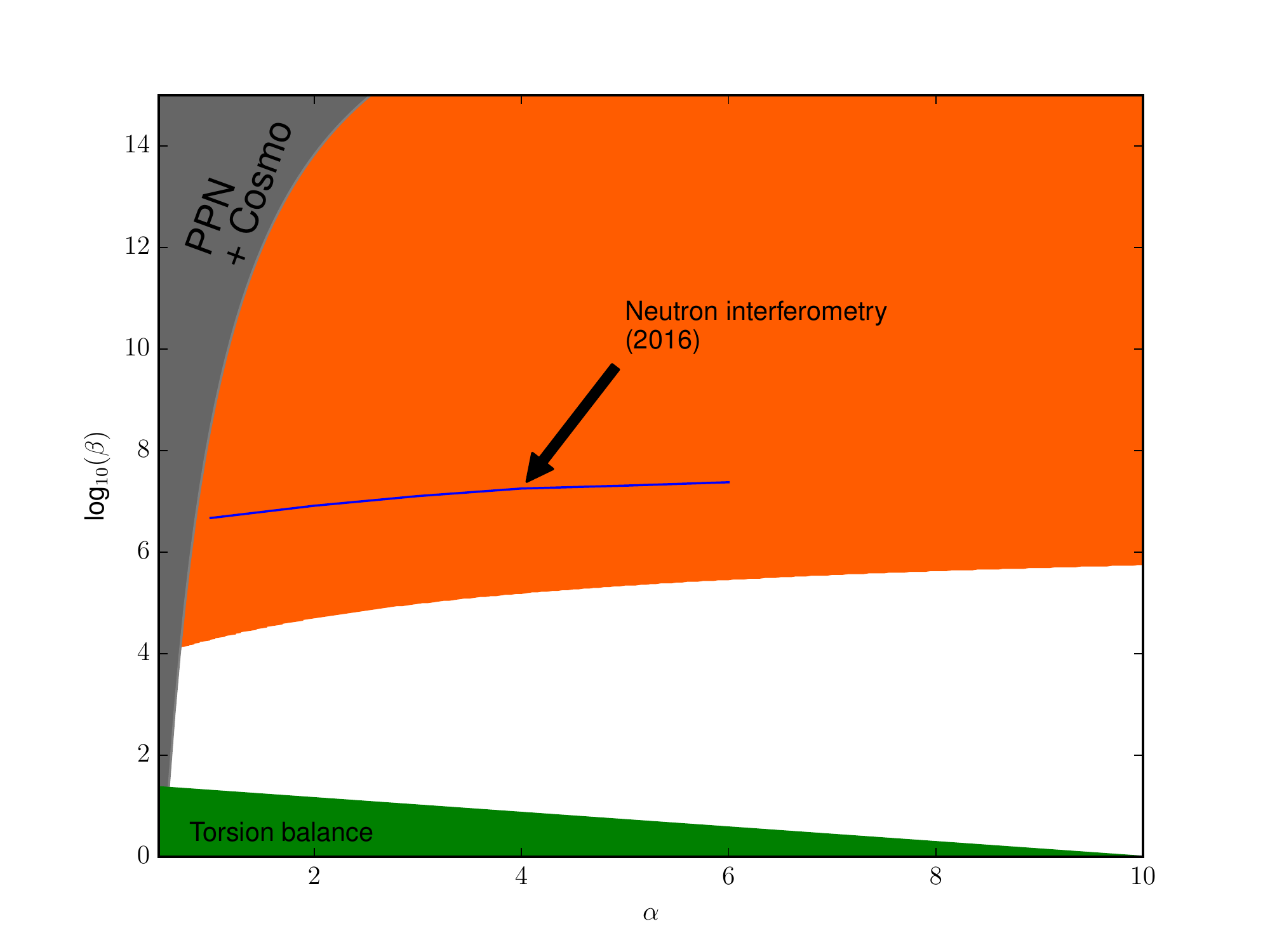}
    \caption{Exclusion contours for the chameleon parameters $\alpha$ and $\beta$ ($\Lambda=2.4~$meV being fixed) for atom interferometry (in orange), combined constraint from PPN parameters and background cosmology (in grey), torsion balance experiment (in green); and neutron interferometry (upper the blue line).} 
    \label{fig:beta_n}

\end{figure}

In this section, we briefly discuss to what extent the exponential chameleon model is viable today in terms of those three parameters, that is the potential ($\alpha,~\Lambda$) and nonminimal coupling ($M$ or equivalently $\beta=\Mp/M$)  parameters. The parameter $M$ is found to be poorly constrained because of the presence of the potential while, as we have already seen in the previous section $\Lambda\sim 1~$meV in order to account for the current cosmic expansion. The current best bounds are represented in Figs.~\ref{fig:exclusion} and \ref{fig:beta_n}. We consider the parameter ranges $10^{-2}~\text{meV}<\Lambda<10^2~$meV, $1<\alpha<10$ and $10^{-15}~\Mp<M<\Mp$.  

In addition to the PPN constraints, stringent constraints on the chameleon parameter space arise from the experimental tests in labs (or in space) and from astrophysical tests of gravity. As a reminder, the larger values of $\beta$ (or small values of $M$) the more efficient is the chameleon mechanism.
%, that is in the so-called thin shell regime, 
For relatively small values of $\beta$, the chameleon field tends to behave like a non-chameleonic field, i.e. quintessence. On the contrary, for large values of $\beta$, the Compton wavelength of the chameleon field $\lambda_\rr{C}$ becomes so small in the presence of massive objects that the range of the fifth force interaction becomes smaller than the size of the objects. It results that the fifth force seems to be sourced by the thin shell of matter at the surface of the  objects only \cite{Amendola:2012ys}. This is the so-called thin-shell regime where the screening mechanism occurs. The precise boundary between screened (i.~e. the thin shell regime) and unscreened objects is determined by the depth of the gravitational potential $\Phi$ (which is related to the density). Assuming spherically symmetric spacetime, the thin-shell parameter $\epsilon$ enables one to quantify the thin-shell effect,
\be
  \epsilon\equiv\frac{\phi_\infty-\phi_\rr{c}}{\Mp \Phi},
\ee
with $\phi_\rr{c}$ the central value of the sourcing object. The screening mechanism is efficient for $\epsilon\ll 1$.

%\tcb{Donner des ordres de grandeur du thin shell. Est-ce réaliste de demander à un champ classique de décroitre sur cette échelle caractéristique de la MQ ?}

\subsection{Constraints from astrophysics}
\label{sec:chamel_astro}
Following \cite{KhouryWeltmanPRD}, all the objects with a sufficiently large compactness, e.g. galaxies like the Milky Way ($s=10^{-6}$), must be screened in order to pass observational tests. However, screened and unscreened objects do not fall at the same rate, leading to a possible violation of the UFF \cite{Hui:2009kc}. The fact that an object is screened or not depends on its density and on its the environment as well as on $\beta$. Indeed, an object can be either self-screened either screened due to the environment (for instance, a dwarf galaxy ($s=10^{-7}-10^{-8}$) can be screened by the local group). Depending on the local environment in the Universe, for instance cosmic voids or above the cosmic mean, dwarf galaxies are thus less or more easily screened respectively since relatively dense environment can be more easily self-screened.
%In order to observe deviations from GR diffuse objects appear to be potentially interesting. 
In addition, when the matter is weakly coupled to the scalar field, i.~e. for small $\beta$, the screening mechanism is less efficient and the UFF is satisfied while if the coupling is strong, i.~e. for large $\beta$, violations can be detectable.
%provided that the external scalar field is primarily sourced by the density rather than the potential, the Yukawa suppression being thus avoided.  
%In the same paper \cite{Hui:2009kc} some observational tests have been proposed, comparing the difference in the acceleration of screened/unscreened galaxies using redshift space distortions or their clustering bias (see Sec.~\ref{sec:LSS}). 
%The chameleon backreaction on the galactic scales have been analyzed by \cite{Pourhasan}.
Several observations have been proposed following this idea. As long as no deviation from GR is detected for less and less compact objects, they have to be unscreened, giving rise to constraint on the $\beta$ parameter.

Some authors studied the effect of the chameleon field inside stars.
The chameleon field is found to leave possible imprints in the mass-radius relationship of both NSs and white dwarfs (see also Sec.~\ref{sec:NS}) by \cite{MotaShaw}. Because of the nonminimal coupling, variations of $\GN$ arise, potentially inside the star itself. Since $G_\rr{eff}$ is larger in unscreened environment, unscreened stars are subjected to a stronger gravitational force, which means that they are brighter and hotter while more ephemeral \cite{Davis:2011qf, Chang:2010xh} compared to screened stars which have (almost) the same properties as predicted by GR \cite{Davis:2011qf} (at the same mass) \cite{Sakstein:2014nfa}. Following \cite{Chang:2010xh}, the stellar evolution of red giants stars is modified, especially their color and luminosity, since the core of those stars ($s\gtrsim 5 \times 10^{-6}$ \cite{Davis:2011qf}) is screened while the envelope is not\footnote{The density of the core is roughly $10^{13} \times$ that in the mantle in red giant stars \cite{casoli2000astronomie}.}. In unscreened galaxies (neither self-screened nor screened by their environment) partially screened red giants stars are found to be hotter than completely screened ones at the same luminosity. This effect is potentially measurable since, looking at Hertzsprung–Russell diagram, that is the classification of stars as a function of their surface temperature and luminosity, there exists a tip in the red giant branch. The chameleon field affects the pattern of this tip, offering a unique signature for modified gravity.

Actually this effect has been found to apply also to the structure of the main sequence stars in the Hertzsprung–Russell diagram by \cite{Davis:2011qf}. In unscreened galaxies (that is dwarf galaxies in cosmic voids), only partially screened stars, which are more luminous and ephemeral than screened ones, can considerably enhance the total galactic luminosity. However, it is difficult to disentangle the chameleon effect to other ones, like the metallicity of the stars. Therefore Davis et al. \cite{Davis:2011qf} proposed  to measure the systematic offsets in luminosity between screened dwarf galaxies in clusters and unscreened galaxies in voids. 

Best bounds using stars have been obtained by comparing distance measurements inferred by the Cepheids and red giant stars observations \cite{Jain}. Since some stars are used as standard candles (at low redshift), a modification of their properties implies a change of distance measurement. In \cite{Jain} authors focused on two specific stages of the giant stars evolution, that is the tip of the red-giant branch and the Cepheids. The key idea is to compare distances inferred using Cepheids and red giants which would agree only in the screened galaxies. As mentioned above the tip is shifted while the period-luminosity relation for Cepheids is also modified (Cepheids pulsate at shorter period at fixed luminosity), both effects adding up to each other \cite{Jain}.  
No deviation has been found up to now, putting strong bounds on the chameleon parameter space \cite{Sakstein:2014nfa}. It results that the only possibly unscreened astrophysical objects in the Universe are isolated gas clouds, the smallest dwarf galaxies and very massive post main sequence stars \cite{Sakstein:2014nfa}.

\begin{figure}[!htb]
    \centering
    \includegraphics[width=0.6\textwidth]{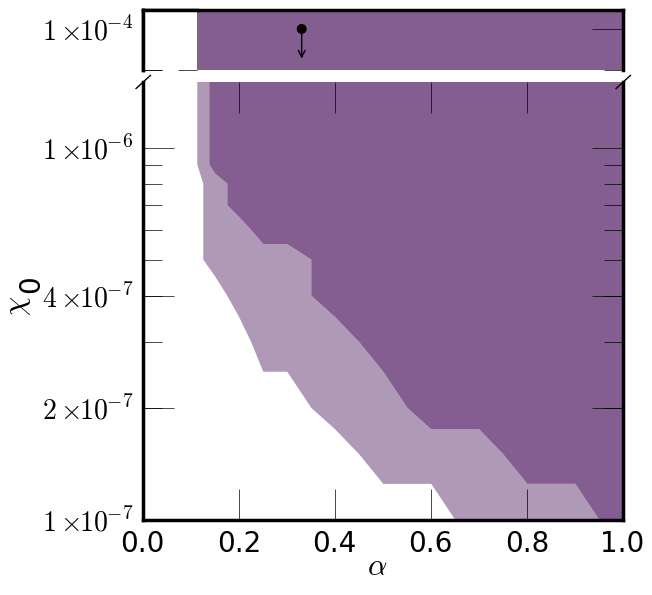}
    \caption{Constraints on the chameleon model using stars as a function of the parameters $\alpha$ and $\chi_0$. The  light  and  dark  shaded  regions  show  the
    regions excluded with 68\% and 95\% C.L. respectively. The black arrow indicates the previous
    constraint coming from galaxy cluster statistic \cite{Schmidt:2008tn}. Reprinted from \cite{Sakstein:2014nfa}.} 
    \label{fig:chamel_astro}
\end{figure}

%They are reported in \cite{Sakstein:2014nfa}.
The current constraints are represented in Fig.~\ref{fig:chamel_astro} for two dimensionless parameters: $\alpha$ which defines the strength of the fifth force interaction outside the thin-shell radius $S$ (see Eq.~\eqref{eq:radius_tt} for a mathematical definition),
\be
  G_\rr{eff}=\GN\left[1+\alpha\left(1-\frac{M(r)}{M(S)}\right)\right],
\ee
with $M(r)$ the mass enclosed into the radius $r$; and $\chi_0$ the self-screening parameter  which defines if an object is completely screened,
\be
  \chi_0<\Phi=\frac{\GN M}{\mathcal{R}}.
\ee

\subsection{Experimental tests of chameleon models}
\label{sec:chamel_exp_cons}
Since the formulation of the chameleon theory \cite{KhouryWeltmanPRL, KhouryWeltmanPRD}, fifth force searches have been found to constrain the chameleon parameter space. In order to compute predictions of the chameleon model, the scalar field profile has to be determined by solving the Klein-Gordon equation \eqref{eq:KG_chamel}. Assuming the spherical symmetry in a Minkowski background $g_{\mu\nu}\simeq \eta_{\mu\nu}$\footnote{More symmetries have been considered in \cite{MotaShaw} for instance.}, an approximation that is valid if the Newtonian gravitational potential is small everywhere and if the backreaction due to the density on $\phi$ remains small \cite{KhouryWeltmanPRD}, the Klein-Gordon equation reads \eqref{eq:KG_chamel},
\bea  \label{eq:KG}
\phi''+\frac{2}{r} \phi'=\frac{{\dd V_{\rm eff}}}{{\dd \phi}}, \hspace{1cm} \frac{{\dd V_{\rm 
eff}}}{{\dd \phi}}=\frac{{\dd V}}{{\dd \phi}} + 
\tilde{\rho} A^3 \frac{{\dd A}}{{\dd \phi}},
\eea
the prime denoting a radial coordinate derivative. Since the relativistic effects are negligible in lab experiments, metric potentials arising in static and spherically symmetric spacetime (see $\nu(r)$ and $\lambda(r)$ in Eq.~\eqref{eq:metric_schwa}) and pressure are assumed to be negligible. The radial profile of the chameleon field $\phi(r)$ can be solved either analytically (see App.~\ref{sec5}) or numerically (see Sec.~\ref{sec4} for an example of such numerical method) provided two boundary conditions\footnote{To be more precise it is necessary to take into account the Yukawa suppression of the fifth force, the second border condition being rather given by Eq.~\eqref{asympt}.}, 
\bea
  \phi'(r=0)&=&0, \\
  \phi(r\longrightarrow\infty)&=&\phi_\rr{min}(\rho_\infty),
\eea
the chameleon field being settled to its attractor that is the minimum of its effective potential, at spatial infinity. 

The fifth force mediated by the chameleon then reads,
\be
  \mathbf{F}_\phi=-\frac{m_\rr{test}}{M}\mathbf{\nabla}\phi
  \hspace{1cm}\Rightarrow\hspace{1cm}
  \mathbf{a}_\phi\equiv\frac{\mathbf{F}_\phi}{m_\rr{test}}=-\frac{1}{M}\mathbf{\nabla}\phi,
\ee
where $m_\rr{test}$ is the mass of the test particle and $a_\phi$ the acceleration induced by the chameleon.  If the scalar field is weakly coupled to matter, the scalar field profile varies like $a_{\phi}\propto 1/(r^2 M^2)$ outside the source mass (see Eq.~\eqref{eq:acc_phi_weak} for the analytical formula) whereas $a_{\phi}\propto 1/(r^2 M)$ in the strongly coupled case (see Eq.~\eqref{eq:acc_phi_strong} for the analytical formula). 
%In the latter case, the chameleon profile is thus independent on $\beta$. 
Computing the $\phi$ profile allows to compare the fifth force mediated by the chameleon to the experimental bounds on the strength of the fifth force $\alpha$ and its length of interaction $\lambda$ (see Eq.~\eqref{eq:fifthforce} where $\mathbf{F}_\phi\equiv -\mathbf{\nabla}V$ for the definition of the parameters $\alpha$ and $\lambda$). At fixed length of interaction, roughly $\lambda~\sim$~10~cm - 1~m for lab experiments, the strength of interaction $\alpha$ is constrained \cite{KhouryWeltmanPRD} (see Fig.~\ref{fig:fifth_force}).

%While compact objects enables to rule out the weakly perturbing regime, strongly coupled chameleons evade such bounds since the chameleon mechanism is much more efficient \cite{Mota, MotaShaw}, the lab experiments being more efficient in this case.
Fifth force test between two macroscopic objects like the Eöt-Wash experiment using torsion pendulum \cite{Kapner, Upadhye:2012qu} gave rise to lower bound $|\alpha|<1$ up to $\lambda=56~\mu$m at $95~\%$ C.L., enabling to rule out a part of the parameter space \cite{Adelberger:2006dh, Upadhye} for $\beta\sim1$. The experimental bounds obtained by Upadhye \cite{Upadhye} with a torsion pendulum are reported in Fig.~\ref{fig:beta_n}. Casimir forces experiment has been also used in order to test the fifth force by measuring the chameleon pressure between two parallel plates in the presence of a medium, e.g. a gas between the plates, affecting the chameleon field \cite{BraxCasimir, Brax:2010xx}. However, the separation between the two plates have to be relatively large, around 10~$\mu$m so that the electrostatic potential is not uniform between the plates. The experiment is thus not straightforward and the total force between the two plates is rather measured as a function of the gas density inside the cavity \cite{Brax:2010xx}.

%\textbf{Casimir force measurements: Measure of the Casimir force between two plates or one plate and a sphere ; the Casimir force is computed analytically and simulated numerically ; ratio of the chameleon force to the Casimir one is reachable... ! Something like $M\leq 10^{16}$ GeV.} REF BIBLIO??
However, such experiments testing fifth force between two macroscopic objects are not able to probe $\beta\gg1$. 
%Fifth force searches between two microscopic bodies allow, among others, to test the existence of light weakly coupled bosons, mediating short-range fifth forces. Such experimental test is complementary to collider experiments looking at heavy strongly interacting particles. While Casimir forces probe the interaction range of $\lambda\sim1$~nm, studying  neutrons and anti-protons allow to probe interaction range of $1~\text{pm}<\lambda<5~\text{nm}$ \cite{Nesvizhevsky:2007by} and $\lambda<1$~pm respectively.
In order to probe the extremely strongly coupled chameleons the fifth force searches between a macroscopic body and a microscopic one, e.g. cold neutrons, appear to be powerful \cite{Brax_neutron, Brax_neutron2, Jenke}. In this case, only the macroscopic body is screened while the cold neutrons may have no thin shell. In the first experiment, ultracold neutrons are bouncing above a mirror. Considering bouncing of the order of mm, neutrons exhibit quantum behavior, their energy levels being discrete, and appear to be unscreened \cite{Brax_neutron}. The chameleon introduces a shift in the quantum gravitational potential, possibly detected by neutron bouncing experiment \cite{Brax_neutron}. While fifth force searches between two macroscopic bodies give rise to upper bound on $\beta$, neutron experiments to lower bounds. The current experimental constraint for $\Lambda\sim2.4\times10^{-12}$~GeV are \cite{Jenke},
\be
    \beta<5.8\times10^{8} \hspace{1cm} \text{for} \hspace{1cm} -2\leq\alpha\leq2 \hspace{0.5cm}  (95~\%~ \text{C.L.}). 
\ee
A second set-up has been proposed using neutron interferometry \cite{Brax_neutron2, Li:2016tux}. Their experimental constraints are even more stringent. They are reported in Tab.~\ref{tab:neutron} and represented in Figs.~\ref{fig:beta_n} and \ref{fig:exclusion}.
%$\beta<10^{7}$ for $\alpha=1$ and $\Lambda\sim2.4\times10^{-12}$~GeV at 95~$\%$ C.L. 

Many other experimental tests of the chameleons have been proposed so far \cite{Ivanov, BraxBurrage, Shih1, Shih2, Haroche,Sukenik, Baum, Harber, Kasevich, Cronin, Harber}. Some of them were realized in space where the ambient density is weaker, the thin shell being thus easier to reach \cite{Joyce:2014kja, KhouryWeltmanPRD, Elder:2016yxm}. The list of experiments presented in this section is not exhaustive, though leaving a part of the parameter space unconstrained.

In the rest of this chapter, we will focus on one lab experiment based on atom interferometry, proposed by Burrage et al. \cite{Burrage} in 2014 and realized by \cite{khoury} in 2015. This experiment offers the best bounds on the chameleon parameter space from now, as reported in Figs.~\ref{fig:exclusion} and \ref{fig:beta_n}.
%,  where the main part of the remaining parameter space is found to be readily accessible. 
Like the neutrons, individual atoms are sufficiently small to let the scalar field unscreened even if their nuclei are dense. Cold atom interferometry experiments were developed recently for measuring the Newton's constant $\GN=6.67\times 10^{-11}~\rr{m^3}~\rr{kg^{-1}~\rr{s}^{-2}}$ with very good accuracy \cite{Fixler, Lamporesi}, the statistical uncertainty being given by $\pm 0.011 \times10^{-11}~\rr{m^3}~\rr{kg^{-1}~\rr{s}^{-2}}$ while the systematic uncertainty $\pm0.003\times10^{-11}~\rr{m^3}~\rr{kg^{-1}~\rr{s}^{-2}}$ \cite{Lamporesi}. Using laser-cooled atoms in a vacuum tube, the acceleration of the atoms due to the presence of a source mass was measured outside the tube. From the knowledge of the mass distribution of the source mass, $\GN$ was determined according to the Newtonian gravitational force.

The experimental setup proposed by Burrage et al. \cite{Burrage} is based on similar atom interferometry experiments while the source mass is now inside the vacuum chamber. It consists in measuring the additional acceleration on individual atoms, due to the scalar field gradient induced by the presence of a source mass at the center of the chamber (see also Sec.~\ref{sec3}). Forecasts provided in \cite{Burrage} and the first experimental results obtained by \cite{khoury}, highlight the fact that most of the remaining part of the chameleon parameter space corresponds to the case where the chameleon field is weakly coupled to matter, i.e. for $M/\mpl\sim 1$ (see Figs.~\ref{fig:exclusion} and \ref{fig:beta_n}). 

In the rest of this chapter, we will provide numerical simulations for the chameleon profile and acceleration measured by the Berkeley experiment. The experimental setup is briefly reviewed in Sec.~\ref{sec3} and the numerical strategy is detailed in Sec.~\ref{sec4}. Numerical results are presented in Sec.~\ref{sec:strong} for the thin shell regime (we report the reader to \cite{Schlogel:2015uea} for a discussion about the weak field regime) where they are compared to analytical results reviewed in App.~\ref{sec5}.  
We finally discuss our results and draw some conclusions and perspectives in Sec.~\ref{sec_CCL}.

%Considering experimental bounds on the chameleon parameter space the range $10^{-5}~\Mp\lesssim M \lesssim \Mp$ has not been experimentally ruled out yet by the atom interferometry experiment \cite{Elder:2016yxm} for $\Lambda=2.4~$meV\footnote{In the following we only consider this value corresponding to the cosmological constant}. The contour plots (reprinted from \cite{Elder:2016yxm}) are reproduced in Figs.~\ref{fig:exclusion} and \ref{fig:beta_n} (see also \cite{Brax_neutron2} for the bounds obtained by neutron interferometry). In Fig.~\ref{fig:exclusion} both neutron interferometry and atom interferometry exclusion bounds are represented for $\alpha=1$ while in Fig.~\ref{fig:beta_n} the parameter space $M-\alpha$ is represented at fixed $\Lambda=2.4~$meV for the same experiments and the torsion pendulum one \cite{Upadhye:2012qu}. Depending on the assumed geometry (spherical or infinite cylindrical, see Eq.~\eqref{phibg_khoury}) the constraints are in light or dark blue. The experimental bounds depicted in Figs.~\ref{fig:exclusion} and \ref{fig:beta_n} have been obtained analytically by \cite{khoury} (see Sec.~\ref{sec:param_space} for details). After describing further the experimental set-up in Sec.\ref{sec3} we will propose a numerical method able to refine those constraints in Sec.~\ref{sec4}. Analytical results are reminded in Sec.~\ref{sec5} and are compared to the numerical results in Sec.~\ref{sec:strong} for the two considered models, refined constraints being established on their parameters.  

\begin{table}[]
\centering
\begin{tabular}{|c|c|c|c|c|c|c|}
\hline
$\alpha$            & $1$ & $2$ & $3$ & $4$ & $5$ & $6$ \\ \hline
$\beta_\rr{lim} \times 10^6$ 	  & $4.7$  & $8.2$  & $12.7$  & $17.9$ & $20.4$ & $23.8$ \\  \hline
\end{tabular}
\caption{Experimental bounds obtained using neutron interferometry. $\beta_\rr{lim}$ corresponds to the upper bounds on $\beta$ at $95~\%$ C.L. \cite{Li:2016tux}.}
\label{tab:neutron}
\end{table}

\section{Experimental setup of the Berkeley experiment}\label{sec3}
%\tcb{Analyse ou commentaire de synthèse pour mettre en évidence l'intérêt de modéliser l'expérience de Berkeley???}
In the last decade, the chameleon model has been tested thanks to cosmological and astrophysical observations, as well as lab experiments, using neutron and atom interferometry. As we can see in Figs.~\ref{fig:exclusion} and \ref{fig:beta_n}, the atom interferometry experiment performed in Berkeley provides the most stringent constraints on the parameter space of the chameleon model today, for large $\beta$ or small $M$. Those results were obtained by measuring the acceleration induced by the chameleon field on cesium-133 atoms inside a ultra-high vacuum chamber in the presence of a source mass. We will here explain in more detail this experimental setup \cite{khoury}. 
% 
% Laboratory experiments measuring the acceleration induced by a source mass can be used to probe and constrain 
% modifications of gravity.  
% As a reminder, the chameleon field is screened in high density environments while it mediates long-range force in 
% sparse ones.
% Therefore atomic particles in an ultra-high vacuum chamber can mimic cosmos conditions. In the first experiments, the 
% source mass was located outside the vacuum chamber \cite{Fixler, 
% Lamporesi}, an 
% experimental setup which is not ideal given that the chamber wall screens the fifth force on the atoms. New experiments 
% have been proposed in 
% \cite{Burrage, khoury} where the
% source mass is located inside the vacuum chamber, which improves the constraints on the acceleration due to the scalar 
% field. 
%The idea is to probe the fifth force (or equivalently the acceleration) between the atoms and the macroscopic 
%source mass due to chameleon field. 
%Here we focus on a recently proposed atom interferometry experiment \cite{khoury} 
%where one 

\begin{figure}
\begin{center}
\includegraphics[scale=0.4]{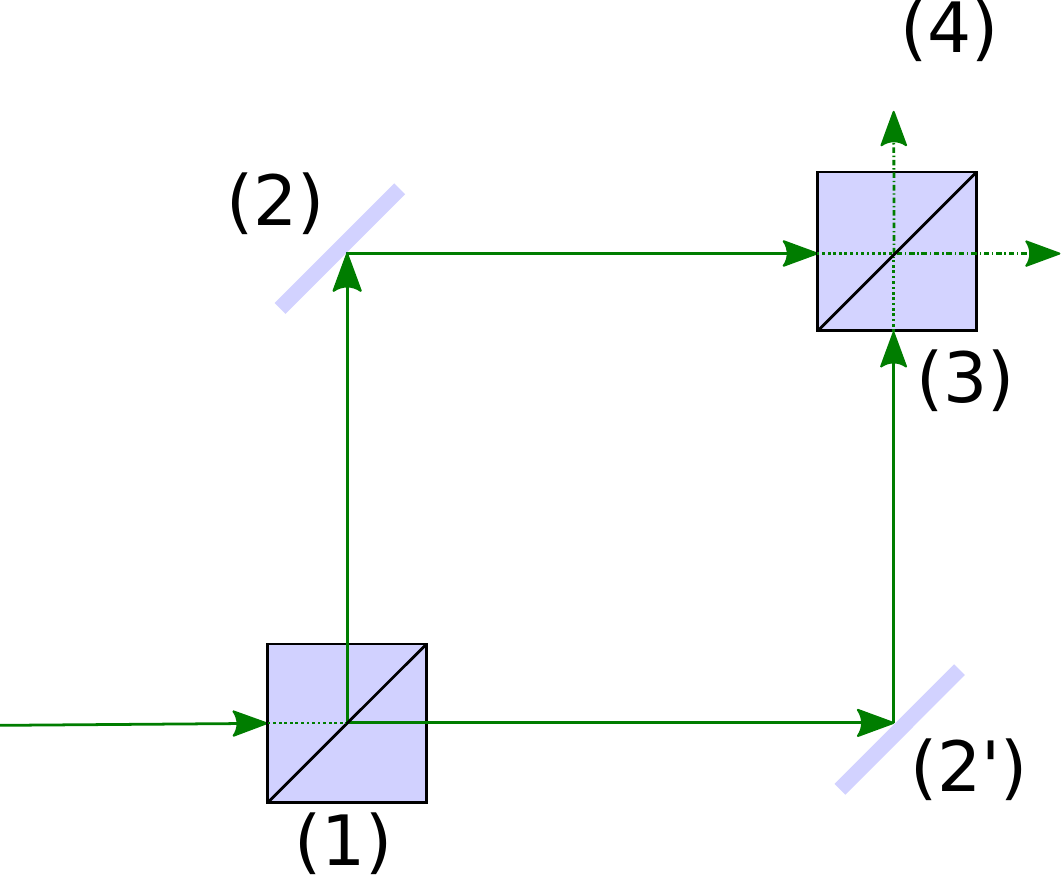}
\caption{Illustration of the Mach-Zehnder interferometer. The incident beam goes through the beamsplitter (1). The two resultant beams are reflected on mirrors (2) and (2') and are recombined by the beamsplitter (3). A detector measures the fringes of interference in (4).}
\centering
\label{fig:interferometer}
\end{center}
\end{figure}

According to quantum mechanics cesium-133 atoms exhibit matter-wave properties 
in a Fabry-Perot cavity. When an atom absorbs/emits a photon, it recoils with a momentum $p=\hbar k$, with $k$ the 
wave number of the absorbed/emitted photon. So, 
one can reproduce the equivalent of a Mach-Zehnder interferometer represented in Fig.~\ref{fig:interferometer} for cold atoms
with three light pulses using counter-propagating laser beams. 
Atoms are initially prepared in a hyperfine state $F=3$ and stored in a 2 dimensional magneto-optical trap.
A first light pulse splits the matter-wave packet in two hyperfine states $F=3$ and $F=4$ (see the beamsplitter $(1)$ in Fig.~\ref{fig:interferometer}) and gives an impulse of $\hbar 
k_{\rm eff}$ to the atoms. The effective wave number $k_{\rm eff}$ depends on the two counterpropagating beam 
wave numbers. The probability of hyperfine transition can be 
controlled by the intensity and duration of both 
laser beams.
The second pulse reverses the relative motion of the beams like the mirror of Mach-Zehnder interferometer (see the mirrors $(2)$ and $(2')$ in Fig.~\ref{fig:interferometer}) and  
the third pulse acts like a beam splitter which allows overlap of partial matter wave packets (see the beamsplitter $(3)$ in Fig.~\ref{fig:interferometer}). 
Because of the recoil of the atoms, the phase difference between the two arms of the interferometer $\Delta \phi$ is 
a function of the acceleration $a$ of atoms,
\be
\Delta \phi=k_{\rm eff} a T^2,
\ee
where $T\sim 10 \,\rm{ms}$ in general, is the time interval between two pulses. 
To alleviate some systematics effects, counterpropagating laser beams are reversed 
and the aluminum sphere can be positioned in two 
places: a \textit{near} and a \textit{far} positions (the source mass surface is respectively located 8.8 mm and 3 cm far from the atoms), 
which allows to disentangle the 
contribution from chameleon force to Earth's gravity.
One measurement consists thus of four interference fringes, corresponding to reversed counterpropagating laser beams 
and both positions of the source mass.
Using this setup, the acceleration induced by the chameleon has been excluded up to
\be
a_\rr{exp}<5.5 \,\rr{\mu m/s^2} \hspace{0.5cm} \Leftrightarrow \hspace{0.5cm} \frac{ a_\rr{exp} }{g}<5\times 10^{-7} \hspace{1cm}\text{at 95$\%$ C.L.},
\label{eq:acc_exp}
\ee
with $g$ the Earth's acceleration of free fall. As a comparison, the Newtonian gravitational attraction due to the source mass is $a_\rr{N}/g(r=8.8~\rr{mm})=\left.(\GN \mA/r^2)/g\right|_{r=8.8~\rr{mm}}=2.25\times10^{-10}$ at the position where the acceleration of the atoms is measured. Since it is more than two orders of magnitude below the current experimental sensitivity, the gravitational acceleration due to the source mass is neglected.

\begin{figure}
\begin{center}
\includegraphics[scale=0.28]{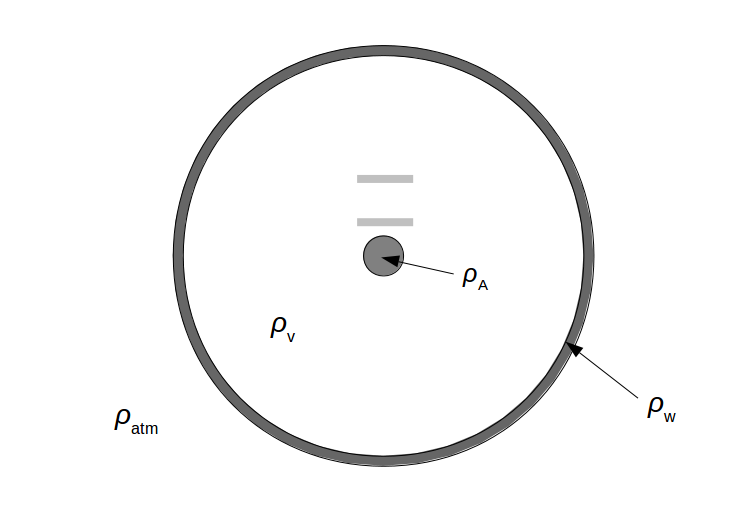}
\caption{Outline of the atom-interferometry experiment, simulated by a four-region model including the source mass, the vacuum chamber, its wall and the exterior environment. In light gray, the \textit{near} and \textit{far} 
positions where the acceleration on atoms is measured (note that we consider a fixed source mass to keep spherical symmetry whereas in the real experimental setup the source mass is moved \protect\cite{khoury}) .}
\centering
\label{plot_exp}
\end{center}
\end{figure}

The experimental setup proposed in \cite{Burrage} is similar, except that they plan to use cooled rubidium atoms 
launched in a small fountain located 1 cm far from the source mass.
Our numerical simulations can be easily adapted for such a configuration. 
Details of the considered experimental setup %we use for our numerical simulations 
are reported in Table~\ref{tab:expconf}. 
%Parameters are \textcolor{red}{those of \cite{khoury, Burrage}, except the wall 
%thickness and density which were not specified in previous papers since the effects of the wall were assumed to be 
%negligible, contrary to what our numerical results show}.
The size and density of the central mass, the geometry of the chamber and the vacuum density are those of \cite{khoury, Burrage}.  
In addition we consider the thickness and density of the vacuum chamber wall, as well as the exterior density.
In Fig.\ref{plot_exp}, we draw the experimental setup considered
in our numerical simulations. The four regions are labeled by their densities%\footnote{In the remainder of this chapter, 
%$\rho$ refers to the density in the Jordan frame.}
: (1) 
the source mass made of aluminum (${\rho}_A$), (2) the vacuum where the acceleration due to the chameleon is measured 
(${\rho}_{\rm v}$), (3) the wall of the chamber (${\rho}_{\rm w}$) made of stainless steel, (4) the exterior of the chamber, mostly filled by air 
at atmospheric pressure (${\rho}_{\rm 
atm}$). 

\begin{table} 
\begin{center}
\begin{tabular}{|c|c|c|} 
\hline
$\RA$ & Radius of the source mass  &  $1 \rm{cm}  / 5.1\times10^{13} \rm{GeV}^{-1} $  \\
$\RL$ & Radius of the chamber & $10 \rm{cm}  / 5.1\times10^{14} \rm{GeV}^{-1} $  \\
$\WT$ & Wall thickness & $1 \rm{cm}  / 5.1\times10^{13} \rm{GeV}^{-1} $   \\
$\mA$ & Test mass  &  11.3g /$6.7\times 10^{24}\rm{GeV}$  \\
$\rhoA$ & Test mass density  & $1.2\times 10^{-17}\rm{GeV}^4$ \\
$\rhoW$ & Wall density &  $3.5 \times 10^{-17} \rm{GeV}^4$ \\
$\rhoV$ & Vacuum density & $5.0 \times 10^{-35} \rm GeV^4$ \\
$\rhoATM $ & Air density $\left(P_\rr{atm}\right)$ & $ 5.2\times 10^{-21} \rm GeV^4$ \\
\hline
\end{tabular}
\caption{Fiducial experimental parameters, corresponding to the 
setup of \protect\cite{khoury} .}
\label{tab:expconf}
\end{center}
\end{table}

\section{Numerical strategy}\label{sec4}
Analytical approaches have been considered so far \cite{Burrage, 
khoury}, which are valid under some theoretical assumptions like static approximation\footnote{A numerical method for studying screening mechanisms in cosmology beyond the quasistatic approximation has been proposed in \cite{Llinares:2013qbh}.}, Minkowski background spacetime, the linearization of the solution (see also \cite{MotaShaw} for a discussion), the coupling function $A(\phi)\sim 1$, and experimental ones like negligible chamber 
wall effects. Numerical methods enable to (in)validate those assumptions and to refine analytical results, by including the effects due to the experimental setup, like the 
thickness and the density of the wall as well as the exterior environment. In the future, numerical results will be also helpful to study 
more realistic situations where the vacuum chamber is not exactly spherical or cylindrical (see e.g. \cite{Elder:2016yxm}).

We consider two methods for solving the KG equation \eqref{eq:KG}: a singular and multipoint boundary
value problem (BVP) solver with unknown parameter 
and a non-linear BVP solver implementing up to sixth order a mono-implicit Runge-Kutta method 
with an adaptative mesh refinement, working in \texttt{quad} precision\footnote{For this purpose we have used the 
Matlab function \texttt{bvp4c} which deals with singular BVP's and a modified version of the \texttt{mirkdc} 
BVP solver with adaptative mesh in Fortran.}.
In the latter case, the density in the four regions was made continuous by considering arctan profiles with negligible widths. The requirement of the \texttt{quad} precision in order to solve the Klein-Gordon equation reveals the strongly fine-tuned nature of chameleon physics.

%\textcolor{red} {In the case where the scalar field where the source mass perturbs the scalar field profile only weakly, 
%the so-called weak-field regime, 
We take the minimal assumption which states that the scalar field is settled to its 
attractor at spatial infinity, i.e. 
$\phi_\infty=\phi_{\rm min}(\rho_{\rm atm})$ as \cite{khoury}.
Then, the asymptotic scalar field profile is obtained by linearizing the KG equation up to first order around spatial infinity,
\bea
\phi''+\frac{2}{r} \phi'= \mathcal{M}^2 \left(\phi-\phi_{\infty}\right),
\eea
with $\mathcal{M}^2=\left.\dd^2 V_{\rm eff}/\dd\phi^2\right|_{\phi=\phi_{\infty}}$, which admits the Yukawa profile 
solution $(\mathcal{M}^2>0)$
\be
\phi=\phi_{\infty}+\frac{\mathcal{C}{\rm e}^{-\mathcal{M}r}}{r},
\label{asympt}
\ee
with $\mathcal{C}$ the constant of integration. Since the KG equation is of second order and the parameter 
$\mathcal{C}$ is to be determined, three boundary conditions are needed.
They are provided by the regularity condition on the scalar field derivative at the origin $\phi'(r=0)=0$ and by the 
asymptotic behavior of $\phi$ and $\phi'$ given by Eq.\eqref{asympt} at the end of the 
integration interval. For the multipoint BVP method, the continuity of $\phi$ and $\phi'$ are imposed at 
the interfaces of each region (6 conditions) while the profile is guaranteed to be continuous for arctan profiles of density.
The density and size of each region are reported in Table~\ref{tab:expconf}.
The two numerical methods have been checked to be in agreement with each other.
Their applicability to the various regimes and their limitations in the deep thin-shell regime 
will be discussed in Sec.~\ref{sec:strong}.  We already point out that this numerical method enables one to properly account 
for the effect of neighboring matter on the chameleon fields and can be easily generalized to other experiments, possibly more sensitive (in the limit of spherical symmetry).

\section{Four-region model: numerical results}\label{sec:strong}
As we can see in Figs.~\ref{fig:exclusion} and \ref{fig:beta_n}, the atom interferometry experiment enables to probe the chameleon models for relatively small $M$ (or large $\beta$), i.e. in the thin shell regime or the so-called strongly perturbing regime (see App.~\ref{sec5} for the mathematical definition). In this regime, the chameleon field is screened in the source mass. On the contrary, the chameleon is unscreened inside the source mass is the so-called weak field regime.
The analytical solutions in those two limit cases are reported in App.~\ref{sec5}. As we will see, the numerical treatment allows one also to probe the transitory regime between the weakly and strongly perturbed cases in addition to refine analytical constraints.  

Since the original chameleon model has been already ruled out by the combined constraints on cosmological and PPN parameters \cite{Hees}, we will mainly focus on the exponential chameleon. We report thus the reader to our paper \cite{Schlogel:2015uea} for the complete numerical results, in particular in the weak-field regime. 

However, probing the deep thin-shell regime, i.e. for very small values of $M$,
%when $(R_{\rr {A,w}} - S_{\rr {A,w}})\ll R_{\rr {A,w}} $ (thin-shell radius of the source mass or of the chamber wall), 
is very challenging numerically because the problem becomes stiff, the chameleon physics being strongly fine-tuned. Up to some point, it is nevertheless possible to track the solution and to check the validity of the analytical estimations, typically using mesh refinement methods. Since the original model is easier to probe we briefly discuss our results in the thin-shell regime.

\subsection{The original model}
\label{sec:strong_field_cham_1}
\begin{figure}[!tbp]
  \centering
  \subfloat[Scalar field profile $\phi$.]{\includegraphics[width=0.5\textwidth, trim= 240 0 250 0, clip=true]{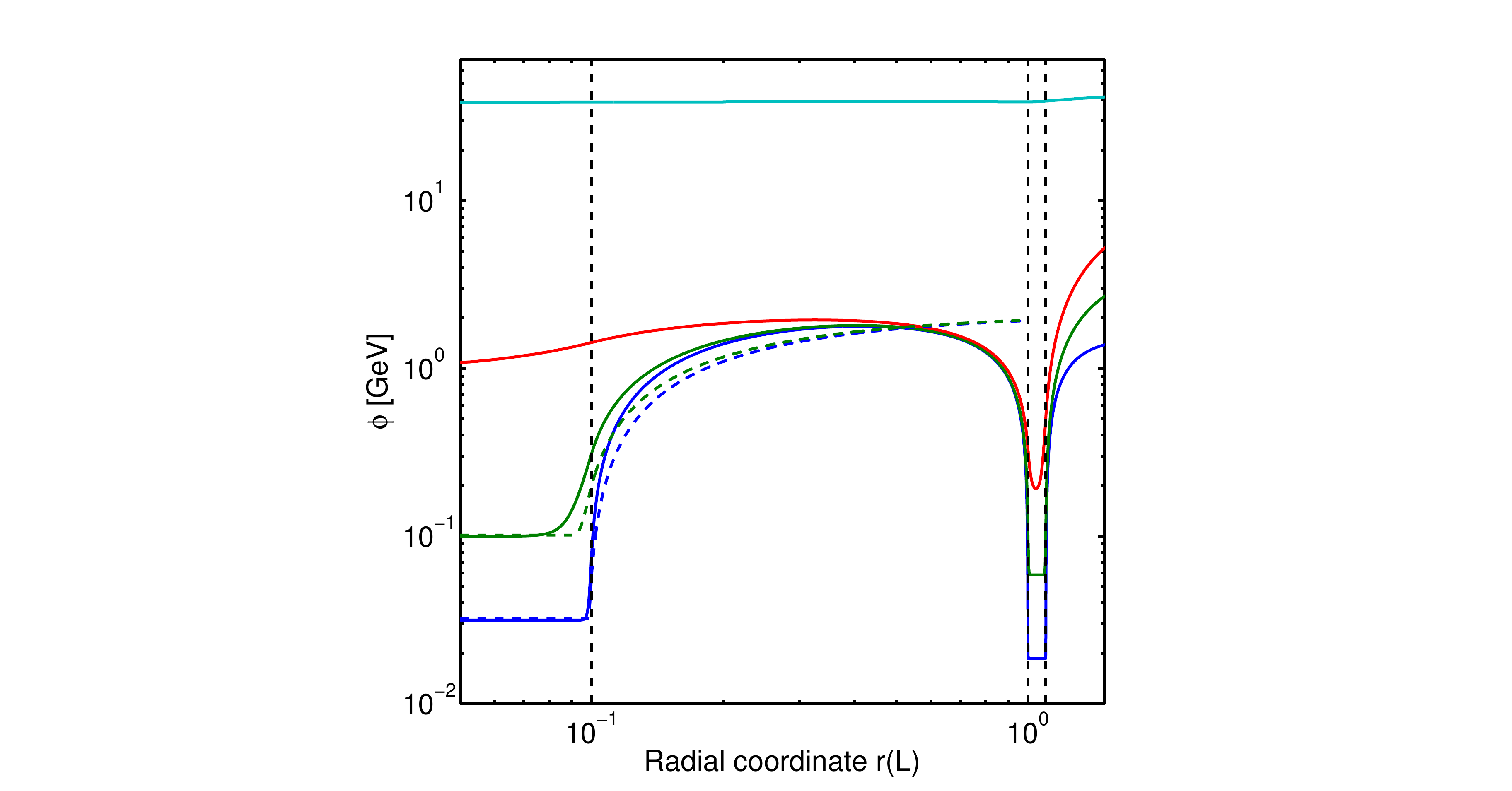}
  \label{plot:profile_field_thinshell1}}
  \hfill
  \subfloat[Acceleration profile  $|a_\rr\phi| / g$. The numerical profile for the four-region model shows that from the middle of the chamber to the wall, the acceleration becomes negative and increases in magnitude. The Newtonian gravitational attraction due to the test mass is plotted in black dashed line.]
  {\includegraphics[width=0.5\textwidth, trim= 340 0 350 0, clip=true]{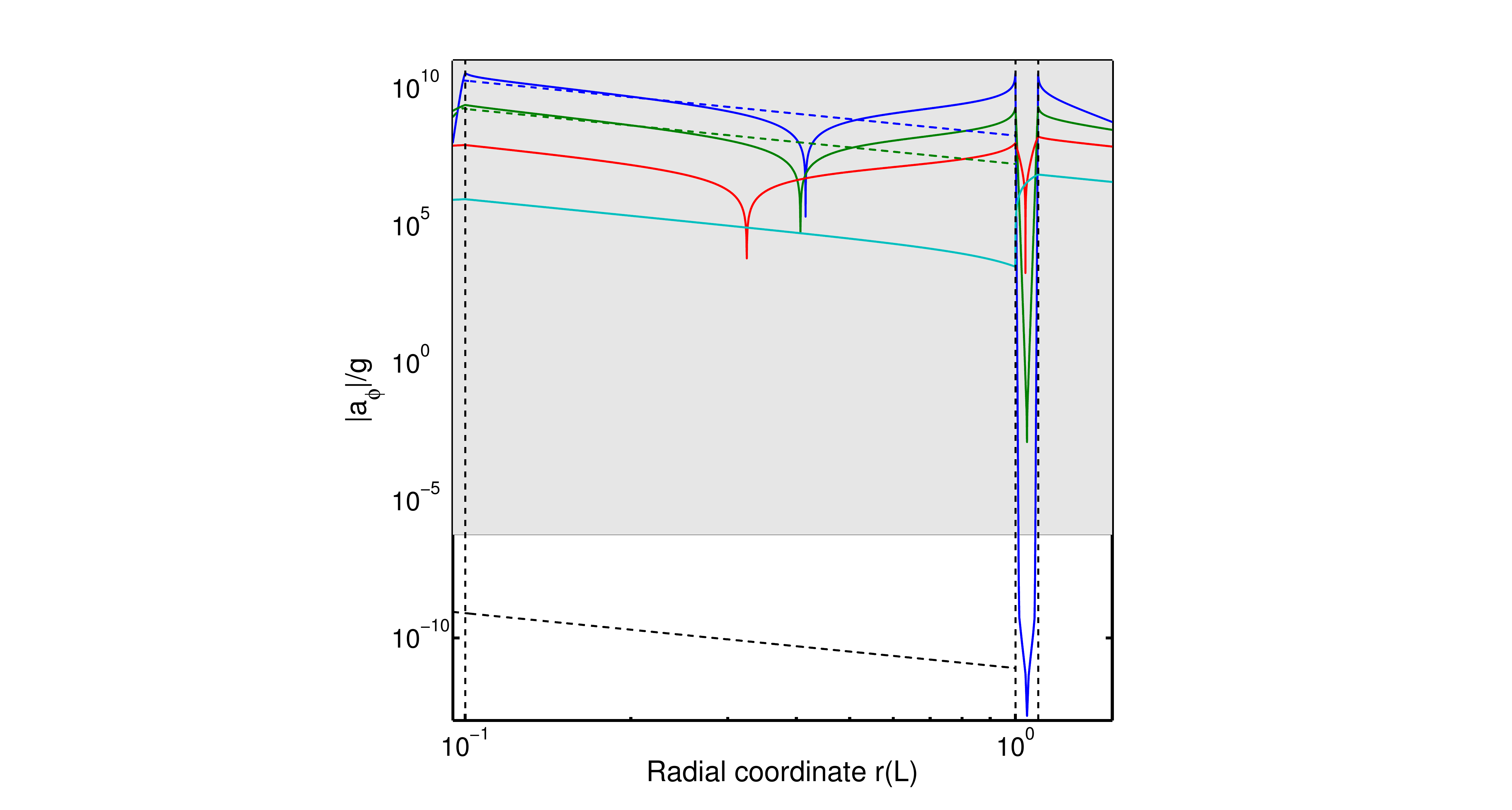}
  \label{plot:profile_acc_thinshell1}}
  \caption{Numerical results (solid lines) and analytical approximation (dashed lines) for the original model, in the \textit{strongly perturbing} (thin shell) regime, for $\Lambda = 2.6 \times 10^{-6} \GeV$ and values of the coupling $M$ listed in Table~\ref{tab:profiles}.  Differences between the two-region and four-region models are non-negligible inside the chamber, especially in the vicinity of the wall.
  Vertical lines mark out the four regions (source mass, chamber, wall and exterior).}
\end{figure}

The numerical scalar field and acceleration profiles for several values of $M$ considering the original chameleon model are represented in solid lines in Figs.~\ref{plot:profile_field_thinshell1} and~\ref{plot:profile_acc_thinshell1}, the dashed lines corresponding to the analytical calculations reported in App.~\ref{sec5}. The shaded region indicate the upper bound on the acceleration \eqref{eq:acc_exp}, the atom interferometry being thus able to rule out the original chameleon model in the strong field regime. This result is in agreement with the analytical analysis (see the Case (3) in App.~\ref{sec5}).

%Our numerical method also allows one to probe the transitory regime.

\begin{table} 
\begin{center}
\begin{tabular}{|c|c|c|c|} %c| 
\hline
Color & M [GeV] & $a_\phi/g$ (\textit{near}) & $a_\phi/g$ (\textit{far}) \\ % $\phi_\infty$ [GeV] &
\hline
% \multicolumn{4}{|c|}{Chameleon-1, \textit{weakly perturbing}: Figs.~\ref{plot_profile_field_k},~\ref{plot_profile_acc_k}  }  \\
% \hline
% %Blue & $10^{11}$  & 48 & $1.3 \times 10^{5}$ & $5.2 \times 10^{4}$\\
% %\hline
% Blue & $10^{13}$&  $1.3 \times 10^{1}$ & $2.8\times 10^0$ \\ %$4.8\times 10^2$ &
% %\hline
% Green & $10^{15}$  &  $1.3\times 10^{-3}$ & $2.8\times 10^{-4}$\\ %$4.8 \times 10^3$ &
% %\hline
% Red & $10^{17}$ &  $1.3\times 10^{-7}$ & $2.8\times 10^{-8}$\\ %$4.8 \times 10^4$ &
% %\hline
% Light blue & $10^{19}$ &  $1.3\times 10^{-11}$ & $2.8\times 10^{-12}$\\ %$4.8 \times 10^5$ &
% \hline
\multicolumn{4}{|c|}{Original chameleon, \textit{thin-shell}: Figs.~\protect\ref{plot:profile_field_thinshell1},~\protect\ref{plot:profile_acc_thinshell1}  }  \\  \hline
Blue & $10^{8}$ &  $5.8 \times 10^{9}$ & $1.4 \times 10^{8}$ \\ %$1.5$ &
Green & $10^{9}$ &  $5.2 \times 10^{8}$ & $5.7 \times 10^{6}$   \\%$1.5$ &
Red & $10^{10}$ &    $1.9 \times 10^{7}$ & $-4.4 \times 10^{6}$  \\ %$4.8 $ &
Light blue & $10^{11}$ &  $2.5 \times 10^{5}$  & $5.5 \times 10^{4}$  \\ %$ 15$  &
\hline
\multicolumn{4}{|c|}{Exponential chameleon, \textit{thin-shell}: Figs.~\protect\ref{plot:cham2_field},~\protect\ref{plot:cham2_acc}  }  \\ \hline
Blue & $10^{14}$&  $ 5.2 \times 10^{-7}$  & $ 1.5 \times 10^{-8}$ \\ %$3.9 \times 10^{10} $ &
Green & $10^{15}$ & $ 5.2 \times 10^{-8}$ & $ 1.5 \times 10^{-9}$ \\ %$3.9 \times 10^{10} $ &
Red &$10^{16}$ & $ 5.2 \times 10^{-9}$ & $ 1.5 \times 10^{-10}$  \\ %$3.9 \times 10^{10} $&
Light blue & $10^{17}$ &  $ 5.2 \times 10^{-10}$ &$ 1.5 \times 10^{-11}$  \\ %$3.9 \times 10^{10} $ &
Purple &$10^{18}$ & $ 5.3 \times 10^{-11}$ &$ 2.4 \times 10^{-12}$  \\ %$3.9 \times 10^{10} $ &
Beige&$10^{19}$ & $ 4.6 \times 10^{-12}$ &$ 6.8 \times 10^{-14}$  \\
\hline
\end{tabular}
\caption{Properties of the numerical scalar field and acceleration profiles
% plotted in 
%Figs.~\ref{plot_profile_field_k} and \ref{plot_profile_acc_k}
for the two models in the different regimes.}
\label{tab:profiles}
\end{center}
\end{table}

As expected from Eq.~\eqref{eq:phimin} given that $\rho_{\rr A} \lesssim \rho_{\rr w}$ with similar source mass radius and wall thickness, when lowering M, the field reaches first the potential minimum $\phi_{\rr{min}}(\rho_{\rr w})$ inside the wall, and then $\phi_{\rr{min}}(\rho_{\rr A})$ within the source mass, over a very thin radius.  
Analytical results appear to be in good agreement with the numerical ones, especially close to the source mass where $a_\phi$ is measured whereas important deviations are found close to the wall. In particular, one observes that the field roughly reaches
the amplitude of the central value of the scalar field in the absence of the source mass $\phibg$ given by Eq.~(\ref{phibgSP}) inside the vacuum chamber, which validates the analytical calculation of~\cite{Burrage} (see App.~\ref{sec5} for the details). In the vicinity of the chamber wall, however, the acceleration changes its sign and becomes negative, with a comparable magnitude with the acceleration close to the source mass.   
%This effect can only be probed by the four-region model, it changes importantly the observational predictions
%in a large part of the chamber and therefore it should be considered in future investigations. 
This effect could be helpful experimentally to discriminate between a signal of modified gravity and systematic errors, by performing measurements of the acceleration at several key positions of the chamber.

 %Its amplitude reaches values comparable to the acceleration at a position close to the source mass, which is a potentially measurable prediction that could be useful to discriminate between experimental systematic effects and an acceleration induced by the presence of some scalar field.

\subsection{The exponential model}  \label{sec:strong_field_cham_2}

\begin{figure}[!tbp]
  \centering
  \subfloat[Scalar field profile $\phi$.]{\includegraphics[width=0.6\textwidth, trim=240 0 260 0, clip=true]{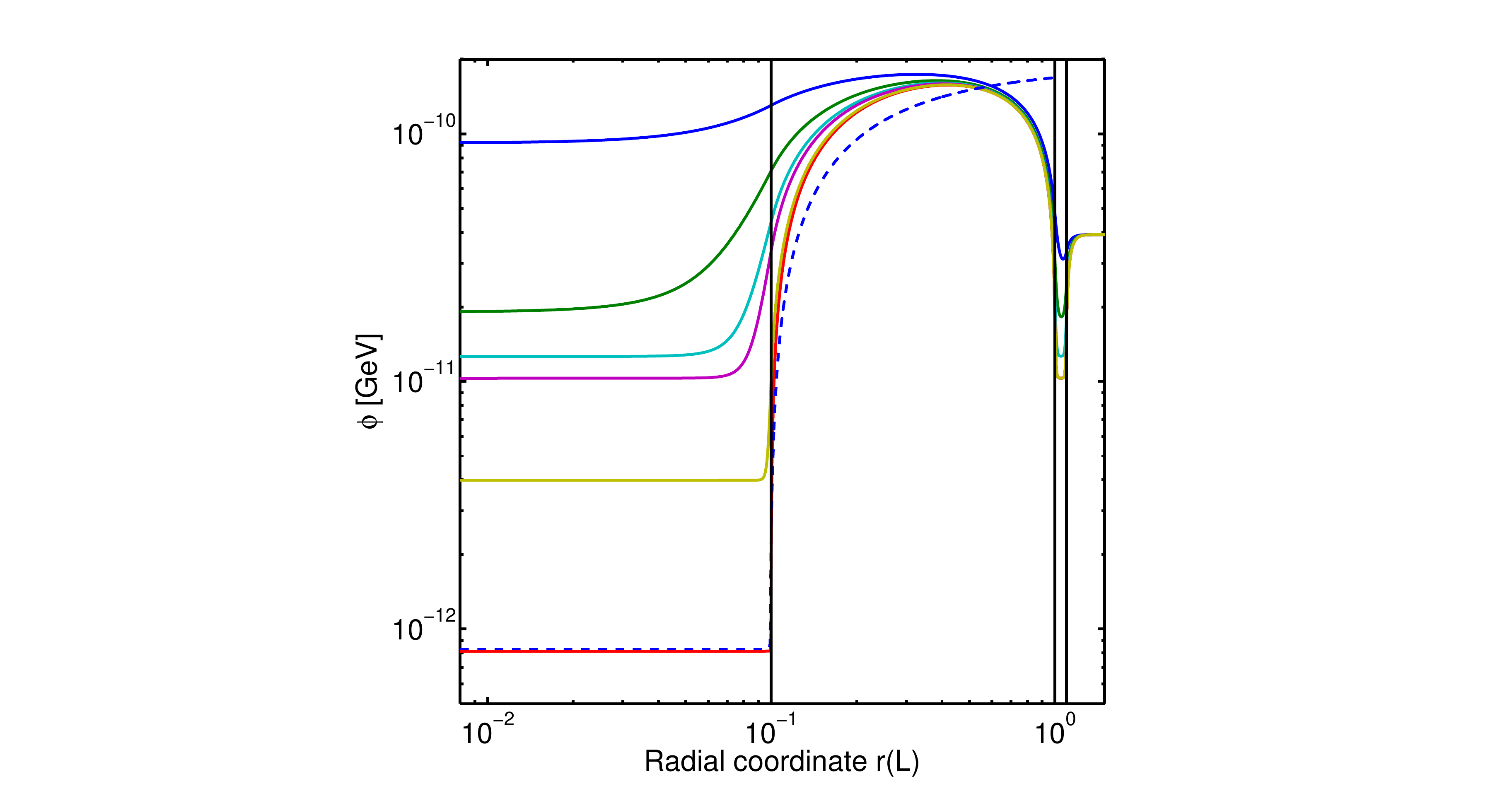}
   \label{plot:density_field}}
  \hfill
  \subfloat[Acceleration profile  $|a_\rr\phi| / g$.]
  {\includegraphics[width=0.6\textwidth, trim= 340 0 350 0, clip=true]{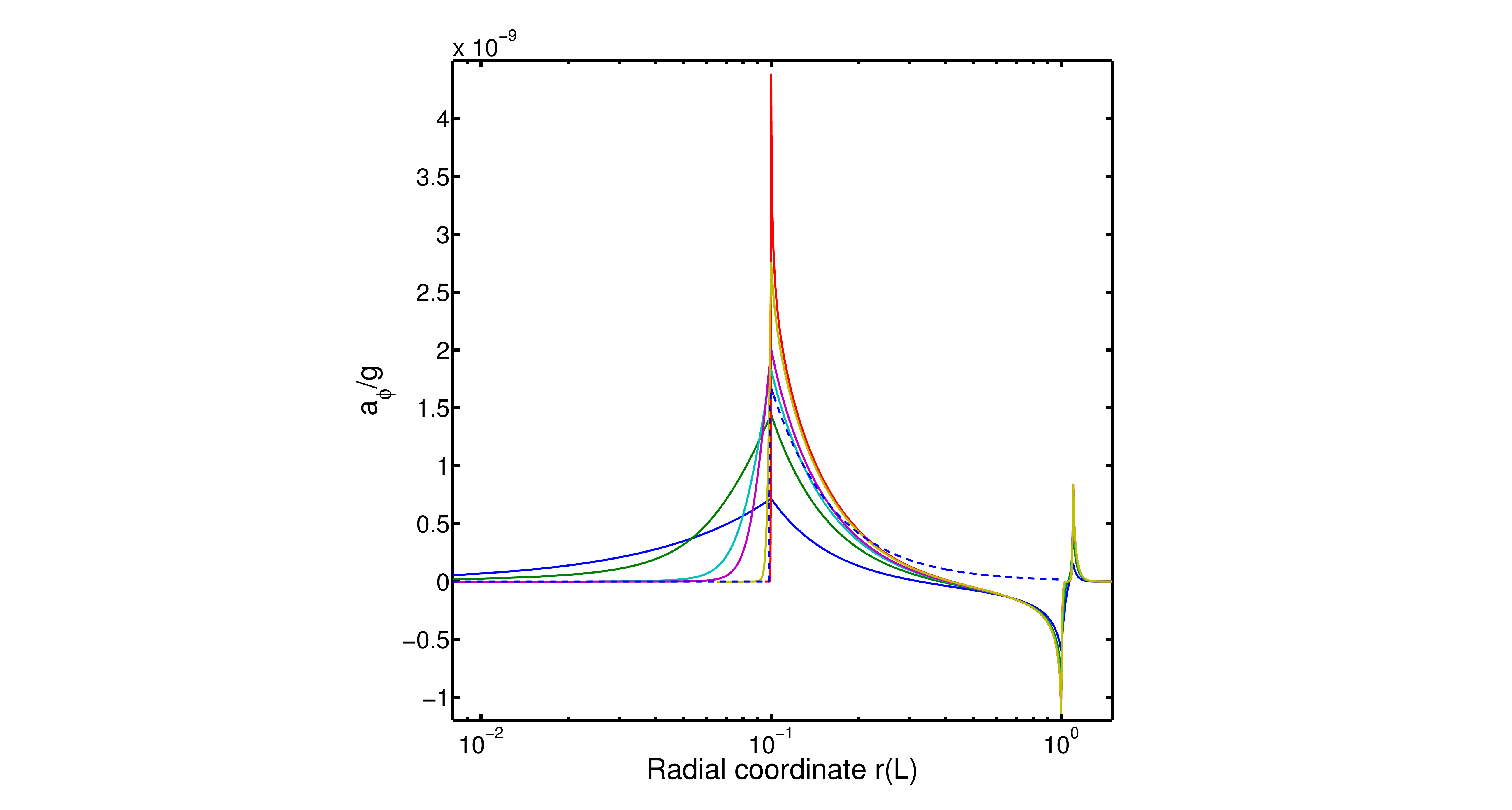}
   \label{plot:density_acc}}
  \caption{Numerical results (solid lines) and analytical approximation (dashed line) for the exponential model, for various $\rhoA$ and $\rhoW$ reported in Tab.~\ref{tab:density}, $M=10^{17}$ GeV being 
  fixed.}
\end{figure}

For the exponential model and the considered experimental set-up, 
it has been impossible to track numerically the thin-shell regime up to the point where the acceleration would have been large enough to be observed in laboratory experiments.  
Nevertheless, the field and acceleration profiles are represented in Figs.~\ref{plot:density_field} and~\ref{plot:density_acc}, for $M=10^{17} \rr{GeV}$ and increasing values of $\rho_{\rr w}$ and $\rho_{\rr A}$.   The attractor field values within the source mass and the wall are reached progressively and the field variations at the borders between the four regions become more steep, as expected given that 
$(R_{\rr {A,w}} - S_{\rr {A,w}}) / R_{\rr {A,w}}  \propto M \rho_{\rr{A,w} }^{-1} R_{\rr {A,w}}^{-2}  $  (see Eq.~(\ref{eq:radius_tt})).  In the case $M=10^{17} \rr{GeV}$,  the attractor is reached inside the source mass for $\rho_{\rr A} \simeq 5 \times 10^{-20} \rr{GeV}^4$, i.e. about 1000 times lower than the aluminum density,  whereas inside the wall, it is reached for $\rho_{\rr w} \simeq 7.5 \times 10^{-20} \rr{GeV}^4 $.  This slight difference is explained by the fact that the central source mass has a diameter two times larger than the wall thickness. 

\begin{table} 
\begin{center}
\begin{tabular}{|c|c|c|} 
\hline
Color & $\rhoA~\left[\rr{GeV}^4\right]$ & $\rhoW~\left[\rr{GeV}^4\right]$  \\
\hline
Blue & $1.0\times 10^{-20}$& $1.0\times10^{-20}$  \\
%\hline
Green & $2.5\times 10^{-20}$  & $2.5\times 10^{-20}$\\
%\hline
Light blue & $5.0\times 10^{-20}$ & $5.0\times 10^{-20}$ \\
%\hline
Purple & $7.5\times 10^{-20}$ & $7.5\times 10^{-20}$ \\
%\hline
Beige  & $5.0\times 10^{-19}$ & $7.5\times 10^{-20}$ \\
%\hline
Red & $1.2\times 10^{-17}$ & $7.5\times 10^{-20}$ \\
\hline
\end{tabular}
\caption{Densities inside the source mass $\rhoA$ and the wall $\rhoW$ for the numerical scalar field and acceleration 
profiles of Figs.~\ref{plot:density_field} and \ref{plot:density_acc}.}
\label{tab:density}
\end{center}
\end{table}

% 
% \begin{figure}
% \begin{center}
%   \includegraphics[width=0.6\textwidth, trim=220 0 240 0, clip=true]{./chapters/chapter4/Img_v2/fig7_density_field_SP.pdf}
%   %\includegraphics[scale=0.42, trim=220 0 240 0, clip=true]{./Img_v2/fig7_density_field_SP.eps}
%   \caption{Numerical results (solid lines) and analytical approximation (dashed line) for the scalar field profile of 
%   the exponential model, for various $\rhoA$ and $\rhoW$ reported in Tab.~\ref{tab:density}, $M=10^{17}$ GeV being 
%   fixed.}
%    \label{plot:density_field}
% \end{center}
% \end{figure}
% 
% \begin{figure}
% \begin{center}
%   %\includegraphics[scale=0.42, trim=220 0 240 0, clip=true]{./Img_v2/fig8_density_acc_SP.eps}
%   \includegraphics[width=0.6\textwidth, trim=220 0 240 0, clip=true]{./chapters/chapter4/Img_v2/fig8_density_acc_SP.pdf}
%   \caption{Numerical results (solid lines) and analytical approximation (dashed line) for the acceleration $a_\phi/g$ 
%   profile of the exponential model, for various $\rhoA$ and $\rhoW$ reported in Tab.~\ref{tab:density}, $M=10^{17}$ GeV 
%   being fixed.}
%    \label{plot:density_acc}
% \end{center}
% \end{figure}

Inside the vacuum chamber, the analytical estimation is roughly recovered in the first half of the chamber.  Once in the thin-shell regime, one can also observe that the field and acceleration profiles inside the chamber are independent of the wall and mass densities,  except at their immediate vicinity. Therefore, in the deep thin-shell regime, the scalar field and acceleration both at the \textit{near} and \textit{far} positions of the interferometer do not depend on the source mass and wall densities and sizes, neither on the exterior environment. In order to obtain the numerical solution inside the chamber, down to low values of $M$, one can therefore use the trick to set the wall and mass densities high enough to be in the thin-shell regime but low enough for the field profile to be numerically tractable through the borders of the four regions. As an example, for $M=10^{17}~$GeV, the numerical solution is tractable for the real source mass density whereas the wall density has been adapted ($\rhoW=7.5\times 10^{-20}~\rr{GeV^{4}}$ instead of $3.5\times 10^{-17}~\rr{GeV^{4}}$) (see the red curves in Figs.~\ref{plot:density_field} and \ref{plot:density_acc}). 
  
\begin{figure}[!tbp]
  \centering
  \subfloat[Scalar field profile $\phi$. Noticeable deviations from the analytical estimation are observed inside the chamber, due to the wall effects. ]{\includegraphics[width=0.48\textwidth, trim= 240 0 250 0,clip=true]{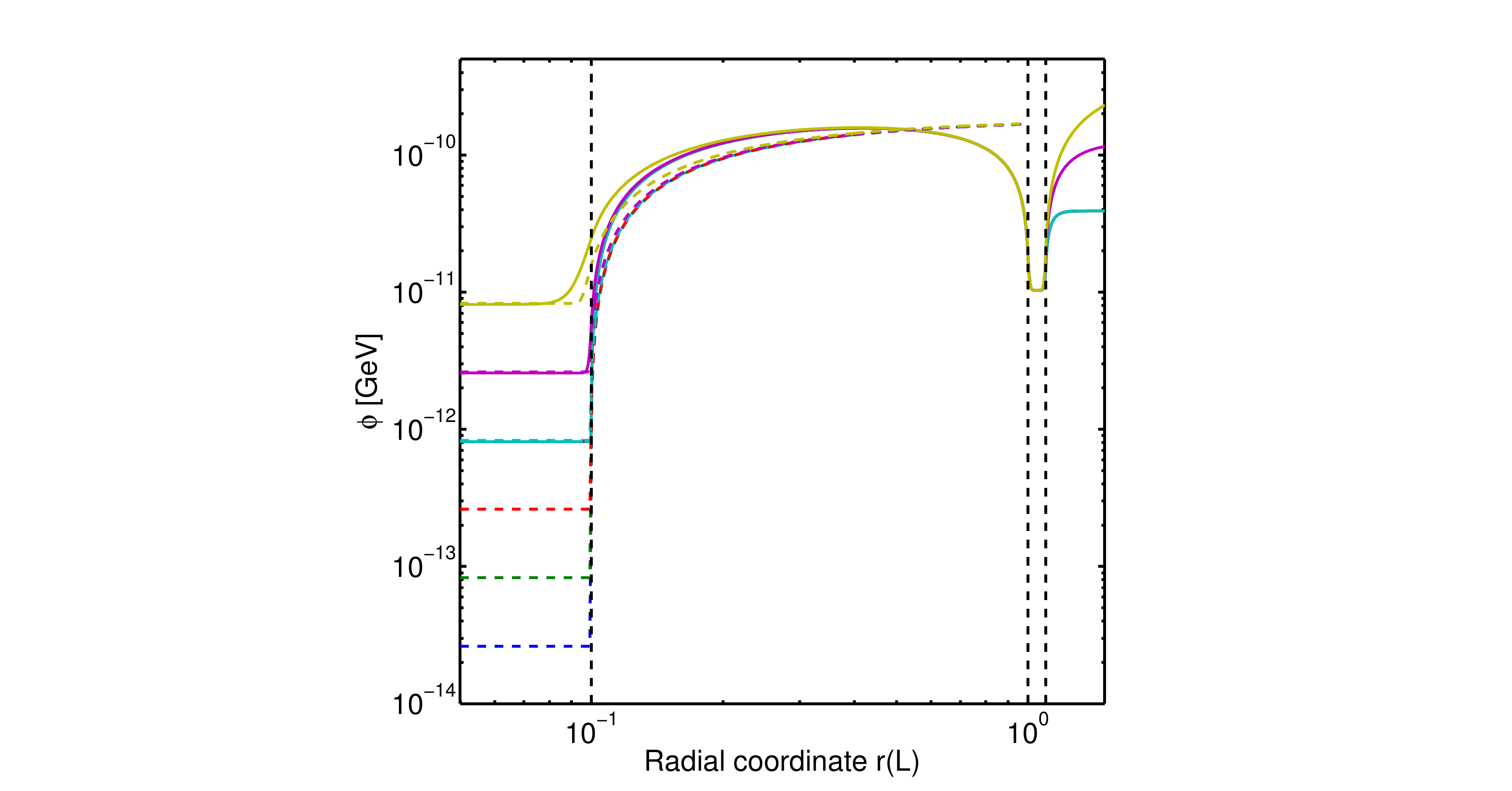}
   \label{plot:cham2_field}}
  \hfill
  \subfloat[Acceleration profile  $|a_\rr\phi| / g$. Strong discrepancies are observed between the four-region (numerical) and the two-region (analytical) models in the vicinity of the wall. The Newtonian gravitational attraction due to the test mass is plotted in black dashed line.]
  {\includegraphics[width=0.48\textwidth, trim= 340 0 350 0, clip=true]{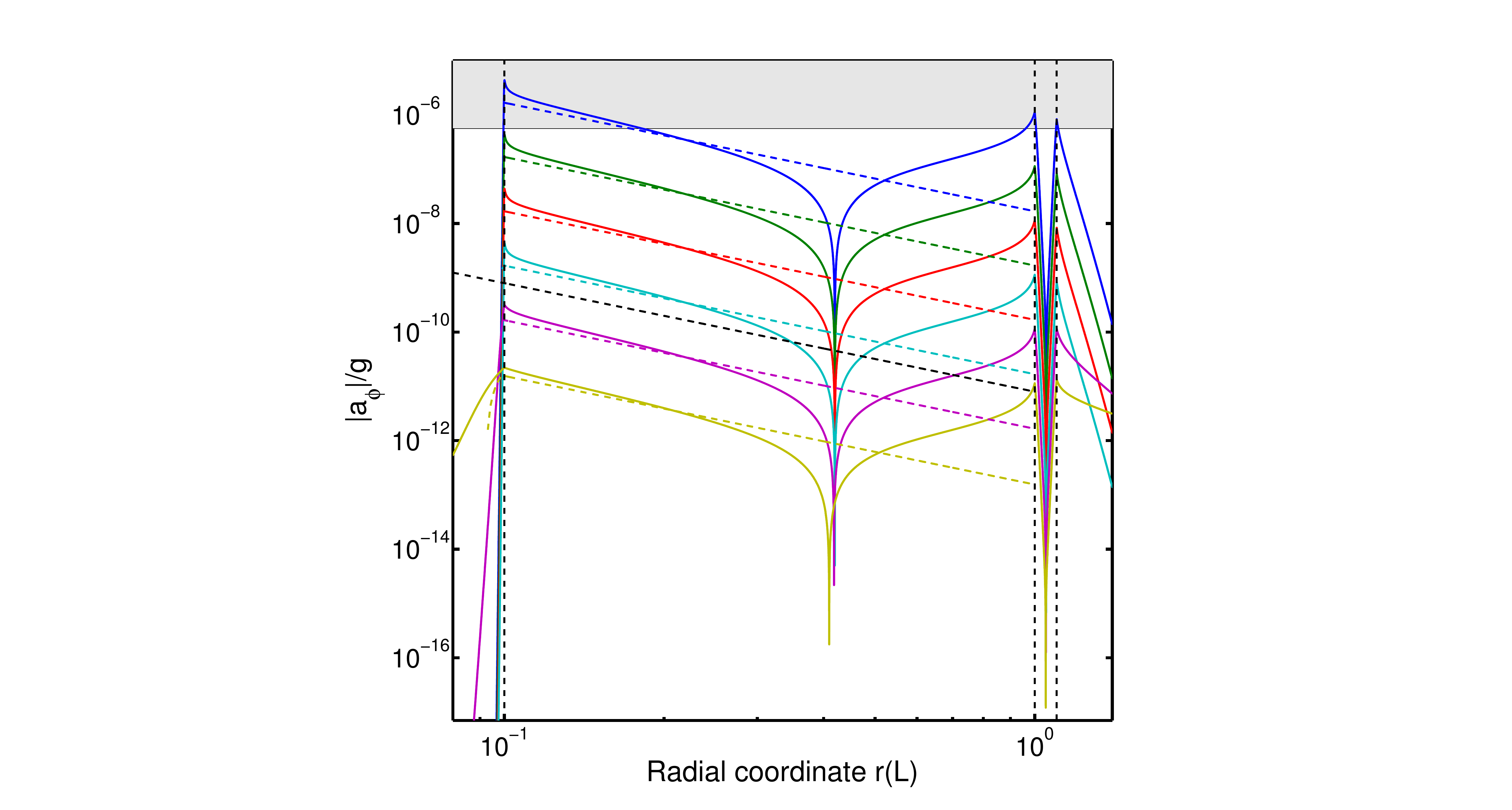}
   \label{plot:cham2_acc}}
  \caption{Numerical results (solid lines) and analytical approximation (dashed lines) for the exponential model, in the \textit{strongly perturbing} (thin shell) regime, for $\Lambda = 2.4 \times 10^{-12} \GeV$ and values of the coupling $M$ listed in Table~\ref{tab:profiles}.  The source mass, wall and exterior densities have been adapted for making the profile numerically tractable, with no effect inside the vacuum chamber (apart in the immediate vicinity of the borders), as explained in Sec.~\ref{sec:strong_field_cham_2}.  The ratios $M/ \rho$ were kept constant (with the same value as for the red curve of Fig.~\ref{plot:density_field}), which fixes the thin-shell radius, apart for $M=10^{18} \rr{GeV}$ (purple) and $M=10^{19} \rr{GeV}$ (beige) for which only the wall density was adapted. Vertical lines mark out the four regions (source mass, chamber, wall and exterior).}
\end{figure}

The field and acceleration profiles have been calculated numerically and compared to the analytical results, for 
several values of $M$ and $\Lambda \simeq 2.4$ meV.   These are represented in Figs.~\ref{plot:cham2_field} 
and~\ref{plot:cham2_acc}.  As expected the profiles have the same behavior as for the original model (see 
Figs.~\ref{plot:profile_field_thinshell1} and~\ref{plot:profile_acc_thinshell1}).   Close to the source mass, one recovers the analytical estimation but one can nevertheless notice differences higher than 20\%. 

%\textcolor{red}{But important differences with the analytical results arise in the chamber, where the field profile is 
%enhanced by more than one order of magnitude when going deeper in the thin-shell regime.NOT TRUE ANYMORE}
Close to the 
wall, the acceleration becomes negative, and its amplitude reaches values comparable to the acceleration at a position 
close to the source mass, which is a potentially measurable prediction that could be useful to discriminate between 
experimental systematic effects and an acceleration induced by the presence of some scalar field.  

In conclusion, we 
find that the 
atom-interferometry experiment of~\cite{khoury} has already excluded values of the coupling parameter $M \lesssim 10^{14} 
\GeV $ at 95\% C.L.\footnote{In \cite{Elder:2016yxm} they found $M<2.3\times 10^{-5}~\Mp$.}.   Moreover, if the experimental sensitivity could be reduced down 
to $a_\phi / g \sim 10^{-11}$ (as it is claimed to be feasible in~\cite{Burrage}), the model would be probed nearly up to the 
Planck scale.  Finally, note that the typical field values reached inside the chamber are too low to induce large 
deviations from $A(\phi) \simeq 1$, which implies that our results are roughly independent of the power-law index $\alpha$. 
%As represented in Fig.~\ref{fig:beta_n}, small values of the $\alpha$ parameter are constrained by the PPN parameters as well as torsion balance experiment whereas large $\alpha$ values are poorly constrained.   

\subsection{Chamber geometry effects}
The numerical method developed in this chapter takes into account the effects of the chamber geometry, in the limit where 
the vacuum chamber is spherical. Exploring various chamber size and wall density, we propose to consider the possibility to 
realize the same atom interferometry experiment in a vacuum room in order to make the test of $M$ values up to the Planck scale possible in the near future.  The largest vacuum rooms have a radius larger than $R=10$~m and their walls
made of concrete are sufficiently large such that the field reaches its attractor inside the walls.  One can thus neglect the exterior of the 
chamber (see Sec.~\ref{sec:strong}).  The vacuum room can sustain a vacuum around $10^{-6}$ Torr (we assume 
$\rhoV= 5 \times 10^{-31} \rr{GeV}^{4}$), low enough to prevent $\phi_{\rr{bg}}$ to reach its effective potential minimum in vacuum.  

\begin{figure}[!tbp]
  \centering
  \subfloat[Scalar field profile $\phi$.]{\includegraphics[width=0.6\textwidth, trim=220 0 240 0, clip=true]{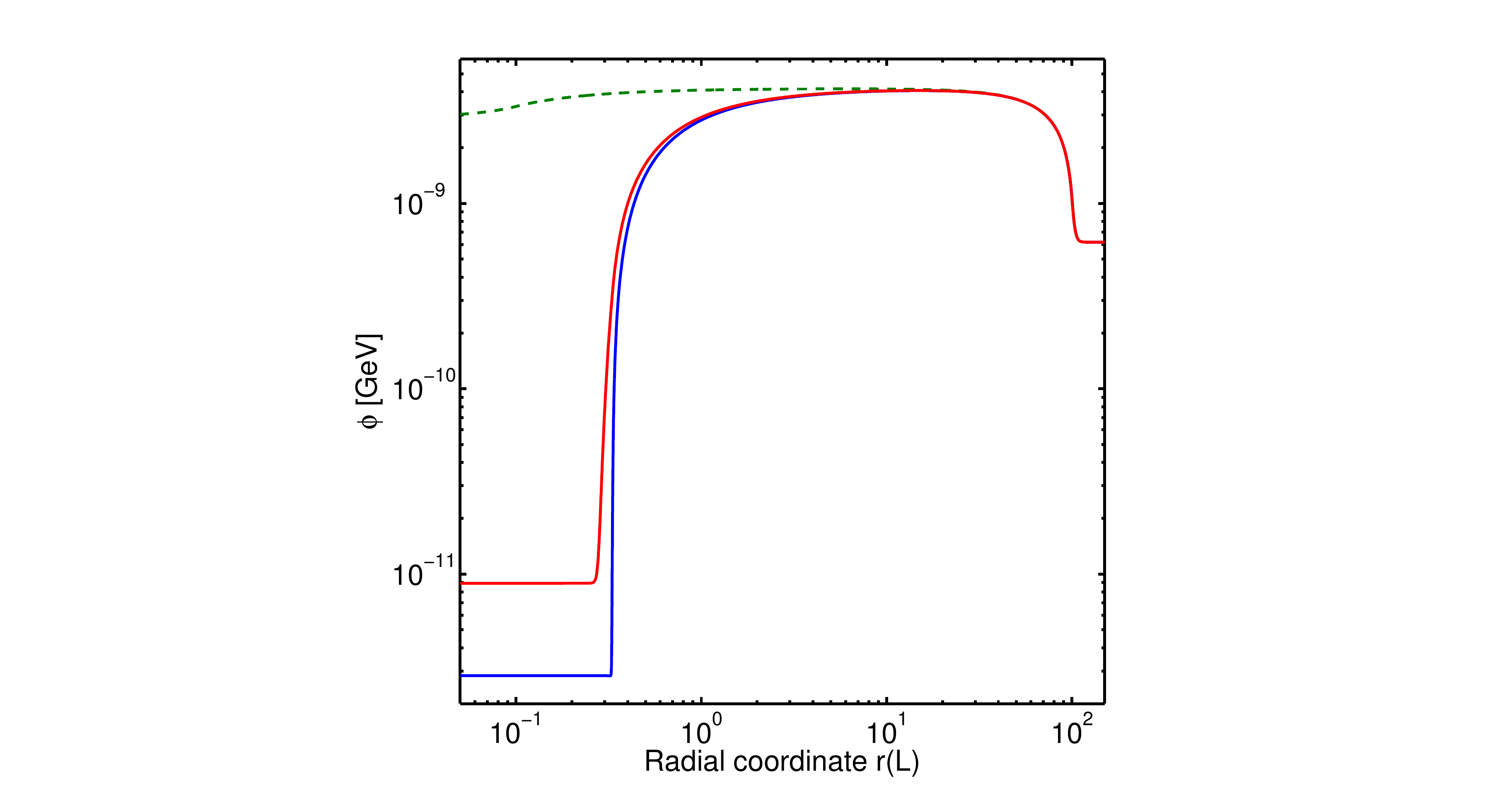}
   \label{plot:vac_room_field}}
  \hfill
  \subfloat[Acceleration profile  $|a_\rr\phi| / g$.]
  {\includegraphics[width=0.47\textwidth, trim= 240 0 250 0, clip=true]{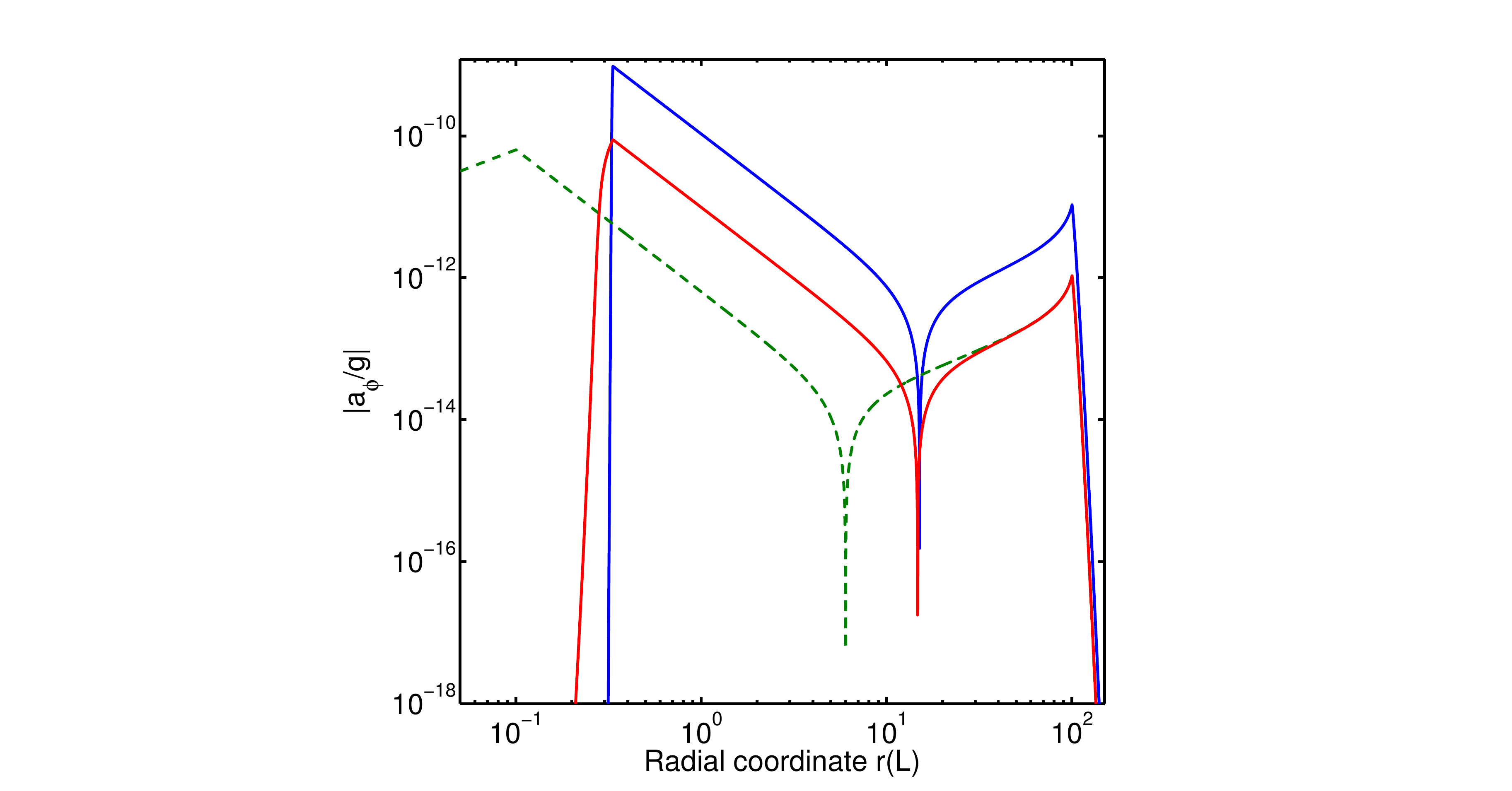}
   \label{plot:vac_room_acc}}
  \caption{Numerical profiles for a vacuum room ($L=10$~m). The green dashed curve is obtained for a test 
   mass of $\RA=$1~cm with $M=m_\rr{p}$ ($\rhoW=2.5\times10^{-21}\,\rr{GeV}^4$) while the blue and the red ones 
   are obtained for $\RA=3.3$~cm with $M=0.1\times m_\rr{p}$ ($\rhoW=2.5\times10^{-22}\,\rr{GeV}^4$) and $M=m_\rr{p}$   
   ($\rhoW=2.5\times10^{-21}\,\rr{GeV}^4$) respectively. We only consider a three regions model, neglecting the effect 
  of the exterior of the vacuum room (see discussion in Sec.~\ref{sec:strong}).}
\end{figure}

Numerical field and acceleration profiles are reported in Figs.~\ref{plot:vac_room_field} and \ref{plot:vac_room_acc} 
respectively. Assuming as before $\rhoA=1.2\times 10^{-17}~\rr{GeV}^4$, it results that 
a source mass of 1~cm radius only enables to probe the regime where the field does not reach $\phi_{\rr A}$ inside the source mass (see dashed green lines on 
Figs.~\ref{plot:vac_room_field} and \ref{plot:vac_room_acc}), the acceleration being thus poorly constrained.  However, 
provided that the source mass radius is larger (e.g. $\RA=3.3$~cm), the strongly perturbing regime is reached and the 
acceleration is large enough to be measurable in the near future for $M$ of the order 
of $m_\rr{pl}$. As a result, for $M=m_\rr{pl}$, $|a_\phi|/g=2.4\times 10^{-10}$ at 8.8~mm from the 
surface of the source mass (the previously called \textit{near} position in Sec.~\ref{sec3}) while $|a_\phi|/g=5.7\times 
10^{-10}$ for $M=0.1~m_\rr{pl}$. In comparison, the source mass of 1~cm gives rise to $|a_\phi|/g=1.7\times 10^{-11}$ for $M=m_\rr{pl}$. 

Similarly to what was obtained in Secs.~\ref{sec:strong_field_cham_1} and \ref{sec:strong_field_cham_2}, the thin shell regime cannot be tracked numerically if the wall density is of the order of the concrete $\rho\sim 10^{-17}\,\rr{GeV}^4$.  But one can safely consider lower values of 
$\rhoW$ (see Fig.~\ref{plot:vac_room_field}) without any significant change of the results inside the vacuum room.

\section{Conclusion} \label{sec_CCL}

The chameleon screening mechanism is able to suppress the fifth force induced by a scalar field in locally dense environment, while allowing the scalar field to be responsible for the current cosmic acceleration on large astrophysical scales and thus to affect significantly the LSS formation. Since chameleon models have not been already ruled out for their entire parameter space, they are still viable candidates for explaining the current cosmic acceleration. 
By the way, they will be tested by future cosmology-dedicated experiments, such as Euclid~\cite{Amendola:2012ys, Amendola:2016saw} 
or the next generation of giant radio-telescopes dedicated to 21~cm cosmology~\cite{Brax:2012cr}.  
Chameleon theories are also well constrained by local tests of gravity in the Solar System, in the galaxy, 
as well as in laboratory.   Recently it has been proposed to use an atom-interferometry experiment 
to constrain chameleon models with an unprecedented accuracy by probing the acceleration induced by the presence of the scalar field on cold atoms.  The experiment is realized inside a vacuum chamber in order to reduce the screening effect, and a central mass is used to source some field gradient.   Forecasts were calculated in~\cite{Burrage} and a first experimental setup was built and used to establish new constraints on the chameleon model~\cite{khoury}.   However, the calculations of the field and acceleration profiles rely on several approximations, and until now had not fully considered the effects of the vacuum chamber wall and of the exterior environment.

The purpose of this work was to refine and eventually validate previous calculations, by using a numerical approach consisting in solving the Klein-Gordon equation in the static and spherically symmetric case for a four-region model representing the central source mass, the vacuum chamber, its wall, and the exterior environment.   Three boundary conditions are imposed:  the field must be regular at the origin and reach the minimum of the effective potential with a Yukawa profile, at large distance in the exterior environment.  Our method allows one to probe the transition between the regime where the central source mass only weakly perturbs the field configuration, and the thin-shell regime where the field inside the central mass and inside the chamber wall reaches the minimum of the effective potential over a very small distance.   Two typical chameleon potentials were considered, in inverse power-laws and allowing varying powers, as well as a standard exponential form for the coupling function.  

In the weakly perturbing regime, it is found that the chamber wall enhances significantly the scalar field inside the vacuum chamber and reduces the induced acceleration, by up to one order of magnitude compared to previous analytical estimations and with a maximal effect close to the wall (see \cite{Schlogel:2015uea} for details). 

Going to the thin-shell regime, for our fiducial experimental setup, the field reaches the 
attractor inside the chamber wall and the exterior environment becomes thus irrelevant.  However, for reasonable value of the 
induced acceleration, the field inside the vacuum chamber does not reach the minimum of the effective potential and is 
instead related to the size of the chamber, as first noticed in~\cite{Burrage}.  Our analysis refines the field and acceleration profiles in the chamber and highlights noticeable deviations 
from the analytical estimation, which is nevertheless roughly recovered close to the central source mass.  
%However, \textcolor{red}{we observe 
%important deviations from the analytical estimations STILL TRUE??}, especially 
Close to the chamber wall, the 
acceleration becomes negative, with a magnitude similar to the one close to the central mass.  We argue that this 
prediction could be useful to distinguish between systematic effects and fifth-force effects which should be 
\textit{maximal and opposite close to the central mass and to the wall}, and should vanish roughly at the 
middle distance between the source mass and the wall.   

Refined constraints have been derived on the coupling parameter $M$ from the atom-interferometry experiment of~\cite{khoury}.  For the chameleon potential $V(\phi) = \Lambda^{4+\alpha} / \phi^\alpha$ and a coupling function $A(\phi) = \exp(\phi /M)$, one finds $M \gtrsim 7 \times 10^{16} \GeV$, independently of the power-law.   For the bare potential $V(\phi) = \Lambda^4 (1+ \Lambda/ \phi)$, we find that 
%$M \gtrsim 4\times 10^{16} \GeV$.   
$M \gtrsim  10^{14} \GeV$.   
We have also confirmed that a future experiment reducing its sensitivity down to $a \sim 10^{-10} \rr{m/s^2}$ would be able to rule out most of the parameter space of the latter model, nearly up to the Planck scale. %\textcolor{red}{CHECK NUMERICAL VALUES}.  

Finally, we have proposed to realize a similar atom-interferometry experiment inside a vacuum room.  The density inside such rooms is low enough for the field profile and the induced acceleration to depend only on the size of the room.  If the room radius is larger than about $10$ meters, we find that the chameleon model could be probed up to the Planck scale.  Nevertheless, further work is needed to implement realistic non-spherical geometries of the room (or of the vacuum chamber).  

We conclude that numerical results will be helpful in the future in order to establish accurate bounds on various modified gravity models. In particular, the effects of the vacuum chamber wall and its exterior environment cannot be neglected. Our numerical method is easily extendable to study other forms of the field potential and other modified gravity models requiring a screening mechanism, such as the symmetron, dilaton and f(R) models. Finally, it can be easily adapted to other experiments.

We point out that relaxation numerical methods have also been developed for modeling the same experiment. We developed a code based on \cite{Ringeval:2004ju} and Elder et al. also proposed a relaxation method on a uniform grid \cite{Elder:2016yxm}. The advantage of this numerical method is its adaptability to various geometry (whereas the mesh refinement method is based on the spherical symmetry) requiring to solve partial differential equations rather than ordinary differential equations. Both results are compatible. In addition Elder et al. propose a space-based experiment in order to prolong the time spent by the atoms near the source mass \cite{Elder:2016yxm}. The acceleration at 2~mm far from the source mass is then of the same order of magnitude as on the Earth,
\be
  a_\phi=2.7\times 10^{-3}\frac{\Mp}{M}\rr{\mu m/s^2},
\ee
with a longer interaction time between the source mass and the atoms. Following \cite{Elder:2016yxm}, such a set-up would in principle be sensitive to the entire parameter space $M\lesssim \Mp$, the same order of magnitude as that probed in a vacuum room.

Even more recently a novel experiment has been proposed using atom-interferometry where atoms are trapped between two parallel plates with different densities but the same total mass \cite{Burrage:2016lpu}. Such a set-up enables to subtract the gravitational background because, even if the field profile for the gravitational profile is symmetric, the chameleon field profile is not. This experimental setup could also probe the chameleon fifth force for smaller $\beta$ (larger $M$) than the Berkeley one.

\renewcommand{\chaptermark}[1]{\markboth{\small\textsc{Chapter \thechapter.\ #1}}{}}
\cleardoublepage

\chapter{The Higgs monopoles} % Main chapter title

\label{chap:Higgs} % For referencing the chapter elsewhere, use \ref{Chapter1} 

\lhead{Chapter 5. \emph{Awesome chapter 5}} % This is for the header on each page - perhaps a shortened title

%%Version electronique
\begin{center}
\textit{based on}\\
\end{center}
\begin{center}
A.~Füzfa, M.~Rinaldi, S.~Schlögel, 
\\\textit{Particlelike distributions of the Higgs field\\ nonminimally coupled to gravity}\\ 
\href{http://journals.aps.org/prl/abstract/10.1103/PhysRevLett.111.121103}{Phys.~Rev.~Lett.~111,~121103 (2013)}, \href{http://arxiv.org/abs/1305.2640}{\texttt{arXiv:1305.2640}}\\
\end{center}
\begin{center}
\textit{and}\\
\end{center}
\begin{center}
S.~Schlögel, M.~Rinaldi, F.~Staelens, A.~Füzfa, \\
\textit{Particlelike solutions in modified gravity: the Higgs monopole}\\
\href{http://journals.aps.org/prd/abstract/10.1103/PhysRevD.90.044056}{Phys.~Rev.~D90:~044056 (2014)}, \href{http://arxiv.org/abs/1405.5476}{\texttt{arXiv:1405.5476}}\\
\end{center}

\vspace{1cm}
%----------------------------------------------------------------------------------------
\noindent
%In this chapter, we will focus on the Higgs inflation model and provide combined constraints for such model in the Solar system and around compact objects. 
In this chapter, we first discuss the possible relation between the Higgs field and gravity. Then we focus on the Higgs inflation model where the Higgs field is responsible at once for inflation and the elementary particle mass generation through the $SU(2)\times U(1)$ spontaneous symmetry breaking at electroweak scale. Because of the nonminimal coupling appearing in Higgs inflation, this model predicts nontrivial distribution of the Higgs field around compact object rather than the Higgs field settled to its vev everywhere. The underlying deviations from GR could rule out Higgs inflation, for instance if it does not fulfill Solar System constraints. We will see that Higgs inflation is indistinguishable from GR in the Solar System and around astrophysical compact objects, even if the distribution of the Higgs field is non trivial. Moreover, we will highlight the existence of a novel amplification mechanism of the Higgs field at the center of compact objects due to the combined effect of the nonminimal coupling and the Higgs potential. % \cite{Fuzfa:2013yba, Schlogel:2014jea}. 

\section[Higgs field and gravity]{Higgs field and gravity} \label{sec:higgs_gravity}
A fundamental question still open today is the nature and the origin of the mass. In particle physics, gauge bosons and fermions have to be massless in order to preserve gauge invariance. In particular, the W and Z gauge bosons mediating the weak interactions appear to be massive since the weak interactions are short-range. In the SM, the Higgs field is responsible for the mass generation of elementary particles\footnote{Note that QCD interactions provide the additional mass arising from bounded microscopic states.}. The most general potential which is renormalizable and gauge invariant under the electroweak symmetry $SU(2)_\rr{L}\times U(1)_\rr{Y}$ where $L$ refers to the left chirality and $Y$ to the hypercharge, is,
\be \label{eq:renorm_pot}
  V(\mathcal{H}^\dagger \mathcal{H})=\mu^2 \mathcal{H}^\dagger \mathcal{H}+\lambda\left(\mathcal{H}^\dagger \mathcal{H}\right)^2, 
\ee
with $\mathcal{H}$ the Higgs $SU(2)$ isospin doublet parametrized by,
\bea
  \mathcal{H}=\begin{pmatrix}
     H^+ \\ H^0
    \end{pmatrix}
  =\frac{1}{\sqrt{2}}\begin{pmatrix}
     H_1+iH_2 \\ H_3+iH_4
    \end{pmatrix},
\eea 
$\mu$ the mass and $\lambda$ the self-interaction parameters. Before the symmetry breaking arising in the early Universe, all elementary particles were massless and the local gauge invariance of the electroweak model was preserved.
%with $H$ the complex Higgs SU(2) doublet, $\lambda\sim0.1$ the strength of the self-interaction and $v=246$ GeV the vev, is spontaneously broken when $H=\pm v$. Before the spontaneous symmetry breaking arising in the early Universe all elementary particles are massless and gauge symmetry $SU(2)_\rr{L}\times U(1)_\rr{Y}$ where $L$ refers to the left chirality and $Y$ to the hypercharge, is preserved. 
By allowing $\mu^2<0$, there is an infinite number of vacua connected together through the residual $U(1)$ symmetry such that the symmetry $SU(2)_\rr{L}\times U(1)_\rr{Y}$ is spontaneously broken while $U(1)_\rr{EM}$ remains invariant. By fixing the gauge to the unitary one ($H_1=H_2=H_4=0$), the complex Higgs doublet reduces to,
\bea
  {\cal{H}}=\frac{1}{\sqrt{2}}\begin{pmatrix}
     0 \\ v+H(x)
    \end{pmatrix},
\eea 
and the potential \eqref{eq:renorm_pot} reads,
\be
  V(H)=\frac{\lambda_\rr{sm}}{4}\left(H^2-v^2\right)^2,
\ee
with $v=\sqrt{-\mu^2/\lambda_\rr{sm}}$ the Higgs vev and $\lambda_\rr{sm}$ the SM Higgs field self-interaction parameter. When the Higgs field is settled to its vev, elementary particles start to behave as they would have acquired a mass because of their coupling to the Higgs field, W and Z gauge bosons via covariant derivatives (the photon remains massless since $U(1)_\rr{EM}$ is unbroken), and fermions, through the Yukawa coupling. Since their masses are proportional to the vev, they depend on the local value of the scalar field.

This is not dissimilar to the Brans-Dicke theory (see Sec.~\ref{sec:STT}) where the scalar field $\Phi$ is responsible for the variation of $\GN$ over spacetime \cite{KaiserScientificA}. In the late 1970s, Zee \cite{PhysRevLett.42.417, PhysRevLett.44.703} and Smolin \cite{Smolin:1979uz} proposed (independently) that a spontaneous symmetry breaking could be incorporated into the theory of gravity.
%to identify the Brans-Dicke and the Higgs field. 
In \cite{PhysRevLett.42.417}, Zee studied the action now referred to as induced gravity\footnote{Induced gravity was first proposed by Sakharov in 1967 \cite{Sakharov:1967pk}.},
\be \label{eq:action_induced_gravity}
  S=\int \dd^4 x\sqrt{-g} \left[\frac{\epsilon \phi^2}{2}R-\frac{1}{2}(\df\phi)^2+V(\phi)\right]+S_\rr{M}\left[\psi_\rr{M};g_{\mu\nu}\right],
\ee
where $\phi$ is a Higgs-like real scalar field responsible for the relevant symmetry group breaking\footnote{Zee considered the breaking of SU(5) to $SU(3)\times SU(2)\times U(1)$.}, $\epsilon\lesssim1$ denotes a dimensionless nonminimal coupling and $V$ the Mexican-hat potential,
\be
  V(\phi)=\frac{\lambda}{4}\left(\phi^2-v^2\right)^2,
\ee
which ensures that $|\phi_\rr{min}|=v$. After the spontaneous symmetry breaking, the scalar field is anchored to its minimum and it generates the Newton's gravitational constant,
\be\label{eq:GN_induced_g}
  \GN=\frac{1}{8\pi \epsilon v^2}.
\ee
The only dimensional constant $\GN$ becomes dynamical and its weakness is explained provided that $v\sim \mpl$. 
Assuming that $\phi$ is rigorously equal to the vev, no deviation from GR would be noticeable and the model passes all observational constraints. However, if $\dd^2 V/\dd\phi^2\neq 0$, then the scalar field acquires a mass and it can affect the dynamics. 
The Zee's theory looks like the Brans-Dicke one, excepted for the self-interaction term $V(\phi)$ which is precisely responsible for the symmetry breaking, and both SM and GR emerge as low-energy effective theories.

%Since $G_\rr{eff}\propto 1/\varphi^2$, the gravitational constant should be settled to $\GN\propto 1/v^2$, the weakness of the gravitational force being thus explained. 
Zee noticed also that a similar theory invoking symmetry breaking with several scalar fields could explain the horizon problem \cite{PhysRevLett.44.703} by a weakening of $\GN$ in the early Universe. One year later, Guth proposed the first inflationary model with a scalar field modelled on the Higgs one \cite{1981PhRvD..23..347G}. In addition, since Linde's work about chaotic inflation \cite{1983PhLB..129..177L}, we know that the Higgs field cannot lead to a viable inflationary model if the Higgs field is minimally coupled to gravity\footnote{No interaction between gauge bosons and inflaton is assumed in chaotic inflation. We will see that this assumption remains in Higgs inflation even if it is questionable.}.

Induced gravity \eqref{eq:action_induced_gravity} has been studied a lot in the framework of cosmology, notably for inflation  (see e.g. \cite{PhysRevD.41.1792, Kaiser:1994vs}), considering $\lambda$ and $\epsilon$ as free parameters. In such models, the scalar field (not yet identified to the Higgs field) is settled to its vev at the end of inflation in order to get the EH action with $\GN$ fixed according to \eqref{eq:GN_induced_g}. Actually, if the SM Higgs field is considered as the inflaton in the framework of induced gravity, its mass is far too large  to solve the horizon and flatness problems \cite{CervantesCota:1995tz}. In order to show this result explicitly, let us start from the induced gravity Lagrangian with the SM Higgs field,
\be
  L=\sqrt{-g}\left[\frac{\epsilon}{2}\mathcal{H}^\dagger \mathcal{H} R-\frac{1}{2}D_\mu\mathcal{H}^\dagger D^\mu\mathcal{H}-V\left(\mathcal{H}^\dagger\mathcal{H}\right)+ \mathcal{L}_\rr{SM}\right],
\ee
where $D_\mu \mathcal{H}$ is the covariant derivative of $\mathcal{H}$ for the internal gauge group $SU(3)_\rr{C}\times SU(2)_\rr{L}\times U(1)_\rr{Y}$, the subscript $\rr{C}$ denoting the color quantum number. $\mathcal{L}_\rr{SM}$ contains the fermionic and massless bosonic fields of the SM, included the Yukawa coupling. Eq.~\eqref{eq:GN_induced_g} where $v$ is the Higgs vev then yields to $\epsilon\sim10^{40}$, highlighting the huge difference between the Planck scale and the vev. In order to get a viable inflationary scenario in terms of the scalar and tensor density perturbations amplitude, the Higgs mass must be very large, $m_\rr{H}\gg G_\rr{F}^{-1/2}\sim300~$GeV, $G_\rr{F}$ being the Fermi constant, in disagreement with the observations, $m_\rr{H}=125$ GeV \cite{PDG-2008}. Moreover, the Higgs field decouples from bosonic and fermionic masses after the symmetry breaking, interacting mostly via gravity \cite{CervantesCota:1995tz}. In conclusion, this model is able to predict viable inflation but is inconsistent with high energy physics.  

Since its discovery at the Large Hadron Collider (LHC) in 2012 \cite{Chatrchyan:2012xdj, Aad:2012tfa}, the Higgs boson has been the first elementary scalar field ever detected in nature (even though it could still be composite \cite{GEORGI1984216}). 
However, several questions remain today like the origin of the spontaneous symmetry breaking of the electroweak interaction during the cosmological history, the (classical) stabilization mechanism of the vev (which cannot vary significantly since it would change the strength of the nuclear interactions) as well as the hierarchy problem between $m_\rr{H}$ and $\mpl$ mentioned above. Recently a novel scenario has been proposed where $m_\rr{H}$ depends on the value of an additional scalar field, the relaxion  \cite{Graham:2015cka}. This model solves the hierarchy problem since in the early Universe, the Higgs field is naturally large and then  decreases gradually to zero. It becomes then unstable and as a result is fixed to its current value through the spontaneous symmetry breaking. On the other hand, the relaxion gives rise to inflation. 
Maybe LHC experiments will allow to reveal a part of the history by discovering the signature of a new particle. Anyway the Higgs sector is probed today by high-energy physics and cosmology at once. 

%It is tempting to tie the Higgs field with gravity. Indeed, the first is responsible for the mass generation of elementary particles while the latter describes the interaction of massive objects, 
\section{Higgs inflation}\label{sec:higgs_inflation}
In 2007, Bezrukov and Shaposhnikov proposed that the SM Higgs boson could be the inflaton provided that it is nonminimally coupled to gravity \cite{Bezrukov:2007ep}. In this section, we briefly expose their model and review their results for inflation.
\subsection{The model}\label{sec:modelHI}
In Higgs inflation, the SM Higgs boson is responsible for the elementary particles mass generation and inflation at once, provided that it is nonminimally coupled to gravity, 
\be 
  \mathcal{L}_\rr{tot}=\mathcal{L}_\rr{SM}-\frac{\Mp^2}{2}\left(1+\xi \mathcal{H}^\dagger \mathcal{H} \right)R,
\ee
with $\mathcal{L}_\rr{SM}$ the Lagrangian density of the SM, including the Higgs sector,  $\xi$ the nonminimal coupling parameter and $\mathcal{H}$ the SM Higgs doublet [GeV]. The shape of the nonminimal coupling function differs from the one of induced gravity (assuming the SM Higgs potential) such that Higgs inflation is able to predict a viable inflation scenario while preserving high energy physics as we will see in the following.

%Let us remind that a minimal coupling preserves the high-energy physics (see Sec.~\ref{sec:higgs_gravity}), whereas it does not give rise to an inflationary phase. In Higgs inflation the nonminimal coupling renders inflation viable in the large field regime since it looks like induced gravity. 

By fixing the SU(2) gauge to the unitary one, 
\bea
  \mathcal{H}=\frac{1}{\sqrt{2}}\begin{pmatrix}
     0 \\ v+H(x)
    \end{pmatrix},
\eea 
the action for Higgs inflation becomes,
\be \label{eq:action_Higgs_inflationJF}
  S_\rr{HI,\,JF}=\int\dd^4x \sqrt{-g}\left[F(h)\frac{R}{2\,\kappa}-\frac{\Mp^2}{2}\left(\df h\right)^2-V(h)\right],
\ee
with the Higgs field normalized as $H=\Mp ~h$. The potential $V(H)$ is assumed to be the SM Mexican-hat,
\be \label{mexican}
  V(h)=\frac{\lambda_\rr{sm}\Mp^4}{4}\left(h^2-\frac{v^2}{\Mp^2}\right)^2,
\ee
with SM parameters $\lambda_\rr{sm}\sim0.1$ \cite{PDG-2008} and $v=246~$GeV, and the nonminimal coupling function reads,
\be \label{eq:nonminHI}
 F(h)=1+\xi h^2.
\ee
The form of this coupling function is further justified by invoking the (semiclassical) renormalization of the energy momentum tensor
associated to the scalar field on a curved background, which needs terms like $H^{2}R$ in the Lagrangian
\cite{1970AnPhy..59...42C}.

By applying the usual conformal transformation \cite{wald},% (see also App.~\ref{app_fR}),
\bea
  \tilde{g}_{\mu\nu}&=&\Omega^2 \, g_{\mu\nu}\hspace{1cm} \text{with}\hspace{1cm}
  \Omega^2=1+\xi h^2, \\
  \tilde{R}&=&\Omega^{-2}\,R-6\Omega^{-3} g^{\alpha\beta}\nabla_{\alpha}\nabla_{\beta}\Omega,
\eea
it is possible to write the action~\ref{eq:action_Higgs_inflationJF} in the Einstein frame,
\be \label{eq:action_Higgs_inflationEF}
  S_\rr{HI,\,EF}=\int\dd^4 x \sqrt{-\tilde{g}}\left[\frac{\tilde{R}}{2\,\kappa}-\frac{\Mp^2}{2}\tilde{g}^{\mu\nu}\tilde{\df}^\mu\chi\tilde{\df}_\nu\chi-U\left(\chi\right)\right],
\ee
with $\chi$ the dimensionless scalar field defined by,
\be \label{eq:implicit_chi}
  \frac{\dd \chi}{\dd h}=\sqrt{\frac{\Omega^2+6\xi^2 h^2}{\Omega^4}},
\ee
and $U$ the potential in the Einstein frame,
\be \label{eq:U_HI}
  U(\chi)=\Omega^{-4}\,V[h(\chi)].
\ee

We can discuss the low energy and the high energy limits in the Einstein frame:
\begin{itemize}
 \item $\xi h^2\ll1 \Rightarrow \Omega\simeq1$: In this case, the Jordan and the Einstein frames are equivalent since $h\simeq \chi$ and $U(\chi)\simeq V(h)$, so that the Higgs field appears to be minimally coupled to gravity in the Jordan frame, giving rise to no inflationary phase. The SM model is thus preserved at low energy scales. 
 \item $\xi h^2\gg1 \Rightarrow \Omega^2\simeq \xi h^2$: In this limit, the Higgs inflation looks like induced gravity in the Jordan frame, leading to an inflationary phase. The implicit equation for $\chi$ \eqref{eq:implicit_chi} reduces to,
  \be \label{eq:chi_large_field}
    \frac{\dd \chi}{\dd h}=\frac{\sqrt{6}}{h}, 
  \ee
 the integration yielding,
  \be
    h=\mathcal{C}\exp\left(\frac{\chi}{\sqrt{6}}\right),
  \ee
 with $\mathcal{C}$  the integration constant. Similarly to the induced gravity scenario, the Higgs vev is determined by,
  \be
    1=\frac{\xi v^2}{\Mp^2} \hspace{1cm} \Rightarrow \hspace{1cm}
    v=\frac{M_\rr{p}}{\sqrt{\xi}},
  \ee
  leading to \cite{Kaiser:1994vs},
  \be \label{eq:h_large_field}
    H=\frac{M_\rr{p}}{\sqrt{\xi}}\exp\left(\frac{\chi}{\sqrt{6}}\right).
  \ee
  The potential in the Einstein frame then reads,
  \bea 
    U(\chi)&=&\frac{\lambda_\rr{sm} \Mp^4}{4 \xi^2}\left(1-\frac{v^2}{H^2}\right)^2, \\
    &=&\frac{\lambda_\rr{sm} \Mp^4}{4 \xi^2}\left[1-\exp{\left(-\frac{2\chi}{\sqrt{6}}\right)}\right]^2.
    \label{eq:U_HI_largeh}
  \eea
  Because of the flatness of $U(\chi)$ for large field $\chi$ values, it results that slow-roll inflation is efficient.
\end{itemize}

\begin{sloppypar}
\subsection{Equivalence between the Starobinsky model and the Higgs inflation} \label{sec:equiv_HI_andStaro}
\noindent
The Planck results for inflationary models depicted in Fig.~\ref{fig:planck_inflation} reveal that the Starobinsky model given by,
\be
  S=\frac{1}{2\kappa}\int\dd^4 x \sqrt{-g} f(R),
\ee
with,
\bea
  f(R)=R+\frac{R^2}{6 M^2},
\label{L_staro}
\eea
$M$ being an energy scale, and the Higgs inflation are equivalent in terms of $n_\rr{s}$ and $r$. The equations of motion for the Starobinsky model are of second order. Indeed, $f(R)$ theories avoid the Ostrogradsky instability (see Sec.~\ref{sec:Ostro}) provided that ${\dd^2 f}/{\dd R^2}=1/(3M^2)\neq 0$\footnote{Actually in order to ensure the stability of FLRW solutions, we must rather impose ${\dd^2 f}/{\dd R^2}> 0$ \cite{Capozziello:2006dj}.} (see e.g. \cite{DeFelice:2010aj}).
In this section the equivalence between both models is explicitly shown \cite{WHITT1984176}.

The Legendre transform of $f(R)$ to another function of an auxiliary field $V(\Phi)$,
\bea
  V(\Phi)=\Phi R(\Phi)-f\left[R(\Phi)\right],
  \label{def_VPHI}
\eea
enables one to rewrite the Starobinsky model as a STT in the Jordan frame in the absence of any kinetic term. Indeed, by defining,
\bea
  \Phi = {\frac{\dd f}{\dd R}}=1+\frac{R}{3M^2},
\eea
the potential function reads,
\bea
  V(\Phi)&=&3M^2 \Phi\left(\Phi-1\right)-3M^2\left(\Phi-1\right)-\frac{3}{2}\left(\Phi-1\right)^2, \\
  &=&\frac{3M^2}{2}\left(\Phi-1\right)^2,
  \label{eq:pot_JF_HI}
\eea
and the action \eqref{L_staro} becomes,
\bea
  S&=&\frac{1}{2\kappa}\int\dd^4 x \sqrt{-g} \left[\Phi R(\Phi)-V(\Phi)\right], \label{eq:action_HI_JF}\\
  &=&\frac{1}{2\kappa}\int\dd^4 x \sqrt{-g}\left[\Phi R-\frac{3M^2}{2}\left(\Phi-1\right)^2\right].
  \label{action_Leg}
\eea
The Starobinsky model corresponds thus to a STT in the Jordan frame in the absence of a kinetic term for the scalar field $\Phi$.

$f(R)$ theory can also be expressed in the Einstein frame by performing the conformal transformation \eqref{eq:Ric_conformal_transfo} with $\Omega=\sqrt{\Phi}$ (see also Sec.~\ref{app:EF}), which requires to compute,
\bea
\tilde{\nabla}_{\alpha}\sqrt{\Phi}&=&\frac{1}{2\sqrt{\Phi}}\tilde{\nabla}_{\alpha}\Phi, \\
\tilde{\nabla}_{\alpha}\tilde{\nabla}_{\beta} \sqrt{\Phi}&=&\frac{1}{2\sqrt{\Phi}}
\left(-\frac{1}{2\Phi}\tilde{\nabla}_{\alpha}\Phi \tilde{\nabla}_{\beta}\Phi
+\tilde{\nabla}_{\alpha}\tilde{\nabla}_{\beta}\Phi\right),
\eea
hence, the Ricci scalar eventually transforms as (see Eq.~\eqref{eq:Ric_conformal_transfo}),
\bea
  R=\Phi \tilde{R} + 3 \tilde{\Box}\Phi-\frac{9}{2} \frac{\left(\tilde{\nabla}\Phi\right)^2}{\Phi}.
\eea
Using Eqs.~\eqref{eq:conformal_det_g}, it results that the action \eqref{eq:action_HI_JF} becomes,
%After the conformal transformation, the action becomes (as a reminder $\sqrt{-\tilde{g}}=\Phi^2\sqrt{-g}$):
\bea
  \hspace{-0.3cm}S&=&\frac{1}{2\kappa}\int \dd^4x \sqrt{-\tilde{g}} \Phi^{-2}
  \left[\Phi^2 \tilde{R} + 3\Phi \tilde{\Box}\Phi-\frac{9}{2} \left(\tilde{\nabla}\Phi\right)^2-V(\Phi)\right],\\
  &=&\frac{1}{2\kappa}\int \dd^4x \sqrt{-\tilde{g}} \left[\tilde{R}+3 \frac{\tilde{\Box}\Phi}{\Phi}-\frac{9}{2}\left(\frac{\tilde{\nabla}\Phi}{\Phi}\right)^2
-\frac{V(\Phi)}{\Phi^2}\right].
\eea
Integrating the second term by parts,
\be
\frac{\tilde{\square}\Phi}{\Phi}=-\tilde{\nabla}_{\alpha}\left(\frac{1}{\Phi}\right)\tilde{\nabla}^{\alpha}\Phi 
=\left(\frac{\tilde{\nabla}\Phi}{\Phi}\right)^2,
\ee
the action yields,
\be
  S=\frac{1}{2\kappa}\int \dd^4x \sqrt{-\tilde{g}} \left[\tilde{R}-\frac{3}{2}\left(\frac{\tilde{\nabla}\Phi}{\Phi}\right)^2
-\frac{V(\Phi)}{\Phi^2}\right].
\ee
Eventually, by rescaling of the scalar field,
\bea \label{eq:rescaled_SF}
\varphi=\sqrt{\frac{3}{2}} \ln\Phi \hspace{1cm} \Rightarrow \hspace{1cm} \Phi=\rr{e}^{\sqrt{2/3}\varphi},
\eea
the potential yields (see Eq.~\eqref{eq:pot_JF_HI}),
\bea
  V(\varphi)=\frac{3 M^2}{2} \left(\rr{e}^{\sqrt{\frac{2}{3}}\varphi}-1\right)^2.
\eea
Hence, the action in the Einstein frame for the Starobinsky model reads,
\bea
  S_\rr{S,\,EF}=\frac{1}{2\kappa}\int \dd^4 x \sqrt{-\tilde{g}} \left[\tilde{R}- \tilde{\nabla}_{\mu}\varphi\tilde{\nabla}^{\mu}\varphi
  -\frac{3 M^2}{2} \left(1-\rr{e}^{-{\sqrt{\frac{2}{3}}\varphi}}\right)^2\right].
\eea
in agreement with the Higgs inflation \eqref{eq:U_HI_largeh},
\bea
S=\int \dd^4 x \sqrt{-\tilde{g}} \left[\frac{\tilde{R}}{2\kappa}-\frac{\Mp^2}{2} \tilde{\nabla}_{\mu}\varphi\tilde{\nabla}^{\mu}\varphi
-\frac{3 M^2 \Mp^2}{4} \left(1-\rr{e}^{-{\sqrt{\frac{2}{3}}\varphi}}\right)^2\right],
\eea
in the limit $\xi h^2\gg1$ provided that $\chi=\varphi$ (see Eq.~\eqref{eq:U_HI_largeh}) and,
\be \label{eq:MequivHI}
  M^2=\frac{\lambda_\rr{sm}\Mp^2}{3 \xi^2}.
\ee

Another way to show the equivalence between the Starobinsky model and the Higgs inflation in the large field limit $(H\gg v)$  consists of starting from the Higgs inflation in the Jordan frame \eqref{eq:action_Higgs_inflationJF} assuming the slow-roll conditions at the action level, $(\df H)^2/2\ll V(H)$ \cite{Kehagias:2013mya}. Starting from Eq.~\eqref{eq:action_Higgs_inflationJF}, the Euler-Lagrange equation for the scalar field yields $\left(h\neq0\right)$, 
\be
  \frac{1}{\sqrt{-g}}\frac{\df S_\rr{HI,\,JF}}{\df h}=0
  \hspace{1cm}\Leftrightarrow\hspace{1cm} h^2=\frac{\xi R}{\lambda_\rr{sm}\Mp^2}. 
\ee
By replacing $h$ into Eq.~\eqref{eq:action_Higgs_inflationJF},
\be
  S=\frac{1}{2\kappa}\int \dd^4 x\sqrt{-g}\left(R+\frac{\xi^2 R^2}{2\Mp^2\lambda_\rr{sm}}\right),
\ee
the Higgs inflation and the Starobinsky model are thus equivalent if Eq.~\eqref{eq:MequivHI} holds.

In summary, both models are equivalent for large Higgs field values ($\xi h^2\gg1$), that is during the inflationary phase only, and at leading order. Radiative corrections as well as the reheating temperature predicted by both models differ \cite{Bezrukov:2011gp} (see also Sec.~\ref{sec:high_energy_higgs}).  However, Planck results do not enable one to distinguish both models.
\end{sloppypar}

\subsection{Constraints from inflation}
In order to confront the Higgs inflation with the Planck observations, the slow-roll analysis in the large field limit ($\xi h^2\gg1$) allows to compute the spectral index $n_\rr{s}$ and the tensor-to-scalar ratio $r$ (see also Sec.~\ref{sec:inflation}). Computing the first and second derivatives of Eq.~\eqref{eq:U_HI_largeh} and using Eq.~\eqref{eq:h_large_field}, the slow-roll parameters read,
\bea \label{eq:slow_roll_HI}
  \epsilon_\rr{V}&\equiv&\frac{1}{2}\left(\frac{\dd U/\dd \chi}{U}\right)^2
  \simeq \frac{4}{3\xi^2 h^4},\\
  \eta_\rr{V}&\equiv&\frac{\dd^2 U/\dd\chi^2}{U}\simeq -\frac{4}{3\xi h^2}.
\eea
Slow-roll inflation ends when $\epsilon_\rr{V}\simeq1$ or equivalently, for the Higgs field value, $h^4_\rr{end}\simeq 4/(3\xi^2)$. The number of e-folds \eqref{eq:e_fold} from the onset $t_\rr{i}$ to the end of inflation $t_\rr{end}$ is then given by,
\bea
  N(t)\equiv\int_{t_\rr{i}}^{t_\rr{end}} H(t) \dd t=\int_{\chi_\rr{i}}^{\chi_\rr{end}} H(\chi) \frac{\dd \chi}{\dot{\chi}}. 
\eea
In the slow-roll conditions \eqref{eq:slow_roll1} and \eqref{eq:slow_roll2}, the Friedmann and the Klein-Gordon Eqs.~\eqref{eq:FR1_SF}, ~\eqref{eq:KG_SF} and \eqref{eq:rho_phi} reduce to\footnote{Note that the slow-roll conditions are given for $U$ rather than $V$ since the Einstein frame looks like GR.},
\bea  
  H(\chi)=\sqrt{\frac{U}{3\Mp^2}},  \hspace{1.5cm}
  \dot{\chi}=-\frac{1}{\sqrt{3}\Mp}\frac{U_\chi}{\sqrt{U}},
\eea
where the subscript $\chi$ denotes a derivative with respect to $\chi$. Using the definition of $\epsilon_\rr{V}$ \eqref{eq:slow_roll_HI}, the number of e-folds is then given by,
\bea
  N(\chi)=-\int_{\chi_\rr{i}}^{\chi_\rr{end}}\frac{U}{U_\chi}\dd \chi
  =-\int_{\chi_\rr{i}}^{\chi_\rr{end}}\frac{1}{\sqrt{2\epsilon_\rr{V}}}\dd \chi,
\eea
which finally reads in terms of the Higgs field $h$ using \eqref{eq:chi_large_field} and \eqref{eq:slow_roll_HI},
\be \label{eq:N_HI}
  N(h)=-\int_{h_\rr{i}}^{h_\rr{end}}\frac{1}{\sqrt{2\epsilon_\rr{V}}}\frac{\dd \chi}{\dd h}\dd h
  =\frac{3\xi}{4}\left(h_\rr{i}^2-h_\rr{end}^2\right).
\ee
Using the definition of $h_\rr{end}$, $h^2_\rr{i}$ yields, 
\be
  h^2_\rr{i}=\frac{4}{3\xi}N+\frac{2}{\sqrt{3}\xi},
\ee
the slow-roll parameters at the onset of inflation reading then approximately,
\bea
  \epsilon_\rr{V}(N)&\simeq&\frac{12}{\left(4N+3\right)^2},\\
  \eta_\rr{V}(N)&\simeq&-\frac{4}{4N+3}.
\eea
The scalar spectral index $n_\rr{s}$ and the tensor-to-scalar ratio $r$ for $N=60$ corresponding to the pivot scale $k_*=0.002/$~Mpc then follow from Eqs.~\eqref{eq:inflation_param},
\bea
  n_\rr{s}&\simeq&0.97,\\
  r&\simeq&0.0033.
\eea
Those results are represented on Fig.~\ref{fig:planck_inflation} in agreement with the Starobinsky model as expected.  

Preheating and reheating were analyzed in detail by \cite{GarciaBellido:2008ab}, Higgs inflation offering the advantage that the couplings between the Higgs field and the other fields of the SM sector are known thanks to particle accelerator experiments, which is not the case for other inflationary models. The dependence of the Higgs inflation predictions on the reheating temperature at which inflation ends has been analyzed numerically by \cite{Martin:2013tda}. The spectral index is found to in good agreement with the data while the contribution of gravity waves is small, whatever reheating temperature in the range $[10^{-2}-10^{14}]$~GeV.

Eventually, the parameter $\xi$ is constrained thanks to the normalization of the CMB power spectrum Eq.~\eqref{eq:power_spectrum_slow_roll} for the pivot scale $k=k_*$ yielding,
\be \label{eq:CMB_norm_HI}
  \mathcal{P}_\zeta(k=k_*)\simeq \frac{H^2_*}{\pi \Mp^2 \epsilon_\rr{V,*}}\simeq\frac{U_*}{3\pi\Mp^4 \epsilon_\rr{V,*}},
\ee
where the asterisk denotes a quantity evaluated at the pivot scale $k_*$ and the Friedmann equation \eqref{eq:FR1_SF} in the slow-roll approximation \eqref{eq:slow_roll1} has been used. According to the COBE satellite measurements $\mathcal{P}_\zeta(k=k_*)=A_\rr{s}/(2\pi^2)\sim 10^{-10}$ \cite{Lyth:1998xn}. Using the expression for $\epsilon_\rr{V}$ \eqref{eq:slow_roll_HI} for $h_*\equiv h_\rr{COBE}$, the definition of $U$ \eqref{eq:U_HI} in the large field limit ($\Omega^2\simeq \xi h^2$) and the expression \eqref{eq:N_HI} for $N_\rr{COBE}$ corresponding to $k=k_*$ gives,
\bea 
  \frac{U_*}{\epsilon_*}&\simeq& \frac{3 \lambda_\rr{sm} \Mp^4}{16} h^4_\rr{COBE}, \\
  &\simeq& \frac{\lambda_\rr{sm} \Mp^4}{3\xi^2}N^2_\rr{COBE}.
\eea
By definition of the normalization of the CMB power spectrum \eqref{eq:CMB_norm_HI} $U_*/\epsilon_*\sim 3\pi 10^{-10} \Mp^4\sim \left(0.027 \Mp\right)^4$, yielding,
\be
  \xi\simeq \sqrt{\frac{\lambda_\rr{sm}}{3}} \frac{N_\rr{COBE}}{(0.027)^2} 
  \simeq 49,000 \sqrt{\lambda_\rr{sm}}\simeq 10^4-10^5,
  %\simeq 49,000 \frac{m_\rr{H}}{\sqrt{2}v}
\ee
since $N_\rr{COBE}\simeq 62$ and $\lambda_\rr{sm}\sim0.1$.

\subsection{High energy physics and extensions of the Higgs inflation} \label{sec:high_energy_higgs}
Higgs inflation appears to be favored by latest cosmological observations provided the nonminimal coupling $\xi\sim10^4-10^5$. However, such a model involves quantum corrections (somewhat flawed by the non-renormalizable character of GR), either from quantum gravity or from loop corrections of the SM fields (among them the Higgs field itself) \cite{Bezrukov:2007ep}. The crucial point is the flatness of the effective potential in the Einstein frame for large $\chi$ which must be preserved. The one-loop and two-loop corrections have been studied assuming that the SM is valid up to the Planck scale \cite{Bezrukov:2008ej, Bezrukov:2009db} (see also \cite{Martin:2013tda}). Following \cite{Bezrukov:2009db} it results that the SM Higgs inflation is viable for Higgs mass values $m_\rr{H}\in\left[126,~194\right]~$GeV depending on the mass of the top quark and the coupling constant of strong interactions $\alpha_\rr{s}$. Their analysis is nevertheless controversial (see e.g. \cite{Barvinsky:2008ia, DeSimone:2008ei} where $m_\rr{H}\sim120-135~$GeV in the latter) notably because the Jordan and the Einstein frames are equivalent at tree level only \cite{Steinwachs:2013tr}. The slow-roll analysis of the radiatively corrected Higgs inflation depending on the potential parameter responsible for the radiative corrections (and on the reheating temperature) has been presented in \cite{Martin:2013tda}. They found that, in agreement with \cite{Barvinsky:2008ia}, radiatively corrected Higgs inflation model appears to be disfavored when cosmological and particle physics data are taken into account altogether.

Moreover, some authors argued that Higgs inflation is an effective theory valid up to the scale $\Lambda_0=\mpl/\xi$ only, below the Higgs field value during the slow-roll inflation, $H\gg\mpl/\sqrt{\xi}$, since above $\Lambda_0$ the Higgs field becomes strongly coupled to the Ricci scalar \cite{Barbon:2009ya} (see Sec.~\eqref{sec:MG_issues}). A similar result was derived by \cite{Burgess:2009ea} where it is shown that the semiclassical treatment of Higgs inflation is correct for energy scale, $\mpl/\xi\gg\Lambda_0\gg\sqrt{\lambda_\rr{sm}}\mpl/\xi$. Otherwise unitarity at the quantum level could be lost for processes like the graviton-Higgs and Higgs-Higgs scattering (in flat space). This means that above the ultra-violet cutoff $\Lambda_0$ the SM should be replaced by a more fundamental theory. In \cite{Bezrukov:2010jz} authors claimed that the cutoff scale depends to the background value of the Higgs field leading to the validity of the semiclassical treatment during inflation where $\Lambda_0\sim \mpl$. Moreover, the effect of the couplings to fermions does not spoil this result while those to gauge bosons lead to a slightly lower cutoff.

In order to avoid the loss of unitarity some modifications of the Higgs inflation have been proposed, either via additional interactions due to the term with covariant derivatives of the Higgs doublet in the action \cite{Lerner:2010mq} or by including additional scalar field like the dilaton \cite{GarciaBellido:2011de, Bezrukov:2012hx} which can lead to the late-time accelerating phase. In addition a model involving a nonminimal derivative coupling of the Higgs field to gravity has been proposed \cite{Germani:2010gm, Germani:2010JCAP},
\be
  S=\int \dd^4 x\sqrt{-g} \left[\frac{R}{2\kappa}
  -\frac{1}{2}\left(g_{\mu\nu}-w^2 G_{\mu\nu}\right)\df^\mu H \df^\nu H
  -\frac{\lambda}{4}H^4\right],
\ee
with $w^2$ a coupling constant in $[\rr{GeV}^{-2}]$. This model also preserves unitarity and leads to viable inflation if $w$ is around the geometric mean of the electroweak and the Planck scale \cite{Germani:2010JCAP}. In Chap.~\ref{chap:FabFour} we will come back on this nonminimal coupling function.
%\cite{Bezrukov:2011sz}
% Spontaneous scalarization \cite{Pani:2010vc} cf.PRL

%\tcb{est-ce que tu parles du modèle de Max où il utilise la Higgs inflation pour faire de la DE? Je pense que ça serait intéressant de mettre un paragraphe là-dessus ici car cela renforce l’intérêt pour le modèle de higgs non minimalement couplé.}

\begin{sloppypar}
\section[Higgs monopoles]{Particlelike distributions of the Higgs Field nonminimally coupled to gravity}

\noindent
Since the nonminimal coupling $\xi$ for a viable inflation model is very large (see Sec.~\ref{sec:higgs_inflation}), of the order of $10^{4}$, it naturally raises concerns about static configurations: how a such strongly coupled Higgs field reacts in the presence of gravitationally bound matter? What does the vacuum look like in the vicinity of a compact object? Since the works of Damour and Esposito-Far\`ese \cite{PhysRevLett.70.2220}, we know that a non-minimally coupled scalar field can give rise to spontaneous scalarization in compact objects (see Sec.~\ref{sec:spont_scala}). In this section, we will show that all spherically symmetric distributions of matter carry a classical Higgs charge, whose magnitude depends on their mass, their compactness, and the strength of $\xi$. However, contrary to spontaneous scalarization, only one particlelike distribution of the Higgs field, that is globally regular and asymptotically flat distribution with finite energy, does exist. This solution is characterized by the radius and baryonic energy density of the compact object as well as the nonminimal coupling. Finally we highlight the existence of a mechanism of resonant amplification of the Higgs field inside the so-called Higgs monopoles that comes into play for large nonminimal coupling. This mechanism might degenerate into divergences of the Higgs field that reveal the existence of forbidden combinations of radius and baryonic energy density.
\end{sloppypar}

%%%%%%%%%%%%%%%%%%%%  SEC II %%%%%%%%%%%%%%%
\subsection{The Model}\label{sec:model_monop}
%%%%%%%%%%%%%%%%%%%%%%%%%%%%%%%%%%%%%%%%
\noindent We start from the same action as for the Higgs inflation \eqref{eq:action_Higgs_inflationJF}, including the matter part of the action,
\bea
\mathcal{L}&=&\sqrt{g}\left[{F\left(H\right)\over {2\kappa}}{R}-\frac{1}{2}\left(\partial H\right)^2-V\left(H\right)\right]
+\mathcal{L}_\rr{M}\left[g_{\mu\nu};\Psi_\rr{M}\right],
\label{eq:lag_monop}
\eea
where $H=\mpl h$ is the Higgs scalar field in the unitary gauge\footnote{Notice that the gauge symmetry does not appear explicitly in the Lagrangian \eqref{eq:lag_monop}. Following \cite{Bezrukov:2007ep} the effect of gauge bosons is neglected according to chaotic inflation scenario.}. The potential $V$ is given by Eq.~\eqref{mexican} with the usual SM model parameters\footnote{We use the SM values for the parameters of the potential  $\lambda_{\rm sm}$ and $v$. However, it would be interesting to study this theory as a generic STT to see in which range these parameters are compatible with the current observations.}
and the nonminimal coupling function by Eq.~\eqref{eq:nonminHI}.
As reminded in Sec.~\ref{sec:higgs_inflation}, this model yields a successful inflation provided $\xi$ is large, of the order $10^{4}$ \cite{Bezrukov:2007ep}.  We will consider only positive values of $\xi$ to avoid the possibility that the effective reduced Planck mass (that can be identified with $(\mpl^{2}+\xi H^{2})^{1/2}$) becomes imaginary during its dynamical evolution. 

A similar Lagrangian for compact objects was already considered by \cite{Salgado:1998sg}, where, however, the potential was neglected. As we will see below, this is an important difference as the presence of the Higgs potential prevents the
solution from converging smoothly to GR. In other words, the solution $H=0$ does not yield the
Schwarzschild solution but, rather, a de Sitter black hole with a cosmological constant proportional to $v^{4}$.

It should also be kept in mind that the Higgs field is in general a complex doublet and, here, it is reduced to a
single real component by choosing the unitary gauge \cite{Bezrukov:2007ep}. However, the other components, also known as
Goldstone bosons, can have physical effects, especially at high energy, when renormalizability imposes a different gauge choice (e.g.\ the so-called $R_{\xi}$-gauges, see for example \cite{peskin}). 
In cosmology, the effects of the Goldstone boson in a toy $U(1)$ model was  investigated by
\cite{Rinaldi:2013lsa, Rinaldi:2014yta}. In the context  of compact object, some results can be found in \cite{vanderBij:1987gi} although the
potential is not of the Higgs type.        

The equations of motion obtained from the Lagrangian
\eqref{eq:lag_monop}  by variation with respect to the metric read,
\be
 \left(1+{\xi\over \mpl^2} H^2\right)G_{\mu\nu}=\kappa\left[T_{\mu\nu}^{(H)}+T_{\mu\nu}^{(\xi)}+T_{\mu\nu}^{(\rr{M})}\right],
\label{eom_tensor}
\ee 
where,% $G_{\mu\nu}$ is the Einstein tensor, 
\bea\label{tmunuH}
T_{\mu\nu}^{(H)}&=&\partial_{\mu} H\partial_{\nu} H - g_{\mu\nu}\left[ \frac{1}{2} \left(\partial H\right)^2 + V\left(H\right)\right],
\eea
is the part of the stress-energy tensor associated to the Higgs field, and, 
\be
T_{\mu\nu}^{(\xi)}=-{\xi\over 4\pi} \left[g_{\mu\nu} \nabla^\lambda\left(H \nabla_{\lambda} H \right)-\nabla_{\mu}\left(H \nabla_{\nu} H\right) \right],
\ee
is the stress-energy tensor induced by the nonminimal coupling $\xi$. Finally, 
%\be
%T_{\mu\nu}^{(\rr{M})}={2 \over \sqrt{-g}} {\delta \mathcal{L}_\rr{M} \over \delta g^{\mu\nu}},
%\ee
the stress-energy tensor of the baryonic matter fields $T_{\mu\nu}^\rr{(M)}$ is given by Eq.~\eqref{eq:def_T} that we assume to have the form of a perfect fluid given by Eq.~\eqref{eq:perfect_fluid} with the specific energy density $\Pi=0$. We point out that we do not introduce any coupling between the Higgs field and baryonic matter.
%, so that
%\bea
%T_{\mu\nu}^{(\rr{M})}&=& \left(\rho+ p \right) u_\mu u_\nu +g_{\mu\nu} p.
%\eea
%where $u_\mu$ is the four-velocity, $\rho$ is the density and $p$ the pressure. 

Here, we adopt the splitting of the energy momentum tensor proposed in \cite{Salgado:1998sg} 
as each part will give distinct contributions, as we will see in Sec.~\ref{sec:num_monop}. 
The set of equations of motion is completed by the Klein-Gordon equation,
\be
\square H+\frac{\xi HR}{8\pi}=\frac{\dd V}{\dd H},
\label{eq:KG_monop}
\ee 
from which we can understand in a qualitative way the main characteristics of the solution, as we show in the next section.

%%%%%%%%%%%%%%%%%%%%  SEC III %%%%%%%%%%%%

\subsection{Effective dynamics}\label{sec:effective_dyn_monop}

%%%%%%%%%%%%%%%%%%%%%%%%%%%%%%%%%%%%%

\noindent Our first goal is to assess whether spherically symmetric and asymptotically flat solutions to the equations of motion exist. 
The term in the Klein-Gordon equation \eqref{eq:KG_monop} that tells us if this is possible, is the one proportional to $\xi HR$. 
For a start, it is clear that the trivial function $H(r)=0$ is always a solution of Eq.~\eqref{eq:KG_monop} even with $\xi\neq0$ and in the presence of matter, i.e. when $R\neq0$. 
%a trivial solution of the Klein-Gordon equation. 
If we consider a static and spherically symmetric spacetime in the Schwarzschild coordinates, described by the metric \eqref{eq:metric_schwa},
% \be
% ds^{2}=-{e}^{2\nu\left(r\right)}dt^{2}+{e}^{2\lambda\left(r\right)} dr^{2}+r^{2} d\Omega^{2},
% \label{schwar}
% \ee
we see from \eqref{eq:lag_monop} that, for $H=0$ and in the absence of matter, we obtain a de Sitter black hole solution since $V(H=0)=\lambda_\rr{sm}v^4/4$.
%because of the term proportional to $v^{4}$ in the potential \eqref{mexican}. 
Therefore, this solution is not asymptotically flat and has infinite energy. 
In the absence of nonminimal coupling $(\xi=0)$, the only asymptotically flat solution of finite energy, namely $H=\pm v$, leads to the usual
Schwarzschild metric (with or without internal matter). On the other hand, with  a nonminimal
coupling and in the absence of matter, there are no-hair theorems that force the solution to be the Schwarzschild one, i.e. again $H(r)=\pm v$ everywhere \cite{Sotiriou:2011dz}. Therefore, the only non-trivial case is the one with nonminimal coupling and nonvanishing baryonic matter density, 
which, as we will show, has indeed finite energy and is asymptotically flat unless $H=\pm v$ is a global solution.  

To examine in detail the dynamics, we rewrite the Klein-Gordon equation \eqref{eq:KG_monop} as,
\be
\square H=-{\dd V_{\rm eff}\over \dd H},
\ee
where,
\bea
V_{\rm eff}=-V+\frac{\xi H^2R}{16\pi}+\mathcal{C},
\label{eq:Veff_monop}
\eea
$\mathcal{C}$ being a constant of integration. Note that the form of the effective potential in a time-dependent inflationary background has the opposite sign with respect to the one in a static and spherically symmetric background. In fact, if the metric has the flat FLRW form given by \eqref{eq:metric_FLRW} with $k=0$,
% \be
% ds^2=-dt^2+a(t)^2 \left(dr^2+r^2 d\Omega^2\right),
% \ee
the scalar field rolls down (in time) into the potential well since the Klein-Gordon equation has the form,
\be
{\dd^{2}H\over \dd t^{2}}+{3\over a}\frac{\dd a}{\dd t}\frac{\dd H}{\dd t} = {\dd V_{\rm eff}\over \dd H}.
\ee
On the other hand, with the static and spherically symmetric metric \eqref{eq:metric_schwa} the Klein-Gordon equation becomes,
\bea
H''-H'\left(\lambda'-\nu'-\frac{2}{r}\right)&\equiv&-{\dd V_{\rm eff}\over \dd H},\\
&=&\left[-\frac{\xi R}{8\pi }+\lambda_{\rm sm}(H^2-v^2)\right]H,\nonumber
\eea
where the prime denotes a derivative with respect to the radial coordinate $r$. For minimal coupling $\xi=0$, while $H=\pm v$ ($H=0$) corresponds to local minima (maximum) in the cosmological case, it corresponds to local maxima (minimum) in the spherical symmetric static configuration.  In addition, for nonminimal coupling, $H=0$ is a stable equilibrium point while $H=v$ is an unstable one. In order to fully characterize the stability of these points in the nonminimal coupling case, it is necessary to compute $R$. 
\begin{figure}
\begin{center}
\includegraphics[width=0.6\textwidth]{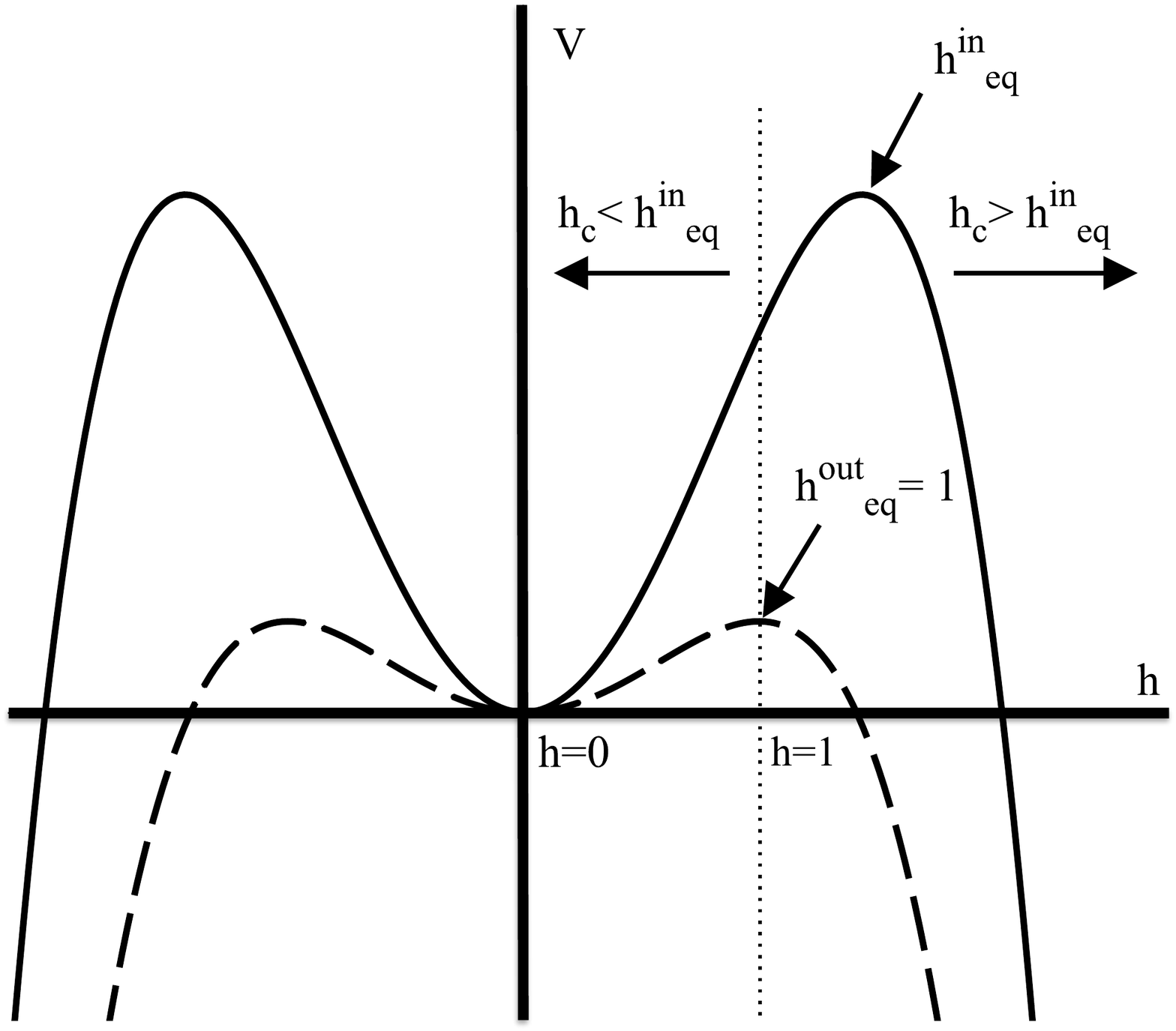}
\caption{Qualitative plot of the potential inside (solid line) and outside (dashed line) the body. The effective potential corresponds to the Higgs one outside the body while the local maxima (see $h^{\rm in}_{\rm eq}$ with $H=\mpl v h$) are displaced from the vev inside the body.}
\centering
\label{fig:plot_Veff_monop}
\end{center}
\end{figure}

For simplicity, a top-hat distribution is assumed of baryonic matter from now on, namely,
\bea\label{eq:rho_monop}
\rho(r)=\left\{\begin{array}{ll}\rho_{0} & 0<r<{\cal R}, \\0 & r>{\cal R},\end{array}\right.
\eea
where ${\cal R}$ is the radius of the spherical body. In this case, the effective potential shows a sharp transition 
between the interior and the exterior of the body. Indeed, if $\xi\neq 0$, the second term in the right-hand side of Eq.\ \eqref{eq:Veff_monop} comes 
into play and we can show that the Ricci scalar satisfies the inequality,
\be
R(r<\mathcal{R})\gg R(r>\mathcal{R}). 
\ee
The reason is that, inside the body, the Higgs field turns out to be almost constant, as shown in Sec.\ \ref{sec:num_monop}. 
Therefore, all the derivatives in the trace of the stress-energy tensor vanish and the only consistent 
contribution to $R$ comes from the potential, as one can easily check by calculating the trace of Eq.\  \eqref{eom_tensor}. If the 
Higgs field is not too much displaced from its vev inside the body, the greatest contribution to the curvature then comes from the baryonic matter, provided the density is sufficiently large.  Outside the body, the Higgs field drops very rapidly towards its vev and $R$ vanishes at large $r$ to match the Schwarzschild solution $R=0$ everywhere. For practical purposes, this means that we can approximate $R$, inside the body, as there was no Higgs field but just matter. To show this property a bit more rigorously, it is sufficient to calculate the trace of Eq.\ \eqref{eom_tensor} and recall that, at the center of the body, we must have $\dd H/\dd r=0$ according to the regularity conditions. Therefore, near the center 
of the body, the trace of the Einstein equations is approximate by,
\bea\label{apptrace}
\left(1+{\xi H^{2}\over \mpl^{2}}\right)R\simeq -\kappa\left(2V+T^{(\rr{M})}-{3\xi\over 4\pi}H\square H\right).
\eea
In addition, for energies much lower than the Planck scale, $H\lll \mpl$, all the terms like $\xi H^{2}/\mpl^{2}$ can be safely neglected, even when $\xi\sim 10^{4}$. Finally by also using the Klein-Gordon equation \eqref{eq:KG_monop}, Eq.\ \eqref{apptrace} can be accurately approximated by,
\bea\non
{1\over v^{2}}\left(R+\kappa T^{(\rr{M})}\right)=-{\kappa\over v^{2}}\left(2V-{3\xi H\over 4\pi}{\dd V\over \dd H}\right),\qquad\qquad\qquad\qquad\\
=-{4\pi\lambda v^{2}\over \mpl^{2}}\left({H^{2}\over v^{2}}-1\right)\left[{H^{2}\over v^{2}}\left(1-{3\xi\over 2\pi}\right)-1\right].
\label{approx_dyn}
\eea

Now, since $(v/\mpl)^{2}\sim 10^{-34}$, we need a very large ratio $H/v$ to yield a non-negligible right hand side, 
even for $\xi$ of the order of $10^{4}$. Therefore, unless we consider planckian energies for the Higgs field, the left hand side of the above equation is negligibly small, at least near the center of the body. This means that, inside the body, the Einstein equations are indistinguishable from the standard GR equation $R=-\kappa T^{(\rr{M})}$. In the App.~\ref{app:num_higgs_monop}, the validity of this approximation is established numerically, so it can be used to investigate the particlelike solutions for a very large range of parameters. 

Let us now study the equilibrium points of $V_{\rm eff}$. Outside the body, where $T^{(\rr{M})}=0$, the scalar curvature is almost vanishing $R\simeq 0$. Therefore, $\dd V_{\rm eff}/\dd H$ vanishes at,
\be \label{eq:eq_points_out}
{H^{\rm out}_{\rm eq}\over v}=0,\,\pm 1.
\ee 
Inside the body, where $R\simeq -\kappa T^{(\rr{M})}$, %we find instead
\be \label{eq:eq_points_in}
{H^{\rm in}_{\rm eq} \over v}=0,\,\pm \sqrt{1+{R \xi \over 8\pi\lambda_{\rm sm} v^2}},
\ee
and the crucial role of a nonvanishing $\xi$ becomes evident. 

The solutions that we are looking for, must interpolate between the value of the Higgs field at the center of the body $H_\rr{c}$  and at the spatial infinity $H_\infty=\pm v$. Furthermore, since  $H'$ must vanish at the origin, $H$ rolls down into the effective potential from rest. Suppose that $\vert H_\rr{c}\vert$ is greater than the nonzero value of $\vert H^{\rm in}_{\rm eq}\vert$ (see the solid line curve in Fig.\ \ref{fig:plot_Veff_monop}). Then, the Higgs field will roll outwards increasing boundlessly its value without any possibility of reaching an equilibrium point outside the body (see the dashed line curve in Fig.\ \ref{fig:plot_Veff_monop}), leading to infinite energy configurations. On the contrary, if $\vert H_\rr{c} \vert $ is smaller than the nonzero value of $\vert H^{\rm in}_{\rm eq}\vert$, the Higgs field rolls down inward, towards the equilibrium at $H_\infty=0$, see Fig.\ \ref{fig:plot_Veff_monop}, a solution of infinite energy too. 
By requiring a Higgs distribution which is globally regular and an asymptotically flat metric, only one
initial value $H_\rr{c}$ converges to the vev at spatial infinity, i.e. $\vert H^{\rm out}_{\rm eq}\vert$, for a given nonminimal coupling $\xi$ as well as energy density $\rho$ and radius $\mathcal{R}$ of the body (or equivalently its compactness $s$ and baryonic mass $m$). 
%, there is only one initial value $H_\rr{c}$ such that $H(r)$ smoothly rolls towards $H_{\infty}=v$ and the geometry is asymptotically flat. 
All the other trajectories lead to either an asymptotically de Sitter solution or to a divergent Higgs field at infinity. These particular solutions, with finite energy and asymptotically flat geometry, are dubbed \emph{Higgs monopoles} since they behave like isolated SM scalar charges. Numerical solutions will be obtained with a specifically designed shooting method in the following sections.

%%%%%%%%%%%%%%%%%%%%%%%%%%%%%%%%%%%%%%%%

\subsection{Analytic properties}\label{sec:analyt_monop}

%%%%%%%%%%%%%%%%%%%%%%%%%%%%%%%%%%%%%%%%
\noindent Before exploring the numerical solutions to the equations of motion, it is worth investigating their analytical properties  in order to obtain information able to target  more efficiently the numerical analysis. 
%We find more convenient, 
For this section, it is more convenient to write the Lagrangian \eqref{eq:lag_monop} in the standard Brans-Dicke form,
\bea\label{eq:BDaction_monop}
{\cal L}_\rr{BD}={\sqrt{-g}\over 2\kappa}\left[\phi R-{\omega\over \phi}(\partial \phi)^{2}-\bar V(\phi)\right]+{\cal L}_\rr{M}.
\eea
where
\bea
\phi=1+{\xi H^{2}\over \mpl^{2}},\quad \omega(\phi)={2\pi\phi\over \xi(\phi-1)},
\eea
and
\bea
\bar V(\phi)={\kappa\lambda_{\rm sm}\over 2}\left[{8\pi\over \xi\kappa}(\phi-1)-v^{2}\right]^{2}.
\eea
The Einstein equations now read,
\bea\label{eq:ee_monop}
R_{\mu\nu}-{1\over 2}g_{\mu\nu}R&=&{1\over \phi}\nabla_{\mu}\nabla_{\nu}\phi+{\omega\over\phi^{2}}\nabla_{\mu}\phi\nabla_{\nu}\phi \non\\
&-&{1\over\phi}\left[\square\phi+{\omega\over 2\phi}(\partial\phi)^{2}+{\bar V\over 2}\right]g_{\mu\nu}+{\kappa\over\phi}T_{\mu\nu},
\eea
where $T_{\,\,\mu}^{\nu}={\rm diag}(-\rho,p,p,p)$ is the energy momentum tensor of the fluid. Using the trace of this 
equation, the Klein-Gordon equation \eqref{eq:KG_monop} becomes,
\bea
(2\omega+3)\square\phi+\frac{\dd\omega}{\dd\phi}(\partial\phi)^{2}-\phi{\dd\bar V\over \dd\phi}+2\bar V= \kappa T.
\eea
With the metric \eqref{eq:metric_schwa}, the $tt-$ and $rr-$components of the Einstein equations are, respectively,
%\begin{widetext}
\bea
\lambda'\left({2\over r}+{\phi'\over\phi}\right)-{\kappa\rho\over \phi }\rr{e}^{2\lambda}+{1\over r^{2}}\left(\rr{e}^{2\lambda}-1\right)-{\phi''\over\phi}-{2\phi'\over r\phi}\non\\
-{\bar V \rr{e}^{2\lambda}\over 2\phi}
-{\omega\over 2}\left( \phi'\over \phi\right)^{2}=0,\\
\nu'\left({2\over r}+{\phi'\over\phi}\right)-{\kappa p\over\phi }\,\rr{e}^{2\lambda}-{1\over r^{2}}\left(\rr{e}^{2\lambda}-1\right)+{2\phi'\over r\phi}\non\\
+{\bar V \rr{e}^{2\lambda}\over 2\phi}
-{\omega\over 2}\left( \phi'\over \phi\right)^{2}=0,
\eea
while the angular component is,
\bea
\nu''+(\nu')^{2}+\nu'\left({1\over r}+{\phi'\over \phi}\right)-\lambda'\left(\nu'+{1\over r}+{\phi'\over\phi}\right)+{\phi''\over\phi}+{\phi'\over r\phi}\non\\
+{\omega\over 2}\left( \phi'\over \phi\right)^{2}+{\rr{e}^{2\lambda}\over\phi}\left({\bar V\over 2}-\kappa p\right)=0.
\eea
Finally, the Klein-Gordon equation becomes,
\bea \label{eq:KG_monop_BD}
(2\omega+3)\left(\phi''+\phi'\nu'-\phi'\lambda'+{2\over r}\phi'\right)+(\phi')^{2}{\dd\omega\over \dd\phi}\non\\
+\rr{e}^{2\lambda}\left(2\bar V-\phi{\dd\bar V\over \dd\phi}\right)+\kappa \rr{e}^{2\lambda}(\rho-3p)=0.
\eea
%\end{widetext}
To these we must add the TOV equation \eqref{eq:TOV} obtained by the usual Bianchi identities. The total energy momentum tensor is identified with the right hand side of Eq.\ \eqref{eq:ee_monop}. Therefore, the total energy density, given by $\rho_{\rm tot}=-T^{0}_{\rm tot\,\,0}$, reads,
\bea
\rho_{\rm tot}=\rr{e}^{-2\lambda}\!\!\left({\phi''\over \phi}-{\phi'\lambda'\over \phi}+{2\phi'\over r\phi }+{\omega 
\phi'^{2}\over 2\phi^{2}}\right)+{\bar V\over 2\phi}+{\kappa\rho\over\phi}.\label{rhotot}
\eea
As mentioned in the previous section, if $\phi$ (and hence, $H$)  varies very slowly with $r$, the energy density is dominated by the baryonic matter. This is certainly true near the center, as there $\phi'(r=0)=0$ is imposed, as required by symmetry arguments.  As a consequence, all the derivatives are negligible and we are left with,
\bea
\rho_{\rm tot}\simeq {\bar V_\rr{c}\over 2\phi_\rr{c}}+{\kappa\rho\over\phi_\rr{c}},
\eea
where a subscript ``$c$'' indicates the value of a quantity at the center of the body. If $\bar V_\rr{c}$ is not too large, that is the Higgs field is not displaced too much from its vev, then the energy density can be taken as the one of GR, as already mentioned in Sec.~\ref{sec:effective_dyn_monop}.

Now, consider the Klein-Gordon equation \eqref{eq:KG_monop_BD} and suppose that there exists a point $\bar r$ such that $\phi'(\bar r)=0$. Suppose also that the energy density is constant inside the body, $\rho=E$. It follows that, at that point,
\bea
\bar\phi''= {\rr{e}^{2\bar\lambda}\over (2\bar \omega+3) }\left[ {64\pi^{2}\lambda_{\rm sm} \phi_{v}(\bar\phi-\phi_{v})\over\kappa \xi^{2}}-\kappa(E-3\bar p)\right],
\eea
where the bar denotes quantities calculated at $\bar r$ and $\phi_{v}=1+\xi\kappa v^{2}/(8\pi)$ is the vev  of 
$\phi$. Outside the body, where $p=E=0$ everywhere and $\rr{e}^{2\lambda}$ is positive definite, we have two cases:
\begin{itemize}
\item $\bar\phi''>0$, i.e. a local minimum, which implies that $\bar\phi>\phi_{v}$,
\item $\bar\phi''<0$, i.e. a local maximum, which implies that $\bar\phi<\phi_{v}$.
\end{itemize}
This shows that if there is a local maximum or a local minimum for $\phi$ at a point outside the body, then the field 
cannot converge to its vev $\phi_{v}$ at infinity. This is possible only if $\phi$ is a monotone and decreasing 
function of $r$ (or if $\phi=\phi_{v}$ everywhere). As we will see further, 
this property allows to approximate the Higgs field outside the body with a Yukawa function and an associated scalar charge (see Eq.~\eqref{eq:scalar_charge} for a general definition). This is no longer true inside the body, as $E-3\bar p>0$ and the displacement of $\phi$ from its vev can be 
compensated by contributions from the energy density and the pressure. Thus, we can have both minima and maxima of the field inside the body. In other words, $\phi$ can oscillate only inside the body. The monopole solution that is reported in App.~\ref{sec5}, illustrates this analytical property.

Let us look at the equations of motion in the absence of matter, i.e. with $\rho=p=0$. 
From the Klein-Gordon equation \eqref{eq:KG_monop_BD}, we see that, in this case, asymptotic flatness (namely $\lambda'= \nu'= \phi'\simeq0$ for large $r$) is consistent with $\phi(r\rightarrow\infty)=\phi_{v}\approx 1$ (since $\kappa\xi v^2\ll 1$)  only if,
\be
  \left. 2 V(\phi_{\infty})-\phi_{\infty} \frac{\dd V}{\dd \phi}\right|_{\phi=\phi_\infty}=0. 
\ee
As discussed in \cite{Sotiriou:2011dz}, this 
implies that the only asymptotically flat solution to the equations of motion is the one that coincides with GR, namely the Schwarzschild metric with constant scalar field.

\subsubsection{Classical energy}

\noindent In STT, it is customary to calculate the binding energy of the system and compare it to the GR value in order to see if a solution is energetically favored. The binding energy (see also Sec.~\ref{sec:schwa_in}) is defined by the difference between the baryonic energy (the energy of the baryons if they were dispersed) and the ADM energy $E_{\rm bin}=E_{\rm bar}-E_{\rm ADM}$. The baryonic energy is defined by,
\bea
E_{\rm bar}&\equiv&\int_\rr{V} \dd^3 x \sqrt{^{(3)}g} T^0_{0,(\rr{M})},\\
&=&\int_\rr{V}\dd^{3}x\sqrt{{}^{(3)}g}\,n(r)m_\rr{b},\\
&=&{4\pi\over \phi }\int_{0}^{\cal R} \dd r\, r^{2}\rho(r)\rr{e}^{\lambda(r)},
\eea
where $n(r)$ is the density number, $m_\rr{b}$ is the average mass of a baryon, $\rho(r)$
is the density profile and $\sqrt{{}^{(3)}g}$ is the proper volume measure. The ADM energy is defined as (see also Eq.~\eqref{eq:ADMmass}),
\bea
E_{\rm ADM}&\equiv&\int_\rr{V} \dd^3 x \sqrt{^{(3)}g} \left[T^0_{0,(\rr{M})}+T^0_{0,(\rr{H})}\right],\\
%&=&-\!\!\int \dd^{3}x\sqrt{{}^{(3)}g}\Bigg|_{r=\infty}\!\!\!\!g_{tt}T^{tt}_{\rm tot},\\
&=&-4\pi\!\int_{0}^{\infty}\!\!\dd r\, r^{2}\rho_{\rm tot},
\eea
where $\rho_{\rm tot}$ is the total energy density, including the scalar field contributions. In our case, it is given by Eq.\ \eqref{rhotot}. In general, when the potential is such that the scalar field vanishes at its minimum, there are always two types of solutions. The first has a vanishing scalar field everywhere and coincides with standard GR solutions. In the absence of matter and angular momentum, this solution is the Schwarzschild metric. The second solution has a varying scalar field and it approaches the Schwarzschild solution only at spatial infinity. In that case, the compact object carries a scalar charge whose effects are asymptotically vanishing \cite{PhysRevLett.70.2220, Salgado:1998sg}. The important point is that the two families of solution are smoothly connected and this allows to compare the binding energy of the two configurations and to determine the stable one, or at least the one that minimizes the energy\footnote{The stability under small perturbations of the metric is a different and much more complicated issue that will not be considered in this work.}. In our case, however, this comparison  is meaningless since the monopole solution cannot smoothly reduce to the Schwarzschild one because of the nonminimal coupling. In fact, the monopole is the unique solution (for a given energy density and radius) with finite energy. All other solutions have either a diverging or vanishing scalar field $H$ at spatial infinity, 
%a result we have already established 
as highlighted in Sec.\ref{sec:effective_dyn_monop}. In the first case, the potential term diverges so $E_{\rm ADM}$ is infinite. In the second case, if $H\rightarrow 0$ then $\phi\rightarrow (2\kappa)^{-1}$ so the term $r^{2}V/\phi$ diverges, yielding again an infinite ADM energy \footnote{For a correct calculation of the mass associated to a de Sitter black hole see Ref.\ \cite{PhysRevD.15.2738}.}.

\subsubsection{The TOV equation}

\noindent We now find an approximate formula for the pressure as a function of the energy density and the value of the scalar field at the center $\phi_\rr{c}$, in analogy with the usual TOV equation (see also Sec.~\ref{sec:schwa_in}). To do so, it is sufficient to expand and solve the equations of motion around $r=0$. It should be kept in mind that, for the monopoles, the value of $\phi_\rr{c}$ (or, equivalently, of $\bar V(r=0)$) is not arbitrary. As we explained in the previous sections, the value of $\phi_\rr{c}$ for a given mass and body radius is determined by the condition that $\phi=\phi_{v}$ at spatial infinity and, therefore,  cannot be fixed by a local expansion. This is the reason why the best we can do, analytically, is to find the central pressure $p_\rr{c}=p(r=0)$ as a function of $\phi_\rr{c}$. As before, we assume that $\rho=E=$ const. At the center, owing to spherical symmetry, the scalar field can be approximated by $\phi\simeq \phi_\rr{c}+\phi_{2}r^{2}$ so that we can solve the $tt$-component of the Einstein equations and find,
\bea
\rr{e}^{2\lambda(r)}=\left[1-{2m(r)\over r}\right]^{-1},
\eea
where
\bea\label{mofr}
m(r)\simeq {\left(2\kappa E+\bar V_\rr{c}\right)r^{3}\over 12\phi_\rr{c}},
\eea
with $\bar{V}_\rr{c}=\bar{V}(\phi_\rr{c})$.
This result can be inserted in the $rr$-component of the Einstein equations, which, together with the usual TOV equation \eqref{eq:TOV}, gives the modified TOV equation,
\bea
{\dd p(r)\over \dd r}\simeq-{\left[p(r)+E\right]\left[\kappa E+3 \kappa  p(r)- \bar V_\rr{c}\right]~r\over 6\phi_\rr{c}-(2\kappa  E+\bar V_\rr{c})~r^{2}},
\eea
that can be solved by separation of variables with the boundary condition that $p=0$ at $r={\cal R}$. The result reads,
\bea\non
p(r)={E(\kappa E-\bar V_\rr{c})\left(\sqrt{1-{2m/ r}}-\sqrt{1-{2m{\cal R}^{2}/ r^{3}}}\right)\over 3\kappa 
E\sqrt{1-{2m{\cal R}^{2}/r^{3}}}- (\kappa E-\bar V_\rr{c})\sqrt{1-{2m/r}}},\\
\eea
where $m$ is the function \eqref{mofr}. This equation reproduces the relativistic expression in the limit  
$\bar V_\rr{c}\rightarrow0$. At the center of the body we have,
\bea
p_\rr{c}={E(\kappa E-\bar V_\rr{c})\left[1-\sqrt{1-{2m({\cal R})/ {\cal R}}}\right]\over {3 \kappa E}\sqrt{1-{2m({\cal R})/{\cal R}}}- \kappa E+\bar V_\rr{c}},
\eea
where $m({\cal R})$ is the mass function \eqref{mofr} calculated at $r={\cal R}$. As for the ordinary relativistic stars, there is a maximum value of the energy density, at which the pressure diverges, given by 
\bea
E_\rr{max}={{12\phi_\rr{c}-\bar V_\rr{c}{\cal R}^{2}}\over 2\kappa {\cal R}^{2}}.
\eea
However, in contrast with the GR case, there exists also a critical value of the energy density, below which the pressure becomes negative, that is
\bea
E_\rr{min}={\bar V_\rr{c}\over \kappa}&=&{\lambda_{\rm sm}\over 2}\left[ {8\pi\over \xi \kappa}(\phi_\rr{c}-1)-v^{2} \right]^{2}, \\\nonumber
&=&{\lambda_{\rm sm}\over 2}\left( H_\rr{c}^2-v^{2} \right)^{2}.
\eea
The interpretation is that the Higgs field potential contributes with a negative pressure at the center of the body, at least in the linearized regime considered in this section. When this approximation is no longer valid, we need to resort to numerical tools to calculate the central pressure and verify in which part of the parameter space it is negative and an eventual threat to the stability of the spherical body.
 %%%%%%%%%%%%%%%%%%%%%
\subsubsection{Discussion about astrophysical compact objects}\label{sec:PPN_monop}
\noindent In App.~\ref{sec:PPN_BD}, the PPN formalism (see also Sec.~\ref{sec:PPN}) for the Brans-Dicke theory is reviewed. It leads straightforwardly to the PPN analysis for the Higgs monopoles, which tells us the amount of deviations from GR outside a body of the size of the Sun. However, because of the presence of the potential, PPN parameters only give upper bounds. According to the PPN prescriptions, we assume that far outside the Sun, the Higgs field is close to its vacuum value so that $V\simeq 0$ and the Newton's constant coincides with its bare value. The PPN parameters follow immediately from Eqs.~\eqref{eq:gamPPN} 
and \eqref{eq:betaPPN2} \cite{Damour:1992we},
\bea
\gamma_\rr{PPN}={\omega+1\over \omega+2},\hspace{1.7cm} \beta_\rr{PPN}-1={1\over(2\omega+3)^{2}(2\omega +4)}{\dd\omega\over \dd\phi}.
\eea
When $\phi\rightarrow \phi_{v}$, $\omega(\phi=\phi_{v})\simeq 2\pi \mpl^2/(\xi^2 v^2) \simeq 1.5\times 10^{26}$ and $(\dd\omega/\dd\phi) (\phi=\phi_{v})\simeq -(2\pi/\xi^3)(\mpl^4/v^4) \simeq -3\times 10^{55} (-2.2\times 10^{-20})$ for $\xi=10^4$ according to Higgs inflation. It results that the PPN parameters $\beta_\rr{PPN}-1=\gamma_\rr{PPN}-1=0$ with a precision far larger than the current observational constraints.
%, as . 
%$\gamma-1 \ll \frac{\xi^2 v^2}{\pi}\approx 10^{-26}$ and
%$\beta-1\ll \frac{\xi^3 v^2}{2\pi^2}\approx 10^{-23}$ for $\xi=10^4$. 
Moreover, these are upper bounds  since the Higgs field is massive ($V\neq0$) and thus decays as a Yukawa function outside the matter distribution at a much faster rate than  $1/r$ (typical in the case of a vanishing scalar potential, \cite{Damour:1992we}). The scalar charge is thus almost completely screened over a distance of a few Schwarzschild radii. 

Following the discussion in Sec.~\ref{sec:effective_dyn_monop}, the equilibrium point of the effective potential inside the compact object \eqref{eq:eq_points_in} gives an upper bound on the central value of the Higgs field $H_\rr{c}$, $|H(r)|\le |H_\rr{c}|\le |H_{\rm eq}^{\rm in}|$ $\forall\,\, r\ge 0$. Inside the matter distribution, the Ricci scalar is found to be nearly constant and well approximated by $R\approx R(r=\mathcal{R})=3 s^3/r_\rr{s}^2$ where $s$ is the compactness of the compact object and $r_\rr{s}=8\pi\rho_0/(3\mpl^2) \mathcal{R}^3$ is its standard Schwarzschild radius\footnote{The physical Schwarzschild radius should take into account also the contribution of the Higgs field. Here, we define it instead as a scale of the theory, uniquely determined by the baryonic mass of the monopole as in GR.} (see Eq.~\eqref{scalR} in the next section). This allows one to give the order of magnitude of $H^\rr{in}_\rr{eq}$,
\be
  {H^\rr{in}_\rr{eq}}-v\simeq\frac{3 s^3 \xi}{16\pi r_\rr{s}^2\lambda_\rr{sm} v}.
\ee
Considering the Sun ($s=10^{-6}$ and $m\sim10^{30}~$kg) and viable nonminimal parameter for Higgs inflation ($\xi=10^4$) yields, 
\be
  H^\rr{in}_\rr{eq}-v\sim 10^{-58} v,
\ee
such that no observable effect can be detected. The effect of the Higgs field in the Sun ($s=10^{-6}$ and $m\sim10^{30}~$kg) remains smaller than $H_{\rm eq}^{\rm in}-v\ll 10^{-2} v$ provided that $\xi<10^{58}$.
Therefore, we can conclude that deviations from GR around astrophysical objects like the Sun are outstandingly small, provided that the nonminimal coupling parameter $\xi$ is not extremely large.
%The theory is thus undistinguishable from GR at the level of Solar System experiments, even for large values of $\xi$. 
%This result is consistent with the fact that, for a mass of the size and compactness of the Sun, the spontaneous scalarization is extremely small (see also Sec.~\ref{sec:num_monop}). 
%An alternative weak-field analysis is discussed in Ref.~\cite{Schlogel:2014jea}. %appendix \ref{appendix1}.

Considering NSs ($s=0.2$ and $m\sim10^{30}~$kg), the variation of the Higgs field cannot be larger than,
\be
  H^\rr{in}_\rr{eq}-v\sim 10^{-41} v,
\ee
for $\xi=10^4$. For example,  if $\xi=10^4$ we find that
$H_\rr{c}/v<1.01$ for a mass $m>3\times 10^{10}$ kg with $s=0.2$ and $H_\rr{c}/v<1.01$  for 
a compactness of $s<10^{-5}$ and a mass of $10^4$ kg. We conclude that no effect is measurable in astrophysical compact objects.

%%%%%%%%%%%%%%%%%%%%%%%%%%%%%%%%%%%%%%%%

\subsection{Numerical results}\label{sec:num_monop}

%%%%%%%%%%%%%%%%%%%%%%%%%%%%%%%%%%%%%%%%
\noindent After discussing the dynamics of the model and some generic analytical results, we now study numerically  the properties of the solutions. We report the reader to App.~\ref{app:num_higgs_monop} for the complete set of equations of
motion, the system of units and the numerical methods that we used. In the previous sections, we have shown that the metric components inside the compact object are almost the same as in GR when we choose the SM values for the parameters of the potential.
Therefore, we follow a simplified procedure, which consists in using the GR solution (with the top-hat matter distribution \eqref{eq:rho_monop}) for the metric components and focus solely on the non-trivial dynamics of the Higgs field. We provide for a proof of this approximation in App.~\ref{app:num_higgs_monop} through the comparison between this approach and the numerical integration of the unaltered system of equations of motion.  With these assumptions, the problem  essentially reduces to solving the Klein-Gordon equation,
\bea
h_{uu}+h_u \left(\nu_u-\lambda_u+\frac{2}{u}\right)
=\rr{e}^{2\lambda}\left(-\frac{R \xi h}{8\pi}+\frac{r_\rr{s}^2}{\mpl^2
\tilde v^2}\frac{\dd V}{\dd h}\right),
\label{kg_simp}
\eea
where $h=H/(\mpl\tilde{v})$, $\tilde{v}=v/\mpl$ being the dimensionless vev, and a subscript $u$ denotes a derivative with respect to $u=r/r_\rr{s}$. The metric fields and the scalar curvature are approximated by the interior and exterior Schwarzschild metric, and read respectively,
\bea
\label{nu}\non
\rr{e}^{2\nu}(u)&=&\left\{\begin{array}{lr}\frac{3}{2}\sqrt{1-s}-\frac{1}{2}\sqrt{1-s^3u^2}, &  \;0<u<s^{-1}, \\ \\ 1-u^{-1}, &
u\ge s^{-1},\end{array}\right.\\\\
\label{lam}
\rr{e}^{-2\lambda}(u)&=&\left\{\begin{array}{lr} 1-s^3u^2, &  \qquad 0<u<s^{-1}, \\\\ 1-u^{-1}, & \qquad u\ge s^{-1},\end{array}\right.\\\non\\
\label{scalR}\non
R(u)&=&\left\{\begin{array}{lr} -\frac{6s^3}{r_\rr{s}^2}\left(\frac{2\sqrt{1-s^3u^2}-3\sqrt{1-s}}{3\sqrt{1-s}-\sqrt{1-
s^3u^2}}\right), &  \;0<u<s^{-1}, \\\\ 0, & u\ge s^{-1},\end{array}\right.\\
\eea
%where $s=r_\rr{s}/\mathcal{R}$ is the compactness. 
Regularity at the origin requires that $h_u|_{u=0}=0$ leaving $h_\rr{c}=h(u=0)$ as the only initial condition for Eq. (\ref{kg_simp}).

\begin{figure}[ht]
\begin{center}
\includegraphics[width=0.6\textwidth,trim=320 0 350 0,clip=true]{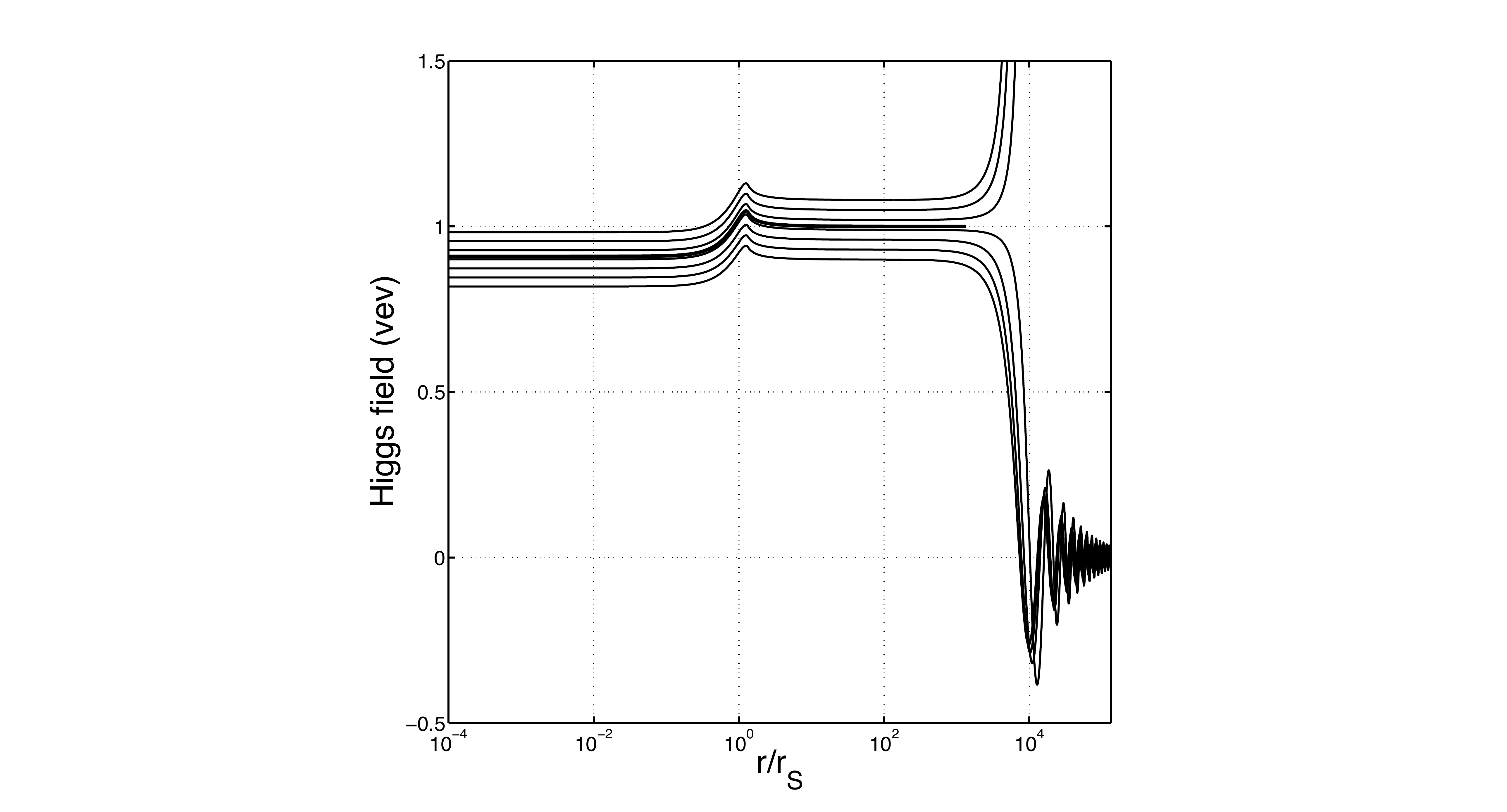}
\caption{Numerical solutions of Eq.\ \eqref{kg_simp} with varying initial conditions $h_\rr{c}=h(r/r_\rr{s}=0)$ for $\xi=10$, $m=10^6\; \rm kg$, and $s=0.75$. The thicker line represents the unique solution that converges to $h=1$ at large $r/r_{s}$.}
\centering
\label{different_hcs}
\end{center}
\end{figure}

In Fig.\ \ref{different_hcs} we plot the numerical solutions of Eq.\ (\ref{kg_simp}) for different values of the initial condition $h_\rr{c}=h(u=0)$. We see that, for fixed mass and compactness, there exists only one value for the initial condition $h_\rr{c}=h_0$ that yields a solution that tends to $h=1$ at spatial infinity (marked by a thicker line). This  solution corresponds to the non-trivial, asymptotically flat, and spherically symmetric distribution of the Higgs field, dubbed ``Higgs monopole'' in \cite{Fuzfa:2013yba, Schlogel:2014jea}. For slightly different initial conditions $h_\rr{c}\ne h_{0}$, the field either diverges (if $h_\rr{c}>h_0$) or tends to zero after some damped oscillations (if $h_\rr{c}<h_0$). This result confirms the analytic treatment of Sec.\ \ref{sec:effective_dyn_monop}.

For each choice of mass $m$, compactness $s$, and coupling strength $\xi$, there exists only one solution of the kind depicted in Fig.~\ref{different_hcs}. Its form varies a lot in function of the parameters, as we show in Fig.~\ref{plot_monop_family} where we plotted several solutions, corresponding to the parametrization listed in Tab.~\ref{table}. We notice that the value of the Higgs field at the center of the monopole 
can be lower than the vev for typically large compactness $s$. For small or moderate compactness, the 
central value of the Higgs field is generically larger than the vev. 

%\afterpage{%    % defer execution until the next page break occurs anyway
   
   \begin{table}[h!] % not t or b or p
      \centering
      \begin{minipage}{\textwidth}
       
%[t!] % not "pt"
      \centering
      \includegraphics[width=0.6\textwidth, trim=340 0 350 0,clip=true] {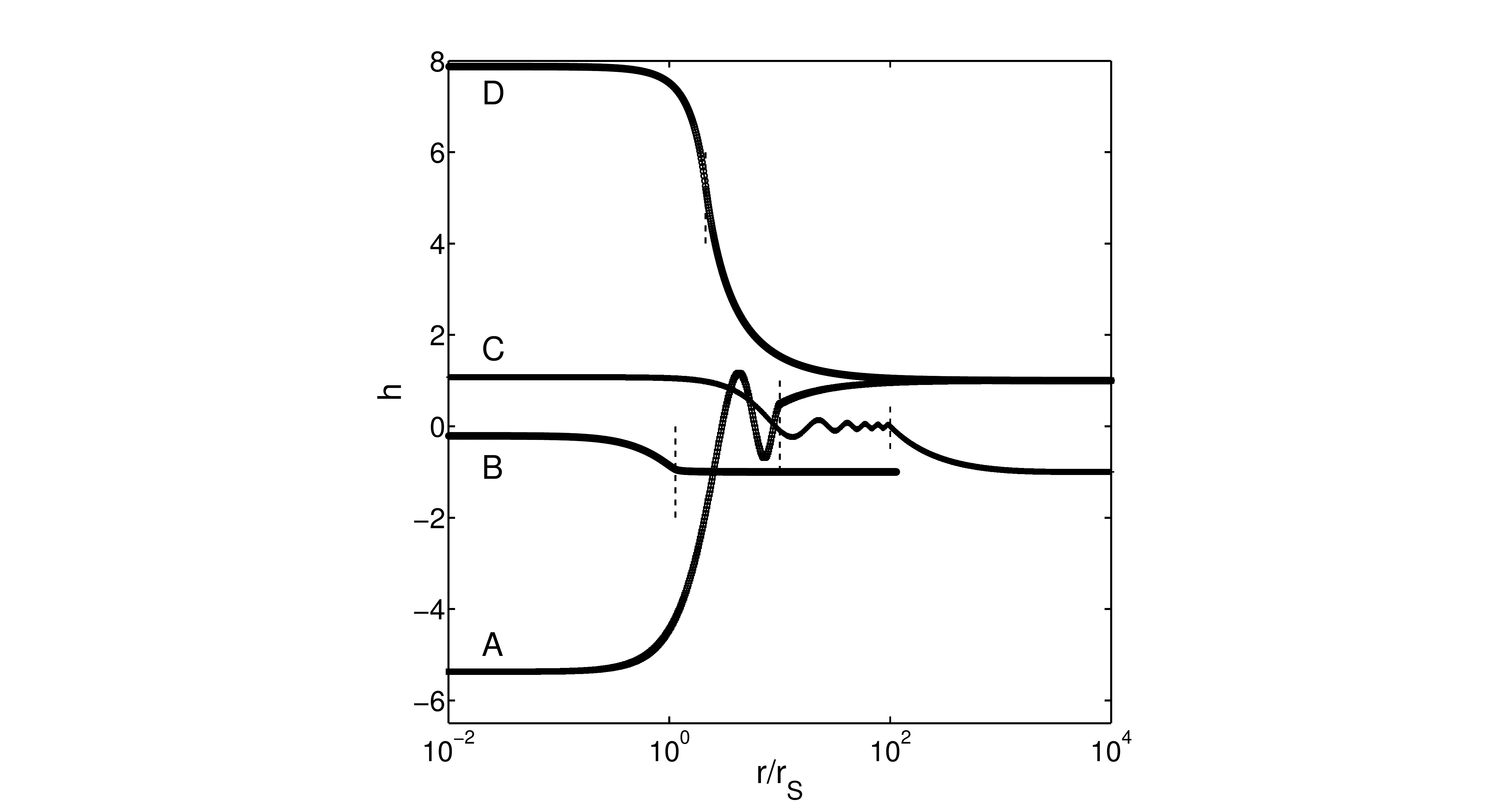}
      \captionof{figure}{Plots of the Higgs field with the parameters listed in Tab.~\ref{table}. The vertical dashed lines mark the radius of the body for each monopole.}
      \label{plot_monop_family}
      \vspace{1cm}
   \end{minipage}
   
      \begin{tabular}{|c|c|c|c|c|c|}
	\hline
	& $h_\rr{c}$ & $\xi$ & m &  s \\
	\hline
	F & 0.91 & 10 & $10^6$ kg & 0.75  \\
	\hhline{|=|=|=|=|=|}
	A & - 5.37 & $10^4$ & $10^3$ kg & $0.1$  \\
	\hline
	B &- 0.21 & $10$ & $10^6$ kg & $0.88$   \\
	\hline
	C & 1.077 & $10^6$ & $10^6$ kg & $0.01$  \\
	\hline
	D & 7.88 & $60$ & $10^4$ kg & $0.47$  \\
	\hline
      \end{tabular}
      \caption{Properties of the Higgs monopoles plotted in Fig.\ \ref{different_hcs} (curve F) and Fig.\ \ref{plot_monop_family} (curves A,B,C,D).}\label{table}
   \end{table}
%} % end of argument of `\afterpage` command

% \begin{figure}[!htbp]
% \begin{center}
% \includegraphics[width=0.6\textwidth, trim=340 0 350 0,clip=true] {./chapters/chapter5/fig3_monop_family-eps-converted-to.pdf}
% \caption{Plots of the Higgs field with the parameters listed in Tab.~\ref{table}. The vertical dashed lines mark the radius of the body for each monopole.}
% \centering
% \label{plot_monop_family}
% \end{center}
% \end{figure}
% %%%%%%%%%%%% TABLE %%%%%%%%%%
% \begin{table}[!htbp]
% \begin{center}
% \begin{tabular}{|c|c|c|c|c|c|}
% \hline
%  & $h_\rr{c}$ & $\xi$ & m &  s \\
%  \hline
%  F & 0.91 & 10 & $10^6$ kg & 0.75  \\
%  \hhline{|=|=|=|=|=|}
% A & - 5.37 & $10^4$ & $10^3$ kg & $0.1$  \\
% \hline
% B &- 0.21 & $10$ & $10^6$ kg & $0.88$   \\
% \hline
% C & 1.077 & $10^6$ & $10^6$ kg & $0.01$  \\
% \hline
% D & 7.88 & $60$ & $10^4$ kg & $0.47$  \\
% \hline
% \end{tabular}
% \caption{Properties of the Higgs monopoles plotted in Fig.\ \ref{different_hcs} (curve F) and Fig.\ \ref{plot_monop_family} (curves A,B,C,D).}\label{table}
% \end{center}
% \end{table}
%%%%%%%%%%%%%%%%%%%%%%%%%%%

\begin{figure}
\begin{center}
\includegraphics[width=0.6\textwidth,trim=280 0 310 0,clip=true]{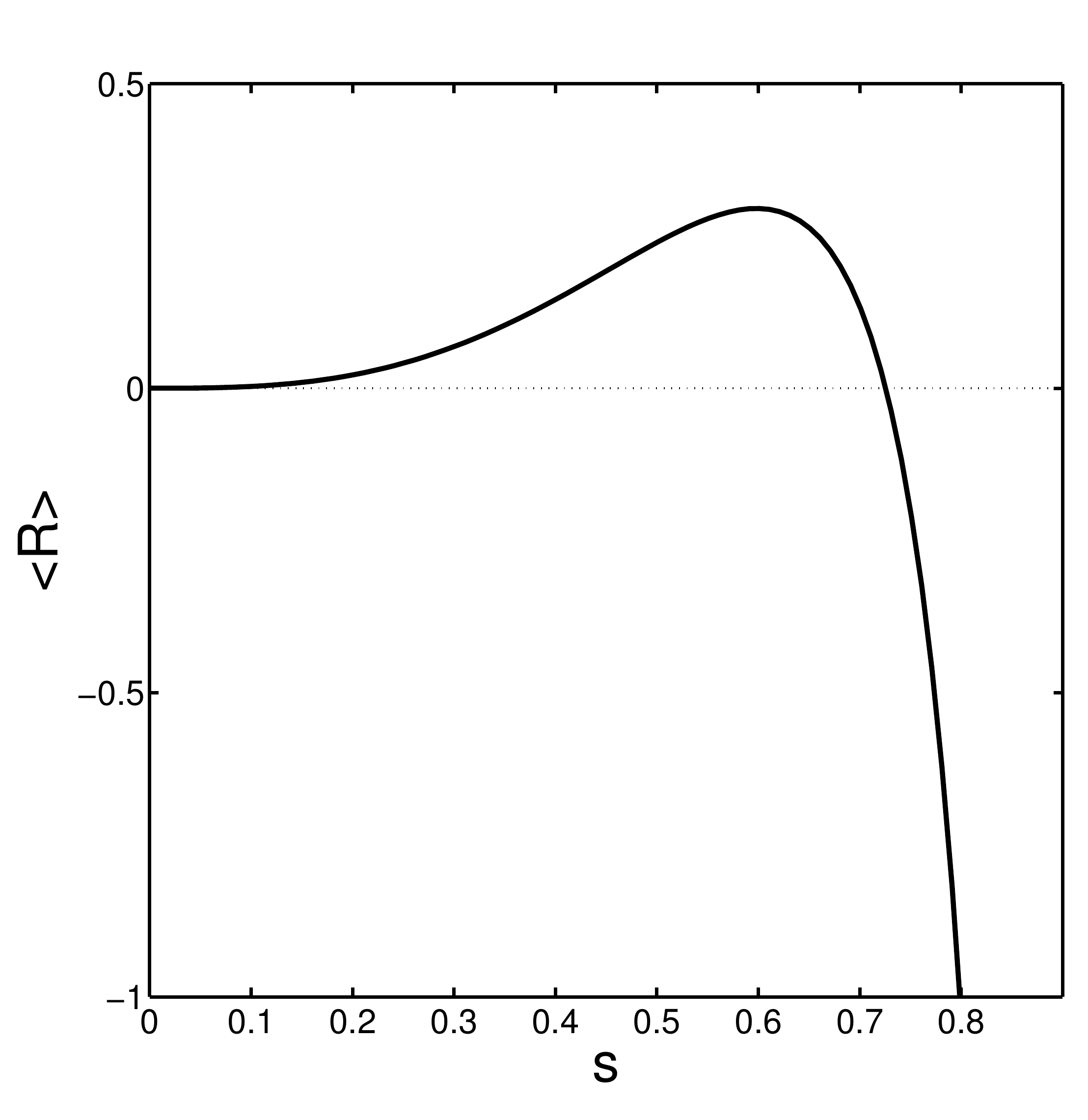}
\caption{Plot of $\langle R\rangle$ as a function of the compactness in the interval $[0, \cal R]$.}
\label{Rmean}
\includegraphics[scale=0.3,trim=280 0 300 0,clip=true]{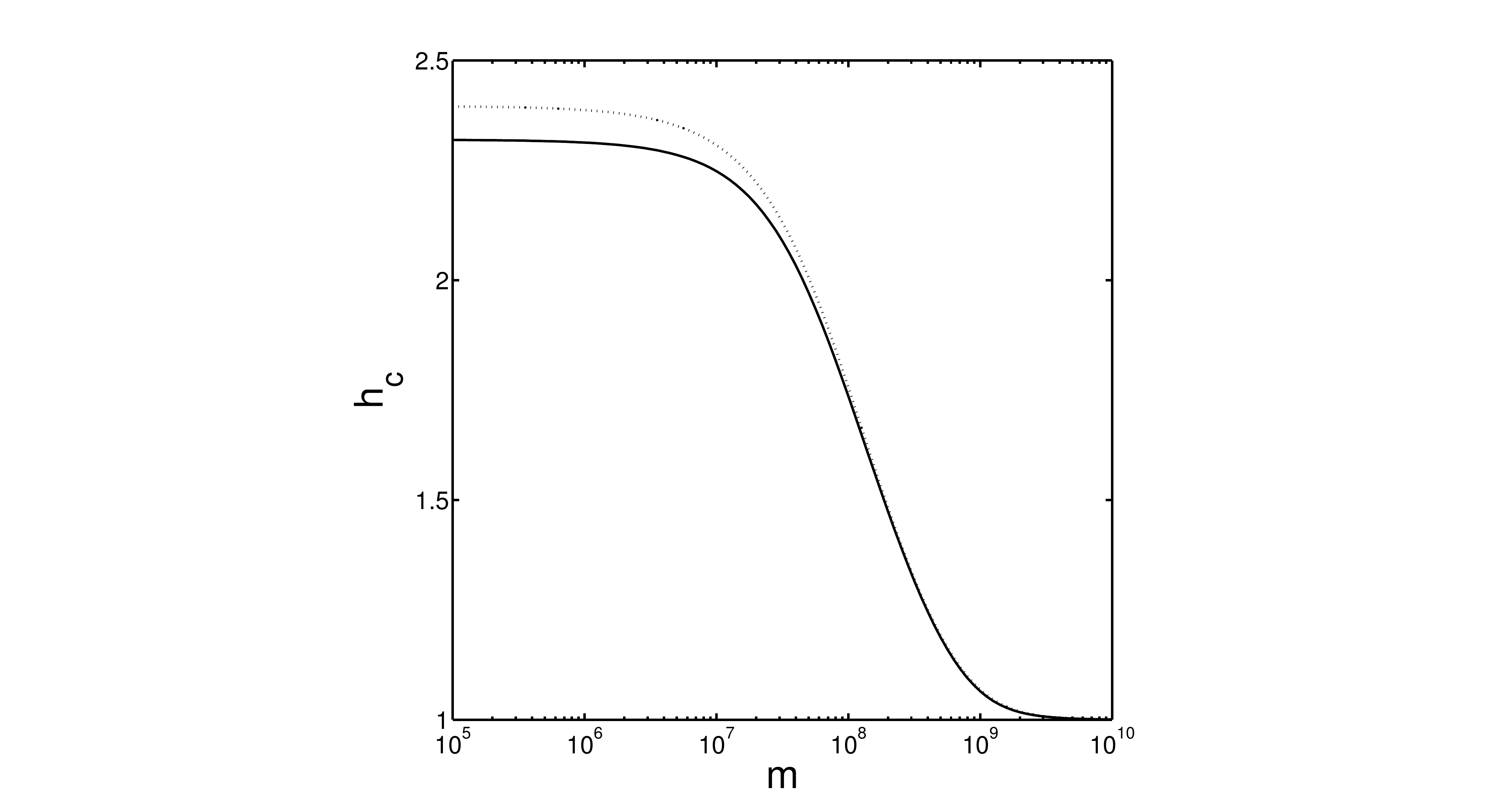}
\caption{Plot of $h_\rr{c}$ as a function of the mass for $s=0.2$ and  $\xi = 60$. The solid line is the result of the numerical analysis, the dotted one is obtained with the analytical approximation described in Sec.\ \ref{sec6}.}
\centering
\label{hc_vs_m2}
\end{center}
\end{figure}

Such behavior can be easily understood by considering the upper bound for $|h_\rr{c}|$ introduced in
section \ref{sec:effective_dyn_monop} and discussed in Sec.~\ref{sec:PPN_monop}. If we work in GeV units, it is expressed as,
\bea
h^{\rm in}_{\rm eq}=0,\pm \sqrt{1+{R \xi \over 8\pi\lambda_{\rm sm} \tilde{v}^2}}=0,\pm \sqrt{1+{R \xi \over 8\pi m_\rr{H}^2}},
\label{h_eq_in}
\eea
where $m_\rr{H}$ is the mass of the Higgs field. Since $R$ depends on the radial coordinate (see Eq.\ \eqref{scalR}) so does the effective potential. In order to show that we may have $|h_\rr{c}|<1$, 
we approximate $R$ in Eq.\ \eqref{h_eq_in} by its spatial average,
\bea
\langle R\rangle=\frac{\int R(u)\sqrt{g}~\dd^3x}{\int\sqrt{g}~\dd^3x}=\frac{\int_0^{1/s} R(u) u~ \rr{e}^{\lambda}\dd u}{\int_0^{1/s} u~ \rr{e}^{\lambda}\dd u}.
\label{eqRmean}
\eea
In Fig.\ \ref{Rmean} we plot the value of  $\langle R\rangle$ in function of the compactness. We see that, for  $s\gtrsim 0.72$, $\langle R\rangle$ becomes negative so that $h^{\rm in}_{\rm eq}<1$, which implies  $|h_\rr{c}|<1$. This happens, for instance, for the monopole represented by the curve B in Fig.\ \ref{plot_monop_family}.
In this plot we also notice that, for large $\xi$, oscillations 
are present only inside the compact body (see monopoles A and C), confirming the analytical results found in Sec.\ \ref{sec:analyt_monop}. 

\begin{figure}
\begin{center}
\includegraphics[scale=0.3,trim=270 0 310 0,clip=true]{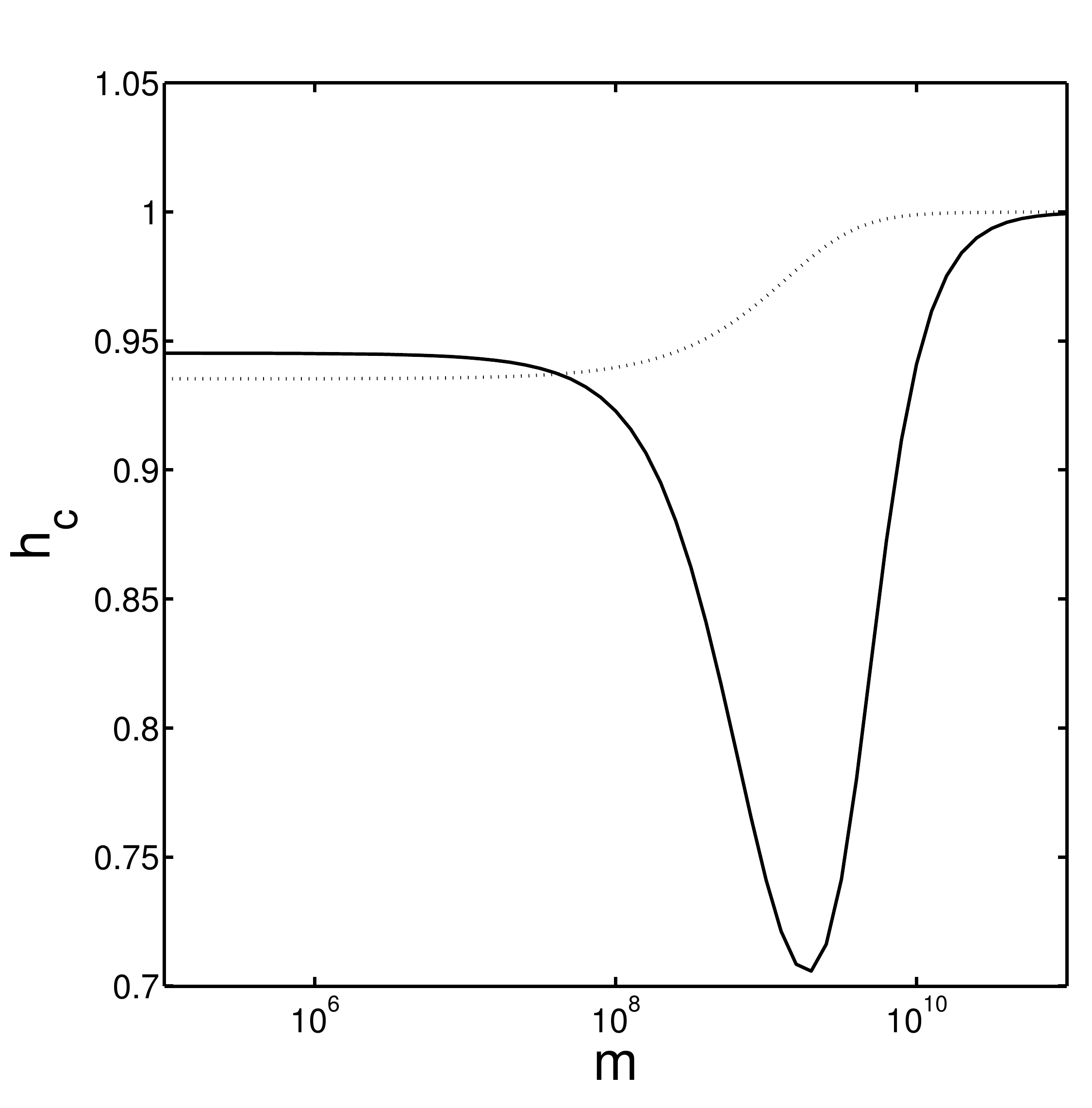}
\caption{Plot of $h_\rr{c}$  in function of the mass 
for  $s=0.73$ and  $\xi = 60$. The solid line is the result of the numerical analysis while the dotted one is obtained with the analytical approximation described in Sec.\ \ref{sec6}. We see that the analytical approximation does not work well with large $s$.}
\label{hc_vs_m1}
\includegraphics[width=0.6\textwidth,trim=410 10 200 10,clip=true]{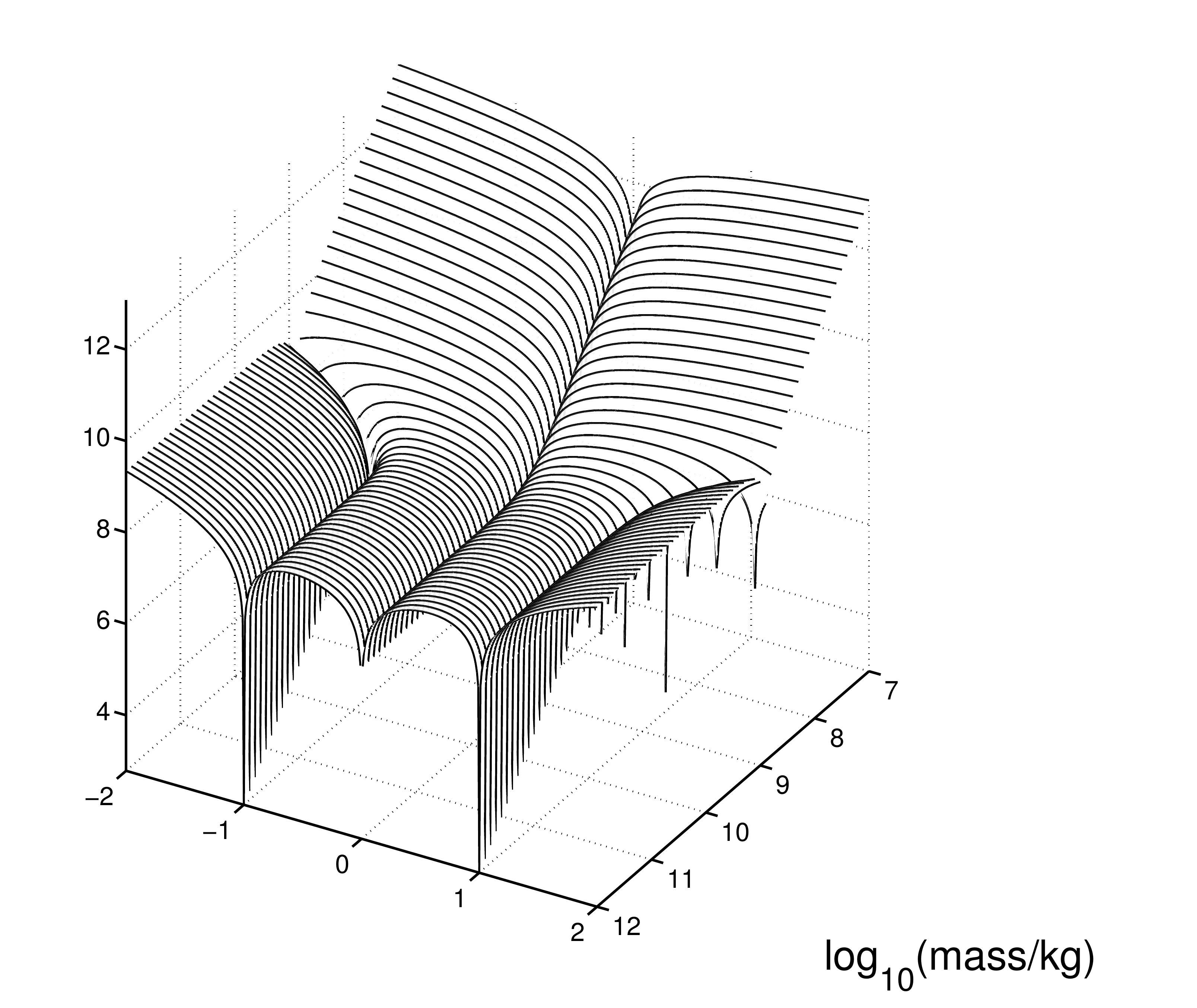}
\caption{Derivative of the effective potential  $V_{\rm eff}$ of  Eq.\ \eqref{eq:Veff_monop} as a function of the mass of the monopole for fixed nonminimal coupling and compactness ($\xi=60$ and $s=0.5$). }
\label{dVeffdh}
\centering
\end{center}
\end{figure}

Finally, we point out that the central value of the Higgs field can be significantly larger than the vev (see e.g.\ the monopole D). We will see below that there is a novel amplification mechanism that explains these large values.
The numerical relation between the mass of the monopole and the value of $h_\rr{c}$ is depicted in  Fig.\ \ref{hc_vs_m2} for a fixed compactness $s=0.2$ and a nonminimal coupling parameter $\xi=60$. The plot shows an interpolation 
between two asymptotic values at small and large masses. For large masses, the value of $h_\rr{c}$ is bounded from above by 
$h^{\rm in}_{\rm eq}\,$ in Eq. (\ref{h_eq_in})
which converges to $h^{\rm in}_{\rm eq}=1$ for $m\approx 10^9 \rm \,\,kg$ (with $s=0.2$
and $\xi=60$). At small masses, the central value $h_\rr{c}$ is independent of the mass because the Higgs potential 
contributes very little to the effective potential inside the matter distribution (see also Fig. \ref{dVeffdh}). 

In Fig.\ \ref{hc_vs_m1} we show that this behavior is present also for large compactness ($s=0.73$), which 
yields  $|h_\rr{c}|<1$, as seen above.
In Fig.\ \ref{dVeffdh} we represent the derivative of the effective potential $V_{\rm eff}$ given in Eq.\ \eqref{eq:Veff_monop} inside the matter 
distribution as a function of the mass of the monopole for fixed $\xi$ and $s$. Local maxima and minima, where $\dd V_{\rm eff}/\dd h=0$, are marked by the peaks appearing on the plot. We see that $h=0$ is always a minimum while there are two maxima at $h^{\rm in}_{\rm eq}$ 
(see Eq.(\ref{h_eq_in})), whose value converges to one for large masses. From the expression of the effective potential \eqref{eq:Veff_monop} (with averaged Ricci scalar),
\bea\label{Veffav}
V_{\rm eff}=-V+\frac{\xi H^2\langle R\rangle}{16\pi},
\eea
and the behavior of $\langle R\rangle$ (see Fig.~\ref{Rmean}) we deduce that the  term $\xi H^{2}\langle R\rangle/(16\pi)$ is dominant  for small masses and becomes negligible compared to the Higgs 
potential for large masses. Thus, for small masses, the field behaves inside the matter distribution as if there was no potential, in a way 
similar to that in spontaneous scalarization \cite{PhysRevLett.70.2220, Salgado:1998sg} where the field inside the body is almost constant. This explains why a plateau appears for small masses in Figs.~\ref{hc_vs_m2} and \ref{hc_vs_m1}. However, outside the body $R\approx0$ and the Higgs potential can no longer be neglected compared to the nonminimal coupling term. As a result, the Higgs field decreases faster than in \cite{PhysRevLett.70.2220, Salgado:1998sg} because of the quartic potential. 
%the Higgs potential can no longer be neglected compared to the nonminimal coupling term. 

\begin{figure}[h]
\begin{center}
\includegraphics[width=0.6\textwidth,trim=300 0 300 0,clip=true]{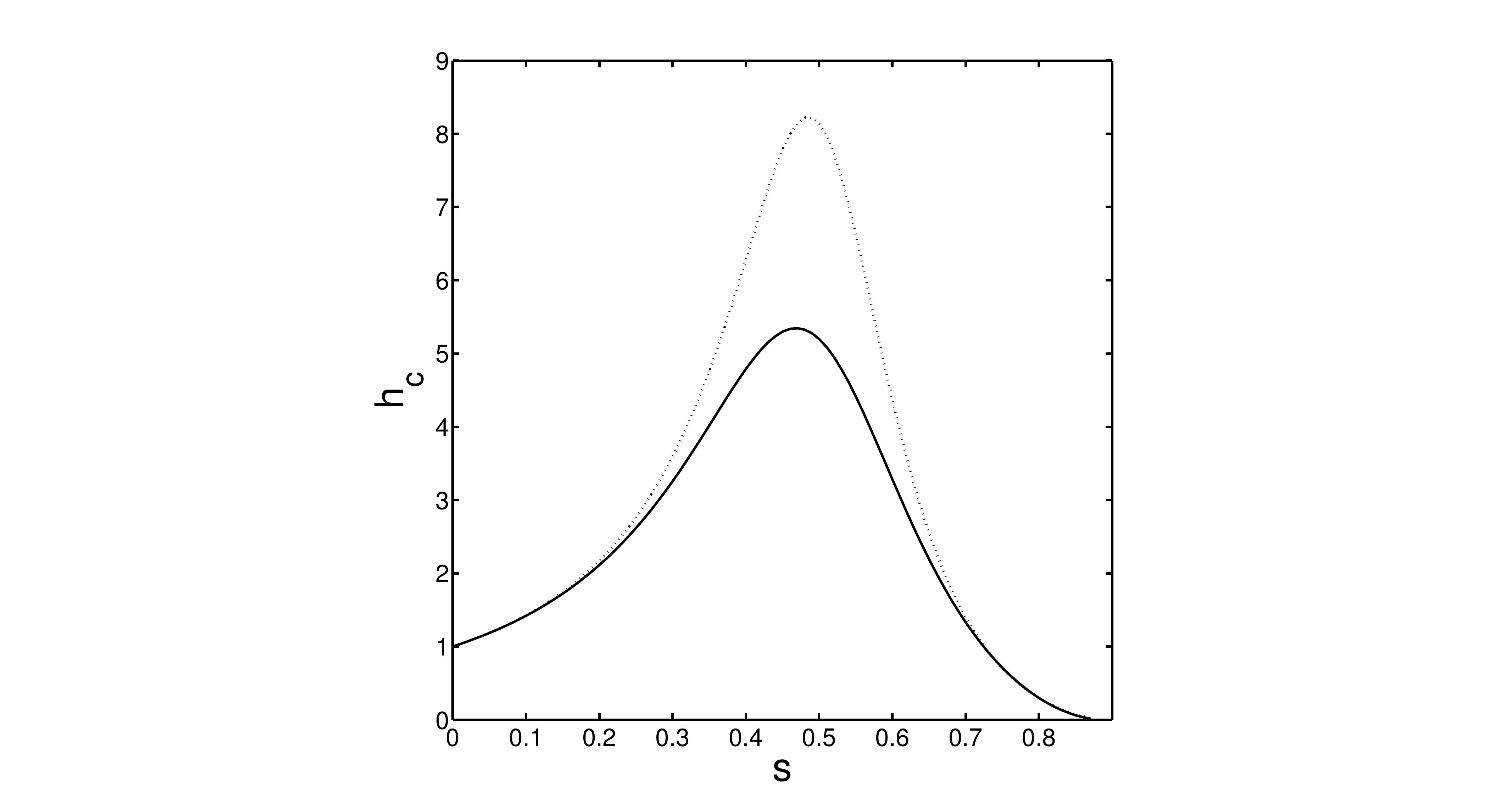}
\caption{
The plot of $h_\rr{c}$ shows a peak at some value of  $s$ and for fixed $\xi$. In the case plotted here (solid line) we choose $m=100\rm \;kg$ and $\xi = 55$. The dotted line is obtained with the analytical approximation described in Sec.\ \ref{sec6}.}
\centering
\label{hc_vs_s}
\end{center}
\end{figure}

\begin{sloppypar}
What fixes the central value $h_\rr{c}$ of the monopoles is a non-linear phenomenon  of classical resonance. In
Fig.\ \ref{hc_vs_s} we show an example, where $h_\rr{c}$ increases around a specific value of the compactness.  For small values of $s$, $h_\rr{c}$ is close to one and the monopole distribution is pretty close to the 
homogeneous GR solution $h(r)=1$. We find that, for astrophysical objects like the Sun, the 
combination of low compactness and large mass  makes the Higgs field extremely close to its vev everywhere, yielding negligible deviations from GR. This is in line with the PPN analysis presented in Sec.\ \ref{sec:PPN_monop}.  On the other hand, we have seen that, for $s>0.7$, $|h_\rr{c}|$  is smaller than one, since $\langle R\rangle$ is negative. Between these two extreme cases, there exists a specific 
value of $s$ that maximizes $h_\rr{c}$. This is a new result due to the combined action of the nonminimal coupling and the field potential. In fact, it is absent if the potential vanishes as in \cite{Salgado:1998sg}. 
\end{sloppypar}

%%%%%%%%%%%%%%%%%%%%%%
\subsection{Amplification mechanism}
\label{sec6}
%%%%%%%%%%%%%%%%%%%%%

\noindent In this section, we present an analytical model of the scalar field resonant amplification found numerically in the previous section. As before, we consider the Klein-Gordon equation \eqref{kg_simp} in dimensionless units and we suppose that the metric fields $\lambda$ and $\nu$ are the ones given by GR. The combination of Eqs.\ \eqref{nu} and \eqref{lam} gives,
\bea
\nu_u-\lambda_u+\frac{2}{u}=
 \left\{\begin{array}{ll} 
	  -{{uR}\over {6\left(1-s^3u^2\right)}}+\frac{2}{u}  & 0<u<1/s,
	  \\{{2u-1}\over{u\left(u-1\right)}}\approx {2\over{u-1}} & u>1/s,\end{array}\right.
\eea
where the top (bottom) line corresponds to the internal (external) solution.

We now expand $\dd V/\dd h$ around  $h=h_*$, where $h_*=1$ outside and $h_*=h_\rr{c}$ inside the body. The function  to be expanded has the form,
\bea
f(h)=\alpha h \left(h^2-1\right),
\eea
where $\alpha=2 \lambda_{\rm sm} r_\rr{s}^2 \mpl^2\tilde v^2$, thus, up to the first order, 
\bea
f(h)\approx \alpha \left[ h_* \left(h_*^2-1\right)+\left(3h_*^2-1\right)\left(h-h_*\right) \right].
\eea
We now examine more carefully the external and internal solutions of the Klein-Gordon equation.

\subsubsection{External solution}
\noindent For the external solution, we can assume that $u\gg 1$, $R\simeq0$ (like for the numerical treatment), $\nu_u-\lambda_u+\frac{2}{u}\simeq \frac{2}{u}$, and $\rr{e}^{2\lambda}\sim1$ and the Higgs field is essentially driven by the quartic potential with extrema at $h\simeq \pm 1$
since the Higgs field settles to its vev at large distance. After the expansion, upon the change of variables $h=Y/u$ and $Z=Y-u$,  Eq.\ \eqref{kg_simp} can be written as,
\bea
Z_{uu}=\alpha Z,
\eea
which has the general solution,
\bea
h_{\rm ext}={\mathcal{C}_1\over u} \rr{e}^{\sqrt{\alpha}u}+{\mathcal{C}_2\over u} \rr{e}^{-\sqrt{\alpha}u}+1,
\label{h_ext}
\eea
for arbitrary constant $\mathcal{C}_1$ and $\mathcal{C}_2$. The requirement  $\lim_{u\rightarrow \infty}h=1$ yields $\mathcal{C}_1=0$, so we find the Yukawa distribution for the Higgs field outside the compact object given by,
\bea
h_{\rm ext}= \frac{Q}{u} \rr{e}^{-u/L}+1.
\eea
\begin{sloppypar}
The parameters $Q=\mathcal{C}_2$ and $L=1/\sqrt{\alpha}$ can be identified as the  scalar charge and the characteristic length respectively, which further justifies the term ``monopole'' used to name these solutions.  To fix $\mathcal{C}_2$, we will use the continuity condition of the Higgs field at the boundary of the compact object given by $h_{\rm ext}(1/s)=h_{\rm int}(1/s)$. In addition,  the continuity condition of the derivative, $h'_{\rm ext}(1/s)=h'_{\rm int}(1/s)$, will lead to an implicit equation for $h_\rr{c}$.
\end{sloppypar}

\subsubsection{Internal solution}

\noindent We now derive the analytical Higgs field profile for the internal region. We make the same assumption as before for the terms involving $\nu$ and $\lambda$, excepted $u\simeq0$ and $R\sim \langle R\rangle \neq0$.
%=\rm{const}
We now expand $f(h)$ around $h_*=h_\rr{c}$ and change the variables according to $h=Y/u$ as well as,
\bea
Z=Y+\frac{B(h_\rr{c})}{A(h_\rr{c})} u,
\eea
where 
\bea
A(h_\rr{c})&=&\frac{\alpha}{2} \left(3h_\rr{c}^2-1\right) - {{ \langle R\rangle\xi}\over{8\pi}}, 
\label{const_A} \\
B(h_\rr{c})&=&-\alpha h_\rr{c}^3.
\label{const_B}
\eea
We then obtain the differential equation,
\bea
Z_{uu}=A\left(h_\rr{c}\right) Z,
\eea
for which it is sufficient to discuss the solution for $A\left(h_\rr{c}\right)<0$, the positive case being basically the same. The case $A\left(h_\rr{c}\right)=0$   is not considered as it corresponds to a fine-tuning of the parameters. The solution reads,
\bea
h_{\rm int}=\frac{\mathcal{D}_1}{u} \rr{e}^{\sqrt{A}u}+\frac{\mathcal{D}_2}{u} \rr{e}^{-\sqrt{A}u}-\frac{B}{A},
\eea
where $\mathcal{D}_1$ and $\mathcal{D}_2$ are constants of integration. The condition of regularity of the Higgs field at the origin, $h_{\rm int}(u=0)=h_\rr{c}$ implies that $\mathcal{D}_1=-\mathcal{D}_2$. In addition, the limit $u\rightarrow0$ enables to fix $\mathcal{D}_1$,
\bea
\mathcal{D}_1=\frac{1}{\sqrt{|A|}} \left(h_\rr{c}+\frac{B}{A} \right),
\label{const_int}
\eea
so, the linearized expression for the Higgs field inside the compact object is given by,
\bea
h_{\rm int}=\frac{\mathcal{D}_1}{u} \sin \left(\sqrt{|A|}u\right)-\frac{B}{A}.
\label{h_int}
\eea
As mentioned above, the conditions of continuity of the Higgs field and its derivative allow to fix  $\mathcal{C}_2$
and to derive an implicit equation for determining $h_\rr{c}$. Indeed, by imposing $h_{\rm ext}\left(1/s\right)=h_{\rm int}\left(1/s\right)$, we find that,
\bea
\mathcal{C}_2=\frac{1}{s} \rr{e}^{\frac{\sqrt{\alpha}}{s}} \left[\mathcal{D}_1 s \sin\left(\frac{\sqrt{|A|}}{s}\right)-\frac{B}{A}-1 \right],
\label{const_ext}
\eea
while the regularity condition $h'_{\rm ext}\left(1/s\right)=h'_{\rm int}\left(1/s\right)$ yields the implicit equation,
\bea\non
\left(h_\rr{c}+{B\over A}\right)\left[ \sqrt{\alpha\over |A|}\sin \left(\sqrt{|A|}\over s\right)+\cos \left(\sqrt{|A|}\over s\right)   \right]\qquad\qquad\qquad\\
\hspace{2.5cm}=\left(1+{B\over A}\right)\left(1+{\sqrt{\alpha}\over s}\right).
\label{simplifed_Ahc}
\eea
The solution for the case $A\left(h_\rr{c}\right)>0$ can be found by replacing the sine and cosine by hyperbolic sine and hyperbolic cosine. 
However, the condition $A\left(h_\rr{c}\right)<0$ is  necessary  for the resonant amplification.
The expression \eqref{simplifed_Ahc} greatly simplifies when $\alpha$ is small as for macroscopic bodies \footnote{As an example, for an object of the mass range of an asteroid ($M\simeq 10^{7}$), $\alpha\simeq 10^
{-25}$.}. In fact,  since $B/A\simeq 0$ when $\alpha$ is negligible, the implicit equation for $h_\rr{c}$ \eqref{simplifed_Ahc} reduces to, 
\bea
h_\rr{c}=\left| \cos\sqrt{{\xi \langle R\rangle}\over {8\pi s^2}}\right|^{-1},
\label{approx_alpha_small}
\eea
where the absolute value is necessary when the positive $h_\rr{c}$ branch is chosen. 
%We see that, 
In this approximation, the central value of the Higgs field $h_\rr{c}$ has periodic divergences corresponding to certain values of $s$, $\xi$ and $m$. %of the mass of the monopole.
E.g. for asteroids, the compactness is very small ($s\sim 10^{-12}$) and one finds that $h_\rr{c}=1$ to great accuracy. Notice that, for small $s$, the condition $A<0$ is no 
longer true and $\cos$ to $\cosh$ must be switched, which yields, however, the same result. We thus confirm the results obtained in the previous section:  for small values of the compactness, the central value of the Higgs field $h_\rr{c}$ is very close to the Higgs vev. 
For larger values of the compactness, the approximate formula \eqref{approx_alpha_small} shows that, for a given $m$ and  $\xi$,  $h_\rr{c}$ has peaks corresponding to critical values of $s$. These are the resonances that we have also seen numerically. The number of peaks depends on the nonminimal coupling $\xi$ as we will see on the next section.
Note that the condition $A\left(h_\rr{c}\right)<0$ is favored by a large nonminimal coupling (see Eq.\ \eqref{const_A}), and so the approximate equation \eqref{approx_alpha_small}, is even more accurate in the large $s$ regime. As we will see in the next section, there exists a critical value of $\xi$ for which one peak splits into two separate peaks. Another interesting limit is $\sqrt{A}/s\ll 1$. In this case, the formula reduces to,
\bea
\left(h_\rr{c}+{B\over A}\right)\left(1+{\sqrt{\alpha}\over s}\right)
\simeq\left(1+{B\over A}\right)\left(1+{\sqrt{\alpha}\over s}\right),
\eea
which implies that $h_\rr{c}\sim 1$. 
The regime $A/s^{2}\sim 0$ corresponds to,  
\bea
r_{s}={16\pi\lambda v^{2} \mathcal{R}^{3}\over 3\xi},
\eea
where $\mathcal{R}$ is the radius of the compact object (assuming $\langle R\rangle\approx 3 s^3/r_\rr{s}^2$). This relation can be written again as $s\xi\simeq(10^{18}\mathcal{R})^{2}$ with $\mathcal{R}$ expressed in meters. 
It is then obvious that this regime is totally unphysical unless $\xi$ is very large \footnote{A very large $\xi$ is not excluded by LHC experiments, see \cite{Atkins:2012yn}.}.

\subsubsection{Analysis of the parameter space}
\noindent 
As we saw in Sec.\ \ref{sec:num_monop},  Figs.\ \ref{hc_vs_m2}, \ref{hc_vs_m1}, and \ref{hc_vs_s}, 
the value of $h_\rr{c}$ as a function of the parameters can be qualitatively reproduced thanks to the analytical model presented in last section. There are some discrepancies (see for instance Fig.\ \ref{hc_vs_m1} for large compactness) but the analytical model expressed by Eq.\ \eqref{simplifed_Ahc} is sufficient to understand the amplification mechanism. For example, in Fig.\   \ref{hc_vs_s} we see a good agreement between our analytical model and the full solution for the position of the resonance, although there is an overestimation of 
its amplitude, up to a factor two.

In the rest of this section, we will use the analytical model to explore the parameter space of the  monopole, given by mass, compactness, and nonminimal coupling. Once these  are fixed,  the 
central value of the Higgs field is uniquely determined by the implicit equation \eqref{simplifed_Ahc}. 

\begin{figure}
\begin{center}
\includegraphics[width=0.6\textwidth, trim=380 0 380 0,clip=true]{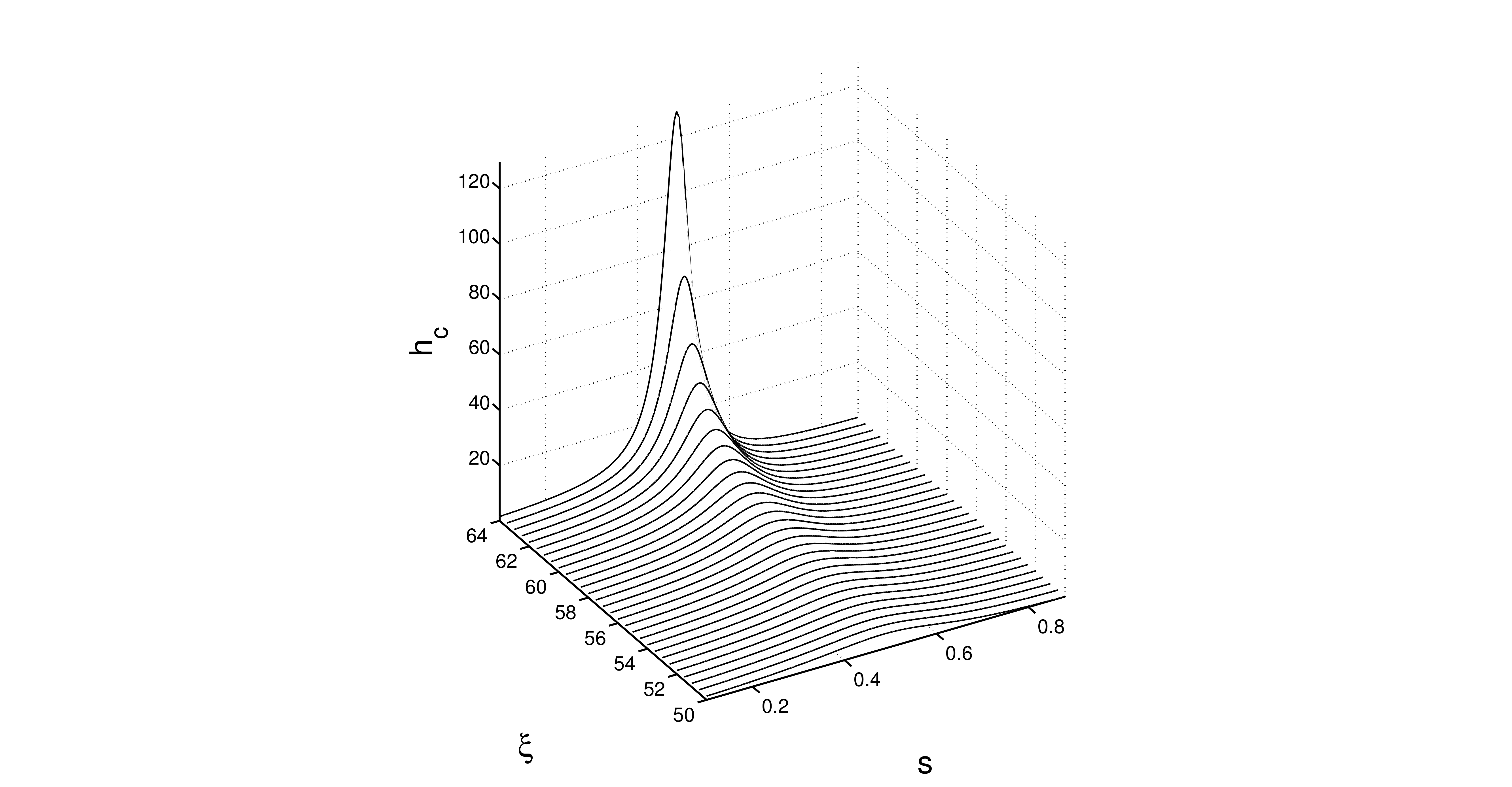}
\caption{Plot of $h_\rr{c}$ in function of $s$ and $\xi$ as given by the implicit relation \eqref{simplifed_Ahc} for a fixed mass $m=10^3\rm kg$.
We see that the peak sharpens for increasing $\xi$.}
\centering
\label{hc_vs_sxi1}
\end{center}
\end{figure}

\begin{figure}
\begin{center}
\includegraphics[width=0.6\textwidth,trim=325 0 380 0,clip=true]{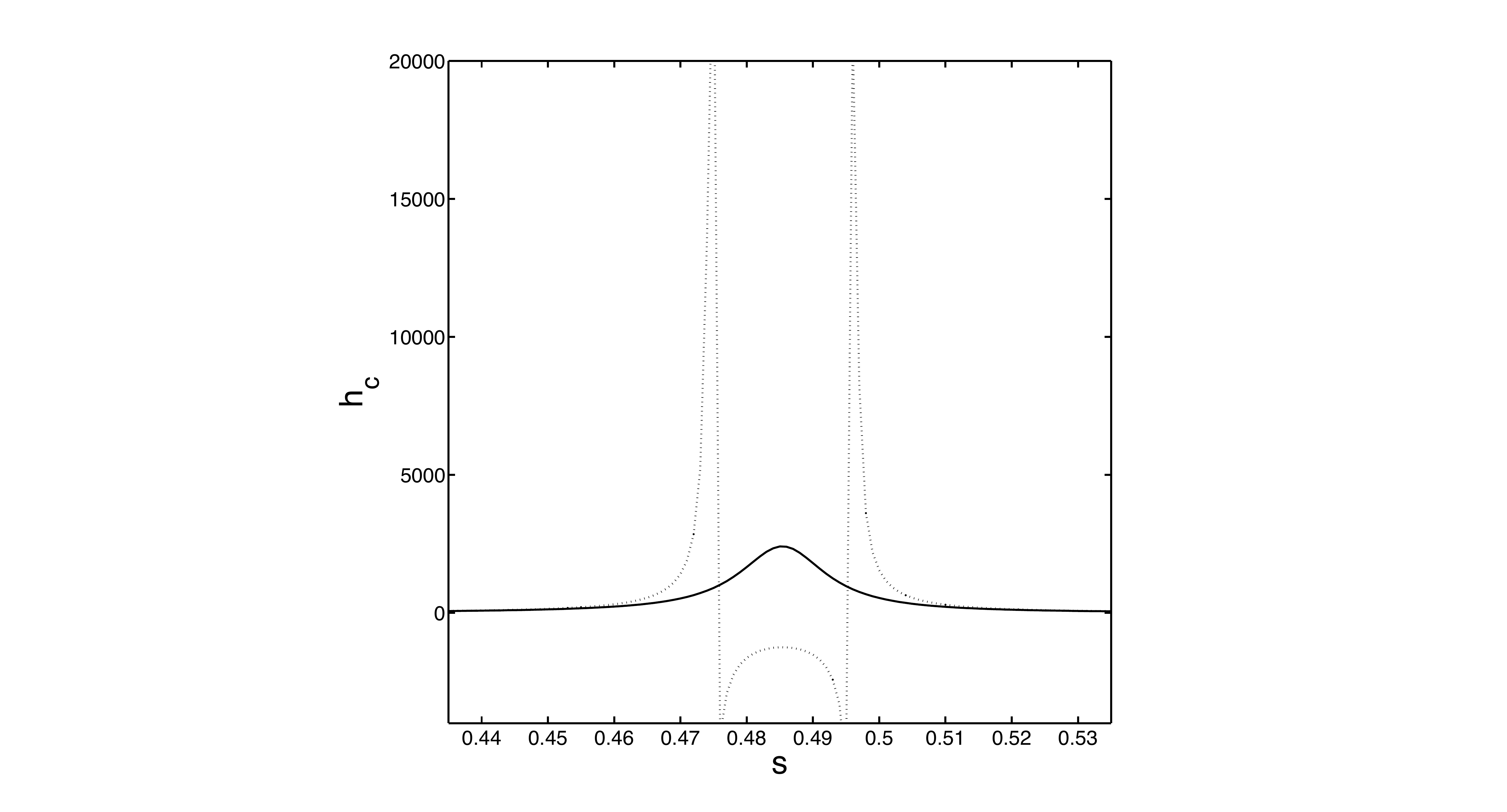}
\caption{Plot of $h_\rr{c}$ given by the implicit Eq.\ \eqref{simplifed_Ahc}
in function of the compactness for $\xi=64.6$ (solid line) and $\xi=64.7$
(dashed line). The monopole mass is fixed at $m=10^3\rm kg$.}
\label{hc_vs_sxi2}
\end{center}
\end{figure}

In Fig.\ \ref{hc_vs_sxi1} we show how the resonance in $h_\rr{c}$ evolves as a function of the compactness and of the nonminimal coupling. 
As $\xi$ increases, the peak grows and sharpens. The question is then how large the resonance can be. It seems that there exists a critical value of $\xi=\xi_{\rm cr}$ above which  $h_\rr{c}$ diverges. This is illustrated in Fig.\ \ref{hc_vs_sxi2} where we plotted $h_\rr{c}$ 
for  both $\xi<\xi_{\rm cr}$ and $\xi=\xi_{\rm cr}$. The two vertical asymptotes in $h_\rr{c}$ appear when the nonminimal coupling becomes larger than $\xi=\xi_{\rm cr}$ and they correspond to a phase transition, in which $h_\rr{c}$ switches sign. We recall in fact that there are two branches corresponding to $v=\pm 246$ GeV. Even though we chose $v$ to be the positive root, there is still the possibility that $h(r)$ jumps to the negative branch, which is a perfectly valid mathematical solution of the Klein-Gordon equation \footnote{This problem could be avoided by considering a Higgs multiplet with an Abelian $U(1)$ symmetry. The amplitude and the phase of the Higgs field would be under a much better analytical and numerical control.}.

\begin{figure}
\begin{center}
\includegraphics[width=0.6\textwidth,trim=350 0 380 0,clip=true]{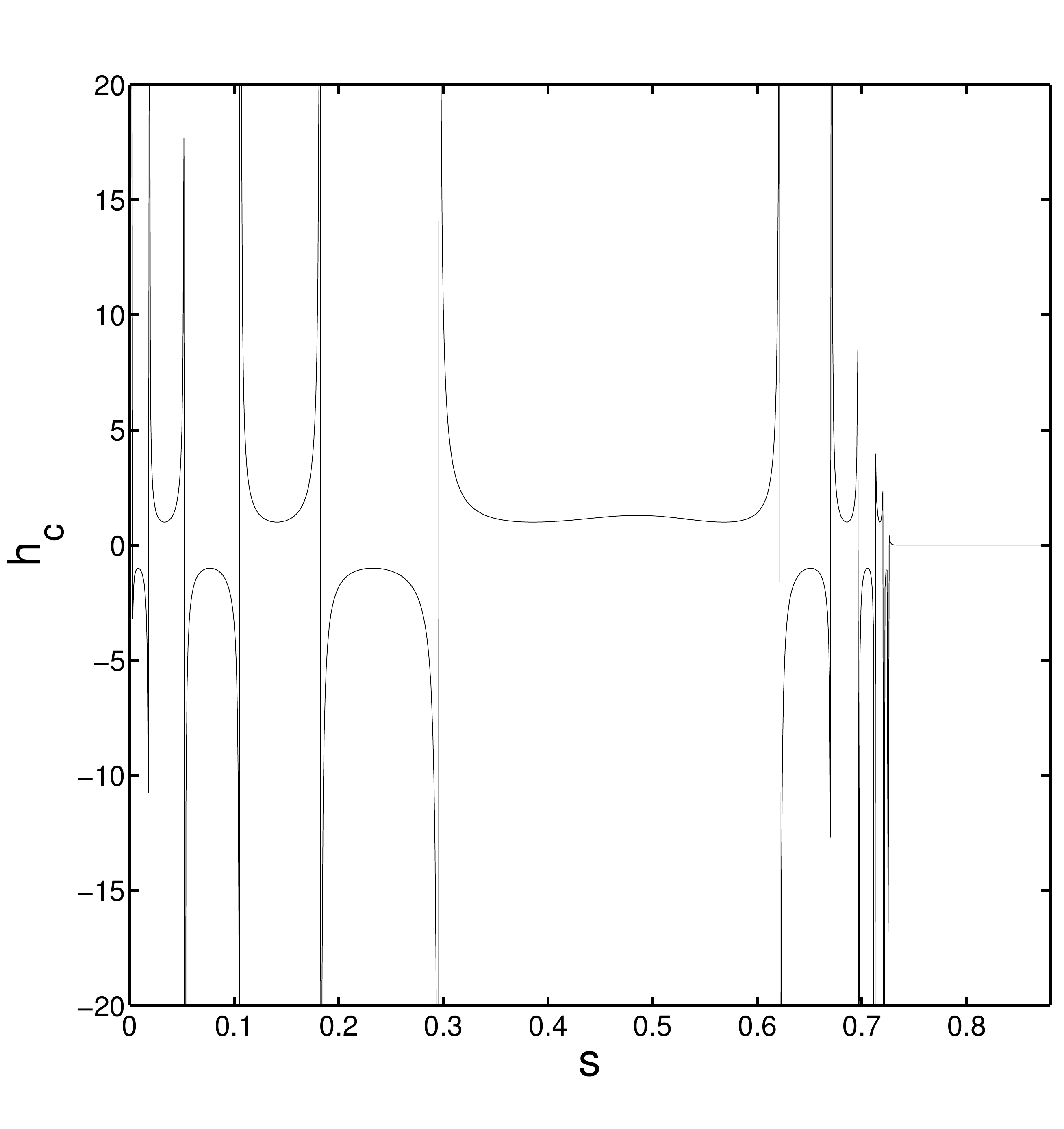}
\caption{Plot of $h_\rr{c}$ in function of $s$ for $\xi=10^{4}$ and $m=10^2\rm kg$ obtained from the expression \eqref{simplifed_Ahc}.}
\centering
\label{hc_vs_sxi1e4}
\end{center}
\end{figure}

\begin{figure*}
 \begin{minipage}{.5\linewidth}
  \centering
  \subfloat[]{\label{hc_ana_ifo_xi:a} \includegraphics[width=0.7\textwidth,trim=415 0 240 0,clip=true]{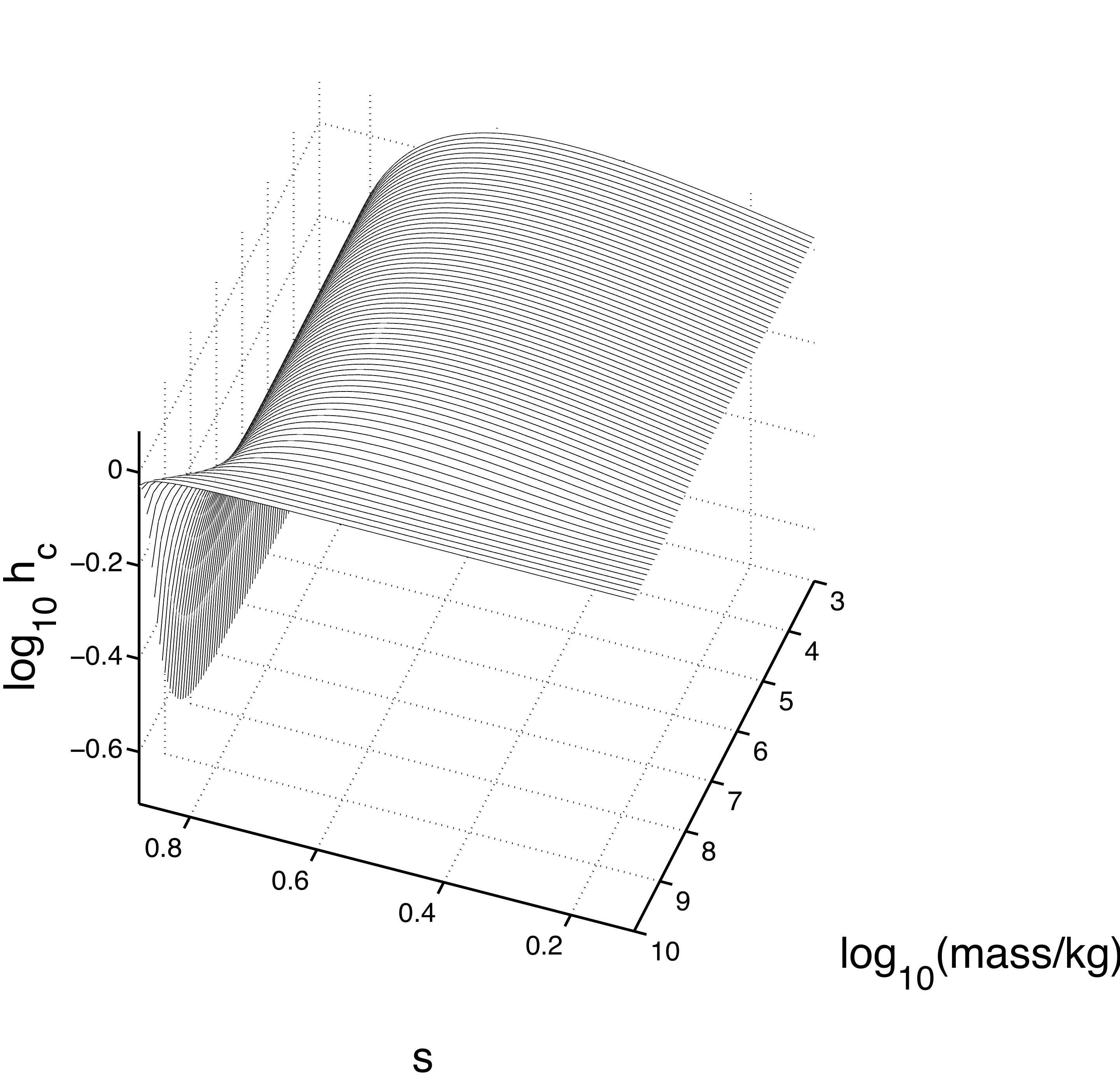}}
 \end{minipage}
 \begin{minipage}{.5\linewidth}
  \centering
  \subfloat[]{\label{hc_ana_ifo_xi:b} \includegraphics[width=0.7\textwidth,trim=415 0 240 0,clip=true]{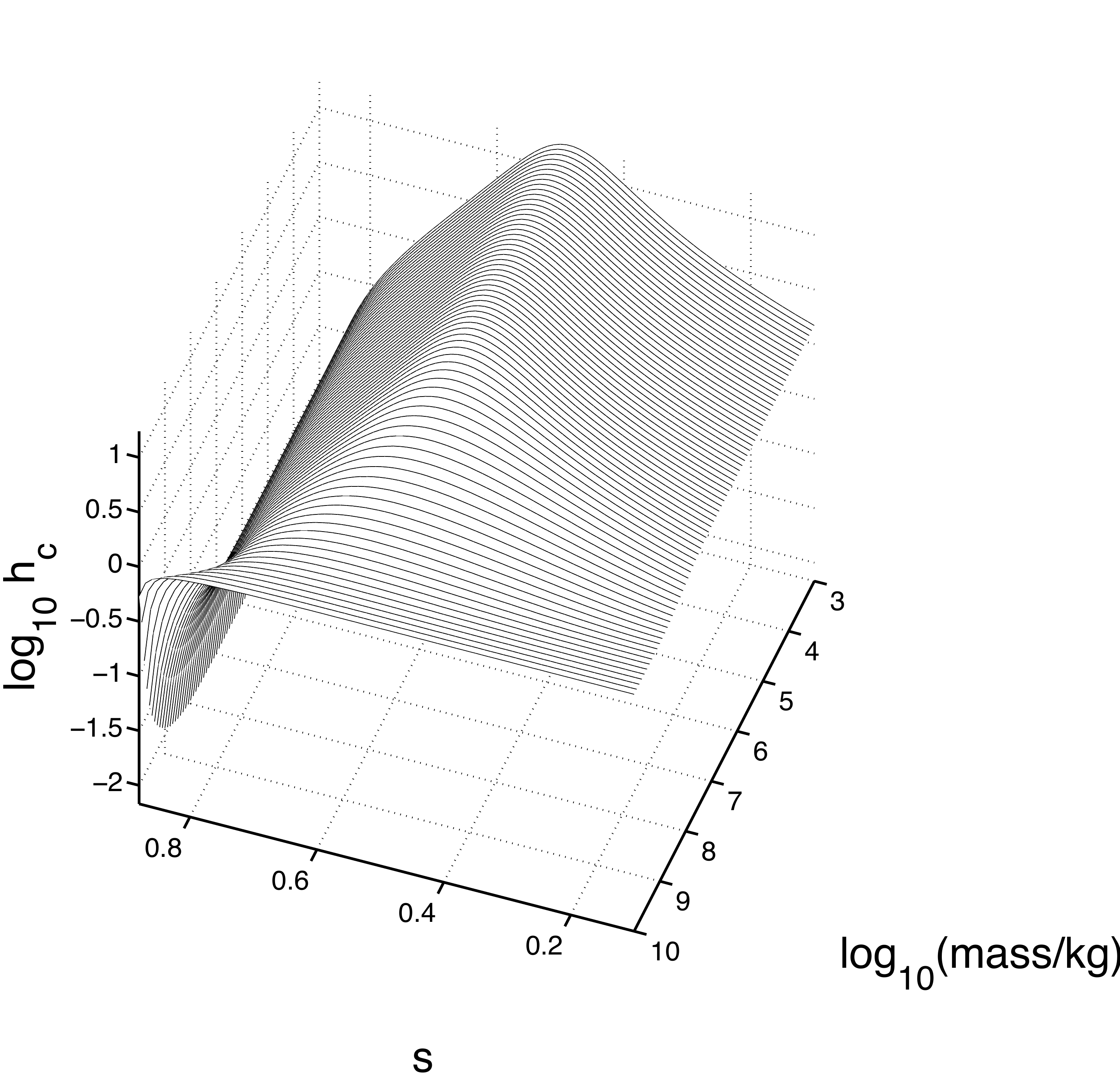}}
 \end{minipage}\par\medskip
 \centering
 \subfloat[]{\label{hc_ana_ifo_xi:c}\includegraphics[width=0.35\textwidth,trim=415 0 240 0,clip=true]{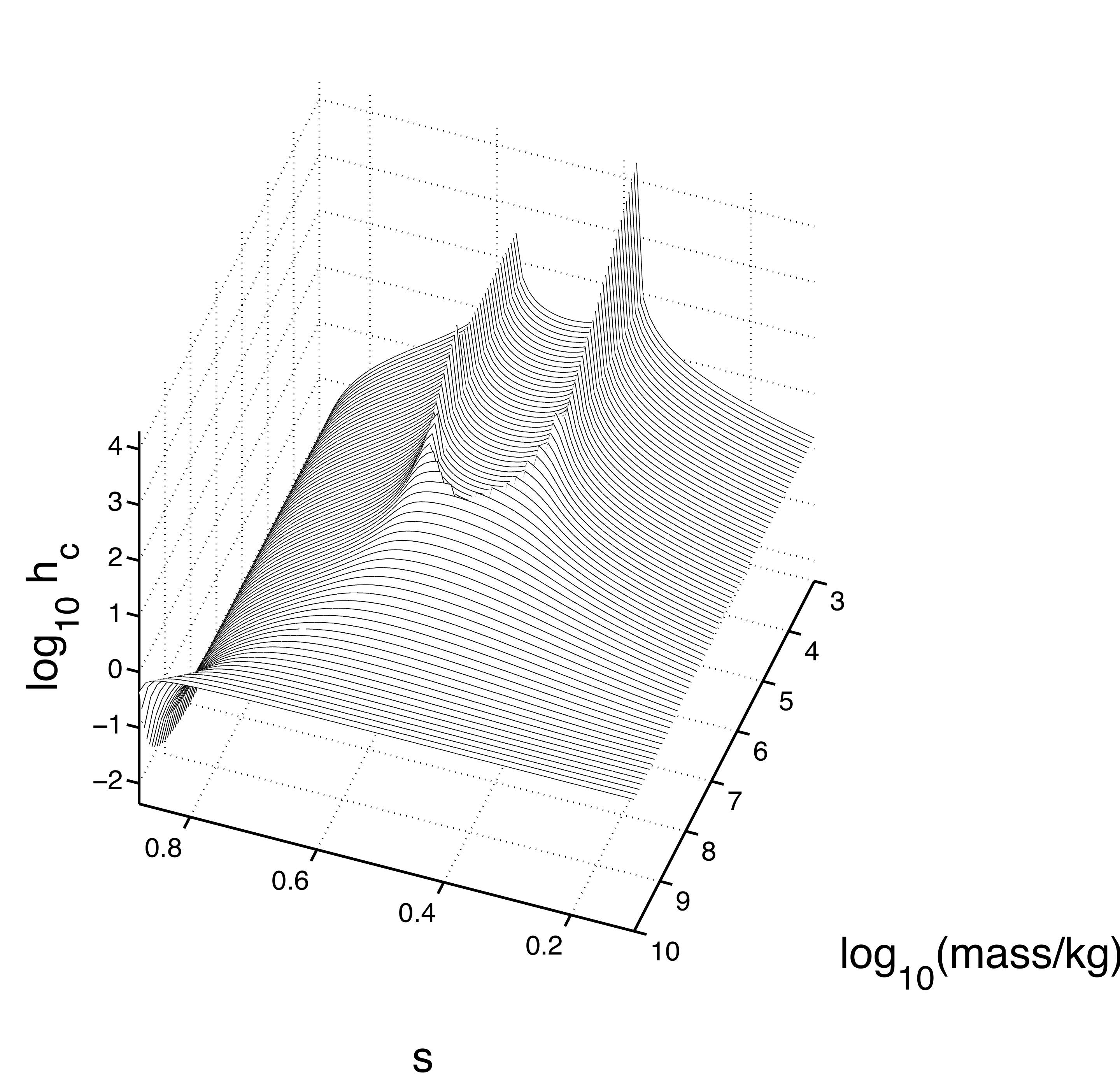}}
 \caption{Plot of $h_\rr{c}$ (obtained with the analytical approximation) in function of the mass of the monopole and its compactness for $\xi=10, 60, 70$ (for Figs.~\ref{hc_ana_ifo_xi:a}, \ref{hc_ana_ifo_xi:b}, \ref{hc_ana_ifo_xi:c} respectively).}                    
\label{hc_ana_ifo_xi}  
\end{figure*}

% 
% \begin{figure*}[ht]
%   \begin{center}
%     \begin{tabular}{ccc}
%       \includegraphics[scale=0.2,trim=415 0 240 0,clip=true]{./chapters/chapter5/fig121_hc_vs_s_m_xi10-eps-converted-to.pdf} &
%       \includegraphics[scale=0.2,trim=415 0 240 0,clip=true]{./chapters/chapter5/fig122_hc_vs_s_m_xi60-eps-converted-to.pdf} &
%       \includegraphics[scale=0.2,trim=415 0 240 0,clip=true]{./chapters/chapter5/fig123_hc_vs_s_m_xi70-eps-converted-to.pdf} 
%     \end{tabular}
%   \end{center}      
% \caption{Plot of $h_\rr{c}$ (obtained with the analytical approximation) in function of the mass of the monopole and its compactness for $\xi=10, 60, 70$ (from left to right).}                    
% \label{hc_ana_ifo_xi}  
% \end{figure*}   

This also implies that, when the nonminimal coupling is larger than  $\xi_{\rm cr}$, there can be forbidden values for $s$ (or, equivalently, for $\mathcal{R}$) in the parameter space. As an example, we plot in Fig.\ \ref{hc_vs_sxi1e4} $h_\rr{c}$ in function of the compactness for $m=10^2$ kg and $\xi=10^4$, which corresponds to the value predicted by Higgs inflation \cite{Bezrukov:2007ep, Bezrukov:2010jz, Bezrukov:2008ej}. We see that there are multiple divergences, also for relatively small values of $s$.
However, this does not prevent the nonminimal coupling parameter to be arbitrarily large since   $\xi_{\rm cr}$ basically depends on the mass of the monopole.

%%%%%%%%%%%%%

On Fig.\ \ref{hc_ana_ifo_xi}, $h_\rr{c}$ is plotted in function of $m$ and $s$ for three values of $\xi$. We see that $h_\rr{c}$ generically settles to its vev,  $h_\rr{c}=1$ for small compactness and large mass. For sufficiently large $\xi$, the peaks appear for small masses (see Fig.\ \ref{hc_ana_ifo_xi:c}) and tend to $h_\rr{c}=1$ at large $m$ values. The peaks sharpen as $\xi$  increases, until $h_\rr{c}$ eventually diverge at some $\xi_{\rm cr}$. In the small mass regime the Higgs potential is much smaller than the coupling term, which is proportional to $\xi$, see Eq.\ \eqref{Veffav}. In the large mass regime, however, the upper bound imposed by Eq.\ \eqref{h_eq_in} becomes closer and closer to one. These two competing effects explain qualitatively the presence of the peaks in the small mass region rather than in the large mass one, provided the compactness $s$ is not too small, otherwise $\langle R \rangle $ is too small and always smaller than the Higgs potential. In such a case, scalar amplification is negligible, no matter the monopole mass. 

In summary, for large values  of the nonminimal coupling, monopoles with small masses cannot exist for certain values of the compactness for which the Higgs field at the center of the body diverges. On the opposite, large mass monopoles always exist but the scalar amplification is much smaller.

\section{Conclusion and perspectives}\label{conclusions}
\begin{sloppypar}
\noindent 
%In this work we have studied in detail the static and spherically symmetric solutions, originally found in \cite{monopole}, of the equations of motion of a theory where the Higgs field in unitary gauge is nonminimally coupled to gravity. 
The Higgs monopole is inspired by Higgs inflation, where the coupling strength $\xi\gtrsim 10^4$ in order to get a viable inflationary model. Such a large coupling could imply a strong deviation from GR, possibly in conflict with astrophysical observations. We show that the deviations are so small ($h_\rr{c}-1<10^{-58}$ for the Sun and $10^{-41}$ for typical neutron stars parameters) that there are not far below the observational sensitivity today. This is due to the hierarchy between the vev and the Planck scale, which remains unexplained. 

However, the main result of \cite{Fuzfa:2013yba, Schlogel:2014jea} is the existence of particlelike solutions that are asymptotically flat and have finite classical energy and that cannot smoothly reduce to GR as they only exist because of the violation of the equivalence principle. The Higgs inflation predicts deviations from the Higgs vev inside any compact objects, the Schwarzschild solution being never recovered. Nevertheless, since the deviations from the Higgs vev depends on the compactness and the baryonic mass of the compact object, the Higgs monopoles solution is indistinguishable from the Schwarzschild one in the case of physical objects. Whereas other particlelike solutions  exist only in the context of exact and unbroken gauge symmetry, as in the Einstein-Yang-Mills system \cite{Bartnik}, the Higgs monopole solution is compatible with spontaneously broken gauge symmetry at the price of a nonminimal coupling to gravity.

In particular, we have found a new non-linear mechanism of resonant amplification that is not present in the models with vanishing potentials studied so far since spontaneous scalarization has been studied for relatively small nonminimal coupling (see e.g. \cite{Salgado:1998sg}). We explored this amplification mechanism numerically and found an analytical approximation that shows that, in the large coupling regime, there are forbidden combinations of radius and baryonic density, at which the value of the Higgs field at the center of the spherical body tends to diverge. As for spontaneous scalarization, the amplification mechanism found here is a general feature that can be applied to cases with different parameters and/or potential shapes.

In principle, the shift in the Higgs expectation value inside the compact object leads to a change in the mass of the $W$ and $Z$ bosons that, in turn, has an impact on the mass of decay products, decay rates and so on. In the case at hand, however, we have seen that the shift in objects like neutron stars is negligible. Larger effects are possible only in ranges of mass and density that are very unphysical, as shown in Tab.~\ref{table}. Therefore, in realistic compact objects we do not expect any observable modification. Similarly, in the case of a Yukawa coupling to fermions we do not expect any dramatic effect for the same reasons. Technically, the addition of a Yukawa coupling to fermions would introduce in the Klein-Gordon equation new terms (one for each fermion) proportional to $H$, which will compete in the dynamics with the non-minimal coupling term $\xi R$, see Eq.\ \eqref{eq:KG_monop}. However, for realistic objects, this contribution will be much larger than the gravitational one so we do not expect significant deviations from the case with ordinary EoS. We point out that this is a very different situation as in previous models, like \cite{Salgado:1998sg}, where the amount of spontaneous scalarization was much larger. In our case, the presence of Higgs potential in the action effectively prevents fundamental interactions to change inside compact objects. 

From the quantum field theory point of view, a $r$-dependent vacuum leads to a non-local effective action whose effect are very small in the regime of small curvature considered here but might be important in the primordial Universe or in strong field configurations \cite{Gorbar:2003yp}. 

We remark that the top-hat profile \eqref{eq:rho_monop} is a simplifying hypothesis that saves some computational effort. Introducing a different profile, and a specific equation of state, would be more realistic but no substantial changes are expected in our results. This claim is supported by some previous work (e.g. \cite{Salgado:1998sg,Damour:1996ke}) and by some numerical tests that we performed by smoothing out the step function.

About the stability of our solutions, we point out that any change of the value of the Higgs field at the center of the body leads to a change in the geometry of the spacetime at infinity: while particlelike solutions are asymptotically flat,  any other solution is asymptotically de Sitter, as discussed in Sec.~\ref{sec:effective_dyn_monop}. In a realistic scenario of a spherical collapse in an asymptotically flat spacetime we expect that the Higgs monopoles are the only solutions and are stable. A formal proof of this statement would require the study of perturbations around the (numerical) monopole solution (see e.g. \cite{Volkov:1998cc}) and goes beyond the scope of this thesis, although it is a very interesting question.

In Higgs inflation, these monopoles could form and if they are not washed out by the exponential expansion, they could constitute a candidate for DM, with a mass range similar to the one of primordial BHs below the evaporation limit  \footnote{Roughly, for  $h(r=\mathcal{R})\approx 1.1v$ we have $m<10^{11}$ kg.}. However, as they also interact through their Higgs external field, the phenomenology is expected to be distinct from the one of BHs. 
We also point out that there exists an intriguing possibility that the formation of these monopoles is related to the semiclassical instability found in \cite{Lima:2010na} and discussed in terms of spontaneous scalarization in\footnote{Note, however, that the stability analysis presented in \cite{Pani:2010vc} cannot be applied to our model because the GR solution does not coexist with the monopole.} \cite{Pani:2010vc}. Although for astrophysical bodies we do not expect that this instability plays a significant role, as the scalarization is negligible, it could be crucial for the formation of inflationary remnants.

There are several aspects that deserve further analysis. For instance, we assumed that the characteristic parameters are the ones of the SM (in particular the coupling $\lambda_{\rm sm}$ and the vev $v$, see Eq.~\eqref{mexican}). As a result, the deviations from GR are negligible. It would be interesting to find to what extent these parameters can vary without violations of the current observational constraints.  

Moreover, we believe that also the symmetry structure of the Higgs field and its influence on the solutions should be studied, relaxing the assumption of the unitary gauge which appears to be restrictive. Indeed, the Higgs field should be treated as a complex multiplet with $SU(2)$ gauge symmetry rather than as a real scalar singlet. Imposing the unitary gauge is possible only if the expansion of the Lagrangian around a classical, time-independent vacuum state. In the case of a nonminimal coupling to gravity, the classical vacuum state is time-dependent since $H=v$ is not a solution anymore. In the case of a complex multiplet, the Higgs multiplet drives rather a multifield inflation due to the presence of Goldstone components \cite{Greenwood:2012aj}, such models being in agreement with current observations from Planck \cite{Kaiser:2013sna}. Moreover, Goldstone bosons might also play a role at low energy where they lead to an acceleration of the expansion rate due to the displacement of the Higgs field from its vev (either in the Abelian and non-Abelian cases) \cite{Rinaldi:2013lsa}, even in the absence of the nonminimal coupling \cite{Rinaldi:2014yta} and if the effect of the coupling of the Higgs field to gauge bosons is taken into account \cite{Rinaldi:2015iza}. Eventually, the Higgs vacuum state has been found to be metastable that is temporarily stable on cosmological time scales, assuming that the SM is valid up to the Planck scale, this result strongly depending on the measure of the top quark mass and the Higgs mass \cite{Degrassi:2012ry, PhysRevLett.115.201802}. The implications of the metastability of the electrovacuum state should be also investigated for compact objects and when the Higgs field is nonminimally coupled to gravity.

Static spherically-symmetric solution for non-Abelian Higgs field have also been studied by \cite{Brihaye:2014vba} for self-gravitating system assuming a nonminimal coupling of the Higgs to gravity. They show that the monopole and the sphaleron solutions  \cite{Volkov:1998cc, vanderBij:2000cu}, that is classical and non-perturbative solutions of Einstein-Yang-Mills-Higgs theory, remain in the presence of the nonminimal coupling. 

The effect of the Yukawa coupling has not been studied yet in this context. It would imply either to build an effective action for matter fields where the Yukawa coupling to fermions is explicit (and possibly the QCD contribution for the energy density too \cite{Shifman1978443}) either to write a field theory where the gauge invariance is explicit and the coupling of the Higgs to fermions is introduced through spinors fields \cite{Boehmer:2007dh}. 

In relation to this, we also recall that there exist exact solutions for Abelian and non-Abelian configuration in Minkowski space called Q-balls \cite{PhysRevD.39.1665, 1985NuPhB.262..263C, 1986NuPhB.269Q.744.}. In the baryonic massless limit, but with the gauge symmetry restored, Higgs monopoles could be generalized to describe gauged Q-balls in curved space. This direction remains to be explored as it might lead to discover solutions with physical properties that are compatible with DM. If not, it would nevertheless be interesting to see if these solutions are excluded by precise Solar System tests.

%Finally the effect of the coupling of the Higgs field to matter through the Yukawa coupling has not been studied yet.
\end{sloppypar}

\renewcommand{\chaptermark}[1]{\markboth{\small\textsc{Chapter \thechapter.\ #1}}{}}
\cleardoublepage

\chapter[When John and George play inflation and gravitation]{Fab Four: When John and George play inflation and gravitation} % Main chapter title

\label{chap:FabFour} % For referencing the chapter elsewhere, use \ref{Chapter1} 

\lhead{Chapter 6. \emph{Awesome chapter 6}} % This is for the header on each page - perhaps a shortened title

%----------------------------------------------------------------------------------------
%% Version electronique
\begin{center}
\textit{based on}\\
\end{center}
\begin{center}
J.-P.~Bruneton, M.~Rinaldi, A.~Kanfon, A.~Hees,  S.~Schlögel, A.~Füzfa, 
\\\textit{Fab Four:\\When John and George play gravitation and cosmology}\\ 
\href{http://www.hindawi.com/journals/aa/2012/430694/}{Advances in Astronomy, Volume 2012 (2012) 430694}, \href{http://arxiv.org/abs/1203.4446}{\texttt{arXiv:1203:4446}}\\
\end{center}

% \begin{center}
% \textit{based on}\\
% \end{center}
% \begin{center}
% J.-P.~Bruneton, M.~Rinaldi, A.~Kanfon, A.~Hees,  S.~Schlögel, A.~Füzfa, 
% \\\textit{Fab Four:\\When John and George play gravitation and cosmology}\\ 
% \textit{Advances in Astronomy, Volume 2012 (2012) 430694}
% \end{center}

\vspace{1cm}

\noindent
In the last two chapters, two STT "à la Brans-Dicke" in the presence of a potential have been studied. We now turn to a more sophisticated model, that is a subclass of the generalized Galileon model dubbed the "Fab Four" in reference to the Beatles, introduced in Sec.~\ref{sec:horndeski}. In particular, we focus on the "John" Lagrangian which exhibits a nonminimal derivative coupling between the scalar field and the Einstein tensor. This model is referred to as "purely kinetic gravity" since no potential is invoked in order to predict inflation and/or dark energy.

First the phenomenology of inflation predicted by this model is analyzed in terms of number of e-folds as well as the no-ghost and causality conditions. Since a kinetically driven inflationary phase requires highly transplanckian values for the initial field velocity, which basically rule out the model, the considerations are extended to a more general model including a coupling of the scalar field to the Ricci scalar, or "George" in the Fab Four terminology. We then study the John plus George model, establishing how far inflation is viable (for background cosmology), provided that the no-ghost and causality conditions are satisfied. Finally, the deviations from GR around compact objects predicted by George and John are studied and the Solar System constraints are derived.

\section{The Fab Four model}
The Fab Four is a subclass of Horndeski gravity \cite{horndeski1974} (see also Sec.~\ref{sec:horndeski}) justified by cosmological considerations, assuming FLRW background. More precisely the Fab Four model contains the four Lagrangians able to alleviate the cosmological constant problem, assuming that the WEP is not violated. In the Fab Four scenario, even if the vacuum energy density $\rho_\rr{vac}$ is large at all time during the Universe history, the vacuum energy is "screened" by the scalar field such that the cosmic expansion is not accelerated \cite{Copeland:2012qf}. As a result, the vacuum energy does not affect significantly the evolution of the scale factor and the inflation/radiation/matter dominated evolution could be recovered for some combination of the Fab Four Lagrangians. The cosmological constant problem might be solved since the vacuum energy $\rho_\rr{vac}$ is allowed to have a much larger value than the cosmological constant one $\rho_\Lambda$.  This solution evades the Weinberg no-go theorem (see Sec.~\ref{sec:DE}) by breaking the Poincaré invariance in the scalar sector \cite{Charmousis:2011bf}. 

The resulting theory reads\footnote{The Fab Four Lagrangians also appear in the Kaluza-Klein reduction of Lovelock gravity\cite{VanAcoleyen:2011mj}.},
\begin{eqnarray}
\label{eq:john}
{\cal L}_\rr{john} &=& V_\rr{john}(\phi)G^{\mu\nu} \nabla_\mu\phi \nabla_\nu \phi, \\
\label{eq:paul}
{\cal L}_\rr{paul} &=&V_\rr{paul}(\phi) P^{\mu\nu\alpha \beta} \nabla_\mu \phi \nabla_\alpha \phi \nabla_\nu \nabla_\beta \phi, \\
\label{eq:george}
{\cal L}_\rr{george} &=&V_\rr{george}(\phi) R,\\
\label{eq:ringo}
{\cal L}_\rr{ringo} &=& V_\rr{ringo}(\phi) \cal{G},
\end{eqnarray}
where $V(\phi)$'s are arbitrary potential functions, $\varepsilon_{\mu\nu\alpha \beta}$ is the Levi-Civita tensor and 
$P^{\mu\nu\alpha \beta} =-\frac{1}{4}\varepsilon^{\mu\nu \lambda \sigma } \;R_{\lambda \sigma \gamma \delta  } \; \varepsilon^{\alpha\beta \gamma \delta}$
is the double dual of the Riemann tensor. GR is recovered considering George only with $V_\rr{george}=$const and the Brans-Dicke model is recovered when $V_\rr{george}\neq$const with the parameter $\omega(\phi)=0$.

The covariant Galileons might pass local tests of gravity thanks to the non-linearities appearing in the kinetic term of the scalar field. This theory thus relies on the Vainshtein screening mechanism (see Sec.~\ref{sec:screening} for a definition) in order to be possibly allowed to reproduce inflation and/or the late-time cosmic acceleration while passing the local tests of gravity. Therefore, considering the Fab Four Lagrangians, the Vainshtein mechanism is expected to work for the John and/or Paul Lagrangian(s). 

In the following we focus first on the John Lagrangian\footnote{George and Ringo must not be considered in isolation for theoretical and phenomenological reasons respectively \cite{Charmousis:2011ea}. In addition, only John and Paul might exhibit the Vainshtein screening mechanism.} and analyze what are the viable inflationary solutions. As we will see, the John Lagrangian is not able to play alone gravitation in a static and spherically symmetric spacetime: its solution is trivial since it is the Schwarzschild one. In the rest of the chapter, we thus study the combination of George and John. The George Lagrangian is reminiscent of the Brans-Dicke theory (with the parameter $\omega=0$) or GR ($V_\rr{george}=\rr{cst}$) such that the phenomenology should allow for minimal modifications of gravity.

\section{The John Lagrangian}
\label{sec:John}
\noindent In order to study the phenomenology predicted by John, we start from the John Lagrangian \eqref{eq:john} where the potential reduces to a constant $V_\rr{john}(\phi)=\rr{cst}$, combined with the EH action with a minimally scalar field, 
\begin{eqnarray}
  \label{actionjohn}
  S=\int \dd^4 x \,\sqrt{-g} \left[\frac{R}{2 \kappa} - \frac{1}{2}\left(g^{\mu\nu}+ \kappa \gamma G^{\mu\nu}\right) \partial_{\mu}\phi\partial_{\nu}\phi\right]
  + S_{\textrm{M}}[\psi_{\textrm{M}};~g_{\mu\nu}],
\end{eqnarray}
where 
$\gamma$ is a dimensionless parameter whereas $\phi$ has the dimension of a mass in natural units. This action is a special case of the generalized Galileon one presented in \cite{Kobayashi:2011nu} (see Eqs.~\eqref{eq:horn1}-\eqref{eq:horn_end}), where $K(X)=X, G_3=0, G_4=1/(2\kappa), G_5 = \kappa \gamma \phi/2, G_{4X}=0, G_{5X}=0$.

The modified Einstein equations are then given by (see App.~\ref{sec:eom_fabfour} for the detailed computations) \cite{Sushkov:2009hk},
\bea
  G_{\mu\nu}=\kappa\left[T^{(\rr{M})}_{\mu\nu}+T^{(\phi)}_{\mu\nu}
  +\kappa\gamma \Theta_{\mu\nu}\right],
\eea
with,
\bea \label{eq:EOM_fabtwo}
\Theta_{\mu\nu} & = & -\frac{1}{2}R\,\phi_{\mu}\phi_{\nu}+2\,\phi_{(\mu}R_{\nu)\alpha}\,\phi^{\alpha}-\frac{1}{2}G_{\mu\nu}\left(\nabla\phi\right)^{2}
  +R_{\mu\alpha\nu\beta}\phi^{\alpha}\phi^{\beta}
  +\phi_{\alpha\mu}\phi_{\;\nu}^{\alpha}
  \non\\& & \qquad-\phi_{\mu\nu}\square\phi
  +\frac{g_{\mu\nu}}{2}\left[ -\phi^{\alpha\beta}\phi_{\alpha\beta}+\left(\square\phi\right)^{2}-2\,\phi^{\alpha}\phi^{\beta}\, R_{\alpha\beta}\right], \\
T^{(\phi)}_{\mu\nu}&=&\df_{\mu}\phi\df_{\nu}\phi-\frac{1}{2}g_{\mu\nu}\left(\df\phi\right)^2,
\eea
while the Klein-Gordon equation reads,
\be \label{eq:KGjohn}
  %\nabla^{\mu}\left[\left(g_{\mu\nu}+\kappa\gamma G_{\mu\nu}\right)\nabla^{\nu}\phi\right]&=&0,\\
  \left(g_{\mu\nu}+\kappa\gamma G_{\mu\nu}\right) \nabla^{\mu}\nabla^{\nu}\phi=0.
\ee
  
\subsection{Inflation with John}
\label{sec:John_inflation}
\noindent As it was realized in \cite{Sushkov:2009hk}, this model allows for a quasi de Sitter  inflation with a graceful exit without the need for any specific scalar potential. Inflation is essentially driven by the non-standard kinetic term of the scalar field and it crucially depends on the initial high velocity of the field, as we will shortly see. Although, in principle, the  inflationary solutions begin at $t=-\infty$ such that this theory does not suffer from Big Bang singularity (see \cite{Sushkov:2009hk}), we will  consider the action as an effective  model only  valid from few Planck times after an unknown transplanckian phase. Our first concern is to establish  whether the model accommodates an inflationary phase together with reasonable assumptions for the initial conditions at that time. This section thus completes the analysis found in \cite{Sushkov:2009hk} by providing the number of e-folds as a function of the free parameters of the theory. The equations of motion derived from Eq.~(\ref{actionjohn}) considering the metric ansatz \eqref{eq:metric_FLRW},
\beq
\label{lineelement}
\dd s^2= -\dd t^2 + \rr{e}^{2 \alpha(t)} \dd\mathbf{x}^2,
\eeq 
read in the absence of matter,
\begin{subequations}\label{eom}
\begin{eqnarray}
&&3 \dot{\alpha}^2 = \frac{ \kappa\dot{\phi}^2}{2} \left( 1- 9 \kappa \gamma \dot{\alpha}^2\right), \label{eom1} \\
&&2\ddot{\alpha}+3\dot{\alpha}^{2}=-\frac{\kappa\dot{\phi}^{2}}{2}\left[1+\kappa \gamma\left(3\dot{\alpha}^{2}+2\ddot{\alpha}+4\dot{\alpha}\ddot{\phi}\dot{\phi}^{-1}\right)\right],
\label{eom2}
\\
&&\frac{1}{a^3} \frac{\dd}{\dd t} \left[a^3 \dot{\phi}\left(1-3 \kappa \gamma \dot{\alpha}^2\right) \right]=0,
\end{eqnarray}
\end{subequations}
with $\alpha = \ln a$, $a$ being the scale factor, and $\dot{\alpha}=H$, $H$ being the Hubble parameter.
In order to solve the equations of motion numerically (see Sec.~\ref{sec:John_num_res} for the numerical results), the system is partially decoupled isolating the second order derivatives $\ddot{\alpha}$ and $\ddot{\phi}$,
\begin{subequations}\label{eomdecoupled}
\begin{eqnarray}
\ddot{\alpha}&=&\frac{\left( 3 \kappa \gamma \dot{\alpha}^2 -1\right)}{2}\frac{3 \dot{\alpha}^2 +  \frac{\kappa \dot{\phi}^2}{2}\left(1-9\kappa \gamma \dot{\alpha}^2 \right) } {1-3\gamma \kappa \dot{\alpha}^2+ \frac{\kappa^2 \gamma \dot{\phi}^2}{2} \left(1+9\kappa \gamma \dot{\alpha}^2 \right)}, \label{eomd1}
\\
\ddot{\phi}&=&\frac{-3 \dot{\alpha} \dot{\phi} \left(1+\kappa^2 \gamma \dot{\phi}^2\right)}{1-3\gamma \kappa \dot{\alpha}^2+ \frac{\kappa^2 \gamma \dot{\phi}^2}{2} \left(1+9\kappa \gamma \dot{\alpha}^2 \right)}.\label{eomd2}
\end{eqnarray}
\end{subequations}
This system can be solved as an initial value problem (IVP) by fixing the initial conditions $\alpha_\rr{i},~\dot{\alpha}_\rr{i},~ \phi_\rr{i},~ \dot{\phi}_\rr{i}$.

\begin{figure}[ht]
\begin{center}
\includegraphics[width=0.6\textwidth,trim=320 0 350 0,clip=true]{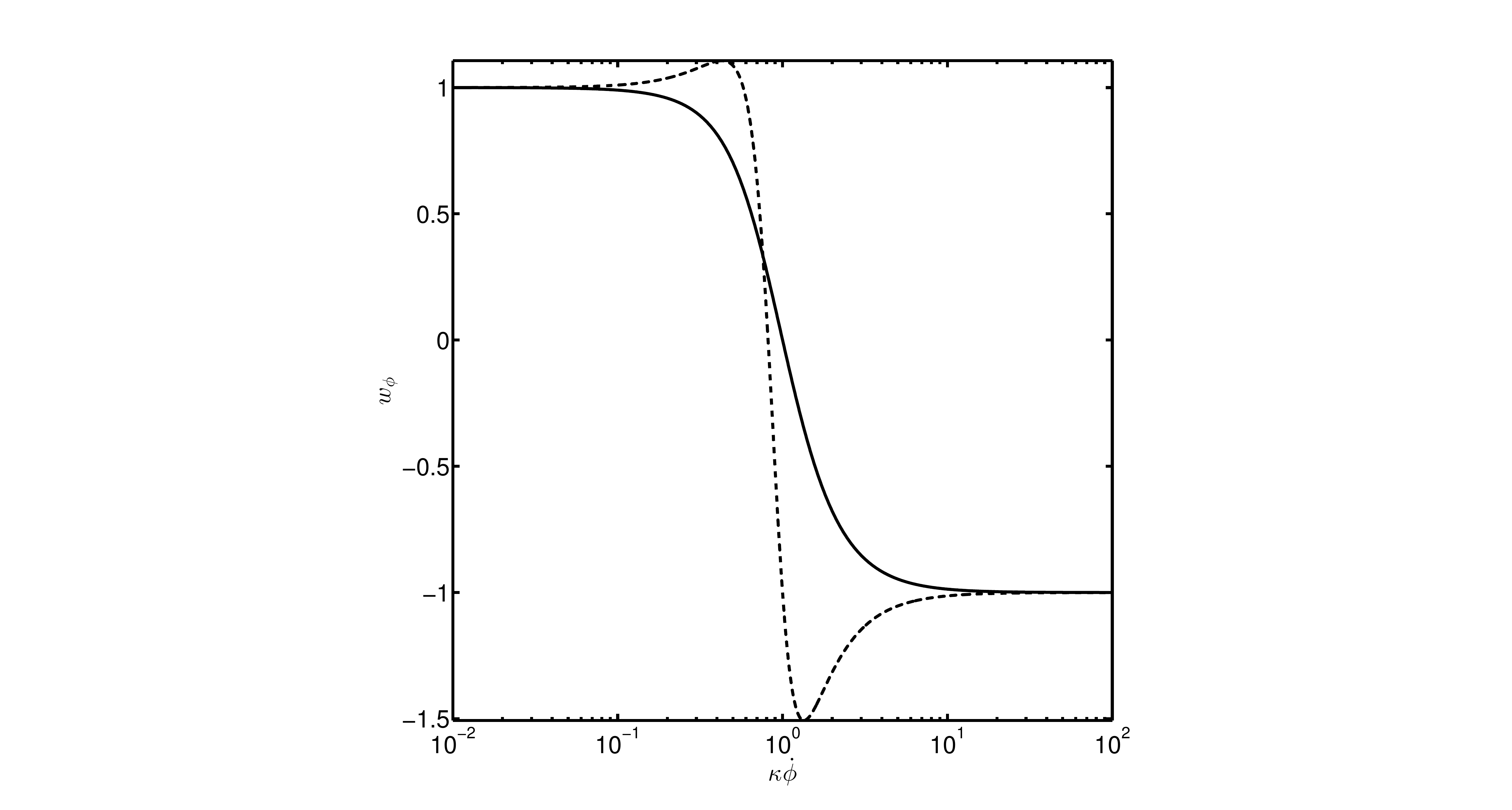}
\end{center}
\caption{EoS of the scalar field $w_\phi$ as a function of its velocity defined by the dimensionless variable $\kappa \dot{\phi}$ for $\gamma=1$ (solid line) and $\gamma=-1$ (dashed line). In the high energy limit ($\kappa \dot{\phi}\gg1)$, the EoS results in a de Sitter phase (the EoS can even be phantom-like $w_\phi<-1$ for $\gamma<0$) where it is of stiff matter in the low energy limit ($\kappa \dot{\phi}\ll1)$. However, the inflationary phase is viable for $\gamma>0$ only.}
\label{figwphi}
\end{figure}

The effective EoS for the scalar field can be obtained from its stress-energy tensor or, more simply, by comparing the equations of motion \eqref{eom1} and \eqref{eom2} directly to the standard Friedmann equations \eqref{eq:FR1_SF} and \eqref{eq:FR2_SF} ($V=0$):
\begin{enumerate}
 \item From Eqs.~\eqref{eom1} and $H^2\equiv \dot{\alpha}^2={(\kappa}/{3})\rho_\phi$ \eqref{eq:FR1_SF}, the energy density of the scalar field reads,
\bea
  \dot{\alpha}^2\left(1+\frac{3}{2}\kappa^2\gamma\dot{\phi}^2\right)=\frac{ \kappa\dot{\phi}^2}{6},&\\
  \dot{\alpha}^2=\frac{\kappa \dot{\phi}^2}{3(2+3 \kappa ^2\gamma \dot{\phi}^2)}&
  \hspace{0.7cm} \longrightarrow \hspace{0.7cm}
  \rho_\phi=\frac{\dot{\phi}^2}{\left(2+{3}\kappa^2\gamma\dot{\phi}^2\right)} \label{eom1_rewritten},
\eea
\item The EoS of the scalar field is identified by comparing the standard Friedmann  equation $\ddot{\alpha}+\dot{\alpha}^2=-(\kappa/6)\rho_\phi(1+3 w_\phi)$  \eqref{eq:FR2_SF} with the modified one \eqref{eom2}. After some algebra, we obtain,
\be
\label{wphi}
w_{\phi}=\frac{\left(2+3\kappa^2 \gamma\dot{\phi}^{2}\right)\left(1-\kappa^2\gamma\dot{\phi}^{2}\right)}{2+3\kappa^2\gamma\dot{\phi}^{2}+3\kappa^{4}\gamma^{2}\dot{\phi}^{4}}.
\ee
\end{enumerate}
The EoS of the scalar field $w_\phi$ is plotted in Fig.~\ref{figwphi} for $\gamma=+1$ and $\gamma=-1$. For both signs of $\gamma$, the EoS tends to $-1$ in the high energy limit ($\kappa |\dot{\phi}| \gg 1$), so that a large initial velocity for the scalar field will result in a quasi de Sitter phase.  However, only the case of positive $\gamma$ can lead to inflation. Indeed,  $\dot{\alpha}$ in Eq.~\eqref{eom1_rewritten} needs to be positive since the Hubble parameter is a real number. Thus, $\gamma <0$ implies that $\kappa\left|\dot{\phi}\right|< \sqrt{-2/3\gamma}$, which, in turn, means that $w_{\phi}>0$ always ($1-\kappa^2\gamma\dot{\phi}^{2}>0$ always). Therefore, the scalar field cannot even start when $w_{\phi} <0 $ if $\gamma <0$. As a result, accelerated phases driven by a scalar field in this model require $\gamma>0$.

Therefore, we focus on the case where $\gamma$ is positive. In that case, we can show that the sign of $\ddot{\phi}$ is always negative (see Eq.~(23) in \cite{Sushkov:2009hk} for instance). 
Hence, the velocity of the field decreases with time and $w_{\phi}$ is driven towards $w_\phi =+1$. 
Assuming that inflation ends at the instant $t_{\text{end}}$, at which  $w_\phi =-1/3$ (the condition $\ddot{a}>0$ is no more satisfied), %\footnote{Throughout this paper we assume a vanishing cosmological constant.}.
$\kappa \gamma \dot{\phi}$ is initially large and leads to the inflationary phase. One may derive an analytical (approximate) solution for the scale factor $a(t)=\rr{e}^{\alpha(t)}$ and the scalar field at early times in this regime  ($\kappa \gamma \dot{\phi}^2\gg1$), considering first Eq.~(\ref{eom1_rewritten}),
\beq
\label{Hhigh}
  H=\dot{\alpha} \simeq \frac{1}{3 \sqrt{\kappa \gamma}}.
\eeq
Integration yields the approximate scale factor,
\beq
\label{scalefactor}
  a(t) \simeq  a_\rr{i} \exp\left( \frac{t -t_\rr{i}}{3 \sqrt{\kappa \gamma}}\right),
\eeq
where the subscript $\rr{i}$ denotes the initial condition.
Inserting Eq.~(\ref{Hhigh}) into Eq.~(\ref{eomd2}), and expanding according to $\kappa^2 \dot{\phi}^2 \gg 1$, gives $\ddot{\phi}/\dot{\phi} \simeq -1/\sqrt{\kappa \gamma}$ and, after integration, yields,
\beq
\label{phidot}
\dot{\phi}(t) \simeq \dot{\phi}_\rr{i} \exp\left(-\frac{t-t_\rr{i}}{\sqrt{\kappa \gamma}}\right).
\eeq
Assuming that $w_{\phi}(t_{\text{end}})=-1/3$ and defining $ \zeta_{\text{end}}\equiv\kappa^2 \gamma \dot{\phi}_\rr{end}^2>0$ with $\phi_\rr{end}=\phi(t_{\text{end}})$, Eq.~\eqref{wphi} yields, 
\be
  w_{\phi}(t_{\text{end}})=\frac{(2+3 \zeta_{\text{end}})(1- \zeta_{\text{end}})}{2+3 \zeta_{\text{end}}+3 \zeta_{\text{end}}^2}=-\frac{1}{3},
\ee
whose solution reads,
\beq
  \zeta_\rr{end} = \frac{1}{6} \left(3+\sqrt{57}\right) \approx 1.76,
\eeq
since only the positive solution verifies $\gamma>0$. Then Eq.~(\ref{phidot}) reduces to, 
\beq
  \kappa^2\gamma\dot{\phi}_\rr{i}^{2}\exp\left[-\frac{2 (t_\rr{end} - t_\rr{i})}{\sqrt{\kappa \gamma}}\right] \simeq \zeta_{\text{end}},
\eeq
leading to,
\beq
\label{deltat}
t_\rr{end}-t_\rr{i}=\frac{\sqrt{\kappa \gamma}}{2}\ln \left(\frac{\kappa^2\gamma\dot{\phi}_\rr{i}^{2}}{\zeta_{\text{end}}}\right).
\eeq
Inserting this expression into Eq.~(\ref{scalefactor}) yields,
\beq
\label{fraca}
\frac{a_{\text{end}}}{a_\rr{i}} \simeq \left(\frac{\kappa^2 \gamma \dot{\phi}_\rr{i}^2}{\zeta_{\text{end}}}\right)^{\frac{1}{6}}.
\eeq
Imposing that inflation lasts for a  number of e-folds $N= \ln (a_{\text{end}}/a_\rr{i})$ larger than $60$, we obtain,
\beq
\label{maineq}
  \dot{\phi_\rr{i}}^2 \gtrsim \frac{\zeta_{\text{end}}}{\kappa^2 \gamma} \exp(360),
\eeq
which is the crucial condition for a successful (purely kinetic-driven) inflationary phase. We see that it involves a rather unusual very large pure number. The Eq.~(\ref{Hhigh}) is also relevant in order to discuss naturalness\footnote{This argument about naturalness is questionable since the scale at which the Fab Four model breaks down might differ from the GR one.}, as it fixes the Hubble parameter at the beginning of the inflationary phase $H_\rr{i} \simeq 1/ (3\sqrt{\kappa \gamma})$. Therefore, Eq.~\eqref{maineq} might be rewritten as,
\beq
\kappa \frac {\dot{\phi_\rr{i}}^2}{H_\rr{i}^2} \gtrsim 9 \, \zeta_{\text{end}}\exp(360) \sim 10^{157}.
\eeq
It follows that the "natural" initial conditions $H_\rr{i} = \mathcal{O}(1)\sim\Mp$ and $\dot{\phi}_\rr{i}  = \mathcal{O}(1)\sim \Mp^2$ in Planckian units are not allowed. On the contrary, a natural value for the initial expansion  $H_\rr{i} =\Mp$ (and thus  $\gamma \approx 0.11$) requires an extremely high transplanckian value for the initial velocity of the field $\dot{\phi}_\rr{i} \gtrsim 10^{78}\Mp^2$. 

It is not even possible to obtain a Planckian value for the initial velocity in this model. Indeed, assuming $\dot{\phi}_\rr{i}\sim\Mp^2$, the initial Hubble parameter will be smaller than the one today, $H_0\simeq2.1\,h~\times 10^{-42}~\rr{GeV}\sim10^{-61}\Mp$. This implies that in such an inflationary scenario, $H_\rr{i}^{-1}\sim\sqrt{\kappa \gamma}$ must be less than the Hubble radius today $H_0^{-1}$ whereas the inflation predicts a huge expansion of the Universe. 

\subsection{Theoretical constraints}
\label{sec:John_theo}
\noindent In this section, we investigate if there exist metric backgrounds for which the propagation of the scalar field becomes pathological, that is non hyperbolic, and thus possibly non causal (see also Sec.~\ref{sec:causality}), or carrying negative energy degrees of freedom, i.e. ghosts (see also Sec.~\ref{sec:MG_issues})). In the following, the cosmological background is assumed to be flat \eqref{lineelement} and we explore the conditions for the theory to be well-defined, for both scalar and tensor metric perturbations.

Conditions for the avoidance of ghosts in scalar, vector and tensor perturbations of the metric have been derived in full generality in a very wide class of Galileon models by \cite{DeFelice:2011bh}. Let us first introduce the reduced dimensionless variables,
\bea
  x(t)&=& \kappa \dot{\phi}, \\
  y(t)&=&\sqrt{\kappa} \dot{\alpha}.
\eea
The no-ghost conditions $Q_\rr{S,~T}>0$ where the subscripts $S$ and $T$ are for scalar and tensor metric perturbations, are given by Eqs.~(23)--(25) in \cite{DeFelice:2011bh} while the conditions for the avoidance of Laplacian instabilities $c_\rr{S,~T}^2\geq0$ are given by Eqs.~(27)--(28) in \cite{DeFelice:2011bh}. Those conditions reduce to rather simple algebraic constraints in our case, after the necessary manipulation using the equations of motion (\ref{eom}) and (\ref{eomdecoupled}),
\begin{subequations}
\begin{eqnarray}
Q_\rr{T}>0  \hspace{1cm} \Leftrightarrow \hspace{1cm} 1+\frac{\gamma x^2}{2} >0, \\
c_\rr{T}^2 \geq 0 \hspace{1cm} \Leftrightarrow \hspace{1cm} 1-\frac{\gamma x^2}{2}  \geq 0,
\end{eqnarray}
\end{subequations}
for the tensor metric perturbations, and,
\begin{subequations}
\begin{eqnarray}
Q_\rr{S}>0 \hspace{0.2cm} &\Leftrightarrow& \hspace{0.2cm} \frac{4+6 \gamma x^2+6 \gamma^2 x^4}{2+3 \gamma x^2} >0,\\
c_\rr{S}^2 \geq 0 &\Leftrightarrow& \frac{12+36 \gamma x^2+19 \gamma^2 x^4-12 \gamma^3 x^6-3 \gamma^4 x^8}{2+3 \gamma x^2+3 \gamma^2 x^4} \geq 0,
\end{eqnarray}
\end{subequations}
for the scalar metric perturbations.

\begin{figure}[!tbp]
  \centering
  \subfloat[The field velocity must be $|x|<1.4$ for $\gamma =1$.]{\includegraphics[width=0.6\textwidth, trim= 240 0 250 0,clip=true]{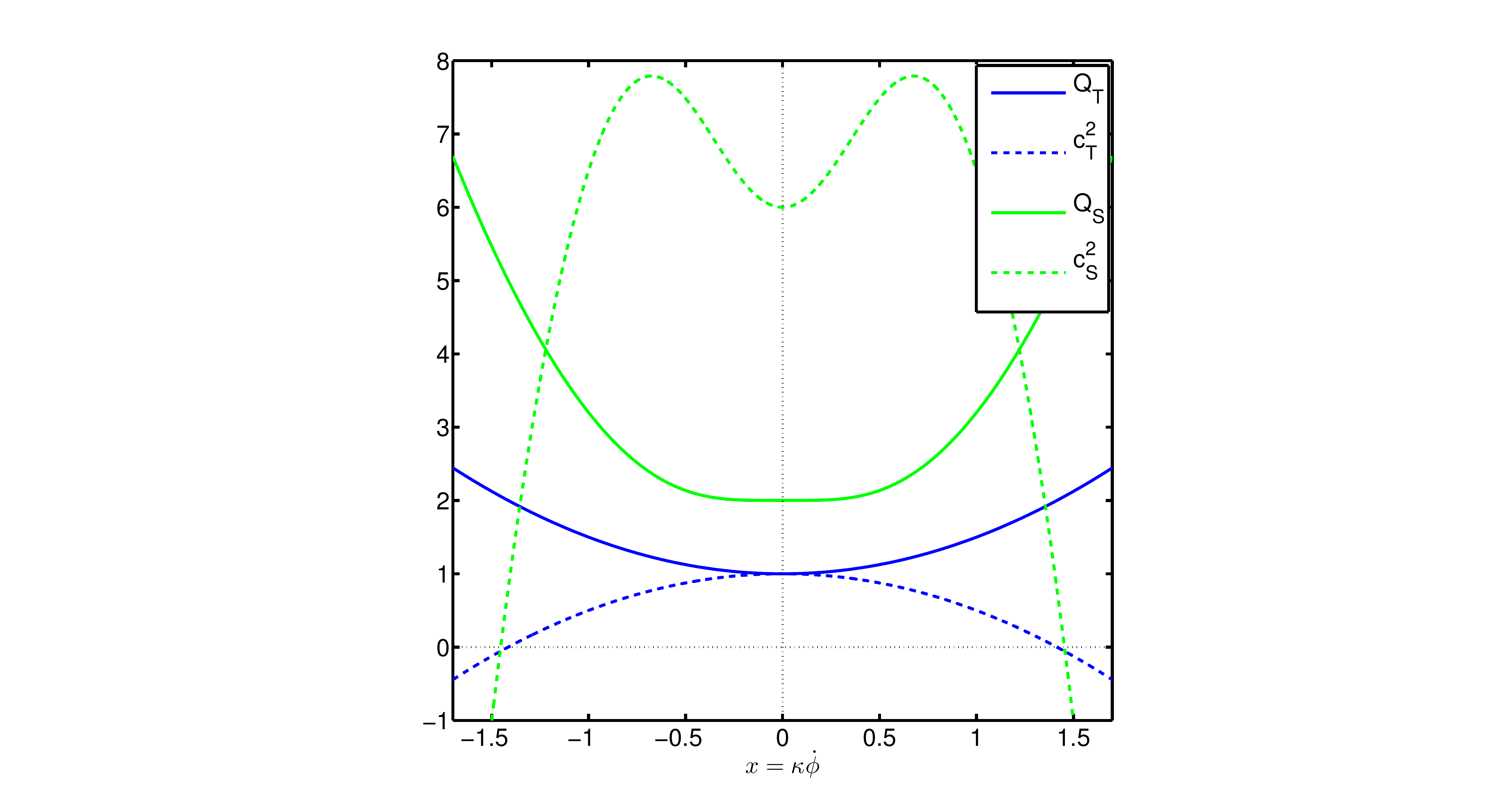}
   \label{ghost1}}
  \hfill
  \subfloat[The field velocity must be $|x|<0.7$ for $\gamma =-1$.]
  {\includegraphics[width=0.6\textwidth, trim= 240 0 250 0, clip=true]{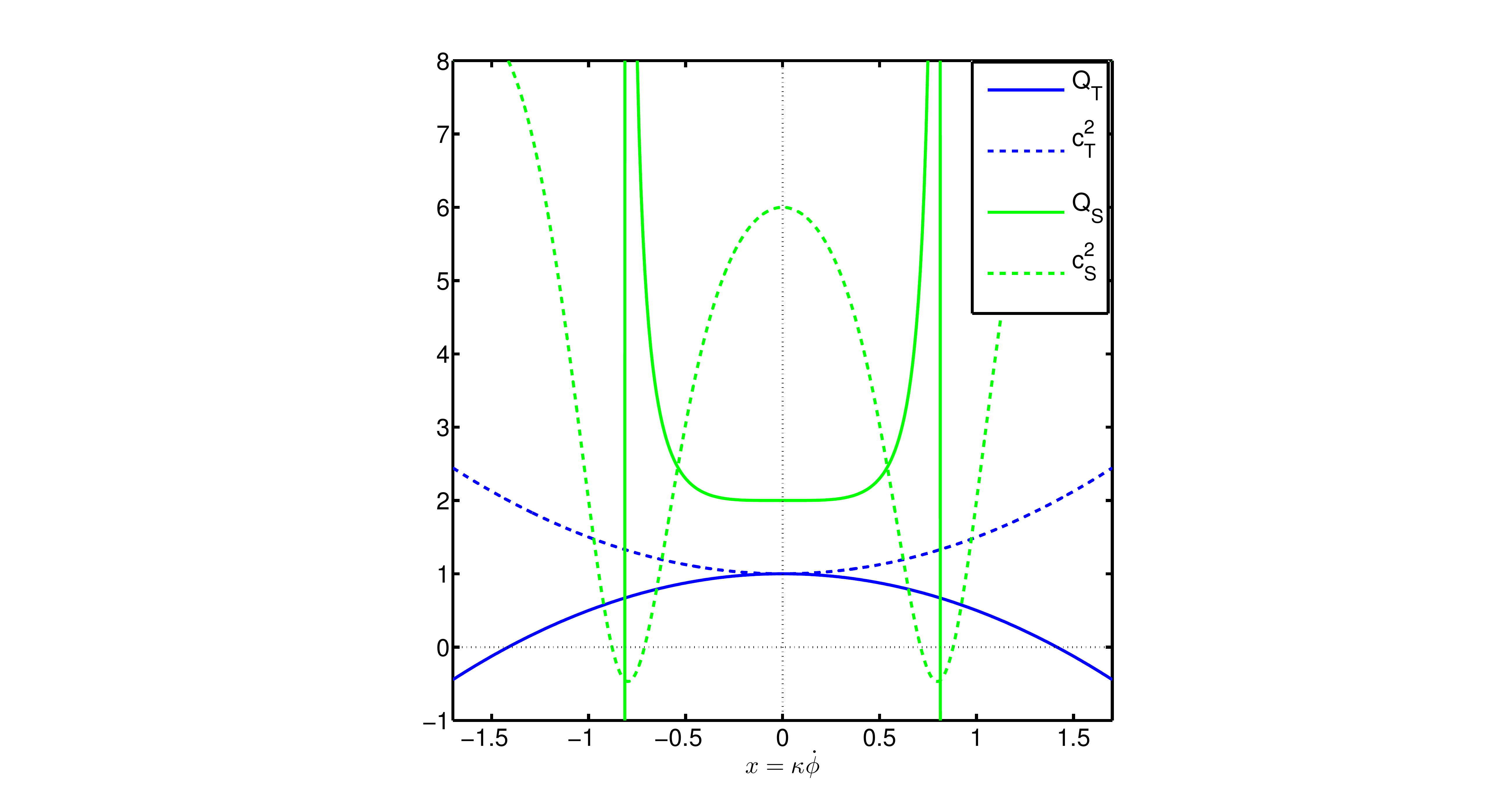}
   \label{ghost2}}
  \caption{Analysis of the causal behavior for and the metric (scalar and tensor) perturbations as a function of the scalar field velocity $x=\kappa\dot{\phi}$. In order to avoid ghost and Laplacian instabilities, $Q_\rr{S}$, $Q_\rr{T}$ and $c^2_\rr{S}$, $c^2_\rr{T}$ must be positive. Allowed values for the field velocity are typically $|x|<\xi_\gamma\sim\mathcal{O}(\gamma^{-1/2})$ in order to preserve causality.}
\end{figure}

This whole set of equations is difficult to reduce algebraically because of the last one. However, one might easily plot the four functions of $x$ defined above, and one typically finds that both  positive and negative values for $\gamma$ are allowed on a given range $|x|< \xi_{\gamma}$, where typically $\xi_{\gamma}$ behaves as $\mathcal{O}\left(1/\sqrt{|\gamma|}\right)$, see the Figs. \ref{ghost1} and \ref{ghost2} for $\gamma=1$ and $\gamma=-1$. Hence,  large (transplanckian) values for $|x|$ are only allowed for small $|\gamma| \ll 1$. This means that the space for possible velocities of the field $x=\kappa \dot{\phi}$ needs to be typically subplanckian, unless $\gamma$ is vanishingly small. This will be linked to the results found earlier, where transplanckian initial velocity were required for a successful inflation,  leading to negative squared speeds $c_\rr{S}^2$ and $c_\rr{T}^2$ in that epoch. This is further discussed in Sec.~\ref{sec:John_num_res}.

Meanwhile, we note that the claim made in the literature (see e.g. \cite{Tsujikawa:2012mk, Germani:2010gm, Germani:2010hd}) according to which only the subclass $\gamma<0$ is a ghost-free theory is wrong (at least in the background considered here).  Notice that the scalar field is well-defined although being a phantom, i.e. $w_\phi<-1$, in a certain regime (in the case $\gamma<0$) (see Fig.~\ref{figwphi}), a situation reminiscent of the one discussed in \cite{Creminelli:2010ba}. However, as shown previously, the Friedmann equations actually prevent the scalar field from entering this regime. 

\subsection{Numerical results}
\label{sec:John_num_res}
\begin{figure*}[bht]
\begin{center}
\includegraphics[width=0.4\textwidth, trim= 210 0 220 0,clip=true]{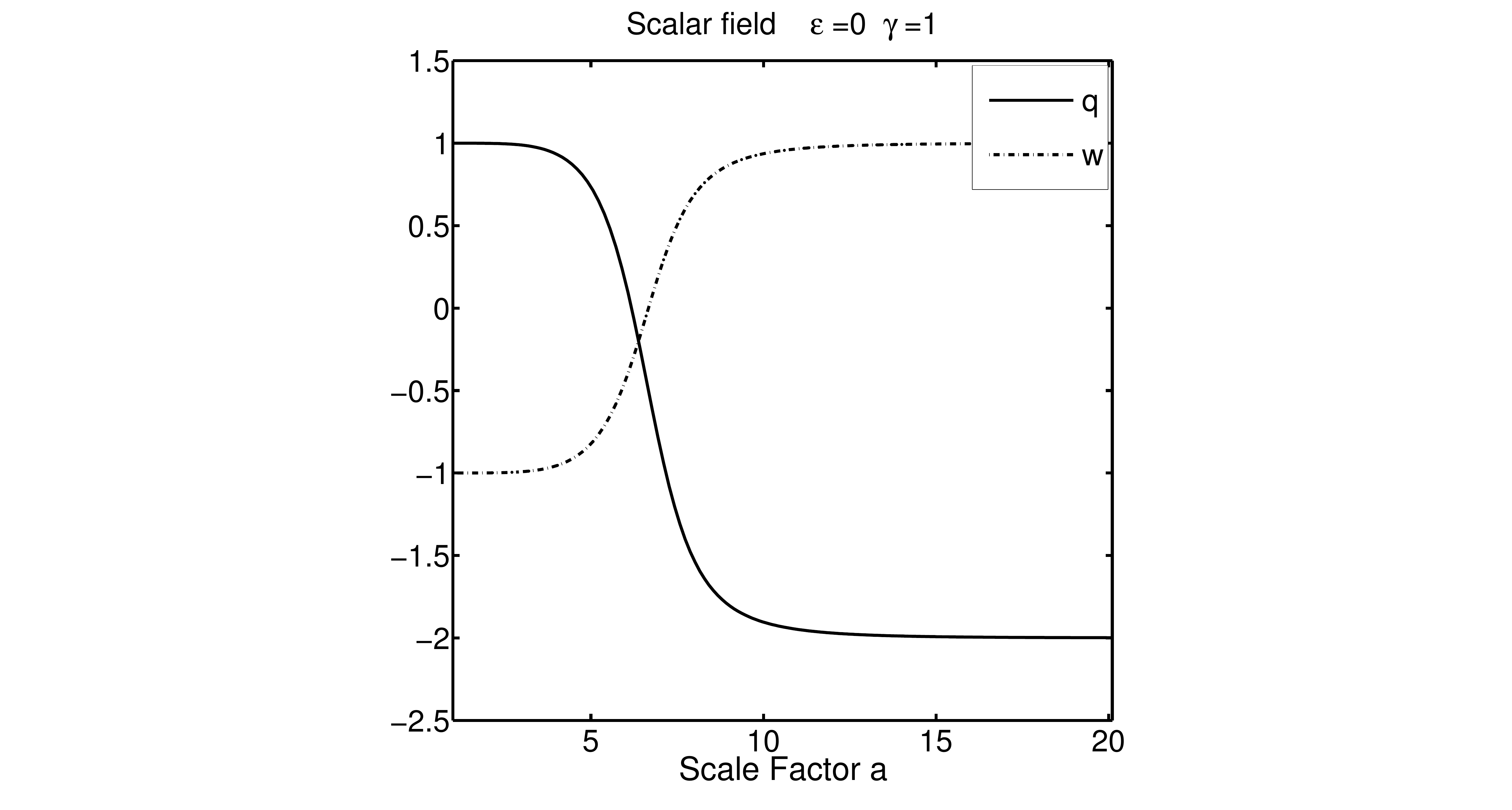}\hspace{1.5cm}
\includegraphics[width=0.4\textwidth]{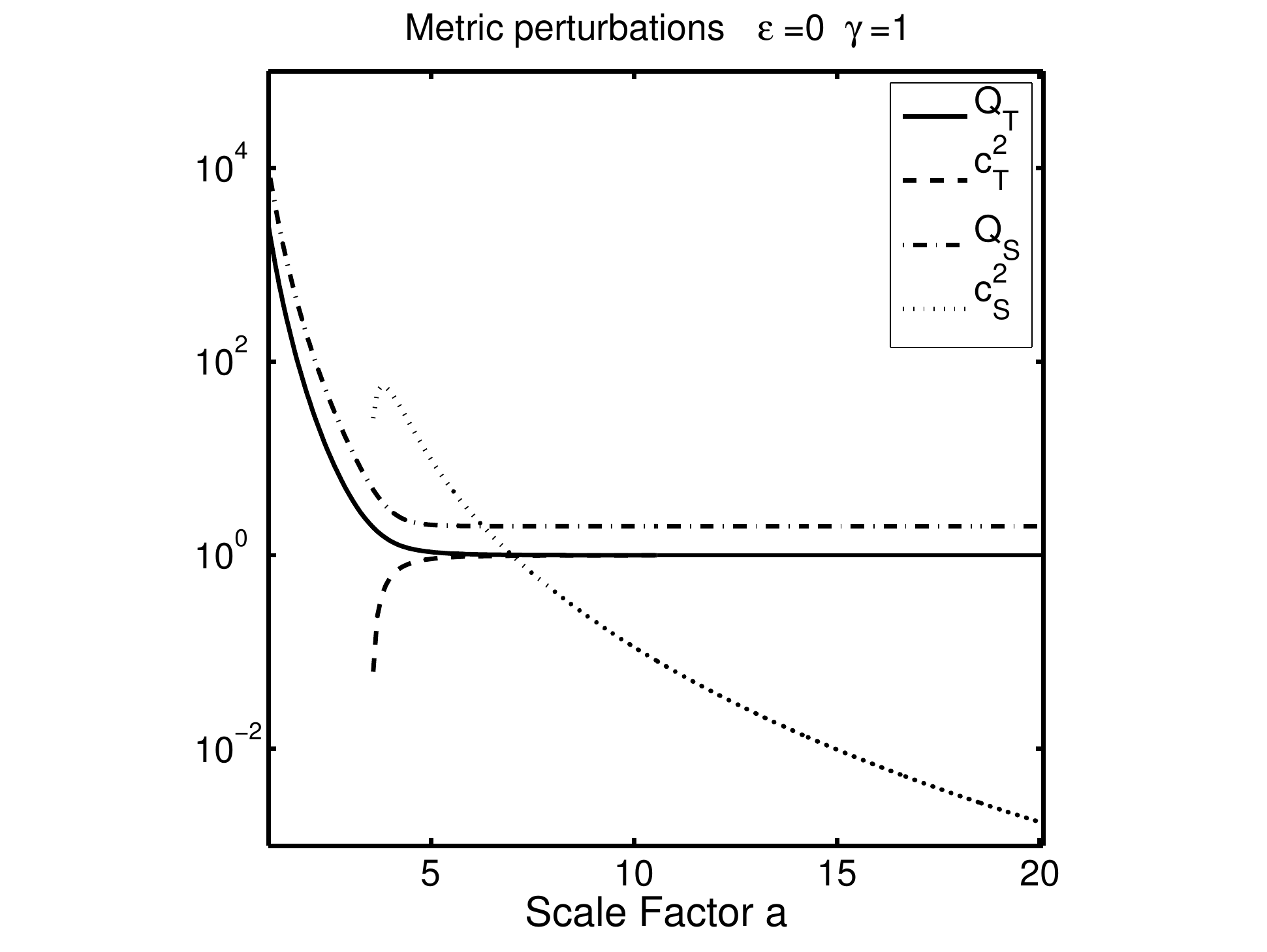}   \\
\includegraphics[width=0.4\textwidth, trim= 210 0 220 0,clip=true]{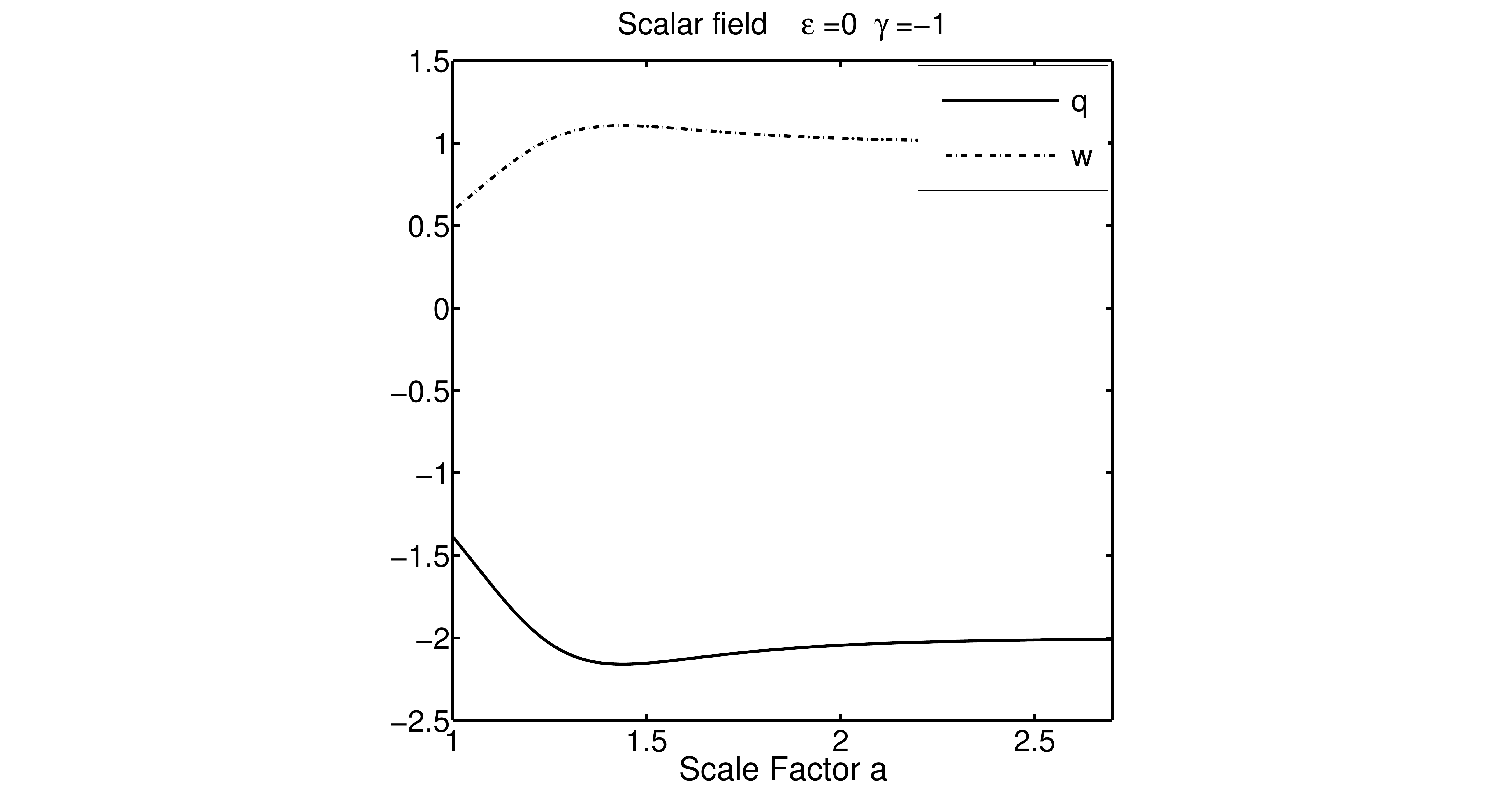}\hspace{1.5cm} 
\includegraphics[width=0.4\textwidth]{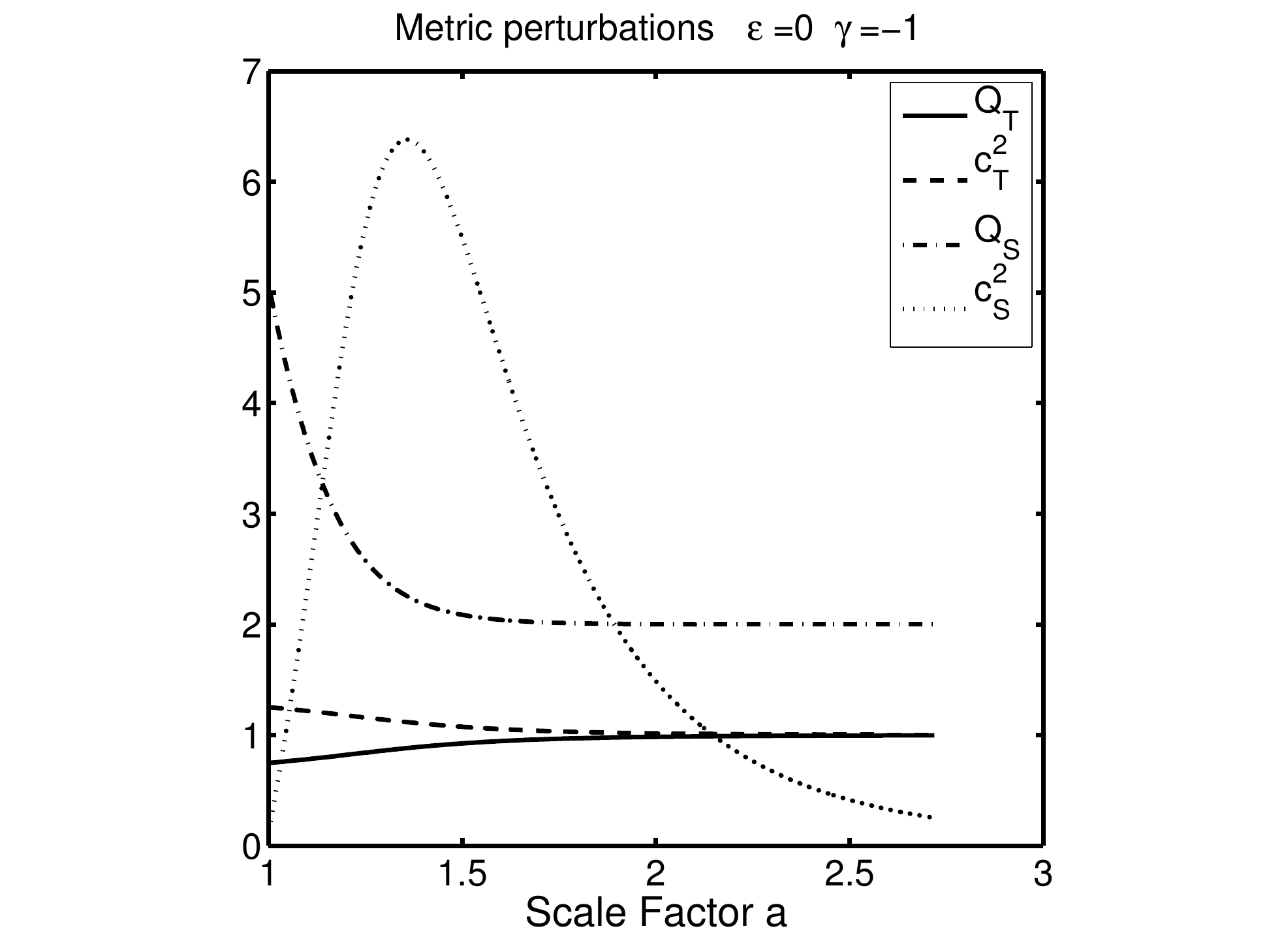} 
\end{center}      
\caption{ Cosmological evolution of $q, Q_\rr{S}, c_\rr{S}^2, Q_\rr{T}, c_\rr{T}^2$ as a function of the scale factor. (Top): This model leads to an accelerating expansion ($q>0$). However, the large initial velocity of the scalar field drives the speed of both scalar and tensor metric perturbations to imaginary values ($c^2_\rr{S}<0$ and $c^2_\rr{T}<0$, the corresponding curves terminate since the scale of the $y$ axis is logarithmic), thus signaling a breakdown of hyperbolicity for metric perturbations. (Bottom): The field starts with an EoS $w_{\phi} \sim 0.5$ and the Universe only decelerates ($q<0$). This model with $\gamma<0$ is well-behaved but does not accommodate inflation.}                    
\label{johncosmofigs}  
\end{figure*}

\noindent In this section, the cosmological evolution predicted by the John model is quickly discussed, for both positive and negative $\gamma$. The equations of motion \eqref{eomdecoupled} were solved by numerical integration as an IVP. The initial conditions for the scalar field velocity were fixed to $\kappa\dot{\phi}_\rr{i} =10$ for $\gamma=1$, and $\kappa\dot{\phi}_\rr{i}=0.1$ for $\gamma=-1$ in order to satisfy the condition  $|x_\rr{i}|=0.1<1.4$ required by the stability conditions (see Fig.~\ref{ghost1}).  In Fig.~\ref{johncosmofigs}, the evolution of the EoS $w_\phi$ as well as the acceleration parameter,
\be
  q=+\frac{\ddot{a}a}{\dot{a}^2},
\ee
are represented depending on the scale factor. The evolution of the conditions to avoid ghost $Q_\rr{S,~T}>0$ and Laplacian instabilities $c^2_\rr{S,~T}\geq0$ are also shown. 
As discussed before, the negative $\gamma$ case leads only to a decelerating Universe: the phantom regime is not an acceptable initial condition (as it entails an imaginary Hubble parameter), and neither can be reached. Only a positive $\gamma$ leads to an accelerated phase of the expansion, and to an inflationary phase in the early Universe, a drawback being the presence of non-causal behavior for the scalar and tensor perturbations of the metric.

\subsection{Discussion}
\label{sec:discuss_John}
We have established that kinetically driven inflation in the Galileon theory involving the simplest coupling to the Einstein tensor, that is the John Lagrangian, is not viable. It requires unnatural transplanckian values for the initial velocity of the field, which, in turn, implies various instabilities. 

This model has anyway another serious drawback. In the absence of any direct coupling to the Ricci scalar, there is no reason why the scalar field should be generated at all since there is no source term in the Klein-Gordon equation \eqref{eq:KGjohn} (even in the presence of a cosmological matter fluid). In other words, $\phi =0$ is always a solution in this class of models, whatever the matter content is. 

This statement is further justified by considering the prediction of the John Lagrangian \eqref{actionjohn} at local scales, in a static and spherically symmetric spacetime. Using the metric ansatz \eqref{eq:metric_schwa}, the Klein-Gordon equation yields,
\bea
  &&\hspace*{-0.7cm}\kappa\gamma \, {r^{2}}\rr{e}^{-2\,\lambda}\left( 3\,\frac{\nu'\phi'}{r^{2}}-6\,\frac{\phi'\nu'\lambda'}{r}+2\,\frac{\nu'^{2}\phi'}{r}-3\,\frac{\phi'\lambda'}{r^{2}}+2\,\frac{\phi''\nu'}{r}+2\,\frac{\phi'\nu''}{r}+\frac{\phi''}{r^{2}}\right)  \non\\
  &&+ {r^{2}}\left[\phi''+\phi'\left(\nu'+\frac{2}{r}-\lambda'\right)\right] +{\kappa\gamma}\left(\phi'\lambda'-\nu'\phi'-\phi''\right)
  =  0,
\eea
the prime denoting derivative with respect to the radial coordinate. By imposing the regularity condition  at the origin $\phi'(r=0)=0$, the solution of the Klein-Gordon equation is trivial,
\be
  \phi''=0 \hspace{1cm} \Rightarrow \hspace{1cm} \phi'=\rr{cst}=0~\hspace{0.5cm}\forall~ r.
\ee
Imposing that the solution is asymptotically flat at spatial infinity, i.e. $\phi(r\longrightarrow \infty)=0$, the solution for a relativistic star is the GR one, i.e. the Schwarzschild solution ($\phi=0\,\forall r$), even in the presence of matter.

To conclude, the model considered so far is trivial in the sense that it cannot be different than GR, except if one imposes non-vanishing initial conditions for the scalar field at early times.
In order to obtain non-trivial solutions, we consider the combination of the John and the George Lagrangians, the latter introducing a direct coupling to the Ricci scalar, $V_\rr{george}(\phi) R$ \eqref{eq:george}.

\section{The John and George Lagrangian}
\label{sec:FabTwo}
We now consider the extended model given by, 
\begin{eqnarray}
  \label{actionjohngeorges}
  S=\int \dd^4 x \sqrt{-g}  \left[\frac{R}{2 \kappa}\left(1+\epsilon \sqrt{\kappa}\phi\right) \right.\qquad\qquad\qquad\qquad\non\\ 
  \left. - \frac{1}{2}\left(g^{\mu\nu}+ \kappa \gamma G^{\mu\nu}\right) \partial_{\mu}\phi\partial_{\nu}\phi\right]+ S_{\textrm{M}}[\psi_\rr{M};~g_{\mu\nu}] ,
\end{eqnarray}
where the nonminimal coupling function is fixed to $V_\rr{george}(\phi)=1+\epsilon \sqrt{\kappa}\phi$ from now on, and $\epsilon$ is a dimensionless, free parameter. The modified Einstein equations then read (see Sec.~\ref{sec:eom_JF} and App.~\ref{sec:eom_fabfour} for the calculations of the equations of motion),
\begin{equation}
G_{\mu\nu}\left(1+\epsilon\sqrt{\kappa}\phi\right)+\epsilon\sqrt{\kappa}\,\left(g_{\mu\nu}\square\phi-\nabla_{\mu}\nabla_{\nu}\phi\right)=\kappa\left(T_{\mu\nu}^{\left(\phi\right)}+\kappa\gamma\Theta_{\mu\nu}\right),
\label{eq:EoM_JG}
\end{equation}
with $T_{\mu\nu}^{\left(\phi\right)}$ and $\Theta_{\mu\nu}$ defined in Eqs.~\eqref{eq:EOM_fabtwo}, while the Klein-Gordon equation reads,
\be
  \left(g_{\mu\nu}+\kappa\gamma G_{\mu\nu}\right) \nabla^{\mu}\nabla^{\nu}\phi+\frac{\epsilon R}{2\sqrt{\kappa}}=0.
\ee
Notice that one can argue that the effective gravitational constant $G_{\textrm{eff}} = G/(1+\epsilon \sqrt{\kappa} \phi)$ might easily become negative in this model, meaning that the action chosen here shall trivially lead to dynamical pathologies for $\epsilon \phi$ sufficiently large and negative\footnote{In fact, what matters in the case $\gamma=0$, is that the scalar field propagates positive energy in the Einstein frame. Performing a conformal transformation, this is equivalent to the usual Brans-Dicke condition $2 \omega + 3 >0$, where $\omega= \epsilon^{-2}(1+ \epsilon \sqrt{\kappa} \phi)$ here. Then, our model with $\gamma =0$ would indeed be pathological if $\phi \leq -[3/(2\epsilon^2)+1]/(\epsilon \sqrt{\kappa})$ . However, there are new terms in the equations of motion for the scalar field due to the presence of the John terms, which invalidate such a conclusion in the general case $\gamma \neq 0$.}. 

Such an argument would call in favor of defining a better coupling function $V_\rr{george}(\phi)$. However, this would be a misleading conclusion here, since the John term introduces a derivative coupling between the metric and the scalar field, thus impacting the propagation of the metric and scalar degrees of freedom. Therefore, only the entire set of stability conditions for both the scalar and the metric perturbations (i.e. positivity of the squared velocities $c^2_\rr{S}\geq0$ and $c^2_\rr{T}\geq0$) can decide which regions of the configuration space are well-behaved. The results are presented in the following sections assuming a cosmological background, based on the conditions derived in App.~\ref{app:causality}. In this light, the function $V_\rr{george}(\phi)$ chosen above is just one of the simplest that we can choose, and might furthermore be understood as retaining only the first term in a series expansion of a more general function $V_\rr{george}(\phi)=\rr{e}^{\epsilon\sqrt{\kappa}\phi}$.

The cosmological evolution predicted by John and George is typically a function of four parameters: the initial value of the field $\phi_\rr{i}$, its velocity $\dot{\phi}_\rr{i}$, as well as the two dimensionless parameters $\gamma$ and $\epsilon$. It goes beyond the scope of this thesis to provide a comprehensive study of this parameter space. However, the numerical results presented in the next section, highlight some essential features in the case where John and George are playing cosmology together. As for the case where John plays alone, the cosmological evolution is studied in terms of $w_\phi$ and $q$ for particular combination of the parameters $\gamma$ and $\epsilon$. The conditions for causality and for energy positivity are analyzed, depending on the signs of $\gamma$ and $\epsilon$.

\subsection{Cosmological behavior}
\label{sec:FabTwo_cosmo}
The equations of motion in a flat, empty Universe, derived from Eq.~(\ref{actionjohngeorges}) are given in App.~\ref{app:cosmo_Fab2} (see Eqs.~\eqref{Eomjg1}-\eqref{Eomjg3}). The analysis of the no-ghost and causality conditions to this more general framework is reported in App.~\ref{app:causality} while the derivation of the scalar field EoS is given in App.~\ref{app:cosmo_Fab2}.

\begin{figure}[!tbp]
  \centering
  \subfloat[We observe the same transition from inflation to stiff matter for the scalar field.]
%   {\includegraphics[width=0.47\textwidth, trim= 240 0 250 0,clip=true]{./chapters/chapter6/scalar_e1_g1}
  {\includegraphics[width=0.4\textwidth, trim= 360 0 370 0,clip=true]{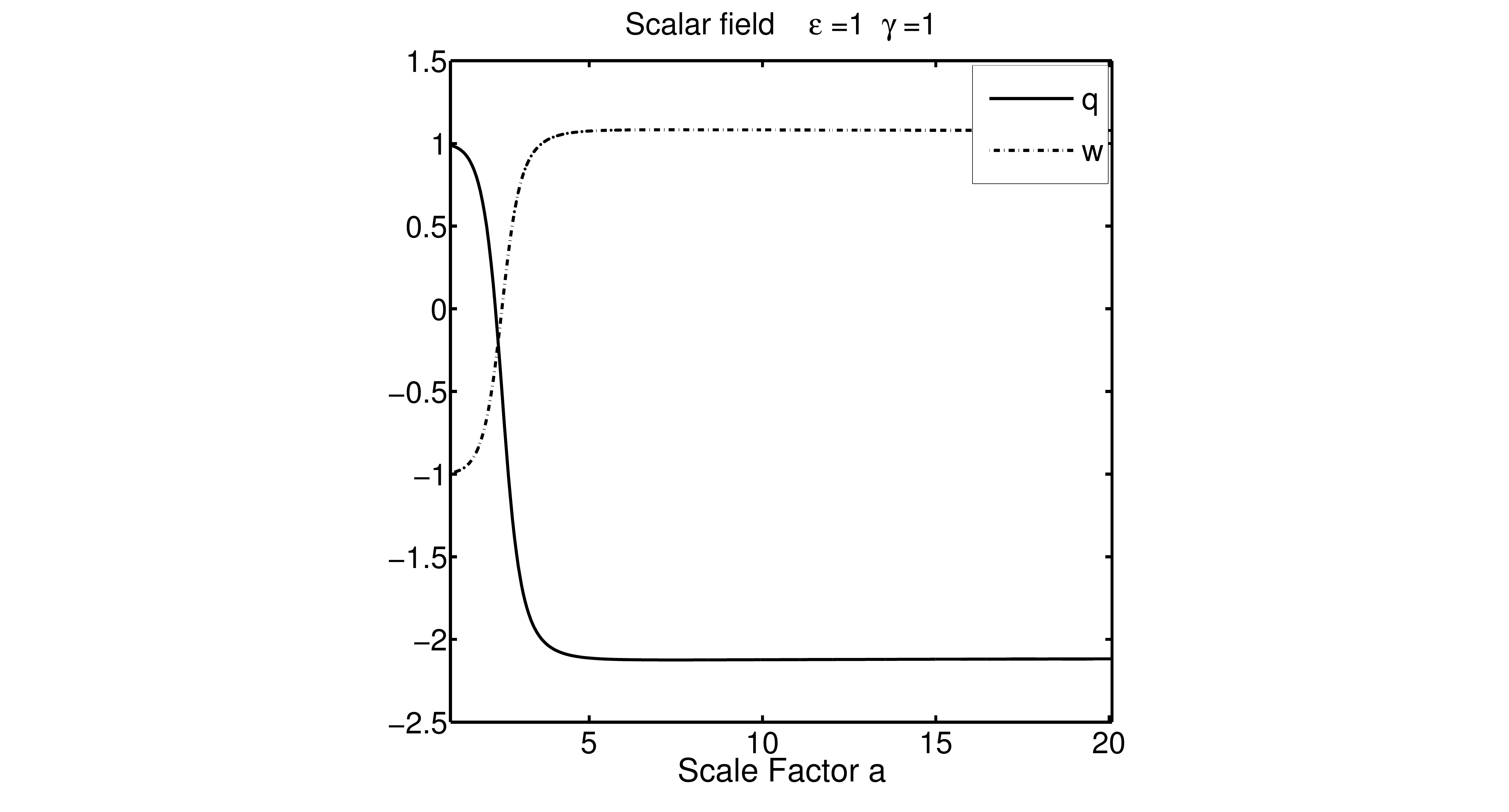}
   \label{jgs11}}
  \hfill
  \subfloat[The "sound" speeds of scalar and tensor metric perturbations are negative in the early Universe.]
  {\includegraphics[width=0.4\textwidth]{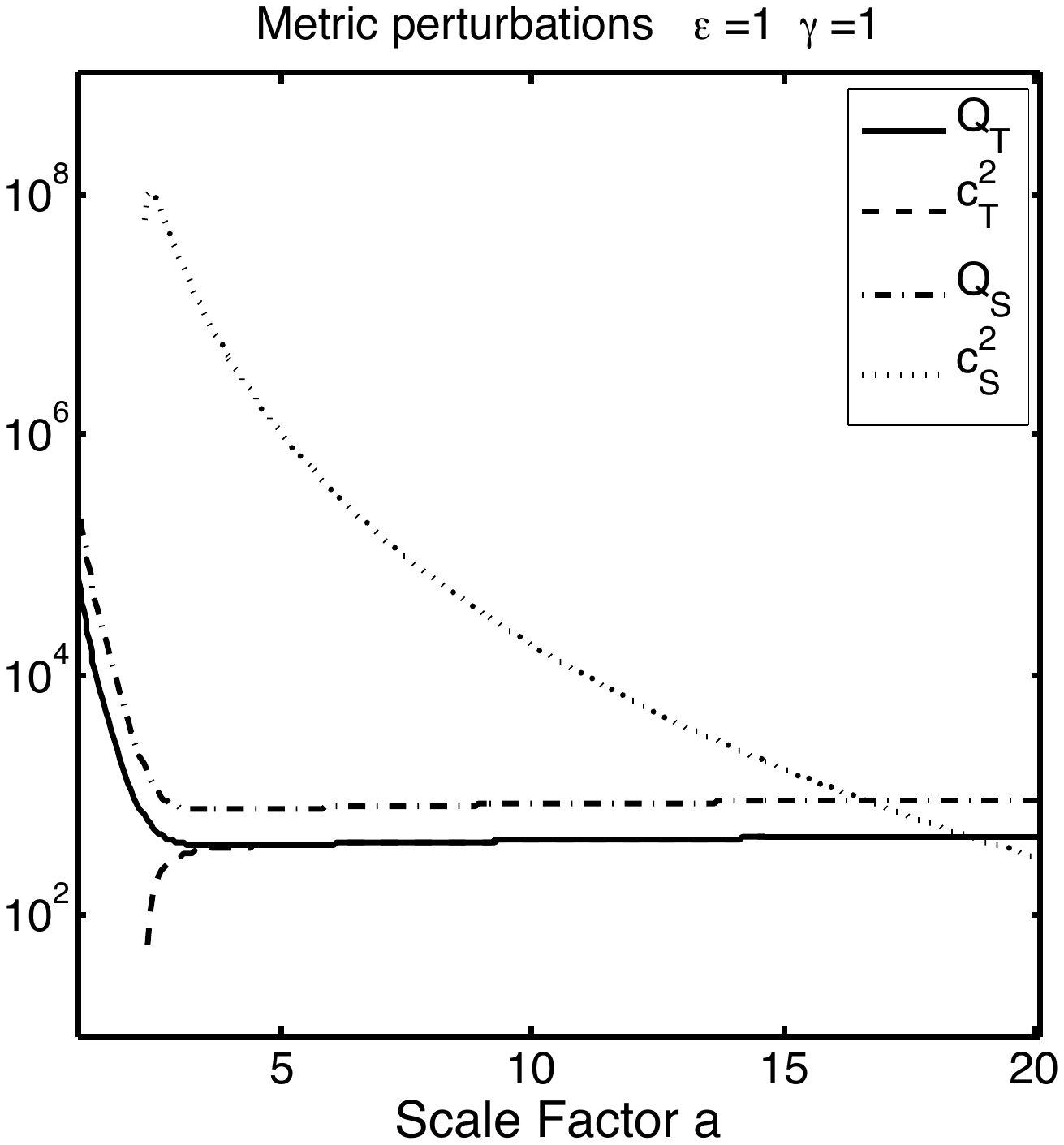}
   \label{jgm11}}
  \caption{Evolution of $q$, $w_{\phi}$ (on the left) as well as $Q_\rr{S}$, $Q_\rr{T}$ and $c_\rr{S}^2, c_\rr{T}^2$ (on the right), as a function of the scale factor $a$, assuming the initial conditions  $\dot{\phi}_\rr{i}=100$ and $\phi_\rr{i}=1$.}
\end{figure}

The numerical results are the following. The case $\epsilon=1$ and $\gamma=1$ is pretty similar to the case John alone ($\gamma>0$), see Figs.~\ref{jgs11} and~\ref{jgm11}. 
Inflation thus occurs in the case $\epsilon>0$ and $\gamma>0$, but the acausal behavior still shows up in the very early Universe. The number of e-folds is a function of the two initial conditions for the field and its velocity, as well as of the dimensionless parameters $\epsilon$ and $\gamma$. A further analysis would determine whether the addition of the George term helps in solving the naturalness problem encountered with John alone in Sec.~\ref{sec:John}.

\begin{figure*}[bht]
\begin{center}
\includegraphics[width=0.35\textwidth, trim=370 0 380 0, clip=true]{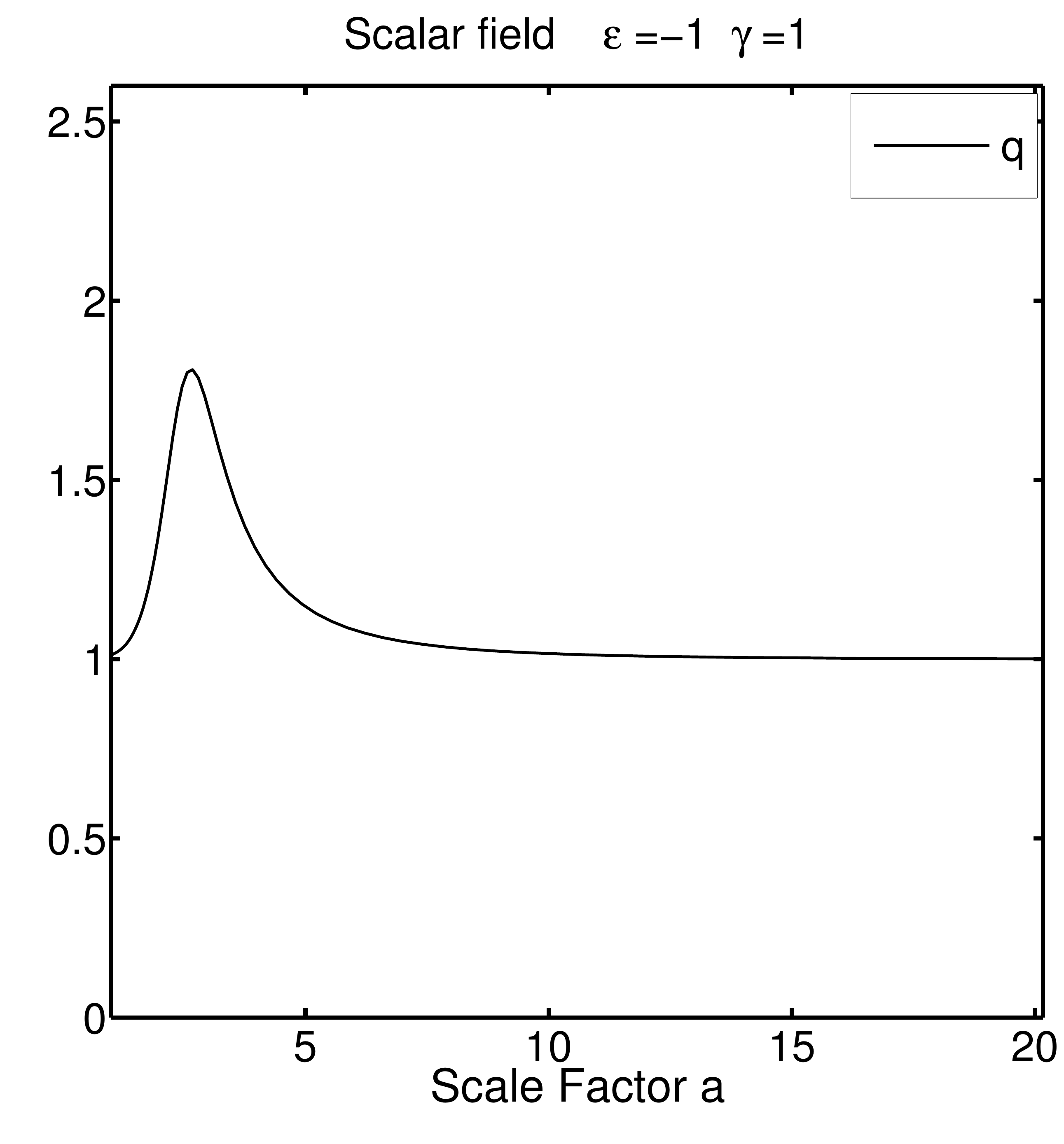}\hspace{1.5cm}
\includegraphics[width=0.35\textwidth]{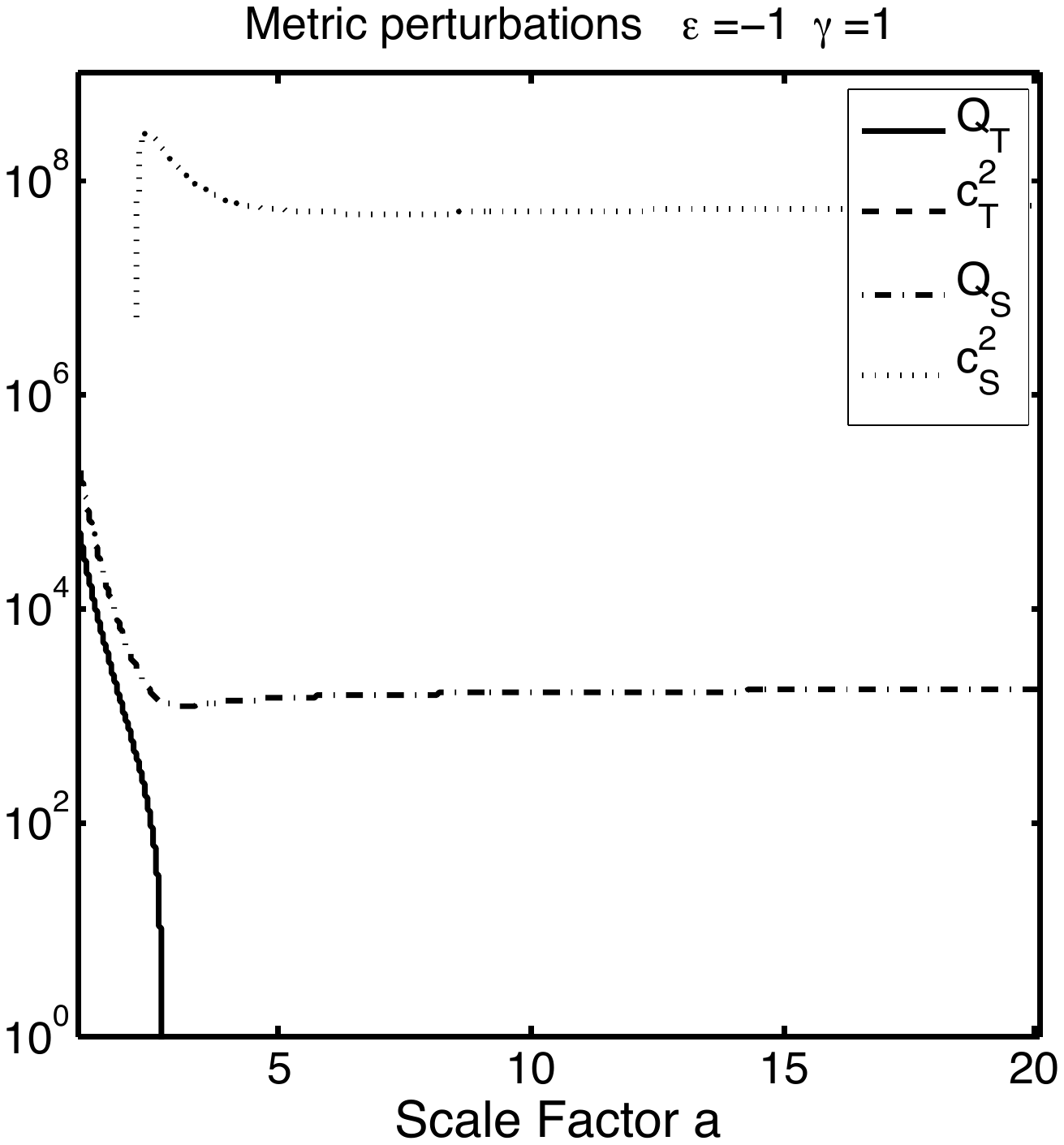}   \\
\includegraphics[width=0.35\textwidth, trim=370 0 380 0, clip=true]{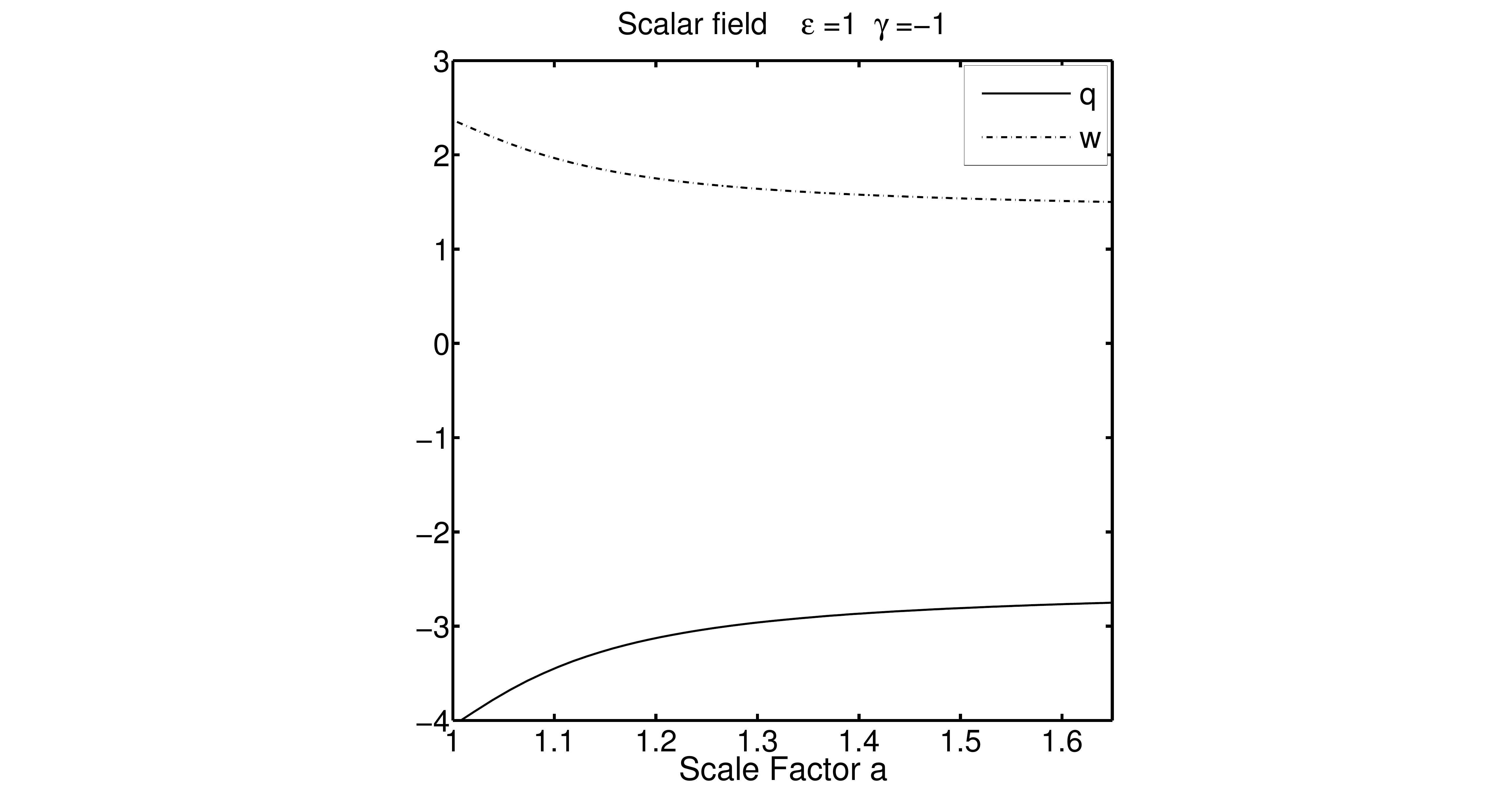}\hspace{1.5cm} 
\includegraphics[width=0.35\textwidth]{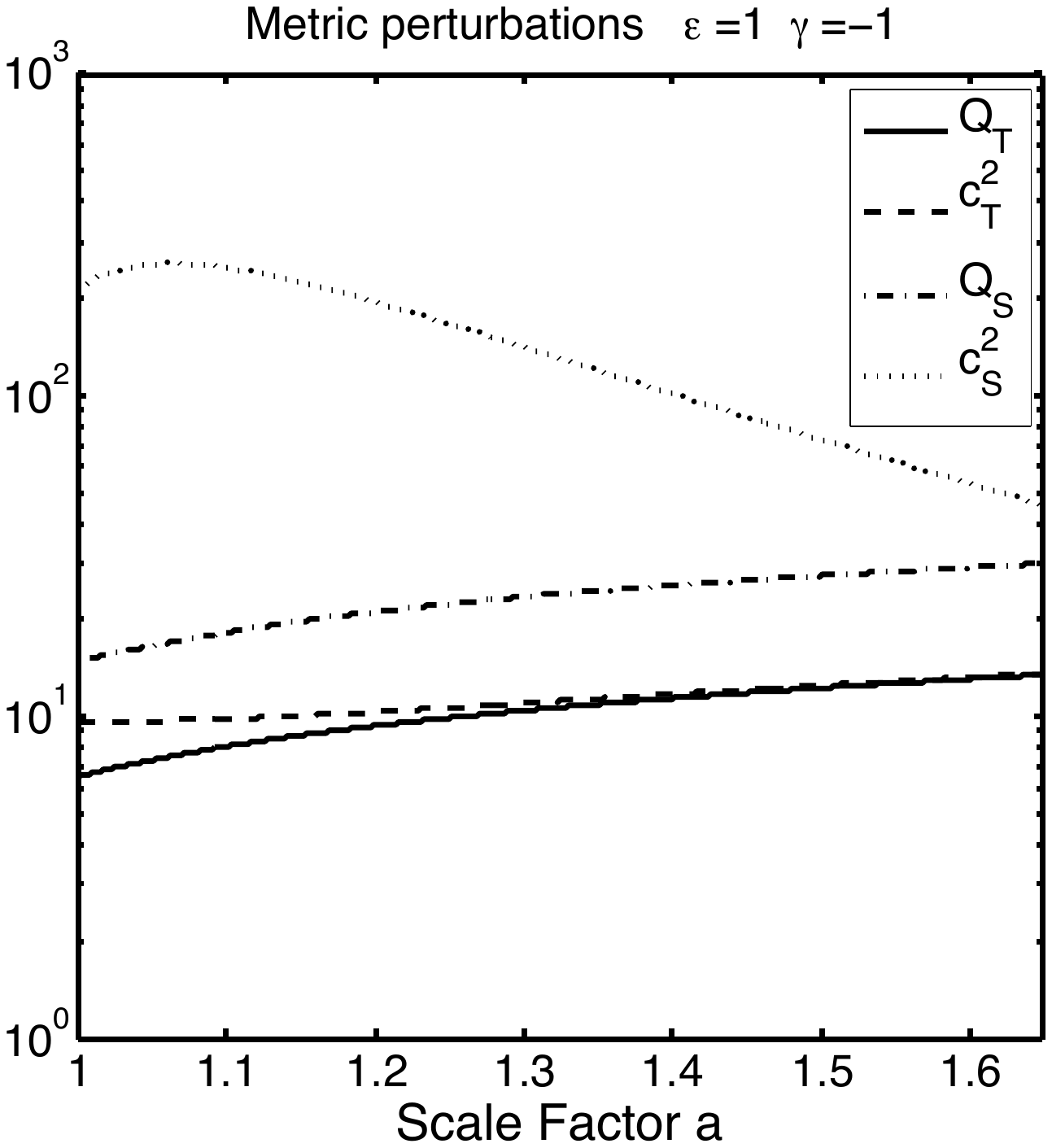} 
\end{center}      
\caption{Cosmological evolution of $q, w_\phi, Q_\rr{S}, c_\rr{S}^2, Q_\rr{T}, c_\rr{T}^2$ with the scale factor. (Top): In the case $\epsilon=-1$ and $\gamma=1$, a "double inflation" scenario is predicted, i.e. $q$ comes twice close to 1. This model is pathological in many respects: $w_\phi$ is imaginary (consequently it is not plotted), $c_\rr{T}^2 <0$ (the $y$ axis is logarithmic such that it is not represented), and there are periods for which $c_\rr{S}^2<0$ and $Q_\rr{T}<0$ (see top right). (Bottom): In the case $\epsilon=1$ and $\gamma=-1$, the model is well behaved but the expansion is not accelerated as in the John alone model. The universe is actually in a super-stiff regime, and hence, in a highly decelerating phase.}                    
\label{jgfigs}  
\end{figure*}

The case $\epsilon =-1$, $\gamma=1$ is pathological since $c^2_\rr{T}<0$ and $w_\phi$ becomes imaginary as seen on Fig.~\ref{jgfigs}. Actually this theory leads to a double inflation scenario (see the acceleration parameter): the Universe transits from one de Sitter phase to another one, and experiences in between a super-acceleration phase. 

Finally, the case with negative $\gamma$ is similar to the John alone model ($\gamma<0$): the theory is well-defined, ghost free and causal, but fails to exhibit any acceleration, see Fig.~\ref{jgfigs}.

\subsection{Compact objects}
\label{sec:compact_objF2}
\noindent
In the last two sections, we explore the phenomenology predicted by the John and George Lagrangians around compact objects and, in particular, in the Solar System. As we will see, the tests of GR in the Solar System put severe constraints on the parameter space of the model. 

In order to study the Fab Two model around compact objects, the full system of equations of motion for a static and spherically symmetric spacetime \eqref{cacaboudin1}--\eqref{cacaboudin2} reported in App.~\ref{app:sphericEOM}, is solved numerically inside and outside the compact object, using a boundary value problem. 
Inside the compact object, the TOV equation~\eqref{eq:TOV} characterizes the pressure profile. For the sake of simplicity, we assume a top-hat profile $\rho=\rho_0~\, \forall~ r<\mathcal{R}$, $\mathcal{R}$ being the radius of the compact object, and a perfect fluid \eqref{eq:perfect_fluid} inside the compact object. 
%, the TOV equation admits an analytical solution (see App.~\ref{app:num_higgs_monop}) such that the pressure profile around the compact object is determined by the solution for the metric field $\nu$. 
Three of the four Einstein equations as well as the TOV equation are integrated numerically, the fourth Einstein equation serving to validate the numerical  results\footnote{One of the Einstein equations is redundant to the others because of the Bianchi identities.}. The boundary conditions corresponding to the dynamical variables read,
\bea
  \nu(0)=0,\hspace{1cm} & \nu'(0)=0, &\hspace{1cm} \lambda(0)=0,\\
  \phi(0)=\phi_\rr{c}, \hspace{1cm} &\phi'(0)=0,& \hspace{1cm} \frac{p}{\rho_0}(\mathcal{R})=0,
\eea
the conditions at the origin being justified by the same arguments as in App.~\ref{app:num_higgs_monop}. 
%\tcb{WHY DO WE NOT HAVE TO SPECIFY THE MASS OF THE COMPACT OBJECT? - la masse de l’objet compact est une grandeur dérivée de rho0 et de la métrique, par intégration sur tout le domaine. Cela pourrait être donné en input mais alors tu devrais ajuster rho0 pour avoir la valeur donnée en input} 
The value of the scalar field at the center of the compact object $\phi_\rr{c}$ is the only remaining unknown and is determined by a shooting method, imposing that the spacetime is asymptotically flat at spatial infinity, namely $\phi\left(r\longrightarrow r_\rr{max}\right)=0$, $r_\rr{max}$ being the maximal value of the integration interval. Outside the compact object, the equations of motion are solved as an IVP, the initial conditions being given by the inner solution at $r=\mathcal{R}$.

\begin{figure}
\begin{center}
\includegraphics[width=10cm,trim= 350 0 380 0 ,clip=true]{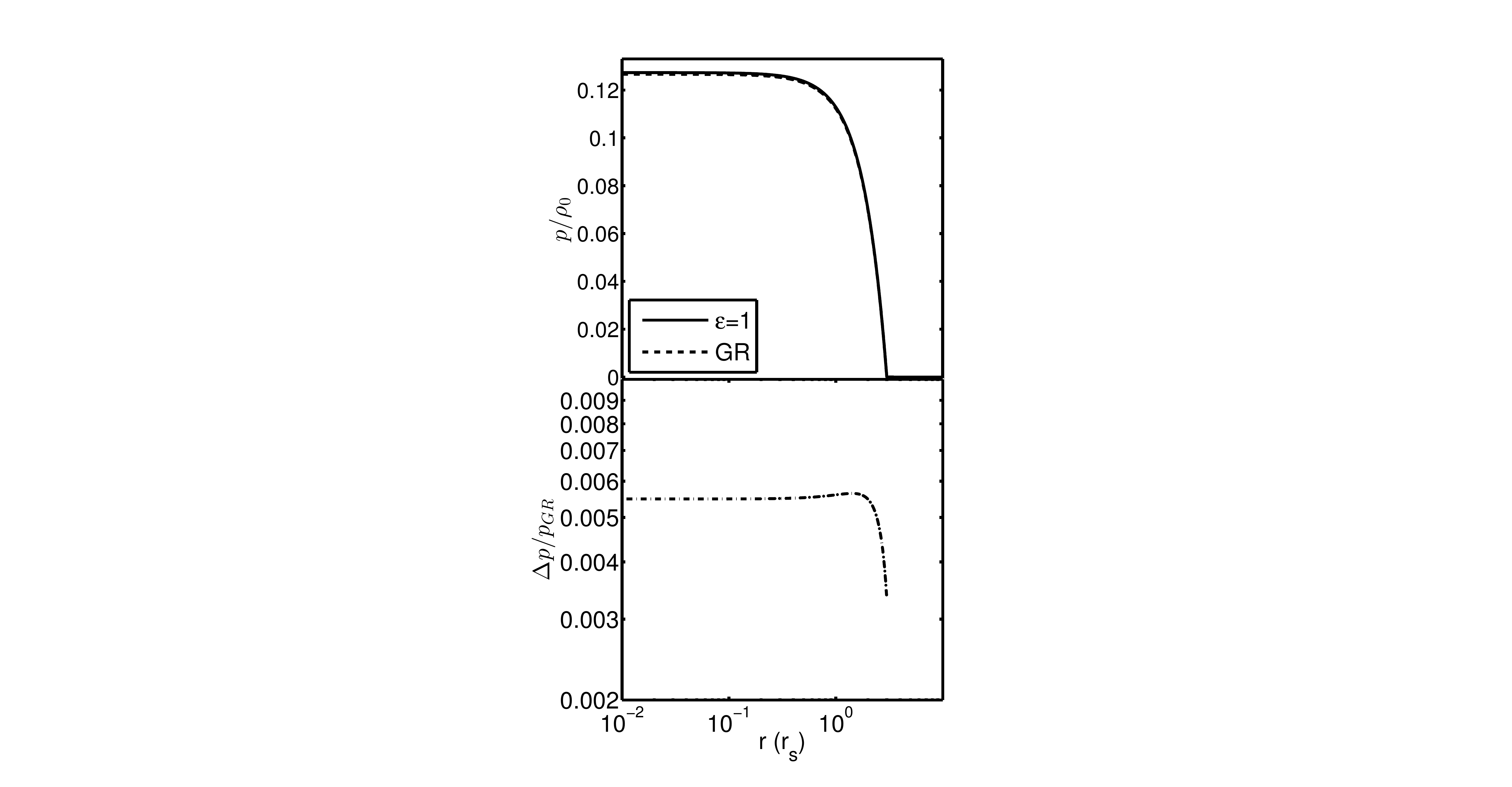}
\caption{Pressure profile (in solid line) predicted by GR and the George model ($\epsilon=1$) in a NS ($s=0.3$). The relative error at the center of the star $\left[p_\rr{c} (\epsilon=1)-p_\rr{c}(GR)\right]/p_\rr{c}(GR)$ (in dashed line) is found to be of the order of $5\times 10^{-3}$. John has negligible influence on this result.}
\centering
\label{fig:george_pressure}
\end{center}
\end{figure}

Contrary to the case where John is playing alone, if George is included, deviations from GR arise. As an example, the  pressure profile for a NS ($s=0.3$) is plotted on Fig.~\ref{fig:george_pressure} for $\epsilon=1$ and $\gamma=0$. Allowing $\gamma\neq0$ affects negligibly the solution such that the Vainshtein mechanism possibly arising in the presence of John, seems to be not that efficient in order to hide the George's effect. Depending on the compactness, the pressure at the center of the NS is expected to be larger than GR ($s=0.3$, see Fig.~\ref{fig:george_pressure}) or smaller ($s=0.5$) if George plays alone ($\epsilon=1$). The relative error $\left[p_\rr{c} (\epsilon=1)-p_\rr{c}(GR)\right]/p_\rr{c}(GR)=5\times10^{-3}$ (see Fig.~\ref{fig:george_pressure}) and $-0.08$ respectively. 

\begin{figure}
\begin{center}
\includegraphics[width=8cm,trim= 230 0 240 0 ,clip=true]{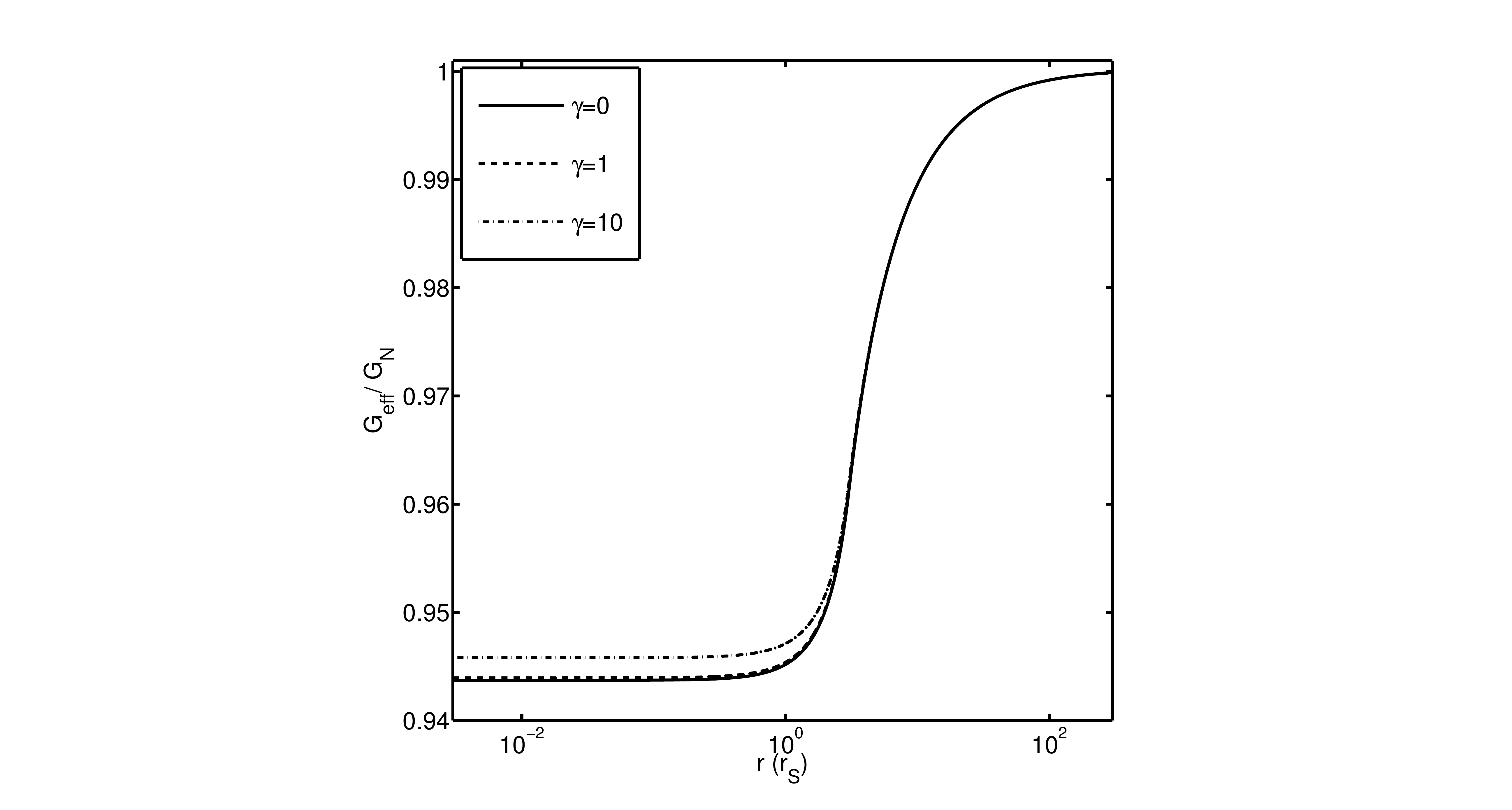}
\caption{Profile of the variation of the gravitational constant $G_\rr{eff}$ for a NS ($s=0.3$) with $\epsilon=1$ and $\gamma=0,~1,~10$. The variations at the center of the compact object are among $5\%$, a result which should be compared to the NSs physics.}
\centering
\label{fig:george_john_Geff}
\end{center}
\end{figure}

\begin{figure}
\begin{center}
\includegraphics[width=8cm,trim= 230 0 240 0 ,clip=true]{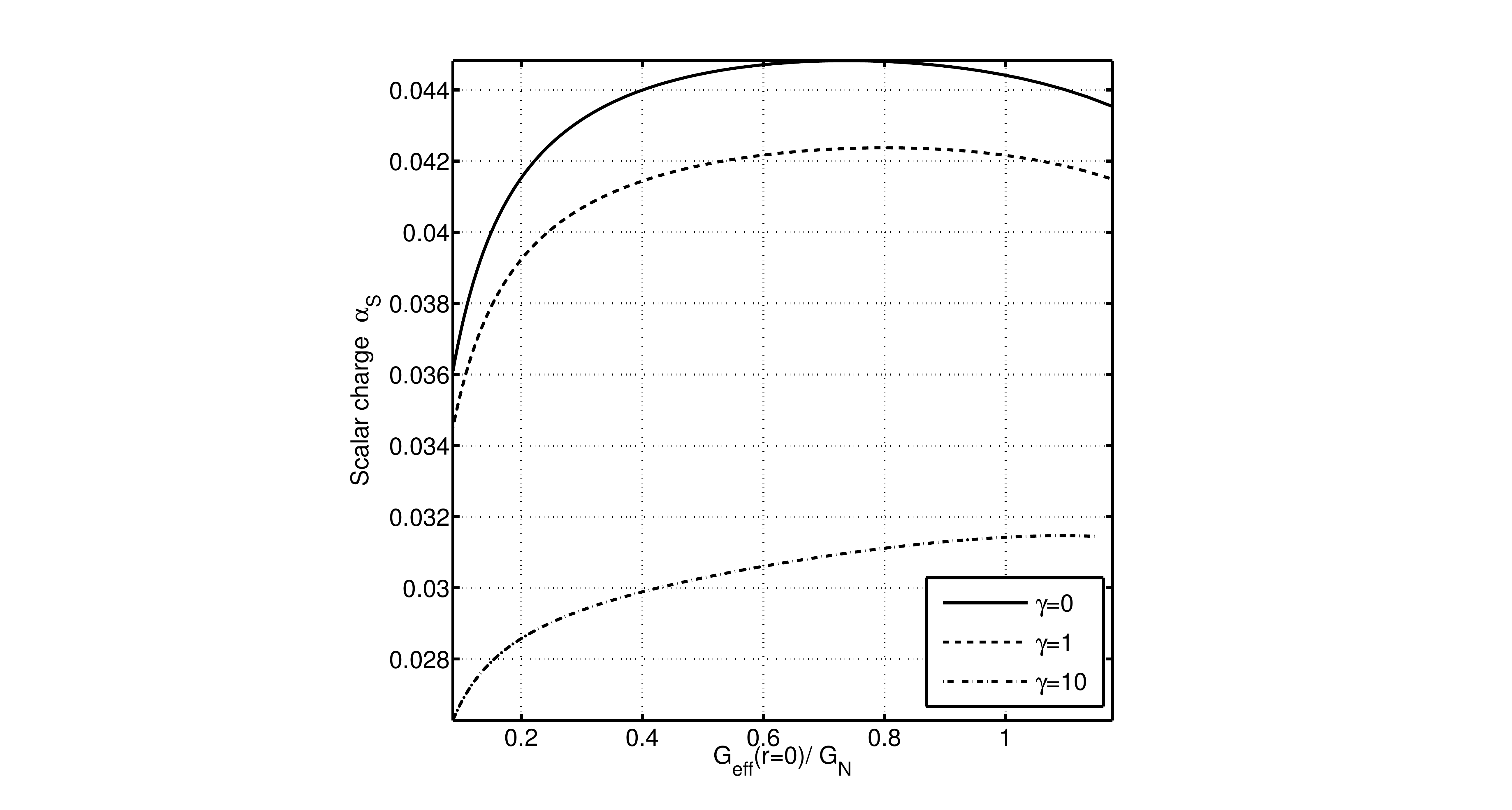}
\caption{Plot of the scalar charge  $\alpha_\rr{s}$ of the NS ($s=0.3$) as a function of $G_\rr{eff}(r=0)/\GN$ for $\epsilon=1$ and $\gamma=0,~1,~10$. The variation of the scalar charge depending on the effective gravitational coupling is not large provided that the deviation of $G_\rr{eff}$ from $\GN$ are about of few percents.}
%\tcb{est-ce qu’on a une interprétation ou un argument pour la valeur maximale de la charge scalaire en fonction de G?}
\centering
\label{fig:george_john_scalar_charge}
\end{center}
\end{figure}

A second physical quantity to be computed is the effective gravitational constant $G_\rr{eff}$ defined as, 
\be
  G_\rr{eff}=\frac{\GN}{1+\epsilon\sqrt{\kappa}\phi(r)},
\ee
which tells one to what extent the SEP is violated. Its profile for a NS ($s=0.3$) is represented on Fig.~\ref{fig:george_john_Geff} for $\epsilon=1$ and various $\gamma$. As a result, the spontaneous scalarization arises for the George model as in other STT (see Sec.~\ref{sec:spont_scala}). Imposing that $\phi=0$ at spatial infinity, variations of the gravitational constant are more than $5\%$ at the center. The larger the John coupling is, the smaller the variation, while of the same order of magnitude. The spontaneous scalarization is further modelled by the scalar charge $\alpha_\rr{s}$ \eqref{eq:scalar_charge} given by,
\be
  \phi(r)=\phi_\infty+\alpha_\rr{s}\frac{\rs}{r}.
\ee
The scalar charge is numerically determined by,
\be
  \alpha_\rr{s}=-\frac{\mathcal{R}^2 \phi'(r=\mathcal{R})}{\rs},
\ee
and is plotted on Fig.~\ref{fig:george_john_scalar_charge} for $\epsilon=1$ and $\gamma=0,~1,~10$ assuming $s=0.3$. We observe that the scalar charge does not vary significantly for values of $G_\rr{eff}$ deviating from $\GN$ by a few percents, $\gamma$ being fixed. However, $\gamma$ has a non negligible influence. Depending on the compactness, $\alpha_\rr{s}$ increases ($s=0.5$) or decreases ($s=0.3$) for increasing $\gamma$. Further analysis would reveal if those values are compatible with the current NSs observations. 

\subsection{Solar system}
\label{sec:fabtwo_sun}
The reader is reported to our paper \cite{Bruneton:2012zk} for the detailed analysis of the John and George model using Solar System observables. The results obtained are briefly exposed in this section. 

The Solar System constraints are usually derived using the isotropic coordinates, the line element of the metric reading,
\begin{equation} \label{sun:metric}
 \dd s^2= -A(r)^2 \dd t^2 + B(r)^2\left( \dd r^2 + r^2 \dd \Omega^2   \right),
\end{equation}
$A$ and $B$ being the metric fields. The equations of motion for the John and George model in the isotropic gauge are reported in App.~\ref{app:sphericEOM}.
The post-Newtonian analysis requires first to expand the metric components  $A(r)$, $B(r)$ as well as the scalar field $\varphi(r)$ in the equations of motion depending on the powers of $1/r$,
\bea \label{eq:expansPPN1}
  A^2(r)&=&1 + \sum_\rr{i} a_\rr{i}/r^\rr{i},\\
  B^2(r)&=&1+\sum_\rr{i} b_\rr{i}/r^\rr{i},\\
  \varphi(r)&=&p_0 + \sum_\rr{i} p_\rr{i}/r^\rr{i},
  \label{eq:expansPPNend}
\eea 
with $p_0=\phi_\infty$ and $p_1=\alpha_\rr{s}\rs$. 
By inserting the field expansions (\ref{eq:expansPPN1}-\ref{eq:expansPPNend}) into the equations of motion reported in App.~\ref{app:sphericEOM}, and equating the coefficients of corresponding powers of $r$, we find,
\begin{subequations}              \label{sun:metcoeff}
	\begin{eqnarray}
		A^2 & = & 1-\frac{r_\rr{s}}{r}+\frac{r_\rr{s}^2}{2r^2} +\frac{\epsilon^2 p_1^2}{4\Mp^2 z^2 r^2}+\frac{p_1\epsilon r_\rr{s}^2}{24 \Mp z r^3}\\
		&&-\frac{p_1^2 r_\rr{s}}{24\Mp^2z r^3} + \frac{3}{4}\frac{\bar\gamma\epsilon^2}{\Mp^4z^2r^4}-\frac{r_\rr{s}\bar \gamma }{8\Mp^4zr^5},\non \\
		B^2&=&1+\frac{r_\rr{s}}{r} - 2\frac{\epsilon p_1}{\Mp z r} -\frac{p_1^2 }{4\Mp^2zr^2}-\frac{\bar \gamma}{4\Mp^4zr^4},
	\end{eqnarray}
\end{subequations}
where $z=1+\epsilon p_0/\Mp$, and $\bar \gamma=\gamma p_1^2$. In the expansion above, we neglected  higher order terms in $r_\rr{s}/r$, $\epsilon$, $p_1/(r \Mp)$ and $\bg/(r \Mp)$ (which means we suppose these terms to be smaller than 1). We recall that, in our conventions,  $\gamma$ and $\epsilon$ are dimensionless parameters. The asymptotic scalar field value  $p_0=\phi(r\to \infty)$ (in GeV) is a free parameter that can eventually be connected to the cosmological evolution of the scalar field since there is no additional scale in the theory in the absence of a potential term for the scalar field. Note that the dimensionless parameter $p_1$ can be related to the scalar charge of the central body derived numerically in Sec.~\ref{sec:compact_objF2}. 

In principle, we could identify the PPN parameters from the metric expansion \eqref{eq:metric_PPN},
\be
  \gamma_\rr{PPN}=-2\frac{\epsilon p_1}{z \Mp r_\rr{s}},\hspace{2cm}
  \beta_\rr{PPN}=\frac{\epsilon^2 p_1^2}{2 z^2 \Mp^2 r_\rr{s}^2}.
\ee
However, the terms $1/r^3$, $1/r^4$ and $1/r^5$ can be larger than the ones $1/r$, $1/r^2$ depending on the nonminimal coupling parameters $\epsilon$ and $\gamma$. As a result, the PPN expansion is not relevant in order to test the Fab Four in the Solar System, which rather requires other tools for computing the observable effects from the metric. In our paper \cite{Bruneton:2012zk}, the anomalous perihelion shifts of the planets and radioscience observables, i.e. using the propagation of light rays in the Solar system \cite{Hees:2012nb} were computed. The four parameters of the John and George model which must be constrained, are ${\bg}/{z}$, ${\bg\epsilon^2}/{z^2}$, ${p_1^2}/{z}$ and ${p_1 \epsilon}/{ z}$. 

The secular perihelion precession rates were computed for the planets of the Solar System (excepted Uranus and Neptune) in \cite{Bruneton:2012zk}. The most stringent constraints for the John and George model are obtained by the data from Mercury and read,
\begin{subequations}   \label{const_per}
\begin{eqnarray}
-3.12 \times 10^{31} \ \rr{m^{4}}   & < \frac{\bg}{\Mp^4z} <     & 6.25 \times 10^{30} \  \rr{m^{4}}\vspace*{5mm}\label{constgamma}, \\
-2.06 \times 10^{23} \ \rr{m^{4}}   & < \frac{\bg \epsilon^2}{\Mp^4z^2} <  & 4.12 \times 10^{22} \ \rr{m^{4}}\vspace*{5mm}, \\ 
-1.13 \times 10^{10} \ \rr{m^{2}}  & < \frac{p_1^2}{\Mp^2 z} <                                & 2.26 \times 10^{9}  \ \rr{m^{2}}\vspace*{5mm}, \\ 
-5.16 \times 10^{-2} \  \rr{m}       & < \frac{p_1\epsilon}{z\Mp} <                        & 1.03 \times 10^{-2} \ \rr{m}.
\end{eqnarray}     
\end{subequations}

The constraints on the John and George Lagrangians were also computed using radioscience simulations (see \cite{Bruneton:2012zk} for technical details). The Doppler effect, i.e. the ratio of frequencies between the emitted and the received signals, perturbing the propagation of light between the Earth and the Cassini spacecraft in orbit around Saturn has been measured with very good accuracy \cite{Cassini}. Requiring the residuals, i.e. deviation between the John and George predictions and the observations, to be lower than the Cassini accuracy yields,
\begin{subequations}
\begin{eqnarray}
  \left| \frac{\bg}{\Mp^4z} \right|= \left|\frac{\gamma p_1^2}{\Mp^4(1+\frac{\epsilon p_0}{\Mp})} \right|   & <  & 3.65 \times 10^{26} \ \rr{m^4}, \\ 
\left| \frac{\bg \epsilon^2}{\Mp^4z^2} \right|= \left|\frac{\gamma p_1^2 \epsilon^2}{\Mp^4(1+\frac{\epsilon p_0}{\Mp})^2} \right| & < &  1.15 \times 10^{26} \ \rr{m^4}, \\
\left| \frac{p_1^2}{\Mp^2 z} \right|= \left|\frac{p_1^2}{\Mp^2(1+\frac{\epsilon p_0}{\Mp})} \right|   &<& 3.53 \times 10^{8} \ \rr{m^2}, \\
\left|\frac{p_1\epsilon}{z\Mp} \right|= \left|\frac{p_1\epsilon}{\Mp(1+\frac{\epsilon p_0}{\Mp})} \right| &<& 5.56 \times 10^{-2} \ \rr{m}.
\end{eqnarray}	
\end{subequations}

It should be noted that radioscience constraints are significantly better for ${\bg}/{(\Mp^4z)}$ and ${p_1^2}/{(\Mp^2 z)}$ compared to the Mercury perihelion advance ones. On the other hand, the constraint from the Mercury perihelion advance on ${\bg \epsilon^2}/{(\Mp^4z^2)}$ is significantly better than the radioscience one, the constraint on ${p_1\epsilon}/{(z\Mp)}$ being of the same order of magnitude.

\section{Conclusions}

\noindent In this chapter, we explored the phenomenology associated to a subset of the Fab Four Lagrangians in cosmology, around compact objects and in the Solar System. The philosophy behind this preliminary work was that we cannot forget about Solar System constraints on the parameter space, even when we deal with inflationary solutions. In general, inflationary models rely upon the fact that the inflaton field decays, at some stage, into ordinary matter through some reheating mechanism. Therefore, the scalar-tensor nature of inflationary gravity is lost very soon in the evolution of the Universe. On the opposite, in the John and George models, the scalar field should live and show its effects until nowadays. Therefore, the parameter space determined by the constraints from cosmological observations must overlap with the one determined by Solar system tests. The Fab Four theory has many parameters with a very rich phenomenology and the entire parameter space must still be further studied.
In this work, we restricted ourselves to the cases John and John plus George. 

When John, i.e. a theory with a nonminimal derivative coupling between the scalar field and the Einstein tensor  whose strength is parametrized by $\gamma$, plays inflation and gravitation:
\begin{itemize}
 \item It was already known that the John Lagrangian admits inflationary solutions with a graceful exit without any ad hoc potential \cite{Sushkov:2009hk}.
 \item If the nonminimal derivative coupling parameter $\gamma$ is positive, John drives an inflationary phase (the acceleration parameter is positive $q>0$). However, very unnatural initial conditions are required. In particular, the field velocity, which is related to the energy density, must be huge compared to the Planck scale, rendering the theory no longer trustworthy. Moreover, the analysis of the scalar field and metric perturbations reveals that $\kappa\dot{\phi}_\rr{i}\lesssim 1/\sqrt{\gamma}$ in order to preserve causality such that $\gamma$ must be very small.
 \item Negative values for $\gamma$ are permitted in the sense that the scalar field and metric perturbations preserve causality. However, this model, while admitting a solution with accelerated expansion, does not allow for inflation since the Hubble parameter would become imaginary. It results that the EoS for the scalar field must be positive.
 \item Finally, the most serious problem comes from the fact that the model turns out to be trivial when one tries to solve the equations of motion inside a compact object. Indeed, we found that the only solution allowed by the regularity conditions is  $\varphi=0$ everywhere.
\end{itemize}

These facts have convinced us to extend the theory to include the term named George, which is nothing but a coupling between the scalar field and the Ricci scalar whose strength is parametrized by $\epsilon$. When George and John are playing together:
\begin{itemize}
 \item The numerical solutions of the equations of motion in cosmology (in the absence of stress-energy-momentum sources) highlight that the sign of the two coupling constants $\gamma$ and $\epsilon$ must be positive in order to have an inflationary phase with graceful exit. However, a non-causal behavior of metric perturbations is expected in the very early Universe. The analysis of the naturalness of the initial conditions for inflation, i.e. the need for extreme initial conditions, is still an open question. 
 \item By solving the Einstein equations for a static and spherically symmetric spacetime, we show that there are non-trivial solutions inside compact objects, like the Sun and NSs, provided that $\epsilon\neq 0$. The John's effect tends to lower the deviation from GR induced by George. However, the John's effect is negligible with respect to the George's one. The efficiency of the Vainshtein mechanism is thus questionable and requires a careful analysis of the complete parameter space.  
 \item The variation of the gravitational coupling $G_\rr{eff}/\GN$ due to George is up to $5\%$ at the center of the NSs, the effect of the John being negligible. 
 \item The George model predicts that NSs are spontaneously scalarized whereas the scalar charge only slightly depends on the deviation of $G_\rr{eff}$ from $\GN$ at the center of the compact objects even for deviations of a few percents. Spontaneous scalarization is decreasing for increasing values of $\gamma$.
 \item In order to provide combined constraints with the tests of GR in the Solar System, a PN analysis of the theory was performed. It required to solve the Einstein equations in a static and spherically symmetric spacetime order by order by expanding the fields to large radial distance. The results still allow for a large parameter space, therefore future work is necessary in order to improve the constraints.
\end{itemize}

Systematic study of the Fab Four phenomenology is still to be investigated, in order to isolate which part of the parameter space is viable from the theoretical and phenomenological point of view. 
%\tcb{un commentaire sur les termes les plus importants à regarder serait le bienvenu. En particulier, je m’attends à ce que certains termes soient triviaux en cosmo par exemple (notamment à cause des opérateurs duaux).} 
Besides these aspects, there are several issues that deserve further analysis:
\begin{itemize}
 \item If the Fab Four model truly leads to inflation with a graceful exit, an alternative to the reheating mechanism is needed since the scalar field is not expected to decay in the early Universe.
 \item The question arises if the Fab Four can lead to late-time acceleration. This would include, at the background level, the study of tracking solutions with a convergence mechanism towards GR, if any (see e.g. \cite{Copeland:2012qf}). The study of cosmological perturbations, in particular CMB spectra and LSS, might then further reduce the parameter space (see \cite{DeFelice:2011bh, Barreira:2013eea} and references therein). 
 \item Astrophysical constraints for Horndeski gravity have been derived in the last few years. While no-hair theorems put severe constraints on the existence of BH solutions, there exists an exact static BH solution considering John provided that the action reads,
 \be \label{eq:john_modif}
  S=\int \dd^4 x \,\sqrt{-g} \left[\frac{R}{2 \kappa} - \frac{\kappa \gamma}{2} G^{\mu\nu} \partial_{\mu}\phi\partial_{\nu}\phi\right],
  %+ S_{\textrm{M}}[\psi_{\textrm{M}};~g_{\mu\nu}],
 \ee
 in the absence of the cosmological constant \cite{Rinaldi:2012vy}. This result was further generalized \cite{Babichev:2013cya}.
 \item Finally, the existence of NSs has been verified in the static and slowly rotating regime (see \cite{Cisterna:2016vdx, Maselli:2016gxk} and references therein). John appears to be the most interesting term (Paul does not give rise to viable stars \cite{Maselli:2016gxk}), the maximal NS mass $M_\rr{max}$ predicted by John being generally smaller than in GR. The limit of $M_\rr{max}=2\,M_\odot$ (see Sec.~\ref{sec:NS}) is reachable for specific EoS only.  
 Provided that the action for gravitation is given by Eq.~\eqref{eq:john_modif} (in the absence of the cosmological constant), cosmological and astrophysical configurations are found to be consistent with each other \cite{Cisterna:2016vdx}.
\end{itemize}

\part{Conclusion}

\renewcommand{\chaptermark}[1]{\markboth{\small\textsc{Chapter \thechapter.\ #1}}{}}
\cleardoublepage

\chapter*{Conclusion} % Main chapter title
\markboth{\textsc{Conclusion}}{\textsc{Conclusion}}

\label{CCL} % For referencing the chapter elsewhere, use \ref{Chapter1} 

\lhead{Chapter 1. \emph{Awesome chapter 1}} % This is for the header on each page - perhaps a shortened title

%----------------------------------------------------------------------------------------
GR has opened the way to precision cosmology. According to Einstein’s theory of gravitation, the cosmological observations today converge towards the $\Lambda-$CDM concordance picture. However, cosmology requires understanding the nature of the  matter-energy sources in the Universe. Within the $\Lambda-$CDM concordance picture, only baryonic matter and radiation which constitute $5\%$ of the Universe’s content, are described by the SM. While the question of the nature of DM has shifted to (astro-)particle physics today, the nature of the late-time cosmic acceleration is still debated. It could reveal the existence of new fields like DE, or modifications of GR at large scales.

The $\Lambda-$CDM concordance picture also suffers from the fine-tuning problem of the initial conditions in the early Universe. The current paradigm assumes an inflationary phase. The simplest models relying on the assumption of a single scalar field responsible for the huge accelerated expansion, are still favored today by the latest CMB observations. Again, the nature of this scalar field is still unknown.

In this thesis the assumption stating that late-time cosmic acceleration and inflation come from modifications of gravity, was investigated. In particular, we considered models where the Einstein metric field has a scalar field counterpart. However, as developed in Chap.~\ref{chap:math_fundations}, GR has a privileged status. Indeed, if the Schlögel-Füzfa conjecture applies, GR is the only theory of gravity in four spacetime dimensions which satisfies the SEP, i.e. the existence of a gravitational field cannot be detected locally whatever observations or experiments are undertaken (see Sec.~\ref{sec:chap1ccl}).

Modifications of gravity are thus challenging, first from the theoretical point of view: modifications of gravity must be well-defined like GR. Further, modified gravity models have to be confronted with the observations: in cosmology, if those models are dedicated to late-time cosmic acceleration or inflation; with local tests of gravity, in the Solar System or in labs; and in astrophysics, looking at compact objects (black holes, gravitational waves and neutron stars). In Chap.~\ref{chap:test_GR}, the tests of GR are reviewed and classified depending on the regime the tests investigate (strong or weak field, in the presence of sources or not, the sources being relativistic or not).

In the rest of this thesis we focused on three different modified gravity models: the chameleon model, the Higgs gravity and the Fab Four. Those models appear to be well-posed, even if the chameleon model possibly suffers from strong coupling (see Sec.~\ref{sec:MG_issues}) and the Higgs gravity could suffer from a loss of unitarity (see Sec.~\ref{sec:high_energy_higgs}). The predictions of these three models were studied at different scales: in the lab, in compact objects and in cosmology. For these cases, the equations of motion were solved numerically.

\section*{Laboratory experiment: the chameleon model}
In Chap.~\ref{chap:chameleon}, we focused on the chameleon model which has been extensively studied in the last decade. Initially, this model was built in order to reproduce the late-time cosmic acceleration. This model exhibits a screening mechanism due to the combined effect of the potential and the nonminimal coupling to gravity. It can pass the Solar System constraints provided that the potential is exponential and the value of the potential parameter $\Lambda$ is of the same order of magnitude as the cosmological constant.

Recently, this model was tested with an unprecedented accuracy in the laboratory, using an atom interferometry experiment inside a vacuum chamber. Analytical forecasts were derived by \cite{Burrage, khoury}, assuming negligible chamber wall effects. The first experimental bounds were obtained in Berkeley \cite{khoury, Elder:2016yxm}. In this thesis we provide numerical simulations for the Berkeley experiments. They lead to the following results:
\begin{itemize}
 \item The numerical method we developed in this thesis allowed us to take the minimal assumption that the chameleon field settled to the minimum of its effective potential far away from the vacuum chamber. Moreover, the effects of the experimental set-up were taken into account using the limit of spherical symmetry, contrary to analytical calculations. 
 \item In the strongly perturbing regime, the numerical method enabled validation and refinement of the analytical calculations close to the test mass where the acceleration induced by the chameleon field is measured. In addition, we highlighted that the acceleration becomes negative close to the wall, this effect being of the same order of magnitude as close to the test mass and thus possibly measurable. 
 \item The effect of the size and the density of the test mass were analyzed. We found that a larger test mass gives rise to a larger induced acceleration. Moreover, the experimentalists in Berkeley now use a test mass made of tungsten rather than aluminum. 
 \item The numerical method confirmed that the chameleon model would be ruled out up to the Planck scale if the induced acceleration is measured with about 3 orders of magnitude more sensitivity. This limit should be reachable in the near future. Indeed, the control of systematics will allow the experimentalists in Berkeley to improve the sensitivity of their experimental set-up. We also provided a forecast for probing the chameleon in a very large vacuum chamber, i.e. vacuum room of 10 m radius, where the chameleon acceleration is found to be (almost) measurable for values of the nonminimal coupling parameter up to the Planck scale with the current experimental sensitivity. 
 \item The numerical method developed in this thesis is easily adaptable to various experiments, in the limit of spherical symmetry, the general modeling of the experiment requires other numerical methods like relaxation. 
 \item Eventually, the experimental set-up developed in Berkeley should also offer the opportunity to test other screening mechanisms like the symmetron, the dilaton and $f(R)$. 
\end{itemize}

Such an experimental set-up reveals that stringent bounds on modified gravity can be obtained in the laboratory, at least for some particular models exhibiting screening mechanisms.

\section*{Compact objects: the Higgs gravity}
The Higgs inflation appears to be a promising model today since it is still favored by the latest observations, among them the Planck satellite, provided that there is nonminimal coupling $\xi>10^4$ \cite{Ade:2015lrj}. 
The Higgs inflation predicts that the tensor-to-scalar ratio is very small $r\simeq0.0033$ such that as long as $r$ is not detected, for instance by the future space mission COrE+, this model will be favored.

In Chap.~\ref{chap:Higgs}, the numerical solutions for the same STT were derived around compact objects, in the presence of baryonic matter. Indeed, no-hair theorems guarantee the Schwarzschild solution in the vacuum, i.e. for black holes. The solution that we derived numerically has the following characteristics:
\begin{itemize}
  \item The distribution of the Higgs field around compact objects is particlelike. The Higgs field converges to the vev at spatial infinity. Its distribution is globally regular, i.e. there is no singularity, it is asymptotically flat (the spacetime is Minkowski at spatial infinity, the Higgs field being settled to its vev) and of finite energy. The Higgs field distribution is characterized by the nonminimal coupling as well as the baryonic mass and the compactness of the object. 
  \item Contrary to the case of spontaneous scalarization (see Sec.~\ref{sec:spont_scala}) there is only one solution which always differs from that of GR, i.e. when the scalar field is minimally coupled to gravity. Indeed, in GR, there exist only unrealistic homogeneous distributions of the Higgs field around compact objects while Higgs inflation predicts non-trivial distributions of the Higgs field. Whatever the compact objects, the vev will be shifted to the center of the objects. 
  \item However, no measurable deviations were found for astrophysical objects. At the center of the Sun, the variation of the Higgs field around its vev is less than $1\%$ provided that the nonminimal coupling is $\xi<10^{58}$, thus, far above the critical value for the inflation $\xi>10^4$. Inside neutron stars, the variation of the Higgs field cannot be higher than $10^{-41}\times$ the vev, assuming $\xi=10^4$. We concluded that this effect was not measurable, either with gravitational or nuclear physics experiments. 
  \item Considering nonphysical values of the compactness and the mass, we highlighted the existence of a mechanism of amplification of the central value of the Higgs field in compact objects. This means that there exists a critical nonminimal coupling above which some compactnesses – or radii - of compact objects are forbidden, since the central value of the Higgs field diverges. This is due to the combined effects of the Higgs potential and the nonminimal coupling function. 
  \item This amplification mechanism could possibly be generalized to other STT. 
  \item The fact that divergence from the central values of the Higgs field appeared, is related to the assumption of the unitary gauge. Indeed, in Higgs inflation, they consider one real scalar field only rather than the Higgs doublet, the positive and negative asymptotic values of the vev no longer being equivalent. In a realistic model, the $\rr{SU}(2)$ gauge symmetry of the electroweak interaction should be included. It shows that the Higgs inflation model should rather be considered as a multifield model due to the presence of the Goldstone bosons \cite{Greenwood:2012aj}. In that case, the Yukawa coupling between the Higgs field and elementary particles could be also considered in-side compact objects. 
  \item Lastly, the question of the stability of this particlelike solution has not been investigated yet. The only argument in favor of this solution is that it is the only one with a finite energy. Indeed, if the scalar field is not settled to its vev at spatial infinity, the potential energy is non-vanishing and becomes infinite at large distances. Also, the question of the formation of the Higgs monopoles as a result of gravitational collapse is still open. 
\end{itemize}

STT provide a framework to model the nonminimal coupling of the Higgs field to gravity. The realistic modeling of the nonminimally coupled Higgs field around compact objects, that is in agreement with the $\rr{SU}(2)$ gauge of the electroweak interactions, requires further investigation. It has already been studied for dark energy \cite{Rinaldi:2015iza}. However, in the presence of baryonic matter, the Yukawa coupling of the Higgs field to elementary particles should also be taken into account.

\section*{From inflation to compact objects: the Fab Four model}
STT consist of a subclass of Horndeski gravity, i.e. the theory of gravity invoking a scalar field counterpart to the metric and leading to second order equations of motion.

In Chap.~\ref{chap:FabFour}, we focused on the Fab Four model which belongs to the Horndeski gravity theory while not being a STT. More precisely we studied the phenomenology predicted by two of the four: John, a nonminimal derivative coupling between the scalar field and the Einstein tensor, and George, a nonminimal non-derivative coupling between the scalar field and the Ricci scalar. In particular, this model does not exhibit any potential term in the Lagrangian.

In this thesis, we explored the predictions of this model for inflation, in the Solar System and for compact objects:
\begin{itemize}
  \item In the case where John plays alone, it can succeed in reproducing inflation provided that the non-minimal coupling parameter is positive. The number of e-folds generated during inflation is sufficient for solving the horizon and the flatness problems and this model predicts a graceful exit. However, the initial conditions, in particular the kinetic energy of the scalar field, were found to be super-Planckian. 
  \item In compact objects, John only predicts trivial solutions: imposing regularity conditions at the center of the compact objects, the scalar field vanishes everywhere. 
  \item In order to obtain a richer phenomenology, we included the George term. In that case, inflation exhibits a graceful exit provided that both nonminimal coupling parameters are positive. Considering that the Solar System constraints are obtained by radio-simulations, a large part of the parameter space appears to still be viable. When John and George play together, spontaneous scalarization arises around compact objects. 
\end{itemize}
The work presented in Chap.~\ref{chap:FabFour} is prospective in the sense that a careful analysis of the parameter space of two of the Fab Four is missing, either in cosmology, in the Solar System or around compact objects. The analysis of two of the Fab Four around NSs has been presented \cite{Cisterna:2016vdx, Maselli:2016gxk}. It has been found that John can predict the existence of NSs of the maximal mass observed today, i.e. $2~M_\odot$, assuming a realistic equation of state. Finally, we only considered two of the Fab Four Lagrangians, the phenomenology predicted by Paul and Ringo being still unexplored.

\phantomsection
\addcontentsline{toc}{chapter}{Conclusion}

% \clearpage
% \newpage
% \phantomsection
% \addcontentsline{toc}{chapter}{List of Tables}
% \listoftables	
% \newpage
% \addcontentsline{toc}{chapter}{List of Figures}
% \listoffigures

\part*{Appendices}
\addcontentsline{toc}{chapter}{Appendices} 
\appendix
\renewcommand{\chaptermark}[1]{\markboth{\small\textsc{Appendix \thechapter.\ #1}}{}}

\chapter[General covariance: a variational approach]{General covariance: \\ a variational approach} % Main chapter title

\label{app:gen_cova_math} % For referencing the chapter elsewhere, use \ref{Chapter1} 

\lhead{Appendix B. \emph{Awesome appendix B}} % This is for the header on each page - perhaps a shortened title

In this appendix, the implications of active diffeomorphism invariance on the EH action are analyzed. As we will see, it results that the second Bianchi identity holds in GR assuming the Levi-Civita connection.

Infinitesimal active diffeomorphisms $\phi_{\Delta\lambda}$, with $\Delta \lambda$ the infinitesimal shift along the vector field $\xi$ from the point $P=\left\{x^\mu\right\}$ to the point $\phi_{\Delta\lambda}(P)=\left\{y^\mu\right\}$ where $y^\mu=x^\mu+\xi^\mu \Delta\lambda$, are generated by the \textbf{Lie derivative} $\mathcal{L}_\mathbf{\xi}$. Applied to a tensor field $\mathbf{T}$, this linear operator is defined as,
\be
  \mathcal{L}_\mathbf{\xi}\mathbf{T}\equiv \lim_{\Delta \lambda\longrightarrow 0}\frac{\phi_{\Delta\lambda} \mathbf{T}(x)-\mathbf{T}(x)}{\Delta\lambda}.
\ee
%Lie derivative tells us how a tensor field $T$ changes when it is pushed forward along the integral curve $\xi$, $x^\mu\longrightarrow x^\mu+\xi^\mu d\lambda$. 
%In GR, the only tensor field involved in the vacuum is $g_{\mu\nu}$ and its Lie derivative reads (up to first order),
The only tensor field appearing in the EH action is the metric $g_{\mu\nu}$. Its Lie derivative reads (up to first order),
\bea
\mathcal{L}_\mathbf{\xi} g_{\mu\nu}~ \Delta \lambda&=&\bar{g}_{\mu\nu}(x)-g_{\mu\nu}(x),\non\\
&=&g_{\alpha\beta}(x+\mathbf{\xi} \Delta\lambda)\frac{\partial x^{\alpha}}{\partial \overline{x}^\mu}\frac{\partial x^{\beta}}{\partial \overline{x}^\nu}
-g_{\mu\nu}(x),\non\\
&\simeq&\left(g_{\alpha\beta}+\df_\rho g_{\alpha\beta}~ \xi^\rho \Delta\lambda\right)
(\delta^\alpha_\mu+\df_\mu \xi^\alpha \Delta\lambda)
(\delta^\beta_\nu+\df_\nu \xi^\beta \Delta\lambda)-g_{\mu\nu},\non\\
%+\mathcal{O}\left(\Delta\lambda\right)^2
&\simeq& \left[\xi^\alpha \partial_\alpha g_{\mu\nu}(x)+g_{\alpha\nu} (x) \partial_\mu \xi^\alpha+g_{\mu\alpha} (x) \partial_\nu \xi^\alpha\right] \Delta\lambda,\non\\ %+\mathcal{O}\left(\Delta\lambda\right)^2,\non\\
\mathcal{L}_\mathbf{\xi} g_{\mu\nu}
&\simeq&\xi^\alpha \nabla_\alpha g_{\mu\nu}+g_{\alpha\nu}\nabla_\mu \xi^\alpha
+g_{\mu\alpha}\nabla_\nu \xi^\alpha + T^\alpha_{\mu\beta} g_{\alpha\nu} \xi^\beta
+ T^\alpha_{\nu\beta} g_{\mu\alpha} \xi^\beta, \non 
%+\mathcal{O}\left(d\lambda\right)^2,
\label{eq:lie_metric}
\eea
where $\bar{g}_{\mu\nu}(x)=\phi_{\Delta\lambda}{g}_{\mu\nu} (x)$ is the metric obtained by applying the infinitesimal active diffeomorphism to $g_{\mu\nu} (x)$, and $T^\alpha_{\nu\beta}$ is the torsion defined by Eq.~\eqref{eq:torsion}. We also used, 
\bea
  \xi^\alpha \nabla_\alpha g_{\mu\nu}&=&
  \xi^\alpha \left(\df_\alpha g_{\mu\nu}-\Gamma^\rho_{\alpha\mu}g_{\rho\nu}
  -\Gamma^\rho_{\alpha\nu}g_{\mu\rho}\right),\\
  g_{\alpha\nu}\nabla_\mu\xi^\alpha&=&g_{\rho\nu}
  \left(\df_\mu\xi^\rho+\Gamma^\rho_{\mu\alpha}\xi^{\alpha}\right).
\eea
Assuming the Levi-Civita connection, the Lie derivative of the metric tensor reads\footnote{This equation reduces to the Killing equation when $\mathcal{L}_\mathbf{\xi} g_{\mu\nu}$ is required to vanish, defining the Killing vectors is the direction of spacetime isometries.},
\be
   \mathcal{L}_\mathbf{\xi} g_{\mu\nu}=\nabla_\mu \xi_\nu + \nabla_\nu \xi_\mu.
\ee

The variation of the EH action (in the absence of the cosmological constant) under an infinitesimal active diffeomorphism, i.e. with $\delta g^{\mu\nu}=\mathcal{L}_\mathbf{\xi} g^{\mu\nu}$ yields,
\bea
  \delta S_\rr{EH}&=&\frac{1}{2\kappa}\int \dd^4x 
	\left(R~ \delta\sqrt{-g}+\delta g^{\mu\nu}\sqrt{-g}~ R_{\mu\nu}
	      +\sqrt{-g}~ g^{\mu\nu}~\delta R_{\mu\nu}\right),\non\\
	&=&\frac{1}{2\kappa}\int \dd^4x \sqrt{-g}
	    \left(G_{\mu\nu}~\delta g^{\mu\nu} + g^{\mu\nu}~ \delta R_{\mu\nu}\right),
	%&=&\frac{1}{2\kappa}\int d^4x\sqrt{-g} G_{\mu\nu}\delta g^{\mu\nu} 
	\label{eq:EH_diffeo}
\eea
using Eq.~\eqref{eq:variation_det-metric}.
The last term of this equation is vanishing. Indeed, the variation of the Ricci tensor reads (see e.g. \cite{hobson2006general}),
\be
  \delta R_{\mu\nu} =\nabla_{\rho}\delta\Gamma_{\nu\mu}^{\rho}-\nabla_{\nu}\delta\Gamma_{\rho\mu}^{\rho},
\ee
such that,
\begin{eqnarray*}
\int \dd^4x \sqrt{-g}\,g^{\mu\nu}\, \delta R_{\mu\nu}
& = & \int\dd^{4}x\,\sqrt{-g}\, g^{\mu\nu}\,\left( \nabla_{\rho}\delta\Gamma_{\nu\mu}^{\rho}-\nabla_{\nu}\delta\Gamma_{\rho\mu}^{\rho}\right), \\
 & = & \int \dd^{4}x\,\sqrt{-g}\,\left[ \nabla_{\rho}\left(g^{\mu\nu}\,\delta\Gamma_{\nu\mu}^{\rho}\right)-\left(\nabla_{\rho}g^{\mu\nu}\right)\;\delta\Gamma_{\nu\mu}^{\rho}\right] \\
 &  & \hspace{0.5cm}-\int \dd^{4}x\,\sqrt{-g}\,\left[ \nabla_{\nu}\left(g^{\mu\nu}\,\delta\Gamma_{\rho\mu}^{\rho}\right)-\left(\nabla_{\nu}g^{\mu\nu}\right)\;\delta\Gamma_{\rho\mu}^{\rho}\right], \\
 & = & \int \dd^{4}x\,\sqrt{-g}\,\left[ \nabla_{\rho}\left(g^{\mu\nu}\,\delta\Gamma_{\nu\mu}^{\rho}\right)-\nabla_{\nu}\left(g^{\mu\nu}\,\delta\Gamma_{\rho\mu}^{\rho}\right)\right], \\
 & = & \int \dd^{4}x\,\sqrt{-g}\,\nabla_{\nu}\left(g^{\mu\rho}\,\delta\Gamma_{\rho\mu}^{\nu}-g^{\mu\nu}\,\delta\Gamma_{\rho\mu}^{\rho}\right), 
\end{eqnarray*}
assuming the Levi-Civita connection.
% \bea
%    g^{\mu\nu} \delta R_{\mu\nu}=\nabla_\mu v^\mu, \hspace{0.7cm}\text{with}\hspace{0.7cm}
%    v^\mu\equiv \nabla_\nu \left(-\delta g^{\mu\nu}+g^{\mu\nu} g_{\alpha\beta}~\delta g^{\alpha\beta}\right),
% \eea
Because of the total derivative, the covariant Gauss-Ostrogradsky theorem applies 
(see e.g. \cite{wald} for a careful treatment of this term),
\be \label{eq:gauss}
  \int_\mathcal{M} \dd^4 x \sqrt{-g} \left(\nabla_\mu V^\mu\right)
  =\int_{\df\mathcal{M}} \dd^3 y \sqrt{-\gamma} n_\mu V^\mu,
\ee
where $\mathcal{M}$ is the spacetime manifold and $\df\mathcal{M}$ its 3 dimensional hypersurface border with $\gamma (y^\mu)$ the induced metric on the border and $n_\mu$ a unit vector normal to the border. Fixing the boundary conditions, it results that the contribution of $\delta R_{\mu\nu}$ in Eq.~\eqref{eq:EH_diffeo} vanishes,
\begin{eqnarray*}
\int \dd^4x \sqrt{-g} g^{\mu\nu} \delta R_{\mu\nu} & = & \int_{\partial\mathcal{M}}\dd^{3}y\,\sqrt{-\gamma}\, n_{\nu}\left(g^{\mu\rho}\,\delta\Gamma_{\rho\mu}^{\nu}-g^{\mu\nu}\,\delta\Gamma_{\rho\mu}^{\rho}\right), \\
 & = & 0,
\end{eqnarray*}
and the variation of the EH action finally yields,
\bea
  \delta S_\rr{EH}&=&
	\frac{1}{2\kappa}\int \dd^4x \sqrt{-g}~
	    G_{\mu\nu}~\delta g^{\mu\nu}.
	%&=&\frac{1}{2\kappa}\int d^4x\sqrt{-g} G_{\mu\nu}\delta g^{\mu\nu} 
\eea

Considering now that the variation of the metric $\delta g^{\mu\nu}$ is generated by the Lie derivative \eqref{eq:lie_metric},
\be \label{eq:infinitesimal_diffeo}
 \delta g_{\mu\nu}=\mathcal{L}_\mathbf{\xi} g_{\mu\nu}\simeq\nabla_\mu \xi_\nu + \nabla_\nu \xi_\mu,
\ee
and using Eq.~\eqref{eq:variation_metric}, the variation of the EH action reads,
\bea
    \hspace*{-0.8cm} \delta S_\rr{EH}&=&-\frac{1}{\kappa}\int_\mathcal{M} \dd^4 x \sqrt{-g} \,G^{\mu\nu}\nabla_\mu \xi_\nu,\\
  &=& \frac{1}{\kappa}\left[\int_\mathcal{M} \dd^4 x \sqrt{-g} \left(\nabla_\mu G^{\mu\nu}\right) \xi_\nu 
  - \int_{\df\mathcal{M}} \dd^3 \Sigma_\mu \sqrt{-g} \left(G^{\mu\nu} \xi_\nu\right) \right], \\
    &\equiv&0.
  \label{eq:EH_diffeo_fin}
%\frac{1}{16\pi G}\int d^4 \sqrt{-g} G_{\mu\nu} \left(\mathcal{L}_\xi g_{\mu\nu}\right), \\
\eea
assuming the Levi-Civita connection, using Eq.~\eqref{eq:variation_metric} and applying the Gauss-Ostrogradsky theorem \eqref{eq:gauss}. Since the vector field $\xi^\nu$ is arbitrary, it results that the second Bianchi identity is a consequence of the active diffeomorphism-invariance, assuming the Levi-Civita connection.

% 
% Including the matter part of the action, the diffeomorphism-invariance must be preserved in order to guarantee the general covariance. Imposing the infinitesimal diffeomorphism Eq.~\eqref{eq:infinitesimal_diffeo} to $S_\rr{M}$,
% % \be \label{eq:gauge_transfo_diffeo}
% % g_{\mu\nu}\longrightarrow g_{\mu\nu}+\mathcal{L}_\mathbf{\xi} g_{\mu\nu}=g_{\mu\nu}+\nabla_\mu \xi_\nu +\nabla_\nu \xi_\mu,
% % \ee
% its variation reads \cite{Bertschinger},
% \bea
%   \delta S_\rr{M}&=& \frac{\delta S_\rr{M}}{\delta g_{\mu\nu}}\mathcal{L}_\mathbf{\xi} g^{\mu\nu},\\
%   &=&-\int \dd^4x~ \sqrt{-g}~ T^{\mu\nu}~ \nabla_\mu \xi_\nu,\\
%   &=&\int \dd^4x~ \sqrt{-g} \left(\nabla_\mu T^{\mu\nu}\right) \xi_\nu,
% \eea
% using the definition of $T_{\mu\nu}$ \eqref{eq:def_T} and neglecting the boundary term. In order to impose the gauge invariance to $S_\rr{M}$, the action has to be extremized, yielding $\nabla_\mu T^{\mu\nu}=0$.

\chapter[Application of PPN formalism to the Brans-Dicke theory]{Application\\ of the PPN formalism\\ to the Brans-Dicke theory} % Main chapter title

\label{sec:PPN_BD} % For referencing the chapter elsewhere, use \ref{Chapter1} 

\lhead{Appendix B. \emph{Awesome appendix B}} % This is for the header on each page - perhaps a shortened title

%----------------------------------------------------------------------------------------
\begin{sloppypar}
The Brans-Dicke formalism leads straightforwardly to the PPN analysis for the Brans-Dicke theory or for any STT in the absence of a potential (see also Sec.~\ref{sec:PPN}).
In this appendix, the Brans-Dicke formalism is briefly reviewed and the expressions for $\gamma_\rr{PPN}$ and $\beta_\rr{PPN}$ are derived.

The Lagrangian density in the standard generalized Brans-Dicke form reads\footnote{The original Brans-Dicke theory does not admit any potential and the function $\omega$ is independent of $\Phi$ \cite{Brans:1961sx}. For the sake of generality, the potential is included here even if it does not appear in the derivation of the PPN parameters. The function $\omega(\Phi)$ is considered since it is relevant for the derivation of $\beta_\rr{PPN}$.}, 
\bea\label{BDaction}
{\cal L}_\rr{BD}={\sqrt{-g}\over 2\kappa}\left[\Phi R-{\omega(\Phi)\over \Phi}(\partial \Phi)^{2}-\bar V(\Phi)\right]+{\cal L}_\rr{M}[\psi_\rr{M};g_{\mu\nu}].
\eea
It is equivalent to Eq.~\eqref{eq:actionSTT} with $F(\phi)=\Phi$, $Z(\phi)=\omega(\Phi)/(2\kappa\Phi)$ and $\bar V (\phi)=V/(2\kappa)$. 
The modified Einstein equations now read (see Sec.~\ref{sec:eom_JF}),
\bea\label{ee}
R_{\mu\nu}-{1\over 2}g_{\mu\nu}R&=&
{1\over \Phi}\nabla_{\mu}\nabla_{\nu}\Phi+{\omega\over\Phi^{2}}\nabla_{\mu}\Phi\nabla_{\nu}\Phi \non\\
&&\qquad-{1\over\Phi}\left[\square\Phi+{\omega\over 2\Phi}(\partial\Phi)^{2}+{\bar V\over 2}\right]g_{\mu\nu}+{\kappa\over\Phi}T_{\mu\nu},
\eea
where the stress-energy tensor is assumed to be a perfect fluid \eqref{eq:perfect_fluid}\footnote{In this appendix, we use the notation $T_{\mu\nu}\equiv T_{\mu\nu}^\rr{(M)}$}. The
% \be
% T_{\mu\nu}=\left(\rho+\rho \Pi+p\right) u_\mu u_\nu +p g_{\mu\nu}. 
% \ee
Ricci scalar then yields,
\bea \label{eq:R_BD}
R=\frac{1}{\Phi}\left[-\kappa T+\frac{\omega(\Phi)}{\Phi}(\partial\Phi)^{2}+3\square\Phi+2\bar V(\Phi)\right].
\eea
The Klein-Gordon equation follows from Eq.~\eqref{eq:KGtensor} which after replacing $R$ according to Eq.~\eqref{eq:R_BD} reads,
\bea \label{eq:KG_PPN_expansion}
(2\omega+3)\square\Phi+{\dd\omega\over \dd\Phi}(\partial\Phi)^{2}-\Phi{\dd\bar V\over \dd\Phi}+2\bar V= \kappa T,
\eea
and the modified Einstein equations becomes,
\bea \label{Einstein_BD}
R_{\mu\nu}={8\pi G\over \Phi}\left(T_{\mu\nu}-{\om +1\over 2\om +3}T g_{\mu\nu}\right)+{\om\over \Phi^{2}}\df_{\mu}\Phi\df_{\nu}\Phi+{1\over \Phi}\nabla_{\mu}\nabla_{\nu}\Phi\qquad \non\\
-\frac{\om_\Phi}{2\Phi}\frac{\left(\df\Phi\right)^2}{3+2\om} g_{\mu\nu}
+\frac{1}{2\Phi\left(3+2\om\right)}\left[\Phi \frac{\dd \overline{V}}{\dd \Phi}+\overline{V}(2\om+1)\right] g_{\mu\nu},
\eea
where the subscript $\Phi$ denotes a derivative with respect to $\Phi$. As previously shown in Sec.~\ref{sec:PPN}, in order to compute $\gPPN$ and $\bPPN$, the modified Einstein equations must be solved up to $\mathcal{O}(2)$ or 1PN and $\mathcal{O}(4)$ or 2PN for $g_{00}$, up to $\mathcal{O}(3)$ or 1.5PN for $g_{0i}$ and up to $\mathcal{O}(2)$ or 1PN for $g_{ij}$. The bookkeeping of the different quantities has already been introduced in Sec.~\ref{sec:PPN} excepted for the scalar field which is expanded as,
\be
  \Phi\left(x^\mu\right)=\Phi_0+\zeta\left(x^\mu\right),
\ee
where $\Phi_0$ is the constant background value and $\zeta$ is at least of order $\mathcal{O}(2)$. In the rest of this appendix, we assume $\overline{V}=0$ since the PPN derivation for STT is exact for vanishing potential only. The steps of the computations then follow:

\begin{enumerate}
 \item Solution for the scalar field $\zeta$ up to $\mathcal{O}(2)$,

The expansion of the Klein-Gordon equation \eqref{eq:KG_PPN_expansion} up to $\mathcal{O}(2)$ enables one to determine $\zeta$ as a function of the gravitational potential $U$ \eqref{eq:U_PPN}. Indeed,
\be
  \square \Phi \equiv|g|^{-1/2}\df_{\mu}(|g|^{1/2}\df^{\mu}\Phi)
  \sim \nabla^{2}\Phi-\df^{2}_{0}\Phi\sim \nabla^{2}\zeta+\mathcal{O}(4),
\ee
where $g_{\mu\nu}=\eta_{\mu\nu}$, the derivative of the scalar field being at least of order $\mathcal{O}(2)$. Only the trace of $T_{\mu\nu}$ further contributes to $\mathcal{O}(2)$ and reads,
\bea \label{T_order2}
T=g_{\mu\nu}T^{\mu\nu}=-\rho\left(1+3\frac{p}{\rho}\right)\simeq -\rho\left[1+\mathcal{O}(2)\right],
\eea
according to the bookkeeping rules given by Eqs.~\eqref{eq:bookkeping}.
The Klein-Gordon equation \eqref{eq:KG_PPN_expansion} up to $\mathcal{O}(2)$ finally yields,
\bea
  \nabla^{2}\zeta^{(2)}=-{8\pi G\over 3+2\om}\rho,
\eea
the superscript denoting the order of the expansion.
Replacing $\rho$ according to the Poisson equation \eqref{eq:poisson} with $U\equiv-\Phi/G$,
the solution for $\zeta$ reads,
\bea\label{scalar}
\zeta^{(2)}-\zeta_{0}={2G U\over 3+2\om}.
\eea
where $\zeta_0$ is the constant of integration.

\item Solution for $h_{00}$ up to $\mathcal{O}(2)$,

Given the expansion of the Levi-Civita connection up to $\mathcal{O}(2)$,
\be \label{levi_exp}
  \Gamma^{\lambda,~(2)}_{\mu\nu}=
  \frac{1}{2}\eta^{\lambda\rho}\left(\df_\mu h_{\rho\nu}+\df_\nu h_{\mu\rho}-\df_\rho h_{\mu\nu}\right),
\ee
the expansion of the Ricci tensor up to $\mathcal{O}(2)$ reads\footnote{Notice that only the terms involving the derivative of the Christoffel symbols are relevant here contrary to order $\mathcal{O}(4)$.} \cite{Will1993},
\be \label{Ric_order2}
R^{(2)}_{\mu\nu}=\frac{1}{2}\left(-\square h_{\mu\nu}-\df_\mu \df_\nu h +\df_\alpha \df_\mu h^\alpha_\nu 
+\df_\nu \df_\alpha h^\alpha_\mu\right),
\ee
with $h=h^\mu_\mu$, hence,
\bea
R^{(2)}_{00}\simeq -{1\over 2}\nabla^{2}h_{00} +\mathcal{O}(>2).
\eea
The only additional terms up to order $\mathcal{O}(2)$ of the $00-$component of Eqs.~\eqref{Einstein_BD} involve the stress-energy tensor, with $T_{00}^{(2)}=\rho$ and $T^{(2)}=-\rho$ (see Eq.~\eqref{T_order2}). The expansion  up to $\mathcal{O}(2)$ thus yields,
\bea
  -{1\over 2}\nabla^{2}h_{00}={8\pi G\rho\over \Phi_{0}}\left(1-{\om+1\over 2\om+3}\right)
  +\mathcal{O}(4).
\eea
Given the Poisson equation~\eqref{eq:poisson}, the solution for $h_{00}$ up to $\mathcal{O}(2)$ is,
\bea \label{h00_order2}
  h^{(2)}_{00}={4G\over \Phi_{0}}{\om +2\over 2\om +3}U\equiv 2 \bar G U,
\eea
where we used Eq.~\eqref{eq:metric_PPN}, $\bar G\equiv G_\rr{Cav}$ (see Sec.~\ref{sec:varying_cst}) being the measured gravitational constant, for instance by Cavendish experiments, which differs from $G$ in STT, 
\bea\label{geff}
  \boxed{\bar G= {2G\over \Phi_{0}}{\om +2\over 2\om +3}}
\eea
In the limit $\omega\longrightarrow\infty$, the measured gravitational constant is the Newton's constant,
\be
  \bar G= {G\over \Phi_{0}}=G,
\ee
assuming that $\Phi_0$ corresponds to the value of $\Phi$ at large distance from the central body.
Considering Eq.~\eqref{geff}, the perturbation around the scalar field background Eq.~\eqref{scalar} yields,
\bea\label{zeta}
  {\zeta\over \Phi_{0}}={\bar G U\over \om +2},
\eea
where $\zeta_{0}$ has been absorbed in $\Phi_{0}$.

\item Solution for $h_{ij}$ up to $\mathcal{O}(2)$,

The expansion of $R_{ij}$ to order $\mathcal{O}(2)$ reads (see Eq.~\eqref{Ric_order2}),
\bea
R^{(2)}_{ij}=-{1\over 2}\left( \nabla^{2}h_{ij}-\df_{i}\df_{j}h_{00}+\df_{i}\df_{j}h^{k}_{\,\,k}-2\df_{k}\df_{j}h^k_{\,\,i}   \right).
\eea
Because of the diffeomorphism-invariance (see Sec.~\ref{sec:gen_cova}), four gauge conditions must be imposed to the modified Einstein equations for fixing the gauge. The {\bf first three gauge conditions} are given by\footnote{Note that this gauge condition is valid to all orders.},
\bea\label{spatialgauge}
\df_{\mu}h^{\mu}_{\,\,\,i}-{1\over 2}\df_{i}h^{\mu}_{\,\,\mu}={1\over \Phi_{0}}\df_{i}\zeta.
\eea
The derivative with respect to spatial component of these conditions expanded up to $\mathcal{O}(2)$ then yields, 
\bea
\df_{j}\df_{k}h^{k}_{\,\,i}-{1\over 2}\left(\df_{i}\df_{j}h^{k}_{\,\, k}-\df_{i}\df_{j}h_{00}\right)={1\over \Phi_{0}}\df_{i}\df_{j}\zeta,
\eea
leading to,
\bea
R^{(2)}_{ij}=-{1\over 2}\nabla^{2}h_{ij}+{1\over \Phi_{0}}\df_{i}\df_{j}\zeta.
\eea
Since $T_{ij}$ is at least of $\mathcal{O}(4)$ according to Eq.~\eqref{eq:bookkeping}, the only term up to $\mathcal{O}(2)$ involving the stress-energy tensor is the one involving $T$ (see Eq.~\eqref{T_order2}). The $ij-$component of Eqs.~\eqref{Einstein_BD} thus reads,
\bea
  \nabla^{2}h_{ij}=-{16\pi G\over \Phi_{0}}{\om+1\over 2\om +3}\rho \delta_{ij},
\eea
which is solved using the Poisson equation~\eqref{eq:poisson} and Eq.~\eqref{geff} yielding,
\bea \label{hij}
  h^{(2)}_{ij}=2\,{\om+1\over \om+2}\,\bar G\, U\, \delta_{ij}\equiv 2\, \gamma_\rr{PPN}\, \bar G\, U\, \delta_{ij}.
\eea
According to the standard PPN metric expansion \eqref{eq:metric_PPN} the parameter $\gamma_\rr{PPN}$ for the Brans-Dicke theory is thus,
\bea \label{eq:gamPPN}
\boxed{\gPPN={\om+1\over \om+2}}
\eea

\item Solution for $h_{0j}$ up to $\mathcal{O}(3)$,

Since $h_{0j}$ is at least of $\mathcal{O}(3)$, the expansion of $R_{0j}$ up to $\mathcal{O}(3)$ reads (see also Eq.~\eqref{Ric_order2}),
\bea
R^{(3)}_{0j}=-{1\over 2}\left(  \nabla^{2}h_{0j}-\df_{j}\df_{k}h^{k}_{\,\,0}+\df_{0}\df_{j}h^{k}_{\,\, k}-\df_{0}\df_{k}h^{k}_{\,\, j}  \right).
\eea
The {\bf fourth gauge condition} is now useful\footnote{Note that this fourth gauge condition cannot be compactified with the three first ones \eqref{spatialgauge} in a covariant way.},
\bea \label{fourth_gauge}
\df_{\mu}h^{\mu}_{\,\,0}-{1\over 2}\df_{0}h^{\mu}_{\,\,\mu}+{1\over 2}\df_{0}h_{00}={1\over \Phi_{0}}\df_{0}\zeta,
\eea
and yields up to order $\mathcal{O}(3)$,
\bea
\df_{i}h^{i}_{\,\,\,0}-{1\over 2}\df_{0}h^{i}_{\,\,i}={1\over \Phi_{0}}\df_{0}\zeta.
\eea
Combining the four gauge conditions  ($\partial_0$~[Eq.~\eqref{spatialgauge}]$\times$ $\partial_j$~[Eq.~\eqref{fourth_gauge}]) up to order $\mathcal{O}(3)$ yields,
\bea
R^{(3)}_{0j}=-{1\over 2}\nabla^{2}h_{0j}-{1\over 4}\df_{0}\df_{j}h_{00}^{(2)}+{1\over \Phi_{0}}\df_{0}\df_{j}\zeta.
\eea
Since $g_{0j}$ is at least of order $\mathcal{O}(3)$, the $0j-$component of the modified Einstein equations \eqref{Einstein_BD} reads to order $\mathcal{O}(3)$,
\bea
R^{(3)}_{0j}={8\pi G\over \Phi_{0}} \left[T^{(3)}_{0j}-\mathcal{O}(>3)\right]+\mathcal{O}(>3)+{1\over \Phi_{0}}\df_{0}\df_{j}\zeta,
\eea
where $T^{(3)}_{0j}=-\rho v_{j}$ assuming a perfect fluid \eqref{eq:perfect_fluid}. Using Eq.~\eqref{h00_order2} it finally leads to,
\be
\nabla^{2}h_{0j}={16\pi G\over \Phi_{0}}\rho v_j-\bar G\df_{0}\df_{j}U.
\ee
Defining two additional potentials, $V_j$ and $\chi$ \cite{Will1993},
\bea
\nabla^{2}V_{j}=-4\pi\rho v_{j}, \qquad \nabla^{2}\chi=-2U,
\eea
and using Eqs.~\eqref{zeta} and \eqref{h00_order2}, $h^{(3)}_{0j}$ reads,
\bea
h^{(3)}_{0j}=-{4\om + 6\over \om +2}\bar G V_{j}+{1\over 2}\bar G\df_{0}\df_{j}\chi.
\eea

\item Solution for $h_{00}$ up to $\mathcal{O}(4)$

The expansion of the Levi-Civita connection up to $\mathcal{O}(4)$ yields,
\be
  \Gamma^{\lambda,~(4)}_{\mu\nu}=
  \frac{1}{2}\left(\df_\mu h_{\nu}^\lambda+\df_\nu h_{\mu}^\lambda-\df^\lambda h_{\mu\nu}\right)
  +\frac{1}{2}h^{\lambda\rho}\left(\df_\mu h_{\rho\nu}+\df_\nu h_{\mu\rho}-\df_\rho h_{\mu\nu}\right),
\ee
the expansion of $R_{00}$ up to $\mathcal{O}(4)$ thus reading \cite{Will1993}, 
\bea
R^{(4)}_{00}&=&-{1\over 2}\nabla^{2}{h_{00}^{(4)}}-{1\over 2}\left(  \df_{0}\df_{0}h^{j\,\,(2)}_{\,\, j}-2\df_{0}\df_{j}h^{j\,\,(3)}_{\,\,0}  \right)
-{1\over 4}\df^{i}{h_{00}^{(2)}}\df_{i}{h_{00}^{(2)}}
\non\\
&&+{1\over 2}\df_{j}h_{00}^{(2)}\left(\df^{k}h^{j\,\,(2)}_{\,\, k}-{1\over 2}\df^{j}h^{k\,\,(2)}_{\,\, k}  \right)
+{1\over 2}h^{jk\,\,(2)}\df_{j}\df_{k}{h_{00}^{(2)}}. \non\\
\eea
Using the fourth gauge condition \eqref{fourth_gauge} and Eq.~\eqref{zeta},
\bea
  \df_0\df_i h^i_0-\frac{1}{2}\df_0^2 h^k_k&=&\frac{1}{\Phi_0}\df_0^2\zeta, \\
  &=& \frac{\bar G}{\om+2}\df_0^2 U,
\eea
as well as the expressions for ${h^{(2)}_{00}}$ \eqref{h00_order2}, $h^{(2)}_{ij}$ \eqref{hij}, and $\zeta$ \eqref{zeta}, the expansion for $R^{(4)}_{00}$ finally reads,
\bea
R^{(4)}_{00}&=&-{1\over 2}\nabla^{2}{h^{(4)}_{00}}+{\bar G\over \om+2}\df^{2}_{0}U-\bar G^{2}{2\om+3\over \om+2}(\nabla U)^{2} \non\\
&&\qquad\qquad\qquad\qquad+~2\bar G^{2}{\om+1\over \om+2} U\nabla^{2} U.
\eea

The modified Einstein equations \eqref{Einstein_BD} expanded to $\mathcal{O}(4)$ then yield,
\bea \label{eq:tmpPPN}
  R^{(4)}_{00}=\frac{8\pi G}{\Phi_0+\zeta}\left[T_{00}^{(4)}-\frac{\om+1}{2\om+3}T^{(4)} g_{00}\right]
		+\mathcal{O}(6)\non\\
		+\frac{1}{\Phi_0}\nabla_0\df_0 \zeta 
		-\frac{\om_\Phi}{2\Phi}\frac{(\df\Phi)^2}{3+2\om} g_{00},
\eea
We first focus on the term involving $T_{\mu\nu}$. In order to expand $T_{\mu\nu}$ up to $\mathcal{O}(4)$, the four-velocity given by,
\bea
g_{\mu\nu}u^{\mu}u^{\nu}=-1,
\eea
is expanded as,
\bea
u^{0}u^{0}&=&-{1+g_{ij}u^{i}u^{j}\over g_{00}},\\
&\simeq& -{1+v^{2}\over -1+2\bar G U},\\
&\simeq& 1+2\bar G U+v^{2} +\mathcal{O}(4),
\eea
where $u^i u_i=v^2$. Thus, assuming a perfect fluid \eqref{eq:perfect_fluid}, the stress-energy tensor $T_{ij}$ and its trace read up to $\mathcal{O}(4)$,
\bea
T^{(4)}_{00}&=& \rho (1+\Pi-2\bar GU+v^{2}), \\
T^{(4)}_{ij}&=& \rho (u_i u_j +\frac{p}{\rho} \delta_{ij}), \\
\qquad T^{(4)}&=&-\rho\left(1+\Pi-{3p\over \rho}\right),
\eea
with $\Pi$ the specific energy density (see Sec.~\ref{sec:PNformalism}).
Moreover the multiplicative term $\Phi^{-1}$ needs to be expanded as,
\bea
{1\over \Phi}\simeq {1\over \Phi_{0}}\left[ 1-{\zeta\over \Phi_{0}}+{\cal{O}}(4)  \right].
\eea
The first term of the right-hand side of Eq.~\eqref{eq:tmpPPN} thus reads,
\bea
  \hspace{-0.4cm}\frac{8\pi G}{\Phi_0+\zeta}\left[T_{00}^{(4)}-\frac{\om+1}{2\om+3}T^{(4)} g_{00}\right] 
  ={8\pi G \rho\over \Phi_{0}}\left(1-{\zeta\over \Phi_{0}}\right)\times \qquad\qquad \non \\
  \left[1+\Pi-2\bar GU+v^{2}+\frac{\om+1}{2\om+3} \left(1+\Pi-{3p\over \rho}\right) \left(-1+2\bar G U\right)\right].
\eea
The second term in Eq.~\eqref{eq:tmpPPN} to be expanded involves the second derivative $\nabla_{0}(\df_{0}\zeta)$ that requires the expansion of the Levi-Civita connection \eqref{levi_exp}. Since $\Gamma^{0}_{\,\,00}\df_{0}\zeta > \mathcal{O}(4)$ the only relevant term is,
\bea
  \Gamma^{i}_{\,\,00}= -{1\over 2}\df^{i}h_{00}+\mathcal{O}(4)\simeq-\bar G\df^{i}U,
\eea
hence,
\bea
\nabla_{0}(\df_{0}\zeta)=(\df^{2}_{0}-\Gamma^{\mu}_{\,\,00}\df_{\mu})\zeta=(\df_{0}^{2}+\bar G\df^{i}U\df_{i})\zeta.
\eea
In the case where $\om=\om(\Phi)$, the last term involving $\om_\Phi$ has also to be expanded,
\bea
-{\om_\Phi\over 2\Phi}\frac{(\partial\Phi)^{2}}{2\om+3} g_{00}&\simeq&-{\om_\Phi\over 2(2\om+3)}{1\over\Phi_{0}}\left( 1-{\zeta\over\Phi_{0}} \right)(-1+h_{00}^{(2)}) \qquad \non\\
&& \times \left[(-1+h_{00}^{(2)})(\df^{0}\zeta)^{2}+(\delta_{ij}+h_{ij}^{(2)})\df^{i}\zeta\df^{j}\zeta\right],\non\\
&=&{\om_\Phi \bar G^{2}\Phi_{0}\over 2(2\om+3)(\om+2)^{2}}(\nabla U)^{2}+\mathcal{O}(6),
\eea
where Eq.~\eqref{zeta} has been used. 
Therefore, the $00$-component of Eqs.~\eqref{Einstein_BD} becomes up to order $\mathcal{O}(4)$,
\bea\label{4thexp}
  \hspace{-0.4cm}-{1\over 2}\nabla^{2}{h^{(4)}_{00}}+{\bar G\over \om+2}\df^{2}_{0}U-\bar G^{2}{2\om+3\over \om+2}(\nabla U)^{2}+2\bar G^{2}{\om+1\over \om+2} U\nabla^{2} U=
  \non\\
  {4\pi \bar G \rho}\frac{2\om+3}{\om+2}\left(1-\frac{\bar G U}{\om+2}\right)\times
  \qquad\qquad\qquad\qquad\qquad\non\\
  \left[\left(1+\Pi -2\bar GU\right)\frac{\om+2}{2\om+3}+v^{2} +{\om+1\over 2\om +3}{3p\over \rho}\right]\non\\
  +\left(\df_{0}^{2}+\bar G\df^{i}U\df_{i}\right)\frac{\bar G U}{\om+2}+{\om_\Phi \bar G^{2}\Phi_{0}\over 2(2\om+3)(\om+2)^{2}}(\nabla U)^{2},
\eea
where $\Phi_0$ has been removed using Eqs.~\eqref{geff} and~\eqref{zeta}. For the term involving $\om_\Phi$, we assume that $\Phi_{0}$ corresponds to the value of $\Phi$ at large distance from the central body, where $\bar G \simeq G$. Therefore, from Eq. \eqref{geff},
\bea
\Phi_{0}\simeq {2\om+4\over 2\om+3}.
\eea
After some algebra, we obtain,
\bea\label{4thexp_bis}
-{1\over 2}\nabla^{2}{h^{(4)}_{00}}&=&4\pi  \bar G \rho\left[ 1+\Pi-{2\om+5\over \om+2}\bar G U+{2\om+3\over \om +2}v^{2}+{3\om +3\over \om+2}{p\over \rho} \right]\non\\
&&\qquad+2\bar G^{2}(\nabla U)^{2}-{2\om+2\over \om+2}\bar G^{2}U\nabla^{2}U \non\\
&&\qquad\qquad+\left[{\om_\Phi\over (2\om+3)^{2}(\om+2)}\right]\bar G^{2}\left(\nabla U\right)^{2}.
\eea
By using the identity,
\be
2(\nabla U)^{2}=\nabla^{2}(U^{2})-2U\nabla^{2}U,
\ee
in order to remove the terms proportional to $(\nabla U)^2$, and by defining four potentials in addition to the one given by Eq.~\eqref{eq:poisson},
\bea\label{Phi}
\nabla^{2}\Phi_{1}=-4\pi \rho v^{2},&\qquad& \nabla^{2}\Phi_{2}=-4\pi \rho U,\\
\nabla^{2}\Phi_{3}=-4\pi \rho \Pi, &\qquad& \nabla^{2}\Phi_{4}=-4\pi p,
\eea
one can solve the equation for $h^{(4)}_{00}$ \eqref{4thexp_bis} and find, 
\bea
{h^{(4)}_{00}}&=&2\bar G U-2 \left[1+{\om_\Phi\over (2\om+3)^{2}(2\om+4)}\right]\bar G^{2}U^{2}+{6+4\om\over 2+\om}\bar G \Phi_{1}\non\\
&&\qquad+\left[{2+4\om\over 2+\om}+{2\om_\Phi\over (2\om+3)^{2}(\om+2)}\right]\bar G^{2}\Phi_{2}+2\bar G\Phi_{3}\non\\
&&\qquad\qquad\qquad+{6+6\om\over 2+\om}\bar G\Phi_{4}.
\eea
In the limit where $\om\rightarrow\infty$ and $\om_\Phi=0$, the GR result is recovered as expected \cite{Will1993},
\be
h^{(4)}_{00}=2\bar G U -2 \bar G^2 U^2 + 4\bar G\Phi_{1}+4\bar G^2\Phi_{2}+2\bar G\Phi_{3}+6\bar G \Phi_{4}.
\ee
In the vacuum ($\Phi_{1}=\Phi_{2}=\Phi_{3}=\Phi_{4}=0$), the expansion becomes $(\om_\Phi\neq0)$,
\bea
{g^{(4)}_{00}}=-1+2\bar G U-2\left[1+{\om_\Phi\over (2\om+3)^{2}(2\om+4)}\right]\bar G^{2}U^{2},
\eea
that must be compared to the standard PPN expansion \eqref{eq:metric_PPN},
\bea
g_{00}=-1+2\bar G U-2 \beta_\rr{PPN}\bar G^{2} U^{2}+{\cal{O}}(6),
\eea
yielding, 
\bea \label{eq:betaPPN2}
\boxed{\bPPN=1+{\om_\Phi\over (2\om+3)^{2}(2\om+4)}}
\eea 
These results are consistent with \cite{Nutku1969, 1972ApJ...176..769N} reported by \cite{Will1993}.
\end{enumerate}
\end{sloppypar}

\chapter[The chameleon model: an analytical approach] {The chameleon model: \\ an analytical approach} % Main chapter title

 % For referencing the chapter elsewhere, use \ref{Chapter1} 

\lhead{Appendix A. \emph{Awesome appendix A}} 
% This is for the header on each page - perhaps a shortened title

%-------------------------------------------------------------------------------

%\label{sec5}
In this appendix we reproduce the main steps of \cite{Burrage} and derive analytically the chameleon field profile in the spherically symmetric and static Minkowski spacetime for a two-region model (the source mass and the vacuum chamber). 
%In the next section, the validity of the various assumptions will be analyzed and the analytical approximations will be compared to the exact numerical results, for the original and exponential chameleon potentials. 
For the sake of simplicity, we assume in this appendix that $\alpha=1$. Those analytical calculations are compared to the numerical computations in Sec.~\ref{sec4} where the analytical analysis is found to be reliable close to the source mass where the induced acceleration is measured. In the second part of this appendix we use the acceleration profiles derived analytically in order to compute the viable parameter space for the Berkeley experiment, as represented in Figs.~\ref{fig:exclusion} and~\ref{fig:beta_n}.

\section{Four different regimes}\label{sec5}
Assuming $A(\phi)=\rr{e}^{\phi/M}\simeq1$, the minimum of the effective potential and its effective mass around it are respectively given by (see Eqs.~\eqref{eq:phimin}--\eqref{eq:m_eff_chamel} with $\alpha=1$),
\bea
\phi_{\rr{min}} = \left( \frac{\Lambda^5 M}{\rho} \right)^{1/2}, \hspace{0.8cm} m_{\rr{min}} = \sqrt 2 \left( 
\frac{\rho^3}{\Lambda^5 M^3} \right)^{1/4}.
\eea
The case where the effect of $A(\phi)$ becomes important, is discussed in our paper \cite{Schlogel:2015uea} for the original model.
For a two-region model the density $\rho$ is either the source mass density $\rho_{\rr A}$ or the density in 
the vacuum chamber $\rho_{\rr{v}}$.  

Four different regimes can be identified, depending on whether the field reaches the effective potential minimum or not: 
(1) the field does not reach the minimum of the effective potential in any region, (2) the field reaches the minimum 
in the vacuum chamber but not in the source mass, (3) the field reaches the minimum in the 
source mass but not in the vacuum chamber,  (4) the field reaches the minimum both inside the test 
mass and the vacuum chamber. The Cases (1) and (2) were 
referred to as the \textit{weakly perturbing regime} in \cite{Burrage}, whereas the Cases (3) and (4) were referred to as 
\textit{strongly 
perturbing}. Below we consider those four cases separately, as in \cite{khoury}.   In principle, one should also distinguish between the cases where the field reaches $\phi_{\rr{min}} $ inside the chamber wall, or not.  When lowering $M$, depending on the central mass density and size, on the chamber wall density and thickness, $\phi_{\rr{min}} $ can be reached first inside the central mass or inside the chamber walls.  Nevertheless, for the considered experimental set-up, the wall and the central mass have similar densities and sizes, and so those two cases will not be distinguished in the following. 

\begin{itemize}
\item{Case~(1): $\phi(r=0) \neq  \phi_{\rr{min}}(\rho_A) $ and $\phi(R_A <r <L) \neq  \phi_{\rr{min}}(\rho_{\rr 
v}) $}

Within the source mass the field does not reach the attractor that is the minimum of the effective potential.  Since $\rho_{\rr v} < 
\rho_{\rr{atm}} 
< \rhoA$,  the second term in the effective 
potential \eqref{eq:KG} dominates, $V_{\rr{eff}} \simeq \phi \rhoA /M$.   The Klein-Gordon equation inside the source mass then reads,
\be
  \phi''+\frac{2}{r}\phi'=\frac{\rhoA}{M}.
\ee
By setting $\phi=Z/r$, the Klein-Gordon equation reads $Z''=(\rho/M) r$, whose solution is given by,
\be \label{eq:solKGtmp}
  Z=\frac{\rhoA}{6M}r^3+ \mathcal{C} r + \mathcal{D},
\ee
with $\mathcal{C}$ and $\mathcal{D}$ the two constants of integration. Imposing that the field profile is regular at the origin, implies that $\mathcal{C}=0$,
\bea \label{eq:KGinA}
\phi = \mathcal{D} + \frac{\mA r^2}{8 \pi M \RA^3}.
\eea
The constant of integration $\mathcal{D}$ is fixed by matching $\phi$ and $\phi'$ to the field solution in the vacuum chamber at $ r=R_A $.  Inside the vacuum chamber the 
field does not reach the attractor value.   Let us denote $\phibg$ the value that the field would take at the center of the chamber in the absence of the source.  Then one 
can consider a harmonic expansion of the potential, 
\bea \label{eq:Veff}
\Veff(\phi) \simeq \Veff(\phibg) + \frac {\mbg^2}{2} (\phi - \phibg)^2~,
\eea
higher order terms being subdominant, the Klein-Gordon equation in the vacuum chamber then reading,
\be
  \phi''+\frac{2}{r}\phi'=\mbg^2\left(\phi-\phibg\right).
\ee
By setting $Y=\phi-\phibg$ and $Y=Z/r$, the Klein-Gordon equation becomes,
\be
  Z''=\mbg^2 Z,
\ee
whose solution reads,
\be
  Z=\mathcal{A}~\rr{e}^{|\mbg| r}+\mathcal{B}~\rr{e}^{-|\mbg| r},
\ee
with $\mathcal{A}$ and $\mathcal{B}$ the two constants of integration.
Assuming that the field profile 
decays at spatial infinity implies that $\mathcal{A}=0$, the scalar field profile thus yielding, 
\bea \label{eq:KGinvacuum}
\phi(r) = \phibg + \frac{\mathcal{B}}{r}~ \rr e^{- |\mbg| r}~.
\eea
Note that at $r=R_A$, one has $\mbg R_A \ll 1$ for typical experimental parameters and thus $\phi(R_A )\simeq \phibg + \mathcal{B}/R_A$.  

By matching the solutions \eqref{eq:KGinA} and \eqref{eq:KGinvacuum} at $r=\RA$, we obtain,
\bea
  \mathcal{B}&=&-\frac{1}{4\pi}\frac{\mA}{M}\rr{e}^{\mbg \RA}\frac{1}{1+\mbg\RA}
  \simeq -\frac{1}{4\pi}\frac{\mA}{M}, \\
  \mathcal{D}&=&\phibg-\frac{1}{8\pi\RA}\frac{\mA}{M}-\frac{1}{4\pi\RA}\frac{\mA}{M}\frac{1}{1+\mbg\RA},\\
  &\simeq& \phibg-\frac{3}{8\pi\RA}\frac{\mA}{M},
\eea
the second equality being obtained assuming $\mbg \RA\ll 1$.
Eventually the field profile in the Case (1) reads,
\bea \label{eq:case1}
\phi^{(1)}(r) = \phibg - \frac{\mA}{8 \pi \RA M} \times \left[ \left( 3 - \frac{r^2}{\RA^2}  \right) \Theta(\RA - 
r)\right.\qquad\qquad \non\\
+\left. \left( 2 \frac{\RA}{r} \rr e^{-\mbg r} \right) \Theta(r- \RA) \right],
\eea
where $\Theta$ is the Heaviside function.  Therefore the effect of the source mass is to deepen the field profile, by a quantity $3 \mA /(8 \pi \RA M) \ll \phibg$ 
at $r=0$.  By definition, the Case (1) is valid as 
long as $|\phibg - \phi^{(\rr 1)}(r=0)| \ll \phibg$.
Outside the source mass, the difference $|\phibg - \phi|$ decreases like $\propto 1/r$ for realistic experimental 
configurations where the exponential decay factor can be 
neglected. 

A subtlety arises in the evaluation of $\phibg$, which in \cite{Burrage} was either the attractor in the vacuum or related to the chamber 
size\footnote{$\rho_{\rr v}$ is 
much lower than the wall density $\rho_{\rr w}$ where the field was assumed to reach its attractor 
$\phi_{\rr{min}}\left(\rho_\rr{w}\right)$.  Thus the 
first term of $\Veff$ in Eq.~(\ref{eq:Veff}) dominates  
the Klein-Gordon equation inside the chamber, which can be solved to get $\phibg $ as a function of 
the size of the vacuum chamber.  However, behind 
this calculation is hidden the assumption that the field reaches $\phi_{\rr{min}}\left(\rho_\rr{w}\right)$ in the 
wall, which is not valid in the Case (1) in most of the parameter space.},
under the assumption that the scalar field reaches its attractor inside the vacuum chamber wall.
This assumption is actually not valid in the Case (1) 
because $\rhoW\sim\rhoA$, and because the wall 
thickness is about the radius of the source mass.  
So in most of the parameter space corresponding to the Case (1), the scalar field does not reach its attractor inside the wall.
 %Actually, the scalar field reaches its attractor inside the wall before 
%inside the source mass with decreasing $M$, while almost at the same time since $\rhoW\sim\rhoA$. }
As a result, $\phibg$ is better approximated by $\phi_\rr{min}(\rhoATM)$, as highlighted by our numerical results which take the effects of the chamber wall on the scalar field profile into account (see our paper \cite{Schlogel:2015uea} for the comparison between analytical and numerical results).
Even if the background field value 
has no effect on the acceleration itself, this result is important because it 
changes the region in the parameter space in which the Case (1) applies: it is extended to lower values 
of $M$, as developed thereafter. 

%Comparison of analytical and numerical profiles, for the chameleon field and the underlying acceleration have been plotted in our paper \cite{Schlogel:2015uea}.
% The analytical field profile and the induced acceleration $a_\phi = \partial_r \phi / M$ have 
% been plotted in Figs.~\ref{plot_profile_field_k} and \ref{plot_profile_acc_k} respectively for various
% values of $M$ reported in Table~\ref{tab:profiles}. 

%before Eq.3 of Clare's draft, $\phi_{lab} \gtrsim \phi(\rho_v) $ false (corrected just before Eq.9).  Eq. 4 false 
%because the field does not reach the attractor in the 
%wall.  But Eq. 4 can be modified and applied to the air.

The acceleration induced by the scalar field gradient inside the vacuum chamber is well approximated by,
\be \label{eq:acc_phi_weak}
 a_\phi \approx \frac{\mA}{4 \pi  M^2 r }  \left( \frac{1}{r} + \mbg  \right). 
\ee
Since $\mbg r \ll 1$ for realistic laboratory experiments, the acceleration is independent of $\Lambda$ and thus one can constrain directly the 
value of $M$. This is the reason why the 
power-law of the potential has no effect on the acceleration as long as $|A(\phi)-1|\ll1$ (see \cite{Schlogel:2015uea} for a discussion about the original chameleon model).

\item{Case~(2): $\phi(0) \neq  \phi_{\rr{min}}(\rho_A) $ and $\phibg = \phi_{\rr{min}}(\rho_{\rr v}) $}

When the size of the vacuum chamber is larger than the characteristic distance over which the field reaches its attractor, that is when,
\bea \label{eq:case2condition}
L \gg \frac{1}{m_{\rr{min}}(\rho_{\rr v})} = \left(\frac{\Lambda^5 M^3}{4 \rho_{\rr v}^3}\right)^{1/4}~,
\eea
 the field profile is still governed by Eq.~(\ref{eq:case1}).  However, the value of $\phibg$ is now simply 
$\phi_{\rr{min}} (\rho_{\rr v}) $.   
In the case of the original chameleon potential $V(\phi) = \Lambda^5 / \phi$, one has $ \Lambda \simeq 2.6 \times 10^{-6} \GeV $ in order to reproduce the late-time cosmic acceleration.  For typical vacuum densities and chamber sizes, e.g. those reported in Table~\ref{tab:expconf}, one finds that this regime would occur when $M\lesssim 10^{-6} \GeV$.  This does not correspond anymore to the weakly perturbing regime requiring $\phibg \gtrsim m_A/ (4\pi R_A M)$, yielding $M\gtrsim 2 \times 10^9 \GeV$ in our fiducial experimental setup.  
In the case of the exponential potential  $V(\phi) = \Lambda^4 
(1+ \Lambda / \phi)$, $\Lambda \simeq 10^{-12}\GeV $ 
is the cosmological constant.   It results that the field 
in the chamber is expected to reach $\phi_{\rr{min}}(\rho_{\rr v})$ only if $M\lesssim 10^5 \GeV$.   This is far from 
the regime where the source mass perturbs only weakly the field, valid when $M \gtrsim 10^{20} \GeV$, i.e. in the 
super-Planckian regime.

 \item{Case~(3): $\phi(0) = \phi_{\rr{min}}(\rho_A) $ and  $\phi(R_A <r <L) \neq  \phi_{\rr{min}}(\rho_{\rr 
v}) $}
 
 In the Case (3) the field reaches $\phi_{\rr A} \equiv \phi_{\rr{min}}\left(\rho_\rr{A}\right)$ inside 
the source mass.  One can 
define a radius $S$ such that $\phi(S) = \phi_{\rr A} 
(1+ \epsilon) $ with $0<\epsilon \ll1 $, so that, for $r<S$,
\be \label{eq:phi_thin_shell}
  \phi\simeq\phi_{\rr A}.
\ee
For $S < r < \RA$, the density term dominates in 
$V_{\rr {eff}}$ and the solution of the linearized Klein-Gordon equation is given by Eq.~\eqref{eq:solKGtmp} (with $\phi=Z/r$), the scalar field profile reading,
 \bea \label{eq:phi_int_case3}
 \phi = \mathcal{D} + \frac{\mathcal{C}}{r} + \frac{\mA r^2}{8 \pi M \RA^3}~,
 \eea
 which is the same as Eq.~(\ref{eq:KGinA}) but with a non-vanishing integration constant $\mathcal{C}$.  Outside the test 
mass, the field still obeys Eq.~(\ref{eq:KGinvacuum}). 
The constants of integration $\mathcal{C}$ and $\mathcal{D}$ are fixed by matching the solutions for $\phi$ and $\phi'$ given by Eqs.~\eqref{eq:phi_thin_shell} and \eqref{eq:phi_int_case3} at $r=S$, yielding,
\bea
  \mathcal{C}&=&\frac{1}{4\pi}\frac{\mA}{M} \frac{S^3}{\RA^3}, \\
  \mathcal{D}&=&\phi_{\rr A}-\frac{3}{8\pi}\frac{\mA}{M}\frac{S^2}{\RA^3}.
\eea
By matching the solutions for $\phi$ and $\phi'$ given by Eqs.~\eqref{eq:phi_int_case3} and \eqref{eq:KGinvacuum} at $r=\RA$, the last constant of integration $\mathcal{B}$ is given by,
\be
  \mathcal{B}=\frac{1}{4\pi}\frac{\mA}{M} \left(\frac{S^3}{\RA^3}-1\right),
\ee
assuming $\mbg\RA\ll1$.
%After matching $\phi$ and $\phi'$ at $r = \RA$ and $\phi$ at $r=S$,  the integration constants $\alpha$, $D$ and $C$ can be fixed. 
The resulting field profile in the Case (3) corresponding to the thin shell regime reads~\cite{Burrage},

%\begin{widetext}
 \bea \label{eq:case3}
 \phi^{(3)}(r) = \left\{
 \begin{matrix*}[l]
 & \phi_{\rr A}~, & 
 &  r < S, & \\
 &  \phi_{\rr A}+  \frac{\mA}{8 \pi \RA^3 M r} \left( r^3 - 3 S^2 r + 2 S^3 \right) ~, &   
 &  S<r<\RA, &  \\
 &   \phibg - \frac{\mA }{4 \pi M r} \rr e^{-m_{\rr{bg}} r} \left( 1 - \frac{S^3}{\RA^3}  \right)~, &
 &  r>\RA, &     
 \end{matrix*}
 \right.\non\\
 \eea
%\end{widetext}

 with the so-called thin-shell radius,
 \bea  \label{eq:radius_tt}
 S \equiv \RA \sqrt{1 - \frac{8 \pi M \RA \phibg}{3 \mA}},
 \eea

 being such that one has typically $(\RA - S) /\RA  \ll 1$.   
 The induced acceleration is well approximated ($\mbg\RA\ll1$) by,
 \be \label{eq:acc_phi_strong}
  a_\phi \approx \frac{\mA}{4 \pi  M^2 r^2 }  \left( 1 - \frac{S^3}{\RA^3}  \right) \simeq \frac{ \RA \phibg}{M r^2 },
 \ee
 and contrary to the Case (1), it is related to the value of $\phibg$. If the wall 
is sufficiently large, then the field reaches $\phi_\rr{min}(\rhoW)$ and so the calculation of $\phibg$ for a spherical chamber in 
\cite{Burrage} is valid,
\be  \label{phibgSP}
\phibg \simeq 0.69 \left(\Lambda^5 L^2\right)^{1/3}.
\ee
Following \cite{khoury}, $\phibg$ is rather given by,
\be \label{phibg_khoury}
  \phibg=\aleph\left[\alpha\left(\alpha+1\right)\Lambda^{4+\alpha}\rhoV\right]^\frac{1}{\alpha+2},
\ee
with $\aleph=1.6,~1.8$ if the vacuum chamber is assumed to be spherical or an infinite cylinder respectively.
Compared to the Case (1), the induced acceleration \eqref{eq:acc_phi_strong} does not only depend on $M$ but also on $\Lambda$ and on the size of the vacuum chamber $L$ (see Eq.~\eqref{phibgSP}). 
When $\Lambda$ is set to the cosmological constant and $L$ to the fiducial value reported in Table~\ref{tab:expconf}, one finds that the Berkeley experiment \cite{khoury} constrains the coupling parameter down to $M \sim 10^{15} \GeV$. The above calculation does not involve the power-law index $\alpha$ (apart indirectly via $m_{\rr{bg} }$, but there is no effect in the limit $m_{\rr{bg} } r \ll1$).  Therefore it is expected that the predictions are independent of $\alpha$, as long as $|A(\phi) - 1 | \ll 1$.  
%\textbf{SC: discussion on the effect of $\alpha$?}. 

% Analytical field profiles and induced    
% accelerations for case (3) are represented in Figs.~\ref{plot:profile_field_thinshell1} and \ref{plot:profile_acc_thinshell1} for the original chameleon model and in Figs.~\ref{plot:cham2_field} and \ref{plot:cham2_acc} for the exponential chameleon,  
% for several values of $M$ reported in Table~\ref{tab:profiles}.   Those are found to be in good agreement with the 
% numerical results, except close to the wall where important deviations are found.

\emph{Remark:} In the strongly perturbing regime, the reliability of the theory is questionable. Indeed the quantum corrections, either in the matter or in the chameleon sector must remain small. Most of the parameter space reachable by the Berkeley experiment \cite{khoury} belongs to this regime. Following \cite{Upadhye:2012vh} the underlying instabilities are harmless and the classical analysis is reliable, keeping in mind that quantum corrections can become large at very small scales.
However, since our aim consists of modeling how the environment can affect the analytical results derived for the classical field, we also provide numerical forecasts in the questionable strongly perturbing regime. 
Nevertheless we do not explore the deeply strongly perturbing regime but focus on the transition between the strongly and the weakly perturbing regimes, where the numerical computations allow one to follow the smooth evolution of the field and acceleration profiles whereas analytical assumptions break.  Our computations show that the analytical estimations are recovered once in the strongly perturbing regime, and that they are quite reliable, at least classically.  The underlying quantum aspects are not discussed in this thesis.

\item{Case~(4): $\phi(0) =  \phi_{\rr{min}}(\rho_A) $ and $\phibg = \phi_{\rr{min}}(\rho_{\rr v}) $}

In the Case (4) the field profile is governed by Eq.~(\ref{eq:case3}) since the field reaches the effective potential 
minimum at the center of the source mass. However, as long as the 
condition Eq.~(\ref{eq:case2condition}) is satisfied, $\phibg = \phi_{\rr{min}}\left(\rho_{\rr v}\right)$.  For the 
original chameleon potential $ 
V(\phi) = \Lambda^5 /\phi $, the Case (4) takes place when $M \lesssim 10^{-3 } \GeV$, whereas for the exponential potential $ V(\phi) 
= \Lambda^4 (1+ \Lambda /\phi) $ one needs $M 
\lesssim 10 \GeV$ in order to reach the strongly perturbing regime inside the source mass.   Therefore the Case (4)  is irrelevant 
for values of $\Lambda$ compatible with cosmology and realistic experimental configurations. 

\end{itemize}

\section{Parameter space}\label{sec:param_space}
It is possible to understand the shape of the viable parameter space depicted on Fig.~\ref{fig:exclusion} in the light of the analytical computations. Following \cite{Burrage, khoury}, it is possible to rewrite the acceleration $a_\phi$ given by Eqs.~\eqref{eq:acc_phi_weak} and \eqref{eq:acc_phi_strong} as,
\be
  a_\phi=\frac{2 \GN \mA}{r^2}\lambda_\rr{at}\lambda_\rr{A}\left(\frac{\Mp}{M}\right)^2
  =\frac{8\pi \GN\rhoA}{3}\frac{\RA^3}{r^2}\lambda_\rr{at}\lambda_\rr{A}\left(\frac{\Mp}{M}\right)^2,
\ee
with $\GN=(8\pi\Mp^2)^{-1}$, $\mA=\rhoA (4/3)\pi \RA^3$ and,
\bea
  \lambda_{i}\simeq\left\{
      \begin{array}{ll}
       1 & \hspace{1cm} \rho_i r_i^2< 3 M \phibg,\\
       1-\frac{S_i^3}{r_i^3} & \hspace{1cm} \rho_i r_i^2 >3 M \phibg,
      \end{array}
  \right.
\eea
corresponding to the weakly and strongly perturbing regimes respectively, the $i$ subscript denoting the species under consideration (atoms or the source mass). 

Four regimes are distinguishable in Fig.~\ref{fig:exclusion}, depending on whether the source mass and/or the atoms are screened as well as on $\phibg=\phi_\rr{min}(\rhoV)$ or given by Eq.~\eqref{phibg_khoury}. In Fig.~\ref{fig:exclusion}, the acceleration is normalized with respect to the Earth's acceleration of free fall $g=\GN M_\oplus /R_\oplus^2=(4/3)\pi \GN R_\oplus \rho_\oplus$, $M_\oplus$, $R_\oplus$ and $\rho_\oplus$ denoting the Earth mass, radius and density respectively. The normalized acceleration then reads,
\be
  \frac{a_\phi}{g}=\frac{2\rhoA}{R_\oplus\rho_\oplus}\frac{\RA^3}{r^2}\lambda_\rr{at}\lambda_\rr{A}\left(\frac{\Mp}{M}\right)^2.
\ee
For large values of $M$ the source mass is unscreened (and a fortiori the atoms are), so $\lambda_\rr{at}=\lambda_\rr{A}=1$, the normalized acceleration reading,
\be
  \frac{a_\phi}{g}^{\rr{(i)}}=2 \left(\frac{\Mp}{M}\right)^2\frac{\rho_A \RA^3}{r^2 R_\oplus \rho_\oplus}.
\ee
This regime corresponds to the Case~(1) in App.~\ref{sec5} and is marked by the vertical line at the top right of Fig.~\ref{fig:exclusion}. If the source mass strongly perturbs the chameleon field whereas the atoms remain unscreened then $\lambda_\rr{at}=1$ while $\lambda_\rr{A}\simeq 3 M\phibg /(\rho_A \RA^2)$ according to Eq.~\eqref{eq:radius_tt}. This regime corresponds to the Case~(3) in App.~\ref{sec5} and the acceleration then yields (see Eq.~\eqref{eq:acc_phi_strong}),
\bea
  \frac{a_\phi}{g}^{\rr{(ii)}}&=&\frac{6 \Mp^2}{r^2 R_\oplus \rho_\oplus}\frac{\phibg \rho_A}{M}, \\
  &=&\frac{6 \Mp^2}{r^2 R_\oplus \rho_\oplus}\frac{\rho_A}{M} \aleph\left[\alpha\left(\alpha+1\right)\Lambda^{4+\alpha}\rhoV\right]^\frac{1}{\alpha+2},
\eea
with $\phibg$ given by Eq.~\eqref{phibg_khoury}. In the previous section we implicitly assumed that the atoms remain unscreened, which is true only as long as $\rho_\rr{at} R_\rr{at} < 3 M \phibg$. Otherwise $\lambda_\rr{at}\simeq 3 M\phibg /(\rho_\rr{at} R_\rr{at}^2)$ and the acceleration becomes,
\bea
  \frac{a_\phi}{g}^{\rr{(iii)}}&=&\frac{18 \Mp^2}{\rho_\rr{at} R_\rr{at}^2 R_\oplus \rho_\oplus}\frac{\RA}{r^2}\phibg^2, \\
  &=&\frac{18 \Mp^2}{\rho_\rr{at} R_\rr{at}^2 R_\oplus \rho_\oplus}\frac{\RA}{r^2}\aleph\left[\alpha\left(\alpha+1\right)\Lambda^{4+\alpha}\rhoV\right]^\frac{2}{\alpha+2}.
\eea
In this case the acceleration is independent of $M$.
If the vacuum chamber is larger than the Compton wavelength of the chameleon (see Eq.~\eqref{eq:case2condition}), then the chameleon reaches its attractor inside the vacuum chamber such that $\phibg$ is given by $\phi_\rr{min}\left(\rhoV\right)$, yielding,
\bea
  \frac{a_\phi}{g}^{\rr{(iv)}}&=&\frac{18 \Mp^2}{\rho_\rr{at} R_\rr{at}^2 R_\oplus \rho_\oplus}\frac{\RA}{r^2}\phibg^2, \\
  &=&\frac{18 \Mp^2}{\rho_\rr{at} R_\rr{at}^2 R_\oplus \rho_\oplus}\frac{\RA}{r^2}
  \left(\frac{\alpha \Lambda^{\alpha+4}M}{{\rhoV}}\right)^{\frac{2}{\alpha+1}}.
\eea
In this latter case, $a_\phi$ depends on $M$ (see on the bottom left of Fig.~\ref{fig:exclusion}). This latter regime is not tractable by our numerical simulations, as discussed in Sec.~\ref{sec4}. Notice also that atom interferometry is not able to test the strongly coupled chameleon ($M>\mpl$).

\chapter{Numerical methods for the Higgs monopoles} % Main chapter title

\label{app:num_higgs_monop} % For referencing the chapter elsewhere, use \ref{Chapter1} 

\lhead{Appendix B. \emph{Awesome appendix B}} % This is for the header on each page - perhaps a shortened title

%----------------------------------------------------------------------------------------

\noindent In Sec.~\ref{sec:num_monop} we present monopole solutions obtained by a simplified integration method, considering the Klein-Gordon equation only. To show that this method is accurate, here we present the results obtained by integrating the full set of equations of motion, namely the Klein-Gordon together with the Einstein equations \eqref{eom_tensor}. This result confirms  that we can safely replace the metric inside the compact object by the standard GR metric as explained in Sec.~\ref{sec:effective_dyn_monop}.

\section{Equations of motion}
\noindent We first list the full set of equations to be solved. The explicit  $tt-$, $\theta\theta-$ and $rr-$ components of the Einstein equations \eqref{eom_tensor} are, respectively,
%\begin{widetext}
\bea
\nu''+\nu'^{2}-\lambda'\nu'+(\nu'-\lambda')\left({1\over r}+{H'\over F }{\dd F\over \dd H}  \right)+{1\over F}{\dd F\over 
\dd H}\left(H''+{H'\over r}\right)\non\\
+{H'^{2}\over F}\left({\kappa\over 2}+{\dd^{2}F\over \dd H^{2}}\right)+\left(\kappa 
V-{3p\over \mathfrak{R}^{2}\rho}\right){\rr{e}^{2\lambda}\over F}=0,
\label{eom_r}
\eea
\bea
\lambda'\left({2\over r}+{H'\over F}{\dd F\over \dd H}\right)-{H''\over F}{\dd F\over \dd H}
-{H'^{2}\over 2F}\left(\kappa+2{\dd^{2}F\over \dd H^{2}}\right)-{2H'\over rF}{\dd F\over \dd H}
\non\\
-{1\over r^{2}}\left(1-\rr{e}^{2\lambda}\right)-{\rr{e}^{2\lambda}\over F}\left(\kappa V+{3\over \mathfrak{R}^{2}}\right)=0,
\label{eom_t}
\eea
\bea
\nu'\left({2\over r}+{H'\over F}\frac{\dd F}{\dd H}\right)-\frac{\kappa H'^{2}}{2F}+{1\over r^{2}}\left(1-\rr{e}^{2\lambda}\right)
+{2H'\over rF}\frac{\dd F}{\dd H}\non\\
-{\rr{e}^{2\lambda}\over F}\left(\frac{3p}{\mathfrak{R}^{2}\rho}-\kappa V\right)=0,
\label{eom_theta}
\eea
%\end{widetext}
where the prime denotes a derivative with respect to $r$ and $\mathfrak{R}^2={\mathcal{R}^3/ r_\rr{s}}$, $r_\rr{s}$ being the standard Schwarzschild radius. We assume a top-hat density profile \eqref{eq:rho_monop} so that $\mathfrak{R}=3/(\kappa\rho_0)$. Finally, the Klein-Gordon equation reads,
\bea\non
H''-H'\left(\lambda'-\nu'-{2\over r} \right)+\rr{e}^{2\lambda} \left({R\over 2\kappa} \frac{\dd F}{\dd H} - \frac{\dd V}{\dd H}\right)=0, \\
\label{KG_static}
\eea
where the Ricci scalar $R$ is given by,
\bea\non
R\!=\!{2\over r^2}-\rr{e}^{-2\lambda}\!\left(\!2\nu''-2\nu'\lambda'  +{4\nu'\over r}+{2\over r^{2}} -{4\lambda'\over r}+2\nu'^{2}\!\right)\!.\\
\label{analyticalR}
\eea

\section{Dimensionless system}
\noindent  To implement the numerical integration we need to write the above equations in a convenient dimensionless units system. This step is actually crucial in order to extract significant numerical results because of the involved scales, like the Planck mass. 
We first rescale the Higgs field as,
\bea
H[\mathrm{GeV}]= \mpl \tilde{v} h= \mpl \tilde{v} \left(1+\chi\right),
\label{change_H}
\eea 
where $h$ and $\tilde{v}=246\; \mathrm{GeV}/\mpl$ are dimensionless. The quantity  $\chi$ characterizes the dimensionless displacement of the Higgs scalar field around its vev. 
We express the radial coordinate in term of $r_\rr{s}$, 
\be
u={r\over r_\rr{s}},
\label{change_r}
\ee
and we remind that the Schwarzschild radius in Planck units is,
\be
r_\rr{s} [\mathrm{GeV^{-1}}]=\frac{2 m}{\mpl^2} \times 5.61 \times 10^{26} [\mathrm{GeV\; kg^{-1}}],
\ee
where $m$ is the baryonic mass of the monopole in [kg]. The numerical factor converts mass from [kg] to 
[GeV] in such a way that the units are consistent. We also define the dimensionless potential,
\be
\mathbb{V}=V {r_\rr{s}^2\over \mpl^2},
\ee
which becomes according to the definition \eqref{change_H},
\be
\mathbb{V}\left(\chi\right)=\frac{\lambda_{\rm sm}}{4} \mpl^2 r_\rr{s}^2 \tilde{v}^4 \chi^2 \left(2+\chi\right)^2.
\ee
Finally, we define the dimensionless coupling function in an analogous way, as,
\be
\mathbb{F}\left(\chi\right)=1+\xi \tilde{v}^2 \left(1+\chi\right)^2.
\ee

\section{Numerical integration method}

\noindent There exist different ways to perform the numerical integration of the Eqs.\ \eqref{eom_r}-\eqref{KG_static}. We choose to treat them like an IVP, by integrating from the center of the body. We first find the internal solution and then use it at the boundary of the compact object to fix the initial conditions for the external solution. We choose to solve the system of  
equations with respect to the variables $\lambda$, $\nu$, $h$ and $p$ since $\rho=\rho_0$ is constant. In addition to the equations of motion, we must consider the TOV equation,
\be
p_u=-\nu_u \left(p+\rho_0\right), 
\ee
where a subscript $u$ denotes a derivative with respect to $u$. 
In the top-hat profile approximation, this equation admits the exact solution,
\be
\frac{p}{\rho_0}=\mathcal{C} e^{-\nu} -1,
\label{TOV_int}
\ee
where $\mathcal{C}$ is a constant of integration to be fixed by the numerical shooting method. In order to find a guess for the shooting method, we use the standard GR expression for the pressure \eqref{eq:pressureGR}. This is a very good approximation since a small discrepancy between the GR solution and the numerical one is expected, as explained in Sec.~\ref{sec:analyt_monop}. In our units, Eq.\ \eqref{TOV_int} becomes (see also Sec.~\ref{sec:schwa_in}),
\bea
{p(u)\over \rho_0}=\left[{\sqrt{1-s}-\sqrt{1-s^3 u^2}}\over \sqrt{1-s^3 u^2}-3\sqrt{1-s} \right].
\label{pressure_GR}
\eea
Imposing the initial condition $\nu(u=0)=0$ \footnote{Note that $\nu(u=0)=0$  is not a regularity condition, which is instead given by asymptotic flatness, namely
 $\nu\left(u\rightarrow \infty\right)=0$. Since we solve an IVP, we prefer to fix $\nu(u=0)=0$ and then shift the solution to
$\nu(u)-\nu_\rr{end}$, without loss of generality.}, $\mathcal{C}$ reads, 
\be
\mathcal{C}={p_\rr{c}\over \rho_0}+1,
\ee
where $p_\rr{c}=p(u=0)$ is the pressure at the center. Then we optimize the value of $\mathcal{C}$ in such a way that it satisfies also the boundary condition for the pressure at the boundary $p(u=1/s)=0$. In addition, this method has the advantage that it allows to test the limit for the central pressure  coming from GR \eqref{eq:lim_pressureGR},
\be
p(u=0)\longrightarrow\infty \Leftrightarrow \mathcal{R}={9m\over 4}.
\ee
It turns out that the difference between $\mathcal{C}$ in GR and in our model is so small that the two solutions are undistinguishable so this step can be safely neglected.
 
Therefore, we are left with the four equations  Eqs.~\eqref{eom_r},~\eqref{eom_t},~\eqref{eom_theta} and \eqref{KG_static}. 
Of these, we keep Eq.~\eqref{eom_theta} as the Hamiltonian constraint and we integrate the other three as an IVP.
The initial conditions for $\lambda$, $\nu_u$ and $h_u$ are obtained from the regularity conditions of the solution at the center of 
the Higgs monopole,
\bea
\lambda(0)=0,
\\
\nu_u(0)=0,
\\
h_u(0)=0.
\eea
In addition, we need to choose a value for  $h_\rr{c}$ to begin the integration. We know that this value is an irrational number that can be determined numerically with finite accuracy only. Thus,  the basic idea of our algorithm consists of incrementing the value of $h_\rr{c}$ digit by digit for a given number of digits. We also have an indication that makes integration easier. Indeed, we saw in Sec.~\ref{sec:effective_dyn_monop} that if $h_\rr{c}$ is larger than  $h^{\rm in}_{\rm eq}$ (see Eq.\ \eqref{h_eq_in}) then it never reaches the vev at spatial infinity. So, we can stop the integration as soon as $h>h_{\rm eq}^{\rm in}$ and reject the chosen value of $h_\rr{c}$. Therefore, we begin by integrating from the 
approximate value of $h_{\rm eq}^{\rm in}$ (we recall that this value is calculated in the approximation that the internal solution is the same as GR) and if $h$ becomes larger than $h_{\rm eq}^{\rm in}$ we stop the integration, we keep the previous digit and perform once again numerical integration with a value of $h_\rr{c}$ incremented by one less significant digit. Otherwise, namely when the 
Higgs field does not become higher than $h_{\rm eq}^{\rm in}$ and is trapped into the local minimum of the effective 
potential $h=0$, we increment the same digit. With this algorithm, we are able to maximize the precision on $h_\rr{c}$ in 
the limit of the precision we impose or, in other words, we are able to push back the radial distance from the center of the body 
at which the scalar field is trapped into the local minimum of the effective potential $h=0$. 

We have also to take care of the ``degeneracy" of the  solution at spatial infinity. Indeed, 
the scalar field can tend to $\pm v$. So, when we perform numerical integration for different values of the parameters, we have to choose between the positive and  the negative solution.

\begin{figure}
\begin{center}
\includegraphics[width=0.6\textwidth, trim=280 0 310 0,clip=true]{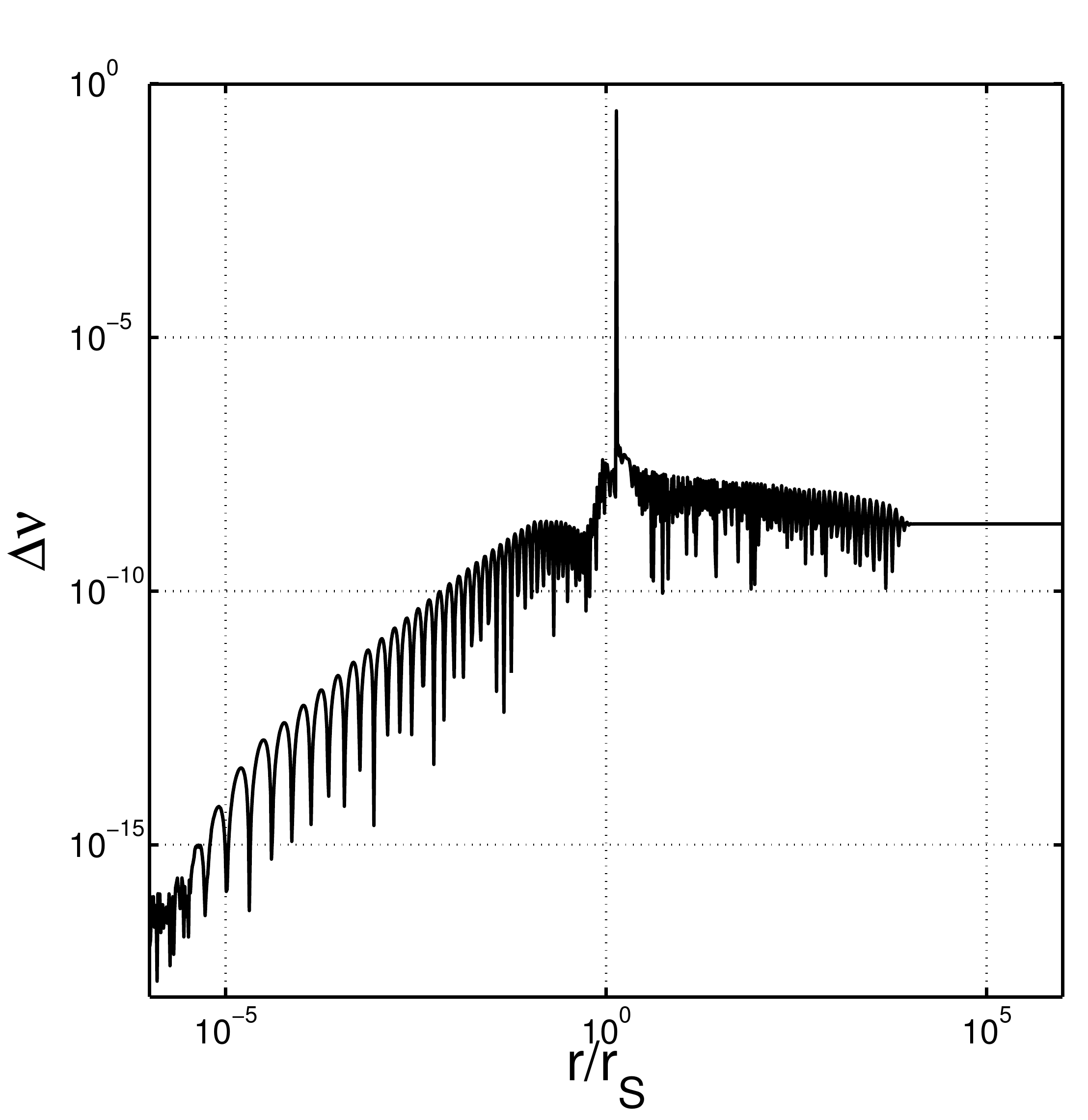}
\caption{Hamiltonian constraint for the same monopole solution as represented in Fig.~\ref{different_hcs} ($\xi=10$, $m=10^6$~kg and $s=0.75$) obtained with the full numerical method.}
\centering
\label{plot_hamil_cons}
\end{center}
\end{figure}

In order to check the validity of our numerical code, we plot in Fig.\ \ref{plot_hamil_cons} the Hamiltonian constraint for the same monopole solution as represented in Fig.~\ref{different_hcs} and obtained with the full numerical integration. Here, the Hamiltonian constraint is defined as the absolute value of the difference between $\nu_u$ coming from Eq.\ \eqref{eom_theta},
where we replaced the values of all fields with the ones found numerically, and the value of $\nu_u$
determined numerically by solving the system of equations. The divergence appearing at the boundary of the monopole comes from the abrupt transition of energy density due to the top-hat approximation. Otherwise, the order of magnitude of the Hamiltonian constraint corresponds to what we expect from the numerical precision.

\begin{sloppypar}
\section{Comparison between the full integration method and the simplified one}
\begin{figure}
\begin{center}
\includegraphics[width=0.6\textwidth, trim=280 0 310 0,clip=true]{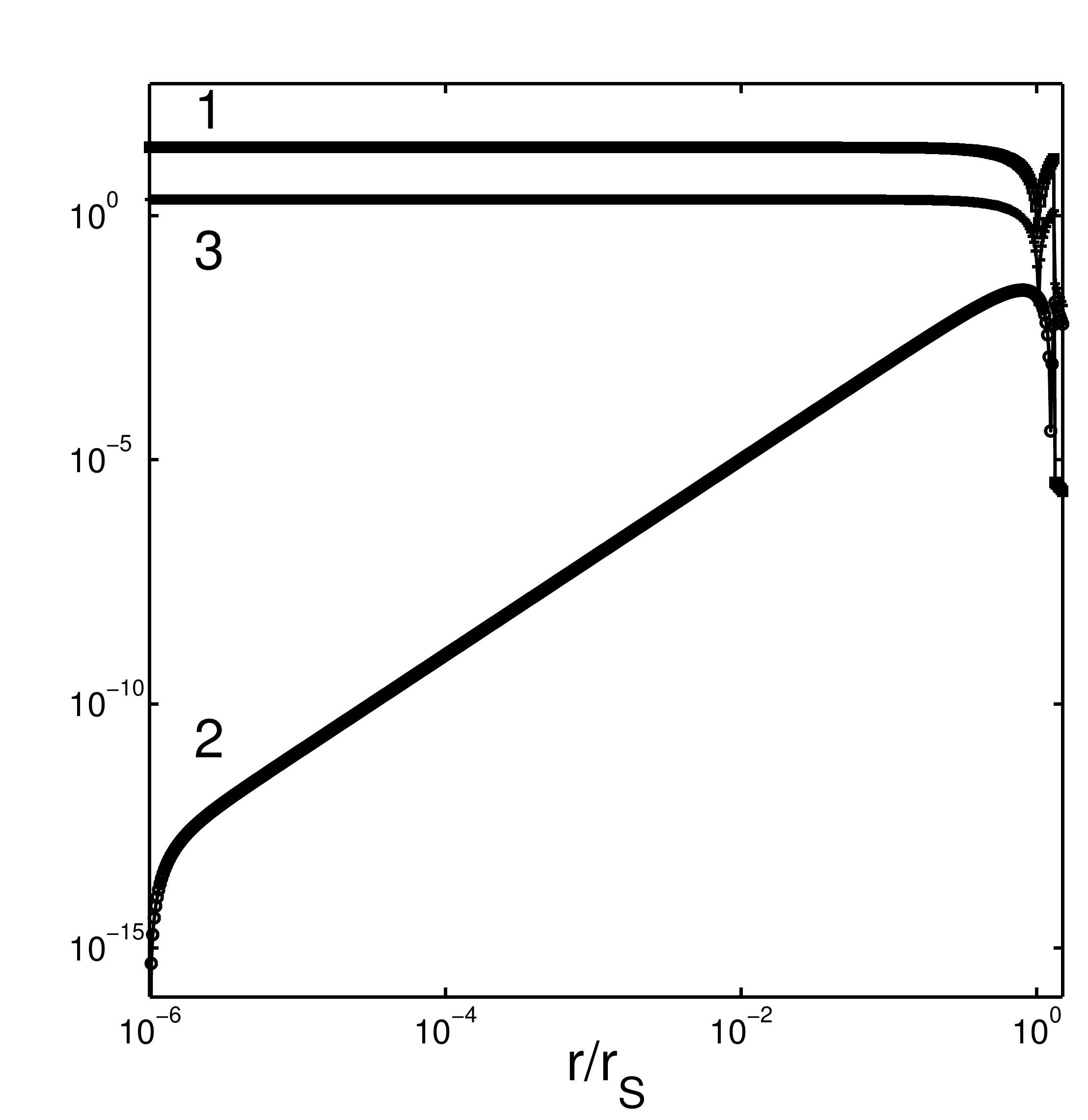}
\caption{Plot of  the trace of the stress-energy tensor contributions $T^{(h)}$ (curve 2),  $T^{(\xi)}$ (curve 3) and of the left-hand side of Eq.~\eqref{eom_tensor} (curve 1). The parameters are chosen as $m=10^6~\rm{kg}$, $s=0.75$, and $\xi=10$.}
\centering
\label{plot_trace_energy_momentum}
\end{center}
\end{figure}

\begin{figure}
\begin{center}
\includegraphics[width=0.6\textwidth, trim=280 0 310 0,clip=true]{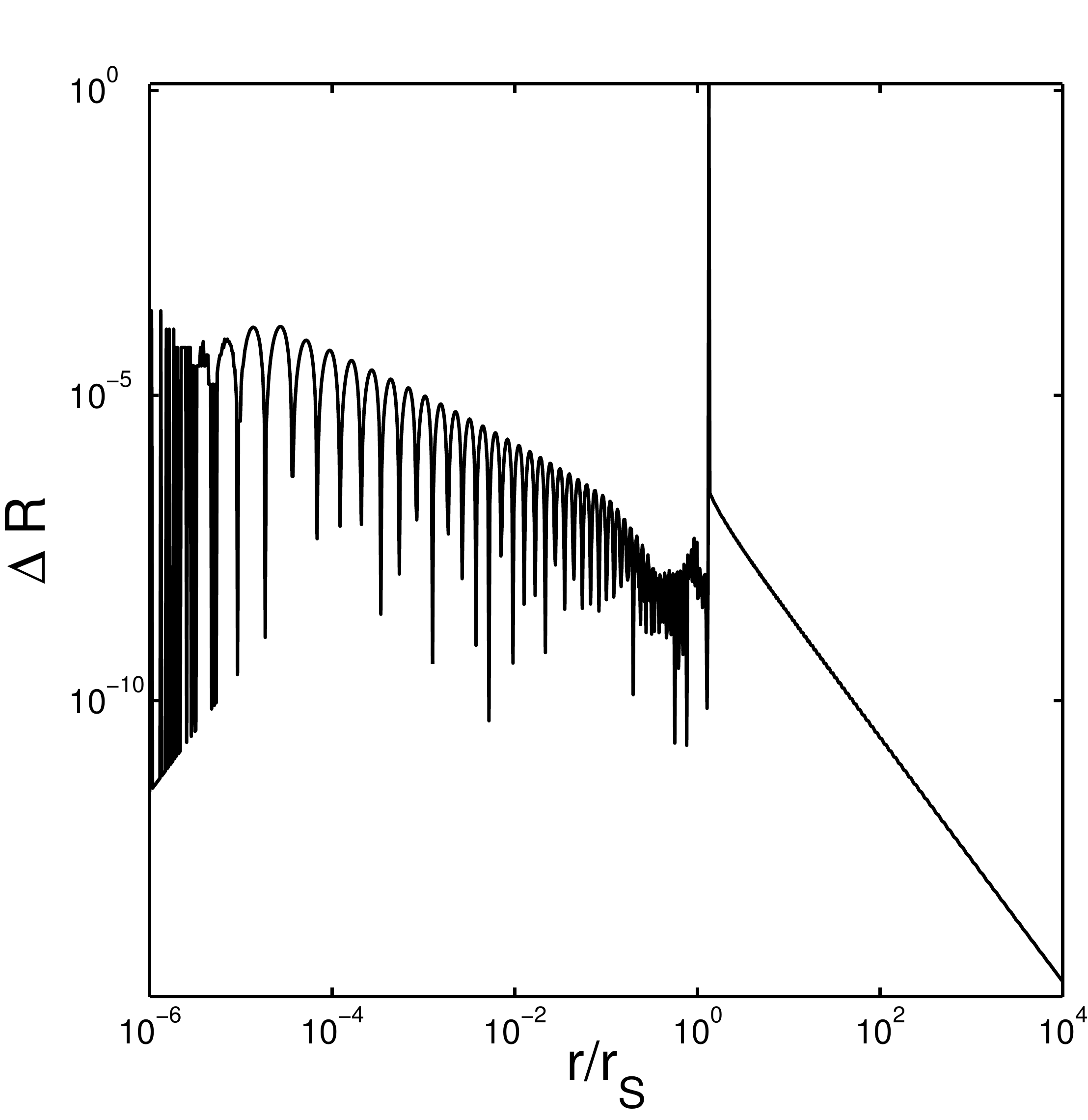}
\caption{Absolute value of the difference between the standard GR curvature scalar and its value calculated with our numerical algorithm for $m=10^6~\rm{kg}$, $s=0.75$ and $\xi=10$.}
\centering
\label{plot_compare_scalar_curvature}
\end{center}
\end{figure}

\end{sloppypar}
\noindent In Sec.\ \ref{sec:analyt_monop} we argue that we can safely neglect the Higgs field inside the body as long as it is sufficiently constant 
and not too much displaced from its vev. This means that, instead of integrating the whole system of equations, we could use the inner Schwarzschild expressions for 
$\lambda$ and $\nu$ (Eqs.\ \eqref{nu} and \eqref{lam}) and integrate only the Klein-Gordon equation as an IVP. 
We now demonstrate the correctness of this claim by comparing our results with a complete numerical integration. We first plot on Fig.\ \ref{plot_trace_energy_momentum} the contribution of each term appearing in the trace of Eq.\ \eqref{eom_tensor} obtained with the full numerical integration in the case when $m=10^6~\rm{kg}$, $s=0.75$ and $\xi=10$. 
In the dimensionless unit system, the contributions of the trace of the stress-energy tensor are given by,
\bea
\frac{r_\rr{s}^2}{\mpl^2} T^{(h)}=-h_u^2+2 V(h), 
\eea
and,
\bea\non
\frac{r_\rr{s}^2}{\mpl^2} T^{(\xi)}&\!\!=\!\!&\frac{3\xi}{4\pi} \Bigg\{h_u^2+h \rr{e}^{-2\lambda}\Bigg[h_{uu}-h_u\left(\lambda_u-\nu_u-\frac{2}{u}\right)\!\!\Bigg]\!\Bigg\}.\\
\eea
We observe in Fig.\ \ref{plot_trace_energy_momentum} that the geometric part is clearly dominant while the contribution coming from the stress-energy tensor components of the scalar field is negligible. This result is confirmed by the comparison of the Ricci scalar given by Eq.\ \eqref{scalR} and Eq.\ \eqref{analyticalR} evaluated numerically. In Fig.\ \ref{plot_compare_scalar_curvature} we plotted the absolute value of the difference between the two expressions  in function of the radial distance for the same parameters as in Fig.\ \ref{plot_trace_energy_momentum}. The difference is clearly negligible while the peak at the boundary of the body is caused only by the top-hat approximation for the energy density.

\begin{figure}
\begin{center}
\includegraphics[width=0.6\textwidth, trim=280 0 310 0,clip=true]{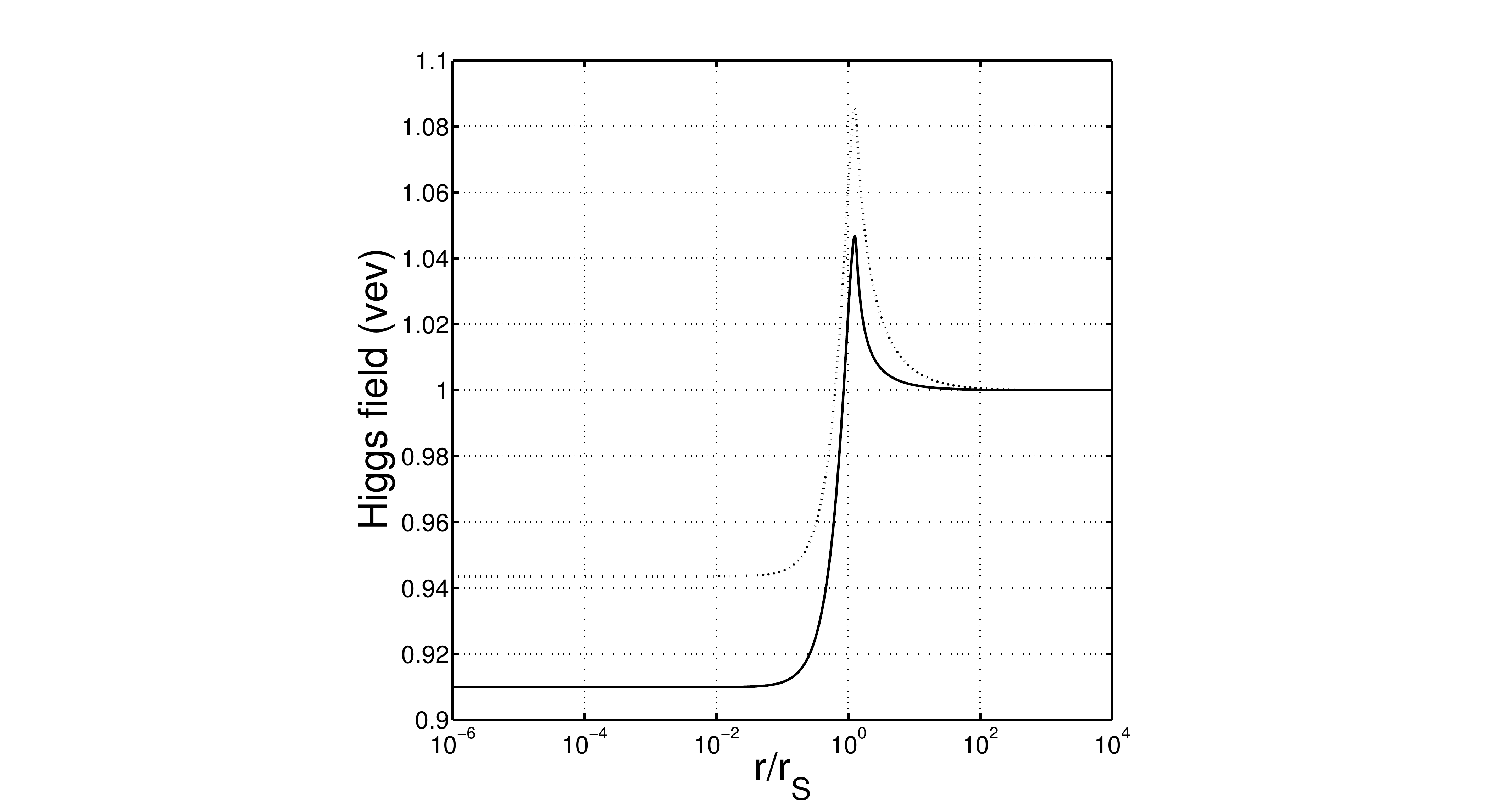}
\caption{Numerical solutions for the monopole with $m=10^6~\rm{kg}$, $s=0.75$, and $\xi=10$ obtained with the full numerical integration and the simplified one.
The difference between the two solutions becomes apparent only inside the body and is negligible outside the body.}
\centering
\label{plot_compare_num_int}
\end{center}
\end{figure}

\begin{figure}
\begin{center}
\includegraphics[width=0.6\textwidth, trim=340 0 340 0,clip=true]{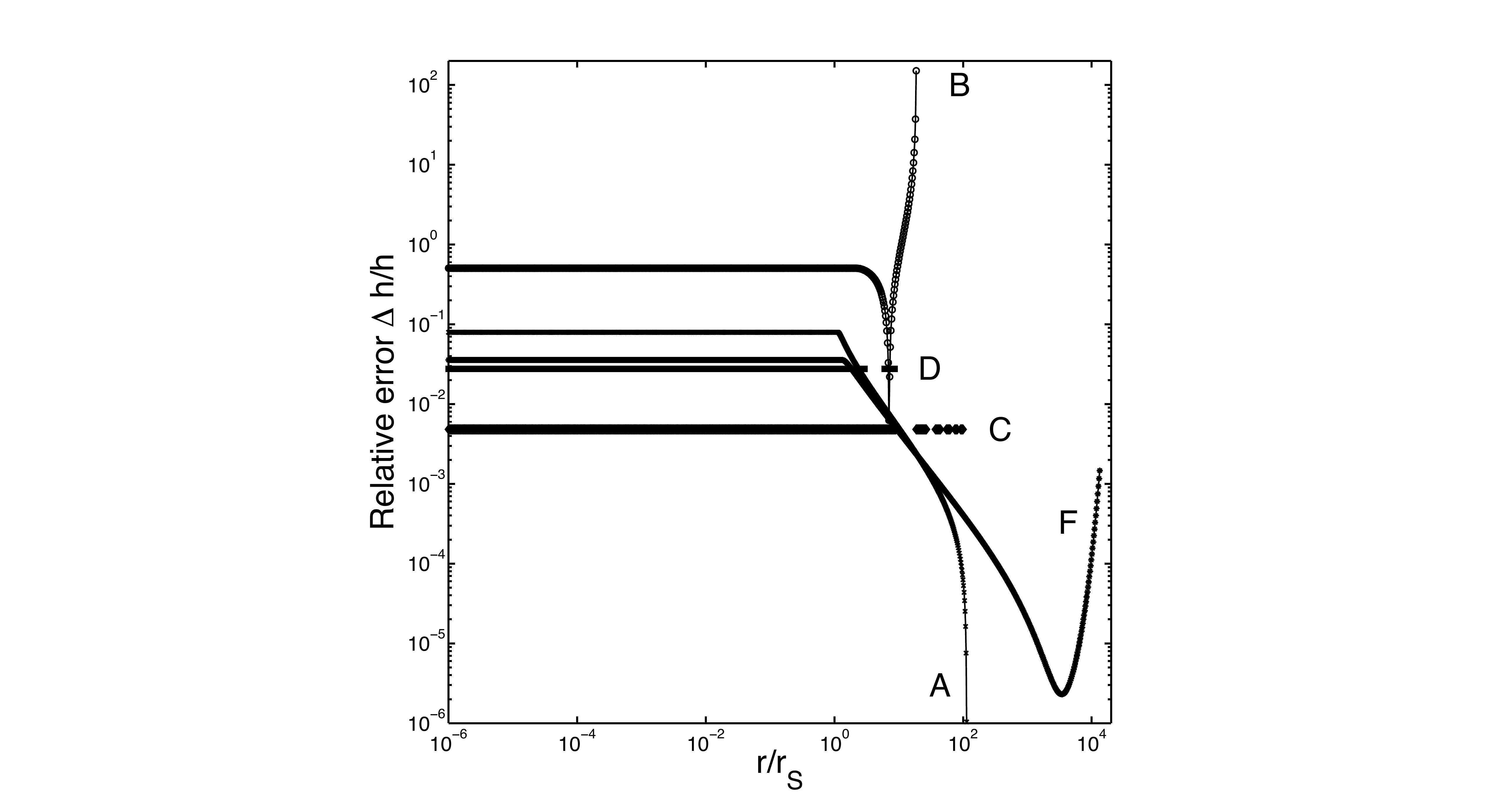}
\caption{Relative error between the Higgs field solutions obtained with the full numerical method and the simplified one. Labels refer to Higgs monopoles listed in Tab.~\ref{table}.}
\centering
\label{plot_rel_error}
\end{center}
\end{figure}

As a further check,  we plot the Higgs field profiles obtained with the two numerical methods in Fig.\ \ref{plot_compare_num_int}  for $\xi=10$, $m=10^6~\rm{kg}$, and  $s=0.75$. The discrepancy inside the body appears only because  the scalar field contribution is neglected in the simplified model. 
In order to get a quantitative result, we plot on Fig.\ \ref{plot_rel_error} the relative errors between the Higgs field solutions obtained with the full numerical method and the simplified one
for various monopole solutions. In general, we see that there is a very good agreement between numerical and approximate solutions only for small compactness.

\chapter{Fab Two: equations of motion and ghosts conditions} % Main chapter title

\label{app:fabtwo} % For referencing the chapter elsewhere, use \ref{Chapter1} 

\lhead{Appendix A. \emph{Awesome appendix A}} % This is for the header on each 
%page - perhaps a shortened title

%-------------------------------------------------------------------------------
We report in this appendix some of the calculations used in Chap.~\ref{chap:FabFour}.

\section{Equations of motion for the Fab Two}\label{sec:eom_fabfour}
In this section, we focus on the non-standard kinetic term of the Fab-Four Lagrangian,
\begin{eqnarray}
  S_\rr{John} & = & \int \dd^{4}x\,\sqrt{-g}\,\phi^{\alpha}\phi^{\beta}\, G_{\alpha\beta},\nonumber \\
  & = & \int \dd^{4}x\,\sqrt{-g}\, g^{\rho\alpha}g^{\sigma\beta}\,\phi_{\rho}\phi_{\sigma}\,\left(R_{\alpha\beta}-\frac{1}{2}R\, g_{\alpha\beta}\right),
 \label{eq:action_non-min}
\end{eqnarray}
where we adopted the convention $\phi_\alpha=\nabla_\alpha \phi$. The variation of $S_\rr{John}$ reads,
\begin{eqnarray}
  E_{\mu\nu}\equiv\frac{1}{\sqrt{-g}}\frac{\delta S_\rr{John}}{\delta g^{\mu\nu}} & = & 			\phi^{\rho}\phi^{\sigma}G_{\rho\sigma}\frac{\delta\sqrt{-g}}{\delta g^{\mu\nu}}
  +\phi_{\rho}\phi^{\beta}G_{\alpha\beta}\frac{\delta g^{\rho\alpha}}{\delta g^{\mu\nu}}
  +\phi^{\alpha}\phi_{\sigma}G_{\alpha\beta}\frac{\delta g^{\sigma\beta}}{\delta g^{\mu\nu}} \non\\
  &&+\phi^{\alpha}\phi^{\beta}\frac{\delta R_{\alpha\beta}}{\delta g^{\mu\nu}}
  -\frac{1}{2}\phi^{\alpha}\phi_{\alpha}\frac{\delta R}{\delta g^{\mu\nu}}
  -\frac{1}{2}\phi^{\alpha}\phi^{\beta} R \frac{\delta g_{\alpha\beta}}{\delta g^{\mu\nu}}, \nonumber \\
  & = & -\frac{1}{2}g_{\mu\nu}\,\phi^{\rho}\phi^{\sigma}\, G_{\rho\sigma}+2\,\phi_{(\mu}R_{\nu)\alpha}\,\phi^{\alpha}\nonumber \\
  &  & \qquad -\frac{1}{2}\, R\,\phi_{\mu}\phi_{\nu}
  \underbrace{+\phi^{\rho}\phi^{\sigma}\frac{\delta R_{\rho\sigma}}{\delta g^{\mu\nu}}}_{(\mathcal{T}_1)}
  \underbrace{-\frac{1}{2}\phi_{\alpha}\phi^{\alpha}\frac{\delta R}{\delta g^{\mu\nu}}}_{(\mathcal{T}_2)}
\label{eq:variation_non-min_initiale},
\end{eqnarray}
using Eqs.~\eqref{eq:variation_metric}--\eqref{eq:variation_det-metric} and adopting the convention on the symmetrization of indices,
\be
  T_{(\mu\nu)}\equiv\frac{1}{2}\left(T_{\mu\nu}+T_{\nu\mu}\right),
\ee
for any tensor $\mathbf{T}$. The variation of the Ricci scalar is given by Eq.~\eqref{eq:variation_scalR} while the one of the Ricci tensor reads, 
\bea
  \frac{\delta R_{\rho\sigma}}{\delta g^{\mu\nu}}=
  \frac{1}{2}\left[g_{\rho\mu}g_{\sigma\nu}\square
  +g_{\mu\nu}\nabla_{\rho}\nabla_{\sigma}
  -2\, g_{\mu(\rho}\nabla_{\sigma}\nabla_{\nu)}\right].
  \label{eq:variation_ricci}
\eea
The term $(\mathcal{T}_1)$ becomes,
\bea
  (\mathcal{T}_1)&=&\frac{1}{2}\,\phi_{\mu}\phi_{\nu}\square +\frac{1}{2}\phi^{\rho}\phi^{\sigma}g_{\mu\nu}\nabla_{\rho}\nabla_{\sigma}
  -\phi_{(\mu}\phi^{\sigma}\nabla_{\sigma}\nabla_{\nu)},\\
  & = & \frac{1}{2}\square\left(\phi_{\mu}\phi_{\nu}\right) + \frac{1}{2}\, g_{\mu\nu}\,\nabla_{\sigma}\nabla_{\rho}\left(\phi^{\rho}\phi^{\sigma}\right)
  -\nabla_{\sigma}\nabla_{(\nu}\left(\phi_{\mu)}\phi^{\sigma}\right),
\eea
by integrating by parts twice the expression and neglecting the boundary terms. Using the commutation relations between the covariant derivatives,
\bea
  \left[\nabla_{\mu},\nabla_{\nu}\right]\phi & = & 0,\\
  \left[\nabla_{\mu},\nabla_{\nu}\right]V^{\rho} & = & R_{\kappa\mu\nu}^{\rho}V^{\kappa},
  \label{eq:defaut_commut}
\eea
for any vector field $\mathbf{V}$, the term $(\mathcal{T}_1)$ eventually yields,
\bea
  (\mathcal{T}_1)=-\phi_{\mu\nu}\square\phi -\, \phi_{\mu\nu\sigma}\phi^{\sigma}
  \qquad\qquad\qquad\qquad\qquad\qquad\qquad\non\\
  + \frac{1}{2}g_{\mu\nu}\left[\left(\square\phi\right)^{2}+\phi^{\rho\sigma}\phi_{\rho\sigma}
  +\phi^{\rho}\left(\phi_{\;\rho\sigma}^{\sigma}+\phi_{\;\sigma\rho}^{\sigma}\right)\right].
\eea
In the same way, the term $(\mathcal{T}_2)$ becomes,
\bea
  (\mathcal{T}_2)&=&-\frac{1}{2}\left[ R_{\mu\nu}\left(\partial\phi\right)^{2}+\phi_{\alpha}\phi^{\alpha}\left(g_{\mu\nu}\square-\nabla_{\mu}\nabla_{\nu}\right)\right], \\
  &=&-\frac{1}{2}\left[R_{\mu\nu}\left(\partial\phi\right)^{2} +g_{\mu\nu}\square\left(\phi_{\alpha}\phi^{\alpha}\right)
  -\nabla_{\nu}\nabla_{\mu}\left(\phi_{\alpha}\phi^{\alpha}\right)\right],\\
   &=&-\frac{1}{2}\Big[ R_{\mu\nu}\left(\partial\phi\right)^{2}
  +2\, g_{\mu\nu}\left(\phi_{\rho\zeta}^{\;\;\zeta}\phi^{\rho}+\phi_{\rho\zeta}\phi^{\rho\zeta}\right)\Big.\non\\
  & & \qquad \Big.-2\,\left(\phi_{\rho\mu\nu}\phi^{\rho}+\phi_{\rho\mu}\phi_{\;\nu}^{\rho}\right)\Big].
\eea
By recombining all the terms we finally obtain,
\bea
%\begin{array}{ccc}
%\frac{1}{\sqrt{-g}}\frac{\delta S_\rr{John}}{\delta g^{\mu\nu}}
  E_{\mu\nu} & = & -\frac{1}{2}R\,\phi_{\mu}\phi_{\nu}+2\,\phi_{(\mu}R_{\nu)\alpha}\,\phi^{\alpha}-\frac{1}{2}G_{\mu\nu}\left(\nabla\phi\right)^{2}
  +R_{\mu\alpha\nu\beta}\phi^{\alpha}\phi^{\beta}
  +\phi_{\rho\mu}\phi_{\;\nu}^{\rho}
  \non\\& & \qquad-\phi_{\mu\nu}\square\phi
  +\frac{g_{\mu\nu}}{2}\left[ -\phi^{\alpha\beta}\phi_{\alpha\beta}+\left(\square\phi\right)^{2}-2\,\phi^{\alpha}\phi^{\beta}\, R_{\alpha\beta}\right].
%\end{array}
\eea

The Klein-Gordon equation is derived by computing the variation of $\mathcal{L}_\rr{John}$ with respect to the scalar field,
\bea
  \frac{\df \mathcal{L}_\rr{John}}{\df \phi}&=&0,\\
  \nabla_\rho\frac{\df \mathcal{L}_\rr{John}}{\df \phi_\rho}&=&
  \nabla_\rho\left(2 G^{\rho\nu}\phi_\nu\right)=2 G^{\rho\nu}\phi_{\nu\rho},
\eea
using the second Bianchi identity for the last equality.

\section{Cosmological equations}
\label{app:cosmo_Fab2}
\noindent  In terms of the reduced variables $x(t)=\kappa \dot{\phi}$, $y(t)=\sqrt{\kappa} \dot{\alpha}$ and $z(t)= 1 + \epsilon \sqrt{\kappa} \phi(t)$, the equations of motion for a flat and empty universe derived from action~(\ref{actionjohngeorges}) are,
\bea
  &&6 \epsilon x y+x^2 \left(-1+9 \gamma y^2\right)+6 y^2 z=0, 
  \label{Eomjg1} \\
  &&4 x \left(\epsilon+\gamma \sqrt{\kappa} \dot{x}\right) y+x^2 \left(1+3 \gamma y^2+2 \gamma \sqrt{\kappa} \dot{y}\right)\qquad\qquad\qquad\non\\
 &&\hspace{3cm}+6 y^2 z+2 \sqrt{\kappa} (\epsilon \dot{x}+2 \dot{y} z)=0,  
 \label{Eomjg2} \\
 &&3 y \left(x-2 \epsilon y-3 \gamma x y^2\right)+\sqrt{\kappa} \left(\dot{x}-3 \gamma \dot{x} y^2-3 (\epsilon+2 \gamma x y) \dot{y}\right)=0, 
\label{Eomjg3}
\eea
% \bea
% 6 \epsilon x y+x^2 \left(-1+9 \gamma y^2\right)+6 y^2 z=0, 
% \label{Eomjg1} \\
% 4 x \left(\epsilon+\gamma \sqrt{\kappa} \dot{x}\right) y+x^2 \left(1+3 \gamma y^2+2 \gamma \sqrt{\kappa} \dot{y}\right)\non\\\qquad\qquad\qquad\qquad\qquad
%  +6 y^2 z+2 \sqrt{\kappa} (\epsilon \dot{x}+2 \dot{y} z)=0,  
%  \label{Eomjg2} \\
% 3 y \left(x-2 \epsilon y-3 \gamma x y^2\right)+\sqrt{\kappa} \left(\dot{x}-3 \gamma \dot{x} y^2-3 (\epsilon+2 \gamma x y) \dot{y}\right)=0, 
% \label{Eomjg3}
% \eea
which can be decoupled in the following way,
\begin{eqnarray}
&& \hspace*{-0.5cm} \dot{x}=\frac{-3 x \left[\epsilon x+4 \left(\epsilon^2+\gamma x^2\right) y+7 \epsilon \gamma x y^2\right]+6 y (-2 x+\epsilon y) z}{2 \kappa^{1/2} \left(3 \epsilon^2+12 \epsilon \gamma x y+\gamma x^2 \left(1+9 \gamma y^2\right)+\left(2-6 \gamma y^2\right) z\right)},\\
&& \hspace*{-0.5cm}
\dot{y}=\frac{2 \epsilon x y \left(1-15 \gamma y^2\right)+x^2 \left[-1+3 \gamma y^2 \left(4-9 \gamma y^2\right)\right]-6 y^2 \left(2 \epsilon^2+z-3 \gamma y^2 z\right)}{2 \kappa^{1/2} \left[3 \epsilon^2+12 \epsilon \gamma x y+\gamma x^2 \left(1+9 \gamma y^2\right)+\left(2-6 \gamma y^2\right) z\right]}.\non\\
\end{eqnarray}
The scalar field EoS is given by,
\beq
w_{\phi}= - x \left(3 \gamma x^2+2 z\right) \frac{N}{D},
\eeq
with,
\begin{eqnarray}
N&=&6 x \left(\gamma x^2-z\right) \left(3 \gamma x^2+2 z\right)^2+6 \sqrt{3} \epsilon^3 \sqrt{x^2 \left(3 \epsilon^2+3 \gamma x^2+2 z\right)} \left(7 \gamma x^2+2 z\right)\non\\
&&\hspace{0.5cm}-2 \sqrt{3} \epsilon \sqrt{x^2 \left(3 \epsilon^2+3 \gamma x^2+2 z\right)} \left(33 \gamma^2 x^4+16 \gamma x^2 z-4 z^2\right)\non\\
&&\hspace{1cm}+9 \epsilon^2 \left(15 \gamma^2 x^5+4 \gamma x^3 z-4 x z^2\right)
-18 \epsilon^4 \left(7 \gamma x^3+2 x z\right),\\
D&=&\left[-3 \epsilon x+\sqrt{3} \sqrt{x^2 \left(3 \epsilon^2+3 \gamma x^2+2 z\right)}\right]^2 \times \non\\
&&\quad\Big[18 \gamma^3 x^6+30 \gamma^2 x^4 z+24 \gamma x^2 z^2+8 z^3 \non\\
&&\qquad+6 \sqrt{3} \epsilon \gamma x \left(\gamma x^2+2 z\right) \sqrt{x^2 \left(3 \epsilon^2+3 \gamma x^2+2 z\right)}
+3 \epsilon^2 \left(3 \gamma^2 x^4+4 z^2\right)\Big].
\nonumber\\
\end{eqnarray}

\section{Stability conditions}
\label{app:causality}
\noindent 
%The coupling to the Ricci scalar, "George", does not change the analysis made for the scalar field sector of the theory (see Sec.~\ref{sec:John_theo}). Thus the two following conditions still hold,
%\begin{eqnarray} \label{eq:ghostJG1}
%Q_{\phi}=\frac{1}{2}\left(1-3 \gamma y^2 \right) &>&0, \\
%c_{\phi}^2= \frac{1 - \gamma \left(3 y^2 +2 \sqrt{\kappa} \dot{y} \right)}{1-3  \gamma y^2 } &\geq& 0.
%\end{eqnarray}
We derive the metric perturbations based on  Eqs.~(23) and (25-27) of \cite{DeFelice:2011bh},
\begin{eqnarray}
Q_\rr{T}>0 &\Rightarrow& z + \frac{\gamma x^2}{2} >0, \\
c_\rr{T}^2 \geq 0 &\Rightarrow& z - \frac{\gamma x^2}{2} \geq 0,
\end{eqnarray}
for the tensorial part, and,
\beq
Q_\rr{S}\,\,\,>0\,\,\, \Rightarrow 3 \epsilon^2+12 \epsilon \gamma x y+9 \gamma^2 x^2 y^2+2 z+\gamma \left(x^2-6 y^2 z\right)>0,
\eeq
for the scalar part of the metric perturbations, while the condition on the squared speed $c_\rr{S}^2 \geq 0$ leads to,
\begin{eqnarray}
\hspace{-0.5cm}&&\hspace{-0.5cm}
2 x \left(\epsilon+\gamma \sqrt{\kappa} \dot{x}\right) \left(\gamma x^2+2 z\right) \left(\epsilon x+3 \gamma x^2 y+2 y z\right) \qquad\qquad\qquad\nonumber \\
&&+2 y \left(\frac{\gamma x^2}{2}+z\right)^2 \left(\epsilon x+3 \gamma x^2 y+2 y z\right)
+\frac{1}{2} \left(\gamma x^2-2 z\right) \left(\epsilon x+3 \gamma x^2 y+2 y z\right)^2 \nonumber \\
&& \hspace{0.3cm}-2 \sqrt{\kappa} \left(\frac{\gamma x^2}{2}+z\right)^2 \left[\epsilon \left(\dot{x}+\frac{2 x y}{\sqrt{\kappa}}\right)+3 \gamma x (2 \dot{x} y+x \dot{y})+2 \dot{y} z\right] \geq 0.
\label{eq:ghostJGend}
\end{eqnarray}

\section{Spherically symmetric equations of motion}
\label{app:sphericEOM}

\noindent We derive the equations of motion for the action (\ref{actionjohngeorges}) in the vacuum for a spherically symmetric and static field configuration, assuming the metric ansatz (\ref{sun:metric}),
%, and  we replace its components in the Lagrangian. With the Noether theorem,  we find the equations of motion for the fields $A$, $B$, $R$, and $\varphi$. Finally, we impose the gauge  $R=r$, and we find three equations plus a Hamiltonian constraint that read,
\begin{eqnarray} 
\non0&=&\Big( 2\varphi'^{2}B{r}^{2}B'' + 4\varphi'  B^2r\varphi'' +4 \varphi' B{r}^{2}  \varphi''  B' -5 \varphi'^{2}B'^{2}{r}^{2}+2\varphi'^2B^2 \Big) \gamma\kappa^{2}\\\non
&-& \Big(4B^{3}\varphi   B''  {r}^{2}+2 B^{3} B'{r}^{2}\varphi' +4B^{4} \varphi' r+8B^{3}\varphi   B' r+2B^{4}{r}^{2}\varphi'' 
\\\non
&-&2 B^2\varphi   B'^{2}{r}^{2} \Big) {\epsilon \kappa^{1/2}}
-8B^{3}rB' -4B^{3}{r}^{2}B'' +2 B^2{r}^{2} B'^{2}- \varphi'^{2}{\kappa}^{2}B^{4}{r}^{2},\\\non\\
\non0&=&\Big( 4 \varphi'^{2} B' A rB +2\varphi'^2B^2A  -3 \varphi'^{2} B'^{2}A  {r}^{2}+4 \varphi' B {r}^{2}\varphi''A B'  +8\varphi'^2B^2rA'  \\\non
&+&2\varphi'^2B'  A' B {r}^{2}+2 \varphi'^{2} A'' {r}^{2} B^2+4 \varphi' B^{2}r \varphi'' A  +4 \varphi'  A' {r}^{2} \varphi''  B^2
\\\non
&+&2 \varphi'^{2}A B  {r}^{2}B''  \Big) \gamma\kappa^{2}-\Big(4B^{3}A   B' {r}^{2}\varphi' -2 B^{2}A \varphi   B'^{2}{r}^{2}+8B^{4}A  \varphi' r\\\non
&+&8\varphi  r A' B^{4}+8B^{3}A \varphi   B'r+4 B^{3}A \varphi   B'' {r}^{2}
+4\varphi   B'  A' {r}^{2}B^{3}+6 A'{r}^{2} \varphi' B^{4}\\\non
&+&4\varphi A'' {r}^{2}B^{4}+4B^{4}A  {r}^{2}\varphi''   \Big) \epsilon\kappa^{1/2}
-4B^{3}A  {r}^{2}B'' -4{r}^{2} A'' B^{4}-8r A'B^{4}\\\non
&-&8B^{3}A  rB' -4 A' {r}^{2} B'B^{3}+2 B^2A  {r}^{2}  B'^{2}- \varphi'^{2}{\kappa}^{2}B^{4}A {r}^{2},\\\non
\end{eqnarray}
\begin{eqnarray}
\non0&=&\Big( 4 \varphi'  A'  B^{3}+4 \varphi' A  B'  B^2+4 \varphi'  B' A {r}^{2} B''B +4 \varphi'  B''A' {r}^{2}B^{2}\\\non
&+&4 \varphi'' r A'  B^{3}+4 \varphi' r A'' B^{3}+4 \varphi'' B'A' {r}^{2}B^{2}+8 \varphi'  B'A' r B^2+4 \varphi'  B''A  r B^2\\\non
&+&4\varphi'' B'A r B^{2}-4 \varphi'  B'^{2}A rB -6 \varphi'  B'^{2} A'{r}^{2}B -6 \varphi'  B'^{3}A {r}^{2}\\\non
&+&4 \varphi'  B' A'' {r}^{2} B^2+2\varphi'' B'^{2}A {r}^{2}B   \Big) \gamma\kappa-\Big(8 B' A r B^{4}+4B^{5}rA' \\\non
&+&4{r}^{2}A   B''B^{4}+2 B'A' {r}^{2}B^{4}+2B^{5}{r}^{2}A'' -2 B'^{2}A {r}^{2} B^{3} \Big) \epsilon\kappa^{-1/2}
\\\non
&+&2 A'{r}^{2} \varphi' B^{5}+2B^{5}{r}^{2}\varphi''A+4 \varphi' B^{5}A r+2 B' {r}^{2} \varphi'A B^{4},\\\non\\
\non0&=&\Big( 2 \varphi'^{2} A'  B^{2}+2r \varphi'^{2} A''  B^2-6 \varphi'^{2} B'^{2}A r+4 \varphi'A  \varphi''  B^2-2 \varphi'^{2} B' A B\\\non
&-&4\varphi'^2A'B'rB + 4 \varphi' r B'A  \varphi'' B +4r \varphi' A'  \varphi''  B^2+2 \varphi'^{2}r B''A B \Big) \gamma\kappa^{2}\\\non
&-& \Big(4\varphi   B'A B^{3}-4r\varphi   B'^{2}A   B^2+4 B^{4}rA' \varphi' +4 B^{4}\varphi  A'' r+4 B^{4}\varphi''A r\\\non
&+&4 B^{4}\varphi  A' +4B^{4}\varphi' A  +4\varphi   B''A r B^{3} \Big) \epsilon\kappa^{1/2}
+4r B'^{2}A  B^{2}-4rA  B'' B^{3}\\\non
&-&4 B' A B^{3}-4B^{4}A' -4B^{4} A'' r-2 B^{4} \varphi'^{2}A r{\kappa}^{2}.
\end{eqnarray}

\begin{landscape}
For the numerical integration the metric ansatz \eqref{eq:metric_schwa} is preferred and the matter sector must be included. The three Einstein equations and the Klein-Gordon equation then read respectively,
\bea \label{cacaboudin1}
  &&\nu''\left[-\frac{\left(1+\epsilon\sqrt{\kappa}\phi\right)}{\kappa}r+\frac{\kappa\gamma}{2}\, \rr{e}^{-2\lambda}\phi'^{2}r\right]
  +\phi''\left(-\frac{\epsilon}{\sqrt{\kappa}}r+\kappa\gamma \rr{e}^{-2\lambda}\phi'+\kappa\gamma \rr{e}^{-2\lambda}\phi'\nu'r\right)\non\\
  && \qquad+\lambda'\left[\left(-\frac{3\,\kappa\gamma}{2}r\rr{e}^{-2\lambda}\phi'^{2}+\frac{\left(1+\epsilon\sqrt{\kappa}\phi\right)}{\kappa}r\right)\nu'-\frac{3\,\kappa\gamma}{2}\rr{e}^{-2\lambda}\phi'^{2}
  +\frac{1+\epsilon\sqrt{\kappa}\phi}{\kappa}+\frac{\epsilon}{\sqrt{\kappa}}\phi'r\right]\non\\
  && \qquad=\nu'^{2}\left[\frac{\left(1+\epsilon\sqrt{\kappa}\phi\right)}{\kappa}r-\frac{\kappa\gamma}{2}\phi'^{2}\rr{e}^{-2\lambda}r\right]%\qquad\qquad\qquad\non\\
  +\nu'\left(\frac{1+\epsilon\sqrt{\kappa}\phi}{\kappa}+\frac{\epsilon}{\sqrt{\kappa}}r\phi'-\frac{\kappa\gamma}{2}\rr{e}^{-2\lambda}\phi'^{2}\right)
  +\frac{\epsilon}{\sqrt{\kappa}}\phi'+\frac{\phi'^{2}r}{2}+\rr{e}^{2\lambda}r\left(-\frac{3}{\tilde{R}^{2}}\frac{p}{\rho}\right),\non\\
  \\
 &  &\phi''\left[-\frac{\epsilon}{\sqrt{\kappa}}\, \rr{e}^{2\lambda}r^{2}+2\,\kappa\gamma\phi'r\right]
+\lambda'\left[\frac{\epsilon}{\sqrt{\kappa}}\, \rr{e}^{2\lambda}\phi'r^{2}+2\, re^{2\lambda}\frac{\left(1+\epsilon\sqrt{\kappa}\phi\right)}{\kappa}-3\,\kappa\gamma\phi'^{2}r\right]
\non\\
 &  & \qquad\qquad\qquad\qquad =\frac{\phi'^{2}}{2}\left[\rr{e}^{2\lambda}r^{2}-\kappa\gamma\left(1+\rr{e}^{2\lambda}\right)\right]
 +2\,\frac{\epsilon}{\sqrt{\kappa}}\rr{e}^{2\lambda}\phi'r+\left(\rr{e}^{2\lambda}-\rr{e}^{4\lambda}\right)\frac{\left(1+\epsilon\sqrt{\kappa}\phi\right)}{\kappa}+r^{2}\rr{e}^{2\lambda}\left(\frac{3}{\tilde{R}^{2}}\right),
\\
&&   \nu'\left[-2\, r\frac{\left(1+\epsilon\sqrt{\kappa}\phi\right)}{\kappa}+3\,\kappa\gamma \rr{e}^{-2\lambda}\phi'^{2}r-\frac{\epsilon}{\sqrt{\kappa}}\,\phi'r^{2}\right]
 +\frac{\phi'^{2}}{2}\left(3\,\kappa\gamma \rr{e}^{-2\lambda}+r^{2}-\kappa\gamma\right)
 -2\,\frac{\epsilon}{\sqrt{\kappa}}\phi'r\non\\
 && \qquad\qquad\qquad\qquad-\left(1-\rr{e}^{2\lambda}\right)\frac{\left(1+\epsilon\sqrt{\kappa}\phi\right)}{\kappa}+\frac{3}{\tilde{R}^{2}}\frac{p}{\rho}\rr{e}^{2\lambda}r^{2}=0,\\
&&\nu''\left(2\,\kappa\gamma r\phi'-\frac{\epsilon}{\sqrt{\kappa}}\, \rr{e}^{2\lambda}r^{2}\right)+\phi''\left(2\,\kappa\gamma r\nu'+\rr{e}^{2\lambda}r^{2}+\kappa\gamma-\kappa\gamma \rr{e}^{2\lambda}\right)\nonumber \\
 &  &\qquad +\lambda'\left[\nu'\left(-6\,\kappa\gamma r\phi'+\frac{\epsilon}{\sqrt{\kappa}}\, \rr{e}^{2\lambda}r^{2}\right)+\phi'\left(-\rr{e}^{2\lambda}r^{2}-3\,\kappa\gamma+\kappa\gamma \rr{e}^{2\lambda}\right)+2\,\frac{\epsilon}{\sqrt{\kappa}}\rr{e}^{2\lambda}r\right]\nonumber \\
 &  &\qquad =-\nu'^{2}\left(2\,\kappa\gamma r\phi'-\frac{\epsilon}{\sqrt{\kappa}}\, \rr{e}^{2\lambda}r^{2}\right)-\nu'\left[\phi'\left(3\,\kappa\gamma-\kappa\gamma \rr{e}^{2\lambda}+\rr{e}^{2\lambda}r^{2}\right)-2\,\frac{\epsilon}{\sqrt{\kappa}}\rr{e}^{2\lambda}r\right]
 -2\,\phi'\rr{e}^{2\lambda}r-\frac{\epsilon}{\sqrt{\kappa}}\rr{e}^{2\lambda}\left(\rr{e}^{2\lambda}-1\right),%\label{eq:J+G_schwa_motion4}
\label{cacaboudin2}
\eea
with $\tilde{R}=3/(\kappa\rho)=\mathcal{R}^3/r_\rr{s}$.

%The third equation is used as Hamiltonian constraint.
\end{landscape}

\bibliographystyle{boudin}
\renewcommand{\tocetcmark}[1]{\markboth{\small\textsc{#1}}{\small\textsc{#1}}}
\bibliography{biblio_PUN}%,biblio_hybridLF}

\end{document}